\documentclass[a4paper,twoside, 11pt]{memoir}

\usepackage{phdstyle}
\usepackage{import}
\usepackage{etoolbox}

\title{Quantum Geometrodynamics of \\ Higher Derivative Theories \\ [10pt] with and without Conformal Symmetry}

\newcommand{\thesistype}{\huge{Inaugural-Dissertation} \\[10pt] \LARGE zur\\[10pt] Erlangung des Doktorgrades \\[10pt] der Mathematisch-Naturwissenschaftlichen Fakult\"at\\[10pt]der Universit\"at zu K\"oln\\[10pt]
vorgelegt von} 

\newcommand{\thesisauthor}{Branislav Nikoli\'c} 
\newcommand{\geburtsort}{Po\v zarevac, Serbien} 

\begin{document}


        
        

        
        

        


\thispagestyle{empty}
\begin{center}

        \begin{figure}[htbp]
             \centering
             \includegraphics[width=.3\linewidth]{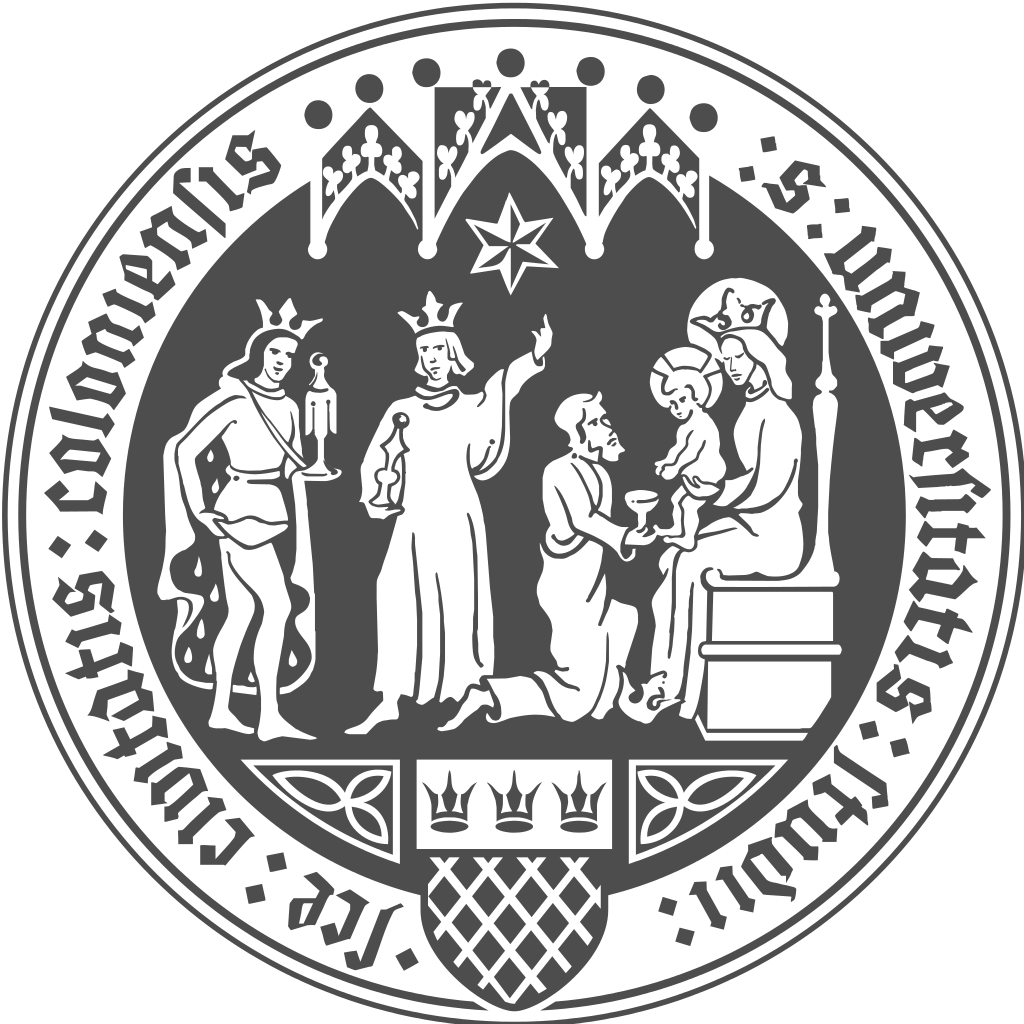}
        \end{figure}
        
            \vspace{1cm}
        
        {\huge
        \textbf{Quantum Geometrodynamics of \\ Higher Derivative Theories \\[10pt] with and without Conformal Symmetry}}

            \vspace{1.5cm}
        
        \thesistype
        
            \vspace{2cm}

        {\huge
        \thesisauthor} \\
        {\LARGE
        aus \geburtsort}\\[10pt]
        
        \vfill
        sedruck K\"oln, 2019

    \end{center}

\newpage
\thispagestyle{empty}
\vspace*{2cm}

\begin{center}
    Diese Arbeit wurde im Jahr 2019 \\
    von der Mathematisch-Naturwissenschaftlichen Fakult\"at \\
    der Universit\"at zu K\"oln angenommen.
\end{center}

\vspace*{10cm}

{\large{1. Gutachter: \hspace{0.5cm}\textbf{Prof. Dr. Claus Kiefer}\\
\hspace*{3.6cm} Institut f\"ur Theoretische Physik\\
\hspace*{3.6cm} Universit\"at zu K\"oln}}\\

{\large{2. Gutachter: \hspace{0.5cm}\textbf{Prof. Dr. Domenico Giulini} \\
\hspace*{3.6cm} Institut f\"ur Theoretische Physik \\
\hspace*{3.6cm} Leibniz Universit\"at Hannover}}\\

{\large{3. Gutachter: \hspace{0.5cm}\textbf{Prof. Dr. Anupam Mazumdar} \\
\hspace*{3.6cm} Van Swinderen Institute\\
\hspace*{3.6cm} University of Groningen, The Netherlands}}

\vspace{1.5cm}

{\large{Tag der m\"undlichen Pr\"ufung:}} 16.09.2019.


\newpage
\thispagestyle{empty}
\begin{center}
    \textbf{\Large Abstract}
\end{center}
{\small This thesis concerns with a framework of canonical quantization of gravity based on the Einstein-Hilbert action extended by terms quadratic in curvature. The aim is to investigate the semiclassical limit of such a theory and compare it with the semiclassical limit of the canonical quantization of the Einstein-Hilbert action alone, the latter of which is the usual approach in this framework.

General Relativity has passed the tests from the length scales of micrometers up to the cosmological scales. The classical evolution of our Universe seems to be described by the so-called $\Lambda$CDM model, which was recently tested by the Planck satelite with success. The recent discovery of gravitational waves seems to confirm also the linearized, long-range behavior of vacuum General Relativity. However, the behavior of gravity at short scales and relatively high energies, i.e. in the regimes where quantum effects of matter fields and spacetime become relevant, remains so far within the many possible theoretical approaches to its understanding. It is expected that near the initial singularity of our Universe --- the Big Bang --- the description of gravity drastically deviates from General Relativity and a theory of quantum gravity is necessary. But already near the theoretical limit of the highest observable energy scale (energy per excitation of a quantum field) --- the Planck energy scale --- it is expected that the effects of quantum field-theoretical description of matter propagating on classical curved spacetimes play a significant role. Because of this, General Relativity changes in two ways. First, the energy-momentum tensor is replaced by the expectation value of the energy-momentum tensor operator. Second, since the latter diverges, the regularization of these divergences has shown that it is necessary to modify General Relativity by adding to the Einstein-Hilbert action, among others, terms quadratic in curvature such as the square of the Ricci scalar and the square of the Weyl tensor. Since these terms generate fourth order derivatives in the modified Einstein equations, the doors were opened for investigating modified classical theories of gravity, in order to provide alternative interpretations of dark matter and the accelerated expansion of the Universe. However, an often neglected fact in these classical approaches is that these terms are suppressed at the present, classical scales. This is also reflected in the fact that the respective coupling constants of these new terms are proportional to the Planck constant and are thus of perturbative nature. Therefore they are only relevant at high energy/strong curvature regimes, typical for the very early universe. At extremely high energy scales, i.e. near and above the Planck energy scale, it is expected that the perturbative description breaks down and that a full quantum theory of gravity --- which assumes that the spacetime itself is quantized as well --- is necessary.

The main goal of this thesis is to quantize the Einstein-Hilbert action extended by the quadratic curvature terms is within the canonical quantization approach, thus formulating quantum geometrodynamics of the higher derivative theories. The motivation is to provide an alternative to the standard canonical quantization based on the Einstein-Hilbert action alone, because the latter does not generate the quadratic curvature terms in the semiclassical limit.  A particular formulation of a semiclassical approximation scheme is employed which ensures that the effects of the quadratic curvature terms become perturbative in the semiclassical limit. This leaves the classical General Relativity intact, while naturally giving rise to its first semiclassical corrections.

Another topic of interest is a classical theory where the quadratic Ricci scalar and the Einstein-Hilbert term are absent from the action, which then enjoys the symmetry with respect to the conformal transformation of fields (local Weyl rescaling). We pay a special attention to this case, because near and beyond Planck scales it is expected that conformal symmetry plays a
very important role, since it provides a natural setting for the absence of the notion of a physical length scale. Certain useful model-independent tools are also constructed in this thesis. Firstly, it is shown that if coordinates are treated as dimensionless and if a set of variables based on the unimodular decomposition of the metric is introduced, the only conformally variant degree of freedom becomes apparent. This makes the geometrical origin of the physical length scale apparent as well, which is especially important in the interpretations of conformally invariant quantum theories of gravity. With such an approach several earlier results become much more transparent. Secondly --- which naturally follows from the application of the set of these new variables --- a model-independent generator of conformal field transformations is constructed in terms of which a reformulation of the definition of conformal invariance is given.  Thirdly, it is argued that a canonical quantization scheme makes more sense to be based on the quantization of generators of relevant transformations, than on the first class constraints.
The thesis thus attempts to combine several minor but important aspects of a theoretical approach and use them to pursue the main goal.}
\thispagestyle{empty}

\newpage
\thispagestyle{empty}
\begin{center}
    \textbf{\Large Kurzzusammenfassung}
\end{center}

{\small Diese Dissertation befasst sich mit einem Modell der kanonischen Quantentgravitation basierend auf der Einstein-Hilbert-Wirkung, die um Terme mit quadratischer Kr\"ummung erweitert wurde. Ziel ist es, die semiklassische Grenze einer solchen Theorie zu untersuchen und mit der semiklassischen Grenze der kanonischen Quantisierung der Einstein-Hilbert-Wirkung allein zu vergleichen, wobei die letztere in diesem Rahmen der \"ubliche Ansatz ist.

Die Allgemeine Relativit\"atstheorie hat die Tests in L\"angenskalen von Mikrometern bis zu  kosmologischen Skalen bestanden. Die klassische Entwicklung unseres Universums scheint durch das sogenannte $\Lambda$CDM-Modell beschrieben zu werden,  das k\"urzlich vom Planck-Satelliten erfolgreich getestet wurde. Die j\"ungste Entdeckung der Gravitationswellen scheint auch das linearisierte weitreichende Verhalten der allgemeinen Relativit\"atstheorie im Vakuum zu best\"atigen. Das Verhalten der Gravitation auf kurzen L\"angenskala und bei relativ hohen Energien, d. h. in den Regimen, in denen Quanteneffekte von Materiefeldern und der Raumzeit relevant werden, bleibt jedoch innerhalb der vielen m\"oglichen theoretischen Ans\"atze unseres Verst\"andnisses. Es wird erwartet, dass in der N\"ahe der anf\"anglichen Sigularit\"at unseres Universums – dem Big Bang - die Beschreibung der Gravitation drastisch von der Allgemeinen Relativit\"atstheorie abweicht und eine Theorie der Quantengravitation erforderlich ist. Aber bereits nahe der theoretischen Grenze der h\"ochsten beobachtbaren Energieskala (Energie pro Quantenfeldanregung) - der Planck-Energieskala - wird erwartet, dass die Effekte der  quantenfeldtheoretischen Beschreibung der Ausbreitung von Materie auf klassische gekr\"ummte Raumzeiten eine bedeutende Rolle spielen. Aus diesem Grund \"andert sich die Allgemeine Relativit\"atstheorie auf zwei Arten. Zuerst wird der Energie-Impuls Tensor durch den Erwartungswert des Energie-Impuls Tensor Operators ersetzt. Zweitens , da dieser divergiert, hat die Regularisierung dieser Divergenzen gezeigt, dass es notwendig ist, die Allgemeine Relativit\"atstheorie zu modifizieren, indem der Einstein-Hilbert-Wirkung unter anderem Terme mit quadratischer Kr\"ummung hinzugef\"ugt werden, wie beispielsweise das Quadrat des Ricci-Skalars und das Quadrat des Weyl-Tensors. Da diese  Terme Ableitungen vierter Ordnung in den modifizierten Einstein-Gleichungen erzeugen, wurden die T\"uren f\"ur die Untersuchung modifizierter klassischer Gravitationstheorien ge\"offnet. Diese erlauben alternative Interpretationen der dunklen Materie und die beschleunigte Expansion des Universums. Eine oft vernachl\"assigte Tatsache in diesen klassischen Ans\"atzen ist jedoch, dass diese Ausdr\"ucke auf der gegenw\"artigen klassischen Skala unterdr\"uckt werden. Dies spiegelt sich auch in der Tatsache wider, dass die jeweiligen Kopplungskonstanten dieser neuen Terme proportional zur Planck-Konstante sind und somit st\"orenden Charakter haben. Daher sind diese nur f\"ur Regime mit hohen Energien, beziehungsweise starker Kr\"ummung relevant, die f\"ur das sehr fr\"uhe Universum typisch sind. Bei extrem hohen Energieskalen, das heißt in der N\"ahe und oberhalb der Planck-Energieskala, wird erwartet, dass die st\"orende Beschreibung zusammenbricht und dass eine vollst\"andige Quantengravitationstheorie erforderlich ist, die davon ausgeht, dass auch die Raumzeit selbst quantisiert wird.

Das Hauptziel dieser Dissertation ist die Quantisierung der Einstein-Hilbert-Wirkung, die durch die quadratischen Kr\"ummungsterme erweitert wird.  Dies  geschieht innerhalb des kanonischen Quantisierungsansatzes um somit die Quantengeometrodynamik der Theorien der h\"oheren Ableitungen zu formulieren. Die Motivation besteht darin, eine Alternative zu der kanonischen Standardquantisierung basierend auf der Einstein-Hilbert-Wirkung allein bereitzustellen, da letztere nicht die quadratischen Kr\"ummungsterme in der semiklassischen Grenze erzeugt. Es wird eine bestimmte Formulierung eines semiklassischen N\"aherungsschemas verwendet, das sicherstellt, dass die Auswirkungen der quadratischen Kr\"ummungsterme in der semiklassischen Grenze st\"orungsfrei werden. Dadurch bleibt die klassische Allgemeine Relativit\"atstheorie erhalten, w\"ahrend auf nat\"urliche Art und Weise die ersten semiklassischen Korrekturen eingef\"uhrt werden.

Ein weiteres Thema von Interesse ist eine klassische Theorie, bei der der quadratische Ricci-Skalar und der Einstein-Hilbert-Term in der Wirkung fehlen.  Die resultierende Wirkung weißt dann die Symmetrie bez\"uglich der konformen Transformation von Feldern (lokales Weyl-Skalieren) auf. Wir widmen diesem Fall besondere Aufmerksamkeit, denn es wird erwartet, dass in der N\"ahe und außerhalb der Planck-Skalen die konforme Symmetrie eine sehr wichtige Rolle spielt, da sie einen nat\"urlichen Rahmen f\"ur das Fehlen einer physischen L\"angenskala bietet. In dieser Arbeit werden außerdem einige n\"utzliche modellunabh\"angige Werkzeuge bereitgestellt. Zun\"achst wird gezeigt, dass, wenn Koordinaten als dimensionslos behandelt werden und ein Satz von Variablen basierend auf der unimodularen Zerlegung der Metrik eingef\"uhrt wird, der einzige konform variierte Freiheitsgrad sichtbar wird. Dadurch wird auch der geometrische Ursprung der physikalischen L\"angenskala sichtbar, was insbesondere bei der Interpretation konform invarianter Quantengravitationstheorie wichtig ist. Mit einem solchen Ansatz werden einige vorherige Ergebnisse deutlich transparenter. Zweitens --- was nat\"urlich aus der Anwendung der Menge neuer Variablen folgt --- wird ein modellunabh\"angiger Generator f\"ur konforme Feldtransformationen konstruiert, anhand dessen eine Neuformulierung der Definition der konformen Invarianz gegeben wird. Drittens wird argumentiert, dass es sinnvoller ist, die  Quantisierung auf den Generatoren relevanter Transformationen aufzubauen, als auf den Zwangsbedingungen der ersten Klasse.
Diese Dissertation versucht daher, einige kleinere, aber wichtige Aspekte einer theoretischen Herangehensweise zu kombinieren und damit das Hauptziel zu verfolgen.}
\thispagestyle{empty}

\frontmatter

\newpage
\raggedbottom
{\small
\tableofcontents*
}
\pagebreak

\mainmatter

\abnormalparskip{6pt}

 \chapter*[Introduction and motivation]{Introduction and motivation}
\addcontentsline{toc}{chapter}{Introduction and motivation}
 \label{ch:Intro}
 \renewcommand\thechapter{I}
      General Relativity (GR) is a theory describing a classical gravitational field as spacetime curved by classical matter.
It is a valid description of gravitational phenomena ``at present scales'' by which we mean either energy scales or length scales characteristic for the gravitational phenomena we are currently able to observe.
The length scales extend from planetary and Solar system scales to the scales characteristic for the Universe as a whole, e.g. Hubble radius (the proper radius of a
fictitious sphere centered at an observer's position from beyond which light can never reach that observer because there the Universe expands faster than the speed of light).
According to the currently satisfying cosmological model, the $\Lambda$CDM model recently tested by the Planck satelite \cite{Planck2018} designed to measure anisotropies of the cosmic microwave background radiation (CMB), our Universe would have started from a point (the Big Bang) and then expanded at exponential rate through a phase called \textit{inflation}, 
ending up evolving as a flat Friedman model with a cosmological constant and matter, such that nowadays it is in an accelerating expansion phase, dominated by the cosmological constant.
How our Universe emerged into existence is not known.
What we do understand is that GR is not a satisfactory description of gravity at the high-energy scales close to the Big Bang, where the typical length scales of gravitational interaction of matter were much smaller than today.
At these scales quantum effects
of matter are expected to have been just as important as matter's gravitational effects; classical GR describes only interactions of classical matter with classical spacetime.
This necessitates a theory of quantum gravity with a valid semiclassical limit that should recover GR and theory of quantum fields propagating on classical curved spacetime.
The notion of a length scale characteristic for gravitational or quantum phenomena and by which means such a length scale can be defined and measured in a
physically realizable setting becomes a very important part of the question, especially if considered within the context of conformal symmetry.

\section{The effect of quantum fields at high energies/short length scales}
To get an idea of these high energy scales, let us briefly take a point of view of a hypothetical experimenter that lives in the time of such regimes.
If such an entity would use one particle to scatter off another particle in order to investigate the latter's properties, the Compton wavelength (in its reduced version),
    \begin{equation*}
         \lambda_{c}: = \frac{\hbar}{ M c}\ ,
    \end{equation*}
corresponding to the mass equivalent $M$ of their total energy $M c^2$ would have a lower observable limit \cite{Hubsch}.
Namely, if the Compton wavelength of such a system of particles is smaller than the Schwarzschild radius 
    \begin{equation*}
        r_{\ssst Sch}: = \frac{2G M}{c^2}\ ,
    \end{equation*}
corresponding to the mass equivalent of their total energy, a black hole would be formed, from which no information could be extracted via such scattering process.
The energy a particle has to have such that this would happen are finite but very large for a single particle, they are of the order $\sim 10^{18}-10^{19}$~GeV (ultra-relativistic compared to the energy equivalent of even the heaviest elementary particles). 
This is \textit{the (reduced) Planck energy scale} and is derived from the condition $\lambda_{c} = r_{\ssst Sch}$, which results in
    \begin{equation*}
        m_{p} c^2 := c^2\sqrt{\frac{\hbar c}{(8\pi) G}}\sim 10^{19}\, \text{GeV}\ ,
    \end{equation*}
where $m_{p}:=\sqrt{\hbar c/8\pi G}$ is the reduced Planck mass and $G,c,\hbar$ are Newton's gravitational constant, speed of light and the (reduced) Planck's constant, respectively.
The ``reduced'' label is usually added to the definition if the factor of $8\pi$ is present --- but the difference is about one order of magnitude and is therefore fundamentally non-existent.
In this work we use the version \textit{with} the factor of $8\pi$, but omitt the ``reduced'' label in the text.
The corresponding Planck length scale --- the mentioned smallest observable Compton wavelength --- is then of the order of
    \begin{equation*}
        l_{p}:=\sqrt{\frac{8\pi\hbar  G}{c^3}}\sim 10^{-35}\,\text{m}\ .
    \end{equation*}
    
Now, according to $\Lambda$CDM cosmological model (which does not take into account the wave-particle duality of matter in the early universe in a way mentioned above), which is a solution to Einstein's equations of GR,
the Universe has no lower limit on its size and no upper limit on energy density --- the time dependent scale factor $a(t)$, describing the relative size of our Universe, towards the initial point $t=0$ in the past tends to zero and the energy density diverges (which is the point referred to as The Big Bang).
This point is called \textit{the initial singularity}.
But as we mentioned above, there seems to be a natural lower limit for the length --- and therefore, size --- of a region of the Universe within which matter interactions could be described in a physically meaningful way, so this singularity is not reached before effects of Planck scales step onto the stage.
The situation could be understood also in terms of \textit{the Planck time}, i.e. the time it takes a massless particle to travel the Planck length,
    \begin{equation*}
        t_{p}:= \frac{l_{p}}{c} := \sqrt{\frac{8\pi\hbar  G}{c^5}}\sim 10^{-44}\,\text{s}\ .
    \end{equation*}
Namely, physical processes which take place over a period of time shorter than the Planck time are unobservable, according to the discussion above.
This means that the extrapolation of the classical description of the universe backwards in time is meaningful only until $t = t_{p}$, i.e. until Planck scales are reached.
Beyond this point into the past another description of the evolution of our Universe is needed, in order to accomodate the effects of Planck scales.

As a first step towards a description of matter-spacetime interactions near Planck scales,
the high-energy regimes approaching the Planck energy should somehow take into account the effect of quantum matter fields on a \textit{classical} spacetime curved by those very same fields.
This is the aim of \textit{quantum field theory in curved spacetimes} \cite{BD,PT} which treats spacetime as classical, but takes into account the effects of high-energies (short length scales) of \textit{quantum} matter.
An important extension of the $\Lambda$CDM model that takes these effects into account to some extent is \textit{inflation} (see e.g.~\cite{Martin}), which is a relatively short period of rapid expansion of the Universe expected to have taken place at most at $\sim 10^{14}$~GeV.
Inflation takes care of some of the problems of the $\Lambda$CDM model (the horizon and the flatness problems) and in the heart of it is the description of an evolving scalar field that drives the rapid
expansion of the Universe and the evolution of \textit{quantized} perturbations of this field.
The latter give rise to natural initial conditions for \textit{classical} perturbations describing the local inhomogeneities as seeds for the structure formation of the Universe.
The important fact here is that the gauge invariant formulation of these perturbations \cite{Mukh} requires that the perturbations of the scalar field are put together into a specific linear combination with the scalar perturbations of the spacetime metric and only then such a mixture is quantized, with an assumption of an initial vacuum state.
This means that the very early period of the Universe's evolution already seems to necessitate quantization of at least perturbations of the spacetime, in order to give rise to the observable randomness of local anisotropies of the CMB.

But at these and even higher energies another important effect of quantum field theory in curved spacetimes needs to be taken into account.
Namely, Einstein equations (EE) --- arising from the sum of the Einstein-Hilbert (EH) and matter action --- change in two ways if the matter action refers to the quantum matter described by quantum fields, instead of the classical matter.

Firstly, instead of the energy-momentum tensor one has to write down the expectation value of the \textit{operator} corresponding to the energy-momentum tensor evaluated with respect to some quantum state.
If part of the matter is classical then the classical energy-momentum tensor is present as well.
These are then not classical but \textit{semiclassical Einstein Equations} (SEE) for a dynamical spacetime background metric interacting with quantum matter through the expectation value of the energy-momentum tensor operator \cite{Ford}.
The spacetime metric unfortunately cannot be solved for in a closed form because the quantum state is unknown until the background metric is known, but the background metric can in general only be determined by the mentioned expectation value.
This fact --- that the gravitational field of the quantum matter reacts back to matter that produces it --- is called \textit{the backreaction}.
The problem is that the calculation of the backreaction term leads to divergent results which depend on the energy scale \cite{BD, PT, UDW}. In order to deal with these divergences, one uses procedures referred to as \textit{regularization} and \textit{renormalization}; chapter 3 in \cite{PT} presents several methods of these procedures.
The former isolates the divergences from the finite terms and it turns out that these divergences are proportional to terms depending only on derivatives of the metric (in a covariant way), not on the matter fields.
This would all be less concerning if the divergent terms were proportional only to the Einstein tensor, the metric tensor and other terms with coupling constants already present in the matter action --- then they would be taken care of by the redefinition of the Newton gravitational, cosmological and other constants in the matter action using the latter method, remormalization (see further below). 
But it turns out that these terms at the first order of approximation contain up to \textit{four} derivatives (in various combinations) of the metric covariantly disguised either as quadratic curvature tensors or as covariant derivatives of curvature tensors --- objects which do not originally appear in the EE.
This is where one gets to know the second way that the EE change.

Namely, because of these higher-derivative divergent terms, renormalization procedure then requires that one \textit{adds additional terms} to the EH action with their own ``bare'' coupling constants which would produce precisely those terms in the SEE which the mentioned divergences are proportional to.
One calls them ``counter-terms'' and there are more counter-terms necessary as energies are increased. 
These counter terms turn out to be made of various contractions of the Riemann tensor with itself and its covariant derivatives: they are scalar terms such as the quadratic\footnote{There are also other terms such as $\nabla_{\mu}\nabla^{\mu}R$ and certain non-local terms, but for simplicity we do not consider these terms here. 
Note that term $\nabla_{\mu}\nabla^{\mu}R$ is not relevant for equations of motion since it is a total divergence, but it may be relevant for a quantum theory of gravity.
The non-local terms are relevant for long-range behavior \cite{Don} at low energies and their coupling constants are theoretically predictable.} Ricci scalar and Ricci tensor $\beta R^2, \gamma R_{\mu\nu}R^{\mu\nu}$ and squared Weyl-tensor $\alpha C_{\mu\alpha\nu\beta}C^{\mu\alpha\nu\beta}\equiv C^2$, where $\beta, \gamma,\alpha$ are coupling constants with dimensions of $\hbar$.
Then by redefining ``bare'' coupling constants $\beta, \gamma,\alpha$ of these new terms in such a way to \textit{include}\footnote{One usually says ``absorb''.} or \textit{counter} the divergent terms arising from the mentioned procedure one ends up with \textit{finite} terms with energy-dependent couplings (the same happens in high-energy particle physics, see e.g. \cite{Don, Hubsch}), $\beta (E), \gamma (E),\alpha (E)$.
This is how one ends up with \textit{additional curvature terms in the SEE apart from the Einstein tensor}. 
The most important consequence of this is that the SEE become \textit{fourth order}.
The corresponding action, with all coupling constants redefined appropriately, is called \textit{the effective action} \cite{BOS}, but is also referred to as \textit{the higher derivative theory of gravity}, for reasons we state in section \ref{Intro_semiclHD} of this introduction.
It is important to note that the effective action is \textit{perturbative} in nature, where $\hbar$ plays the role of the perturbation parameter, the powers of which the mentioned additional terms are proportional to.
Thus, at low energies --- due to its perturbative nature --- the correction terms do not contribute significantly compared to the EH term \cite{Don}, namely e.g. the term $R^2$ is significant only if $\beta (E) R \gtrsim 10^{70}$m$^{-2}$ or $\beta (E) \nabla_{\mu}\nabla^{\mu}R/R \gtrsim 10^{70}$m$^{-2}$.
On the other hand, again due to its perturbative nature, it is expected that near Planck energies the SEE break down because the mentioned higher-derivative terms become significant.
At these scales one must abandon the effective action with a perturbative approach and find a different description of gravity.

\section{Quantum gravity} 
This is where quantum gravity enters the stage. There are quite a few approaches to quantum gravity \cite{OUP} and we have so far motivated it in one way; there are other reasons to motivate quantum gravity such as the need for unification of matter and gravitational interactions, or consistent description of interaction of black holes with quantum matter \cite{KieferBH}.
Quantum gravity is a general name for a theory which treats \textit{both} gravitational and matter interactions as quantum. 
In such theories the spacetime itself is of quantum nature.
Whatever the final quantum theory of gravity is, it should not only describe the spacetime at the mentioned energy regimes close to (and perhaps beyond) the Planck scales but also have a valid and consistent \textit{semiclassical limit}.
From this semiclassical limit a correct description of the classical world must emerge under certain conditions.
In ordinary quantum mechanics these conditions are achieved by what is usually referred to as the limit of vanishing Planck's constant, $\hbar\rightarrow 0$. 
One could think of this as ``classical mechanics is a regime of scales with respect to which the quantum of action (i.e. $\hbar$) looks negligibly small''; equivalently but somewhat formally, we would like to say that an action $S$ describes classical physics if $S/\hbar \gg 1$.
We prefer the latter, because that statement does not depend on the choice of units (i.e. it is dimensionless) --- a guideline we shall prefer to adopt in this thesis.
Now, just as classical mechanics is a limit of quantum mechanics, classical gravity (assumed to be described by GR) should be the limiting case of a potential quantum gravity theory.
But there must be one intermediate step in this approximation which must arise from any quantum gravity theory: the SEE mentioned in the previous two sections. 
Namely, a full quantum gravity theory has to explain the emergence of \textit{classical} spacetime and its interaction with \textit{quantum matter fields} that propagate on it.
Which parameter serves the role of regulating the semiclassical approximation to a quantum gravity theory?
Since Planck length scale is much smaller than even the lowest observed length scales \cite{EWash} $l_{0}\gtrsim 137 \mu$m where classical gravitational phenomena are still described by GR and Newtonian limit, we could say that the enormous dimensionless ratio of at least $l_{0}/l_{p} \sim 10^{28}$ (independent of chosen units!) is a good parameter which can tell us that any quantum phenomena relevant at Planck scales are negligible with that order of precision at scales described by $l_{0}$.
This ratio could also be interpreted as the ratio of a radius of the presently relevant spacetime curvature with the radius corresponding to the much stronger curvature at Planck scales.
However, this ratio could be smaller for gravitational phenomena involving high mass-energy densities such as the ones in the very early universe where energy per particle approaches Planck energies, or even in very strong gravity regimes in the present-day Universe such as formation of black holes. 
In such regimes a typical curvature radius of the relevant region of spacetime becomes comparable with the Planck length, i.e. $l_{0}/l_{p}\sim 1$.
If we interpret $l_{0}$ as the Compton wavelength of a typical particle in such strong-gravity regions of spacetime and recall the aforementioned example of scattering particles at high energies, we could say that towards Planck energies the Compton wavelength becomes comparable with the Planck length.
These are few of several various ways of interpreting $l_{0}$ and they seem to make $l_{0}/l_{p}\gg 1 $ a good candidate for controlling the semiclassical approximation to a quantum gravity theory. 
Indeed, it is the gravitational coupling constant expressed in terms of the Planck length (or Planck mass $m_{p}$) via $G\sim l_{p}^2c^3/\hbar=\hbar c/m_{p}^2$ which tells one about the strength of gravity, yet only in given units and thus in an ambiguous way.
But since $G$ can be expressed in terms of a fundamental length (or mass) unit, i.e. the Planck scale, then it makes more sense to express the strength of gravity \textit{with respect} to some given length scale, in this case the Planck scale, as $l_{0}/l_{p}$, which is what we do in this thesis.
The semiclassical picture should emerge from a quantum theory of gravity once the limit $l_{0}/l_{p}\gg 1$ is taken and should be able to show that SEE emerge, just as classical mechanics emerges from quantum mechanics in $S/\hbar\gg 1$ limit.
It is thus important to review the SEE in some more detail.

\section{Semiclassical and higher-derivative gravity}
\label{Intro_semiclHD}
The most drastic consequence of the SEE after the procedure of renormalization has taken place is that the presence of quadratic curvature terms in the SEE implies that not only the solution for the metric is different compared to the original EE but also that there are \textit{more} solutions to the resulting differential equations due to their \textit{fourth} order nature;
moreover, some of these new solutions exhibit instabilities in the sense that they diverge as one takes the limit of $\beta, \gamma,\alpha \rightarrow 0$, and thus fail to give a meaningful low energy limit.

An important example of such additional solutions is the Starobinsky inflation based on the work by Starobinsky \cite{Staro}, a solution to the vacuum SEE stemming from the EH action extended by an $R^2$ term.
There, the additional degree of freedom appears due to the $R^2$ term which can be shown to mimic a scalar field with a certain potential (referred to as the Starobinsky potential).
This solution is, however, stable.
The situation is more sever if other curvature terms resulting in four derivatives of the metric are included as the first necessary counter-terms, as mentioned in the previous section; few years before Starobinsky's paper Stelle addressed the most general quadratic curvature effective action containing the EH term in two papers \cite{Stelle1, Stelle2}, 
i.e. the EH action with most general combination of curvature terms containing four derivatives of the metric.
Stelle showed that such an action --- unlike pure EH gravity --- is renormalizable\footnote{It is not possible in some theories to introduce a finite number of counter-terms to absorb the divergencies appearing in the theory as one approaches the high energies. The EH action describing GR is one such an example as shown in \cite{tHoofVelt} and such theories one calls \textit{non-renormalizible}.
Non-renormalizibility of GR is one additional motivation to pursue alternative theories to GR.} \cite{Stelle1}.
Furthermore, in \cite{Stelle2} the same author considered this action as purely \textit{classical} and looked at linearized solution to its \textit{fourth order} differential equations of motion in the context of a static spherically symmetric ansatz.
Apart from the usual Newtonian $1/r$ term in the potential, he obtained a Yukawa-like term as well as terms exponentially increasing and decaying with $r$.
They compete with the Newtonian potential (because some of them have an opposite sign and thus behave as anti-gravity) and at $r=0$ conspire to give a finite result.
Furthermore, if the linearized theory is discussed in the context of general perturbations of the metric, it is found that it has \textit{eight} dynamical degrees of freedom: 
apart from the usual two associated with a massless spin-2 state associated with the gravitational waves in GR, one ends up with five degrees of freedom associated with a massive spin-2 and one degree of freedom associated with a massive spin-0 (scalar) component.
If even higher order terms were included as counter-terms in the action (which is necessary with increasing energies) there would be even more degrees of freedom and one would need to make sense of them.

Now, the problem is not only the increased number of dynamical degrees of freedom.
The problem is that some of these additional solutions are \textit{unstable} and diverge. An example of this phenomenon is given by a theory which is made of $C^2$ term (which we refer to as the \textit{Weyl-tensor term}), whose linearized version gives a wave whose amplitude linearly increases with time, as shown in \cite{Riegert}, which thus diverges for $t\rightarrow \infty$.
This issue is not unique to higher-derivative theories of \textit{gravity}. In a generic (non-gravitational) higher derivative theory that contains interactions, the corresponding Hamiltonian is necessarily unbounded \cite{Wood}, i.e. such a theory contains unstable, run-away solutions.
In particular, for a quite generic higher-derivative theory of gravity it can be shown \cite{Nun} that it necessarily suffers from unstable solutions, thereby representing a serious generic problem of higher-derivative theories of gravity that aim to substitute GR as \textit{exact} classical theories.
Furthermore, if quantization is performed, this pathological feature is manifested as negative norms \cite{HH}, thus breaking unitarity, which may be an important drawback of quantum versions of higher-derivative theories.
These unstable solutions or modes are called ``ghosts'' (not to be confused with Fadeev-Popov ghosts) or ``poltergeists''.
However, it is interesting that in spite of these problems, classical higher-derivative theories of gravity are quite popular and a considerable effort is made to make sense of them (see e.g. \cite{Balach}), mostly because the general hope is that these models can explain dark matter and dark energy beyond the GR \cite{CapFar, Mann1}.
A rather general effective action with non-local terms has been considered by Calmet at al. \cite{Xav} at the linearized level to pave the way for possible methods of measuring the involved coupling constants $\beta, \gamma, \alpha$ individually via gravitational wave experiments.
They found that no fine tuning of coupling constants and parameters could eliminate ghosts, but they also claim that ghosts are not a problem as long as one only speaks of \textit{classical} gravitational fields --- they simply contribute to the repulsive gravitational potential (as was also found by Stelle \cite{Stelle2}).
Indeed, \textit{classical} gravitational waves other than the standard ``cross'' and ``plus'' transversal modes are perfectly acceptable as solutions to the linearized higher-derivative gravity formulated as an extension of GR, as shown in e.g. \cite{CapGW, Hol}, where in the former reference also prospects of their detection in LIGO and VIRGO observatories has been discussed.
These solutions simply stretch the space in several additional ways other than ``cross'' and ``plus'' modes of the pure GR.
Furthermore, recently in \cite{Xav1} it is shown that the massless spin-2, massive spin-2 and massive spin-0 modes are a relevant model-independent prediction of the effective action (the same one used in their earlier paper \cite{Xav}) that needs to be taken into account in future simulations of black hole mergers.
They estimate (based on data from \cite{EWash}) that in order for the massive spin-2 mode to be produced (taking into account its constraints which they also discuss) the centers of two black holes would have to be apart from one another at most of the order of $10$ cm, which is well inside any astrophysical black hole's Schwarzschild radius.
This provides an expected length scale at which higher-derivative terms would be relevant.

On one hand, it seems that it is the conflict between the appeal of robustness of classical higher-derivative theories and the plague of their ghost solutions that is usually motivating the methods of ``how to deal with ghosts'' in quantization of higher-derivative theories, e.g. by alternative ways of quantization \cite{BendMan,Bend}.
On the other hand, we think that crucial importance of higher-derivative extensions of GR does not lie in the hope for providing alternatives to dark matter and dark energy but in the hope for bridging the low energy scales (where GR is an appropriate classical theory) and high energy scales at which the full theory of quantum gravity is expected to rule the description of gravity-matter interactions.
For example, if one thinks that it is important to discuss classical gravitational waves in higher-derivative theories then one must admit that it is also important to discuss these theories in the very early universe as well, in the context of inflation, because the tensor (gravitational wave) modes and their quantization are predicted in pure GR with inflation \cite{StaroGW}.
This is why one would eventually \textit{have} to deal with ghosts and issues with higher-derivative theories, if they are taken seriously, as it was recently emphasized by Matsui (see \cite{Matsui} and references therein) in the context of instability of spacetime in the presence of higher-derivative terms. For example, in \cite{Derr} the inflationary power spectrum of quantized ghost gravitational modes in a theory with a $C^2$ term was inspected and it was found that it is indeed relevant and that its behavior, remarkably, depends on a coordinate system employed. Thus it seems that a special care is necessary in order to treat and understand this problem.
That is one reason why we do \textit{not} take higher-derivative theories as \textit{exact classical} theories of gravity seriously in this thesis.
Another reason --- which actually follows from cautiously interpreting the effective action --- is that higher-derivative terms should be treated as \textit{\textbf{perturbations}} of the classical action, as they indeed are, being proportional to the powers of $\hbar$.
This fact seems to have been largely missed in most of the references we have stated so far on the topic, including \cite{Xav,Xav1} (and many other, which can be found therein)\footnote{An exception must be mentioned \cite{Mazm,Maz,Mazum}, which is concerned with formulation of \textit{non-local} theories of gravity, that can be rewritten as an infinite sum of \textit{infinitely} increasing order of derivatives; these theories do not suffer from ghosts or extra degrees of freedom.
We think that infinite-derivative formulations deserve more attention as theories with higher derivatives, especially because they aim to abridge the low energy and high energy end of a theory of gravity in a consistent way.
Also, Donoghue \cite{Don} acknowledges promises of perturbative methods described below.}.
If these terms are \textit{local perturbations} of the EH action, then the corresponding equations of motion (i.e. the SEE) are to be treated as \textit{perturbed} EE.
But that means that the spacetime metric, as the solution to these equations, has \textit{no valid meaning as an exact solution} but only as a \textit{perturbative solution}. 
This simply follows from adopting the perturbative method of solving differential equations.
The essential consequence of this is that the additional solutions arising from the presence of the higher-derivative terms are automatically excluded and thus there are no extra degrees of freedom, no massive or ghost modes, independently of the order of derivative terms included in the action.
The recognition of the perturbative nature of higher-derivative terms in general was first recognized by Bhabha \cite{Bhabha} already in 1946 in the case of the Lorentz-Dirac equation for an electron and what is know as the Abraham-Lorentz force, which describe the influence of the electron's own electromagnetic field back on the electron's own motion. 
This equation, if treated exactly, leads to exponentially increasing acceleration, but if treated perturbatively such a runaway solution is excluded \cite{Nakhleh} and no problems occur.
Furthermore, the perturbative nature of the quadratic terms in the effective gravitational action and their solutions was first emphasized by Simon almost three decades ago in
\cite{Simon1,Simon2} and further boosted in a short series of research during the 1990's starting with \cite{ParSim}, in which the \textit{second order form} of the SEE was derived using \textit{the perturbative reduction} of the fourth order equations.
This method is referred to as \textit{the method of perturbative constraints} (MPC) or \textit{perturbative order reduction}.
MPC has recently been concisely and clearly reviewed by Cheng et al. \cite{Cheng}, who, among their results, showed on a higher-derivative toy model of two masses coupled through two springs that unstable solutions are perturbatively excluded at low energy (one spring much stiffer than the other).
For gravity, this means that MPC enables one to take the $\beta,\gamma, \alpha \rightarrow 0$ limit without any issues.
To quote Bhabha \cite{Bhabha}:
    \begin{quote}
        \textit{``The exact equations of motion of point particles possess two types of solutions; the first type, called the physical solutions, are continous functions of the interaction constants at the point where the values of these constants are zero, and hence can be expanded as series in ascending powers of the constants; the second type, called the non-physical solutions, have an essential singularity at the point where the values of the interaction constants is zero, and hence cannot be expanded as series in ascending powers of the interaction constants.''}
    \end{quote}

Therefore, in this thesis we take the position that higher-derivative actions make sense as classical actions \textit{only if the higher-derivative terms are treated consistently as perturbation terms} thus giving rise only to the solutions of the SEE which are perturbatively expandable (i.e. analytic) in their coupling constants.
At energies where these terms are relevant, one must abandon the perturbative interpretation of the higher-derivative terms and quantize the theory, thereby pushing the additional degrees of freedom to the quantum regime, which then requires a separate analysis that we do not go into here.
This systematically eliminates all problems in the low energy limit mentioned above.
Let us now review how do the SSE arise from a particular approach to quantum gravity. 

\section{On quantum geometrodynamics and its semiclassical limit} The context among the approaches to quantum gravity we put this thesis into is the approach of \textit{quantum geometrodynamics} or QGD, in short.
It was introduced by DeWitt \cite{Witt67} in 1967 and is one of the conservative approaches to quantum gravity because it is based on Dirac quantization of the Hamiltonian formulation of GR \cite{ADM} in an analogous way as Dirac quantization of classical mechanics, without adding any additional mathematical structure.
We shall refer to this theory as \textit{quantum geometrodynamics of GR} or QGDGR in short\footnote{Such more precise nomenclature is necessary because we are concerned in this thesis with quantization of theories based on actions containing quadratic curvature terms in addition to the EH term and we shall refer to ``QGD'' as a \textit{tool} for quantizing an arbitary theory of gravity.}.
In the focus of QGDGR \cite{OUP} is the \textit{Wheeler-DeWitt equation} (WDW), an equation of motion for the wave function\textit{al} of the three-dimensional metric field and non-gravitational fields.
As mentioned before, it is important to have a semiclassical approximation scheme at one's disposal, leading to the SEE and quantum field theory on curved spacetimes, determined by those SEE.
This is achieved in a combination of a Born-Oppenheimer-type and WKB-like approximation which comes with an expansion of the wave functional in powers of $G^{-1}$ (or equivalently $m_{p}^2$ or $l_{p}^{-2}$ \cite{KiefSing,PadSing}).
(As we argued further above, we think it is more meaningful to use dimensionless parameter $l_{0}/l_{p}\gg 1$ as the expansion parameter; the results will not change.)
This was shown on a number of occasions \cite{Banks,KiefSing,PadSing} to lead at the highest order in the approximation to a semiclassical picture of gravity: the Einstein-Hamilton-Jacobi (EHJ) equation \cite{Peres} (which is equivalent to the Einstein equations, as shown by Gerlach \cite{Gerlach}) and the quantum field theory on a fixed curved background spacetime formulated as the functional Schr\"odinger equation in terms of an emerging evolution parameter referred to as the ``semiclassical time'' (which has nothing to do with the coordinate time at first).
However, as we reviewed before, one still must employ regularization and renormalization procedures that will take care of divergences in the emerging SEE and the functional Schr\"odinger equation --- 
these procedures are \textit{not} automatically included in the semiclassical approximation nor QGDGR and this is why one needs to introduce the counter-terms by hand. 
It would be preferable that counter-terms somehow emerge from the full QGDGR so that one simply has to take the $l_{0}/l_{p}\gg 1$ limit leading to the semiclassical approximation and things should take care of by themselves.
But since introducing these counter-terms changes the action, QGDGR --- in its present state --- can no longer be an adequate starting point for a quantum gravity theory that aims to derive a consistent semiclassical limit because its gravitational part is based only on the EH term without the counter-terms.
To investigate the possibility of a quantum gravity theory based on the approach of QGD that is able to give rise to the SEE \textit{with counter-terms}, there are at least two ways of proceeding.
The first is to deal with ill-defined second functional derivatives with respect to the fields evaluated at the same point, since these produce divergencies; according to a recent work by Feng \cite{Feng}, these ill-defined objects can be remedied by a certain procedure which formally produces nothing other than the quadratic curvature terms arising in the SEE; it would be interesting to investigate the interplay of this procedure with the regularized and renormalized SEE and understand the role of these additional terms derived in \cite{Feng}. 
The second way --- which we adopt in this thesis --- is to simply quantize an action that already contains the counter-terms and analyze the consequences to the semiclassical approximation.

Let us thus summarize the discussion in the following two important points that must be taken into account, given the state of matters and our chosen approach in this thesis:
    \begin{itemize}
        \item QGDGR is based on the EH action. However, if we take the point of view that any quantum gravity theory has to recover the SEE in its semiclassical approximation, then QGDGR is expected to produce the mentioned higher order counter-terms, which the standard approach to QGDGR \cite{OUP} fails to achieve.
        One needs to add counter-terms by hand \textit{after} the semiclassical approximation and because of this we think that QGDGR --- in the present state of affairs --- is unlikely to be a valid method of quantizing gravity.
        \item Suppose that one indeed has at one's disposal a potential quantum gravity theory based on the QGD of the EH action extended by the counter-terms.
        Now suppose that a valid semiclassical approximation can be obtained using $l_{0}/l_{p}\gg 1$ as an expansion parameter such that the SEE with all necessary counter-terms arise.
        Then one is faced with the fact that these equations are at least of the fourth order, thereby changing the nature of classical gravity solutions.
        But since the counter-terms are perturbative in nature, the solutions must be treated perturbatively as well.
        This necessarily invites a \textit{modified} semiclassical approximation scheme by means of which the perturbative nature of the quantized counter-terms must be taken into account.
    \end{itemize}
The work in this thesis aims to provide one possible remedy for the above two points. We shall seek a formulation of a QGD based on an action containing the EH term, $R^2$ term and the $C^2$ term, with non-minimally coupled scalar field.
An example of such theory was studied by the author in his Master thesis \textit{Quantum Geometrodynamics of Conformal Gravity} \cite{MSc}, where the EH action extended by the $C^2$ term was considered.
The resulting semiclassical approximation was performed in terms of the \textit{dimensionful} ratio $c^3 m_{p}^2/\hbar\alpha$, where $\alpha$ is the coupling of the $C^2$ term and it was shown that the classical Einstein gravity emerges. 
However, despite the significance of the latter result, two important points were not realized at the time: the fact that conformal and non-conformal degrees of freedom become explicit if one employs the so-called \textit{unimodular decomposition} of the metric, and the fact that a concrete formulation of the MPC in the context of the (quantized) higher-derivative theories is available in the literature and is indeed well-defined line of attacking the problem.
The former is not directly related to the semiclassical approximation scheme but it does considerably help to clearly separate and understand at a deeper level the contributions of the $R^2$ term from the contributions of the $C^2$ term.
It also demonstrates the reward of an effort to seek a relatively more elegant formulation of a theory in terms of symmetry-motivated new set of variables and thus is also of a great pedagogical and inspirational value for a daring young theorist.
The latter fact is crucial for achieving some intermediate steps in this thesis and is motivated not only by the mentioned works of Simon \cite{Simon1, Simon2} but also and especially by the work of Mazzitelli \cite{Mazzit} from 1992.
Mazzitelli was the first to combine the perturbative approach with QGD based on the quadratic curvature extensions of the EH action and he has shown that the correct SEE \textit{with} counter-terms arises in the semiclassical limit to the \textit{perturbed} WDW equation so the only thing one was left to do in addition was to perform the regularization and renormalization of the coupling constants, which he successfully realized.
Thus it seems at first that aims of this thesis repeat the already established results of \cite{Mazzit}.
But this is not the case. Namely, the subtlety of Mazzitelli's result is that he employed the MPC \textit{before} the quantization (which we shall refer to as ``perturbation before quantization'', PbQ), whereas the results of the present author's Master thesis have shown (on a more restricted example of EH plus $C^2$ action) that the same result could be expected if one employs (what is now known to the author as) the MPC formalism \textit{after} the quantization (which we shall refer to as ``quantization before perturbation'', QbP).
The difference is not in the mathematical aspect of the two approaches (which does remind one of the chicken-and-egg question), which prevents one from favoring either of the approaches over the other.
The difference is in the physical aspect of this apparent ambiguity.
Indeed, as argued above, there is a way to motivate the QbP in a very simple way: 
the higher-derivative terms can be allowed to overcome the EH term only at high energies, while at low energies (i.e. in the SEE) they have purely classical but \textit{perturbative} nature; 
that is the reason why it does not seem reasonable to us to quantize the higher-derivative terms \textit{after} they have already been identified as \textit{low-energy} perturbations (as Mazzitelli \cite{Mazzit} did).
That is the point of view we adopt in this thesis and is one of the main motivations for pursueing the quantization of higher-derivative theories of gravity.
Moreover, we would like to show that pure GR does not necessarily arise only in the QGDGR approach or in QGD of the EH plus $C^2$ action, but may arise from the more general local quadratic curvature gravity with the EH term. This may also have significant implications for other (especially canonical) approaches to quantum gravity.

This thesis also has a couple of side-endeavours which seem useful for both classical and quantum contexts of theories of gravity and thus are worth spending few sections on.
Namely, we employ a decomposition of the metric and matter fields based on their conformal properties. 
The decomposition isolates the part of variables invariant under conformal field transformations in a new set of conformally invariant variables, while allowing only one single variable to transform under conformal transformations --- \textit{the scale density}, defined as $(\sqrt{g})^{1/4}$, where $g$ is the absolute value of the metric determinant.
The consequence of this rather simple trick is that any metric theory of gravity reveals its conformal features manifestly: conformally invariant theories --- such as $C^2$ gravity, electromagnetism and conformally coupled scalar field --- take a \textit{manifestly} conformally invariant form, while conformally non-invariant theories --- such as GR, $R^2$ gravity or minimally coupled scalar field --- take a \textit{manifestly} conformally non-invariant form.
Such formulation not only significantly simplifies both the Lagrangian and Hamiltonian formulations of a theory but also provides one with a clear physical insight into conformal degrees of freedom of a theory.
Why is this so important to emphasize? Because, as will be shown in one part of this thesis, if we consider coordinates as \textit{dimensionless} (which is not usually done), then the scale density carries the meaning of a length scale which we introduced above as $l_{0}$ (that one uses as ``rods'' and ``clocks''), that ties the interpretation of a characteristic length scale with the notion of the length defined with the spacetime metric.
Consequently, by defining a generator of conformal field transformation, we shall show that an action is invariant under conformal transformations if it possesses no functional dependence on the scale density variable and is thus unable to give rise to a meaningful notion of the length scale.
Definition of conformal invariance in terms of our generator could provide a very useful tool for studying gravity and matter at high energies since it seems reasonable to expect that conformal symmetry may be unbroken at very high energies both in matter and gravitational sector \cite{tHooft}.
Due to its theory-independent formulation and off-shell validity, its can be envisioned as a very useful tool in other approaches to high-energy formulation of theory of gravity.

The thesis is organized as follows. Chapter \ref{ch:umtocf} is a pedagogical warm-up exercise on coordinate transformations in which we take a relatively novel approach to understanding the basic coordinate transformations and their effect on the metric components. 
This serves to motivate the unimodular decomposition of the metric in a rather smooth way by investigating conformal and non-conformal (shear) coordinate transformations.
We also review some old results on the group of general linear transformations which are not usually mentioned in standard textbooks on GR.
In chapter \ref{ch:umtocf1} we introduce the unimodular decomposition of the metric and extend it to field theory and $3+1$ decomposition of spacetime.
We also introduce the notion of the characteristic length scale $l_{0}$ by demanding the coordinates be dimensionless.
The definition of the generator of conformal transformation and definition of conformal invariance in terms of the scale density are presented in chapter \ref{ch:defcf}.
In chapter \ref{ch:HDclass} the higher-derivative terms are introduced into the EH action and their perturbative nature with consequences on the equations of motion is discussed.
This sets up the stage for chapter \ref{ch:Quant} where a canonical quantization of the action based on the EH term extended by $R^2$ and $C^2$ with non-minimally coupled scalar field is presented.
Such quantum gravity theory is compared to the QGDGR in a general context.
The emphasis will be on the semiclassical approximation and emergence of the SEE.
Each chapter is ended by some final remarks which summarize the main insights and provide some further ideas.
The summary and outlook is presented in Conclusions, and the Appendix gives several calculations or definitions which would otherwise interfere with the flow of the main text.
The references are organized alphabetically and cited by a numerical system.

   {\centering \hfill $\infty\quad$\showclock{0}{30}$\quad\infty$ \hfill}

\newpage

{\centering \hfill \textbf{Notation} \hfill}

\begin{itemize}
    \item $l_{\ssst P} = \sqrt{\frac{(8\pi)\hbar  G}{c^3}}\sim 10^{-35}$~m $\rightarrow$ the (reduced) Planck length;
    \item $t_{\ssst P} = \sqrt{\frac{(8\pi)\hbar  G}{c^5}}\sim 10^{-44}$~s $\rightarrow$ the (reduced) Planck time;
    \item $m_{\ssst P} = \sqrt{\frac{\hbar c}{(8\pi) G}}\sim 10^{19}$~$\frac{\text{GeV}}{c^2}$ $\rightarrow$ the (reduced) Planck mass;
    \item unless otherwise specified, throughout the thesis we adopt $c=1$ units;
    \item $l_{0}$ $\rightarrow$ characteristic length scale measured by the four- or three-dimensional metric;
    \item $l:=\frac{l_{0}}{l_{\ssst P}}$ $\rightarrow$ dimensionless length scale relative to the Planck length scale;
    \item the metric signature convention is $(-, +, +, +)$;
    \item greek indices designate spacetime components and run as $\mu = 0,1,2,3...,d-1$, while latin ones designate spatial components and run as $i = 1,2,3,...,d-1$;
    \item the Riemann tensor convention is $R^{\alpha}{}_{\mu\beta\nu} = \del_{\beta}\Gamma^{\alpha}{}_{\mu\nu} + ...$, and $R_{\mu\nu}=R^{\alpha}{}_{\mu\alpha\nu} = \del_{\alpha}\Gamma^{\alpha}{}_{\mu\nu} + ...$ for the Ricci tensor;
    \item $g := \vert \det g_{\mu\nu}\vert$ $\rightarrow$ the absolute value of the determinant of an $n$-dimensional metric, so the usual minus sign does not appear in the volume element, which we write as $\sqrt{g}$;
    \item $A_{(\mu\nu)}$ and $A_{[\mu\nu]}$ $\rightarrow$ symmetrization and antisymmetrization of the enclosed pair of indices, respectively;
    \item $A_{\mu\nu}^{\srm{T}} := A_{\mu\nu} - \frac{1}{d}g_{\mu\nu}A^{\alpha}_{\alpha}$ $\rightarrow$ the traceless part of $A_{\mu\nu}$;
    \item $\mathbb{1}_{\alpha\beta}^{\mu\nu}:=\delta_{\alpha}^{\mu}\delta_{\beta}^{\nu}$ $\rightarrow$ the identity matrix on the space of second-rank tensors;
    \item $\mathbb{1}_{\alpha\beta}^{{\srm{T}}\mu\nu}:=\delta_{\alpha}^{\mu}\delta_{\beta}^{\nu} - \frac{1}{d}g_{\alpha\beta}g^{\mu\nu}$ $\rightarrow$ the identity matrix on the space of traceless second-rank tensors
\end{itemize}

\renewcommand{\thechapter}{\arabic{chapter}}
 \chapter{A fresh look on general coordinate transformations}
     \label{ch:umtocf}
     The term ``conformal transformations'' can be encountered in several different contexts with various meanings: conformal coordinate transformations, scale transformations, local and global Weyl rescaling, as well as the related symmetries.
Therefore, it is of crucial importance to spend some time elaborating precisely what one means by a ``conformal transformation'' in this thesis, especially in order to avoid misunderstandings.
Independent of which kind of conformal transformations one is referring to, they all have one thing in common: they are such transformations that leave angles and shapes invariant, while affecting only volumes, areas and scales.
This chapter is a plunge into defining features of conformal transformations, offering an alternative, yet more valuable approach (compared to what is usually found in textbooks about them) to intuitive understanding of what conformal transformations actually are.
In short, if one would like to use mathematical language to say ``let observers at each point have their own measure of unit length'' (be that using coordinates or fields) one would use nothing other than conformal transformations to describe the change of units from a point to a point.
But is this somehow related to the underlying geometry? We shall see that a careful inspection of coordinate transformations and thereby induced transformations of the metric reveals that only some pieces of the
geometry are affected by conformal transformations. Much like the discussion above, there is a notion of ``shape'' that can be attributed to the metric describing the part left invariant under any kind of conformal
transformation. Identifying this ``shape'' part of the metric and separating it from what we shall call the ``scale'' part of the metric is what one calls \textit{unimodular decomposition} and the thesis relies heavily on this point of view.

\section{Active and passive coordinate transformations, Lie derivative}
\label{sec_actpas}

We start this chapter by discussing the notion of general coordinate transformations and their interpretation.
General Relativity belongs to a class of theories invariant under reparametrization, i.e. \textit{reparameterization-invariant theories}\footnote{In classical mechanics a Lagrangian which is not explicitly dependent on time belongs to this class. In field theories, the
same holds except there are four parameters (as four coordinates) instead of just one.}.
It is equivalent of saying that all equations describing the laws of interaction of matter with spacetime are written using tensors and therefore do not change their form under any change of coordinates $x^{\alpha}\rightarrow \tilde{x}^{\mu}=\tilde{x}^{\mu}(x^{\alpha})$.
These changes of coordinates are described by the following matrix (and its inverse)\footnote{These are in general functions of coordinates but we suppress the dependence for clarity of notation.},
    \begin{equation}
    \label{transfmatrix}
        \ta^{\mu}{}_{\nu}:=\dd{\tilde{x}^{\mu}}{x^{\nu}}\ , \qquad \tilde{\ta}^{\mu}{}_{\nu}:=\dd{x^{\mu}}{\tilde{x}^{\nu}}\ ,\qquad \ta^{\mu}{}_{\alpha}\tilde{\ta}^{\alpha}{}_{\nu}=\delta^{\mu}_{\nu}\ ,
    \end{equation}
It is obvious that after such an arbitrary change of coordinates the line element, for example, remains invariant\footnote{We write explicitly tensor product $\otimes$ here, but allow ourselves to suppress this explicit notation for simplicity. In the definition of the metric as a symmetric bilinear form it is often left out.}
    \begin{equation}
    \label{tansfg}
        \d s^{2}=g_{\mu\nu}(x)\d x^{\mu}\otimes\d x^{\nu}=\tilde{g}_{\mu\nu}(\tilde{x})\d \tilde{x}^{\mu}\otimes\d \tilde{x}^{\nu}=\d \tilde{s}^2\ ,
    \end{equation}
even though the components of the metric have changed to $\tilde{g}_{\mu\nu}(\tilde{x})=\tilde{\ta}^{\alpha}{}_{\mu}\tilde{\ta}^{\beta}{}_{\nu}g_{\alpha\beta}(x)$ and the expanded line element might not resemble the original one in these new coordinates, $\d \tilde{s}^2$ still refers to one and the same distance.
The same is with any other tensor. For example, components of a vector field $\scb{V}$ change
according to
    \begin{equation}
    \label{tansfV}
        \scb{V}=V^{\mu}(x)\del_{\mu}
        =\tilde{V}^{\alpha}(\tilde{x})\ta^{\mu}{}_{\alpha}\tilde{\ta}^{\beta}{}_{\mu}\tilde{\del}_{\beta}
        =\tilde{V}^{\mu}(\tilde{x})\tilde{\del}_{\mu}
        =\tilde{\scb{V}}
    \end{equation}
where $\tilde{\del}_{\mu}:=\dd{}{\tilde{x}^{\mu}}$ and $\tilde{\scb{V}}$ refers to the same vector field but expressed in different coordinates. 
Similarly with a scalar field $\scb{X}$, except that a scalar field is determined by a single ``component'', so matrix given by eq.~\eqref{transfmatrix} is not involved and ``the only component of a scalar field'' remains unchanged, 
    \begin{equation}
    \label{tansfS}
        \scb{X}=\phi(x)=\tilde{\phi}(\tilde{x})=\phi(\tilde{x})=\tilde{\scb{X}}\ .
    \end{equation}
However, we ought to make statements in eqs.~\eqref{tansfg}, \eqref{tansfV} and \eqref{tansfS} more precise. Namely, a vector field (as an example of a general tensor field) can be thought of as a collection of arrows each attached to one point uniquely
\footnote{This is a very simplified way of referring to a \textit{vector flow}. On a differentiable manifold $M$, at a point $P$ one constructs a tangent space $T_{\ssst P}M$ which hosts all vectors tangent to all smooth curves on $M$ passing through that point. The studied vector field will
always have a representative ``arrow'' that lives in $T_{\ssst P}M$ that is a tangent to some curve through that point. This curve is the \textit{flow} of the vector field that passes through point $P$: along this curve the arrows will change the magnitude but all the arrows that are tangent to that curve belong to the same vector field.
We might as well pick another point $Q$, with another tangent space $T_{\ssst Q}M$ then \textit{the same vector field} will be represented by another flow, this time through point $Q$.
Thus, a vector field is a collection of all
arrows that one attaches to each point on $M$ and is thus an entity independent of which arrow one picks to keep track of via its flow; one can always pick another arrow without disturbing the vector field itself.
The vector field can therefore be thought of as a \textit{distribution} (in a differential geometry context, not in the context of analysis!) of $d$-tuplets (where $d$ is the dimension of the manifold) over points on a manifold.}, each pointing at certain direction and having their own certain magnitude, but if one writes
$V^{\mu}(x)\del_{\mu}$, one refers to a \textit{single arrow}, thereby attached to a \textit{single point}. 
Therefore, in order to remedy the notation, if $\scb{V}$
is evaluated at a point $P$ to which one attaches a set of four numbers $x_{\ssst P}\equiv\{ x^{\mu} \}$ in one coordinate system and a set of some other four numbers $\tilde{x}_{\ssst P}\equiv\{ \tilde{x}^{\mu} \}$ in another coordinate system, then one refers to its components \textit{with respect to a basis defined at point} $P$ and one writes accordingly,
    \begin{equation}
    \label{tansfP}
        \d s^{2}\vert_{\ssst P}=\d \tilde{s}^{2}\vert_{\ssst P}\ ,\qquad \scb{V}\vert_{\ssst P}=\tilde{\scb{V}}\vert_{\ssst P}\ ,\qquad \scb{X}\vert_{\ssst P}=\tilde{\scb{X}}\vert_{\ssst P}\ .
    \end{equation}
In simple words, eq.~\eqref{tansfV} says that \textit{a collection of arrows representing a vector field exists on its own and is independent of the choice of coordinates that one uses to represent these arrows}, which then implies eq.~\eqref{tansfP} according to which 
\textit{a particular arrow (its magnitude and direction) at a particular point is not affected by a change of the coordinate system}. For the example of a vector field, this means that eq.~\eqref{tansfV} is more precisely written as
    \begin{align}
    \label{VfieldP}
        \scb{V}\vert_{\ssst P}=V_{\ssst P}^{\mu}(x_{\ssst P})\del_{\mu}^{\ssst P}
        =\tilde{V}_{\ssst P}^{\mu}(\tilde{x}_{\ssst P})\tilde{\del}_{\mu}^{\ssst P}=\tilde{\scb{V}}\vert_{\ssst P}
    \end{align}
and similarly for other tensor fields.
Based on these conclusions, we say that if we interpret a coordinate transformation which does not ``move the point'' or does not ``move an arrow'' from the point $P$, i.e. does not describe ``picking another arrow at another point'', as the \textit{passive transformation}.

What if we wanted to compare two neighbouring arrows of a vector field located at two infinitesimally close points $P$ and $Q$? Then we are looking for 
    \begin{equation}
    \label{compVPQ}
        \scb{V}\vert_{\ssst P}-\scb{V}\vert_{\ssst Q}=V_{\ssst P}^{\mu}(x_{\ssst P})\del_{\mu}^{\ssst P}-V_{\ssst Q}^{\mu}(x_{\ssst Q})\del_{\mu}^{\ssst Q}\ ,
    \end{equation}
i.e. the difference between the vector field evaluated at $P$ and the same vector field evaluated at $Q$.
However, since coordinate values at points $P$ and $Q$ are related by 
    \begin{equation}
    \label{infPQ}
        x^{\mu}_{\ssst Q}=x^{\mu}_{\ssst P}+\xi^{\mu}_{\ssst P}\ ,\qquad \xi^{\mu}_{\ssst P}\ll 1\ ,
    \end{equation}
where $\xi^{\mu}_{\ssst P}$ is a coordinate-dependent\footnote{We suppress the notation for its dependence on $x^{\mu}$ in order to keep the notation clean.} vector that designates the distance and direction from $P$ to $Q$ and whose components are given with respect to the basis at $P$, this seems to be just an infinitesimal coordinate transformation version of\footnote{Plugging eq.~\eqref{infPQ} into eq.~\eqref{transfmatrix} produces $\delta^{\mu}_{\nu}+\del_{\nu}^{\ssst P}\xi^{\mu}_{\ssst P}$ and $\delta^{\mu}_{\nu}-\del_{\nu}^{\ssst Q}\xi^{\mu}_{\ssst Q}$, respectively.} eq.~\eqref{transfmatrix}. But we saw that eq.~\eqref{transfmatrix} implies eq.~\eqref{tansfV}, i.e. the vector field (the abstract object itself) does not care about which
coordinate system it is represented in, so $\scb{V}\vert_{\ssst P}-\scb{V}\vert_{\ssst Q}$. This might seem a bit odd, but that is only because of a not so ideal notation for certain abstract concepts.
An abstract entity designated by $\scb{V}\vert_{\ssst P}$ and an abstract entity designated by $\scb{V}\vert_{\ssst Q}$ refer to the same distribution of arrows; the suffix
``$\vert_{\ssst P}$'' and ``$\vert_{\ssst Q}$'' only have a meaning once one looks into what this vector field is made of --- and it is made of a bunch of arrows, each attached to a point, each having their own components. Thus, that one arrow is different from
another can be told only by inspecting and comparing the components of each arrow with one another, while these different arrows with their components (with respect to the corresponding basis) encode the information about \textit{the same vector field}, i.e. the same distribution of arrows.
Then, we know that expression in eq.~\eqref{compVPQ} vanishes identically from its LHS. But in order to make this explicit in the RHS as well, we have to evaluate each term \textit{with respect to the same basis}.

Suppose now we are located at point $P$ and we have all the information about the magnitude and components $V_{\ssst P}^{\mu}(x_{\ssst P})$ of the arrow at point $P$ with respect to chosen coordinates and basis we constructed there. Let us then express the second term in basis $\del_{\mu}^{\ssst P}$. Then we see that we need to obtain information about
the arrow at infinitesimally close point $Q$ but in terms of our own coordinate system at $P$. That means that we need to change from $V_{\ssst Q}^{\mu}(x_{\ssst Q})$ to $V_{\ssst Q}^{\mu}(x_{\ssst P})$ 
and from $\del_{\mu}^{\ssst Q}$ to $\del_{\mu}^{\ssst P}$. How do we do that? The former is simply a Taylor expansion around the point $P$, so using eq.~\eqref{infPQ} in the second term in eq.~\eqref{compVPQ}, we can describe the ``motion'' from $P$ to $Q$ and express the value of components $V_{\ssst Q}^{\mu}(x_{\ssst P})$ with respect to $P$,
    \begin{equation}
    \label{compVPQ1}
        0=V^{\mu}_{\ssst P}(x_{\ssst P})\del_{\mu}^{\ssst P}- V^{\mu}_{\ssst Q}(x_{\ssst P})\del_{\mu}^{\ssst Q}-\xi^{\alpha}_{\ssst P}\del_{\alpha}^{\ssst P}V^{\mu}_{\ssst Q}(x_{\ssst P})\del_{\mu}^{\ssst Q}\ .
    \end{equation}
The latter is done by
    \begin{equation}
    \label{delTr}
        \del_{\mu}^{\ssst Q}=\ta^{\alpha}{}_{\mu}\del_{\alpha}^{\ssst P}=\del_{\mu}^{\ssst P} - \del_{\mu}^{\ssst P}\xi^{\alpha}_{\ssst P}\del_{\alpha}^{\ssst P}\ .
    \end{equation}
Keeping only terms up to the first order in $\xi^{\mu}$ and its derivatives, plugging eq.~\eqref{delTr} into eq.~\eqref{compVPQ1} results in
    \begin{align}
    \label{compVPQ2}
        0&=V^{\mu}_{\ssst P}(x_{\ssst P})\del_{\mu}^{\ssst P} - V^{\mu}_{\ssst Q}(x_{\ssst P})\del_{\mu}^{\ssst P}
        +V^{\mu}_{\ssst Q}(x_{\ssst P})\del_{\mu}^{\ssst P}\xi^{\alpha}_{\ssst P}\del_{\alpha}^{\ssst P}
        -\xi^{\alpha}_{\ssst P}\del_{\alpha}^{\ssst P}V^{\mu}_{\ssst Q}(x_{\ssst P})\del_{\mu}^{\ssst P}\ ,
    \end{align}
We can now drop the labels ``$P$'' from coordinates and $\xi^{\mu}_{\ssst P}$ and the following result is obtained component-wise,
    \begin{equation}
        \delta_{\xi}V^{\mu}(x):=V^{\mu}_{\ssst P}(x)-V^{\mu}_{\ssst Q}(x)=\xi^{\alpha}\del_{\alpha}V^{\mu}_{\ssst Q}(x)
        -V^{\alpha}_{\ssst Q}(x)\del_{\alpha}\xi^{\mu}=\mathcal{L}_{\xi}V^{\mu}(x)\ .
    \end{equation}
We can recognize from the above equation that we have just derived the expression for the Lie derivative of the contravariant vector field. Actually, there is a slight abuse of notation when one writes $\mathcal{L}_{\xi}V^{\mu}(x)$, because the Lie derivative acts on the \textit{field} itself and then we pick the $\mu$ component of the result, so an honest notation states
    \begin{align}
        \mathcal{L}_{\xi}\scb{V}
        &=\Bigpar{
        \xi^{\alpha}\del_{\alpha}V^{\mu}(x)
        -V^{\alpha}(x)\del_{\alpha}\xi^{\mu}
        }\del_{\mu}\ ,\\[6pt]
        \mathcal{L}_{\xi}V^{\mu}(x)&\equiv\Bigpar{\mathcal{L}_{\xi}\scb{V}}^{\mu}=\xi^{\alpha}\del_{\alpha}V^{\mu}(x)
        -V^{\alpha}(x)\del_{\alpha}\xi^{\mu}
    \end{align}
and in the last line we wrote the source of imprecise notation.
Due to its common use in physics, we stick to this imprecise notation in this thesis, but must keep in mind the correct reading and writing of the Lie derivative of tensors (and non-tensorial objects such as the connection) as explained above.

To illustrate further more clearly that transformation from $P$ to $Q$ introduced by eq.~\eqref{infPQ} is interpreted differently than the passive coordinate transformation that gives rise to eqs.~\eqref{tansfg}-\eqref{tansfS}, we take a look at the transformation of the scalar field under a ``motion'' given by eq.~\eqref{infPQ}.
As with the vector field, the field $\scb{X}$ itself is one and the same field, be it is expressed at a point $P$ or at a point $Q$, so again we have
    \begin{align}
        0=\scb{X}\vert_{\ssst P}-\scb{X}\vert_{\ssst Q}&=\phi_{\ssst P}(x_{\ssst P})-\phi_{\ssst Q}(x_{\ssst Q})\nonumber\\[6pt]
        &=\phi_{\ssst P}(x_{\ssst P})-\phi_{\ssst Q}(x_{\ssst P})-\xi^{\mu}_{\ssst P}\del_{\mu}^{\ssst P}\phi_{\ssst Q}(x_{\ssst P})\ ,
    \end{align}
where we used the Taylor expansion around $Q$ in the second line. From here it follows that the Lie derivative with respect to $\xi^{\mu}$ is
    \begin{equation}
        \delta_{\xi}\phi(x)=\mathcal{L}_{\xi}\phi=\xi^{\mu}\del_{\mu}\phi(x)\ .
    \end{equation}
Again, note the difference between this result and eq.~\eqref{tansfS}: in the latter the scalar field is evaluated at the same point in two different sets of coordinates (cf. eq.~\eqref{tansfP}), while in the former is the scalar field is evaluated at two different points, which is why $\phi_{\ssst P}(x_{\ssst P})\neq\phi_{\ssst Q}(x_{\ssst P})$.
Furthermore, we can also arrive in the same way at the Lie derivative of the metric. The only difference is that we have one additional term as compared to the vector case because the metric tensor is rank 2 tensor and we have to use eq.~\eqref{delTr} two times. We state it here without the proof
    \begin{align*}
        \d s^2\vert_{\ssst P}-\d s^2\vert_{\ssst Q}=0\quad\rightarrow\quad \mathcal{L}_{\xi}g_{\mu\nu}(x) = \xi^{\alpha}\del_{\alpha}g_{\mu\nu}(x)
        +g_{\mu\alpha}(x)\del_{\nu}\xi^{\alpha}
        +g_{\alpha\nu}(x)\del_{\mu}\xi^{\alpha}\ ,
    \end{align*}
which is just the standard result. 

To get a better feeling of the difference between a passive and an active view of coordinate transformations, compare the coordinate transformation from Cartesian to polar coordinates in two dimensions with rotations in two dimensions,
    \begin{align}
    \label{chpol}
        x & = r \cos\theta  & y & = r \sin\theta\\[6pt]
    \label{chrot}
        x & = \tilde{x} \cos\phi - \tilde{y} \sin \phi  & y & = \tilde{x} \sin\phi + \tilde{y} \cos \phi\ .
    \end{align}
The transformation to polar coordinates in \eqref{chpol} does not require introduction of any parameter: it is enough to know which coordinates we would like to transform to and this transformation replaces one grid of coordinate line with another, globally (of course there are points which cannot be included by the new
system but that is irrelevant now). However, for rotations in \eqref{chrot}, what we basically do is that we not only replace the coordinate lines, but we also give a direction to which they are pointing using parameter $\phi$. Stated as they are, these rotations introduce the new coordinates globally.
Now, these two coordinate transformations indeed look quite different, but let us take a look how do their differentials change\footnote{Similarly for the transformation of $\del_{\mu}$.},
    \begin{align}
    \label{rotx}
        \d x & = \cos\theta\,\d r - \sin\theta (r\,\d\theta)   &  \d y & = \sin\theta\,\d r +\cos\theta (r\,\d\theta)\\[6pt]
    \label{rotds}
        \d x & = \cos\phi\,\d \tilde{x} - \sin \phi\, \d \tilde{y}  & \d y & = \sin\phi \,\d \tilde{x}  + \cos \phi \,\d \tilde{y} \ .
    \end{align}
And now we see that \textit{locally}, i.e. if we focus on the transformation of \textit{(co)frames} which are defined at a point and not globally, these two transformations look the same, provided we introduced a local orthonormal frame $\theta^{1}=\d r\ ,\theta^2 = r\,\d \phi$. That is, they both act like a rotation
of \textit{frames}. Indeed, even though $(r,\phi)$ are curvilinear coordinates this coordinate system is an orthogonal one so the basis vectors at each point are orthogonal to each other, but their orientation depends on $\theta$.
Therefore, if one would like to relate a frame in Cartesian coordinates to an orthonormal frame in $(r,\phi)$ coordinates one would use the rotation of frames by $\theta$. But this is just the same as if we started with rotation
\eqref{rotds} in the first place, except that $\d \tilde{y}$ is integrable but $r\,\d \theta$ is not, so the curvilinear coordinate axes are straight only in a small neighbourhood of a point in which the frame is defined and one can approximate them with a Cartesian coordinate system with Eucliedan
metric only locally. This introduction of locally orthonormal frames can be extended to curved spaces as well in the same way. Then we have that the metric can locally be represented by
    \begin{equation}
        \d s^2 = \eta_{\ssst A\ssst B}\theta^{\ssst A}\theta^{\ssst B}\ ,
    \end{equation}
where $\theta^{\ssst A}=\tilde{E}^{\ssst A}{}_{\mu}\d x^{\mu}$ is the orthonormal coframe, i.e. an arbitrary linear combination of $\d x^{\mu}$ encoded in a matrix $\tilde{E}^{\ssst A}{}_{\mu}$ (that in general has nothing to do with a coordinate transformation in eq.~\eqref{transfmatrix}) called \textit{vielbein},
and $\eta_{\ssst A\ssst B}$ is the constant diagonal metric with $\pm 1$ as its entries (Minkowski metric, if we are talking about spacetime). Then one can always
find a coordinate system valid around a small neighbourhood of a point called \textit{Riemann normal coordinate
system}, whose axes measure geodesic distance and that gives rise to the flat metric $\eta_{\mu\nu}$ and vanishing of the Christoffel symbols at that point. More generally, one can introduce a coordinate system around a timelike geodesic (i.e. at each point \textit{along} a chosen geodesic) such that the metric is $\eta_{\mu\nu}$ and the Christoffel symbols vanish \textit{along} this geodesic (this is called \textit{Fermi normal coordinate system}).
This is a rough mathematical version of what Einstein essentially did in order to formulate his Equivalence Principle: \textit{it is the active view of transformation that describes the switching from a non-inertial to an inertial frame}, describing a freely-falling observer along a timelike geodesic. 
Moreover, one can also formulate Fermi normal coordinate systems for null geodesics \cite{BFW}, the so-called \textit{null Fermi coordinates}, which are suitable for tracking null rays along geodesics; this is the closest as one would get to transforming into ``a frame attached to a photon'' and is more appropriate to think of it as being attached to a wave front.

In summary, the active transformation can distinguish among the observers found in different physical situations. It describes switching among different ``points of view'' (local frames).
This induces a transformation of the frame, meaning that the point of view needs to be updated with information about the new frame, as if the observer has to keep reconstructing their original frame (by means of
eq.~\eqref{delTr}) at each next point as they advance, in order to evaluate this change. As we saw, the resulting change is encoded in the components of the Lie derivative with respect to a single frame, i.e. with respect to a single point of view of choice. This interpretation gives rise to a visualization of ``instantaneous motions'', e.g. one says ``let us boost into a freely falling frame''; what is meant here is that we use a transformation of a frame \textit{at one point}, i.e. \textit{not} globally, which may or may not be associated with a coordinate transformation (in a small neighbourhood around that point) and is thus more fundamental.

In the following sections we are interested in the change of fields's components from an active point of view on coordinate transformations.
It should be noted that since we are working in Riemannian geometry\footnote{In Riemannian geometry the metricity condition $\nabla_{\alpha}g_{\mu\nu}$ is satisfied and torsion (antisymmetric part of the connection) is set to vanish, thus leaving the Levi-Civita
connection (Christoffel symbols).} all the above-stated expressions for the Lie derivatives can be written in terms of the covariant derivatives in place of the partial derivatives, and we do so whenever the need arises in this thesis.

\section{General coordinate transformations}
\label{sec_genCT}

Before we familiarize ourselves with conformal transformations it is of great use to analyze general coordinate transformations.
We shall focus on active \textit{infinitesimal} coordinate transformations (i.e. point transformations) because we would like to inspect the local so-called \textit{physical} change of tensor components expressed by the means of a Lie derivative, as explained in the previous section.
Namely, restating eq.~\eqref{infPQ}, a general infinitesimal coordinate transformation is given by
    \begin{equation}
    \label{gencoordtr}
        \tilde{x}^{\mu}=x^{\mu}+\xi^{\mu}\ ,
    \end{equation}
where $\xi^{\mu}$ is a $d$-dimensional vector with each component being a function of coordinates and $\xi^{\mu}\ll 1$, 
induces a change of the metric components in the form (valid for Riemannian spaces)
    \begin{equation}
    \label{CT_1}
        \delta_{\ssst \xi} g_{\mu\nu}=\mathcal{L}_{\ssst \xi}g_{\mu\nu}=2\nabla_{(\mu}\xi_{\nu)}
    \end{equation}
and the change of the Christoffel symbols of the form\footnote{See the proof in appendix~\ref{App_VarGamma}.}
    \begin{subequations}    
        \begin{align}
        \label{CT_2}
            \delta_{\ssst \xi} {\Gamma^{\alpha}}_{\mu\nu}&=
            g^{\alpha\beta}\Biggpar{
            \nabla_{\mu}\nabla_{(\beta}\xi_{\nu)}
            +\nabla_{\nu}\nabla_{(\beta}\xi_{\mu)}
            -\nabla_{\beta}\nabla_{(\mu}\xi_{\nu)}
            }\nonumber\\[6pt]
            &=g^{\alpha\beta}
            \Biggpar{
            \nabla_{(\mu}\nabla_{\nu)}\xi_{\beta}
            -\xi_{\rho}{R^{\rho}}_{(\mu\nu)\beta}
            }\\[6pt]
        \label{CT_2a}
            &=\nabla_{(\mu}\nabla_{\nu)}\xi^{\alpha}-{R^{\alpha}}_{(\mu\nu)\beta}\xi^{\beta} \ .
        \end{align}
    \end{subequations}
Note that this variation can be derived even if there were no metric --- it too is more generally defined as the Lie derivative of the connection\footnote{As mentioned in the previous section, one keeps in mind that notation $\delta_{\ssst \xi} {\Gamma^{\alpha}}_{\mu\nu}$ means ``$\alpha\mu\nu$-component of the Lie derivative of the connection''.} in the direction of $\xi^{\mu}$; then the covariant derivative is unrelated to the metric and the proof is slightly different.
Equation \eqref{CT_2} is more convenient for our purposes.
One is familiar with isometries, i.e. those coordinate transformations which do not change the metric components, the consequence of which are the following two equations
    \begin{subequations}
        \begin{align}
        \label{Killeq}
            \delta_{\ssst \xi} g_{\mu\nu} =2\nabla_{(\mu}\xi_{\nu)}&=0\ ,\\[12pt]
        \label{KillGamma}    
            \delta_{\ssst \xi} {\Gamma^{\alpha}}_{\mu\nu}=0\quad\Rightarrow\quad
            \nabla_{(\mu}\nabla_{\nu)}\xi_{\beta}
                &=\xi_{\rho}{R^{\rho}}_{(\mu\nu)\beta}\ ,
        \end{align}
    \end{subequations}
then we call eq.~\eqref{Killeq} \textit{the Killing equation} and vector $\xi^{\mu}$ is referred to as \textit{the Killing vector}, while equation\footnote{Note that this equation can be derived even if there were no metric; thus it is a statement independent of eq.~\eqref{Killeq} and is \textit{necessary} for finding all Killing vectors.}
\eqref{KillGamma} is usually referred to as the integrability condition for $\xi^{\mu}$. To warm up for the approach presented below, one can read the above equation as follows: \textit{if the symmetric part of tensor $\nabla_{\mu}\xi_{\nu}$ vanishes, $\xi^{\mu}$ is a Killing vector}.
But conformal transformations, which we are aiming to talk about here, are not isometries; they are simply a class of general coordinate transformations with certain special properties.

Coming back to general coordinate transformations, great insight into various transformations may be gained if one decomposes $\xi^{\mu}$ into directions orthogonal to ``vector''\footnote{Note that this not a tensorial object that transforms as a vector under coordinate transformations. It is just a set of $d$ scalar functions.} $x^{\mu}$ and parallel to it\footnote{To the author's knowledge the following approach to describing coordinate transformations is not introduced in textbooks.}.
To define these directions, introduce projectors $P^{\mu}_{{\ssst \bot}\nu}$ and $P^{\mu}_{{\ssst \parallel}\nu}$ which obey
    \begin{equation}
        P^{\mu}_{{\ssst \bot}\nu}x^{\nu}=0\ ,\quad P^{\mu}_{{\ssst \parallel}\nu}x^{\nu}=x^{\mu}\ ,\quad
        P^{\mu}_{{\ssst \bot}\nu}P^{\nu}_{{\ssst \parallel}\alpha}=0\ ,\quad
        P^{\mu}_{{\ssst \bot}\nu}P^{\nu}_{{\ssst \bot}\alpha}=P^{\mu}_{{\ssst \bot}\alpha}\ ,\quad P^{\mu}_{{\ssst \parallel}\nu}P^{\nu}_{{\ssst \parallel}\alpha}=P^{\mu}_{{\ssst \parallel}\alpha}\ ,
    \end{equation}
such that $\xi^{\mu}$ is split in the following way
    \begin{equation}
    \label{xiDecomp}
        \xi^{\mu}=
        {\xi^{\ssst \bot}}^{\mu}+ g^{\mu\nu}\nabla_{\nu}\sigma\ ,
        \qquad {\xi^{\ssst \bot}}^{\mu}:=P^{\mu}_{{\ssst \bot}\nu}\xi^{\nu}\ , 
        \qquad  g^{\mu\nu}\nabla_{\nu}\sigma := P^{\mu}_{{\ssst \parallel}\,\,\nu}\xi^{\nu}\ ,
    \end{equation}
and
    \begin{equation}
    \label{projcomps}
        \nabla_{\mu}{\xi^{\ssst \bot}}^{\mu}=0 \ ,\qquad {\xi^{\ssst \bot}}^{\mu}\nabla_{\mu}\sigma=0\ .
    \end{equation}
We call ${\xi^{\ssst \bot}}^{\mu}$ and $\nabla_{\nu}\sigma $ transversal and longitudinal component, respectively. This decomposition is encouraged by Presnov \cite{Pres02} where it was introduced in the context of studying chaotic systems. It was noted there that the two conditions: vanishing divergence of ${\xi^{\ssst \bot}}^{\mu}$ is one and its orthogonality to $x^{\mu}$ is another, may or may not imply one another and it is a matter of choice what would one like to do and what kind of situation one has. We choose both because for the matters discussed here it seems to be advantageous for an intuitive understanding of coordinate transformations.
If one does not introduce transverse-longitudinal projectors, but stays with decomposition into divergence and divergence-less parts, one has the usual Helmholtz-Hodge decomposition.

So established point of view will help us understand conformal coordinate transformations in a way that is not found in textbooks, to the best of author's knowledge.
Now, we already have some intuition about transversal and longitudinal components that we can borrow from our understanding of electrodynamic potential $A_{\mu}$ associated with a vacuum electromagnetic field.
There the transversal component carries the two remaining gauge invariant degrees of freedom after the gauge freedom has been used. The longitudinal component is missing because the mass term is missing --- mass, or inertia,
acts like a kind of friction to suppress the propagation of waves, thus its absence means the waves propagate with the maximum possible velocity. 
In other words, mass acts like the spring to which a body is suspended: non-vanishing ellasticity coefficient induces oscillations in the body's position; these oscillating modes are akin to the longitudinal mode of wave propagation.
We could also think of the mass term as being related to field's longitudinal effects on charges: its absence inhibits any changes to charge distributions in the longitudinal direction of the wave propagation.
It is useful to keep in mind this relationship between a mass term and longitudinal degree of freedom for later on.

Let us give an example to obtain some further intuition about the transversal and longitudinal components of $\xi^{\mu}$.
Consider a spatial rotation. Let us write the position vector as $\vec{r}$ instead of $x^{\mu}$ for a moment. Then a spatial rotation of $\vec{r}$ in a certain plane will shift the tip of that vector in the direction \textit{orthogonal} to it, while keeping its length fixed and keeping its stem fixed to the origin. Hence this is an orthogonal transformation --- it adds to $\vec{r}$ an infinitesimal displacement vector\footnote{Iwth conditions $\varphi << 1$,  $\vec{n}\cdot\vec{n}=1$.} $\varphi\vec{n}$ \textit{orthogonal} to it, such that
    \begin{equation}
        \vec{r}^{'}=\vec{r}+\varphi\vec{n}\ ,\qquad \vec{n}\cdot\vec{r}=0\ ,
    \end{equation}
where $\varphi$ is the small rotation angle.
This can be generalized to spaces of any number of dimensions. Wherever $x^{\mu}$ is pointing, a rotation always changes $x^{\mu}$ in the direction orthogonal to it, such that its length remains invariant,
    \begin{equation}
        \eta_{\mu\nu}\tilde{x}^{\mu}\tilde{x}^{\nu}\approx \eta_{\mu\nu}x^{\mu}x^{\nu}+2\eta_{\mu\nu}x^{\mu}\xi^{\nu}\shalleq \eta_{\mu\nu}x^{\mu}x^{\nu}\ ,
    \end{equation}
from which it follows that $\eta_{\mu\nu}x^{\mu}\xi^{\nu}=0$.
Extending this to an example of Lorentz transformation and using the language of transversal-longitudinal decomposition, we may describe $\xi^{\mu}$ associated with a Lorentz transformation as being orthogonal to $x^{\mu}$, that is,
    \begin{equation}
    \label{xixzero}
        \text{if $\xi^{\mu}$ is a Lorentz transformation then}\quad \eta_{\mu\nu}\xi^{\mu}x^{\nu}=0 \quad\Rightarrow\quad \xi^{\mu}=\xi^{\mu}_{\ssst \bot}\ ,
    \end{equation}
that is, Lorentz transformations are described by $\xi^{\mu}$ whose longitudinal component vanishes. Let us examine this statement more closely. Act with $x^{\alpha}\del_{\alpha}$ on $\eta_{\mu\nu}\xi^{\mu}x^{\nu}=0$ in eq.~\eqref{xixzero} to get
    \begin{equation}
    \label{xOrthoXi}
        0=x^{\alpha}\xi_{\alpha}=-x^{\alpha}x^{\mu}\del_{\alpha}\xi_{\mu}=-x^{\alpha}x^{\mu}\del_{(\alpha}\xi_{\mu)}\quad\Rightarrow\quad \del_{\alpha}\xi_{\mu}=\del_{[\alpha}\xi_{\mu]}\quad\Rightarrow\quad\xi_{\mu}=\xi_{\mu}^{\ssst\bot}
    \end{equation}
since partial derivatives on $\sigma$ commute. This proves that general Lorentz transformations are described by the antisymmetric part of $\del_{\alpha}\xi_{\mu}$ and thus by the transversal component $\xi_{\mu}^{\ssst\bot}$ only.
Note, in passing, that condition $\del_{(\alpha}\xi_{\mu)}=0$ is just what follows from eq.~\eqref{Killeq} for Minkowski spacetime, pointing to the equivalence of the two approaches.
The form of the vector which satisfies these conditions is given by
    \begin{equation}
    \label{xiL}
        \xi^{\mu}_{L}=m^{\mu}{}_{\nu}x^{\nu}\ ,
    \end{equation}
where $m_{\mu\nu}$ is an antisymmetric matrix of constant parameters.

What about translations? Translations are described by $\xi^{\mu}=a^{\mu}=const.$ and this means that a vector $V^{\mu}(x)$ can be translated in any direction while its length is preserved, that is,
    \begin{equation}
    \label{Vlength}
        \tilde{V}^{\mu}\tilde{V}_{\mu}\approx V^{\mu}V_{\mu}+2V^{\mu}V^{\nu}\del_{\mu}\xi_{\nu}\shalleq V^{\mu}V_{\mu}\quad\Rightarrow\quad \del_{(\mu}\xi_{\nu)}=0\ ,
    \end{equation}
and we see that this trivially includes the case of $\xi^{\mu}=a^{\mu}=const.$
Hence, we have a choice to say $\xi^{\mu}_{\ssst\bot}=a^{\mu}$, $\xi^{\mu}_{\ssst\parallel}=a^{\mu}$ or that both components contribute to $a^{\mu}$.
Note that condition in eq.~\eqref{Vlength}, in accordance with Minkowski spacetime versions of eq.~\eqref{Killeq} and eq.~\eqref{KillGamma}, also includes Lorentz transformations, i.e. it determines Poincare symmetries of the Minkowski spacetime.

The transverse-longitudinal decomposition used here is based on the Helmholz decomposition theorem which states that any vector can be decomposed into divergence-free part (${\xi^{\ssst \bot}}^{\mu}$) and curl-free part ($\sigma$). Here this theorem is used in the context of a general coordinate system.
The second equation in \eqref{projcomps} can be read as: \textit{derivative of the longitudinal scalar degree of freedom along the transversal direction vanishes}, which is just a consequence of Helmholz decomposition being orthogonal.

What about the interpretation of the longitudinal part? What sort of a change of coordinates may be done \textit{along} the direction of $x^{\mu}$? The simplest example to think of is dilations, while still in Minkowski spacetime. 
dilations are such transformations which change the length of a position vector by some constant factor $\Lambda $. For this to happen, we must have that
    \begin{equation}
    \label{DilatXX}
        \eta_{\mu\nu}\tilde{x}^{\mu}\tilde{x}^{\nu}\approx\eta_{\mu\nu}x^{\mu}x^{\nu}+2\eta_{\mu\nu}x^{\mu}\xi^{\nu}\shalleq \Lambda\, \eta_{\mu\nu}x^{\mu}x^{\nu}
    \end{equation}
from where it follows that
    \begin{equation}
    \label{xilambda}
        \xi^{\mu}=\lambda x^{\mu}
    \end{equation}
such that $\Lambda=1+2\lambda$ and we see that dilations transform the coordinates in the following way
    \begin{equation}
    \label{Xdilat}
        \tilde{x}^{\mu}=(1+\lambda)x^{\mu}\ ,
    \end{equation}
where $\lambda$ is a constant parameter of dilation transformation.
Since $\xi^{\mu}$ is proportional to $x^{\mu}$, it is obvious that dilations cannot be described by the transversal component. Hence,
    \begin{equation}
        \text{if $\xi^{\mu}$ describes dilations then }\quad \xi^{\mu}=\eta^{\mu\nu}\del_{\nu}\sigma
    \end{equation}
and one may even find that $\sigma=\lambda \eta_{\mu\nu}x^{\mu}x^{\nu}/2$ up to a constant, but for our discussion there is no need for such detail. What is more interesting is to realize that $\lambda$ is determined from eq.~\eqref{xilambda} by taking a divergence of both sides, 
    \begin{equation}
        \lambda=\frac{1}{d}\del_{\mu}\xi^{\mu}=\frac{1}{d}\Box\sigma\ ,
    \end{equation}
where $\Box$ is the d'Alambertian. But $\del_{\mu}\xi^{\mu}$ is just the trace of the Minkowski spacetime version of eq.~\eqref{CT_1}! Therefore, one can relate the trace of $\del_{\mu}\xi_{\nu}$ with dilations. Indeed, dilations belong to the class of \textit{conformal coordinate transformations} which are defined with such a $\xi_{\mu}$ which obeys
    \begin{equation}
    \label{defConf}
        \delta_{\xi}g_{\mu\nu}=2\nabla_{(\mu}\xi_{\nu)}=2\omega g_{\mu\nu}\ ,
    \end{equation}
where $\omega$ is \textit{a function of coordinates} and we wrote the most general definition of conformal transformations in arbitrary space (for a moment moving away from the Minkowski spacetime).
In textbooks with standard treatment of conformal transformations eq.~\eqref{defConf} is usually read as: \textit{conformal transformations leave the metric components invariant up to an arbitrary scaling function $\omega(x)$}.
However, such a definition puts somewhat misleading attention to a sort of a deviation from isometry rather than on features of conformal coordinate transformations. 

Instead of such a definition of conformal (or any other non-isometry) coordinate transformations, we would like to look at general coordinate transformations as comprised of three classes (or subgroups) of transformations, described by the following conditions
    \begin{enumerate}
        \item $\delta_{\xi}g_{\mu\nu}=2\nabla_{(\mu}\xi_{\nu)}=0$
        \item $\delta_{\xi}g_{\mu\nu}=2\nabla_{(\mu}\xi_{\nu)}=2\omega g_{\mu\nu}$
        \item $\delta_{\xi}g_{\mu\nu}=2\nabla_{(\mu}\xi_{\nu)}=2S^{\srm T}_{\mu\nu}$, such that $g^{\mu\nu}S^{\srm T}_{\mu\nu}=0$
    \end{enumerate}
Class 1. clearly determines the Killing vectors; isometries do not change the metric components and are described by the remaining part: the antisymmetric part $\nabla_{[\mu}\xi_{\nu]}$. 
Class 2. should be read as: \textit{conformal coordinate transformations are defined by the trace part of $\delta_{\xi}g_{\mu\nu}$}.
We are quite familiar with the first two classes. However, it now becomes clear that Class 3. can be introduced, describing those coordinate transformations that do not fall into the first two classes. These transformations are the remaining set of transformations complementary to the conformal transformations; they are defined by the tracelss part of $\delta_{\xi}g_{\mu\nu}$.
Therefore, we may split the $d^2$-component tensor $\nabla_{\mu}\xi_{\nu}$ into three orthogonal sets of components:
    \begin{subequations}
        \begin{align}
        \label{GenCtAll}
            \underbrace{\nabla_{\mu}\xi_{\nu}}_{d^2}&=\underbrace{M_{\mu\nu}}_{\frac{d(d-1)}{2}}+\underbrace{S^{\srm T}_{\mu\nu}}_{\frac{d(d+1)}{2}-1}+\frac{1}{d}\underbrace{g_{\mu\nu}S}_{1}\ ,\\[12pt]
        \label{GenCtfIso}
             M_{\mu\nu}&:=\nabla_{[\mu}\xi_{\nu]}\ ,\\[12pt] 
        \label{GenCtfTr}
             S &:=\nabla_{\alpha}\xi^{\alpha}\ ,\\[12pt]
        \label{GenCtfTrless}
             S^{\srm T}_{\mu\nu}&:= \nabla_{(\mu}\xi_{\nu)}-\frac{1}{d}g_{\mu\nu}\nabla_{\alpha}\xi^{\alpha}\ .
        \end{align}
    \end{subequations}
In 4 dimensions this split amounts to $16 = 6 + 9 + 1$ components. 
But these components are somehow determined by the transversal and longitudinal parts of $\xi^{\mu}$ and it is interesting to see in which way. To see this, simply apply decomposition given in eq.~\eqref{xiDecomp} to eqs.~\eqref{GenCtfIso}-\eqref{GenCtfTrless}, obtaining the following expressions,
    \begin{align}
        \label{GenCtfIso1}
             M_{\mu\nu}&=\nabla_{[\mu}\xi_{\nu]}^{\ssst\bot}\ ,\\[12pt] 
        \label{GenCtfTr1}
             S &=\Box\sigma\ ,\\[12pt]
        \label{GenCtfTrless1}
             S^{\srm T}_{\mu\nu}&= \nabla_{(\mu}\xi_{\nu)}^{\ssst\bot}+\left[\nabla_{(\mu}\nabla_{\nu)}-\frac{1}{d}g_{\mu\nu}\Box\right]\sigma\ .
    \end{align}
In eq.~\eqref{GenCtfIso1}, which describes 6 parameters of isometries (Class 1.), the longitudinal component drops out because the covariant derivatives commute in Riemannian geometry\footnote{It would be interesting to study isometries and other coordinate transformations in non-Riemannian geometry in terms of transversal and longitudinal components.}.
From eq.~\eqref{GenCtfTr1} we can see that only the longitudinal component contributes to the trace component of the metric variation, describing conformal transformations. Both longitudinal and transversal components feed the traceless part, eq.~\eqref{GenCtfTrless1}.

For the end of this section it is instructive to state what is the variation of Christoffel symbols and curvature tensors induced by general coordinate transformations in terms of eq.~\eqref{GenCtAll}. Plugging this equation in the first line of eq.~\eqref{CT_2} we see that isometries drop out, while the rest can be grouped into two \textit{independent} coordinate variations
    \begin{align}
    \label{varGamma}
        \delta_{\ssst \xi} {\Gamma^{\alpha}}_{\mu\nu}&=
            g^{\alpha\beta}\Biggpar{
            \nabla_{\mu}S_{\beta\nu}^{\srm T}
            +\nabla_{\nu}S_{\beta\mu}^{\srm T}
            -\nabla_{\beta}S_{\mu\nu}^{\srm T}
            }+ 
            \frac{1}{d}\left(\delta^{\alpha}_{\mu}\delta^{\beta}_{\nu}+\delta^{\alpha}_{\nu}\delta^{\beta}_{\mu}-g_{\mu\nu}g^{\alpha\beta}\right)\del_{\beta}S\nonumber\\[6pt]
            &=g^{\alpha\beta}
            \Biggpar{
            \nabla_{(\mu}\nabla_{\nu)}\xi_{\beta}
            -\xi_{\rho}{R^{\rho}}_{(\mu\nu)\beta}
            }\ .
    \end{align}
Besides the above, one can further calculate $\delta_{\xi}R^{\alpha}{}_{\mu\beta\nu}=2\nabla_{[\nu\vert}\delta_{\ssst \xi} {\Gamma^{\alpha}}_{\mu\vert\beta]}$ by using eq.~\eqref{varGamma}.
In particular, for Minkowski spacetime, all Christoffel symbols and curvatures vanish and one may wonder what is the meaning of the above equations and why are they even necessary. The meaning of $\delta_{\xi}R^{\alpha}{}_{\mu\beta\nu}$ in Minkowski spacetime is that no coordinate transformation can change the fact that all curvatures vanish at each point --- this follows from the covariant (tensorial) nature for curvature.
The reason why they are necessary is that they provide differential equations for finding $\xi^{\mu}$ in any spacetime, including Minkowski spacetime, as we shall see in the next section.

Now the question is: which components of the
metric are affected by conformal transformations encoded in the trace $S$ and which by those (yet to be named) transformations encoded in the traceless part $S^{\srm T}_{\mu\nu}$? The question is essentially concerned with the nature of trace and traceless parts of $\delta_{\xi}g_{\mu\nu}$ to which we turn to in the following sections.

\section{Conformal (shape-preserving) coordinate transformations}
\label{conftr}

Apart from the fact that conformal transformations are defined by eq.~\eqref{defConf}, they are also described as those transformations that change the lengths, areas and volumes, but leave the angles invariant. We shall describe what does this mean on an example.
A thorough review with lots of clarifications of subtleties usually omitted elsewhere can be found in a paper by Kastrup \cite{Kas} devoted to all kinds of conformal transformations. 
For all known results regarding conformal transformations one can refer to that paper, as we do here.

We start by visualizing the \textit{Mercator mapping}. Take a flat sheet of paper whose width is equal to the circumference of the globe and height equal to its half. Draw a Cartesian grid, identify the equator line across the
middle of the paper along the width and try to cover a globe with that sheet of paper by aligning the equators --- it is not possible, there is some excess surface of the paper which is a signature that the surface of the globe is positively curved. 

But now imagine that this piece of paper is elastic and can be stretched or contracted at each point however we like along some chosen direction (with even more special ability that once we stretch it or contract it, it stays ``frozen'' that way without returning to its original state).

With this new property of the paper we would like to think of the excess areas of the paper as the
ones which are ``stretched too much'' compared to the corresponding area on the globe and we would like to correct this mismatch.
Let us then contract these areas of this elastic paper in such a way (that means choose the directions of stretching appropriately) that the excess area disappears and the sheet of paper covers
the globe in such a way that the grid on the paper aligns precisely with the grid on the globe. To describe that in some points (infinitesimally close to the equator) we do not need to contract
the paper and that in some other points (more and more as we move to the poles) we need to contract a lot, we use a coordinate-dependent function $\omega(x)$.
What has happened as a result? First of all, all points on the upper edge (parallel to the equator) of the paper had to be identified: the area excess infinitesimally close to that edge was
the greatest, thus maximal contraction had to be performed there with a result that all points of the edge have been identified --- all these points have become the north pole.
The same thing happens with the lower edge --- they are identified with the south pole on the globe.
Furthermore, the side edges of the paper (the ones orthogonal to the equator) are identified one with another --- this enables us to travel around the world, literally, as we are confined to the sheet of paper.
These identifications mean that the same \textit{topology} as the globe's had to be imposed on the finite sheet of paper before identifying it with the globe --- otherwise there would be no
smooth lines across the edges on the paper. This is a topology of a 2-sphere.
Lastly, since the necessity for contraction increases as we move from the equator towards each of the poles, 2 adjacent coordinate lines orthogonal to the equator which were parallel to each other on the uncontracted sheet of paper converge to a single point in
both directions, the north and south pole. To visualize what this means, draw a square with one edge aligned with the equator initially, then drag this square towards the upper edge of the paper such that the mentioned edge is always parallel to the equator. On the uncontracted sheet of paper the square remains of the
same size during the whole dragging process. But on the globe the dragging along the same path towards the pole between the two adjacent meridians makes the square shrink in size, with a greater rate around its edge closer to the pole, only to degenerate into a point at the pole
itself. However, even though the square has been deformed in a certain way, the four inner angles which make this square \textit{a square} have been left unchanged during the process of conformal
transformation. In other words, the angles --- for which we say they characterize \textit{the shape} of a figure --- are \textit{invariant under conformal transformations}.

To see this more clearly, recall the simple dilation we described by eq.~\eqref{DilatXX}, but now generalize $\Lambda$ to a coordinate-dependent function $\Lambda=\Lambda(x)=1+2\omega(x)$,
    \begin{equation}
    \label{ConfXX}
        \eta_{\mu\nu}\tilde{x}^{\mu}\tilde{x}^{\nu}\approx (1+2\omega(x))\, \eta_{\mu\nu}x^{\mu}x^{\nu}\ .
    \end{equation}
Note that we cannot claim that a coordinate-dependent version of eq.~\eqref{xilambda} is valid in this case --- things are a bit more complicated and we shall soon see why.
Now, that the angles are invariant can be witnessed from a more general definition of an ``angle''
    \begin{align}
    \label{confTrangles}
        \frac{\eta_{\mu\nu}\tilde{x}^{\mu}\tilde{y}^{\nu}}{\vert \eta_{\mu\nu}\tilde{x}^{\mu}\tilde{x}^{\nu}\vert^{1/2} \vert \eta_{\mu\nu}\tilde{y}^{\mu}\tilde{y}^{\nu}\vert ^{1/2}}
        &\approx\frac{(1+2\omega)\eta_{\mu\nu}x^{\mu}y^{\nu}}{\vert (1+2\omega)\eta_{\mu\nu}x^{\mu}x^{\nu}\vert^{1/2} \vert (1+2\omega)\eta_{\mu\nu}y^{\mu}y^{\nu}\vert ^{1/2}}\nonumber\\[6pt]
        &=\frac{\eta_{\mu\nu}x^{\mu}y^{\nu}}{\vert \eta_{\mu\nu}x^{\mu}x^{\nu}\vert^{1/2} \vert \eta_{\mu\nu}y^{\mu}y^{\nu}\vert ^{1/2}}\ ,
    \end{align}
where $x^{\mu}$ and $y^{\mu}$ are some arbitrary position ``vectors''. Thus we see that particular ratios of lengths (i.e. the generalization of a cosine of an angle between two vectors in Euclidean space) are invariant under conformal transformations since any change cancels out.
If one repeats the whole procedure described in our example above by using a very fine resolution grid, one would indeed witness \textit{the preservation of shape and inflation/deflation of areas and volumes} under conformal transformations. 
Actually, this fact is where the name \textit{conformal} comes from\footnote{In latin \textit{con} -- same;  \textit{forma} -- shape, form}: \textit{same + shape}.
Therefore, in some way angles (or shapes) and lengths, volumes (or scales) are complementary to each other, much in the same way that eq.~\eqref{GenCtfTrless} is complementary to eq.~\eqref{GenCtfTr}.

It is of crucial interest to move away from a two-dimensional example above to a general $d$-dimensional spacetime, because we would like to generalize the notions of ``shape'' and ``scale'' to some geometric objects that one can refer to instead of cubes or squares or triangles and their sizes.
Since conformal transformations affect only scales and volumes, it is natural to start by looking at how conformal transformations change $\sqrt{g}$, which one usually calls a bit imprecisely ``the volume''\footnote{It is more precise to refer to it as \textit{the component of the invariant volume form}, defined in four dimensions as $\d \rm{vol} =\sqrt{g}\,\d x^{0}\wedge\d x^{1}\wedge\d x^{2}\wedge\d x^{3}$, where $\wedge$ is the antisymmetric exterior product.}.
We immediately see this upon taking the trace of the variation of the metric components given by eq.~\eqref{defConf},
    \begin{subequations}
        \begin{align}
        \label{tracedeltag}
            g^{\mu\nu}\delta_{\xi}g_{\mu\nu}=\frac{2}{\sqrt{g}}\delta_{\xi}\sqrt{g}&=2\nabla_{\mu}\xi^{\mu}=2\omega\, d \ ,\\[6pt]
            \delta_{\xi}\sqrt{g}&=\del_{\mu}\underline{\xi}^{\mu}=\underline{\omega}\, d
        \end{align}
    \end{subequations}
where $\underline{\xi}^{\mu}:=\sqrt{g}\xi^{\mu}$ and $\underline{\omega}:=\sqrt{g}\,\omega$ are vector and scalar density, respectively. The second equation above is more suitable for general spacetimes due to the absence of covariant derivatives.
We can somewhat simplify the above equations by introducing an object which caries the meaning of coordinate-dependent \textit{length scale}, similar to $\sqrt{g}$ that carries the meaning of coordinate-dependent volume,
    \begin{equation}
    \label{scaledef}
        A:=(\sqrt{g})^{\frac{1}{d}}\ ,
    \end{equation}
which we shall call \textit{the scale density}; its weight is $1/d$. 
Then eq.~\eqref{tracedeltag} can be equivalently written as
    \begin{equation}
    \label{deltaAomega}
        \delta_{\xi}A=\omega A\ ,
    \end{equation}
from which we see that $\delta_{\xi}A /A$ is a scalar; also note that eq.~\eqref{deltaAomega} can easily be interpreted as local rescaling of lengths (see section \ref{confFtransf}).
Furthermore, the scale density is a single degree of freedom of the metric tensor and, according to eq.~\eqref{tracedeltag}, is the only degree of freedom affected by conformal transformations. Let us isolate this degree of freedom from the
metric components by decomposing the metric tensor into $A$ and something else, in such a way that a conformal transformation changes only $A$, leaving the remaining part manifestly invariant.
This remaining part then has to have $d(d+1)/2 - 1$ components, which are put into an object defined by\footnote{Also called ``unimodular metric'', for reasons that we shall elaborate more on in \ref{unimodDec}. Note that Fulton et al. \cite{Fult} have noticed the relevance of $\gb_{\mu\nu}$ in the context of conformal transformations and their relevance in physics.}
    \begin{equation}
    \label{shapedef}
        \gb_{\mu\nu}:=A^{-2}g_{\mu\nu}
    \end{equation}
which we shall call \textit{the shape density}.
We can now define conformal coordinate transformations as those \textit{that leave the shape density invariant}, or, equivalently, those \textit{that change only the scale density}. 

One says that \textit{two sets of metric components $\tilde{g}_{\mu\nu}$ and $g_{\mu\nu}$ are conformal to each other if they are related by a conformal transformation}. 
An important well-known fact related to this concerns the Weyl tensor $C^{\mu}{}_{\alpha\nu\beta}$ defined as the totally traceless part of the Riemann tensor
    \begin{align}
    \label{Weyltens}
        C^{\alpha}{}_{\mu\beta\nu}&=R^{\alpha}{}_{\mu\beta\nu}
        -\frac{2}{d-2}\left(\delta^{\alpha}_{[\beta}R_{\nu]\mu}-g_{\mu[\beta}R^{\alpha}{}_{\nu]}\right)
        -\frac{2}{(d-1)(d-2)}\delta_{[\nu}^{\alpha}g_{\beta]\mu}R\ .
    \end{align}
introduced by Herman Weyl in \cite{Weyl18a}.
Namely, Weyl tensor is that part of the Riemann tensor which is invariant under conformal transformations
\footnote{Moreover, Weyl tensor is invariant under a general local rescaling of the form $\tilde{g}_{\mu\nu}(x)=\Omega(x)g_{\mu\nu}(x)$ with no reference to a coordinate transformation but we leave this important detail for section \ref{confFtransf}.} given by eq.~\eqref{defConf}.
Therefore, one may say that Weyl tensor does not ``see'' the conformal factor $\omega(x)$, it simply cancels out. It follows that a Lie derivative of the Weyl tensor along a vector generating conformal transformations vanishes \cite{YanoNag},
    \begin{equation}
    \label{LieConfWeyl}
        \text{if $\xi^{\mu}$ generates conformal trnasformations, then }\qquad \delta_{\xi}C^{\mu}{}_{\alpha\nu\beta}=\mathcal{L}_{\xi}C^{\mu}{}_{\alpha\nu\beta} = 0\ .
    \end{equation}
Then Weyl tensor calculated from $\tilde{g}_{\mu\nu}$ and Weyl tensor calculated from $g_{\mu\nu}$ are equal and one says that the two metrics are conformal to each other.
But since conformal transformations affect only the scale density according to eq.~\eqref{deltaAomega}, we can now make an educated guess that the Weyl tensor is completely determined actually only by the shape density defined above with eq.~\eqref{shapedef}.
We shall come back to this important remark in section~\ref{unimodDec}.

We note here without proof that the variation of $\gb_{\mu\nu}$ is traceless (cf. eq.~\eqref{vargbTrzero}), meaning that whatever coordinate transformation gives rise to a change in $\gb_{\mu\nu}$ it will be encoded in the traceless symmetric part described by eq.~\eqref{GenCtfTrless} and it would be something other than a conformal transformation.
Therefore, substituting eq.~\eqref{GenCtfTr} and eq.~\eqref{GenCtfTrless} into the variation $\delta_{\xi}\gb_{\mu\nu}$ based on eq.~\eqref{shapedef} one can finally deduce
    \begin{align}
    \label{varGbStless}
        A^{2}\delta_{\xi}\gb_{\mu\nu}&=2S_{\mu\nu}^{\srm T}\qquad\text{$=0$ for conformal transformations}\ ,\\[6pt]
    \label{varGbSTr}
        \frac{\delta_{\xi}A}{A}&=\frac{S}{d}\ .
    \end{align}
Since conformal transformations have $S_{\mu\nu}^{\srm T}=0$, clearly some non-conformal transformations describe the variation of the shape density and we come back to them in section~\ref{shearshape}.

\section{Conformal coordinate transformations in Minkowski spacetime}
\label{conftrM}

In Minkowski spacetime the metric tensor $\eta_{\mu\nu}$ does not contain any dependence on coordinates. According to eq.~\eqref{scaledef} and eq.~\eqref{shapedef} the Minkowski metric has constant scale and shape density, i.e. $A=1$ and $\gb_{\mu\nu}=\eta_{\mu\nu}$.
Since conformal transformations change only the scale density, the result will be a new scale density $A'(x)=1+\omega(x)$. It is our task now to find function $\omega(x)$.

In the literature on conformal field theories (see e.g. chapter 2 in \cite{BlumPlau}) 
one usually starts from Minkowski spacetime version of eq.~\eqref{defConf}, that is,
    \begin{equation}
        \del_{(\mu}\xi_{\nu)}=\omega\eta_{\mu\nu}
    \end{equation}
and interprets conformal transformations as ``the ones that leave the Minkowski metric invariant up to an overall function''. Then one proceeds to take a derivative of the above equation and seek a rather specific sum of terms with permuted indices that gives a useful equation relating the second derivatives of $\xi_{\mu}$ and a derivative of $\omega$,
    \begin{equation}
    \label{confTransf_eq1}
        \del_{\mu}\del_{\nu}\xi_{\alpha}=(\eta_{\mu\alpha}\del_{\nu}+\eta_{\nu\alpha}\del_{\mu}-\eta_{\mu\nu}\del_{\alpha})\omega
    \end{equation}
Furthermore, one finds a second-order equation for $\omega$,
    \begin{equation}
    \label{confTransf_eq2}
        (d-2)\del_{\mu}\del_{\nu}\omega +\eta_{\mu\nu}\Box\omega = 0
    \end{equation}
by using the freedom to contract index of the vector with any of the derivative indices acting on it because the Minkowski metric commutes with partial derivatives.
From the above equations one deduces $\Box\omega=0$ and therefore $\del_{\mu}\del_{\nu}\omega=0$ in $d>2$. Now, it is interesting that in the literature on conformal coordinate transformations in Minkowski spacetime one cannot find an explanation of why eqs.~\eqref{confTransf_eq1} and \eqref{confTransf_eq2} need to be sought by taking a very specific combination of derivative terms and why the need for taking another derivative.
But the reason why this specific combination of $\del_{\mu}\del_{\nu}\omega$ works is the same as the one used for deriving the non-isometry integrability condition (see appendix \ref{App_VarGamma}) encountered in eq.~\eqref{CT_2a}. Hence, instead of guessing the specific sum of index permutations of some expressions, simply demand that the variation of the Christoffel symbols has to be proportional to the second term in eq.~\eqref{varGamma}. (In quantum field theory on Minkowski background one is usuall not familiar with Christoffel symbols because one's attention is always on the Minkowski metric. Then it is expected that the only way to arrive at eq.~\eqref{confTransf_eq1} is to guess it as it is usually done, if one wants to avoid introducing geometrical concepts relevant to theories of gravity.)

In any case, one arrives at the conclusion that for\footnote{For $d=2$, term $\del_{\mu}\del_{\nu}\omega$ drops out from eq.~\eqref{confTransf_eq2} and one has that infinitely many coordinate transformations can conformally transform the metric.} $d>2$
    \begin{equation}
    \label{omega}
        \del_{\mu}\del_{\nu}\omega=0\qquad \Rightarrow \qquad \omega = \lambda + 2 b_{\mu}x^{\mu}\ ,
    \end{equation}
where $\lambda=const.$ and $b_{\mu}$ are covariant components of a arbitrary constant vector (or they can be thought of as components of a differential one-form); we chose the factor of 2 for mere convenience. Recalling that $\omega=\del_{\mu}\xi^{\mu}/d$, it follows that $\xi^{\mu}$ itself is at most quadratic (thus non-linear!) in coordinates,
    \begin{equation}
        \xi^{\mu}=\lambda + {c^{\mu}}_{\alpha\beta}x^{\alpha}x^{\beta}
    \end{equation}
where ${c^{\mu}}_{\alpha\beta}$ contains $b_{\mu}$ in some combination that can be deduced by plugging the above equation into eq.~\eqref{confTransf_eq1}, after which one finds
    \begin{align}
        {c^{\mu}}_{\alpha\beta} &= \delta^{\mu}_{\alpha}b_{\beta}+\delta^{\mu}_{\beta}b_{\alpha}-\eta_{\alpha\beta}b^{\mu}\ ,\\[6pt]
    \label{confKillingxi}
        \xi^{\mu}&=\lambda x^{\mu} + 2b_{\alpha}x^{\alpha}x^{\mu}-b^{\mu}x^2
    \end{align}
where $x^2=\eta_{\alpha\beta}x^{\alpha}x^{\beta}$. The above vector describes infinitesimal conformal coordinate transformations\footnote{In the literature it is referred to as \textit{conformal Killing vector}, but we reserve the attribute ``Killing'' only for isometry transformations.}. Note that one could freely add to this vector translations and Lorentz transformations $a^{\mu}+{m^{\mu}}_{\nu}x^{\nu}$, because the former is just a constant vector and the latter has an antisymmetric constant matrix such that eq.~\eqref{confTransf_eq1} is trivially satisfied for $\omega=0$, without contradiction. But we keep the focus on $\omega\neq 0$ transformations only.
We see that eq.~\eqref{confKillingxi} consists of $d+1$ constant parameters and thus there are $d+1$ independent vectors,
    \begin{align}
    \label{xiconfD}
        \xi_{\ssst D}^{\mu}& := \lambda x^{\mu}\ ,\\[6pt]
    \label{xiconfK}
        \xi^{\mu}_{\ssst K}& :=  2b_{\alpha}x^{\alpha}x^{\mu}-b^{\mu}x^2\ ,
    \end{align}
one for $\lambda$, which we see from eq.~\eqref{xilambda} describes dilations, and one for each component of $b^{\mu}$. The latter describes \textit{special conformal transformations}, which are non-linear.

\begin{figure}[h]
    \label{Fig_confCube}
        \includegraphics[width=\textwidth - 1cm]{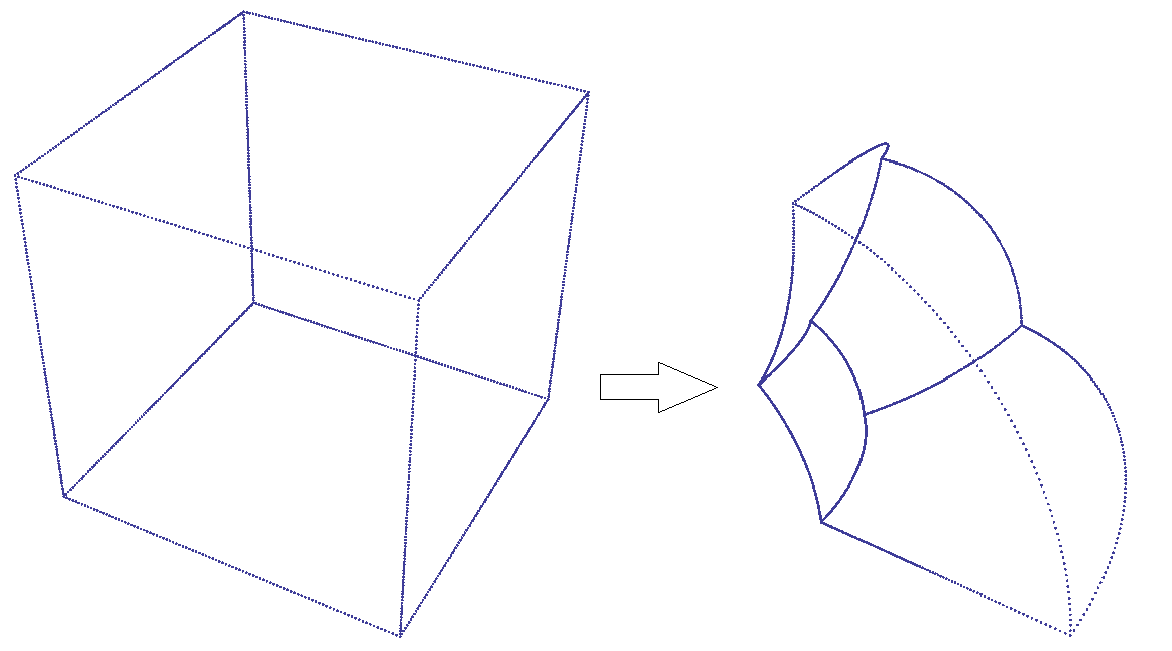}
        {\centering
        \caption[Conforally distorted cube]{\small A cube before (left) and after (right) a special coordinate transformation of the form given by eq.~\eqref{specConfTrX}
        with $b^{\mu}=(0.5,0,0)$. Note how the the edges stretch differently in every point but the right angles are preserved if measured infinitesimally close to the corresponding vertices. [generated in Wolfram Mathematica]}}
    \end{figure}

The special conformal transformations are most easily understood by inspecting their finite version. This is given by (see \cite{Wess1,Wess} for the conformal transformations and symmetry in the context of quantum field theory)
    \begin{equation}
    \label{specConfTrX}
        \tilde{x}^{\mu}=\frac{x^{\mu}-b^{\mu}x^2}{1-2b_{\mu}x^{\mu}+b^2 x^2}\ ,
    \end{equation}
from which one can deduce that the norm of the position vector gets rescaled at each point differently by a local conformal factor (compare with eq.~\eqref{DilatXX}),
    \begin{equation}
    \label{specConfXX}
        \tilde{x}^{2}=\Lambda^2(x)x^2\ ,\qquad \Lambda^2(x)=\frac{1}{1-2b_{\mu}x^{\mu}+b^2 x^2}\ .
    \end{equation}
One can now divide eq.~\eqref{specConfTrX} by the above norm and introduce new coordinates $y^{\mu}=x^{\mu}/x^2$, then eq.~\eqref{specConfTrX} reduces to
    \begin{equation}
    \label{spec_confTranl}
        \tilde{y}^{\mu}=y^{\mu} - b^{\mu}\ ,\qquad y^{\mu}=\frac{x^{\mu}}{x^2}\ ,\quad \tilde{y}^{\mu}=\frac{\tilde{x}^{\mu}}{\tilde{x}^{2}}
    \end{equation}
which is just a translation. If we pay attention to the order of introducing new coordinates, special conformal transformations are just a composition of an inversion, translation, and another inversion. The inversion is the part that makes it non-linear in the original coordinates. 
To get some (relatively!) intuitive picture of what special conformal transformations in 3D Euclidean space do, see Fig. \ref{Fig_confCube}.
    
Another way to understand this is to think of eq.~\eqref{spec_confTranl} as rules for translations which have to be employed if one would like to talk about a special conformal transformation in eq.~\eqref{specConfTrX} of a point at infinity: in order to render the result finite, one needs to divide this equation by another quantity that blows up, the norm of the position vector, which as a result lets one interpret eq.~\eqref{spec_confTranl} as translations of points near infinity. For this reason it is sometimes said that translations are dual to special cofnormal transformations (and vice versa), as well as that the point around at the origin (of a chosen coordinate chart) is dual to the point at infinity\footnote{These are notions that have a rigorous definition and clear geometrical meaning in \textit{projective geometry}, (see \cite{DelpProj} for insightful exposition on relationship with special relativity), into which we do not go into in this thesis.}.
It thus is expected that translations and special conformal transformations have the same number of generators. 
These are well-known results.

The example of special conformal transformations is very illuminating for understanding the meaning of transversal and longitudinal components of the transformation vectors. What is the character of $b^{\mu}$ vector --- transversal or longitudinal? We now have to distinguish between the infinitesimal and finite transformations because the answer will be different. For non-small $b_{\mu}$ we have
    \begin{equation}
        b_{\mu}=b_{\mu}^{\ssst \bot}+b_{\mu}^{\ssst \parallel}\ ,\qquad b_{\mu}x^{\mu}=b_{\mu}^{\ssst \parallel}x^{\mu}\ ,\qquad b^2=b_{\mu}^{\ssst \bot}b_{\ssst \bot}^{\mu}+b_{\mu}^{\ssst \parallel}b_{\ssst \parallel}^{\mu}
    \end{equation}
and we see from eq.~\eqref{specConfXX} that both longitudinal and transversal components of $b^{\mu}$ contribute to the change of the length. 
One can interpret this as: a finite special conformal transformation acts to change the length of a straight distance not only by simple rescaling
along the line but also by \textit{bending the line} in the direction orthogonal to it. This is precisely what one sees with a coarse grid that we used as an example in the previous section to illustrate a conformal transformation --- a square deforms in such a way that its angles are
preserved but its edges are bent and deformed differently on different parts of the map. But if one used a finer grid, having smaller (say, infinitesimally small) squares, an infinitesimal square stays an infinitesimal square, but slightly larger or smaller, which is described by $\delta_{\xi}\eta_{\mu\nu}=2(\lambda+2b_{\mu}x^{\mu})$, based on eq.~\eqref{omega}. The longitudinal and transversal components of the transformation vector are
    \begin{align}
    \label{xiConfLong}
        \xi^{\mu}_{\ssst\parallel} & =
        \lambda x^{\mu}
        +2b^{\ssst\parallel}_{\alpha}x^{\alpha}x^{\mu}
        -b_{\ssst\parallel}^{\mu}x^2\ ,\\[6pt]
    \label{xiConfBot}
        \xi^{\mu}_{\ssst \bot} & = 
        -b_{\ssst\bot}^{\mu}x^2\ ,
    \end{align}
from which we can see that $b_{\ssst \bot}^{\mu}$ \textit{does not contribute} to the infinitesimal change of length along the given direction $x^{\mu}$. (The longitudinal component of the conformal transformation vector is \textit{shared} by dilations and special conformal transformations, while the transversal is determined only by the special conformal transformation.)
One can visualize this in three dimensions\footnote{Which is actually more valid than the two-dimenisonal case because the results derived are valid in $d>2$. However, one could do some naive counting of components: in 2D special conformal transformation vector has 2 parameters; these are distributed as one component to each of the longitudinal and transversal parts, and one can think of them as one component in the given direction to change the length while the other in the direction orthogonal to it to bend the line.} in spherical coordinates in the following way. 
For the given direction choose the radial direction. Then $b_{\ssst\bot}^{i},\ i=1,2,3$ at a point along the radial direction is tangential to a 2-sphere drawn through that point. Thus $b_{\ssst\bot}^{i}=(0,b_{\ssst\bot}^{\theta},b_{\ssst\bot}^{\phi})$. This explanation fits quite nicely to the spirit of \cite{Pres02}.
Hence we can say that the role of changing the length of vectors is played by the longitudinal component $\xi^{\mu}_{\ssst\parallel}$, which confirms the expectations we made in the previous section.

Much more could be said and explored about conformal coordinate transformations, but we wanted merely to offer an intuitive explanation of what the consequences of their action on the components of the metric are.
One could talk about consequences such as conserved currents, but that is not of main interest here. Instead, we turn to shear transformations.

\section{Shear (volume-preserving) coordinate transformations}
\label{shearshape}

Let us now make sense of $S_{\mu\nu}^{\srm T}$, the ``complement''\footnote{In the next section we clarify why is this word under a quote.} of conformal transformations. These transformations are described by a traceless matrix of $\frac{d(d+1)}{2}-1$ \textit{coordinate-dependent} entries. However the problem with volume-preserving transformations is that there are infinitely many generating vectors
$\xi^{\mu}$, because, as we shall see, one is not able to determine these vectors from the procedure that we used to obtain conformal transformations.
There is nothing inconsistent about this fact --- there are infinitely many coordinate transformations and indeed it is expected that if one finds only five of them in conformal transformations, the complement set has infinite number of them.
However, in spite of this fact, we can focus on some special cases which will illuminate the nature of shear transformations as a special case of volume-preserving transformations. This section is motivated by \cite{Kaca1,ShearHehl} and discussions with Ka\'ca Bradonji\'c \cite{KacaB}.

We start with a question, what is $\xi^{\mu}$ such that the Minkowski metric transforms according to
    \begin{equation}
        S=0 \quad\Rightarrow\quad \delta_{\xi}\eta_{\mu\nu}=2S_{\mu\nu}^{\srm T}\ ?
    \end{equation}
The answer is sought in the same way as in the case of conformal transformations: in addition to the above, one demands that Christoffel symbols vary as the first term in eq.~\eqref{varGamma}.
(Note that, again, there is no need to guess the necessary sum of index permutations because a clear geometric statement eq.~\eqref{varGamma} is available.)
For Minkowski metric this gives the following equation for $\xi^{\mu}$
    \begin{equation}
    \label{ddxishear}
        \del_{\mu}\del_{\nu}\xi_{\alpha}=
            \del_{\mu}S_{\alpha\nu}^{\srm T}
            +\del_{\nu}S_{\alpha\mu}^{\srm T}
            -\del_{\alpha}S_{\mu\nu}^{\srm T}\ .
    \end{equation}
Taking a derivative $\del_{\beta}$, then taking two different traces, over $\mu\nu$ and over $\alpha\beta$ indices and using $\del_{\alpha}\xi^{\alpha}=0$ for the definition of the volume-preserving transformations, one obtains
    \begin{align}
    \label{shear1}
        2\del^{\alpha}\del_{(\mu}S_{\nu)\alpha}^{\srm T }-\Box S_{\mu\nu}^{\srm T}=0\ ,\\[6pt]
    \label{shear2}
        \del^{\alpha}\del^{\beta}S_{\alpha\beta}^{\srm T}=0\ .
    \end{align}
Equation \eqref{shear1} is analogue of eq.~\eqref{confTransf_eq2}. So let us take its trace; but as a result one trivially finds eq.~\eqref{shear2}, which cannot be used in eq.~\eqref{shear1} anyhow. This is in drastic contrast the case of conformal transformations: there the trace of eq.~\eqref{confTransf_eq2} provided
some new information which could be used back into that same equation, allowing one to derive eq.~\eqref{omega}. But in the case of volume-preserving transformations this is not the case, since one could have any transversal vector $\xi^{\mu}_{\ssst \bot}$ giving rise to such transformations.

Therefore, one can conclude that while there are only $d$ independent ways of changing only the volume, i.e. the scale $A$, there are infinitely many ways of changing only the shape, i.e. $\gb_{\mu\nu}$. This makes sense, because there are infinitely many general coordinate transformations with respect to which the total metric tensor is covariant.

In spite of this, we can look at some special cases in order to obtain some intuition about these transformations. The first guess is that $S_{\mu\nu}^{\srm T}$ is a matrix of constant parameters, which surely satisfies the above conditions.
This is analogous to Lorentz transformations, where one encounters an antisymmetric matrix $m_{\mu\nu}$ of constant parameters. With the assumption $S_{\mu\nu}^{\srm T}=s_{\mu\nu}$ and remembering that this matrix is traceless, we have
    \begin{equation}
    \label{xishearS}
        \xi^{\mu}_{S}={s^{\mu}}_{\nu}x^{\nu}\ .
    \end{equation}
Since this vector is linear in $x^{\mu}$, we can think of the dilation parameter $\lambda$ as the missing trace piece that complements ${s^{\mu}}_{\nu}$ to form a symmetric matrix of constant parameters. If we added translations and Lorentz transformations, one would obtain a matrix
with symmetric traceless, antisymmetric and trace parts amounting in total to $d^2$ parameters plus $d$ parameters of translations. In four dimensions this is 20 parameters and one has described all possible choices for linear transformations. Thus, it makes sense to consider the matrix of constant elements.

Another example is to impose the following condition,
    \begin{equation}
        \del^{\beta}S_{\alpha\beta}^{\srm T}\sim c_{\alpha}
    \end{equation}
where $c_{\alpha}$ is a constant vector and any proportionality constant is irrelevant and can be absorbed. Then we can show that $S_{\mu\nu}^{\srm T}$ takes the form
    \begin{equation}
    \label{specShearS}
        S_{\mu\nu}^{\srm T}=c_{\mu}\eta_{\nu\alpha}x^{\alpha}+c_{\nu}\eta_{\mu\alpha}x^{\alpha}-\frac{2}{d}\eta_{\mu\nu}c_{\alpha}x^{\alpha}\ ,
    \end{equation}
plus the already introduced matrix constant $s_{\mu\nu}$ which we do not include here. From eq.~\eqref{ddxishear} one then derives
    \begin{equation}
    \label{xishearQ}
        \xi^{\mu}_{Q}:=c^{\mu}x^2-\frac{2}{d}c_{\alpha}x^{\alpha}x^{\mu}+\frac{1}{d}c^{\mu}x^2\ ,
    \end{equation}
which we may call the generating vector of \textit{special shear transformations}, because of its similarity with the special conformal vector given by eq.~\eqref{xiconfK}.
An interesting feature of these transformations is observed when we recall the analysis in eqs.~\eqref{xiConfLong}-\eqref{xiConfBot}. Namely, vector $b^{\mu}$ describing the special conformal transformations has all $d$ components since both longitudinal and transversal components contribute. But special shear transformations given by eq.~\eqref{specShearS} may equally be described by
    \begin{equation}
        S_{\mu\nu}^{\srm T}=c_{\mu}\eta_{\nu\alpha}x^{\alpha}+c_{\nu}\eta_{\mu\alpha}x^{\alpha}\quad\Rightarrow\quad
        \xi^{\mu}_{Q}=c^{\mu}_{\bot}x^{2}\ ,\qquad c_{\mu}^{\bot}x^{\mu}=0\ ,
    \end{equation}
in which case one does not need to worry about dimensional dependence. If described in such a way, special shear transformations are manifestly transversal and are determined by $d-1$ parameters only.

Independently of the fact that volume-preserving transformations can be represented in infinitely many ways, they are all \textit{volume-preserving} and this fact is valid in a general space and dimension, by definition
    \begin{equation}
    \label{delgbarShape}
        A^2\delta_{\xi}\gb_{\mu\nu}=2S_{\mu\nu}^{\srm T}\ ,\quad \delta_{\xi}A = 0\ ,
    \end{equation}
based on eq.~\eqref{shapedef}. Thus, enough evidence is gathered for motivating the split of the metric tensor into scale and shape density. But shape and scale density can be recognized by other transformations than coordinate transformations, see section~\ref{confFtransf}.

\section{Generators of coordinate transformations and their algebra}
\label{groupalg}

In this brief section we talk about group-theoretical observations regarding the general coordinate transformations and their subsets that we mentioned in the previous sections. 

In group-theoretical language one says that in $d$ dimensions matrices given by \eqref{transfmatrix} have $d^2$ independent real elements and they, together with an operation of multiplication among them, form the \textit{general linear group} over real numbers\footnote{Meaning that elements of a matrix are real numbers.} denoted by $GL(d,\mathbb{R})$. In $d=4$ dimensions each matrix has 16 independent elements and we have $GL(4,\mathbb{R})$. So far we have looked at conformal transformations (sections \ref{conftr}-\ref{conftrM}) and shear transformations (section \ref{shearshape}) and only in passing we mentioned translations and Lorentz transformations. We have found their explicit form only in Minkowski spacetime, see eqs.~\eqref{xiconfD}, \eqref{xiconfK} and \eqref{xishearS}, with an important remark that there are infinitely many shear transformations of which one example was given by eq.~\eqref{xishearQ}. In addition to the stated, we have translations and rotations given by
    \begin{align}
    \label{xiLP}
        \xi_{\ssst P}^{\mu}&=a^{\mu}\ ,\\[6pt]
    \label{xiLL}
        \xi_{\ssst L}^{\mu}&=m^{\mu}{}_{\nu}x^{\nu}\ .
    \end{align}
There are two questions to be asked. How do all these transformations fit into $GL(d,\mathbb{R})$ and how do they generalize to curved spaces?

The answer to the first question was given in a remarkable paper by Ogievetsky \cite{Ogi1} (see also \cite{Ogi2} for an important application of this result) as follows and we shall focus now on the relevant case of $d=4$ dimensions. First of all, since matrix $A^{\mu}{}_{\nu}$ can be expanded around the identity, i.e.
    \begin{equation}
    \label{infmatA}
        \ta^{\mu}{}_{\nu}\approx\delta^{\mu}_{\nu}+\del_{\nu}\xi^{\mu}\ ,
    \end{equation}
where $\del_{\nu}\xi^{\mu}$ contains small real parameters, we are in the realm of Lie groups. We shall need the notion of those matrices $\ta^{\mu}{}_{\nu}$ that have unit determinant. Since determinant of the matrix in eq.~\eqref{infmatA} is roughly a fourth power of
its RHS then $ \text{J}\equiv\det \ta^{\mu}{}_{\nu} \approx 1 + \del_{\mu}\xi^{\mu}$ (this is just the Jacobian matrix determinant), so a unit determinant requires $\del_{\mu}\xi^{\mu}=0$ (we shall soon see that this is indeed related to volume-preserving transformations). Then we can decompose the transformation matrix $\ta^{\mu}{}_{\nu}$ into its determinant \text{J} and unit-determinant\footnote{If an object has a unit determinant it is often referred to as ``unimodular''.} $\bar{\ta}^{\mu}{}_{\nu}$ parts,
    \begin{equation}
    \label{Adec}
        \ta^{\mu}{}_{\nu}=\text{J}^{\frac{1}{4}}\bar{\ta}^{\mu}{}_{\nu}\ ,\qquad \bar{\ta}^{\mu}{}_{\nu}:=\text{J}^{-\frac{1}{4}}\ta^{\mu}{}_{\nu}\ .
    \end{equation}
Now, in Lie groups one deals with generators of the corresponding transformations and their algebra (i.e. commutation relations).
The generators of translations $P_{\mu}$, rotations $L_{\mu\nu}$, shears $S_{\mu\nu}$, dilations $D$ and special conformal $K_{\mu}$ transformations can be represented as differential operators as \cite{Ogi1}
    \begin{align}
    \label{poingen}
        L_{\mu\nu} &=-i\left(\eta_{\mu\alpha}x^{\alpha}\del_{\nu}-\eta_{\nu\alpha}x^{\alpha}\del_{\mu}\right)\ ,\qquad P_{\mu} = -i\del_{\mu}\\[6pt]
    \label{sheargen}
        S_{\mu\nu}&=-i\left(\frac{1}{2}\left(\eta_{\mu\alpha}x^{\alpha}\del_{\nu}+\eta_{\nu\alpha}x^{\alpha}\del_{\mu}\right)-\frac{1}{4}\eta_{\mu\nu}x^{\alpha}\del_{\alpha}\right)\ ,\\[6pt]
    \label{confgen}
       D&=-i x^{\mu}\del_{\mu}\ ,\qquad K_{\mu}=-i \left(2\eta_{\mu\alpha}x^{\alpha}x^{\beta}\del_{\beta}-x^{2}\del_{\mu}\right)\ .
    \end{align}
If we act on $x^{\rho}$ with these generators contracted by the correscponding parameters we recover the corresponding coordinate transformations.
This might look complicated but the matter is much simpler than it seems. Namely, as it was shown by  Ogievetsky~\cite{Ogi1}, it can be recognized that the sum of Lorentz and shear transformations is just the sum of the antisymmetric and symmetric traceless parts of the following generator
\footnote{Factor $1/2$ is added because of the definition of antisymmetrization on the Lorentz transformations piece. However, the following definition is independent on how are such factors distributed among the generators.}
    \begin{equation}
    \label{Lbargen}
        \frac{1}{2}L_{\mu\nu}+S_{\mu\nu}=\bar{L}_{\mu\nu}:= -i\left(\eta_{\mu\alpha}x^{\alpha}\del_{\nu}-\frac{1}{4}\eta_{\mu\nu}x^{\alpha}\del_{\alpha}\right)\ ,
    \end{equation}
which is a generator of the \textit{special linear group} $SL(4,\mathbb{R})$. This is a group of all matrices $\ta^{\mu}{}_{\nu}$ with a unit determinant and is a \textit{subgroup} of $GL(4,\mathbb{R})$. For its infinitesimal version in eq.~\eqref{infmatA} this means
    \begin{equation}
    \label{unitdetcond}
        \del_{\mu}\xi^{\mu} = 0\ ,
    \end{equation}
which just means that matrices of $SL(4,\mathbb{R})$ are described by $\bar{\ta}^{\mu}{}_{\nu}$, i.e. the unimodular piece of eq.~\eqref{Adec}.
But now recall the split of $\del_{\mu}\xi_{\nu}$ into antisymmetric, symmetric traceless and trace parts, i.e. the Minkowski spacetime version of eqs.~\eqref{GenCtAll}-\eqref{GenCtfTrless}: eq.~\eqref{unitdetcond} is nothing other than $S=0$, which is the requirement for excluding conformal transformations, leaving us with \textit{volume-preserving} transformations. The generators of $SL(4,\mathbb{R})$ obey the following algebra
    \begin{align}
    \label{algLL}
        \Bigsq{
        L_{\mu\nu},L_{\alpha\beta} }&=-2i\left(\eta_{[\mu\vert\alpha}L_{\vert\nu]\beta}-\eta_{[\mu\vert\beta}L_{\vert\nu]\alpha}\right)\ ,\\[6pt]
    \label{algLS}
        \Bigsq{
        L_{\mu\nu},S_{\alpha\beta} }&=-2i\left(\eta_{[\mu\vert\alpha}S_{\vert\nu]\beta}+\eta_{[\mu\vert\beta}S_{\vert\nu]\alpha}\right)\ ,\\[6pt]
    \label{algSS}
        \Bigsq{
        S_{\mu\nu},S_{\alpha\beta} }&=-2i\left(\eta_{(\mu\vert\alpha}L_{\vert\nu]\beta}+\eta_{[\mu\vert\beta}L_{\vert\nu)\alpha}\right)\ .
    \end{align}
Now observe what happens if we add dilations to the generator of shear transformations,
    \begin{equation}
        S_{\mu\nu}+\frac{1}{4}\eta_{\mu\nu}D= D_{\mu\nu}:=-i\eta_{(\mu\vert\alpha}x^{\alpha}\del_{\vert\nu)}\ ,
    \end{equation}
and check the algebra with Lorentz transformations: the algebra is identical to eq.~\eqref{algLS} because dilations commute with Lorentz transformations
    \begin{equation}
        \Bigsq{L_{\mu\nu},D}=0\ .
    \end{equation}
Furthermore, translations have the following algebra with the mentioned generators:
    \begin{align}
    \label{poinc}
        \Bigsq{P_{\mu},L_{\alpha\beta}}&=-i\left(\eta_{\mu\alpha}P_{\beta}-\eta_{\mu\beta}P_{\alpha}\right)\ ,\\[6pt]
    \label{PDalg}
        \Bigsq{P_{\mu},D_{\alpha\beta}}&=-i\left(\eta_{\mu\alpha}P_{\beta}+\eta_{\mu\beta}P_{\alpha}\right)\ .
    \end{align}
while they obviously commute with itself. 
This means that the algebra of $L_{\mu\nu},D_{\mu\nu}$ and $P_{\mu}$ closes and they all form \textit{the linear realization of the rigid affine group} $A(4,\mathbb{R})$ which is a semidirect product of translation group and the linear group,
$A(4,\mathbb{R})=\mathbb{R}^{4}\rtimes GL(4,\mathbb{R})$, i.e. the group of transformations which acts \textit{on coordinates linearly}\footnote{A remark on wording is of use here: even those transformations which act non-linearly on coordinates
are transformations which act linearly on vectors and these are precisely the matrices $A^{\mu}{}_{\nu}$ of $GL(4,\mathbb{R})$. In a linear realization of the rigid affine group we have only linear coordinate transformations, meaning that the coefficients $a^{\mu}{}_{\nu}$ and $a^{\mu}$ below are constant.}
    \begin{equation}
    \label{linCS}
        \tilde{x}^{\mu}=a^{\mu}{}_{\nu}x^{\nu}+a^{\mu}\ ,
    \end{equation}
where $a^{\mu}{}_{\nu}=m^{\mu}{}_{\nu}+s^{\mu}{}_{\nu}+\lambda \delta^{\mu}_{\nu} $ are 16 constant parameters consisting of Lorentz transformations given by eq.~\eqref{xiLL}, shear transformations given by eq.~\eqref{xishearS} and dilations given by eq.~\eqref{xiconfD}, in addition to four
translations given by eq.~\eqref{xiLP}. Therefore, not only that the Poincar\'e group in eq.~\eqref{poinc} is a subgroup of $A(4,\mathbb{R})$, but dilations and the linear realizaton of $SL(4,\mathbb{R})$ are also subgroups of $A(4,\mathbb{R})$.

What about the special conformal transformations given by eq.~\eqref{xiconfK} and their generators in eq.~\eqref{confgen}? Their algebra with other transformations is given by
    \begin{align}
    \label{algKP}
        \Bigsq{K_{\mu},P_{\nu}}&=-2i\left(\eta_{\mu\nu}D-L_{\mu\nu}\right)\ ,\\[6pt]
    \label{algKL}
        \Bigsq{K_{\mu},L_{\alpha\beta}}&=i\left(\eta_{\mu\alpha}K_{\beta}-\eta_{\mu\beta}K_{\alpha}\right)\ ,\\[6pt]
    \label{algKD}
        \Bigsq{K_{\mu},D}&=-i K_{\mu}\ ,\qquad \Bigsq{D,D}=0\ ,\\[6pt]
    \label{algKSD}
        \Bigsq{K_{\mu},D_{\alpha\beta}}&=-i\left(\eta_{\mu\alpha}K_{\beta}+\eta_{\mu\beta}K_{\alpha}\right)-2i \eta_{\mu\rho}\eta_{\alpha\sigma}x^{\rho}x^{\sigma}\del_{\beta}\ .
    \end{align}
We see from eqs.~\eqref{algKP}-\eqref{algKD} and eqs.~\eqref{poinc}-\eqref{PDalg} that the algebra of Poincare, dilations and special conformal group of transformations closes into algebra of \textit{conformal group} $C(4,\mathbb{R})$ that has 15 generators in total. However, from eq.~\eqref{algKSD} we see something odd: there is a piece $x^{\rho}x^{\sigma}\del_{\beta}$ which does not belong to any of the so far found
generators (see Table \ref{TabAlg}). Therefore, if one takes into account the algebra of special linear group $SL(4,\mathbb{R})$ and conformal group $C(4,\mathbb{R})$ and demands their \textit{closure}, one can produces new kinds of generators and this was Ogievetsky's
main observation. 
    \begin{table}[ht]
    \renewcommand{\arraystretch}{1.8}
    \renewcommand{\tabcolsep}{12pt}
    \centering
    \begin{minipage}{12cm}
        \begin{tabular}{ccccccccccccc}
            \textbf{}                       & \textbf{P}                      & \textbf{L}                      & \textbf{S}                      & \textbf{D}                      & \textbf{K}               &                       & $\bot$                 & $\parallel$            &  &  &  &  \\ \cline{2-6} \cline{8-9}
            \multicolumn{1}{c|}{\textbf{P}} & \multicolumn{1}{c|}{0}          & \multicolumn{1}{c|}{P}          & \multicolumn{1}{c|}{P}          & \multicolumn{1}{c|}{P}          & \multicolumn{1}{c|}{D+L} & \multicolumn{1}{c|}{} & \multicolumn{1}{c|}{+} & \multicolumn{1}{c|}{+} &  &  &  &  \\ \cline{2-6} \cline{8-9}
                                            & \multicolumn{1}{c|}{\textbf{L}} & \multicolumn{1}{c|}{L}          & \multicolumn{1}{c|}{S}        & \multicolumn{1}{c|}{0}          & \multicolumn{1}{c|}{K}   & \multicolumn{1}{c|}{} & \multicolumn{1}{c|}{+} & \multicolumn{1}{c|}{-} &  &  &  &  \\ \cline{3-6} \cline{8-9}
                                            &                                 & \multicolumn{1}{c|}{\textbf{S}} & \multicolumn{1}{c|}{L}          & \multicolumn{1}{c|}{0}          & \multicolumn{1}{c|}{K+$xx\del$}   & \multicolumn{1}{c|}{} & \multicolumn{1}{c|}{+} & \multicolumn{1}{c|}{-} &  &  &  &  \\ \cline{4-6} \cline{8-9}
                                            &                                 &                                 & \multicolumn{1}{c|}{\textbf{D}} & \multicolumn{1}{c|}{0}          & \multicolumn{1}{c|}{K}   & \multicolumn{1}{c|}{} & \multicolumn{1}{c|}{-} & \multicolumn{1}{c|}{+} &  &  &  &  \\ \cline{5-6} \cline{8-9}
                                            &                                 &                                 &                                 & \multicolumn{1}{c|}{\textbf{K}} & \multicolumn{1}{c|}{0}   & \multicolumn{1}{c|}{} & \multicolumn{1}{c|}{+} & \multicolumn{1}{c|}{+} &  &  &  &  \\ \cline{6-6} \cline{8-9}
        \end{tabular}
        {\centering
            \caption[Lie algebra]{\label{TabAlg}\small
            \textit{Left}: Schematic representation of the Lie algebra given by eqs.~\eqref{algLL} - \eqref{algSS}, \eqref{poinc}, \eqref{PDalg}, \eqref{algKP} - \eqref{algKSD} of generators of translations (\textbf{P}), Lorentz transformations (\textbf{L}), linear shear transformations (\textbf{S}), dilations (\textbf{D}) and special conformal transformations (\textbf{K}). Note how \textbf{K}-\textbf{S} commutator extends the algebra of conformal group and shear group to include more general second order generators. \textit{Right}: Presence (+) and absence (-) of transversal $\bot$ and longitudinal $\parallel$ components of the vector corresponding to each generator.}}
    \end{minipage}
    \end{table}
This means that if one takes $x^{\rho}x^{\sigma}\del_{\beta}$ as the generator of some transformations and commutes it with the generator, say, $K_{\mu}$, one obtains a generator proportional to $x^{\alpha}x^{\rho}x^{\sigma}\del_{\beta}$.
He showed by mathematical induction that if one continues with such a procedure one can write \textit{any $n$-th order generator of general covariance group $GL(4,\mathbb{R})$}
    \begin{equation}
    \label{genGen}
        ^{n}Lg^{n_{0},n_{1},n_{2},n_{3}}_{\mu}=-i(x^{0})^{n_{0}}(x^{1})^{n_{1}}(x^{2})^{n_{2}}(x^{3})^{n_{3}}\del_{\mu}\ ,
    \end{equation}
where $n=n_{0}+n_{1}+n_{2}+n_{3}$ is the sum of non-negative integers and denotes the order of non-linearity, i.e. the total power of $x^{\mu}$,
\textit{as a linear combination of commutators of the generators of the special linear $SL(4,\mathbb{R})$ and conformal $C(4,\mathbb{R})$ groups}.
Therefore, \textit{all} coordinate transformations described by eq.~\eqref{transfmatrix} and all kinds of \textit{motions} which we interpreted as active coordinate transformations can be constructed from the generators of the linear coordinate
transformations given by eq.~\eqref{linCS} and the generators of conformal transformations given by eq.~\eqref{algKD}. There are of course infinitely many ways to construct generators in given by eq.~\eqref{genGen} which is expected because there are infinitely many coordinate
transformations at one's disposal to represent physical objects in. But conformal transformations (which contribute only to the scale and volume variation expressed with eq.~\eqref{varGbSTr}) have only 5 parameters, so conformal group is not the place to look for this freedom.
We have already caught a glimpse of the freedom that is ``missing''  --- in ``special shear transformations'' given by eq.~\eqref{xishearQ} in section \ref{shearshape}. Namely, that was only one ``guessed'' example of, as it was stated there, infinitely many volume-preserving transformations. One can see that term
$x^{\rho}x^{\sigma}\del_{\beta}$ that is produced in eq.~\eqref{algKSD} can be related to a part of eq.~\eqref{xishearQ}. It can also be checked that the commutator of $\xi^{\mu}_{\ssst Q}\del_{\mu}$ with $K_{\mu}$ gives terms of third order in
$x^{\mu}$. This example agrees with Ogievetsky's results and therefore we conclude that the infinite freedom is found in the special linear group $SL(4,\mathbb{R})$ represented by non-linear volume-preserving transformations.
This answers our first question.

The second question was how does one generalize these transformations to curved spaces? The problem is that this depends on the metric. Even in flat spacetime, in coordinates other than Cartesian, things become more complicated because the Christoffel symbols no longer vanish. In curved spacetimes, in addition, the
curvatures do not vanish. Therefore one needs to solve equations \eqref{varGamma}, which can be tricky. However, one can still talk about the special linear and conformal groups \textit{locally}. Then one takes the affine group and promotes the
20 constant parameters to functions of coordinates and demands that the matter action in question is invariant under such \textit{local} transformations, leading to various variations of a gauge theory of gravity~\cite{Blag,HMN,HB}. But this goes beyond the scope of this thesis.

The important thing to take away from this section is the understanding of the definitions of the scale density and shape density given by eq.~\eqref{scaledef} and eq.~\eqref{shapedef}, respectively, in terms of the groups we mentioned here.
Namely, metric can be split into irreducible components with respect to the conformal group $C(4,\mathbb{R})$ or with respect to the special linear group $SL(4,\mathbb{R})$; the result is the same, that is, the scale density $A$ is defined up to a conformal transformation and shape density $\gb_{\mu\nu}$ is defined up to a
volume-preserving transformation. 
Perhaps it is illustrative to collect the results in the following,
    \begin{align}
        GL(4,\mathbb{R})&=C(4,\mathbb{R})\ltimes SL(4,\mathbb{R}) \ ,\\[6pt]
    \label{Amatdec}
        \ta^{\mu}{}_{\nu}&=\text{J}^{\frac{1}{4}}\bar{\ta}^{\mu}{}_{\nu} \ ,\\[6pt]
        g_{\mu\nu}&=A^{2}\gb_{\mu\nu}\ .
    \end{align}
Then conformal transformations of $C(4,\mathbb{R})$ are characterized by $\bar{\ta}^{\mu}{}_{\nu}=\delta^{\mu}_{\nu}$, while volume-preserving transformations of $SL(4,\mathbb{R})$ are characterized by $\text{J}=1$.
In the following chapter we shall see that this decomposition can be motivated by means other than with respect to coordinate transformations.

\section{Final remarks}

This chapter offers one way of motivating the separation of the metric into the scale and shape densities: by examining the subsets of general coordinate transformations and how their action affects the metric in several nonequivalent ways.
Our aim was to carefully describe the details around the meaning of the conformal coordinate transformations, because in the next chapter we contrast them with another kind of ``conformal'' transformation and we shall then be able to define clearly which kind of ``conformal transformation'' is the important one in this thesis and why.

It is author's hope that this chapter also has a pedagogical value, because the way that the $GL(d,\mathbb{R})$ group and the algebra of its subgroups are represented motivates the introduction of the scale and shape density parts of the metric, which may be understood in terms of shape-preserving and volume-preserving coordinate transformations, respectively. 
Their introduction was achieved with the aim of painting an intuitive picture with the help of transversal and longitudinal parts of the generating vector, while still smoothly wrapping these concepts into the language of group theory and thereby offering an invitation to a more rigorous considerations if one would like to pursue so further.
This makes the material of the current chapter suitable for those who would otherwise be discouraged from pursuing the mentioned concepts starting from the more abstract mathematics necessary to define them.
It is thus author's opinion that it can provide a good starting point for conceiving a complementary material for a course of General Relativity that seems not to have been encountered in standard textbooks about the topic. 

{\centering \hfill $\infty\quad$\showclock{0}{25}$\quad\infty$ \hfill}

\chapter[Conformal field transformation and unimodular-conformal decomposition][Conformal field transformation and...]{Conformal field transformation and unimodular-conformal decomposition}
\label{ch:umtocf1}
    Unlike coordinate transformations, we can also perform transformations directly on fields. Of main interest in this thesis is not a conformal coordinate transformation but a \textit{conformal field transformation}, also known more precisely as \textit{Weyl transformation} or \textit{Weyl rescaling}. We shall show in this chapter that decomposition of the metric with respect to this transformation leads to an
equivalent definition of the scale density and the shape density that we met in the previous chapter, thus allowing us to investigate conformal properties of a theory in a much more general sense in terms of the scale density $A$, without any reference to any specific kind of conformal transformations --- coordinate or field one.
Of particular importance will be the identification of a \textit{physical length scale} solely through the scale density $A$, which will allow one to keep track of the dimensions of all fields simply by keeping track of the scale density.
We also dive into more detail by looking at the implications of such a decomposition for curvature tensors and apply the decomposition to the 3+1 formalism, thus setting the grounds for the material in the following chapters.
The power of unimodular-conformal decomposition will be demonstrated on the example of a non-minimally coupled scalar field with an arbitrary potential. This also serves as a motivation for reformulating the notion of conformal invariance with respect to the scale density, which is an invitation for the upcoming chapter.

\newpage

\section{Conformal field transformation and a local change of length scale}
\label{confFtransf}

A conformal \textit{field} transformation, also known more precisely as \textit{local Weyl rescaling} or simply \textit{Weyl transformation}, consists of transforming the metric tensor and any field by multiplying them with an arbitrary function of coordinates ($\Omega(x)>0$) to some power, without any reference to a coordinate transformations. It is given by
    \begin{equation}
    \label{eqn:conftrans}
        g_{\mu\nu}(x)\quad\rightarrow\quad \tilde{g}_{\mu\nu}(x) = \Omega^2 (x)g_{\mu\nu}(x)\,, \qquad \phi_{\ssst I}(x)\quad\rightarrow\quad \tilde{\phi}_{\ssst I}=\Omega^{n_{\ssst I}} (x)\phi_{\ssst I}(x)\,,
    \end{equation}
where $n_{\ssst I}$ is usually called ``conformal weight'' of any kind of field (scalar, vector, spinor...) labeled by index $I$ that transforms homogeneously under this transformation.
There are also fields which transform inhomogeneously under conformal transformation. An example is the extrinsic
curvature, Christoffel symbols, Weyl gauge vector. Where necessary, we shall generalize the above definition to such a field, but for scalar fields and some vector fields this definition is enough.

What is the meaning of such a transformation? The meaning can be understood immediately if we make an analogy with dilations, described by eq.~\eqref{Xdilat}. Namely, with respect to the space of coordinates $x^{\mu}$, $\lambda$ is a constant. 
In analogy, we can talk about \textit{configuration space} --- the space of all components of fields in consideration --- and introduce an operation that multiplies each field with a constant
with respect to configuration space, but not constant with respect to spacetime. So if we introduce fields defined on spacetime, then the action behind dilations and all other coordinate transformations are logically extended --- they can now act on fields.
Metric transformations in eq.~\eqref{confFtransf} relate two sets of metric tensor components that describe two \textit{different} geometries --- Riemann tensor ``sees'' the difference between the two metrics, but Weyl tensor does not because Weyl tensor is
conformally invariant (for similar reasons as in eq.~\eqref{LieConfWeyl}, see section~\ref{unimodDec}). One says that two metrics are \textit{conformal to each other} (i.e. belong to the same conformal \textit{class}) if they are related by eq.~\eqref{confFtransf}, except that this conformal correspondence is not generated by coordinate transformations.
To be more precise, the line element itself is transformed under this transformation:
    \begin{equation}
        \d s^2(x)\quad\rightarrow\quad \d \tilde{s}^2(x) = \Omega^{2}(x)\d s^2(x)\ ,
    \end{equation}
from which the metric components transformation in eq.~\eqref{eqn:conftrans} follows.

Since the physical content is inscribed in the space of field configurations, not in the space of coordinates, any transformation in configuration fields is called \textit{internal} and any transformation due to a change in coordinates is
called \textit{external}, since coordinates are parameters which have nothing to do with the features of the field theory in question; then, a symmetry transformation is called internal or external, respectively. 

One should be aware that a general variation of a given field is then a sum of two variations: external and internal, i.e. $\delta\phi_{\ssst I}(x)=\delta_{\xi}\phi_{\ssst I}(x)+\delta_{\epsilon}\phi_{\ssst I}(x)$, where $\xi$ and
$\epsilon$ are infinitesimal parameters of external and internal transformation. Therefore, in general, one takes into account both and then investigates various behavior of a field theory at hand, including derivation of conserved currents that have both internal
and external characteristics. A very detailed treatment of such variations and related symmetries can be found in \cite{Blag, HB}. However, through the rest of this chapter we argue that it is justified for one to not take into account external transformations in the part of the thesis where we use a toy model to study conformal symmetry. This is because
both conformal coordinate transformations and conformal field transformations change only the geometric volume $\sqrt{g}$, or equivalently, the scale density $A$, and leave the shape density $\gb_{\mu\nu}$ invariant; it is for this reason that the two are easily confused under the less precise term ``conformal transformations''.
Since we are interested in conformal invariance in field theory, this ultimately
invites investigation of whether the scale density $A$ is present in the theory or it is not. If it is, a conformal transformation --- be that internal or external --- will affect the theory (as well as the equations of motion) and any resulting dynamics of the fields in question. If it is not present, we expect
the theory to be invariant under conformal transformations of any kind. We find support for this idea in \cite{Fult} (see also section 4.2 of \cite{Blag}), where it was established that the invariance under Weyl rescaling in curved spacetime implies conformal coordinate invariance in flat spacetime. Therefore, when necessary, we shall restrict our reference only to conformal field transformations\footnote{Another reason
for taking into account only internal transformations is that a canonical quantum theory of gravity that we are concerned with in this thesis
explicitly depends only on the three-dimensional metric field and other, non-gravitational fields, so that one is concerned directly with field transformations.} and from now on we refer to them simply as \textit{conformal transformation}. 

Therefore, based on the previous sections and arguments presented above, our tools will comprise of mechanisms of keeping track of the scale density $A$ throughout the calculations, not of a particular conformal transformation that should otherwise be specified.

We shall mainly deal with infinitesimal version of eq.~\eqref{eqn:conftrans} in this thesis. That means that the stated transformation should be expanded around identity $\Omega(x)\approx 1+\omega(x)$ and then we have,
    \begin{equation}
    \label{ConfFieldInf}
        \delta_{\omega}g_{\mu\nu}(x)=2\omega(x)g_{\mu\nu}(x)\ ,\qquad \delta_{\omega}\phi_{\ssst I}(x)=n_{\ssst I}\omega(x)\phi_{\ssst I}(x) \ ,\qquad\omega(x)\ll 1\ .
    \end{equation}
Let us note that the $n_{\ssst I}=2$ for the metric is a choice, but we justify it further below. 
It is by now clear that a conformal transformation of the metric field in eq.~\eqref{eqn:conftrans} produces the same effect as an active conformal transformation of coordinates: the metric is rescaled by a function of coordinates. Recalling eq.~\eqref{deltaAomega}, that a conformal coordinate transformation affects only the scale density $A$, it follows that conformal field transformation affects only the scale density $A$, leaving the shape density $\gb_{\mu\nu}$ invariant. Indeed, if we calculate the determinant of the metric in eq.~\eqref{eqn:conftrans}, we obtain that the $d$-dimensional volume transforms under conformal field transformation as
    \begin{equation}
        \sqrt{\tilde{g}}=\Omega^4\sqrt{g}\ .
    \end{equation}
Taking the fourth root, we can see that a transformation of the metric by an $\Omega^2$ can be ``explained'' as a transformation of the square of the scale density $\tilde{A}^2=\Omega^2 A^{2}$, defined by eq.~\eqref{scaledef}. That is, using the decomposition on both sides of the equation, we have
    \begin{align}
        \tilde{g}_{\mu\nu}&=\Omega^2 g_{\mu\nu}
        \nonumber\\[6pt]
        \tilde{A}^2\bar{\tilde{g}}_{\mu\nu}&=\Omega^2 A^2\gb_{\mu\nu}=(\Omega A)^2\gb_{\mu\nu}
    \end{align}
and it follows that
    \begin{equation}
        \tilde{A}=\Omega A\ ,\qquad \bar{\tilde{g}}_{\mu\nu}=\gb_{\mu\nu}\ ,
    \end{equation}
i.e. the shape density is invariant under conformal transformations. This completes the evidence that the scale density and shape density behave under conformal field transformations in
the same way as under conformal coordinate transformations, as we anticipated. Thus our focus on scale density instead on a conformal transformation is justified.

If the scale density produces a factor $\Omega$ under a conformal transformation, then could one look at other fields in a similar way as on the decomposed metric? There is nothing stopping us from defining some new fields $\chi_{\ssst I}$ such that
    \begin{equation}
    \label{fieldresc}
        \chi_{\ssst I} : = A^{-n_{\ssst I}}\phi_{\ssst I}\ ,
    \end{equation}
and in this way $\chi_{\ssst I}$ are conformally invaraint, if $\phi_{\ssst I}$ transforms homogeneously under conformal transformations\footnote{The new field can be defined even if the old field transforms inhomogeneously, in order to at least
compensate the scaling by $\Omega$.}. In other words, the idea is to introduce a set of new fields rescaled appropriately by the scale density $A$ such that the conformal transformation of the old fields is compensated for. 
We shall refer to such decomposition of fields as \textit{conformal decomposition}.
Note that these new fields are not absolute tensors but tensor \textit{densities} of weight $w^{\ssst I}=n_{\ssst I}/d$, but since we shall not encounter $\sqrt{g}$ explicitly, it is justifiable to introduce a \textit{scale weight}, defined by
    \begin{equation}
    \label{scaleweightdef}
        \wb_{\ssst I}: = w_{\ssst I} d = - n_{\ssst I}
    \end{equation}
such that the scale weight is equivalent to the negative of the conformal weight of the original field and the negative \textit{length dimension} of the original field, that we shall explain shortly. Note that introduction of the scale weight enables one to define $\gb_{\mu\nu}$ as a ``tensor density of scale weight $-2$'' and to define $A$ as a ``scalar density of scale weight 1''. Moreover, one may call fields $\chi_{\ssst I}$ ``tensor density of scale weight $\wb_{\ssst I}$''. 

One could have chosen any other convention for $n_{\ssst g}$ in the conformal transformation of the metric, but $n_{\ssst g}=2$ is convenient because we would like to think of the set of the metric tensor components as a dimensionful object that carries information about the measureable length and size of things, that is,
    \begin{equation}
        \left[g_{\mu\nu}\right]= L^2\ ,\qquad \left[x^{\mu}\right]=1\ ,
    \end{equation}
where $L$ is the unit of length, while coordinates are kept dimensionless.
This is equivalent to the argument that coordinates are only helpful set of labels with no measurable physical meaning and thus they should be dimensionless. One thus says that the length dimension of the metric components is two. Similarly, length dimensions of other fields' components are introduced based on the form of their Lagrangians. 
Here, we related the length dimension to the conformal weight (which is for non-gravitational fields also deduced from their Lagrangians). In that way conformal transformation using $\omega(x)$ means ``let us change the unit length scale at each point in spacetime differently'', while coordinates are kept fixed.
Choosing the metric tensor components as carriers of length units raises a question ``what is the length scale which provides meaningful units to $g_{\mu\nu}$?'' and this question is important to be asked. This actually depends on a context. 
One usually \textit{compares} relevant scales and here it is of interest for the discussion of the quantum-gravitational phenomena to measure physically relevant scales
with respect to the Planck length $l_{p}$. By physically relevant scales we mean those that are measured by observations, \textit{which can take place only with the help of interactions among non-gravitational fields}. 
These observable interactions are essentially events in spacetime that are separated by spacetime distances. These spacetime distances are said to be ``large'' or ``small'' only with respect to some other physically relevant length scale --- any other non-relative notion of ``large'' or ``small'' has no clear meaning. Therefore, if the metric
carries the units of length then the scale density $A$ is the piece of the metric that describes the ``size'' of the region in which the observed physical phenomena are taking place. Nowadays, any physical phenomena that we study in experiments take place across sizes which are much greater than the Planck scale. 
Thus we may say that for relevant non-quantum-gravitational phenomena
    \begin{equation}
    \label{relLength}
        \frac{l_{0}}{l_{p}}\gg 1\ ,
    \end{equation}
where $l_{0}$ is a number measured by the spacetime distance, or equivalently, by $A$, according to our new language.
But in the very early Universe, this ratio was closer to 1 as compared to today, which is why it is important to have it at disposal. Note that because this ratio is \textit{dimensionless}, it is suitable for any approximations that involve the Planck scale.
Having this in mind, we shall make the scale density $A$ \textit{dimensionless}, by formally redefining it as
    \begin{equation}
    \label{Alen}
        A \rightarrow l_{0}A\ ,
    \end{equation}
such that any expression that contains $A$ also has a dimensionful constant accompanying it. Apart from the metric components, non-gravitational fields are also dimensionful in general but this depends on a particular theory.
Separating the length in this way is particularly useful in exposing the dimensions of all coupling constants in a given theory, as we shall see in the next section. 
This may also be important for discussions about the renormalization group equations for quantum fields on curved spacetime, but this is beyond the scope of the thesis.

\section[Unimodular-conformal decomposition: scale and shape \texorpdfstring{\\}{} parts of  geometry][Unimodular-conformal decomposition: scale and shape parts...]{Unimodular-conformal decomposition: scale and shape parts of geometry}
\label{unimodDec}

The definition of the scale density and the shape density which we arrived at in previous sections is equivalent to demanding a decomposition of the metric under the action of the conformal group $C(d,\mathbb{R})$ or under the action of conformal field transformation given by eq.~\eqref{eqn:conftrans}.
There is nothing new in this definition compared to the information given in the previous sections and one can take the previous sections as a pedestrian way of motivating what usually goes under a name \textit{unimodular decomposition} and can be stated as a starting point as
    \begin{equation}
    \label{gsplit1}
        g_{\mu\nu}=A^2\gb_{\mu\nu}\ ,\qquad g^{\mu\nu}=A^{-2}\gb^{\mu\nu}\ ,\qquad A=(\sqrt{g})^{\frac{1}{d}}\ ,\qquad \vert\det \gb_{\mu\nu} \vert= 1\ ,
    \end{equation}
such that $\delta \gb_{\mu\nu}=0$ for variations due to any kind of conformal transformation. Also note that $\gb_{\mu\alpha}\gb^{\alpha\nu}=\delta^{\nu}_{\mu}$.
The new piece of information that we haven't mentioned so far is the unit determinant of $\gb_{\mu\nu}$ (thus the name ``unimodular'' decomposition). It can be checked easily from the definition of $A$ that this is indeed the case.
Alternatively, one could have defined unimodular decomposition by the requirement that $\vert\det \gb_{\mu\nu} \vert= 1$, from which it would follow that the power of $\sqrt{g}$ that enters the definition of $A$ has to be $1/d$.
Consequentially, the shape density is invariant under $C(d,\mathbb{R})$ and conformal field transformation in eq.~\eqref{eqn:conftrans}, while the scale density is invariant under $SL(d,\mathbb{R})$.
Together with eq.~\eqref{fieldresc}, we shall refer to this decomposition as \textit{unimodular-conformal decomposition}.

An important feature of the shape density that follows from here is that its variation is traceless. Namely, a general variation of the metric splits according to
    \begin{equation}
    \label{deltaGbarA}
        \delta g_{\mu\nu} = A^{2}\delta \gb_{\mu\nu} + 2  \gb_{\mu\nu} A\delta A\ ,
    \end{equation}
and after taking the trace with respect to $g_{\mu\nu}$, one arrives at
    \begin{equation}
        2\frac{\delta(\sqrt{g})}{\sqrt{g}} = A^{2}g^{\mu\nu}\delta \gb_{\mu\nu} + 2  d\frac{\delta A}{A}\ ,
    \end{equation}
but because $\delta \sqrt{g}= d A^{d-1}\delta A$ according to the definition in eq.~\eqref{gsplit1}, it must be that
    \begin{equation}
    \label{vargbTrzero}
        g^{\mu\nu}\delta \gb_{\mu\nu}=A^{-2}\gb^{\mu\nu}\delta \gb_{\mu\nu}=0\ .
    \end{equation}
This means that the shape density does not change under a conformal variation defined in eq.~\eqref{ConfFieldInf}, i.e. its conformal variation vanishes. A consequence of this is that since $\delta \gb_{\mu\nu}$ may stand for any derivative of $\gb_{\mu\nu}$, its trace is identically vanishing. For example,
    \begin{equation}
        \gb^{\mu\nu}\del_{\alpha} \gb_{\mu\nu}=0\ ,\qquad \gb^{\mu\nu}\del_{\alpha}\del_{\beta}\gb_{\mu\nu}-\gb^{\mu\nu}\gb^{\rho\sigma}\del_{\alpha}\gb_{\mu\rho}\del_{\beta}\gb_{\nu\sigma}=0\ ,
    \end{equation}
where the second identity follows from the first one by differentiating and lowering indices on $\gb_{\mu\nu}$ within the derivative, as shown in the work by Katanaev \cite{Kat06}. Katanaev calls $\gb_{\mu\nu}$ ``metric density''. They apply unimodular split only to the Ricci scalar and Einstein Equations and emphasize the
latter's resulting polynomial form in $\gb_{\mu\nu}$ and its derivatives, as well as the difference between using $A$ and using a scalar field to model the non-conformal degree of freedom of geometry.
But here we go further than their work and inspect what is the consequence of eq.~\eqref{gsplit1} to other curvature tensors.

\subsection{Scale and shape connection}

First, let us emphasize that both $A$ and $\gb_{\mu\nu}$ are tensor \textit{densities}. We can understand that the determinant is a kind of an object that carries information about conformally non-invariant
properties of spacetime. The shape density, on the other hand, has a fixed determinant \textit{in every coordinate system} and carries information about conformally invariant properties of the spacetime.
But since they transform --- as $g_{\mu\nu}$ does --- under coordinate transformations, they do not uniquely denote a feature of a geometry in a coordinate invarint way.
Namely, one can find a coordinate transformation (which will be a conformal one) which changes $A$ to $A'=1$ for any given metric. Therefore if one would like to make some physically relevant statements in terms of scale and shape density one needs to look into curvature tensors and curvature scalar invariants.

To this purpose, let us plug eq.~\eqref{gsplit1} into Christoffel symbols. 
One then has that they split into two parts\footnote{Note that the last term in eq.~\eqref{Sigma} can also be written in terms of the metric, because $A$ simply cancels out.}
    \begin{align}
    \label{GammaDec}
        \Gamma^{\alpha}{}_{\mu\nu}&=\Gb^{\alpha}{}_{\mu\nu}+\Sigma^{\alpha}{}_{\mu\nu}\ ,\\[6pt]
    \label{Gammabar}
        \Gb^{\alpha}{}_{\mu\nu}&=\frac{1}{2}\gb^{\alpha\rho}\left(\del_{\mu}\gb_{\rho\nu}+\del_{\nu}\gb_{\rho\mu}-\del_{\rho}\gb_{\mu\nu}\right)\ ,\\[6pt]
    \label{Sigma}
        \Sigma^{\alpha}{}_{\mu\nu}&=\left(2\delta_{(\mu}^{\alpha}\delta_{\nu)}^{\beta}-\gb_{\mu\nu}\gb^{\alpha\beta}\right)\del_{\beta}\log A \ ,
    \end{align}
which the following properties (taking into account eq.~\eqref{vargbTrzero} for the third term in eq.~\eqref{Gammabar}),
    \begin{align}
    \label{Gammabarprop}
        \Gb^{\alpha}{}_{\alpha\nu}&=0\ , & \gb^{\mu\nu}\Gb^{\alpha}{}_{\mu\nu}&=-\del_{\mu}\gb^{\mu\alpha}\ ,\\[12pt]
    \label{Clogaprop}
        \Sigma^{\alpha}{}_{\alpha\nu}&=d\,\del_{\nu}\log A\ ,\quad &\gb^{\mu\nu}\Sigma^{\alpha}{}_{\mu\nu}&=(2-d)\gb^{\alpha\nu}\del_{\nu}\log A\ ,\\[6pt]
    \label{tlessSigma}
       \Sigma^{\beta}{}_{(\mu\vert\alpha}&g_{\beta\vert\nu)}-\frac{1}{d}g_{\mu\nu}\Sigma^{\beta}{}_{\beta\alpha}=0\ .& &
    \end{align}
We see that $\Gb^{\alpha}{}_{\mu\nu}$, which we call \textit{the shape connection}, is the traceless in the up-down indices and $\Sigma^{\alpha}{}_{\mu\nu}$, which we call \textit{the scale connection}, seems to be the trace of the Christoffel symbols.
The last identity, eq.~\eqref{tlessSigma}, basically means that the symmetric traceless part with respect to the first two indices of $\Sigma^{\alpha}{}_{\mu\nu}$ vanishes.
\\

{\small (However, it should be noted that splitting the connection into traceless and trace pieces \textit{does not imply} the same split under the unimodular decomposition in the case of non-Riemannian geometry, i.e. if $\nabla_{\alpha}g_{\mu\nu}\neq 0$. Namely, there is a connection called \textit{projective connection} \cite{Thom} denoted by $\Pi^{\alpha}{}_{\mu\nu}$ and defined by
     \begin{equation}
     \label{projGamma}
        \Gamma^{\alpha}{}_{\mu\nu}=\Pi^{\alpha}{}_{\mu\nu}+\frac{1}{d+1}(\delta^{\alpha}_{\mu}\Gamma_{\nu}+\delta^{\alpha}_{\nu}\Gamma_{\mu})\ ,\qquad \Pi^{\alpha}{}_{\alpha\nu}=0\ ,
     \end{equation}
where no metric compatibility has been assumed.
It too is relevant in the context of the group $SL(4,\mathbb{R})$ and in discussions about unparametrized geodesics \cite{Kaca1,KacaB,EPS,Thom} and can be defined independently of the metric. This means that $\Gamma_{\mu}$ above a priori has nothing to do with the trace in the first equation in \eqref{Clogaprop} which is determined from the metric.
One can also define \textit{projective curvature tensor} (see~\cite{Lube} for comparison with the Weyl tensor) based on the projective connection and this curvature is invariant under \textit{projective transformations} that are a subset of $SL(4,\mathbb{R})$, which we do not speak of in this thesis.)}

At this point it is important to relate general variation of the shape and scale densities with general variation of the Christoffel symbols and understand this relationship in the context of variations of Christoffel symbols with respect to coordinate transformations that we derived in eq.~\eqref{varGamma}.
The variation of the Christoffel symbols due to the variation given in eq.~\eqref{deltaGbarA} splits in the following way,
    \begin{align}
    \label{deltaGammaDec}
        \delta\Gamma^{\alpha}{}_{\mu\nu}&=\delta\Gb^{\alpha}{}_{\mu\nu}+\delta\Sigma^{\alpha}{}_{\mu\nu}\nonumber\\[6pt]
        &=\frac{1}{2}\gb^{\alpha\rho}\left(\del_{\mu}\delta\gb_{\rho\nu}+\del_{\nu}\delta\gb_{\rho\mu}-\del_{\rho}\delta\gb_{\mu\nu}\right)
        +\left(2\delta_{(\mu}^{\alpha}\delta_{\nu)}^{\beta}-\gb_{\mu\nu}\gb^{\alpha\beta}\right)\del_{\beta}\log \delta A\ .
    \end{align}
Now compare the above equation with eq.~\eqref{varGamma}. One concludes that conformal transformations (of any kind) give rise to $\delta\Sigma^{\alpha}{}_{\mu\nu}$ and variations with respect to the volume-preserving transformations give rise to $\delta\Gb^{\alpha}{}_{\mu\nu}$, thus agreeing with eq.~\eqref{varGbSTr} and eq.~\eqref{varGbStless}, respectively.
This conclusion nicely fits the content of section \ref{groupalg}.
    
\subsection{Shape covariant derivative}

The split of connection induced by the unimodular decomposition means that the covariant derivative splits as well. It is then of interest to inspect the metricity condition.
Using the fact that $\gb_{\mu\nu}$ is a tensor density of scale weight $\wb=-2$ and definitions in eqs.~\eqref{GammaDec}-\eqref{Sigma} and \eqref{tlessSigma}, the metricity condition on the metric implies
    \begin{align}
    \label{metricity}
        \nabla_{\alpha}g_{\mu\nu}&=A^{2}\nabla_{\alpha}\gb_{\mu\nu}=0\nonumber\\[6pt]
        &=A^{2}\Biggpar{\del_{\alpha}\gb_{\mu\nu}-2\Gb^{\beta}{}_{(\mu\vert\alpha}\gb_{\beta\vert\nu)}
        -2\Sigma^{\beta}{}_{(\mu\vert\alpha}\gb_{\beta\vert\nu)}
        +2\del_{\alpha}\log A \gb_{\mu\nu}
        }\nonumber\\[6pt]
        &=A^{2}\Biggpar{
        \del_{\alpha}\gb_{\mu\nu}-2\Gb^{\beta}{}_{(\mu\vert\alpha}\gb_{\beta\vert\nu)}
        }\nonumber\\[6pt]
        &=A^{2}\bar{\nabla}_{\alpha}\gb_{\mu\nu}=0\ ,
    \end{align}
where we have defined $\bar{\nabla}_{\alpha}$ to be the ``covariant derivative'' built from $\Gb^{\alpha}{}_{\mu\nu}$ only. Note that all derivatives of $A$ cancel out. From this one deduces an interesting conclusion:
the metricity condition with respect to the metric and the
connection is equivalent to the metricity condition with respect to the shape density and shape connection. Thus, the metricity condition is conformally covariant (since $A^{2}$ can be cancelled). This could be important in the context of non-Riemannian geometry with a projective connection \cite{Kaca1}.

A covariant derivative of a covariant vector density $\mathcal{V}_{\alpha}$ of scale weight $\wb$ splits in the following way,
    \begin{align}
    \label{nablaVdec}
        \nabla_{\mu}\mathcal{V}_{\alpha}&= \del_{\mu}\mathcal{V}_{\alpha}-\Gb^{\beta}{}_{\alpha\mu}\mathcal{V}_{\beta}
        -\Sigma^{\beta}{}_{\alpha\mu}\mathcal{V}^{\beta}
        -\wb \mathcal{V}_{\alpha}\del_{\mu}\log A\nonumber\\[6pt]
        &=\bar{\nabla}_{\mu}\mathcal{V}_{\alpha}
        -\left((1+\wb)\delta^{\beta}_{\alpha}\delta_{\mu}^{\rho}+\delta_{\mu}^{\beta}\delta_{\alpha}^{\rho}-\gb_{\mu\alpha}\gb^{\beta\rho}\right)\mathcal{V}_{\beta}\del_{\rho}\log A\ ,
    \end{align}
and upon taking the trace to form a covariant divergence one gets
    \begin{align}
        g^{\mu\alpha}\nabla_{\mu}\mathcal{V}_{\alpha}&=g^{\mu\alpha}\del_{\mu}\mathcal{V}_{\alpha}+(d-\wb-2)g^{\mu\alpha}\mathcal{V}_{\alpha}\del_{\mu}\log A\\[6pt]
    \label{nabladivdec}
        &=\del_{\mu}\mathcal{V}^{\mu}+(d-\wb)\mathcal{V}^{\mu}\del_{\mu}\log A\ ,
    \end{align}
where we used the traceless property of the shape connection in eq.~\eqref{Gammabarprop} and wrote in the second line the expression for the contravariant vector density.
In eq.~\eqref{nablaVdec} we defined \textit{the shape covariant derivative}
    \begin{equation}
        \bar{\nabla}_{\mu}\mathcal{V}_{\alpha}:=\del_{\mu}\mathcal{V}_{\alpha}-\Gb^{\beta}{}_{\alpha\mu}\mathcal{V}_{\beta}\ ,
    \end{equation}
and similarly for the contravariant version. In Bradonji\'c \& Stachel \cite{Kaca1} this is called ``conformal covariant derivative''. The shape covariant derivative therefore ``does not see'' the difference between a vector and a vector density; this definition easily generalizes to a tensor density of arbitrary rank.
From eq.~\eqref{nabladivdec} one concludes that a vector density of scale weight $\wb=d$ eliminates the explicit scale connection from this derivative. It is for this reason that for the special case of divergence of vector density of
weight $w=\wb/d=1$ simplifies, see eq.~\eqref{nabladivdec}. We can also ask what is the traceless part of eq.~\eqref{nablaVdec}? In fact, we shall also impose symmetrization on the two indices since such case appears in this work in section \ref{sec_31conf}; we obtain the following answer
    \begin{align}
    \label{delVtless}
        \left[\nabla_{(\mu}\mathcal{V}_{\nu)}\right]^{\srm T}&=\left[\bar{\nabla}_{(\mu}\mathcal{V}_{\nu)}\right]^{\srm T}-(2+\wb)\left[\mathcal{V}_{(\mu}\del_{\nu)}\log A\right]^{\srm T}\ .
    \end{align}
We see from here that for the special case of covariant vector density of scale weight $\wb=-2$ there is no difference between the usual covariant derivative and the shape covariant derivative because any scale density dependence cancels out in that case\footnote{This does not mean that the
expression is conformally invariant, because this depends on the conformal properties of $\mathcal{V}_{\nu}$. Conformal invariance holds of course in the special case when $\mathcal{V}_{\nu}$ is conformally invariant itself.}. We will encounter one such example in this thesis (the shift vector, see section \ref{sec_31conf}).

Shape covariant derivative is useful only if one makes a restriction from $GL(d,\mathbb{R})$ to $SL(d,\mathbb{R})$, i.e. if one excludes conformal transformations from the original generally covariant theory. Such an example is the case of \textit{unimodular gravity}, see e.g. \cite{Alv, Ellis, Fink, Un} and thereby cited references. In unimodular gravity one imposes a constraint on the metric itself that its determinant is \textit{fixed} $\sqrt{g}=1$ (which can also be thought of as gauge fixing) and one must follow the consequences of this constraint. 
This eventually leads to interesting dynamics which is related to the solutions of the Einstein vacuum equations with cosmological constant \cite{Alv, Ellis,Un}. It can also be used in theories of matter quantum fields on a dynamical curved spacetime background together to study regimes of the very early Universe (energies above
$10^{2}{\rm GeV}$) in which the matter content enjoyed conformal symmetry, see \cite{BuchDrag2}. However, in a more recent paper \cite{PadSal} it has been claimed that locally there is no difference between classical unimodular gravity and classical GR, \textit{since the former is just a locally gauge-fixed version of the latter}.
It is also claimed that the previously claimed ``new perspective'' of the problem of cosmological constant were not formulated carefully because one needs to study the cosmological constant within the context of semiclassical gravity and take into account necessary renormalization requirements. The paper also argues that there is
an equivalence between quantum theory based on unimodular gravity up to an arbitrarily high energies within the framework of path integral approach. Therefore the notion of unimodular gravity as a theory distinct from GR has to be taken with care.

Here, however, we do not impose any constraint on the metric: the number of its independent components is still $d(d+1)/2$ except that with the help of unimodular decomposition we look at them as $1 + (d(d+1)/2 - 1)$ components instead. 
Therefore, no constraint must be added if we would like, for example, to look at the Einstein equations --- one simply implements the consequences of such decomposition, such as eqs.~\eqref{GammaDec}-\eqref{tlessSigma}.
One only needs to be careful not to
interpret objects built from the shape and scale connection as general-covariant tensors, but rather as tensors with respect to the restricted group of volume-preserving coordinate transformations.

\subsection{Curvatures in terms of the scale and shape densities}
\label{subsec_curvAsh}

Scale and shape density, like the metric, take on a different form in different coordinate systems. The same is with scale and shape parts of the Christoffel symbols. For given $A(x)$ and $\gb_{\mu\nu}(x)$ one can always find a coordinate transformations such that the scale density becomes equal to one,
    \begin{equation}
        \tilde{A}(\tilde{x})=\text{J}^{\frac{1}{d}}(x)A(x) = 1\ ,
    \end{equation}
where $\text{J}$ is the determinant of the transformation matrix, recall eq.~\eqref{Amatdec}. But one \textit{cannot} always find a coordinate transformation that transforms the shape density into the constant matrix. This is just a consequence of the fact that if the space is curved then there is no global coordinate transformation that will bring the metric into the Minkowski/Euclidean form.
Thus we see that there is a certain asymmetry between the scale density and the shape density. This is roughly speaking reflected in the semidirect product of $SL(d,\mathbb{R})$ with $C(d,\mathbb{R})$, as discussed in section \ref{groupalg}.
Since the only way to tell if the space is flat or curved is to ask if the Riemann tensor vanishes at every point of space or not, one expects this asymmetry to be reflected in the Riemann tensor as well, once we look at how it decomposes as a consequence of the unimodular decomposition.

Riemann tensor is defined in terms of the Christoffel symbols as
    \begin{equation}
    \label{riem}
        R^{\alpha}{}_{\mu\beta\nu}=\del_{\beta}\Gamma^{\alpha}{}_{\mu\nu}-\del_{\nu}\Gamma^{\alpha}{}_{\mu\beta}+\Gamma^{\alpha}{}_{\beta\rho}\Gamma^{\rho}{}_{\mu\nu}-\Gamma^{\rho}{}_{\mu\beta}\Gamma^{\alpha}{}_{\nu\rho}\ ,
    \end{equation}
and we immediately see that there is going to be cross terms $\Gb\cdot\Sigma$ once one uses eq.~\eqref{GammaDec}. This is a signal that scale density interacts with the shape density and the Riemann tensor cannot be separated as a direct sum of tensors dependent only on $A$ and tensors dependent only on $\gb_{\mu\nu}$.
But since we have learned, as mentioned above, that one can always find a coordinate transformation which eliminates the scale density, there must be at least one tensorial piece of the Riemann tensor which does not care about such transformations (i.e. conformal transformations) because it has to survive to tell us about the curvature of the space.
We know that $\Gb^{\alpha}{}_{\mu\nu}$ is invariant under conformal transformations. We also know that the Weyl tensor, defined by eq.~\eqref{Weyltens} and alternatively by
    \begin{align}
    \label{WeyltensS}
        C^{\alpha}{}_{\mu\beta\nu} &=R^{\alpha}{}_{\mu\beta\nu}
        -  2\left(\delta^{\alpha}_{[\beta}P_{\nu]\mu}  -  g_{\mu[\beta}P^{\alpha}{}_{\nu]}\right)\ ,
    \end{align}
where in the second line we used the Schouten tensor defined by
    \begin{equation}
    \label{Schoudef}
        P_{\mu\nu}=\frac{1}{d-2}\left(R_{\mu\nu}-\frac{1}{2(d-1)}g_{\mu\nu}R\right)\ ,
    \end{equation}
is invariant under conformal transformations. But we could pretend that we do not know about the Weyl tensor and ask what is the tensor that is built solely from $\Gb^{\alpha}{}_{\mu\nu}$? This is the question asked by Thomas \cite{Thom25,Thom26} in 1925-26. They call $\Gb^{\alpha}{}_{\mu\nu}$ ``conformal connection''\footnote{Note that this is the the same as ``trace-free Christoffel symbols'' of \cite{Kaca1}.} and they showed that it is not enough to simply take the structure of the Riemann tensor's
definition in eq.~\eqref{riem} and substitute $\Gamma^{\alpha}{}_{\mu\nu}\rightarrow \Gb^{\alpha}{}_{\mu\nu}$ because such an expression is not a tensor. However, only when one subtracts all the traces from such an expression one obtains a tensorial object, which he showed is equivelant to the Weyl tensor.
We will only sketch this result with the following line of reasoning. Take a look at the first two terms in eq.~\eqref{riem}, pretending for a moment that all $\Gamma^{\alpha}{}_{\mu\nu}\rightarrow \Gb^{\alpha}{}_{\mu\nu}$. In the second term, the $\alpha=\beta$ component is missing because $\Gb^{\alpha}{}_{\beta\alpha}=0$. But
that means that a coordinate transformation will give rise to a non-tensorial term from the $\alpha=\beta$ trace in the first term which will not have a counterpart to be cancelled with. Therefore, $\alpha=\beta$ trace must be subtracted from the potential definition, which from the LHS means that one needs to subtract Ricci
tensor. Repeating this argument for all problematic traces, one can find that the resulting definition is given by eq.~\eqref{Weyltens} or eq.~\eqref{WeyltensS} with all $\Gamma^{\alpha}{}_{\mu\nu}\rightarrow \Gb^{\alpha}{}_{\mu\nu}$ in it.
In other words, all terms containing $\Sigma^{\alpha}{}_{\mu\nu}$ cancel out in those equations.
We shall not prove this, but we shall inspect how Ricci tensor and Ricci scalar look like under unimodular decomposition. Contracting with $\alpha=\beta$ in eq.~\eqref{riem} and using eqs.~\eqref{GammaDec}-\eqref{Clogaprop} one obtains
    \begin{align}
    \label{riccidec}
        R_{\mu\nu}&=\bar{R}_{\mu\nu}+2\,\del_{[\alpha}\Sigma^{\alpha}{}_{\mu]\nu}+2\Sigma^{\beta}{}_{\mu[\nu}\Sigma^{\alpha}{}_{\alpha]\beta}-2\Gb^{\alpha}_{\beta(\mu}\Sigma^{\beta}{}_{\nu)\alpha}+\Gb^{\beta}_{\mu\nu}\Sigma^{\alpha}{}_{\alpha\beta}\nonumber\\[12pt]
        &=\bar{R}_{\mu\nu}-\left(d-2\right)\left(\mathbb{1}_{(\mu\nu)}^{\alpha\beta}+\frac{1}{d-2}\gb_{\mu\nu}\gb^{\alpha\beta}\right)\bar{\nabla}_{\alpha}\del_{\beta}\log A \nonumber\\[12pt]
        &\quad +\left(d-2\right)\left(\mathbb{1}_{(\mu\nu)}^{\alpha\beta}-\gb_{\mu\nu}\gb^{\alpha\beta}\right)\del_{\alpha}\log A\,\del_{\beta}\log A\ ,
    \end{align}
where $\bar{R}_{\mu\nu}$ is defined by\footnote{This is the so-called ``November tensor'' \cite{Renn} that Einstein ended up with in one of his attempts on deriving his field equations, using the gauge condition $\sqrt{g}=1$, which corresponds to $A=1$ here.}
    \begin{equation}
    \label{novR}
        \bar{R}_{\mu\nu}:=\del_{\alpha}\Gb^{\alpha}{}_{\mu\nu}-\Gb^{\rho}{}_{\mu\alpha}\Gb^{\alpha}{}_{\nu\rho}\ .
    \end{equation}
Taking the trace of eq.~\eqref{riccidec}, the Ricci scalar decomposes as
    \begin{align}
    \label{Rdec}
        R=g^{\mu\nu}R_{\mu\nu}&=
        A^{-2}\biggsq{\bar{R}
        -2\left(d-1\right)\gb^{\mu\nu}\biggpar{\bar{\nabla}_{\mu}\del_{\nu}\log A  +\frac{d-2}{2}\del_{\mu}\log A\,\del_{\nu}\log A
        }
        }\nonumber\\[12pt]
        &=A^{-2}\biggsq{\bar{R}
        -2\left(d-1\right)A^{\frac{2-d}{2}}\del_{\mu}\biggpar{A^{\frac{d-2}{2}}\gb^{\mu\nu}\del_{\nu}\log A}
        }\nonumber\\[12pt]
        &=A^{-2}\biggsq{\bar{R}
        -2\left(d-1\right)A^{-2}\biggpar{A\del_{\mu}\left(\gb^{\mu\nu}\del_{\nu} A\right)  +\frac{d-4}{2}\gb^{\mu\nu}\del_{\mu} A\,\del_{\nu} A
        }
        }\nonumber\\[12pt]
        &=A^{-2}\biggsq{\bar{R}
        -2\left(d-1\right)A^{\frac{2-d}{2}}\del_{\mu}\biggpar{A^{\frac{d-4}{2}}\gb^{\mu\nu}\del_{\nu} A}
        }\nonumber\\[12pt]
        &\stackrel{\text{if } d\neq 2}{=}A^{-2}\biggsq{\bar{R}
        -\frac{4\left(d-1\right)}{d-2}A^{\frac{2-d}{2}}\del_{\mu}\biggpar{\gb^{\mu\nu}\del_{\nu} A^{\frac{d-2}{2}}}
        }\ ,
    \end{align}
where $\bar{R}:=\gb^{\mu\nu}\bar{R}_{\mu\nu}$ is the conformaly invariant part of the Ricci scalar.
We have presented above several different ways of writing the decomposed Ricci scalar that may be useful for various purposes. For example, from the expression in the second line it follows that in $d=2$ dimensions $\sqrt{g}R$ is a total divergence. Third and fourth line contain useful expressions for $d=4$ dimensions.
Furthermore, a rule of thumb can be used to quickly determine the conformally transformed Ricci scalar: simply add a term which
is obtained from the $A$-dependent term in the first line of eq.~\eqref{Rdec} by making a substitution $A\rightarrow \Omega$ and $\bar{\nabla}_{\mu}\rightarrow \nabla_{\mu} $, or from the $A$-dependent term in the third line of eq.~\eqref{Rdec} by making a substitution $A\rightarrow \Omega$ and $\del_{\mu}\rightarrow \nabla_{\mu} $. The result is
    \begin{align}
    \label{Rconftransf}
        \tilde{R}&=R-\frac{2\left(d-1\right)}{ \Omega^2}\Biggl(\nabla_{\mu}\left(g^{\mu\nu}\del_{\nu}\log \Omega\right)+\frac{d-2}{2}g^{\mu\nu}\del_{\mu}\log\Omega\,\del_{\nu}\log\Omega\Biggr)\nonumber\\[12pt] 
        &=R-\frac{2\left(d-1\right)}{ \Omega^4}\Biggl(\Omega
        \nabla_{\mu}\left(g^{\mu\nu}\del_{\nu}
        \Omega\right)+\frac{d-4}{2}g^{\mu\nu}\del_{\mu}\Omega\,\del_{\nu}\Omega\Biggr)
    \end{align}
and note that $A^{-2}$ has been absorbed into $g^{\mu\nu}=A^{-2}\gb^{\mu\nu}$. Indeed, this is the correct conformal transformation \cite{Far}. On the other hand, the last line in eq.~\eqref{Rdec} is useful when discussing non-minimally coupled scalar field.
Now, the same manipulation could be done with the Riemann tensor, but for our purposes it is enough to say that the $A$-dependent terms are exactly cancelled by $A$-dependent terms in the Schouten tensor in eq.~\eqref{WeyltensS}, leaving the $A$-independent and \textit{therefore} conformally invaraint Weyl tensor. Based on eq.~\eqref{Schoudef}, eq.~\eqref{riccidec} and eq.~\eqref{Rdec} Schouten tensor decomposes as
    \begin{equation}
    \label{Schoudec}
        P_{\mu\nu}=\bar{P}_{\mu\nu}-(d-2)\Biggpar{
         \bar{\nabla}_{\mu}\del_{\nu}\log A
         -\left(\mathbb{1}_{(\mu\nu)}^{\alpha\beta}-\frac{1}{2}\gb_{\mu\nu}\gb^{\alpha\beta}\right)\del_{\alpha}\log A\,\del_{\beta}\log A
        }\ .
    \end{equation}
where $\bar{P}_{\mu\nu}$ is defined as the conformally invariant part of the Schouten tensor
    \begin{equation}
        \bar{P}_{\mu\nu}:=\frac{1}{d-2}\left(\bar{R}_{\mu\nu}-\frac{1}{2(d-1)}\gb_{\mu\nu}\bar{R}\right)\ .
    \end{equation}
Note that the $A$-dependent part in eq.~\eqref{Schoudec} has the same form in any dimension, unlike Ricci tensor and scalar. Finally, our educated guess takes the form:
    \begin{equation}
    \label{Weylbar}
        C^{\alpha}{}_{\mu\beta\nu}=\bar{R}^{\alpha}{}_{\mu\beta\nu}
        -  2\left(\delta^{\alpha}_{[\beta}\bar{P}_{\nu]\mu}  -  \gb_{\mu[\beta}\bar{P}^{\alpha}{}_{\nu]}\right)
    \end{equation}
where $\bar{R}^{\alpha}{}_{\mu\beta\nu}$ is an object that has the same structure as the Riemann tensor given in eq.~\eqref{riem} with $\Gamma^{\alpha}{}_{\mu\nu}\rightarrow \Gb^{\alpha}{}_{\mu\nu}$. This makes the Weyl tensor manifestly conformally invariant and allows one to call it ``the shape curvature tensor''. Manifest conformal invariance is a guiding principle for the choice of tools in this thesis and we shall use it whenever possible.

There are two more tensors worth mentioning, the traceless part of Ricci tensor and the Einstein tensor. The former is given by
    \begin{equation}
    \label{Riccitlessdec}
        R_{\mu\nu}^{\srm T}:=R_{\mu\nu}-\frac{1}{d}g_{\mu\nu}R=\bar{R}_{\mu\nu}^{\srm T}-\left(d-2\right)\left(\bar{\nabla}_{(\mu}\del_{\nu)}\log A-\del_{\mu}\log A\,\del_{\nu}\log A\right)^{\srm T}\ ,
    \end{equation}
where $\bar{R}_{\mu\nu}^{\srm T}$ is the traceless part of eq.~\eqref{novR}. The Einstein tensor is given by
    \begin{align}
    \label{Einsdec}
        G_{\mu\nu}:=R_{\mu\nu}-\frac{1}{2}g_{\mu\nu}R&=\bar{G}_{\mu\nu}-\left(d-2\right)\Bigg[\left(\mathbb{1}_{(\mu\nu)}^{\alpha\beta}-\gb_{\mu\nu}\,\gb^{\alpha\beta}\right)\bar{\nabla}_{\alpha}\del_{\beta}\log A\nonumber\\[12pt]
        &\qquad\qquad-\left(\mathbb{1}_{(\mu\nu)}^{\alpha\beta}-\frac{3-d}{2}\gb_{\mu\nu}\,\gb^{\alpha\beta}\right)\del_{\alpha}\log A\,\del_{\beta}\log A\Bigg]\ ,
    \end{align}
where $\bar{G}_{\mu\nu}:=\bar{R}_{\mu\nu}-\frac{1}{2}\gb_{\mu\nu}\bar{R}$.

Extracting the $A$-independent parts from $GL(d,\mathbb{R})$-tensors using unimodular decomposition only offers a suitable method for dealing with various cordinate choices, unless one is especially interested in restricting to $SL(d,\mathbb{R})$. 
For example, one may choose a coordinate gauge in which the $A$-dependent term of the Einstein tensor vanishes (the simplest is $A=1$). Or one can look for conformally flat spaces by demanding that $\gb_{\mu\nu}=\eta_{\mu\nu}$. Of course, only for the latter 
one has a generally covariantly expressed condition, i.e. the vanishing of the Weyl tensor. For the former condition there is no generally covariant condition. The only generally covariant thing one could do to make sure the scale density $A$ does not contribute to the curved space is to 
require vanishing of the Ricci tensor. We give here a remarkably simple and intuitive proof of this fact. Namely, consider the metric components expressed in a neighbourhood of a geodesic (measured by $\vert\zeta^{\mu}\zeta_{\mu}\vert < 1$), i.e. in Fermi normal coordinates, given by the usual Taylor expansion \cite{MM},
    \begin{equation}
    \label{RNC}
        g_{\mu\nu}(\zeta)=\eta_{\mu\nu}+\frac{1}{3}R_{\mu\alpha\nu\beta}\zeta^{\alpha}\zeta^{\beta} + \mathcal{O}(\zeta^{\gamma}\zeta^{\tau}\zeta^{\sigma})\ .
    \end{equation}
In the above expression the metric components and the components of the Riemann tensor are evaluated along a chosen geodesic at a point which belongs to it. The meaning of this equation is that the Riemann tensor measures deviation of a metric from the flat one in a small neighbourhood along a geodesic. So which pieces of the
Riemann tensor measure deviation of the shape and scale parts of the metric then\footnote{In \cite{Lov} it is shown what is the interpretation of the Riemann and Ricci tensor and Ricci scalar. However, we find their derivation for the
interpretation of the Ricci tensor is cumbersome and therefore offer here, based on the unimodular decomposition, a much simpler proof of the same claim that follows below.}? Consider the difference $\delta_{0}g_{\mu\nu}(\zeta):=g_{\mu\nu}(\zeta)-\eta_{\mu\nu}$, which has nothing to do with a Lie derivative but we may consider it to be a type of a variation at a point on a manifold due to a coordinate transformation. Using eq.~\eqref{deltaGbarA} and
eq.~\eqref{vargbTrzero} in eq.~\eqref{RNC} and separating the trace from traceless parts one obtains\footnote{We found that eq.~\eqref{RNCtr} agrees with Corollary 2.3 in \cite{Zouq}.}
    \begin{subequations}
    \begin{align}
    \label{RNCtless}
       A^2\delta_{0}\gb_{\mu\nu} & = \frac{1}{3}C_{\mu\alpha\nu\beta}\zeta^{\alpha}\zeta^{\beta}+\frac{1}{3(d-2)}\biggsq{\zeta^2R_{\mu\nu}^{\srm T} -2\Bigpar{\zeta_{(\mu}R_{\nu)\beta}^{\srm T}-\frac{1}{d}g_{\mu\nu}R_{\alpha\beta}^{\srm T}\zeta^{\alpha}}\zeta^{\beta}
       }\nonumber\\[6pt]
       &\quad -\frac{1}{3d(d-1)}R\Bigpar{\zeta_{\mu}\zeta_{\nu}-\frac{1}{d}g_{\mu\nu}\zeta^2}\ ,\\[12pt]
    \label{RNCtr}
        \frac{\delta_{0}A}{A}&=\frac{1}{6d}R_{\mu\nu}\zeta^{\mu}\zeta^{\nu}=\frac{1}{6d}\biggpar{R_{\mu\nu}^{\srm T}\zeta^{\mu}\zeta^{\nu}+\frac{1}{d}R\zeta^2}\ ,
    \end{align}
    \end{subequations}
where $\zeta^2=\zeta^{\mu}\zeta_{\mu}$ and we have additionally split the Ricci tensor into its traceless and trace parts. Equation \eqref{RNCtr} proves that the Ricci tensor is that part of the Riemann tensor which measures the effect of the spacetime curvature on the $d$-dimensional volume (expressed here in terms of scale density).
Therefore, the only covariant statement regarding the constancy of the volume is the vanishing of the Ricci tensor, meaning that all vacuum solutions of GR have the property that along the freely-falling trajectories an observer measures a constant four-dimensional volume.
On the other hand, we see that even for conformally flat spacetimes the absence of the Weyl tensor in eq.~\eqref{RNCtless} does not mean that the shape part of the metric is not curved. As we have stated earlier in this subsection, this is because the
Riemann tensor does not split into scale-independent and shape-independent pieces under unimodular decomposition --- there is ``mixing'' between $A$ and $\gb_{\mu\nu}$ in the Ricci tensor, which is just a consequence of the non-linear nature of the Riemannian curvature.
Furthermore, one can see that for Einstein spaces ($R_{\mu\nu}^{\srm T}=0, R=const.$) we have that both the shape and the scale parts of the metric experience the curvature of spacetime.

Is it possible to have such a metric that in Fermi normal coordinates only its scale density experiences the curvature but not the shape? This is not possible, because, as one can see from eq.~\eqref{RNCtless}, that would mean 
that all $C_{\mu\alpha\nu\beta},R_{\mu\nu}^{\srm T}$ and $R$ have to vanish\footnote{Since $\zeta^{\mu}$ is arbitrary along the geodesic.}, which implies that the space is flat and $\delta_{0}A=0$ as well. 
This is the same asymmetry between the scale and shape parts of the metric that we discussed earlier in this subsection regarding the decomposition of the Riemann tensor under the unimodular decomposition.
It is the property of the Riemannian geometry itself that the roles of the scale and shape parts of the metric are non-trivially rooted in the Riemannian curvature tensor and its traces.
Said simply, the asymmetry could mean that the concept of shape \textit{could} be defined without the concept of scale, while the concept of scale \textit{could not} be defined without the concept of shape\footnote{This might be a part of a more general geometric relationship among $p$-dimensional hypersurfaces. Namely, the basis of 3-forms $\d x^{\mu}\wedge \d x^{\nu}\wedge \d x^{\alpha}$ in $d$ dimensions (i.e. a 3-volume) can only be defined if the bases of 1-forms (lines) and 2-forms (planes) have already been defined, while the definition of the basis of 2-forms does not require the definition of the basis of 3-forms nor any higher forms.}.

Does the flat spacetime have a meaningful notion of a scale density? Even though this kind of question is valid, in the light of discussion presented so far this question needs to be made a bit more precise. As mentioned in the previous paragraph, only flat spacetime gives rise to zero change of both scale and shape parts across the manifold. However, this does not mean that $A=0$ but
it does not even mean $A=const.$ (otherwise metric would not be defined). Indeed, in spherical coordinates the volume is $A^{4}=\sqrt{g}=r^2\sin \theta$, while in Cartesian coordinates $A^4=1$, so it depends on coordinate system. But one would like to
have a coordinate-independent answer to the asked question and that answer is given again by eq.~\eqref{RNCtr}, which vanishes at every point for the flat spacetime case. It simply means that an observer does not measure a curvature-induced deviation of a small volume along a geodesic.
The same is with the shape density: eq.~\eqref{RNCtless} vanishes identically, so the shape density of a
flat spacetime does not deviate from the flat metric as measured by the observer along a geodesic. These two statements are independent of a coordinate system used.
But now we can ask where is $l_{0}$ in the flat spacetime metric? Indeed, this is an important question, especially if one is interested in studying some field theory on a flat background. Since we have just established that even the flat spacetime has a scale, although that is not obvious in Cartesian coordinates, one simply needs to use
eq.~\eqref{Alen} and make coordinates dimensionless. Then in spherical coordinates the volume is $A^{4}=\sqrt{g}=l_{0}^4 r^2\sin \theta$ and in Cartesian coordinates $A^4=l_{0}^{4}$, and with this that the scale density always has the meaning of a length is made clear in our formalism. Then the Minkowski metric takes the form $\d s^2 = l_{0}^2 \eta_{\mu\nu}\d x^{\mu}\d x^{\nu}$.
(This way of thinking might have a lot of interesting consequences for quantum field theories on Minkowski background especially in the context of dilation invariance.)
Note that now eq.~\eqref{RNC} has to be rewritten with $l_{0}$ appearing explicitly, 
    \begin{align}
    \label{RNCL}
        g_{\mu\nu}(\zeta)=l_{0}^2\eta_{\mu\rho}\left(\delta^{\rho}_{\nu}+\frac{1}{3}R^{\rho}{}_{\alpha\nu\beta}\zeta^{\alpha}\zeta^{\beta}\right)
        \ ,
    \end{align}
where we raised the index on the Riemann tensor using $\eta_{\mu\rho}$, which is valid at the approximation order we are considering. With this writing we have that the Riemann tensor and $\zeta^{\mu}$ are manifestly dimensionless and the second term describes corrections to the Minkowski metric (that is, its scale and shape parts) such that dimensionless numbers
    \begin{equation}
        R^{\rho}{}_{\alpha\nu\beta}\zeta^{\alpha}\zeta^{\beta}\ll 1\ ,
    \end{equation}
which means that the approximation in eq.~\eqref{RNC} is valid as long as the size of the neighbourhood of the point of interest is a much smaller fraction of $l_{0}$ compared to the size of the curvature radius (which is roughly the inverse of the curvature squared). Thus, extracting the characteristic length scale naturally gives dimensionless numbers which control ``sizes''. The similar thing will take place with comparisons that involve relative strengths of coupling constants, as we shall see in the next chapter.

\section{Application to the 3+1 decomposition of spacetime}
\label{sec_31conf}

So far we have only referred to the full $d$-dimensional metric and its unimodular-decomposition. But we will also need to consider the three-dimensional unimodular decomposition since the approach to quantum gravity that we are taking in this thesis is based on the 3+1 decomposition of spacetime. Space + time splitting is briefly summarized in appendix \ref{App_31} where the relevant references are also mentioned. The author has also written about it in detail in his Master thesis \cite{MSc}.
However only during the work on the current thesis has the author derived the material presented in this section\footnote{This is a part of the relevant paper \cite{KN17}.}.

\subsection{Unimodular-conformal decomposition and \texorpdfstring{$3+1$}{} decomposition}
\label{subs_umoddec31}

First we take a look at all relevant elementary variables used in $3+1$ decomposition. Conformal transformation of the $3+1$-decomposed metric is given by
    \begin{align}
    \label{g31Om}
        \tilde{g}_{\mu\nu}&=\Omega^2 g_{\mu\nu}=\Omega^2 h_{\mu\nu}-\Omega^2 n_{\mu}n_{\nu}
        =\left(
        \begin{array}{ccc}
        -\Omega^2 N^{2}+\Omega^2N_{i}N^{i}& \Omega^2 N_{i}\\[12pt]
        \Omega^2 N_{i} & \Omega^2 h_{ij} 
        \end{array}
        \right)\ ,\\[12pt]
    \label{detg31Om}
        \sqrt{\tilde{g}}&=\Omega^4 \sqrt{g}= \Omega N \,\Omega^3 \h\ ,
    \end{align} 
based on which one can deduce the following transformation of the individual objects
    \begin{align}
    \label{conf3met}
        h_{ij}\quad\rightarrow &\quad \tilde{h}_{ij}=\Omega^2 h_{ij}\ ,\\[12pt]
    \label{conf3det}
        \sqrt{h}\quad\rightarrow &\quad \sqrt{\tilde{h}}=\Omega^3 \sqrt{h}\ ,\\[12pt]
    \label{confn}
        n_{\mu}\quad\rightarrow &\quad \tilde{n}_{\mu}=\Omega n_{\mu}=\Omega\left(-N,0\right)\ ,\\[12pt]
    \label{confnn}
        n^{\mu}\quad\rightarrow &\quad \tilde{n}^{\mu}=\Omega^{-1} n^{\mu}=\Omega^{-1}\left(\frac{1}{N},\frac{-N^{i}}{N}\right)\ ,\\[12pt]
    \label{confN}
        N\quad\rightarrow &\quad\,\,\tilde{N}=\Omega N\ ,\\[12pt]
    \label{confNi}
        N^{i}\quad\rightarrow &\quad\tilde{N}^{i}=N^{i}\quad  {\rm and}\quad \tilde{N}_{i}=\tilde{h}_{ij}\tilde{N}^{j}=\Omega^2 N_{i}\ .
    \end{align}
Here $N$ is the lapse function, $N^{i}$ is the shift vector, while $n^{\mu}$ is a timelike vector orthogonal to the three-hypersurface whose metric is $h_{ij}$.
Furthermore, based on eqs.~\eqref{conf3met}-\eqref{confNi}, \eqref{Kijdef} and \eqref{Ktrdef}, it can be deduced that the extrinsic curvature transforms non-covariantly under conformal transformation,
    \begin{equation}
    \label{eqn:confK}
        K_{ij}\quad\rightarrow\quad \tilde{K}_{ij}=\Omega K_{ij}+h_{ij}\mathcal{L}_{n}\log\Omega,
    \end{equation}
which is thanks to the inhomogeneous transformation of its trace,
    \begin{equation}
    \label{eqn:confKtr}
        K\quad\rightarrow\quad \tilde{K}=\frac{1}{\Omega}\left( K+3\mathcal{L}_{n}\Omega \right).
    \end{equation}

Now, in complete analogy to eq.~\eqref{gsplit1} we define
    \begin{equation}
    \label{3scalea}
        a:=(\sqrt{h})^{\frac{1}{3}}\ ,
    \end{equation}
such that $a$ is the only geometric variable that transforms under conformal transformation. We call it \textit{the three-scale density} but if the context allows we shall simply refer to it as the scale density and we shall make sure there is no ambiguity.
Then, instead of referring to the four-dimensional scale and shape densities, the focus shifts to the three-dimensional scale and shape densities.
Based on eq.~\eqref{3scalea} the three-metric decomposes as 
    \begin{equation}
    \label{3unimoddec}
        h_{ij}=a^{2}\hb_{ij}\ ,\qquad \hb_{ij} =a^{-2} h_{ij}\ ,\qquad \det \hb_{ij}=1\ .
    \end{equation}
This decomposition is now with respect to the conformal group $C(3,\mathbb{R})$, i.e. with respect to three-dimensional conformal coordinate transformations, or, equivalently, with respect to the same group of the field conformal transformation eq.~\eqref{eqn:conftrans}. Accordingly, the scale density $a$ is invariant under $SL(3,\mathbb{R})$ group of three-dimensional volume-reserving transformations.
It is important to note that the Levi-Civita tensor \textit{density} components $\varepsilon_{ijkl}$ have a conformal weight of $3$ because it represents the three-volume. Therefore we could write
    \begin{equation}
    \label{3eps}
        \bar{\varepsilon}_{ijk}:=a^{-3}\varepsilon_{ijk}
    \end{equation}
as the conformally invariant Levi-Civita \textit{tensor} components and this is just the Levi-Civita \textit{symbol} itself which has, of course, zero conformal weight. This is important to keep in mind for the definition of the magnetic part of the Weyl tensor, see eq.~\eqref{magWresc}.
In order to cancel the effect of conformal transformation in eqs.~\eqref{conf3met}-\eqref{confNi} we define the corresponding rescaled objects as
    \begin{align}
    \label{VarsNbar}
        \Nb  := a^{-1}N\ ,\qquad &\Nb^{i}=N^{i}\ ,\qquad  \Nb_{i}=a^{-2} N_{i}\ ,\\[6pt]
    \label{Varsnbar}
        \bar{n}_{\mu}:= a^{-1}n_{\mu}\ ,& \qquad \bar{n}^{\mu}:= a n^{\mu}\ .
    \end{align}
Due to this rescaling, we have
    \begin{equation}
    \label{LieResc}
        \mathcal{L}_{n}\mathcal{T}=a^{-1}\mathcal{L}_{\bar{n}}\mathcal{T}
    \end{equation}
for a tensor density $\mathcal{T}$ of any weight.

Extrinsic curvature deserves special care.
The time derivative $\dot{h}_{ij}$ will give a term proportional to $\dot{\hb}_{ij}$ and to $h_{ij}\dot{a}$, which immediately reminds us of the split into traceless and trace parts in analogy to eq.~\eqref{deltaGbarA}.
This can be seen once we use eqs.~\eqref{3unimoddec} and \eqref{VarsNbar} in the explicit definition of $K_{ij}$ given by eq.~\eqref{Kijdef},
    \begin{align}
        K_{ij} & =\underbrace{\frac{a}{2\Nb}\left(\dot{\hb}_{ij}-2\left[D_{(i}\Nb_{j)}\right]^{\srm T}\right)} \quad + \quad \underbrace{a^{-1}h_{ij}\frac{1}{\Nb}\left(\frac{\dot{a}}{a}-\frac{1}{3}D_{i}N^{i}\right)}\\[12pt]
         \frac{1}{2}\mathcal{L}_{n}h_{ij}& =\qquad\qquad\frac{1}{2}a\mathcal{L}_{\bar{n}}\hb_{ij}\qquad\quad\,\,\,\, +\qquad\qquad a^{-1}h_{ij}\mathcal{L}_{\bar{n}}a
    \end{align}
where  $\mathcal{L}_{\bar{n}}$ denotes the projected Lie derivative with respect to $\bar{n}^{\mu}$. 
From these we can read off the following expressions
    \begin{align}
    \label{trtrK}
        K_{ij} & = K_{ij}^{\srm T}+\frac{1}{3}h_{ij}K\,\\[12pt]
        K_{ij}^{\srm T}&=\frac{1}{2}a\mathcal{L}_{\bar{n}}\hb_{ij}=\frac{a}{2\bar{N}}\left(\dot{\bar{h}}_{ij}-2\left[D_{(i}\bar{N}_{j)}\right]^{\srm T}\right)\\[12pt]
        K&=3a^{-1}\mathcal{L}_{\bar{n}}a=3a^{-1}\frac{1}{\bar{N}}\left(\frac{\dot{a}}{a}-\frac{1}{3}D_{i}N^{i}\right)\ .
    \end{align}
We can see that in both expressions above there is a factor of $a$ that appears on both sides in the second equality of each equation. This is precisely the source of conformal covariance of extrinsic curvature which is witnessed from eq.~\eqref{eqn:confK}. Factoring this scale density out, we can define
    \begin{align}
        \ktb_{ij} & := a^{-1}K_{ij}^{\srm T} = \frac{1}{2}\mathcal{L}_{\bar{n}}\hb_{ij}\nonumber\\[12pt]
    \label{Kbardef2}
        & =\frac{1}{2\Nb}\left(\dot{\hb}_{ij}-2\left[D_{(i}\Nb_{j)}\right]^{\srm T}\right)
        =\frac{1}{2\Nb}\left(\dot{\hb}_{ij}-2\left[\bar{D}_{(i}\bar{N}_{j)}\right]^{\srm T}\right)\ ,\\[12pt]
        \kb & := \frac{a K}{3}=\mathcal{L}_{\bar{n}}a\nonumber\\[12pt]
    \label{Ktbardef2}
        & = \frac{1}{\Nb}\left(\frac{\dot{a}}{a}-\frac{1}{3}D_{i}N^{i}\right)
        =\frac{\bar{n}^{\mu}\del_{\mu}a}{a}-\frac{\del_{i}N^{i}}{3\Nb}.
    \end{align}
Note that in the second line in eq.~\eqref{Kbardef2} we have used eq.~\eqref{delVtless} and the fact that $\Nb_{i}$ is a vector density of scale weight $\wb=-2$ according to eq.~\eqref{VarsNbar} to write $\left[D_{(i}\Nb_{j)}\right]^{\srm T}=\left[\bar{D}_{(i}\Nb_{j)}\right]^{\srm T}$, thus showing that this expression, and therefore $\kb_{ij}^{\srm T}$, is independent of $a$ and hence manifestly conformally invariant.
With these definitions the extrinsic curvature can now be compactly expressed as
    \begin{equation}
    \label{Kalldec}
        K_{ij}=a\left(\ktb_{ij} + \hb_{ij}\kb\right)\ .
    \end{equation}
Since conformal transformation affects only the scale density $a$, we have that only the following two objects transform under conformal transformation,
    \begin{align}
    \label{31_confa}
        a  \rightarrow \Omega a &\quad\Rightarrow \quad \delta_{\omega} a =\omega a\\[6pt]
    \label{31_confK}
        \kb  \rightarrow \kb + \bar{n}^{\mu}\del_{\mu}\log\Omega & \quad\Rightarrow \quad \delta_{\omega}\kb = \bar{n}^{\mu}\del_{\mu}\log\omega,
    \end{align}
from which the conformal transformation of eq.~\eqref{Kalldec} is rather obvious,
    \begin{equation}
    \label{31_KconfTransfDec}
        K_{ij}\quad\rightarrow\quad \tilde{K}_{ij}=\Omega a\left(\kb_{ij}^{\srm T} + \hb_{ij}\kb\right)+ \bar{n}^{\mu}\del_{\mu}\log\Omega \quad\Rightarrow \quad \delta_{\omega} K_{ij} = \omega  K_{ij} + \bar{n}^{\mu}\del_{\mu}\log\omega
    \end{equation}
where the last term is just $\mathcal{L}_{\bar{n}}\log\omega$, as in eq.~\eqref{eqn:confKtr}, coming from $\kb$. We refer to $\ktb_{ij}$ as the ``shear density'', while we refer to $\kb$ as the ``expansion density''.
Note that $K_{ij}^{\srm T}$ is usually called ``shear'' and $K$ ``expansion''.
With definitions given by eqs.~\eqref{Kbardef2} and \eqref{Ktbardef2} it becomes clear that the shear is the change of the three-dimensional shape, while the expansion is the change of the three-dimenional scale in time: shapes shear and scales expand (or contract).

Of all these newly introduced rescaled variables only two are not conformaly invariant: the scale density $a$ and the expansion density $\kb$, which is built form $a$. Therefore, one can expect great simplifications in investigation of conformal properties of various expressions. In the same way that we proposed that conformal
properties of four-dimensional covariant expressions are encoded in terms dependent on scale density $A$, conformal properties of $3+1$-decomposed expressions is encoded in terms depending on the scale density $a$ and its space and time derivatives, that sit in the expansion density $\kb$. Indeed, these variables will prove very powerful for this purpose. 
Note that some of the variables introduced here have already been used, mostly in studies on numerical relativity in relation to the Cauchy initial value problem, e.g. in the so-called BSSN formalism\footnote{Baumgarte-Shapiro-Shibata-Nakamura formalism.} \cite{BS,SN}. 
One introduces a new metric conformal to the physical one and requires its determinant to be equal to one --- this is analog to $\hb_{ij}$; in \cite{DB,DB1} BSSN formalism has been recast in a conformally invariant form by relaxing the unit determinant condition
before the evolution equations for the metric and traceless part of the extrinsic curvature have been found. These examples show that unimodular decomposition in $3+1$
formalism has a very useful application. It can already be anticipated that in a genuinely conformally invariant theories this decomposition can simplify investigations of their Hamiltonian formulation considerably.

If there are some conformally covariant non-gravitational fields $\phi_{\ssst I}$ present in a theory under consideration, its conformally invariant part is defined in analogy to eq.~\eqref{fieldresc},
    \begin{equation}
    \label{fieldresc3}
        \chi_{\ssst I}:=a^{-n_{\ssst I}}\phi_{\ssst I}\ ,
    \end{equation}
which is a three-dimensional scalar density of scale weight $\wb_{\ssst I}$. Note that the difference between definitions in eq.~\eqref{fieldresc3} and eq.~\eqref{fieldresc} arises from eq.~\eqref{g31mat} and eq.~\eqref{VarsNbar} since
    \begin{equation}
        A=\Nb^{\frac{1}{4}} a\ ,
    \end{equation}
which means that the scale weight is unchanged after unimodular decomposition of $3+1$ variables. The factor of certain power of $\Nb^{\frac{1}{4}}$ that enters definition in eq.~\eqref{fieldresc} only complicates things if one would stick with that four-dimensional definition of rescaled fields and it does not change the interpretation of the rescaled field.
Namely, note that length scale $l_{0}$ that we introduced by redefining $A$ with eq.~\eqref{Alen} is now found in the scale density $a$. Therefore we have
    \begin{equation}
    \label{lenghta}
        a\rightarrow l_{0}a\ ,
    \end{equation}
which makes sense because it tells about the size of the spatial three-dimensional line element as well. Then it is clear that scale density $a$ is the only variable that has physical dimension. All barred variables (including $N^{i}$) and all non-gravitational fields eq.~\eqref{fieldresc3} are \textit{dimensionless}. 
Therefore, in accordance to the analogous conclusion about the scale density $A$, we expect that the physical dimension of coupling constants and fields in a theory which is decomposed in $3+1$ formalism can be uncovered and tracked with the scale density $a$.

\subsection{Electric and magnetic parts of the Weyl tensor}
\label{EBwumod}

Let us use the so far presented formalism to prove that electric and magnetic parts of the Weyl tensor given by eq.~\eqref{elW} and eq.~\eqref{magW} are conformally invariant\footnote{That these objects should be independent not only of $K$ but equivalently of $a$ and $\kb$ was not noticed by authors in any of previous more detailed works \cite{Blw,ILP,Kaku1982,Kluson2014} that contain $3+1$ formulation of the Weyl-tensor action.}. This will be our first direct application of the unimodular-conformal decomposition for the
purpose of exposing conformal invariance of an object in $3+1$ formalism and to our knowledge such a formulation does not exist in the literature.

Since the Weyl tensor (with one upper index) is conformally invariant, we expect it to be independent of $a$ and $\kb$.
For the magnetic part eq.~\eqref{magW} let us first use the traceless-trace split of $K_{ij}$ in eq.~\eqref{trtrK},
    \begin{align}
        C_{ij}^{\srm B}& =\varepsilon^{kl}{}_{(i\vert}D_{k}K_{\vert j)l}=\varepsilon^{kl}{}_{(i\vert}D_{k}K_{\vert j)l}^{\srm T}+\frac{1}{3}\varepsilon^{kl}{}_{(i}h_{j)l}\del_{k}K=\varepsilon^{kl}{}_{(i\vert}D_{k}K_{\vert j)l}^{\srm T}\ ,
    \end{align}
from which we already see that the term with trace $K$ drops out because $\varepsilon^{kl}{}_{(i}h_{j)l}=\varepsilon^{k}{}_{(ji)}=0$. Therefore, the only stem of conformal transformation is now hidden in the Christoffel symbols and in the conformal weight of $K_{j)l}^{\srm T}$. 
Expanding the covariant derivative, using eq.~\eqref{GammaDec} and the rescaling defined in the first line of eq.~\eqref{Kbardef2} we have,
    \begin{align}
    \label{magWresc}
        C_{ij}^{\srm B} & =\varepsilon^{kl}{}_{(i\vert}\del_{k}K_{\vert j)l}^{\srm T}-\varepsilon^{kl}{}_{(i}\Gb^{b}{}_{j)k}K_{bl}^{\srm T}-\varepsilon^{kl}{}_{(i}\Sigma^{b}{}_{j)k}K_{bl}^{\srm T}
        -\varepsilon^{kl}{}_{(i\vert}\Gb^{b}{}_{lk}K_{\vert j)b}^{\srm T}-\varepsilon^{kl}{}_{(i\vert}\Sigma^{b}{}_{lk}K_{\vert j)b}^{\srm T}\nonumber\\[12pt]
        & =a\Bigpar{
        \varepsilon^{kl}{}_{(i\vert}\del_{k}\ktb_{\vert j)l}-\varepsilon^{kl}{}_{(i}\Gb^{b}{}_{j)k}\ktb_{bl}
        }
        -a\Bigpar{
        \varepsilon^{kl}{}_{(i}\Sigma^{b}{}_{j)k}\ktb_{bl}-\varepsilon^{kl}{}_{(i\vert}\ktb_{\vert j)l}\del_{k}\log a
        }\nonumber\\[12pt]
        &=a\varepsilon^{kl}{}_{(i\vert}\bar{D}_{k}\ktb_{\vert j)l}\ ,
    \end{align}
where the entire second term in the middle line vanishes due to antisymmetrization of the symmetric shape and scale parts of Christoffel symbols over indices $kl$; in the second equality we used eq.~\eqref{Kbardef2} to expose the scale density $a$; in transition to the third equality the second parentheses from the second equality cancels out using the definition of the scale part of Christoffel symbols in eq.~\eqref{Sigma}, i.e. this term is equal to
    \begin{align}
        \underbrace{
        \varepsilon^{kl}{}_{(i}\delta^{b}_{j)}\del_{k}\log a\,\ktb_{bl}
        +\varepsilon^{kl}{}_{(i}\del_{j)}\log a\,\ktb_{kl}-\varepsilon^{kl}{}_{(i}h_{j)k}h^{bc}\del_{c}\log a\ktb_{bl}
        }_{=\varepsilon^{kl}{}_{(i}\Sigma^{b}{}_{j)k}\ktb_{bl}}
        -\varepsilon^{kl}{}_{(i\vert}\ktb_{\vert j)l}\del_{k}\log a=0\ .
    \end{align}
The first and the last term above cancel out, the second term vanishes due to the antisymmetrization of the symmetric pair of indices on $\ktb_{kl}$, while the third term vanishes because $\varepsilon^{kl}{}_{(i}h_{j)k}=0$. 
Now, recall eq.~\eqref{3eps} which says that there is $a^3$ hidden in the Levi-Civita tensor density in eq.~\eqref{magWresc} and note that two indices are raised by two inverse three-metric tensors which also hide $a^{-2}$ each. Exposing all this, we have
    \begin{equation}
    \label{magWunimod}
        \bar{C}_{ij}^{\srm B}=C_{ij}^{\srm B}=a\, a^{3}a^{-2}a^{-2}\hb^{kb}\hb^{kc}\bar{\varepsilon}_{bc(i\vert}\bar{D}_{k}\ktb_{\vert j)l}=\hb^{kb}\hb^{kc}\bar{\varepsilon}_{bc(i\vert}\bar{D}_{k}\ktb_{\vert j)l}\ ,
    \end{equation}
which completes our proof that the magnetic part of the Weyl tensor is conformally invariant, since the only fields affected by a conformal transformation cancel out.
We write an overbar in $\bar{C}_{ij}^{\srm B}$ to denote this fact explicitly.

The electric part defined by eq.~\eqref{elW} requires a bit more manipulation. The main problem here is that the traceless part of $\mathcal{L}_{n}K_{ij}$ is not equal to the Lie derivative of the tracless part of $K_{ij}$.
Starting from the split of $\mathcal{L}_{n}K_{ij}$ into
its traceless and trace part, 
    \begin{equation}
    \label{LieKsplit}
        \mathcal{L}_{n}K_{ij}=\left(\mathcal{L}_{n}K_{ij}\right)^{\srm T}+\frac{1}{3}h_{ij}h^{ab}\mathcal{L}_{n}K_{ab}\ ,
    \end{equation}
using the traceless-trace decomposition of the extrinsic curvature in
eq.~\eqref{trtrK} as well as the following identity, 
    \begin{equation}
    \label{LieKtr}
        \mathcal{L}_{n}K=h^{ab}\mathcal{L}_{n}K_{ab}-2K_{ab}K^{ab},
    \end{equation}
where $K^{ab}=-\mathcal{L}_{n}h_{ab}/2$, one can show that the traceless part of $\mathcal{L}_{n}K_{ij}$ can be expressed in terms of the Lie derivative of $K_{ij}^{\srm T}$,
    \begin{equation}
    \label{eqn:tLieKt}
        \left(\mathcal{L}_{n}K_{ij}\right)^{\srm T}=\mathcal{L}_{n}K^{\srm T}_{ij}+\frac{2}{3}K_{ij}^{\srm T}K-\frac{2}{3}h_{ij}K_{ab}^{\srm T}K^{ab\srm T}.
    \end{equation}
Subtracting $K_{ab}^{\srm T}K$ from both sides we get
    \begin{equation}
    \label{tLieKtelW}
        \left(\mathcal{L}_{n}K_{ab}\right)^{\srm T}-K_{ab}^{\ssst \rm T}K=\mathcal{L}_{n}K^{\srm T}_{ij}-\frac{1}{3}K_{ij}^{\srm T}K-\frac{2}{3}h_{ij}K_{ab}^{\srm T}K^{ab\srm T}\ .
    \end{equation}
Note that by taking the trace of eq.~\eqref{tLieKtelW} we obtain
    \begin{equation}
    \label{trLieKt}
        h^{ij}\mathcal{L}_{n}K^{\srm T}_{ij}=2K_{ij}^{\srm T}K^{ij\srm T}\ ,
    \end{equation}
which actually simply follows also from $\mathcal{L}_{n}h^{ij}K_{ij}^{\srm T}=0$. These manipulations allow us to trade $\left(\mathcal{L}_{n}K_{ab}\right)^{\srm T}$ for $\mathcal{L}_{n}K^{\srm T}_{ij}$ in eq.~\eqref{elW} using eq.~\eqref{tLieKtelW}, leaving us with
    \begin{equation}
    \label{elWsmpl}
        C_{ij}^{\srm E}=\mathcal{L}_{n}K_{ij}^{\srm T}-\frac{1}{3}K_{ij}^{\srm T}K-\frac{2}{3}h_{ij}K_{ab}^{\srm T}K^{ab\srm T}-\!\,^{\ssst(3)}\!R_{ij}^{\srm T}-\frac{1}{N}D_{ij}^{\srm T}N \ .
    \end{equation}
This expression is still not manifestly conformally invariant and there is still an explicit dependence on $K$, which should somehow cancel out.
To make conformal invariance evident we apply the unimodular-conformal decomposition by using eqs.~\eqref{VarsNbar}, \eqref{Kbardef2} and \eqref{Ktbardef2} in eqs.~\eqref{LieKdef1} and \eqref{LieKdef2} with  $K_{ij}^{\srm T}$ instead of $K_{ij}$, in order to separate any scale-dependent pieces in the first term in eq.~\eqref{elWsmpl}.
This results in
    \begin{align}
    \label{LieKtdec}
        \mathcal{L}_{n}\ktb_{ij} & =\frac{1}{N}\Big(\dot{K}_{ij}^{\srm T}  - \mathcal{L}_{\vec{N}}K_{ij}^{\srm{T}}\Big) = \nld
        & =\frac{1}{a\Nb}\Big(
        a\dot{\kb}_{ij}^{\srm T}  - 
            a N^{b}\del_{b}\ktb_{ij} - a \ktb_{ik}\del_{j}N^{k} - a\ktb_{kj}\del_{i}N^{k} + \underbrace{\kb_{ij}^{\srm T}\dot{a} - \ktb_{ij}N^{b}\del_{b} a}_{= a\Nb\kb_{ij}^{\srm T} \bar{n}^{\mu}\del_{\mu}a}
        \Big)\nld
        & = \frac{1}{\Nb}\Big(
        \dot{\kb}_{ij}^{\srm T}  - 
             N^{b}\del_{b}\ktb_{ij} - \ktb_{ik}\del_{j}N^{k} - \ktb_{kj}\del_{i}N^{k} \Big) 
             + \kb_{ij}^{\srm T}\frac{\bar{n}^{\mu}\del_{\mu}a}{a}\nld
        & \stackrel{eq.~\eqref{Ktbardef2}}{=} \frac{1}{\Nb}
        \dot{\kb}_{ij}^{\srm T}  - 
             \frac{1}{\Nb}\underbrace{\left(N^{b}\del_{b}\ktb_{ij} + \ktb_{ik}\del_{j}N^{k} + \ktb_{kj}\del_{i}N^{k}  
             - \frac{1}{3}\kb_{ij}^{\srm T}\del_{k}N^{k} \right)}_{= \mathcal{L}_{\vec{N}}\ktb_{ij}} + \kb_{ij}^{\srm T}\kb\nld
        & = \frac{1}{\Nb}
        \dot{\kb}_{ij}^{\srm T} - \frac{1}{\Nb}\mathcal{L}_{\vec{N}}\ktb_{ij} + \kb_{ij}^{\srm T}\kb\ ,
    \end{align}
from which it follows
    \begin{equation}
        \mathcal{L}_{\bar{n}}\ktb_{ij} = \mathcal{L}_{n}\ktb_{ij} - \kb_{ij}^{\srm T}\kb\ ,
    \end{equation}
where 
    \begin{equation}
    \label{LieKtdec1}
        \mathcal{L}_{\bar{n}}\ktb_{ij} := \frac{1}{\Nb}
        \dot{\kb}_{ij}^{\srm T} - \frac{1}{\Nb}\mathcal{L}_{\vec{N}}\ktb_{ij}\ .
    \end{equation}
Note that $\mathcal{L}_{\vec{N}}\ktb_{ij}$ is the Lie derivative of a tensor \textit{density} of weight $-1/3$ (corresponding to scale weight $-1$).
Furthermore, from now on we shall write
    \begin{align}
    \label{Kbold}
        \bktb &\equiv \ktb_{ij}\ ,\\[6pt]
    \label{KKbold}
        \bktb \cdot \bktb &:= \ktb_{ij}\hb^{ia}\hb^{jb}\ktb_{ab}\ ,
    \end{align}
where ``$\cdot$'' denotes a contraction over all available pairs of indices\footnote{This bold-font notation will be used on more occasions in this thesis.}. 
We thus obtain
    \begin{equation}
    \label{elWumod}
        \bar{C}_{ij}^{\srm E}=C_{ij}^{\srm E}=\mathcal{L}_{\bar{n}}\kb_{ij}^{\srm
          T}-\frac{2}{3}\hb_{ij}\bktb \cdot \bktb - \,^{\ssst
          (3)}\!\bar{R}_{ij}^{\srm
          T}-\frac{1}{\Nb}\left[\bar{D}_{i}\del_{j}\Nb\right]^{\srm
          T}\ , 
    \end{equation}      
from which $a$ and $\kb$ have cancelled out, as expected, and we put an overbar as we did for the magnetic.
The last two terms in eq.~\eqref{elWsmpl} still seemingly contain $a$ and its first and second derivatives, however, we have proved in appendix \ref{App_DD} that the scale density cancels out, see eq.~\eqref{DDRT}, resulting in the last two terms in eq.~\eqref{elWumod}. 
(It should be kept in mind that since these two terms separately are not $GL(3,\mathbb{R})$ tensors but only $SL(3,\mathbb{R})$ tensors, they should always be considered together.)
We have thereby exposed the manifest conformal invariance of both the electric and the magnetic parts of the Weyl tensor.
For chapter \ref{ch:HDclass} we will need the square of the Weyl tensor, whose decomposition in $3+1$ formalism can be found in appendix \ref{App_31} resulting in eq.~\eqref{C2dec} in terms of the electric and magnetic parts.
Then, using the results of this subsection, the following holds,
    \begin{equation}
    \label{W2EB}
        \sqrt{g}C_{\mu\nu\lambda\rho}C^{\mu\nu\lambda\rho}
        =2\Nb\left( \bar{\mathbf{C}}^{\srm E}\cdot\bar{\mathbf{C}}^{\srm E} - 2\bar{\mathbf{C}}^{\srm B}\cdot\bar{\mathbf{C}}^{\srm B}\right)
    \end{equation}
where
    \begin{equation}
    \label{EBbold}
         \bar{\mathbf{C}}^{\srm E}\cdot\bar{\mathbf{C}}^{\srm E}\equiv\bar{C}_{ij}^{\srm E}\hb^{ik}\hb^{jl}\bar{C}_{kl}^{\srm E}\ ,\qquad
         \bar{\mathbf{C}}^{\srm B}\cdot\bar{\mathbf{C}}^{\srm B}\equiv \bar{C}_{ij}^{\srm B}\hb^{ik}\hb^{jl}\bar{C}_{kl}^{\srm B}\ ,
    \end{equation}
since the scale density $a$ cancels out, thus showing the manifest conformal invariance of the weighted square of the Weyl tensor.

\subsection{Ricci scalar}
\label{Rumod}

The Ricci scalar is important for the Hamiltonian formulation of not only General Relativity but also of non-minimally coupled scalar field theory and semiclassical gravity in which terms such as $R^{2}$ usually appear in the action.
Its explicit dependence on derivatives of $K$ signals that this object is not conformally invariant.
But we can still simplify it in the spirit of the material presented so far.

Using eqs.~\eqref{VarsNbar}, \eqref{Varsnbar}, \eqref{LieResc} and \eqref{Ktbardef2} the Lie derivative of $K$ along the timelike normal vector $n^{\mu}$ decomposes as follows,
    \begin{align}
    \label{LieKdec}
        \mathcal{L}_{n}K = 3 a^{-1}\mathcal{L}_{\bar{n}}\left(a^{-1}\kb\right) & =
        3 a^{-2} \mathcal{L}_{\bar{n}}\kb - a^{-2}\kb\frac{\mathcal{L}_{\bar{n}}a}{a}\nld
        & = 3 a^{-2} \mathcal{L}_{\bar{n}}\kb - a^{-2}\kb^2\nl
        & = 3 a^{-2} \left(\frac{1}{\Nb}\dot{\kb} - \frac{N^{i}}{\Nb}\del_{i}\kb - \frac{1}{3}\frac{\del_{i}N^{i}}{\Nb} \right) - 3 a^{-2}\kb^2 \nld
        & = 3 a^{-2}\bar{n}^{\mu}\del_{\mu}\kb - \frac{1}{3}\frac{\del_{i}N^{i}}{\Nb} - 3 a^{-2}\kb^2\ ,\\[12pt]
    \label{LieKdecn}
        \mathcal{L}_{\bar{n}}\kb & : = \frac{1}{ \Nb}\dot{\kb} - \frac{1}{\Nb}\mathcal{L}_{\vec{N}}\kb\ .
    \end{align}
The above result can also be written as
    \begin{align}
    \label{LieKdec1}
        \mathcal{L}_{n}K = a^{-2}\left(\frac{3}{ \Nb}\dot{\kb} - \frac{3}{\Nb}\mathcal{L}_{\vec{N}}\kb - 3\kb^2\right) = 3a^{-2}\mathcal{L}_{\bar{n}}\kb - 3a^{-2}\kb^2\ ,
    \end{align}
where $\mathcal{L}_{\vec{N}}$ is the Lie derivative of the scalar density $\kb$ of weight $1/3$ (corresponding to scale weight 1) with respect to shift vector $N^{i}$,
    \begin{equation}
    \label{LieKbarN}
        \mathcal{L}_{\vec{N}}\kb = N^{i}\del_{i}\kb + \frac{1}{3}\del_{i}N^{i}\kb\ .
    \end{equation}
Then using the unimodular-conformal variables and eq.~\eqref{LieKdec} in eq.~\eqref{Rdec31} we obtain
    \begin{align}
    \label{Rdec31umod}
        R =a^{-2}\left( 6 \,\mathcal{L}_{\bar{n}}\kb 
        + 6 \kb^2 
        + \bktb \cdot \bktb
        \right)
        + \,^{\ssst (3)}\! R - \frac{2a^{-2}}{\Nb}\mathbf{D}\cdot\mathbf{D}\Nb\ ,
    \end{align}
where we leave the last two terms undecomposed because $a$ cannot cancel out from there. Therefore, we have exposed manifest conformal non-invariance of the Ricci scalar in $3+1$ formalism.
A more suitable form of the Ricci scalar will be of use for the Hamiltonian formulation of GR in section \ref{sec_HEH}, namely, the one given by eq.~\eqref{Rdec31a} but in unimodular-conformal variables,
    \begin{equation}
    \label{Rdec31umod2}
        R = a^{-2}\Bigl(a^2\,^{\ssst (3)}\! R
        + \bktb\cdot\bktb - 6\kb^{2} + 6\nabla_{\mu}\left(\bar{n}^{\mu}\kb\right)
	    -\frac{2}{\Nb}\mathbf{D}\cdot\mathbf{D}\Nb\Bigr)
    \end{equation}
which is obtained by simply using eqs.~\eqref{VarsNbar} and \eqref{Kalldec} in eq.~\eqref{Rdec31a}.
Other curvature tensors could be dealt with in a similar way but such a complete treatment would take more than intended space of this thesis.

\section{An example: non-minimally coupled scalar field}
\label{sec_xiphi}

The Lagrangian of a non-minimally coupled scalar field with a potential is an excellent example for demonstrating the power of unimodular-conformal decomposition. It is given \cite{BD} by
    \begin{equation}
    \label{Lagnm}
        \mathcal{L}^{\varphi}=-\frac{1}{2}\sqrt{g}\left(g^{\mu\nu}\del_{\mu}\varphi\del_{\nu}\varphi+\xi R\varphi^2+2V(\varphi)\right)\ ,
    \end{equation}
where $\xi$ is a dimensionless non-minimal coupling and $V(\varphi)$ is the potential term (e.g. $V(\varphi)=m^2\varphi^2/2$).
For $\xi=(d-2)/4(d-1)$ and a potential either vanishing or proportional to $\varphi^{2d/(d-2)}$ Lagrangian in eq.~\eqref{Lagnm} is conformally invariant up to a total divergence.
However, this is not at all apparent from the form of eq.~\eqref{Lagnm}.
The same is true for the Klein-Gordon (KG) equation, which is derived by varying the above action with respect to $\varphi$ and has the following form
    \begin{subequations}
    \begin{align}
    \label{KGphi}
        \frac{1}{\sqrt{g}}\del_{\mu}\left(\sqrt{-g}g^{\mu\nu}\del_{\nu}\varphi\right)-\xi R\varphi-V'(\varphi)&=0\\[12pt]
    \label{KGphi1}
        \frac{1}{2\sqrt{g}}\del_{\mu}\left(\sqrt{-g}g^{\mu\nu}\del_{\nu}\varphi^2\right)-\frac{1}{4\varphi^2}g^{\mu\nu}\del_{\mu}\varphi^2\del_{\nu}\varphi^2-\xi R\varphi^2-\varphi V'(\varphi)&=0\ ,
    \end{align}
    \end{subequations}
where the prime denotes its derivative with respect to $\varphi$ and we gave it in another form by eq.~\eqref{KGphi1} as well because some expressions we encounter later simplify if they are expressed in terms of $\varphi^2$.
It is the purpose of this section to show how can conformal features of this action be exposed using the unimodular-conformal decomposition in both $d$-dimensional and $3+1$ formulation.

\subsection{Covariant formulation in \texorpdfstring{$d$}{TEXT} dimensions}

Using eq.~\eqref{fieldresc} with $s=n_{\varphi}=-\wb$ being the negative of the scale weight $\wb$ of the new field $\chi$ yet to be determined:
    \begin{align}
    \label{kinetic}
        \sqrt{g}g^{\mu\nu}\del_{\mu}\varphi\del_{\nu}\varphi
        & = A^{d-2(1-s)}\gb^{\mu\nu}\Biggl[
        \del_{\mu}\chi\del_{\nu}\chi 
        + s A^{-1}\del_{\mu}A\del_{\nu}\chi^2
        + \gruline{s^2 A^{-2}\del_{\mu}A\del_{\nu}A\chi^2}
        \Biggr]\ ,\\[12pt]
    \label{nonmin}
        \xi\sqrt{g} R \varphi^2 & = 
        \xi A^{d-2(1-s)}\Biggl[
        \bar{R} -\reduline{2(d-1) A^{-1}
        \del_{\mu}\left(\gb^{\mu\nu}\del_{\nu}A\right)}\nonumber\\[6pt]
        &\qquad\qquad\qquad\qquad -\gruline{(d-1)(d-4)A^{-2}\gb^{\mu\nu}\del_{\mu}A\del_{\nu}A}
        \Biggr] \chi^2\ ,\\[12pt]
        \sqrt{g}V(\varphi) & =: A^{d-2(1-s)}\bar{V}(\chi,A)\ .
    \end{align}
We may choose to use partial integration either on the second term in eq.~\eqref{kinetic} or in the second term in eq.~\eqref{nonmin}.
Since terms in eq.~\eqref{nonmin} come from the Ricci scalar, it is advisable to stay as close to its original form as possible as we have split it into non-tensorial quantities using the fourth line in eq.~\eqref{Rdec}. 
That means it is better to chose the former possibility, for which we obtain the following form,
    \begin{align}
    \label{chipartInt}
        sA^{d-3+2s}\del_{\mu}A\del_{\nu}\chi^2 & = s\del_{\mu}
        \Bigpar{
        A^{d-3+2s}\chi^2\gb^{\mu\nu}\del_{\nu}A
        }
        -\chi^2\del_{\nu}
        \Bigpar{
        A^{d-3+2s}\gb^{\mu\nu}\del_{\nu}A
        }\nonumber\\[12pt]
        & =s\del_{\mu}
        \Bigpar{
        A^{d-3+2s}\chi^2\gb^{\mu\nu}\del_{\nu}A
        }\nonumber\\[12pt]
        & \quad - s A^{d-2(1-s)}\Bigsq{
        \reduline{A^{-1}\del_{\mu}\left(\gb^{\mu\nu}\del_{\nu}A\right)}
        + \gruline{(d-3+2s)A^{-2}\gb^{\mu\nu}\del_{\mu}A\del_{\nu}A}
        }\chi^2\ ,
    \end{align}
where the first term is a total divergence and will contribute to a boundary term in the action. 
Now, summing \txr{red} and \txg{green} underlined terms together, the Lagrangian of the scalar field $\varphi$ is reformulated as Lagrangian of the scalar density field $\chi$ and is settled into the following form
    \begin{align}
        \mathcal{L}^{\varphi}=\mathcal{L}^{\chi}&=-\frac{1}{2}A^{d-2(1-s)}\Biggl[
        \gb^{\mu\nu}\del_{\mu}\chi\del_{\nu}\chi
        + \xi\bar{R}\chi^2
        + 2\bar{V}(\chi,A)\nonumber\\[6pt]
        &\quad - \txr{a}A^{-1}\del_{\mu}\left(\gb^{\mu\nu}\del_{\nu}A\right)\chi^2
        + \txg{b}A^{-1}\gb^{\mu\nu}\del_{\mu}A\del_{\nu}A\chi^2
        \Biggr]\nonumber\\[12pt]
        &\quad -\frac{s}{2}\del_{\mu}\left(\chi^2 A^{d-3+2s}\gb^{\mu\nu}\del_{\nu}A\right)\ ,
    \end{align}
where the coefficients resulting from the addition of terms are correspondingly marked with red and green colors and they are given by
    \begin{equation}
    \label{coeff}
        \txr{a}=s + 2\xi(d-1)\ ,\quad \txg{b}=s^2 - \xi(d-1)(d-4)-s(d-3+2s)\ .
    \end{equation}
These two terms and the potential are the ones responsible for breaking the conformal symmetry of the non-minimally coupled scalar field, apart from the total divergence term and an overall factor of $d-2(1-s)$ powers of $A$.

Let us now determine the length dimension (and therefore the scale weight) $\wb = - s$. This can be done by demanding that the kinetic term is $A$-independent. This ensures that the kinetic term explicitly has dimension of $[\hbar]$ and is conformally invariant. Such a demand is satisfied if $d-2(1-s)=0$, which sets the scale weight of $\chi$ to be
    \begin{equation}
    \label{sweight}
        s=\frac{2-d}{2}\quad\Rightarrow\quad \wb = \frac{d-2}{2}\ .
    \end{equation}
Then coefficients in eq.~\eqref{coeff} reduce to
    \begin{equation}
        \txr{a}=2(d-1)(\xi-\xi_{cf})\ ,\qquad 
        \txg{b}=-(d-1)(d-4)(\xi-\xi_{cf})\ ,
    \end{equation}
so we can write down the final form of the Lagrangian
    \begin{align}
    \label{nonmin_Lag_decomp}
        \mathcal{L}^{\chi}&=-\frac{1}{2}\Biggl[
        \gb^{\mu\nu}\del_{\mu}\chi\del_{\nu}\chi
        + \xi\bar{R}\chi^2
        + 2\bar{V}(\chi,A)
        \nonumber\\[12pt]
        &\qquad
        -2(d-1)\xi_c\biggpar{A^{-1}
        \del_{\mu}\left(\gb^{\mu\nu}\del_{\nu}A\right)\nonumber +\frac{d-4}{2}A^{-2}\gb^{\mu\nu}\del_{\mu}A\del_{\nu}A
        }
        \chi^2
        \Biggr]
        \nonumber\\[12pt]
        &\qquad
        +\frac{d-2}{4}\del_{\mu}\left(\chi^2 A^{-1}\gb^{\mu\nu}\del_{\nu}A\right)\ ,
    \end{align}
where $\xi_{cf}$ and $\xi_{c}$ are defined by\footnote{We have assumed $d\neq 1$, which is a trivial case of no interest here.}
    \begin{equation}
    \label{xiconf}
        \xi_{cf}:=\frac{d-2}{4(d-1)}\ ,\quad \xi_{c}:=\xi-\xi_{cf} \ .
    \end{equation}
The importance of eq.~\eqref{xiconf} is obvious: all interaction terms between $A$ and $\chi$ disappear for the special case $\xi_{c}=0 \Leftrightarrow \xi=\xi_{cf}$ (except the potential term and the total divergence). This value of $\xi$ is called \textit{conformal coupling} \cite{DB}.
Moreover, compare that $A$-dependent expression with the one given in eq.~\eqref{nonmin}; exposing $A$ in the kinetic term changes the coefficient of these terms in eq.~\eqref{nonmin} from $\xi$ to $\xi_{c}$.

How does the KG equation for the scalar density $\chi$ look like now? We derive it from Lagrangian eq.~\eqref{nonmin_Lag_decomp} to be
    \begin{align}
    \label{KGchi}
        \ddel{S_{\chi}}{\chi}&=0\quad\Rightarrow\quad \del_{\mu}\left( \gb^{\mu\nu}\del_{\nu}\chi\right) -\xi\bar{R}\chi-
        \dd{\bar{V}(\chi,A)}{\chi}\nonumber\\[6pt]
        &
        +2(d-1)\xi_{c}\biggpar{A^{-1}
        \del_{\mu}\left(\gb^{\mu\nu}\del_{\nu}A\right) +\frac{d-4}{2}A^{-2}\gb^{\mu\nu}\del_{\mu}A\del_{\nu}A
        }
        \chi=0\ .
    \end{align}
Comparing eq.~\eqref{KGchi} with eq.~\eqref{KGphi}, as well as eq.~\eqref{nonmin_Lag_decomp} with eq.~\eqref{Lagnm}, we witness the isolation of all conformally-variant terms and complete decoupling of $A$ from the scalar density $\chi$ in the case of conformal coupling and vanishing $\del \bar{V}/\del A$.
Note that any coupling constant related to the interactions with $\chi$ appears only in $\bar{V}$.
Moreover, eq.~\eqref{KGchi} might be simpler to handle in certain models due to the simplification of the d'Alambertian. We shall see the advantages of using this KG equation in the upcoming chapters.

Formulation of the scalar field theory in terms of the unimodular-conformal variables 
shows that \textit{any breaking of conformal symmetry must come from the presence of $A$, the scale degree of freedom of the metric as the only field responsible for conformal transformation.}
In fact, the whole purpose of the unimodular-conformal decomposition could be motivated with the single example of non-minimally coupled scalar field: formulate the theory in terms of such variables that only
the scale $A$ (and objects derived from it) is affected by a  conformal transformation. 
But does this result generalize to other theories as well?

To prepare an answer this question, it is useful to first formalize this result.
How can we formally state the dependence of an action on the scale density $A$? The key is to use the notion of variational derivative of the action or the Lagrangian with respect to the scale $A$,
    \begin{align}
    \label{varLA}
        A\ddel{\mathcal{S}_{\chi}}{A} & = A\ddel{\mathcal{L}_{\chi}}{A} 
        =-A\dd{\bar{V}}{A} -\xi_{c}(d-1)\times \nonumber\\[6pt] 
        &\quad
        \times\Biggpar{
        \del_{\mu}\left( \gb^{\mu\nu}\del_{\nu}\chi^2\right)
        -(d-2)\Bigsq{
        A^{-1}\del_{\mu}\bigpar{
        \gb^{\mu\nu}\del_{\nu}A
        }+\gb^{\mu\nu}A^{-1}\del_{\mu}A
        \del_{\nu}
        }\chi^2
        }\ .
    \end{align}
The reason for multiplying the variational derivative with $A$ will become clear in the next chapter. For now, assume first that the potential is independent of $A$, i.e. the first term in the above expression vanishes. 
Then we notice that variation given by eq.~\eqref{varLA} is \textit{identically zero} if we have conformal coupling $\xi_{c}=0$ and an $A$-independent potential
    \begin{equation}
    \label{Vconf}
        \dd{\bar{V}}{A}\equiv 0 
    \end{equation}
simultaneously.
Some obvious examples where this can be tested are the mass term and the $\varphi^4$ term in $d=4$ dimensions,
    \begin{align}
    \label{Vmass}
        \sqrt{-g}V  =
        \frac{1}{2}\sqrt{-g}\,m^2\varphi^2  & \quad\rightarrow\quad \bar{V}=\frac{1}{2}l_{0}^2m^2A^2\chi^2\ ,\\[6pt]
    \label{V4}
        \sqrt{-g}V  = \frac{1}{4}\sqrt{-g}\,\lambda\varphi^4 &  \quad\rightarrow\quad \bar{V}=\frac{1}{4}\lambda\chi^4\ ,
    \end{align}
whereas a $d>2$-dimensional generalization of eq.~\eqref{V4} is given by
    \begin{equation}
    \label{Vd}
        \sqrt{-g}V  = \frac{1}{n}\sqrt{-g}\,\lambda\varphi^n  \quad\rightarrow\quad \bar{V}=\frac{\lambda}{\frac{2d}{d-2}}\chi^{\frac{2d}{d-2}}\ ,
    \end{equation}
and is $A$-independent for $n=2d/(d-2)$, $\lambda$ being a dimensionless constant.
Potential in eq.~\eqref{Vmass} has explicit \textit{dimensionful} coupling $m$ and thus breaks conformal symmetry and it is important to observe that $l_{0}$ explicitly appears \textit{together} with this dimensionful coupling. 
In contrast to this term, potential in eq.~\eqref{Vd} does not depend on $A$ and thus  $l_{0}$ cancels out, so $\lambda$ is \textit{dimensionless} and this term preserves the conformal symmetry. 
Therefore, if one wants to have a conformally invariant Lagrangian for the scalar field, the potential needs to be conformally invariant, which translates to an \textit{independence on dimensionful coupling constants}, allowing only eq.~\eqref{Vd}. One can anticipate that this conclusion is quite general and we will address this in the next chapter.

\subsection{3+1 formulation}
\label{sec_31chi}

For practical purposes we need also the $3+1$ decomposition of the previous section's result. It is not straightforward to simply apply the results of appendix~\ref{App_31} and section~\ref{sec_31conf} to
Lagrangian in eq.~\eqref{nonmin_Lag_decomp} and KG eq.~\eqref{KGchi}, because one ends up with many derivatives of $\Nb$ due to definition $\varphi=\chi/A^{\wb} = \chi /(a\Nb)^{\wb}$.
In other words, one has to be careful whether $\Nb$ is included in the definition of the new field $\chi$ or not because these two are not the same. 
These two definitions coincide only for $\Nb=1$, which corresponds to the choice of the so-called ``conformal time''.
We shall choose to work with
    \begin{equation}
    \label{chidef3}
        \chi:= a \varphi
    \end{equation}
because such a definition does not depend on the choice of $\Nb$. We use the same letter to designate this new field as in the covariant case, but make sure to make it clear within the context in question (it will be clear which one because we shall not mix covariant with $3+1$ formalism within one section/derivation). 
But we anyway have to go through a tedious but straightforward calculation in order to express conformal features manifestly in $3+1$ formalism. 
We have done so in appendix \ref{app_31chi} where it can be observed that independence of the Lagrangian on the scale density $a$ and the expansion density $\kb$ is achieved precisely for conformal coupling, as these are the only objects which transform under conformal transformation. 

However, the final expression for the Lagrangian presented in eq.~\eqref{eqn:nonx_LagdeSone}, is not so easy to work with and for this reason we here rewrite it in a more familiar and compact form which is
particularly suitable for studying perturbations of $\chi$ on a spatially homogeneous background spacetimes (but we do not assume spatial homogeneity here).
The only difference will be in the second line of eq.~\eqref{eqn:nonx_LagdeSone}, which we trace back to the combination of eqs.~\eqref{eqn:nonx_IIa} and \eqref{eqn:nonx_IIb} on one hand and eq.~\eqref{eqn:nonx_Ricdec} on the other.
Namely, here we do not decompose $a^{2}\,^{\ssst (3)}\!R$ as in eq.~\eqref{eqn:nonx_Ricdec} but only collect $a$-dependent terms from the former two equations.
Doing so, the Lagrangian takes the following form
    \begin{align}
    \label{chi31Lagdec}
        \mathcal{L}^{\varphi}=\mathcal{L}^{\chi} & = \frac{1}{2}\Nb\Bigg[\left(\bar{n}^{\mu}\del_{\mu}\chi
        + 6\xi_{c}\bar{K}\chi
        - \frac{\del_{i}N^{i}}{3\Nb}\chi\right)^2
        - V^{\ssst\chi} \Bigg]
        - \xi\del B
        + \xi\rm{BT}
    \end{align}
where we defined the potential as
    \begin{align}
    \label{Vchi_def}
    V^{\ssst\chi} & := 
        U^{\ssst\chi} + 36\xi\xi_{c}\kb^2\chi^2 + \xi\bktb\cdot\bktb\chi^2 \\[12pt]
    \label{Uchi_def}
    U^{\ssst\chi} & := \xi a^{2}\,^{\ssst (3)}\!R\chi^2
        + \hb^{ij}\del_{i}\chi\del_{j}\chi\nonumber\\[12pt]
    &\quad 
        - \hb^{ij}\del_{i}\log a\,\del_{j}\chi^2
        + \chi^2\hb^{ij}\del_{i}\log a\del_{j}\log a
        + \xi D_{j}\left(\hb^{ij}D_{j}\chi^2\right)\ ,
    \end{align}
and where $\xi_{c}=\xi-1/6$, recalling eq.~\eqref{xiconf}, and the total divergence term now has the form:
    \begin{equation}
    \label{chi31_divs}
        \rm{BT}=\del_{i}\left(\chi^2\hb^{ij}D_{j}\Nb-\Nb\hb^{ij}D_{j}\chi^2\right)\ ,
    \end{equation}
while $\del B$ is given by eq.~\eqref{eqn:nonx_delB}. This total divergence and the last term in the third line of eq.~\eqref{chi31Lagdec} arise from expanding $\sqrt{h}h^{ij}D_{i}D_{j}N\,\varphi^2$ in terms of unimodular-conformal variables but in a different way than in eq.~\eqref{eqn:nonx_IIb} and this term vanishes if $N$ does not depend on spatial coordinates, as it is the case in spatially homogeneous spacetimes.
Therefore, only the first two lines of eq.~\eqref{chi31Lagdec} survive for spatially homogeneous models.

Even though it is not so obvious from eq.~\eqref{chi31Lagdec} that conformal invariance is achieved for conformal coupling $\xi=\xi_{cf}=1/6$, $\xi_{c}=0$, we can use that form of the Lagrangian straightforwardly in calculations for this case.
However, for this thesis only spatially homogeneous models will be relevant and in that case the Lagrangian simplifies significantly and conformal invariance is manifest.

\section{Final remarks}
\label{conc2}

We give a few final general remarks on the conformal symmetry and unimodular-conformal decomposition introduced in the current chapter. 
This chapter was a slow-paced invitation for introducing the unimodular-conformal decomposition in four (see eq.~\eqref{gsplit1}) and 3+1 dimensions (see eqs.~\eqref{3unimoddec}-\eqref{Ktbardef2}) including the recipe for its utilisation in basic geometric objects used in Riemannian geometry.
The main point to take away from this chapter is that separating the scale
density as the geometrical meaning of \textit{dimensionful} ``size'' not only from the metric but also from the other fields exposes any implicit conformal properties of any expression by revealing them as $A$-dependent (in full covariant treatment) or $a$- and $\kb$-dependent 
(in $3+1$ treatment) terms. This also exposes physical length dimension of a field by an appropriate rescaling with a scale density such that the conformal weight (and therefore the length dimension) is compensated for.
Then ``a test'' of conformal invariance of any expression could be formulated as a test of whether or not expressions depend on the only conformally non-invariant fields in a theory: $A$ or $a$ and $\kb$.
The example of non-minimally coupled scalar field presented in section \ref{sec_xiphi} clearly supports this conclusion.
A concrete formulation of such test is precisely the topic of the upcoming chapter. 
Then, based on the fact that the vanishing of eq.~\eqref{varLA} eliminates any $A$ is equivalent to the claim that in such a case the action is confromally invariant, we anticipate that the variational independence of an action on $A$ can be read as: \textit{if an action does not respond to the variations of the scale $A$ then such an
action is conformally invariant.} 
A remarkable consequence of this and the fact that we consider coordinates as dimensionless but the metric dimensionful is that
independence on $A$ clearly implies the absence of dimensionful coupling constants and we shall revisit this important observation as well.
This asks for a concrete definition of conformal invariance that can quite generally be applied to any field theory, as we shall see in the next chapter.
A few more side remarks are given below before we move on.

\textbf{A note on Weyl gauging}.
It should be kept in mind that there is a way of implementing true local invariance under the choice of units and this is referred to as \textit{the Weyl gauging}, initiated by Weyl himself \cite{Weyl18a}. A modern formulation within the context of gauge theory of gravity can be found in \cite{HB}.
This and more general variations of this idea are recently becoming again important~\cite{Esch} and one of the reasons is the search and discovery of the Higgs particle in LHC as the only
known neutral scalar field, which is responsible for giving a definite scale in the Standard Model of particle physics ($\sim 125 \text{GeV}$) and giving mass to other fields through interactions with them. 
At the energies above the scale of the Higgs symmetry breaking mechanism, the formal Lagrangian of the Standard Model enjoyed conformal symmetry with an exception of the formal mass term of the Higgs field.
With the appearance of a definite dimensionful scale, this symmetry is formally broken. Therefore, it is rather important to study the role of local conformal invariance and its breaking by implementing it in a theory in a certain way. 
Weyl gauging is one way to do it and it basically consist of promoting the Riemannian geometry to the so-called \textit{Weyl geometry} \cite{Wheel} in which the affine connection is conformally invariant (unlike the Levi-Civita connection). This ``deviation'' from the
Levi-Civita connection is expressed in terms of the \textit{non-metricity} such that instead of eq.~\eqref{metricity} one has $\nabla_{\alpha}g_{\mu\nu}=-2Q_{\alpha}g_{\mu\nu}$, where $Q_{\alpha}$ is called \textit{the Weyl vector}. Weyl vector serves a similar purpose as the $U(1)$
connection of electromagnetism $A_{\mu}$ --- to establish the \textit{local Weyl gauge symmetry}, i.e. the symmetry under local conformal rescaling, which is ambiguously referred to as the conformal or scale symmetry. 
One thus has the possibility to explore the interactions and relationships of the Weyl vector (especially in a particular case where it is described as the gradient of a scalar field) with the Higgs field and basic ideas are reviewed in \cite{Esch1,Esch2}; see also \cite{Ghil} for a recent and representative treatment of quadratic curvature\footnote{Terms such as $R^2$ and  $C_{\mu\alpha\nu\beta}C^{\mu\alpha\nu\beta}$ enter the action, beside the EH term.
Quadratic curvature gravity in Reimannian geometry context is the topic of this thesis.} gravity within the Weyl geometry in relation to the spontaneous symmetry breaking and the Higgs mechanism.
In contrast to Weyl gauging, in this thesis we talk about conformal symmetry without leaving the Riemannian geometry, but we think that the context of Weyl geometry would be a reasonable next step in which one could study a quantum gravity theory.

\textbf{A note on the use of unimodular decomposition in renormalization methods.}
One could imagine that there is certain hope that unimodular decomposition can simplify calculations not only in classical theories but also in covariantly formulated quantum
field theory as well. Namely, as mentioned in the Introduction, if one advances any classical field theory towards higher energies, one requires perturbative modifications due to quantum corrections \cite{BD}. One is then faced with tools of
renormalization, a method of redefining coupling constants and fields in a theory such that they depend on the energy scale and are able to absorb divergent terms that appear when one takes into account the quantum fields. It turns out that this is a necessary
procedure if the coupling constants and involved fields have non-zero length dimension and this has to do with coordinate dilation and conformal invariances (since this is effectively a change of unit of length).
Now, it is shown in Kalmykov and Kazakov
\cite{KK1} on a general model of quadratic gravity that the use of unimodular decomposition (they call it ``conformal parametrization'') simplifies certain results of renormalization. Namely, 
in the standard approach Newton gravitational constant $G$ needs to be renormalized and this procedure is not gauge-independent, but repeating the procedure with the use of unimodular decomposition it turns out that renormalization of $G$ is not necessary in a spacetime
without a boundary to all orders of the perturbation theory both with and without massless fields interactions. The renormalization procedure shifts to the metric, in particular to the scale density (``conformal mode'' in their paper). Hence, the ``dimensionfulness'' of $G$ is taken care of through the renormalization of the scale density if one uses the unimodular decomposition of the metric.
This makes sense because the scale density is the one that carries \textit{the geometric meaning of a scale and length} in any field theory and we think that by using eq.~\eqref{Alen} this becomes clear because $G$, which sits in front of the Ricci scalar in the EH action, can be rescaled by $l_{0}$ to be dimensionless, meaning that the necessity for
renormalization can be thought of in relation to the dependence on the scale density $A$. It seems suggestive then to push ideas of \cite{KK1} further and rescale also the non-gravitational fields in a given theory
according to eq.~\eqref{fieldresc}, taking into account eq.~\eqref{Alen}, bringing about our full unimodular-conformal formalism. Then we expect that the need for renormalizing all dimensionful coupling constants and fields in the matter sector is completely shifted to the renormalization of the scale density, in a similar way that is suggested by Kalmykov and Kazakov for the case of the metric. A good and simple example to study this would be the non-minimal scalar field presented in section \ref{sec_xiphi}, but this is, however, beyond the topic of this thesis.

{\centering \hfill $\infty\quad$\showclock{0}{20}$\quad\infty$ \hfill}


\chapter{Definition of conformal invariance}
    \label{ch:defcf}
    In this chapter we shall pursue a general definition of conformal invariance  of a field theory in terms of the scale density $A$. This is a different approach than the usual definition which says that a conformally invariant matter field theory (in the Weyl rescaling sense) is that which has an identically vanishing trace of the corresponding Hilbert energy-momentum tensor. Nevertheless, the two definitions do share some important points. 
The new definition will be motivated on the example of a non-minimally coupled scalar field with a general potential that we met in section \ref{sec_xiphi} and then formulated independently of a theory in terms of the variational derivative with respect to the scale density. 
This motivates the introduction of a generator of conformal field transformation much alike the generator of dilations $D$ that we met in eq.~\eqref{confgen}. The generator is formulated independently of a theory in question and we argue why it should be so.
The use of unimoduar-conformal decomposition of the metric tensor and the non-gravitational fields established in chapter \ref{ch:umtocf1} plays a crucial role in establishing these statements.
The new definition of conformal invariance in terms of this generator
is then compared with the standard definition of conformal invariance and the equivalence between the two established.
Its application and consequences are demonstrated on some well-known theories in $d$-dimensions: Einstein-Hilbert action, vacuum electromagnetic field theory (EM) and Weyl-tensor action, the latter of which is an important part of this thesis.

\newpage

\section{Energy-momentum tensor and the definition of conformal invariance}
\label{sec_defconf}

The usual definition of conformal invarince of a given action is given with a reference to the trace of the corresponding variation with respect to the metric $g_{\mu\nu}$. 
A general matter action, expressed either through Lagrangian density $\mathcal{L}^{m}$ or Lagrangian $L^{m}=\mathcal{L}^{m}/\sqrt{-g}$, is defined as
    \begin{equation}
    \label{mattaction}
        S^{m}=\int {\rm d}^d x\, \mathcal{L}^{m}=\int {\rm d}^d x\,\sqrt{-g}\, L^{m}\ .
    \end{equation}
The variation of the action with respect to the metric components (denoted by $\delta_{g}$) defines the energy-momentum tensor\footnote{This is the Hilbert definition of the energy-momentum tensor. The \textit{canonical} energy-momentum tensor is defined as a Noether current but we do not use that definition in this thesis, since we are in the curved Riemannian geometry.},
    \begin{equation}
    \label{Tmndef}
        \delta_{g} S^{m} = \int {\rm d}^d x \,\ddel{ \mathcal{L}^{m}}{ g^{\mu\nu}}\delta g^{\mu\nu}=:-\frac{1}{2}\int {\rm d}^d x\,\sqrt{-g} \,T_{\mu\nu}\delta g^{\mu\nu}\ ,
        \qquad T_{\mu\nu} =:-\frac{2}{\sqrt{-g}} \frac{\delta \mathcal{L}^{m}}{\delta g^{\mu\nu}}\ .
    \end{equation}
Alternatively one could work with energy-momentum tensor \textit{density} defined as
    \begin{equation}
    \label{Tmndensdef}
        \delta_{g} S^{m} = \int {\rm d}^d x \,\ddel{ \mathcal{L}^{m}}{ g^{\mu\nu}}\delta g^{\mu\nu}=:-\frac{1}{2}\int {\rm d}^d x\, \mathcal{T}_{\mu\nu}\delta g^{\mu\nu}\ ,
        \qquad \mathcal{T}_{\mu\nu} =:-2\frac{\delta \mathcal{L}^{m}}{\delta g^{\mu\nu}}\ ,
    \end{equation}
whose weight is one. This density is not only useful for expressing covariant conservation laws in terms of partial derivatives, but it is also remarkably directly related to the variation with respect to the scale and shape, as we shall see in this chapter.

Now, the usual definition of conformal invariance \cite{Wald} states that an action is invariant under conformal transformations iff the trace of the corresponding energy-momentum tensor vanishes, i.e. if $T:=g^{\mu\nu}T_{\mu\nu}=0$, \textit{on-shell}. ``On-shell'' means ``taking into account the equations of motion'', hence, only if one uses the equations of motion in $T$ can one obtain that $T=0$. 
The reason why the trace of the energy-momentum tensor lies in the core of this statement is that conformal variations given by eq.~\eqref{ConfFieldInf} are proportional to the metric itself. Then from eq.~\eqref{Tmndef} we have,
    \begin{align}
    \label{dactionconfzero}
        \delta_{\omega}S  &=
        \int {\rm d}^d x \,\omega\,\ddel{ \mathcal{L}^{m}}{ g^{\mu\nu}} g^{\mu\nu}= 0\ ,\\[12pt]
     \label{vardefConfInv}
        \ddel{ \mathcal{L}^{m}}{ g^{\mu\nu}} g^{\mu\nu}\vert_{\ssst\text{on-shell}}&=0\quad
        \Leftrightarrow\quad   S\text{ is conformally invariant}\ ,
    \end{align}
by using $\delta_{\omega}g^{\mu\nu}=-2\omega g^{\mu\nu}$. If $S$ is a matter action then this just means that the trace $T$ of the energy-momentum tensor vanishes if the action is conformally invariant and we say that we are dealing with \textit{conformal matter}.
Now, there are two challenges in this standard formulation of conformal invariance that we want do pursue:
    \begin{enumerate}
        \item we would like to extend the definition to \textit{any} action by shifting the emphasis from the trace $T$ to the variation of the action with respect to the metric,
        \item statement of conformal invariance of an action should not depend on whether equations of motion are satisfied or not.
    \end{enumerate}
This essentially boils down to reformulating eq.~\eqref{dactionconfzero}, eq.~\eqref{vardefConfInv} and the following statement in terms of the scale density $A$: if for \textit{any} action the trace of its variation with respect to the metric vanishes identically, that action is conformally invariant. 

In order to do that we propose here a general recipe for applying unimodular-conformal decomposition and exposing conformal properties of a given theory.
This recipe is given as follows. We first prepare a given theory in the following way:
    \begin{enumerate}
        \item  Decompose the metric into scale and shape density according to eq.~\eqref{gsplit1}.
        \item Determine the length dimension of all fields (recalling that the dimension of the action is $[\hbar]$) and apply conformal decomposition into appropriately defined densities according to eq.~\eqref{fieldresc}, such that the scale weight equals length dimension.
        \item Use eq.~\eqref{Alen} to extract the length dimension form each $A$-dependent term in the resulting action. The result of this is that it will become obvious that each term with a dimensionful coupling constant is necessarily $A$-dependent. The last step is to redefine the coupling constants into their dimensionless versions, by absorbing factors of $l_{0}$ which appear in the correpsonding terms.
    \end{enumerate}
After an action has been prepared according to these steps we have the following theorem:
    \begin{quote}
        \textit{An action prepared as above is conformally invariant iff its variation with respect to the scale density identically vanishes up to a boundary term.}
    \end{quote}
In what follows, we shall test this theorem on several field theories and in the end propose a concrete formulation of this theorem.

\section{Energy-momentum tensor revisited}

Before we turn to the formulation of the generator, we ask for a more obvious interpretation of the variation in eq.~\eqref{varLA}: since
$A$ is just a degree of freedom of the metric, then isn't expression \eqref{varLA} somehow related to the energy-momentum tensor?
We are therefore motivated to formulate the definition of the energy-momentum tensor in terms of the unimodular-conformal decomposition and to ask what can we learn about its relationship with the variational derivative with respect to $A$.
We shall study this topic again on the example of the non-minimally coupled scalar field that we met in section \ref{sec_xiphi} and understand the meaning of eq.~\eqref{varLA}.

Using unimodular variation in eq.~\eqref{vardecUP}, the variation of the action is split into two parts:
    \begin{align}
        \delta_{g} S^{m}& = \int {\rm d}^d x\,
        \left[ A^{-2}\ddel{ \mathcal{L}^{m}}{ g^{\mu\nu}}\delta \gb^{\mu\nu} - 2  \gb^{\mu\nu} A^{-3}\ddel{ \mathcal{L}^{m}}{ g^{\mu\nu}}\delta A\right]\nonumber\\[12pt]
        &=- \frac{1}{2}\int {\rm d}^d x\,A^d \left[A^{-2}T_{\mu\nu}\delta \gb^{\mu\nu}
        -2A^{-1} T_{\mu\nu}g^{\mu\nu}\delta A 
        \right]\nonumber\\[12pt]
    \label{Tmnvar}
        &=- \frac{1}{2}\int {\rm d}^d x\, \left[A^{-2}\mathcal{T}_{\mu\nu}\delta \gb^{\mu\nu}
        -2A^{-1} \mathcal{T}_{\mu\nu}g^{\mu\nu}\delta A 
        \right]\ ,
    \end{align}
from which definitions of the tracelss and trace part of the energy momentum tensor (density) \textit{follow directly},
    \begin{subequations}
    \begin{align}
    \label{Tmn_dec}
        T\equiv T_{\mu\nu}g^{\mu\nu} & := A^{1-d}\frac{\delta S^{m}}{\delta A}\ ,&
        \qquad T_{\mu\nu}^{\srm T} & := -2A^{2-d}\left(\frac{\delta S^{m}}{\delta \gb^{\mu\nu}}\right)^{\ssst\rm T}\ ,\\[12pt]
    \label{Tmn_decDens}
        \mathcal{T} & := A\frac{\delta S^{m}}{\delta A}\ ,  &  \mathcal{T}_{\mu\nu}^{\srm T} & := - 2A^{2}\left(\frac{\delta S^{m}}{\delta \gb^{\mu\nu}}\right)^{\ssst\rm T}
    \end{align}
    \end{subequations}
where we explicitly indicate with superscript ``${\srm{ T}}$'' that the variation with respect to $\gb^{\mu\nu}$ results in a traceless object, as a direct consequence of eq.~\eqref{tracelessvar}.
Now we can conclude: the variation of an action with respect to the conformally invariant part of the metric defines the \textit{traceless} part of the energy momentum tensor (density) and variation with respect
to the scale part of the metric defines its \textit{trace} part. Note again the theme of unimodular decomposition $\rightarrow$ traceless-trace decomposition.

Let us first calculate the energy-momentum tensor of the non-minimally coupled scalar field based on the usual definition given by eq.~\eqref{Tmndef} with Lagrangian in eq.~\eqref{nonmin_Lag_decomp}.
Calculating its trace and traceless parts as well, we have,
    \begin{align}
    \label{Tmnphinm}
       T_{\mu\nu} &=(1-2\xi)\del_{\mu}\varphi\del_{\nu}\varphi +2\left(\xi-\frac{1}{4}\right) g_{\mu\nu}g^{\alpha\beta}\del_{\alpha}\varphi\del_{\beta}\varphi
       - 2\xi\varphi\left(\nabla_{\mu}\nabla_{\nu} -g_{\mu\nu}\Box\right)\varphi\nonumber\\[12pt]
       &\quad +\xi\left(R_{\mu\nu}
       -\frac{1}{2}g_{\mu\nu}R\right)\varphi^2-g_{\mu\nu}V(\varphi)\ ,
    \end{align}
    \begin{align}
    \label{TmnphinmTr}
        T & = \frac{2-d}{8\varphi^2}\,\xi_{c}\,g^{\alpha\beta}\del_{\alpha}\varphi^2\del_{\beta}\varphi^2
        +2(d-1)\xi\left(\frac{1}{2}\Box\varphi - \xi_{cf} R \varphi^2\right) - d V(\varphi)\nonumber\\[12pt]
        &\stackrel{\eqref{KGphi}}{=}(d-1)\xi_{c}\Box\varphi^2 + \frac{(d-2)}{2}\varphi V'(\varphi)- d V(\varphi)\\[20pt]
    \label{TmnphinmTless}
        T_{\mu\nu}^{\srm T} & = \frac{1}{4\varphi^2}\left(\del_{\mu}\varphi^2\del_{\nu}\varphi^2-\frac{1}{d}g_{\mu\nu}g^{\alpha\beta}\del_{\alpha}\varphi^2\del_{\beta}\varphi^2\right) - \xi\left(\nabla_{\mu}\nabla_{\nu} - \frac{1}{d}g_{\mu\nu}\Box\right)\varphi^2 + \xi R_{\mu\nu}^{\srm T}\varphi^2
    \end{align}
and similarly for they densitized versions.
It is important to observe that in the second line in eq.~\eqref{TmnphinmTr} we have used the KG equation to eliminate $\Box\varphi$ as is usual and necessary. This results in two terms that in general break conformal invariance according to definition in eq.~\eqref{vardefConfInv}. Apart from already familiar condition $\xi_{c}=0$, at the same time one has to have the following condition, 
    \begin{equation}
    \label{Vconfphi}
        \varphi \frac{V'(\varphi)}{V(\varphi)} = \frac{2d}{d-2}
    \end{equation}
such that conformal invariance of the action is established. But eq.~\eqref{Vconfphi} is just previously derived eq.~\eqref{Vd} in disguise, i.e. for a potential of the form $V(\varphi)\sim \varphi^{n}$ it implies $n=2d/(d-2)$. Only in this case can the trace vanish. 

On the other hand, using our unimodular-conformal decomposition the trace and traceless parts defined by eq.~\eqref{Tmn_decDens} are calculated to be\footnote{One has to keep in mind that none of the terms are individually tensorial objects under $GL(d,\mathbb{R})$.}
    \begin{align}
    \label{traceT}
        \mathcal{T} & = -\xi_{c}(d-1)\Biggpar{
        \del_{\mu}\left( \gb^{\mu\nu}\del_{\nu}\chi^2\right)
        - (d-2)A^{-1}\Bigsq{
        \del_{\mu}\bigpar{
        \gb^{\mu\nu}\del_{\nu}A
        }
        + \gb^{\mu\nu}\del_{\mu}A
        \del_{\nu}
        }\chi^2
        }\nonumber\\[6pt]
        &\quad
        - A\dd{\bar{V}}{A}
        \ ,
    \end{align}
    \begin{align}
    \label{tlessT}
        \mathcal{T}^{\srm T}_{\mu\nu}&=A^{2}\Biggbr{
        \frac{1}{4\chi^2}
        \del_{\mu}\chi^2\del_{\nu}\chi^2
        + \xi\biggpar{
        \bar{R}_{\mu\nu}- \bigpar{\delta^{\alpha}_{\nu}\del_{\mu} - {\bar{\Gamma}^{\alpha}}_{\mu\nu}
        }\del_{\alpha}
        }\chi^2
        \nonumber\\[6pt]
        &\qquad\qquad\qquad
        + 2\xi_{c}(d-1)\Biggpar{
        A^{-1}\del_{(\mu}A\del_{\nu)}
        - \frac{d-2}{2}A^{-2}
        \del_{\mu}A
        \del_{\nu}A
        }\chi^2
        }^{\srm T} \ ,
    \end{align}
where we recognize eq.~\eqref{varLA} as the trace of the energy-momentum tensor given by eq.~\eqref{traceT}. 
It is interesting to observe that eq.~\eqref{traceT} contains second order derivatives of the scale density $A$, while eq.~\eqref{tlessT} contains only first derivatives.
For conformal coupling $\xi_c=0$ the energy-momentum tensor density pieces reduce to
    \begin{align}
    \label{traceTc}
        \mathcal{T}&=-A\dd{\bar{V}}{A}
        \ ,\\[6pt]
    \label{tlessTc}
        {\mathcal{T}^{{\srm T}\alpha}}_{\nu}&=\gb^{\alpha\mu}\Biggbr{
        \frac{1}{4\chi^2}
        \del_{\mu}\chi^2\del_{\nu}\chi^2
        +\xi_{cf}\biggpar{
        \bar{R}_{\mu\nu}- \bigpar{\delta^{\alpha}_{\nu}\del_{\mu}-{\bar{\Gamma}^{\alpha}}_{\mu\nu}
        }\del_{\alpha}
        }\chi^2
        }^{\srm T} \ ,
    \end{align}
where we have raised an index to $\mathcal{T}_{\mu\nu}^{\srm T}$ in order to get rid of the factor of $A^2$ in its definition. 
We see that, completely equivalent to the discussion around eq.~\eqref{varLA}, the trace of the energy-momentum tensor (density) in eq.~\eqref{traceTc}
for conformally coupled scalar density field \textit{vanishes identically} if the potential satisfies eq.~\eqref{Vd} (or eq.~\eqref{Vconf}, equivalently), \textit{without using the KG equation}.
Note also that all $A$-dependent terms from eq.~\eqref{tlessTc} have canceled, leaving it manifestly conformally invariant, whereas this is not evident from eq.~\eqref{TmnphinmTless} for conformal coupling $\xi=\xi_{cf}$.
Therefore, we arrive at one of the most important results in this thesis: \textit{variational independence of an action on the scale density $A$ implies its conformal invariance off-shell}:
    \begin{equation}
    \label{varSconfinv}
        A\ddel{S}{A}=0\qquad\Leftrightarrow\quad S\text{ is conformaly invariant}\ .
    \end{equation}
This is the reason why the trace of the energy-momentum vanishes for such matter actions. Since a conformally invariant action does not depend on $A$ up to a boundary term it also does not contain any length scale, i.e. dimensionful coupling constant.
But the converse is not true: an action might have the property that $l_{0}$ cancels out (which would mean that it does not have dimensionful coupling constants) but this does not necessarily imply that it is conformally invariant. An example is the kinetic term of a minimally coupled scalar field, as can be seen from eq.~\eqref{kinetic} for $s=(2-d)/2$.

Comparing eq.~\eqref{varSconfinv} with the generator of dilations $D$ in eq.~\eqref{confgen} it is suggestive to think of eq.~\eqref{varSconfinv} as some kind of generator acting on the space field configurations and functionals that depend on them. This is the topic of the next section.

\section{Generator of conformal field transformation and conformal invariance}
\label{sec_genconf}

Recall that the generator of dilations in conformal coordinate transformations given by eq.~\eqref{confgen} is $\sim x^{\mu}\del_{\mu}$. All coordinates $x^{\mu}$ enter this generator because all of them are affected by dilation \textit{by definition}.
To draw an analogy, lift the general meaning of ``$x^{\mu}$'' to configuration space of metric and matter fields. Since dilations are rescaling of coordinates by a constant, 
this would correspond in field theory precisely to a conformal field transformation in eq.~\eqref{ConfFieldInf} (since the conformal factor $\Omega(x^{\mu})$ is constant in configuration fields, even though it depends on coordinates) with an exception that each field has its own way of rescaling, i.e. conformal weight.
But after applying unimodular-conformal decomposition, the scale density is \textit{the only} field that transforms under conformal transformation and its conformal weight is one, which represents a departure from the analogy with coordinates.
In a later chapter we shall mention the $3+1$ formulation of the generator of conformal transformation first recognized by Irakleidou et al. \cite{ILP} on the example of the Weyl-tensor gravity.
Our formulation of the generator in this chapter is a \textit{covariant} formulation, valid in any dimension.

\subsection{Formulation}
\label{subsec_genform}

Let us propose the form of a generator of conformal field transformations. 
In general, if there are $N$ fields $\phi_{\ssst I}$, $I=1,2,...N$,  in a theory out of which $M$ have conformal weight $n_{\ssst I}$ as assumed in eq.~\eqref{eqn:conftrans} but $N-M$ are conformally
invariant, then we define the generator of conformal transformation in a configuration space as (summation over $J$ implied)
    \begin{equation}
    \label{Gconfdef}
        \hat{\mathcal{G}}\,\cdot := n_{\ssst J}\int\d^{d}x\, \omega(x)\phi_{\ssst J}(x)\, \ddel{\,\cdot}{\phi_{\ssst J}(x)}\ ,\quad J=1,2...M \ ,
    \end{equation}
where the dot ``$\cdot$'' is to be replaced by whatever functional the generator acts on, as an operator, such as an action or a field. This is in almost complete analogy to the generator of dilations, except that we have to consider the integral because we are dealing with functional derivatives; 
we have also included the infinitesimal parameter of conformal transformation $\omega(x)$ \textit{into} the generator\footnote{Otherwise by ``generator'' we would have to call only $\phi_{\ssst J}(x)\, \ddel{\,\cdot}{\phi_{\ssst J}(x)}$ which would not have much meaning without the integral.}. 
To see how would this work, let us produce a conformal transformation of the metric tensor $g_{\mu\nu}$, whose conformal weight is $n_{g}=2$. We could imagine that the exponential of the generator in eq.~\eqref{Gconfdef} is an element of a Lie group, but the problem is that we are dealing with a 
\textit{functional space} and it is not clear to us how to proceed rigorously. Nevertheless, one could imagine that a finite and infinitesimal conformal transformations of the metric by $\Omega(x)=exp(\omega(x))$ can be \textit{defined} using eq.~\eqref{Gconfdef} with a demand $n_{g}=2$ and then proceeding by expanding around the identity transformation as follows,
    \begin{align}
        \left(e^{\hat{\mathcal{G}}} g_{\mu\nu}\right)(x) & := \Omega^2(x)g_{\mu\nu}\nonumber\\[12pt]
        \approx 1 + \left(\hat{\mathcal{G}}g_{\mu\nu}\right)(x)  & = \left(1+2\omega (x)\right)g_{\mu\nu}\nonumber\\[12pt]
        g_{\mu\nu}(x) + 2\int\d^{d}y\, \omega(y)g_{\alpha\beta}(x)\, \ddel{g_{\mu\nu}(x)}{g_{\alpha\beta}(y)}  & = \left(1+2\omega (x)\right)g_{\mu\nu}\nonumber\\[12pt]
        g_{\mu\nu}(x) + 2\int\d^{d}y\, \omega(y)g_{\alpha\beta}(x)\, \mathbb{1}^{\alpha\beta}_{(\mu\nu)}\delta(x-y)  & = \left(1+2\omega (x)\right)g_{\mu\nu}\ ,
    \end{align}
where
    \begin{equation}
        \ddel{g_{\mu\nu}(x)}{g_{\alpha\beta}(y)}=\mathbb{1}^{\alpha\beta}_{(\mu\nu)}\delta(x-y)\ ,
    \end{equation}
and $\delta(x-y)$ is $d$-dimensional delta-function which cancels the integral over $y$ and leaves only $2\omega(x)g_{\mu\nu}(x)$ as it should. Note that in the third line only one term from the sum over $J$ in eq.~\eqref{Gconfdef} has survived --- the metric. 
The action of this generator is similar to the $U(1)$ group of transformations (i.e. the phase transformation in quantum mechanics or the local gauge group of electromagnetism) except that the group element is real, not complex, which is why it is a scale transformation.
This kind of formulation seems to work in principle also for any other field in a similar way. However, if we use unimodular-conformal decomposition, none of the fields except the scale density transforms under infinitesimal conformal transformations,
    \begin{equation}
        \left(\hat{\mathcal{G}} A \right)(x) =  n_{\ssst A} \int\d^{d}y\, \omega(y)A(y)\, \ddel{A(x)}{A(y)}:=\omega(x)A(x)\ ,
    \end{equation}
by demanding $n_{\ssst A}=1$.
Then one could use the generator formalism to define not only conformal field transformation but also the
unimodular-conformal decomposition itself by asking for a set of tensor field densities $\chi_{\ssst I}(x)$ of scale weight $\wb_{\ssst I}$ such that the generator annihilates them, that is,
    \begin{align}
        \left(\hat{\mathcal{G}} \chi_{\ssst I}\right)(x)  & \shalleq 0 \\[12pt]
        n_{\ssst J}\int\d^{d}x\, \omega(y)\phi_{\ssst J}(y)\, \ddel{\left(A^{\wb_{\ssst I}}\phi_{\ssst I}\right)(x)}{\phi_{\ssst J}(y)} & = 0\nonumber\\[12pt]
         n_{\ssst J}\int\d^{d}x\, \omega(y)\left(
         \phi_{\ssst J}(y)\, \ddel{A^{\wb_{\ssst I}}(x)}{\phi_{\ssst J}(y)}\phi_{\ssst I}(x)  + \phi_{\ssst J}(y)\, A^{\wb_{\ssst I}}(x)\ddel{\phi_{\ssst I}(x)}{\phi_{\ssst J}(y)}\right) & = 0 \nonumber\\[12pt]
         \int\d^{d}x\, \omega(y)\biggpar{\wb_{\ssst I}
         A(y)A^{\wb_{\ssst I}-1}(x)\phi_{\ssst I}(x)  + n_{\ssst J}\phi_{\ssst J}(y)\, A^{\wb_{\ssst I}}(x)\delta_{\ssst IJ}}\delta(x-y) & = 0\nonumber\\[12pt]
          \left(\wb_{\ssst I} + n_{\ssst I}\right)A^{\wb_{\ssst I}}(x)\phi_{\ssst I}(x) & = 0\ ,
    \end{align}
from which it follows
    \begin{equation}
        \wb_{\ssst I} =- n_{\ssst I}\ ,
    \end{equation}
i.e. that the scale weight of the conformally invariant tensor density field $\chi_{\ssst I}$ has to be negative of the conformal weight of the original field in order for the new fields to be conformally invariant, which agrees with our original definition given by eq.~\eqref{fieldresc} and eq.~\eqref{scaleweightdef}. This is how one can define all new rescaled and conformally invariant fields, including the shape density.

An important consequence of introducing unimodular-conformal variables for the generator of conformal transformations is that only one term from the sum over $J$ in its definition in eq.~\eqref{Gconfdef} survives: the scale density. This means the generator of conformal transformations in any dimension in any field theory defined on Riemannian geometry can be defined as
    \begin{equation}
    \label{GconfdefA}
        \hat{\mathcal{G}}_{\omega}\,\cdot := \int\d^{d}x\, \omega(x)A(x)\, \ddel{\,\cdot}{A(x)}\ ,
    \end{equation}
which simply comes from
    \begin{equation}
    \label{traceAvar}
        -2g^{\mu\nu}\ddel{}{g^{\mu\nu}}=A\, \frac{\delta }{\delta A}\ .
    \end{equation}
The analogy with dilations can be taken further. We could introduce the notion of the Lie derivative of any functional of scale density, shape density and conformally invariant non-geometric fields $F[q^{\ssst I}]$,\quad $q^{\ssst I}(x)=\left(A(x),\gb_{\mu\nu}(x),\chi_{\ssst I}(x)\right)$ along the ``direction'' of a vector analogous to the generating vector of dilations in eq.~\eqref{xiconfD},
    \begin{align}
    \label{dFgenconf}
        \delta_{\Xi}F[q^{\ssst I}] \equiv \mathcal{L}_{\Xi}F[q^{\ssst I}]:&= \int\d^{d}x\, \Xi^{\ssst J}(x)\, \ddel{F[q^{\ssst I}]}{q^{\ssst J}(x)}\nonumber\\[12pt]
        &=\int\d^{d}x\, \omega (x) A(x)\, \ddel{F[q^{\ssst I}]}{A(x)}\ ,\qquad \Xi^{\ssst J}(x) := \left(\omega (x) A(x), 0, 0...\right)\ ,
    \end{align}
where $q^{\ssst J=A}(x)=A(x)$. This shows that $F[q^{\ssst I}]$ is analogous to a scalar field on spacetime. We see that ``direction'' $\Xi^{\ssst I}(x)$ in the space of fields in which the conformal transformation happens has only the first component non-vanishing --- this is the direction along the sale density $A$ and in the future we shall write $\delta_{\omega}$ instead of $\delta_{\Xi}$ for the variation, just for simplicity. Do we have something similar among cordiante transformations? We have something close to it.
Namely, if one would rewrite the generators of dilations $D$ in \textit{spherical} coordinates in spacetime, one would have left with only two coordinates which are affected by dilations: the time coordinate and the radial coordinate. Angles are, as was explained in chapter \ref{ch:umtocf}, invariant under dilations or special
conformal transformations because they are silent about the notion of size or length. It can be indeed shown (we skip the straightforward proof here) that the generator reduces to $D\sim t\del_{t}+r\del_{r}$. If we further introduced hyperbolic polar
coordiantes (i.e. Rindler coordinates) $t=v \sinh u,\, r=v\cosh u$ it can be shown that $D\sim v\del_{v}$, i.e. only one component\footnote{This situation speaks for itself in favor of using polar coordinates in order to study conformal coordinate
transformations.}, because $u$ is the hyperbolic \textit{angle}. This would roughly correspond to what happened in our case with unimodular-conformal decomposition except that the kind of transformation we are using here is rather different compared to coordinate transformations to polar coordinates. 
To finalize, the discussion of this paragraph paints ``the big picture'' of what conformal field transformation is. In the special case where $F[q^{\ssst I}]$ is an action, this interpretation of conformal transformation and invariance relates to the question of whether $A$ is a dynamical field or not. Namely, if $A$ is dynamical, that means there exists an equation of motion for $A$. An equation of motion for $A$ would arise from extremization of the action with respect to $A$, i.e. from its first order variation. This equation of motion holds for arbitrary variations $\delta A$. On the other hand, conformal invariance requires the vanishing of the first order variation given by eq.~\eqref{dFgenconf} of the action with respect to a specific variation $\delta A = \delta_{\omega} A=\omega A$ and thus needs to hold for arbitrary $\omega$ and arbitrary $A$. The two cases side by side are compared as follows,
    \begin{align}
    \label{confinvS1}
        \ddel{S[q^{\ssst I}]}{A(x)}&=0\ ,\quad\forall \delta A(x) \qquad\Rightarrow\qquad \text{E.O.M. for $A(x)$}\\[12pt]
    \label{confinvS2}
        \ddel{S[q^{\ssst I}]}{A(x)}&=0\ ,\quad\forall  A(x) \qquad\Rightarrow\qquad \text{conformal invariance}\ ,
    \end{align}
where eq.~\eqref{confinvS2} basically means that the vanishing is identical.
This should be kept in mind in order not to confuse validity of equations of motion and conformal invariance; conformal invariance stated by eq.~\eqref{confinvS2} does not require $A$ to obey equations of motion (thus $\forall A(x)$), i.e. it holds off-shell, as we showed in the previous section on the example of the non-minimally coupled scalar
field. This should actually be expected because conformal invariance of an action concerns only the structure of the action itself, not the equations of motion. And if an action is conformally invariant then it follows that $A$ is not dynamical, i.e. it is arbitrary. All information about conformal properties is contained in the action already and we have the restatement of the theorem proposed in section \ref{sec_defconf}:
\textit{An action is invariant under conformal transformation iff it is annihilated by the action of the generator of conformal transformations},
    \begin{equation}
    \label{defGenconfFin}
       \delta_{\omega} S[q^{\ssst I}] = \hat{\mathcal{G}}_{\omega}S[q^{\ssst I}]=\int\d^{d}x\, \omega (x) A(x)\ddel{S[q^{\ssst I}]}{A(x)}=0\ ,
    \end{equation}
 implying eq.~\eqref{confinvS2}. This is one of the main results of this thesis. Condition in eq.~\eqref{defGenconfFin} is valid for any theory in any dimension and essentially completes the formulation of the theorem proposed in section \ref{sec_defconf}. We hope that this formulation might inspire some more rigorous treatment of the notion of generator of conformal transformation in the space of fields.
 We would next like to see this generator in action on some well-known examples.

\subsection{Einstein-Hilbert action}
\label{subsec_EH}

Because of conformal transformation of the Ricci scalar in eq.~\eqref{Rconftransf} the Einstein-Hilbert action
    \begin{equation}
    \label{EHaction}
        S^{\srm{ EH}} = \frac{1}{2\kappa}\int\d^{d}x\, \sqrt{g}(R-2\Lambda)
    \end{equation}
is not conformally invariant unless $d=2$ and $\Lambda=0$. Note that $\Lambda$ has dimensions of $L^{-2}$ and that dimension of $\kappa$ depends on the spacetime dimension.
GR, which is the special case of the above for $d=4$, is also not conformally invariant. Unimodular-conformal decomposition makes this obvious. Namely, let us see under which circumstances eq.~\eqref{confinvS2} holds. Because of eq.~\eqref{traceAvar} and eq.~\eqref{Alen} we have
    \begin{align}
    \label{GonEH}
        \hat{\mathcal{G}}_{\omega}S^{\srm{EH}}\stackrel{?}{=}0\quad\Rightarrow\quad\ddel{S^{\srm{EH}}}{A} & = \frac{-2}{\kappa}\int\d^{d}x\,\sqrt{g}\left[\left( R_{\mu\nu}-\frac{1}{2}g_{\mu\nu}R\right)g^{\mu\nu}+d\Lambda\right]\nonumber\\[16pt]
        & =\frac{-2}{\kappa}\int\d^{d}x\,A^{d}\left[l_{0}^{d-2}\frac{2-d}{2} R + d l_{0}^d\Lambda\right]\stackrel{?}{\equiv}0\ .
    \end{align}
If we demand conformal invariance of this action then the following has to hold
    \begin{equation}
    \label{GonEHconfcond}
       \frac{2-d}{2} R\equiv 0\quad\wedge\quad  \Lambda = 0\ ,
    \end{equation}
for all $A$, but we see that this is possible only if $d=2$, so only in two dimensions the EH theory is conformally invariant. Reacall that we claim that conformal invariance is related to the absence of dimensionful coupling constants from the action. In the EH action in eq.~\eqref{EHaction} the gravitational coupling $\kappa$ has units which depend on dimension. To see this explicitly, observe from the second line in eq.~\eqref{GonEH} that the following \textit{dimensionless} ratio
    \begin{equation}
    \label{ratiokappa}
        l:=\frac{l_{0}^{d-2}}{\kappa [S^{\srm{EH}}]}
    \end{equation}
where $[S^{\srm{EH}}]$ is the unit of action, has to be dimensionless (we avoid referring to $\hbar$ in the case of arbitrary dimension because the relationship between length, time, mass and $G, c, \hbar$ depends on the dimension of spacetime). It follows that $[\kappa]=L^{d-2}/[S^{\srm{EH}}]$ and $\kappa$ may be rewritten in terms of another constant with the meaning of length to the power of $d-2$.
Hence, in $d=2$ dimensions $[\kappa]=1/[S^{\srm{EH}}]$, i.e. dimensionless in the inverse units of action. 

In $d=4$ dimensions, where we have GR, a natural length scale is given by the Planck length\footnote{To be precise, this is the \textit{reduced Planck length}, which is defined with a factor of $8\pi$ hidden in $\kappa$.} $l_{p}$ defined \cite{OUP} as
    \begin{equation}
    \label{PlanckLkappa}
        l_{p}:=\sqrt{\kappa\hbar}\ ,
    \end{equation}
where $\kappa$ has dimensions of $L^2/[\hbar]$ (in $c=1$ units). Then from eq.~\eqref{ratiokappa} it follows that eq.~\eqref{ratiokappa} is given by
    \begin{equation}
    \label{relscale}
        l = \frac{l_{0}}{l_{p}}\ ,
    \end{equation}
and we call it \textit{the relative length scale} or \textit{the relative gravitational coupling constant}. Recalling the discussion around \ref{relLength}, this number has a clear meaning: it measures how big or small the observed physically relevant region of spacetime is as compared to the Planck length or, alternatively,
it can be thought of as the dimensionless measure of the strength of gravity in a given finite region of spacetime. It is thus obvious that it plays a crucial role in the transition from quantum to classical gravity. Using eq.~\eqref{relscale} in eq.~\eqref{GonEH} for $d=4$, we obtain
    \begin{equation}
    \label{EHactionL}
        S^{\srm{ EH}} = \frac{\hbar}{2}\int\d^{4}x\, A^{4}\left(l^2 R-2 l^{2}(l^{2}_{0}\Lambda)\right)
    \end{equation}
and it is now clear that $[\Lambda]=L^{-2}$. Now $l^{2}\hbar$ is what determines ``classicallity'' of the action and we have $l\gg 1$  if the 
EH action is classical, which agrees with the definition in eq.~\eqref{relscale} and the claim in eq.~\eqref{relLength}. This alternative interpretation of $l$ is useful to keep in mind if matter and quantum corrections predicted by quantum field theory are taken into account. 
Furthermore, it is interesting that expression $l^{2}(l^{2}_{0}\Lambda)$ can be given a familiar interpretation. Namely, we can identify $l_{0}$ with a relevant cosmological scale measured by the Hubble horizon as $l_{0}\equiv c/H_{0}$ and then we have
\footnote{Using the values from Table XXXIII in \cite{CODATA} for the Planck length $l_{p}=\sqrt{8\pi\hbar G/c^3} =  1.616 229\cdot 10^{-35}\sqrt{8\pi} \,\text{m}$ and the speed of light $c=299 792 458\, \text{m}\text{s}^{-1}$ and
the value for the Hubble constant $H_{0}=2.1928\cdot 10^{-18}\, \text{s}^{-1}$ from \cite{Planck2018}. } 
     \begin{equation}
     \label{lambdadef1}
         l^{2}_{0}\Lambda 
         = 3\,\Omega_{0,\Lambda} \frac{l^{2}_{0}H_{0}^2}{c^2}
         = 3\,\Omega_{0,\Lambda}\approx 2.1 \ ,
     \end{equation}
where $\Omega_{0,\Lambda}$ is the dimensionless density parameter for the energy density of $\Lambda$. Then using $l^2 \approx 2.8\cdot 10^{120}$ one concludes that $l^{2}l^{2}_{0}\Lambda\sim 10^{120}$. This may be referred to as the ``dimensionless cosmological constant''. 
We think that one should tend to use the such dimensionless, relative coupling constants in calculations and any kind of approximations because it is independent of the choice of units and has a direct physical interpretation.

\subsection{Massive vector field and electromagnetic field}

A massive vector field $V_{\mu}$ (as a prototype of massive weak gauge vector bosons) is described by the following action in $d$ dimensions,
    \begin{equation}
    \label{Vem}
        S^{\srm V}=-\int\d^{d}x\,\sqrt{-g}\Biggpar{\frac{1}{4}F_{\mu\nu}F^{\mu\nu}+\frac{1}{2}m^2 V_{\mu}V^{\nu}}\ ,
    \end{equation}
which is not invariant under gauge $U(1)$ transformation due to the mass term where the \textit{dimensionful} coupling constant $m$ breaks it. The kinetic term is constructed from the field strength:
    \begin{equation}
        F_{\mu\nu}=\del_{\mu}V_{\nu}-\del_{\nu}V_{\mu}\ .
    \end{equation}
Now, using unimodular-conformal decomposition and bearing in mind that $A$ hides the length scale according to eq.~\eqref{Alen}, we use $V_{\mu}= A^{s}\bar{V}_{\mu}$ and the action becomes
    \begin{align}
        S^{\srm V}&=-\int\d^{d}x\, (l_{0}A)^{d-4+2s}\Biggpar{\frac{1}{4}\gb^{\mu\alpha}\gb^{\nu\beta}\bar{F}_{\mu\nu}\bar{F}_{\alpha\beta}+\frac{1}{2}m^2 l^{2}_{0}A^{2} \gb^{\mu\alpha}\bar{V}_{\mu}\bar{V}_{\alpha}}\ ,\\[12pt]
    \label{Fmndec}
        \bar{F}_{\mu\nu}&:=F_{\mu\nu}-s\, l^{s}_{0} A^{s-1}\Bigpar{\bar{V}_{\mu}\del_{\nu}A-\bar{V}_{\nu}\del_{\mu}A}\ .
    \end{align}
We see that the kinetic term has units of action if we set $s=(4-d)/2$. Fixing such an $s$, the action becomes
    \begin{equation}
        S^{\srm V}=-\int\d^{d}x\, \Biggpar{\frac{1}{4}\gb^{\mu\alpha}\gb^{\nu\beta}\bar{F}_{\mu\nu}\bar{F}_{\alpha\beta}+\frac{1}{2}m^2 l^{2}_{0}A^{2} \gb^{\mu\alpha}\bar{V}_{\mu}\bar{V}_{\alpha}}\ ,
    \end{equation}
but we still do not have a conformally invariant $\bar{F}_{\mu\nu}$ as can be seen from eq.~\eqref{Fmndec} and its dependence on $A$. This dependence can be elliminated if $s=0$, which would then imply that conformall invariance is possible only in $d=4$.
But we see that even in four dimensions conformal invariance could only be achieved if the vector field is \textit{massless} $m=0$. If these conditions are assumed, we have the well-known case of electromagnetism and the vector potential does not require any rescaling, i.e. $\bar{V}_{\mu}=V_{\mu}$ and its conformal weight is zero.
This is in accordance with the well-known fact that the trace of the energy-momentum tensor for a massless vector field given in standard formulation
    \begin{equation}
        T= -\frac{d-4}{4}F_{\mu\nu}F^{\mu\nu}
    \end{equation}
vanishes only in $d=4$ dimensions, assuming that $V_{\mu}$ has zero conformal weight. 
It is unnecessary to bother ourselves with calculating the trace of the energy-momentum tensor as defined in our approach with eq.~\eqref{Tmn_dec} because (apart from being a bit tedious and non-illuminating) we only want
to have eq.~\eqref{confinvS2} fulfilled and it is already been deduced that this can happen only if $s=0$, $d=4$ and $m=0$. Only in that case the generator of conformal transformation defined by eq.~\eqref{defGenconfFin} annihilates the action.

Some comments about the comparison of this case with the case of the scalar field treated in section \ref{sec_xiphi}. From the kinetic term of the scalar field given by eq.~\eqref{kinetic} we see that there is only one $g^{\mu\nu}$ since there is only one pair of indices to be contracted, unlike the kinetic term in eq.~\eqref{Vem}, 
which requires \textit{two} $g^{\mu\nu}$. Since in our approach coordinates are dimensionless and inverse metric has units of $L^{-2}$ there is already enough units of length in four dimensions to cancel
$L^{4}$ unit of volume and that is why $V_{\mu}$ is already dimensionless and need not be rescaled, unlike the scalar field $\varphi$.
Note, however, that if coordinates are the ones which are dimensionful, as is usually assumed, then both kinetic terms give $L^{-2}$ dimension from the derivatives, which then implies that $[V_{\mu}]=L^{-1}$ as well.
Therefore, the length dimension of a field depends on whether or not one considers coordinates dimensionful. 
We think that ``length-less'' $V_{\mu}$ rhymes well with its conformal invariance in $d=4$ dimensions and motivates the use of dimensionless coordinates in this case\footnote{Note, however, that under active conformal \textit{coordinate} transformations given by eq.~\eqref{confKillingxi} the kinetic term in eq.~\eqref{Vem} requires $V_{\mu}$ to transform as well, unlike with conformal field transformation discussed here.}.

\subsection{Weyl-tensor gravity}

An action formed by the invariant made of the Weyl tensor,
    \begin{equation}
    \label{WeylactionG}
        S^{\srm W}=-\frac{\aw}{4}\int\d^{d}x\,\sqrt{-g}\,{C^{\mu}}_{\nu\alpha\beta}{C_{\mu}}^{\nu\alpha\beta}\ ,
    \end{equation}
in $d=4$ dimensions will be the topic of a part of this thesis and we shall refer to it \textit{the Weyl-tensor action/theory}. Note that $\aw$ is a coupling constant whose length dimension depends on the actual dimension of spacetime (for similar reasons as $\kappa$ in section~\ref{subsec_EH}) because the square of the Weyl tensor has a fixed dimension of $L^{-4}$. To see this, apply unimodular decomposition with eq.~\eqref{Alen} as before; the action takes the following form
    \begin{equation}
        S^{\srm W}=-\frac{\aw l_{0}^{d-4}}{4}\int\d^{d}x\,A^{d-4}\gb_{\mu\rho}\gb^{\nu\sigma}\gb^{\alpha\tau}\gb^{\beta\delta}{C^{\mu}}_{\nu\alpha\beta}{C^{\rho}}_{\sigma\tau\delta}\ ,
    \end{equation}
from which we conclude that the coupling constant has to have a dimension of $[\aw]=L^{4-d}$. Since we showed in eq.~\eqref{Weylbar} that the Weyl tensor is determined solely by the shape $\gb_{\mu\nu}$ and is thus conformally invariant in its up-down-down-down index version, we can easily see that this action is conformally invariant only in $d=4$ dimensions, as only then the scale density then disappears.
The action of the generator of the conformal transformation in  eq.~\eqref{defGenconfFin} is then
    \begin{equation}
    \label{WeylactionDconf}
        \hat{\mathcal{G}}_{\omega}S^{\srm W}=-\aw l_{0}^{d-4}\int\d^{d}x\,\omega\biggpar{\frac{d-4}{4} A^{d-4}}\gb_{\mu\rho}\gb^{\nu\sigma}\gb^{\alpha\tau}\gb^{\beta\delta}{C^{\mu}}_{\nu\alpha\beta}{C^{\rho}}_{\sigma\tau\delta}
    \end{equation}
which identically vanishes only for $d=4$ and only in this case $\aw$ is dimensionless in units of action. This example is quite similar to the electromagnetic field action, but the fact that the Weyl tensor is $A$-independent in any dimension made things simpler.

Just for amusement, we could ask if there is a higher-dimensional conformally invariant action based on the Weyl tensor. The answer is yes, but the dimension has to be even. In $d=6$ dimensions three Weyl tensors will suffice
    \begin{equation}
        S^{\srm W}=-\frac{\aw}{4}\int\d^{6}x\,\sqrt{g}\,
        C^{\mu\nu}{}_{\alpha\beta}
        C^{\alpha\beta}{}_{\epsilon\zeta}
        C^{\epsilon\zeta}{}_{\mu\nu}\ ,
    \end{equation}
and by counting the number of $g^{\mu\nu}$ we see that there is exactly $A^{-3\cdot 2}$ factor which cancels the six-dimensional volume.
    
\section{Final remarks}

The most important message from this chapter concerns the definition of conformal invariance stated in the theorem given in section \ref{sec_defconf} and further
elaborated on with the definition of the generator of the conformal transformation in eq.~\eqref{defGenconfFin}.
We have shown that conformal invariance of an action is achieved \textit{iff} none of the terms in the action depend on the scale density $A$,
up to a boundary term (a total divergence in the Lagrangian). Apart from this, an important improvement compared to the old definition given by eq.~\eqref{vardefConfInv} is that our definition holds off-shell, i.e. independently of equations of motion.
In the case of the non-minimally coupled scalar field we think that this is because our proof involves partial integration in eq.~\eqref{chipartInt} in order to be able to cancel the $A$ from the Lagrangian which is what happens in the standard approach when one derives the KG equation. Thus, using the KG equation in the old approach to show that $T=0$ for conformally invariant scalar field only appears to be necessary because no partial integration in derivation of $T_{\mu\nu}$ was necessary.
But such partial integration was necessary for our approach in the derivation of eq.~\eqref{traceT} and this was enough for it to vanish for a confomally coupled $\chi$.
One simply needs to apply the unimodular-conformal decomposition according to the steps presented in section \ref{sec_defconf} and inspect whether or not $A$ cancels out.
Such an achievement greatly increases the significance of the unimodular-conformal decomposition and encourages further applications.

It is important to keep in mind that the identification of the physical length scale $l_{0}$ in $A$ allows one to rather evidently relate the conformal invariance with the absence of dimensionful coupling constants in a theory. 
This might have important implications for studying the behavior of a quantum field theory at high energies, as mentioned in section~\ref{conc2}, while the generator of the conformal transformation could have certain relationship with the so-called \textit{beta functions} which are central in renormalization methods used there. 
Furthermore, in the case of conformally non-invariant theories, such as GR, or conformally coupled massive scalar field, we saw that the length dimension of dimensionful coupling constants is ``compensated'' by a
certain power of $l_{0}$ which arises in those terms, which invites a redefinition of these couplings as dimensionless ratios that can be used to distinguish among different regimes of a theory independently of the choice of units.
One example is the dimensionless gravitational coupling in eq.~\eqref{relscale} which has a very useful and clear interpretation: it measures the strength of gravity or the size of the region in which the gravitational phenomena occur, as compared to the Planck length.
Other dimensionless couplings could be introduced based on the mass parameter $m$ of any field and the cosmological constant $\Lambda$. In the former case, the natural length scale associated with $m$ could be the corresponding Compton wavelength $\lambda_{m}=h/m$ which can then be absorbed into a dimensionless coupling constant $\lambda_{m}/l_{0}$. This ratio could be interpreted as the relative size of the region within which the field is localized, for example, and might have useful applications in studying cosmological perturbations. 
In the case of $\Lambda$, the resulting number $l^{2}l^{2}_{0}\Lambda\sim 10^{120}$ can be used to distinguish between matter- and $\Lambda$-dominated era of the Universe, if compared to $l^2 R$ in the EH action.

Even though the generator of conformal transformation introduced in this chapter seems to be formally viable and 
will play an important role in the following chapters (especially when we discuss the quantization procedure), more care would have to be taken in order to make its definition in terms of functional derivatives mathematically rigorous and consistent, but we leave it here as it is.

{\centering \hfill $\infty\quad$\showclock{0}{15}$\quad\infty$ \hfill}

\chapter[Classical higher derivative theories and their \texorpdfstring{\\}{} perturbative interpretation][Classical higher derivative theories and their perturbative...]{Classical higher derivative theories and their perturbative interpretation}
    \label{ch:HDclass}
    Classical GR is modified at higher energies by the presence of quantum matter fields in a way that requires to change the Einstein-Hilbert action by adding terms quadratic in curvatures ($R^2, R_{\mu\nu}R^{\mu\nu}, C_{\mu\alpha\nu\beta}C^{\mu\alpha\nu\beta}\equiv C^2$) and terms which are non-local.
We are interested in the interpretation of such an effective theory and justification for its use as a base for the quantum theory, which we study in the following chapter.
We shall restrict ourselves only to two quadratic curvature terms, $R^2$ and $ C^2$; this will be enough to study the general features of the theory and its implications for the quantum theory.
To investigate the meaning of higher-derivative terms in gravity, a toy model will be presented in which a one-dimensional harmonic oscillator is modified by a term quadratic in second order time derivatives.
It is argued, based on already existing results, that if higher-derivative terms are small corrections becoming relevant towards higher energies of the system then this implies that they are \textit{perturbations} to the first order, low energy theory and should be mathematically treated that way.
The conclusions are directly applicable to the theory of gravity with higher derivatives and they provide hints to formulate the guidelines for the quantization.
We shall also review some important basic features of the quadratic curvature actions with non-minimally coupled scalar field using the approach of the unimodular-conformal variables both in covariant and $3+1$ Hamiltonian formulation, which shall set the stage for the canonical quantization in the following chapter.
A particular attention is paid to the conditions under which the conformal symmetry could be established.
This will necessitate a discussion on $3+1$ formulation of the generator of conformal transformation we defined in chapter \ref{ch:defcf}.
Throughout the chapter we demonstrate a rather natural use of dimensionless coupling constants as introduced in section \ref{subsec_EH}.

\section{Why higher-derivative theories?}
\label{sec_HD}

The shortest answer must not be anything less than ``it depends on what is meant by \textit{higher-derivative theory}''. Namely, if these theories are motivated by the results of an effective theory approach --- a high-energy extension of a purely classical theory --- then such theories are better referred to as ``theories with higher derivatives''.
The terms giving rise to higher derivatives appear as corrections to the purely classical action due to the effects of high energies, while at low energies they are negligible.
If, on the other hand, these theories are aimed to substitute GR as alternative \textit{classical} theories of gravity, usually with an intention to provide alternative understanding of the dark matter problem and accelerating expansion of the Universe, then they deserve a name ``classical higher derivative theories'' and their motivation from the effective approach is irrelevant and outside the context.
The problem is that the latter theories, the purely classical ones --- which dominate the literature --- are almost exclusively motivated by the results of the effective theories and this inconsistency has a price.

\subsection[Semiclassical Einstein equations and higher-derivative \texorpdfstring{\\}{} counter-terms]{Semiclassical Einstein equations and higher-derivative counter-terms}
\label{subs_HDSEE}

Let us sketch the main points of the SEE and its features.
The classical Einstein equations, following from the EH action given by eq.~\eqref{EHaction} supplemented by a matter action, take the following form
    \begin{equation}
    \label{EE_eqn}
        \frac{1}{\kappa}\left( R_{\mu\nu} - \frac{1}{2}g_{\mu\nu}R + g_{\mu\nu}\Lambda\right)= T_{\mu\nu}\ .
    \end{equation}
If the matter action describes quantum matter then instead of $T_{\mu\nu}$ on the RHS of eq.~\eqref{SEE_eqn} we have 
    \begin{equation}
    \label{SEE_eqn}
        \frac{1}{\kappa}\left( R_{\mu\nu} - \frac{1}{2}g_{\mu\nu}R + g_{\mu\nu}\Lambda\right)= \langle \hat{T}_{\mu\nu}\rangle\ .
    \end{equation}
where $\langle \hat{T}_{\mu\nu}\rangle$ is the expectation value of the energy-momentum tensor operator with respect to some state.
If we were in flat spacetime, a simple normal ordering procedure would make $\langle \hat{T}_{\mu\nu}\rangle$ a finite value by eliminating the divergences appearing upon summation of all modes of a given matter field, without any problems encountered.
But since the normal ordering procedure is essentially a subtraction of the \textit{vacuum} contribution from $\langle \hat{T}_{\mu\nu}\rangle$ and because the notion of vacuum in curved spacetime is ambiguous (but already in flat spacetime in some coordinates which do not refer to an inertial observer, see Unruh effect in \cite[chapter 4]{BD}) due to the lack of appropriate symmetries that the Minkowski spacetime enjoys, it is impossible to define the normal ordering procedure in quantum field theory in curved spacetimes and one must work a little harder.

Upon evaluation of the backreaction term \cite{BD,ParSim,PT,Simon2,Wald} it turns out that it can be separated into a finite term and a divergent term in the process called \textit{regularization},
    \begin{equation}
    \label{SEE_tmnfindiv}
        \langle \hat{T}_{\mu\nu}\rangle = \langle \hat{T}_{\mu\nu}\rangle_{fin} + \langle\hat{T}_{\mu\nu}\rangle_{\infty}\ .
    \end{equation}
The issue here is the divergent term $\langle\hat{T}_{\mu\nu}\rangle_{\infty}$, for which it can be shown \cite{Ford,PT}, to the first order in $\hbar$, that it is proportional to a linear combination of covariantly conserved tensors 
    \begin{equation}
    \label{SEE_tmndiv}
        \langle \hat{T}_{\mu\nu}\rangle_{\infty} = a_{0} g_{\mu\nu} + a_{1}G_{\mu\nu} + c_{1}H_{\mu\nu}^{\ssst{(1)}} + c_{2}H_{\mu\nu}^{\ssst{(2)}}\ ,
    \end{equation}
where $G_{\mu\nu}$ is the Einstein tensor; $a_{0},a_{1},c_{1},c_{2}$ are constants proportional to $\hbar$ which depend on the regularization scheme employed and they all diverge upon the completion of such procedure.
The last two terms turn out to be obtainable from the variational principle from the $R^2$ and $R_{\mu\nu}$ terms, respectively\footnote{These expressions were checked with \texttt{xAct} package \cite{xact} in Wolfram Mathematica.},
    \begin{subequations}
    \begin{align}
    \label{SEE_varR2}
        H_{\mu\nu}^{\ssst{(1)}} & =
        -\frac{2}{\sqrt{g}}\ddel{}{g^{\mu\nu}}\intxx\sqrt{g}R^2\nonumber\\[12pt]
        & = 4\left(R_{\mu\nu}^{\srm T}R - \left(\nabla_{\mu}\nabla_{\nu} - g_{\mu\nu}\Box\right)R\right)\ ,\\[12pt]
    \label{SEE_varRic2}
        H_{\mu\nu}^{\ssst{(2)}} & = -\frac{2}{\sqrt{g}}\ddel{}{g^{\mu\nu}}\intxx\sqrt{g}R_{\mu\nu}R^{\mu\nu}\nonumber\\[12pt]
        & = 4\left(
        \left[R_{\mu\alpha}R^{\alpha}{}_{\nu}\right]^{\srm T} 
        + \frac{1}{2}\left(
        g_{\mu\nu}\nabla^{\alpha}\nabla^{\beta} - \delta^{\alpha}_{\mu}\delta^{\beta}_{\nu}\Box\right)R_{\alpha\beta}
        \right)\ ,
    \end{align}
    \end{subequations}
where ``$\srm T$'' denotes the traceless part with respect to the free indices of a tensor.
One can see that these terms contain fourth order derivatives of the metric.
It is remarkable that the divergent terms only depend on the metric and its derivatives, independently of which matter is considered.
This could be understood as an effect of a considerable energy density of quantum matter on spacetime: the spacetime at smaller scales (probed by higher derivatives, just like in a Taylor expansion of a function in a small neighborhood of a point) curves locally because it ``feels'' the presence of high-energy quantum effects of matter fields.
It is then expected that spacetime will be modified at small scales by quantum corrections as a response to the presence of high energy quantum matter.

Now, the first two terms in eq.~\eqref{SEE_tmndiv} can be absorbed into $\kappa$ and $\Lambda$ which are already introduced by the EH action in eq.~\eqref{SEE_eqn}.
This is done by \textit{renormalizing} or redefining the ``bare'' coupling constants $\kappa$ and $\Lambda$ as
    \begin{equation}
    \label{redefconst}
        \frac{1}{\kappa} = \frac{1}{\kappa_{\ssst{phys}}} - a_{1}\ ,\qquad 
        \Lambda = \Lambda_{\ssst{phys}} - a_{0}\ ,
    \end{equation}
after which the divergences in $a_{0}$ and $a_{1}$ are cancelled. 
We can thus say that two ``counter-terms'' --- which \textit{counter} the divergences in the backreaction --- are already included in the Einstein equations.
It is then $\kappa_{\ssst{phys}}$ and $\Lambda_{\ssst{phys}}$ which are the \textit{physical} coupling constants that we measure in experiments.
A priori, $\kappa$ and $\Lambda$ in the EH action have no physical meaning.
Since $a_{0}$ and $a_{1}$ depend on an energy scale, $\kappa_{\ssst{phys}}$ and $\Lambda_{\ssst{phys}}$ shall depend on it too\footnote{The exact form of energy dependence depends on the matter content and the specific spacetime model. However, such details --- which can be found in e.g. \cite{BD, PT} --- are not relevant for this thesis}.
On the other hand, terms in eqs.~\eqref{SEE_varR2}-\eqref{SEE_varRic2} do not appear in the original action and one is thus faced with the following fact: high-energy description of gravity interacting with quantum matter must deviate from the pure EH action in order for divergences in $c_{1}$ and $c_{2}$ to be cancelled.
The cancellation is done by redefining the couplings $\beta_{1},\beta_{2}$ similarly to eq.~\eqref{redefconst} to include $c_{1}$ and $c_{2}$. 
The remaining term $\langle \hat{T}_{\mu\nu}\rangle_{fin}$ is finite and does not depend on geometric terms but on the quantum state in question (which one does not know a priori). Only in certain cases, such as massless minimally or conformally coupled scalar field on conformally flat backgrounds is $\langle \hat{T}_{\mu\nu}\rangle $ determined entirely by the geometric terms. 
This modification rests upon accepting that at high (but considerably lower than Planck) energies the gravitational action takes the following form:
    \begin{equation}
    \label{SEE_effAct}
        S_{g}=\intxx\sqrt{g}\left( \frac{1}{2\kappa}\left(R - 2 \Lambda\right) + \beta_{1}\hbar R^2 + \beta_{2}\hbar R_{\mu\nu}R^{\mu\nu}\right)
    \end{equation}
where $\beta_{1},\beta_{2}$ are dimensionless bare coupling constants. We have made the coupling of the quadratic terms explicit in order to emphasize that they are relevant only at relatively small scales.
Stelle \cite{Stelle1} has estimated that the physical value of these constants are very weakly bounded by Solar system scale observations, i.e. $\beta_{1}^{\ssst phys},\beta_{2}^{\ssst phys} \lesssim 10^{74}$ in $\hbar=1$ units, and this means that these terms have little effect for classical (low energy) gravity \cite{Don,Don1}.
So far we think there is enough evidence to consider these quadratic curvature terms what they literally are: \textit{perturbations} of the EH action relevant at increasing energies.
This point of view agrees with those of \cite{Don,Don1}, who recalled the works of Simon \cite{Simon1, Simon2} where it was clearly shown that classical solutions to the equations of motion based on eq.~\eqref{SEE_effAct} make sense only as solutions to the Einstein equations perturbed by the fourth-order terms.
In the language of our formalism introduced in chapter \ref{ch:umtocf1}, we choose the coordinates to be dimensionless and put the length dimension in the metric tensor components through the characteristic length scale $l_{0}$. Then eq.~\eqref{SEE_effAct} becomes (cf.~eqs.~\eqref{PlanckLkappa}, \eqref{relscale} and \eqref{EHactionL})
    \begin{equation}
    \label{SEE_effActL}
        S_{g}=\intxx\sqrt{g}\left( \frac{l^{2}\hbar}{2}\left( R - 2 \bar{\Lambda}\right) + \beta_{1}\hbar R^2 + \beta_{2}\hbar R_{\mu\nu}R^{\mu\nu}\right)
    \end{equation}
where we wrote $\bar{\Lambda}\equiv l^{2}_{0}\Lambda$. There are no changes to already dimensionless $\beta_{1},\beta_{2}$ couplings because $l_{0}$ cancels out from the corresponding terms.
It should not be confusing that $\hbar$ appears explicitly with the EH term because this is an artefact of our choice of writing the coupling constant as $l^{2}\hbar$: there is an $\hbar$ hidden in the denominator of $l^{2}$ due to the definition of the Planck length in eq.~\eqref{PlanckLkappa}, but its dimension is not visible because $l^{2}$ is dimensionless.
This allows us to compare the terms in the basis of dimensionless constants $l^{2},\beta_{1},\beta_{2}$ instead with respect to $\hbar$, so now the classical GR is recovered if $l^{2}\gg \beta_{1},\beta_{2}$.
We shall spend some time in the next subsection using a toy model to explain how one must deal with theories with higher derivatives in a given energy scale.
This will clarify and motivate our treatment of quantization of such theories in the next chapter.

As explained in the Introduction, in spite of their perturbative nature, it has been overwhelmingly more popular in the literature (starting from \cite{Stelle1, Stelle2} in late 1970's) to treat eq.~\eqref{SEE_effAct} and its variations as an exact classical theory, even though the same exemplary works from the literature motivate such theories from the point of view of quantum field-theoretical corrections, as done here.
If indeed there is any classical signature of these \textit{exact} theories, these terms could only be relevant in strong gravity regimes such as black hole mergers \cite{Xav1,Hol} and it was pointed out in \cite{CapGW} and \cite{Xav} that their signature should be included in future simulations of gravitational waves generation from such events. Note, however, that potential future searches for stochastic gravitational wave background contain model-dependent features which are not yet taken into account \cite{IsiStein}.

\subsection{An example: simple harmonic oscillator with a higher-derivative term}
\label{subs_HDLHO}

We shall first recall the action of the simple harmonic oscillator in one dimension and the corresponding real solutions and then introduce a higher-derivative theory toy model based on that example.
There are a lot of higher-derivative toy models but they all share features which are originally met in the Pais-Uhlenbeck oscillator, the prototype of a higher derivative theory; see \cite{Pav} for a concise overview of its features.
This section is motivated by works of Bhabha~\cite{Bhabha} and Simon~\cite{Simon1,Simon2} who discussed few such examples of a higher-derivative theory in the context of perturbative approach to its solutions.
We construct our own example here which is equally well suited for demonstration of peculiar features of theories with higher derivatives.
This choice does not delete any of the main features shared with other higher-derivative models and only enriches the spectrum of higher-derivative Lagrangians which one could examine in order to understand them. 

The action of a simple harmonic oscillator reads
    \begin{equation}
    \label{LHOaction}
        S = \int \d t \left( 
        m\frac{\dot{x}^{2}}{2} - k\frac{x^2}{2} 
        \right)\ ,
    \end{equation}
where $m$ and $k$ are positive constants.
The Euler-Lagrange equations of motion are given by
    \begin{equation}
    \label{LHOEOM}
        \dd{L}{x} - \frac{\d}{\d t}\dd{L}{\dot{x}} = 0 \quad \Rightarrow \quad \ddot{x} + \omega_{m}^{2} x = 0\ ,
    \end{equation}
where $\omega_{m} := \sqrt{k/m}$.
Note that we have obtained this equation after dividing by $m$, which is why we must assume $m\neq 0$. For this simple harmonic oscillator --- where there would be no dynamics at all if $m=0$ --- it is a trivial condition. 
But as we shall see later, the coefficient in front of the highest order term in the equation of motion needs to be treated with care if it is multiplied by a small parameter and one is interested in an approximate solution.
A general real solution may be written in the form of
    \begin{equation}
    \label{LHOsol}
        x = A \cos (\omega_{m} t - \phi)\ ,
    \end{equation}
where $A, \phi$ are \textit{two} arbitrary constants (amplitude and phase) parametrizing a solution to the \textit{second order} differential equation.
Our discussion will not be affected by limiting ourselves to this real solution.

Now let us modify the oscillator in the following way
    \begin{equation}
    \label{ALHOaction}
        S = \frac{1}{2}\int \d t \left( 
        m\dot{x}^{2} - k x^2 - g \left(\ddot{x}-f x\right)^2
        \right)\ ,
    \end{equation}
where $g$ and $f$ are real positive\footnote{The positivity assumption can be relaxed but for the main point of this section it is enough to assume only positive values.} constants.
The choice of signs in this new term does not affect the conclusion and the choice of the term itself could be different,
as long as, more importantly, we have a Lagrangian which contains second time derivatives in a way that cannot be reduced to depend on the first derivatives only.
Formally, we could say that if the Lagrangian $L=L(x(t),\dot{x}(t),\ddot{x}(t))$ is nonlinear in second and higher derivatives,
    \begin{equation}
    \label{ALHO_HDcond}
        \frac{\del^2 L}{\del (x^{(i)})^2} \neq 0\ ,
    \end{equation}
for at least one $i>1$, where $x^{(k)}$ is the $k$-th time derivative of $x$, then Lagrangian $L$ describes a \textit{higher derivative theory}.
In that case no boundary term can be added to reduce it to a first order Lagrangian
\footnote{An example of a first order Lagrangian which contains second derivative term that can be eliminated by an addition of a boundary term is the Einstein-Hilbert Lagrangian describing GR.}.
In this thesis we shall take an approach to theories containing higher derivatives based on the following discussion. 

Let us imagine a weighing balance (weighing scale) instrument as a metaphor of comparison of two terms in a given action: the kinetic term on one plate and the non-linear second derivative term on the other plate.
The role of weights for each of these terms is played by their respective coupling constants: the coupling constant $m$ for the kinetic term and the coupling constant $g$ for the higher derivative term.
If we choose units such that $[t]=[x]=T$, i.e. units of time, and interpret the action as dimensionless ($[S] = 1$) then we see that the two coupling constants cannot be compared because they have different dimensions: $[m]=T^{-1}$ and $[g]=T$, while we also note that $[k]=T^{-3}$ and $[f]=T^{-2}$.
But we could introduce a certain characteristic physical time scale (in analogy way as we introduced characteristic length scale $l_{0}$ in chapter \ref{ch:umtocf1}) in terms of which $g$ can be expressed.
Then we could extract the characteristic time scale from coordinates as $t\rightarrow t_{0} t$ after which $t$ and $x$ become dimensionless.
We could think of this characteristic time scale as the period $\sqrt{m/k}$ of the simple oscillator, for example.
Now we can see that $t_{0} m \rightarrow m$ and $g/t_{0}\rightarrow g$ can be compared because they are dimensionless (and similar can be done for the other two coupling constants).
Dimensions of the original $g$ suggest that it could be interpreted as a kind of a time scale.
This time scale can be thought of as a characteristic time scale over which the higher-derivative effects are relevant.
Regarding the value of the new, dimensionless $g$ itself, if $g<1$ then this time scale is shorter than the characteristic time scale $t_{0}$ and if $g>1$ then it is longer.
If $m > g$ the balance is in favor of the kinetic term and the effects of the second derivative term in the equation of motion dominate over the fourth derivative term.
In the opposite case $m < g$ the balance is in favor of the fourth derivative term.
Now let us imagine that $m$ and $g$ are non-constant weights, i.e. that their value decreases in time for an unknown and for our discussion irrelevant reason, such that $g$ decreases relatively faster compared to $m$.
Let us also assume that this change happens over a much greater time period compared to $t_{0}$.
We need this last assumption because we cannot implement the unknown time dependence of $m$ and $g$ and we will make sure that Emmy and Richard --- two physicists from two very distant periods of time --- run their experiment over timescales within which both $m$ and $g$ are approximately constants.
Let Emmy and Richard model an oscillator according to the Lagrangian in eq.~\eqref{ALHOaction} and let the values of $m$ and $g$ be known to them at the time of their respective experiments (and let Emmy's and Richard's value of $k$ and $f$ be the same).
Let Emmy know with certainty that $m \ll g$ as a result of some independent set of measurements from her time.
On the other hand, let Richard know with certainty that $g \ll m$ as a result of some independent set of experiments from his time and that $g=0$ is a good approximation to the relevant observed phenomena.
Both Emmy and Richard use the following Euler-Lagrange equation of motion for describing the problem,
    \begin{align}
    \label{ALHOEOMEL}
        \dd{L}{x} - \frac{\d}{\d t}\dd{L}{\dot{x}} + \frac{\d^2}{\d t^2}\dd{L}{\ddot{x}} &= 0\ ,\\[12pt]
    \label{ALHOEOM}
        \Rightarrow \qquad g \left(\frac{\d^2}{\d t^2} -  f \right)\left(\frac{\d^2}{\d t^2} -  f \right) x + m \ddot{x} + k x & = 0\ ,
    \end{align}
where $\ddot{x}$ denotes the second derivative and we shall also use in the future $\ddot{\ddot{x}}\equiv x^{(4)}$ to designate the fourth derivative.
But they will treat this equation differently.
What are the solutions for the equations of motion that Emmy and Richard can use and what assumptions are Emmy and Richard allowed to make in order to find approximate solutions?
This is the most important question that we believe sits in the core of understanding which methods can be used in dealing with a theory containing higher than second derivatives.

Let us first examine what is the solution to Emmy's problem. 
Since the weight balance is tipped in favor of a higher derivative term for Emmy, as she knows that $m \ll g$, she can implement this assumption in eq.~\eqref{ALHOEOM} by dividing by $g \neq 0$ to obtain
    \begin{equation}
    \label{ALHOMeom}
        \ddot{\ddot{x}} - \left(2f - \frac{m}{g}\right) \ddot{x} + \left(f^2 + \omega^{2}_{g} \right) x = 0\ ,
    \end{equation}
where $\omega_{g}:=\sqrt{k/g}$ and solve this equation exactly with an ansatz $x\sim \exp{\tilde{\omega}t}$ to obtain \textit{four} solutions
    \begin{subequations}
        \begin{align}
        \label{ALHOMom12}
            \tilde{\omega}_{1,2} & = \pm \left(\frac{m}{2g} - f -  \sqrt{\frac{m^2}{4 g^2} - \frac{m f}{g} - \omega^2_{g} } \,\right)^{\frac{1}{2}}\ ,\\[12pt]
        \label{ALHOMom34}
            \tilde{\omega}_{3,4} & = \pm \left(\frac{m}{2g} - f + \sqrt{\frac{m^2}{4 g^2}  - \frac{m f}{g} - \omega^2_{g}} \,\right)^{\frac{1}{2}}\ .
        \end{align}
    \end{subequations}
Since $m < g$, it can be inferred from the above that the exponents of Emmy's solutions are always complex.
This means they contain both oscillatory (coming from the imaginary part) and exponentially decaying/increasing (coming from the real part) factor.
This is the main feature of a higher-derivative theory: it always contains more than two independent solutions and always contains solutions which are so called ``runaway'', i.e. the norm of the amplitude diverges with time.
Emmy can expand her four solutions in the extreme case of $m \ll g$, or use this approximation as a tool for finding an approximate solution to her problem.
One says that her solutions are \textit{perturbatively expandable} in perturbation parameter $m/g$; it means that using the limit $m \ll g$ does not produce inconsistency.
Then Emmy could write\footnote{Or $x\approx x_{0} + \frac{m}{g} x_{1}$, but if we assume that all constants except $m$ are of the order 1 then the stated approximation suffices, since $m$ is dimensionless.} $x \approx x_{0} + m x_{1}$ and use this in eq.~\eqref{ALHOMeom}, summing all terms with the same power of $m$ to zero.
She would obtain
    \begin{subequations}
        \begin{align}
        \label{ALHOMeom0}
            \ddot{\ddot{x}}_{0} - 2f\ddot{x}_{0} + \left(f^2 + \omega^{2}_{g} \right) x_{0} = 0\ ,\\[12pt]
        \label{ALHOMeom1}   
            \ddot{\ddot{x}}_{1} - 2f\ddot{x}_{1} + \left(f^2 + \omega^{2}_{g} \right) x_{1} + \frac{1}{g}\ddot{x}_{0}= 0\ .
        \end{align}
    \end{subequations}
The first equation determines $x_{0}$, which is the solution one would obtain if Lagrangian in eq.~\eqref{ALHOaction} did not have the kinetic term (i.e. $m = 0$) to start with.
Emmy would solve for $x_{0}$ and plug this solution into eq.~\eqref{ALHOMeom1} and then solve for $x_{1}$, thus finding the solution to the full equation of motion with precision of up to $\mathcal{O}(m)$.
The same solution could be found by simply Taylor-expanding eq.~\eqref{ALHOMom12} and eq.~\eqref{ALHOMom34} around $m=0$ (or $m/g = 0$); the two methods give identical results, as expected.
In summary, Emmy finds four independent solutions to the fourth order equation of motion for $x$ and this means she needs to impose four initial conditions: position, velocity, acceleration and the first derivative of acceleration.
This is also true for the approximate solution, taking $m \ll g$.

Richard, on the other hand, has a different problem.
The observations from his time give with certainty $g \ll m$ and the weight balance for him takes the opposite position compared to Emmy's.
He thinks of eq.~\eqref{ALHOEOM} as a second order equation of motion which has a small correction in the form of the fourth derivative of $x$ and he tries to proceed by finding a solution to the following equation of motion
    \begin{equation}
    \label{ALHOReom}
        \ddot{x} + \omega^2_{m} x + \frac{g}{m}\left(\ddot{\ddot{x}} - 2f\ddot{x} + f^2 x\right)  = 0\ ,
    \end{equation}
where $\omega_{m} = \sqrt{k/m}$, i.e. he divides eq.~\eqref{ALHOEOM} by $m$.
But now he is in a dilemma: does he treat eq.~\eqref{ALHOReom} as a fourth order equation or does he treat it as a second order equation with a small perturbation proportional to $g/m$?
If he treats it exactly then the solutions are found in the same way as in Emmy's case and lead to four of them, given by eqs.~\eqref{ALHOMom12}-\eqref{ALHOMom12}.
But if one is not careful then one could miss an important fact: solutions in eqs.~\eqref{ALHOMom12}-\eqref{ALHOMom12} are found under the assumption $g \neq 0$ (since one must divide by $g$) and $m \ll g$, so even though the latter can be relaxed, these solutions are thus \textit{not perturbatively expandable in powers of} $g$ \textit{around} $g=0$.
To emphasize: dividing by $g$ is \textit{forbidden} if one is looking for perturbative solutions \cite{Simon1,Simon2}.
Indeed, the limit of $g \rightarrow 0$ in eqs.~\eqref{ALHOMom12}-\eqref{ALHOMom34} diverges.
On the other hand, if he tries to find the solution perturbatively he would expand the solution as 
    \begin{equation}
    \label{ALHORxpert}
        x \approx x_{0} + g x_{1}\ ,
    \end{equation}
and end up with
    \begin{subequations}
        \begin{align}
        \label{ALHOReom0}
            \ddot{x}_{0} + \omega^2_{m} x_{0} & = 0\ ,\\[12pt]
        \label{ALHOReom1}
            \ddot{x}_{1} + \omega^2_{m} x_{1} + \frac{1}{m}\left(\ddot{\ddot{x}}_{0} - 2f\ddot{x}_{0} + f^2x_{0}\right)  & = 0\ .
        \end{align}
    \end{subequations}
The second derivative of eq.~\eqref{ALHOReom0} can be used to eliminate the fourth order derivatives from eq.~\eqref{ALHOReom1}, which results in 
    \begin{equation}
    \label{ALHOReom2}
        \ddot{x}_{1} + \omega^2_{m} x_{1} + \frac{1}{m}\left(\omega^{2}_{m} +  f\right)^2 x_{0}  = 0\ .
    \end{equation}
Now let Richard assume that his solution is of the form $x = A \cos (\tilde{\omega}t) $ (choosing a vanishing phase).
Then according to eq.~\eqref{ALHORxpert} one has
    \begin{align}
    \label{ALHOXexp}
        x &\approx A \cos ((\omega_{m} + g \omega_{1} )t)\nonumber\\[6pt]
        & \approx \underbrace{A \cos (\omega_{m}t)}_{=x_0} - g \underbrace{ A \omega_{1} t \sin (\omega t)}_{=x_{1}}
    \end{align}
with terms of order $\mathcal{O}(g^2)$ and above neglected and with initial conditions
    \begin{equation}
        x_{0}(0)  = A\ ,\quad \dot{x}_{0}(0)  = 0\ ,\quad g x_{1}(0)  = 0\ ,\quad g \dot{x}_{1}(0)  = 0\ ,
    \end{equation}
compatible with $x(0)=A$ and $\dot{x}(0)=0$. In other words, Richard must impose the initial conditions \textit{at each perturbative order}.
It is important to note that the first line in eq.~\eqref{ALHOXexp} is valid for all values of $t$, while the approximation in the second line is valid only if $\vert g \omega_{1} t\vert \ll 1$ is assumed in addition; if the system is observed during a time beyond $t\sim 1/\vert g\omega_{1}\vert $ the second approximation in eq.~\eqref{ALHOXexp} breaks down.
Solving eq.~\eqref{ALHOReom2} with $x_{1}$ ansatz from eq.~\eqref{ALHOXexp} gives $\omega_{1} = (f+\omega_{m}^{2})^{2} / 2 m\, \omega_{m}$.
In this way Richard has found a \textit{perturbative} solution 
    \begin{align}
    \label{ALHOXexpS1}
        x &\approx A \cos \left(\omega_{m}t + \frac{g}{m}  \frac{(f + \omega_{m}^{2})^2}{2 \omega_{m}} t\right)
    \end{align}
which, as expected, reduces to the solution of a simple harmonic oscillator once $g\rightarrow 0$ limit is taken and which he may expand as in the second line of eq.~\eqref{ALHOXexp} if careful about its validity only up to some timescale $t\sim 1/\vert g\omega_{1}\vert$.
The fact that the correction is proportional to $t$ should not worry Richard if he is using the model over a finite period of time.
Otherwise, the solution given by eq.~\eqref{ALHOXexpS1} is valid for all times.

In order to obtain more intuition about the perturbative approach, Richard comes up with another way of deriving his perturbative solution. 
Namely, if he claimed that eq.~\eqref{ALHOReom0} holds before even deriving the full equations of motion, he could use this zeroth order equation in the higher-derivative terms directly in his Lagrangian, i.e. substituting $\ddot{x}\rightarrow \omega_{m}^2 x$ directly in eq.~\eqref{ALHOaction}, obtaining
    \begin{align}
    \label{ALHOactionSubs}
        S & = \frac{1}{2}\int \d t \left( 
        m\dot{x}^{2} - k x^2 - g \left(\omega_{m}^2 + f \right)^2x^2
        \right)\ ,\nonumber\\[12pt]
        & = \frac{1}{2}\int \d t \left( 
        m\dot{x}^{2} - \left(k + g\left(\omega_{m}^2 + f \right)^2\right)x^2
        \right)\ ,
    \end{align}
from which the following equation of motion and its solution in the $g/m \ll 1$ limit can be derived
    \begin{equation}
    \label{ALHOpertsolfin}
        \ddot{x} + \tilde{\omega}^2_{m} x^2 = 0\quad\Rightarrow \quad \tilde{\omega}_{m}\approx \omega_{m}\left(1 + \frac{g}{m}\frac{\left(\omega_{m}^2 + f \right)^2}{2\omega_{m}^2}\right)\ .
    \end{equation}
Comparison of the above solution with eq.~\eqref{ALHOXexpS1} shows that this is an identical result.
This procedure might be the most straightforward one: substitute all higher-derivatives in the Lagrangian by derivatives of the zeroth-order solution.

\label{p_pertg} In summary, we see that Richard's \textit{perturbative} approach is the one which gives him consistent results and he cannot use the exact solution to the fourth order theory.
The consistency is reflected in the fact that Richard's Lagrangian is not exact because assumption $g \ll m$ makes the kinetic term dominate the higher-derivative term; therefore the corresponding equation of motion cannot be exact and the corresponding solution cannot be exact, but they must be treated perturbatively.
This is why Richard has only two degrees of freedom instead of four like Emmy.
Moreover, it can be shown \cite{Simon2} that perturbative solution does not make sense if it is truncated at the order higher than the highest order of the higher-derivative term in the Lagrangian; in other words, if Lagrangian contains higher-derivative terms up to order $g^n$ then the solution makes sense only if it is expanded up to order $n$ and not above, otherwise one obtains again non-perturbative solutions as $g\rightarrow 0 $ is taken.

There is a question of origin of this higher-order perturbation in Richard's case.
Suppose Richard discovers Emmy's theoretical and experimental results in a paper written long before his time.
At first, he is confused because they both used the same Lagrangian but soon he discovers (by investigating the observational data from Emmy's time) that Emmy's constants $m$ and $g$ differ drastically from the same constants measured in his time.
This is based on the fact that (according to the dimensional analysis of the action in eq.~\eqref{ALHOaction} which is discussed at the beginning of this section) 
$m$ couples a term which is proportional to the characteristic time scale, while $g$ couples a term which is inversely proportional to the characteristic time scale: they cannot contribute in the same way for small characteristic time scales as for the large ones and from this one may deduce that the higher-derivative term ``resolves'' effects of a high-frequency (fast oscillator with a low period) oscillator.
The higher the frequency, the shorter the characteristic time scale and the more important the higher-derivative term is; this can be easily seen from eq.~\eqref{ALHOXexpS1}.
Alternatively, one can say that the slower the oscillator, the less important the higher-derivative term is.
This reminds one of a Taylor series: an analytic function in a small neighborhood of a point can be expanded in an infinite Taylor series in powers of a $\epsilon \ll 1$ parameter that measures the size of the point's neighbourhood.
Each next order of the series is of a higher and higher derivative term and gives a finer and finer modification to the value of a function at the point --- and this is what we mean by ``resolving'' (in this case the smaller patches of the point's neighborhood).
Then one can imagine that in Richard's case the fourth derivative term in the Lagrangian is just the first term of an infinite series of higher and higher derivatives which converge to form some non-local contribution.
We will only briefly here mention how Simon \cite{Simon1, Simon2} has shown this very elegantly.
Namely, consider the following equation of motion (we use Simon's notation and come back to ours only after presenting his findings)
    \begin{equation}
    \label{SimonEOM}
        \ddot{x} + \omega^2  \int_{-\infty}^{+\infty}\d s\, \frac{e^{-\vert s\vert}}{2} x(t + \epsilon s) = 0\ .
    \end{equation}
We shall soon explain what this integral term actually means.
If one expands $x(t + \epsilon s)$ in Taylor series around $\epsilon = 0$, by direct integration one obtains an infinite sum of even derivatives of $x$,
    \begin{equation}
    \label{SimonEOM1}
        \ddot{x} + \omega^2 \sum_{n=0}^{\infty}\epsilon^{2n} \frac{\d^{2n}}{\d t^{2n}}x=0\ .
    \end{equation}
The crucial point here is that eq.~\eqref{SimonEOM} is a \textit{second order} equation and eq.~\eqref{SimonEOM1} is also a \textit{second order} equation, since the latter is derived from the former.
That means that there are only \textit{two} degrees of freedom, independently of the fact that eq.~\eqref{SimonEOM1} contains infinite number of terms with forever-increasing number of derivatives!
The next crucial point is that this fact \textit{would not change} if we chose to truncate the series at some order $\epsilon^{k}$.
One may choose to keep terms only up to $\epsilon^{2}$ in which case one obtains\footnote{The following equation is based on Simon's notation and the corresponding Lagrangian $L=\frac{1}{2}\left((1+\epsilon^2\omega^2)\dot{x}^2 - \omega^2 x^2 - \epsilon^2 \ddot{x}^2\right)$.}
    \begin{equation}
        (1+\epsilon^2\omega^2)\ddot{x} + \omega^2 x^2 + \epsilon^2 \ddot{\ddot{x}}=0\ ,
    \end{equation}
which corresponds by analogy to eq.~\eqref{ALHOReom} in our case.
It is also possible, as Simon showed in equation (34) in \cite{Simon1}, to find a Lagrangian from which eq.~\eqref{SimonEOM} can be derived. 
We will not write the Lagrangian here, but only note that it contains a similar integral as in eq.~\eqref{SimonEOM}.
The main point is that these terms containing integrals are \textit{non-local} in the sense of ``action-at-a-distance'': according to the values of integration boundaries in eq.~\eqref{SimonEOM}, contributions from infinitely distant past and infinitely distant future are contributing to the acceleration at time $t$ --- this is the feature of non-locality.
The moral of the story is that one does not need to know which non-local theory a higher-derivative term is derived from in order to solve the problem at the given order.
But if one finds --- like Richard has inferred from the available data on values of $m$ and $g$ --- that a higher-derivative term in the Lagrangian is relatively small compared to the kinetic term, then it needs to be treated that way (i.e. perturbatively) and this ensures that the theory has two degrees of freedom independently of how many small higher-derivative terms contribute to the Lagrangian.

One can indeed deduce that even if terms of order higher than $\ddot{x}$ appear in the Lagrangian they would even more finely ``resolve'' the time scales than the term proportional to $g$ because the corresponding coupling constant would be proportional to a higher power of the $g$-timescale.
Note that in Emmy's case (where $g > 1$) inclusion of higher and higher-order terms would give rise to more and more solutions and degrees of freedom.
It would be hard to motivate increasing number of independent solutions and one would need increasing number of initial conditions in order to solve the problem.
This is, however, not the problem with the \textit{infinite} sum but the price to pay is non-locality.
But the bigger issue is that such a hypothetical theory does not converge and cannot be reformulated as some non-local theory (or its truncation) as is with Richard's case.
This summarizes the most important problem with treating a higher-derivative theory as an exact one.
It is for this reason that we think that Richard's approach is the correct one for treating \textit{classical} higher-derivative theories, which in the case of this thesis refers to a theory of gravity: any theory of gravity containing higher derivatives of the metric tensor components in addition to the EH action is to be treated \textit{perturbatively} with respect to higher derivatives, preserving the second order nature of SEE.
We shall review this approach in the next subsection.

We have made sense of Richard's approach to his Lagrangian but what is then the interpretation of Emmy's Lagrangian? In this thesis we take the approach that if the couplings behave relative to one another to tip the balance towards the higher-derivative term as in Emmy's case, then the Lagrangian does not make sense as a classical one but must describe the quantum version of the theory.
In other words, until the value of $g$ decreases below $m$, the Lagrangian is to be quantized and describes high-frequency (high-energy), short time scale oscillations.
Only when $m\gg g$ Richard may recognize the Lagrangian as describing a classical theory, which describes the low-frequency, long time scale oscillations, while no additional solutions arise; otherwise, Richard interprets Emmy's Lagrangian within the context of a quantum theory.
It should be kept in mind that this line of thought implies that \textit{any classical theory derived from such a quantum theory in a semiclassical approximation would have to involve invoking the assumption of a frequency-dependent (or scale-dependent or energy-dependent) couplings $g$ and $m$, possibly through the methods of renormalization}, as reviewed in the previous section.
(However, we do not seek an implementation of energy-depending couplings in this thesis.)
This is in accordance with our discussion above on the relation between the size of coupling $g$ and the ``resolution'' of time scales: the more influential the higher order derivative terms are, the closer one is to the requirement to shift to the quantum description because the physics of small scales then becomes more important and perturbative approach breaks down for $g\sim m$ and high frequencies.

\subsection{Semiclassical Einstein equations and their perturbative solution}
\label{subsec_SEEpert}

We shall take Richard's situation described in the previous subsection as the analog of the scales and energies we consider today as the domain of validity of GR and its higher-derivative corrections.
Towards higher energies it is required --- as explained in section \ref{subs_HDSEE} --- to include quadratic curvature terms in the EH action and renormalize the coupling constants.
As in the case of the toy model used above, the quadratic curvature terms in the gravitational action give rise to fourth order derivatives.
Such an action is the basis of the \textit{effective approach} \cite{BOS, Don1} and we ought to have learned from Richard's \textit{perturbative} way of going about making sense of the solutions to such a theory. 
In the case of gravity the perturbative treatment of the SEE was introduced by Simon \cite{Simon1, Simon2} and Parker and Simon \cite{ParSim} where it was shown that if the order of the SEE based on eq.~\eqref{SEE_effActL} is reduced perturbatively the theory does not suffer from unstable solutions and spacetime metric is perturbatively expandable in powers of $\hbar$, giving a sensible classical limit.
We shall sketch their procedure here but use a slightly different form of eq.~\eqref{SEE_effActL}.

In principle one could have included the term $R_{\mu\alpha\nu\beta}R^{\mu\alpha\nu\beta}$ in eq.~\eqref{SEE_effActL} in the integral but it turns out that in four dimensions there is an identity among the metric variation of the three curvature terms in the action (see e.g. Appendix B in \cite{tHoofVelt} or a recent review on quadratic gravity by Salvio \cite[section 2.1]{Salv}).
Namely, the Gauss-Bonnet term in four dimensions is a topological invariant, being a total covariant divergence and takes the following form,
    \begin{equation}
    \label{GaussBonnet}
        G=\frac{1}{4}\epsilon^{\mu\alpha\rho\sigma} \epsilon^{\nu\beta\tau\gamma}R_{\mu\alpha\nu\beta}R_{\rho\sigma\tau\gamma} =
        R_{\mu\alpha\nu\beta}R^{\mu\alpha\nu\beta} - 4 R_{\mu\nu}R^{\mu\nu} + R^2 = \text{cov. div.}\ ,
    \end{equation}
where $\epsilon^{\nu\beta\tau\gamma}$ is the Levi-Civita tensor density. 
Its metric variation therefore does not contribute to the equations of motion and one can use it to express the Riemann tensor squared in terms of $R_{\mu\nu}R^{\mu\nu}$ and $R^2$ up to a divergence
    \begin{equation}
    \label{RiemTGauss}
       R_{\mu\alpha\nu\beta}R^{\mu\alpha\nu\beta} =
        4 R_{\mu\nu}R^{\mu\nu} - R^2 + G\ .
    \end{equation}
On the other hand, Riemann tensor squared can be expressed in terms of its irreducible pieces based on eq.~\eqref{WeyltensS}, 
    \begin{equation}
    \label{RiemToWeyl}
        R_{\mu\alpha\nu\beta}R^{\mu\alpha\nu\beta} = C_{\mu\alpha\nu\beta}C^{\mu\alpha\nu\beta} + 2 R_{\mu\nu}R^{\mu\nu} - \frac{1}{3} R^2\ .
    \end{equation}
Using eq.~\eqref{RiemTGauss} in eq.~\eqref{RiemToWeyl} one obtains\footnote{\label{SchWsquared} It is interesting to see that $ \frac{1}{2}C_{\mu\alpha\nu\beta}C^{\mu\alpha\nu\beta}  - \frac{1}{2} G =R_{\mu\nu}R^{\mu\nu} - \frac{1}{3} R^2 = G_{\mu\nu}P^{\mu\nu}$ where $P^{\mu\nu}$ is the Schouten tensor introduced in eq.~\eqref{Schoudef}.}
    \begin{equation}
        R_{\mu\nu}R^{\mu\nu} = \frac{1}{2}C_{\mu\alpha\nu\beta}C^{\mu\alpha\nu\beta} + \frac{1}{3} R^2 - \frac{1}{2} G\ ,
    \end{equation}
which can be used in the action given by eq.~\eqref{SEE_effActL} to  get (using $C^2\equiv C_{\mu\alpha\nu\beta}C^{\mu\alpha\nu\beta}$)
    \begin{equation}
    \label{SEE_TheAction}
        S_{g}=\intxx\sqrt{g}\left( \frac{l^{2}\hbar}{2}\left( R - 2 \bar{\Lambda}\right) + \frac{\br \hbar}{4} R^2 - \frac{\aw\hbar}{4} C^2\right)
    \end{equation}
where we define the new coupling constants $\br := 4\beta_{1} + 4\beta_{2}/3$ and $\aw := -4 \beta_{2}/3$; we subtract the Gauss-Bonnet term since we are not concerned with spacetimes with a boundary and assume there are no topological issues, for simplicity.
Recall that all couplings in the Lagrangian have no physical meaning until renormalization procedure is taken care of.
Only then one could make sensible predictions of the theory both in high-energy and low-energy limits.
We do, however make a constraint that $\aw, \br > 0$ in order to incorporate indications that such choice ensures non-tachyonic modes \cite{Stelle2, Salv}.
It looks like the $R^2$ and $C^2$ terms are the only ones in four dimensions which contribute to the equations of motion to the order of $\hbar$; they are also two independent pieces of the Riemann tensor.
We shall see in section \ref{sec_HHDall} that this has a deeper meaning.

If eq.~\eqref{SEE_TheAction} is supplemented by a matter action containing both quantized and classical matter, its variation gives the following equations of motion
    \begin{equation}
    \label{SEE_full}
        l^2\hbar \left( R_{\mu\nu} - \frac{1}{2}g_{\mu\nu}R + g_{\mu\nu}\bar{\Lambda}\right) + \br \hbar H_{\mu\nu} - 2\aw\hbar  B_{\mu\nu}= T_{\mu\nu}^{\ssst cl} + \langle \hat{T}_{\mu\nu}\rangle
    \end{equation}
where $T_{\mu\nu}^{\ssst cl}$ is the classical energy-momentum tensor and recall that $\langle \hat{T}_{\mu\nu}\rangle=\mathcal{O}(\hbar)$.
Tensors $H_{\mu\nu}$ and $B_{\mu\nu}$ are defined as
    \begin{align}
    \label{R2eom}
        H_{\mu\nu} &= \frac{1}{4}H_{\mu\nu}^{\ssst{(1)}} = R_{\mu\nu}^{\srm T}R - \left(\nabla_{\mu}\nabla_{\nu} - g_{\mu\nu}\Box\right)R\ ,\\[12pt]
    \label{C2eom}
        B_{\mu\nu} &= \left(\nabla_{(\alpha}\nabla_{\beta)} + \frac{1}{2}R_{\alpha\beta}\right)C^{\alpha}{}_{\mu}{}^{\beta}{}_{\nu}\ .
    \end{align}
Tensor $B_{\mu\nu}$ is called the Bach tensor \cite{Bach} and arises from the variation of the Weyl-tensor term $C^2$.
Since the Weyl-tensor term is conformally invariant (cf. eq.~\eqref{WeylactionDconf}), Bach tensor is also conformally invariant, i.e. variation with respect to the scale density vanishes.
Using the unimodular-conformal decomposition this is shown explicitly in eq.~\eqref{BachSmpl}.
Because of this, Bach tensor contribution changes only the traceless part of the Einstein equations, while $H_{\mu\nu}$ changes also its trace:
    \begin{align}
    \label{SEE_full_Tless}
        l^2\hbar R_{\mu\nu}^{\srm T} + \br \hbar H_{\mu\nu}^{\srm T} - 2\aw\hbar  B_{\mu\nu}&= T_{\mu\nu}^{{\srm T}\ssst cl} + \langle \hat{T}_{\mu\nu}^{\srm T}\rangle\, \\[12pt]
    \label{SEE_full_Tr}
        l^2\hbar \left( 4\bar{\Lambda} - R \right) + 3\br \hbar \Box R & = T^{\ssst cl} + \langle \hat{T}^{\mu}_{\mu}\rangle\ ,
    \end{align}
where 
    \begin{equation}
    \label{SEE_full_HTless}
        H_{\mu\nu}^{\srm T} = R_{\mu\nu}^{\srm T}R - \left(\nabla_{\mu}\nabla_{\nu} - \frac{1}{4} g_{\mu\nu}\Box\right)R\ .
    \end{equation}
Depending on a specific spacetime model and type of matter, there could be certain simplifications, but also some additions to eq.~\eqref{SEE_full}.
For example, for conformally flat and Einstein spacetimes $B_{\mu\nu}=0$.
For conformal classical matter one has $T=0$, but for conformal \textit{quantum} matter it turns out that $\langle \hat{T}^{\mu}_{\mu}\rangle\neq 0$, which means that quantum corrections of a massless conformally coupled scalar field or pure electromagnetic field, for example, break conformal symmetry.
The latter is named \textit{conformal anomaly} \cite{BD,BOS,PT} and is a very important subject, especially in relation to the possibility of having a conformal symmetry in a quantum gravity theory.
It is given by
    \begin{equation}
    \label{confanom}
        \langle \hat{T}^{\mu}_{\mu}\rangle \sim C^2 + \left(G - \frac{2}{3}\Box R\right)\ ,
    \end{equation}
with each their own \textit{finite} constants proportional to $\hbar$ that can be calculated for a specific matter field theory.
Where does eq.~\eqref{confanom} come from? It can be shown (see e.g. \cite[section 2]{Mottola}) that certain non-local terms must be added to the action in order to generate the conformal anomaly via the variational principle.
These local terms have their local version \cite[eq. (2.14)]{Mottola} but only if one introduces an additional scalar field --- which represents the dynamical degree of freedom that appears because the broken conformal symmetry in the matter sector introduces a non-vanishing $\langle \hat{T}^{\mu}_{\mu}\rangle$.
The local version of the action is the an action for a \textit{fourth order derivative} theory of a scalar field which is non-minimally and conformally coupled to gravity.
This makes the resulting theory a sort of a scalar-tensor theory but with a lot more complicated interaction terms.
Due to its complicated nature we shall not consider conformal anomaly terms in the action in this thesis, but we stress that they should be included in the further research on the topic in this thesis.
Nevertheless, the conformal anomaly in eq.~\eqref{confanom} would contribute to eq.~\eqref{SEE_full} in addition to the quadratic curvature terms, after calculating $\langle \hat{T}_{\mu\nu}\rangle$ explicitly.
Another special case are classical vacuum spacetimes, i.e. $T_{\mu\nu}=0$.
\textit{In vacuum spacetimes there are no corrections} because both $B_{\mu\nu}$ and $H_{\mu\nu}$ vanish (the latter vanishes because the Bianchi identity for the Riemann tensor reduces to $\nabla_{\alpha}C^{\alpha}{}_{\mu\beta\nu}=0$).

What is the solution to eq.~\eqref{SEE_full}? 
Before we set on this endeavour let us agree that we have already regularized $\langle \hat{T}_{\mu\nu}\rangle$ according to eq.~\eqref{SEE_tmnfindiv} and absorbed $\langle \hat{T}_{\mu\nu}\rangle_{\infty}$ into couplings $l^2,\aw$ and $\br$ by
renormalizing them into physically meaningful and energy-scale dependent couplings  $l^2_{\ssst phys},\aw^{\ssst phys}$
and $\br^{\ssst phys}$, but that we drop the ``$\ssst phys$'' label in order to simplify the notation. So from now on, all couplings are renormalized and in the future equations we write $\langle \hat{T}_{\mu\nu}\rangle_{fin}$ in place of $\langle \hat{T}_{\mu\nu}\rangle_{\infty}$.
As we have argued so far in the current chapter, the SEE should not be solved exactly for the metric because $H_{\mu\nu}$ and $B_{\mu\nu}$ are suppressed by their couplings compared to $l^{2}$.
If one uses $1/\kappa$ instead of our $l^2\hbar$ then one says that the two tensors are suppressed by $\mathcal{O}(\hbar)$, this was the way Parker and Simon approached the problem \cite{ParSim}.
But since we have turned $1/\kappa$ into $l^2\hbar$, we have to work with relative strengths of $l^2,\aw,\br$ instead of $\hbar$
This is required because we have multiplied and divided $1/\kappa$ by $\hbar$ in order to transform to $l^2$ and hence $\hbar\rightarrow 0$ is not a valid thing to do\footnote{This can be understood as a consequence of the fact that $\hbar,c, G$ form one set of independent coupling constants while $l_{p},t_{p},m_{p}$ form \textit{another} set of independent coupling constants, which means their limits cannot be mixed.}. This represents another departure from \cite{ParSim}.
To make a connection to Richard's story form the previous subsection, $l^2$ is analogous to $m$, while $\aw,\br$ are analogous to $g$.
Therefore, we are allowed to divide by $l^2$ but not with $\aw,\br$, since we have to make sure it is possible to take the limit $\aw,\br\rightarrow 0$.

The perturbative solution can be constructed in the following way.
First, assume that $T_{\mu\nu}^{\ssst cl}$ is of the order $l^2\hbar$. 
This is necessary in order to make the perturbative treatment compatible with the classical interpretation of $T_{\mu\nu}^{\ssst cl}$ and it ensures that $\hbar$ is explicitly eliminated from it. 
We have to keep in mind that in the scales at which the geometry of spacetime is classical we have $l^2\gg 1$, which already means that we are in the domain of validity of 
eq.~\eqref{SEE_full}, i.e. well above the Planck length scale, so only classical matter can curve the spacetime.
(Also, note that if $T_{\mu\nu}^{\ssst cl}/l^2\hbar \ll 1$ then this means that we are in nearly vacuum spacetimes, which is a trivial case.
Furthermore, in some cases the entire $T_{\mu\nu}^{\ssst cl}$ could emerge from the $\mathcal{O}(\hbar^{0})$ terms in $\langle \hat{T}_{\mu\nu}\rangle$ in the limit of large number of ``particles'' (as excitations of the quantum fields).
Moreover, the quantum matter could represent the quantized perturbations of a scalar field, whose background component is classical and generates $T_{\mu\nu}^{\ssst cl}$.)
Secondly, divide eq.~\eqref{SEE_full_Tless} and eq.~\eqref{SEE_full_Tr} by\footnote{In \cite{ParSim} it was not allowed to divide by $\hbar$ as this was their perturbation parameter. In our case perturbation parameters are $\aw/l^2$ and $\br/l^2$, so we must refrain from dividing by these parameters.} $l^2\hbar$, and multiply both equations by\footnote{In \cite{ParSim} the equation was multiplied by $\hbar$, their perturbation parameter. As we progress towards the next chapter, we shall obtain an intuition that our choice of perturbation parameters makes things a bit more transparent due to their dimensionless nature and compatibility with the meaning of the semiclassical approximation to the quantum version of the theory, which we give in the following chapter.} $\br/l^2$ to obtain
    \begin{subequations}
    \begin{align}
    \label{SEE_full_TlessB}
         \frac{\br}{l^2}R_{\mu\nu}^{\srm T} + \frac{\br}{l^2}\left[\frac{\br}{l^2} H_{\mu\nu}^{\srm T} - 2\frac{\aw}{l^2}  B_{\mu\nu}\right] & =
         \frac{\br}{l^2}\frac{1}{l^2\hbar}T_{\mu\nu}^{{\srm T}\ssst cl} + \frac{\br}{l^2}\frac{1}{l^2\hbar} \langle \hat{T}_{\mu\nu}^{\srm T}\rangle_{fin}\, \\[12pt]
    \label{SEE_full_TrB}
        \frac{\br}{l^2}\left( 4\bar{\Lambda} - R \right) + 3\left(\frac{\br}{l^2}\right)^2\Box R & =
        \frac{\br}{l^2}\frac{1}{l^2\hbar}T^{\ssst cl} + \frac{\br}{l^2}\frac{1}{l^2\hbar}\langle \hat{T}^{\mu}_{\mu}\rangle_{fin}
    \end{align}
    \end{subequations}
Next, take the following approximation,
    \begin{equation}
        \frac{\aw}{l^2}\ll 1\ ,\quad \frac{\br}{l^2}\ll 1\ ,\quad \frac{\aw}{l^2}\frac{\br}{l^2}\ll 1\ .
    \end{equation}
by which one essentially assumes that higher-order terms do not contribute to the Einstein equations arising from the classical gravitational (EH) action.
Furthermore, note that the last term both in eq.~\eqref{SEE_full_TlessB} and eq.~\eqref{SEE_full_TrB} is of the order of $\hbar$, not $l^{2}\hbar$, because it contains no information about the classical matter (assuming it is all in $T_{\mu\nu}^{\ssst cl}$).

Then one neglects all terms of the order $\mathcal{O}\left(\br/l^2\right)$ and arrives at
    \begin{subequations}
    \begin{align}
    \label{SEE_full_TlessB1}
         \frac{\br}{l^2}R_{\mu\nu}^{\srm T\ssst cl} & =
         \frac{\br}{l^2}\frac{1}{l^2\hbar}T_{\mu\nu}^{{\srm T}\ssst cl} +\mathcal{O}(\frac{\br}{l^2},\frac{\br}{l^4})\, \\[12pt]
    \label{SEE_full_TrB1}
        \frac{\br}{l^2}\left( 4\bar{\Lambda} - R^{\ssst cl} \right) & =
        \frac{\br}{l^2}\frac{1}{l^2\hbar}T^{\ssst cl} + \mathcal{O}(\br/l^2,\br/l^4)
    \end{align}
    \end{subequations}
These are just Einstein equations and the label ``${\sst cl}$'' refers to the fact that these tensors are evaluated with the purely classical metric, the usual metric that one would obtain if there were no higher-derivative terms, which we denote as $g_{\mu\nu}^{\ssst cl}$. 
The same discussion with exact same result is valid for multiplying the equations with $\aw$ instead of $\br$.
What is the meaning of eqs.~\eqref{SEE_full_TlessB1}-\eqref{SEE_full_TrB1}? The meaning is: at the perturbation order of $\aw/l^2$ and $\br/l^2$ impose the classical Einstein equations as a constraint --- this is where the name \textit{perturbative constraint} derives from.
Then the total metric makes sense to be determined only up to $\mathcal{O}(\aw/l^2,\br/l^2)$ and it has to be expanded as
    \begin{equation}
    \label{SEE_gmetexp}
        g_{\mu\nu} = g_{\mu\nu}^{\ssst cl} + \frac{\aw}{l^2}h_{\mu\nu}^{\ssst \alpha} + \frac{\br}{l^2}h_{\mu\nu}^{\ssst \beta}\ .
    \end{equation}
This is completely analogous to eqs.~\eqref{ALHORxpert} and \eqref{ALHOReom0} where the zeroth order solution is the usual simple harmonic oscillator.
The next thing to do is to either plug eq.~\eqref{SEE_gmetexp} back into to eqs.~\eqref{SEE_full_Tless} and \eqref{SEE_full_Tless} (keeping in mind our agreement concerning the renormalized couplings), expand all tensors (including $T_{\mu\nu}^{\ssst cl}$) up to $\mathcal{O}(\aw/l^2,\br/l^2)$, sum one set of all terms with the same power $\aw/l^2$ and another with all terms with the same power $\br/l^2$ and solve for $h_{\mu\nu}^{\ssst \alpha}$ and $h_{\mu\nu}^{\ssst \beta}$, using already solved Einstein equations for $g_{\mu\nu}^{\ssst cl}$.
Somewhat formally, these equations would in principle read
    \begin{subequations}
    \begin{align}
    \label{SEE_full_TlessB2}
         \delta R_{\mu\nu}^{\srm T}\left[g^{\ssst cl},\, \frac{\aw}{l^2}h^{\ssst \alpha},\, \frac{\br}{l^2}h^{\ssst \beta}\right]& + \frac{\br}{l^2}  H_{\mu\nu}^{\srm T}\left[g^{\ssst cl}\right] - 2\frac{\aw}{l^2} B_{\mu\nu}\left[g^{\ssst cl}\right] \nonumber\\[12pt]
         & = \frac{1}{l^2\hbar} \delta T_{\mu\nu}^{{\srm T}\ssst cl}\left[g^{\ssst cl},\, \frac{\aw}{l^2}h^{\ssst \alpha},\, \frac{\br}{l^2}h^{\ssst \beta}\right] + \frac{1}{l^2\hbar}\langle \hat{T}_{\mu\nu}^{\srm T}\rangle_{fin}\, \\[18pt]
    \label{SEE_full_TrB2}
        - \delta R\left[g^{\ssst cl},\, \frac{\aw}{l^2}h^{\ssst \alpha},\, \frac{\br}{l^2}h^{\ssst \beta}\right] & + 3\frac{\br}{l^2}\Box R 
        = \frac{1}{l^2\hbar}\delta T^{\ssst cl}\left[g^{\ssst cl},\, \frac{\aw}{l^2}h^{\ssst \alpha},\, \frac{\br}{l^2}h^{\ssst \beta}\right]  + \frac{1}{l^2\hbar}\langle \hat{T}^{\mu}_{\mu}\rangle_{fin}
    \end{align}
    \end{subequations}
where tensors $\delta R_{\mu\nu}^{\srm T},\delta R,\delta T_{\mu\nu}^{\srm T\ssst cl}$ and $\delta T^{\ssst cl}$ with arguments in the square brackets $[...]$ are to be understood as the \textit{first order perturbations} in $h^{\ssst\alpha}_{\mu\nu}$ and $h^{\ssst\beta}_{\mu\nu}$ of the corresponding tensors, while for the rest of the tensors $[g^{\ssst cl}]$ means that they depend only on the classical solution, found at the previous order.
This procedure ensures that the SEE equations are second-order and thus do not suffer from runaway solutions.
Therefore, all terms in the above two equations are $\mathcal{O}(\aw/l^2,\br/l^2)$, with one exception: $\langle \hat{T}_{\mu\nu}^{\srm T}\rangle_{fin}$ and $\langle \hat{T}^{\mu}_{\mu}\rangle_{fin}$. But this shouldn't be confusing because at this order there is no relative coupling multiplying these terms which would determine whether the size of $\langle \hat{T}_{\mu\nu}^{\srm T}\rangle_{fin}/l^2\hbar$ and $\langle \hat{T}^{\mu}_{\mu}\rangle_{fin}/l^2\hbar$ is comparable to the rest of the terms in the respective equations.
This reflects the high non-linearity of the SEE: perturbations $h^{\ssst\alpha}_{\mu\nu}$ and $h^{\ssst\beta}_{\mu\nu}$ cannot be in general determined unless $\langle \hat{T}_{\mu\nu}\rangle_{fin}$ is known, but the latter is not known until one solves for the quantum state with respect to which it is evaluated; however the quantum state is not known a priori, because it depends on the spacetime geometry on which it propagates and this complicates the issue considerably.
Only in special cases, such as e.g. conformally coupled massless scalar field on conformally flat backgrounds $\langle \hat{T}_{\mu\nu}\rangle_{fin}=0$, $B_{\mu\nu}[g^{\ssst cl}]=0$ could one explicitly write down the solution~\cite{ParSim} and in that case there is another term (cf. eq.~\eqref{confanom}) which contributes to the trace eq.~\eqref{SEE_full_TrB2} that we did not include.

There is another way of solving the SEE which is achieved in direct analogy with the method that led to eq.~\eqref{ALHOactionSubs} in section~\ref{subs_HDLHO}.
Namely, we showed there that Richard could simply substitute the zeroth-order solution directly into the higher-derivative part of the Lagrangian, which lead to the perturbed \textit{second-order} equations of motion.
Richard could do the same with the presently discussed higher-order gravity, since the form of the SEE \eqref{SEE_full_Tless}-\eqref{SEE_full_Tr} is conveniently given in terms of the Ricci curvature: using Einstein equations~\eqref{ttREE}, i.e. $R_{\mu\nu}^{\srm{T}\ssst cl} = T_{\mu\nu}^{\srm{T}\ssst cl}/l^2\hbar$ and $R=(4\bar{\Lambda}-T^{\ssst cl})/l^2\hbar$, eliminate the Ricci curvature appearing in the higher-derivative terms from either the action in eq.~\eqref{SEE_effActL} or the equations of motion~\eqref{SEE_full_Tless}-\eqref{SEE_full_Tr} themselves.
The result is the following gravitational part of the \textit{perturbatively constrained} semiclassical action,
    \begin{align}
    \label{SEE_effActLR}
        S_{g} = \intxx\sqrt{g}\Bigg[ &
        \frac{l^{2}\hbar}{2}\left( R - 2 \bar{\Lambda}\right)\nld
        & + \frac{\beta_{1}}{l^2} \left(4\bar{\Lambda} - T^{\ssst cl}\right)^2
         + \frac{\beta_{2}}{l^2} \left(T_{\mu\nu}^{{\srm T}\ssst cl}T_{\ssst cl}^{{\srm T}\mu\nu} + 4\left(4\bar{\Lambda} - T^{\ssst cl}\right)^2 \right)
        \Bigg]\ .
    \end{align}
Variation of the total action (i.e. with the additional matter contributions) results in perturbed SEE, but their form depends on the form of $T_{\mu\nu}^{\ssst cl}$.
Alternatively, but similarly, the zeroth order solution could simply be substituted into the higher-derivative term in the equations of motion themselves and then proceed by solving the resulting \textit{second order} differential equation.
All these procedures reduce the order of the equations of motion and give the same result \cite{ParSim}.

We shall not go into more details, because the perturbative treatment of the SEE has already been studied extensively in~\cite{ParSim} for several spacetime models with $\hbar$-corrections to the classical solutions to GR.
Let us only copy here the result of one of the models discussed in~\cite{ParSim}, namely the spatially flat Friedman model filled with radiation, without cosmological constant.
The perturbative procedure described in the present subsection leads to the following solution (after suitably rescaling the involved quantities to absorb irrelevant constants) for the scale factor as a function of time,
    \begin{equation}
        a(t) = (t - \tau_{0})^{\frac{1}{2}} - \hbar \alpha_{3} (t - t_{0})^{-\frac{3}{2}} - \hbar \tau_{1} (t - \tau_{0})^{-\frac{1}{2}}\ ,
    \end{equation}
where $\tau_{0}$ and $\tau_{1}$ are integration constants (note that $\tau_{1}$ is relevant only at the order of $\hbar$) while $\alpha_{3}$ is the coupling of the Gauss-Bonnet term in the conformal anomaly in eq.~\eqref{confanom}, which we did not include in our equations.
The first term is the purely classical term, with a familiar $t^{1/2}$ behavior of radiation-dominated universe. Terms proportional to $\hbar$ are semiclassical corrections. 
It can be seen that for late times $t\rightarrow \infty$, away from the initial singularity, one has $t\approx \tau_{0}$ and the scale factor behaves as in the classical case without any abnormalities.
The solution is thus perturbatively expandable in $\hbar$.
But close to the initial singularity the solution diverges.
This is expected, because the perturbative treatment breaks down at very early times and very high energies.

We find it of considerable significance to have reviewed this subject because it has gone almost unnoticed in the classical higher-derivative theory communities for a couple of decades, while research within the classical gravity context has been going mostly in the direction of making sense of actions such as eq.~\eqref{SEE_TheAction} in the purely classical non-perturbative approach and trying to tackle the issue of runaway solutions. 
In the light of the recent detection of the gravitational waves we think that the perturbative approach to the SEE should be revived as a physically more meaningful treatment of strong gravity regimes in black hole mergers.
It was concluded in \cite{Xav1} that \textit{exact} effects of the higher-derivative terms in the gravitational sector should be implemented into numerical simulations of merges of compact objects since these terms could have observational signatures. 
However, we think that one should at least in parallel try to implement the perturbative treatment of higher-derivative terms into not only numerical simulations of these mergers but also the physics of the primordial fluctuations of spacetime and matter and the inflationary universe.



\section[Hamiltonian formulation of a simple harmonic oscillator with higher derivatives][Hamiltonian formulation of a simple harmonic oscillator...]{Hamiltonian formulation of a simple harmonic oscillator with higher derivatives}
\label{subs_Ham_osc}

In the rest of this chapter we shall present the Hamiltonian formulation of the action in eq.~\eqref{SEE_TheAction} in unimodular-conformal variables introduced in section~\ref{sec_31conf}.
This is necessary for the canonical quantization.
But since canonical quantization of an action implies its quantum interpretation instead of the classical one and thus assumes its validity at short length scales/high energy scales, this contradicts the initial assumption that the action is \textit{perturbative} and semiclassical.
This motivates us to make a clear distinction between two approaches to the Hamiltonian formulation of theories with higher derivatives. 
The perturbative nature of the action must be taken into account if the action is treated as a classical one.
On the other hand, if the action is treated as a quantum action, such perturbative nature cannot be assumed.
In these two cases one ends up with two qualitatively different Hamiltonian formulations.
In the present section we clarify this difference on the example of our model of a simple harmonic oscillator with higher derivatives from section \ref{subs_HDLHO}.
We explain the Hamiltonian formulation in two ways. First we assume that Emmy treats action in eq.~\eqref{ALHOaction} as an exact action. Then we show how Richard must treat the same action if he takes the perturbative approach.
The comparison of the two cases will give us guidelines to choose the approach which is the appropriate one for quantization of theories of gravity. 

\subsection{Exact interpretation of the higher derivative theory} 
\label{subs_HD_exact}

Let us find Emmy's momentum. Since eq.~\eqref{ALHOaction} is currently treated as an exact action, i.e. recall that in this case $g/m > 1$ makes sense, the system has four genuine degrees of freedom.
In the Hamiltonian approach this means that one needs two have \textit{two} pairs of canonical variables.
Furthermore, Hamiltonian formalism is a first order formalism, which means that the extra variables must be utilized in a way to reduce the higher order nature of the theory to the second order (but not by means of perturbative approach).
This is usually referred to as the method of \textit{Ostrogradski order reduction} and is well explained in \cite{Wood}, whose line of thought we incorporate in our toy model.
The method consists of defining a new set of variables
    \begin{align}
    \label{H_alho_q1}
        q_{1} & = x\ ,\qquad& p_{1} & = \dd{\mathcal{L}}{\dot{x}} - \frac{\d }{\d t}\dd{\mathcal{L}}{\ddot{x}} \ ,\\[12pt]
    \label{H_alho_q2}
        q_{2} & = \dot{x}\ ,\qquad& p_{2} & = \dd{\mathcal{L}}{\ddot{x}}
    \end{align}
and performing a Legendre transform to find the Hamiltonian.
Equation \eqref{H_alho_q1} for the momentum is simply the variation of the Lagrangian with respect to velocity $\dot{x}$ --- another term appears due to the higher-derivative nature of the theory.
The momenta are given as follows,
    \begin{align}
    \label{H_alho_q1a}
        p_{1} & = m \dot{x} + g\left(\dot{\ddot{x}} - f\dot{x}\right)= m q_{2} + g\left(\ddot{q}_{2} - f q_{2}\right)\ ,\\[12pt]
    \label{H_alho_q2a}
        p_{2} & = - g\left(\ddot{x} - f x\right) = - g\left(\dot{q}_{2} - f q_{1}\right)\ .
    \end{align}
Now, the important thing is to be able to invert for the velocities, i.e. the highest appearing derivatives. This can be done easily form the above equations,
    \begin{align}
    \label{H_alho_q1b}
        \ddot{q}_{1} & = \frac{1}{g}\left( p_{1} - m q_{2} \right) + f q_{2}\ ,\\[12pt]
    \label{H_alho_q2b}
        \dot{q}_{2} & = - \frac{1}{g}p_{1} + f q_{1}\ .
    \end{align}
The condition for inverting the velocities is that at least one of the highest order variables satisfy
    \begin{equation}
        \dd{^2\mathcal{L}}{(q^{\ssst A}_{N})^2}\neq 0
    \end{equation}
where $N$ is the highest order derivative (non-linear) term appearing in the Lagrangian and $A$ runs from $1...M$ where $M$ is the dimension of the configuration space.
In our case this reduces to
    \begin{equation}
    \label{H_cond}
        \dd{p_{2}}{\dot{q}_{2}} = - g \neq 0\ ,  
    \end{equation}
which indeed is the case. 
Let us at this point pause for a moment and reflect on what happens to the condition given by eq.~\eqref{H_cond} if one takes the limit $g\rightarrow 0$. 
We immediately see that this is inconsistent with the assumption that the phase space has four degrees of freedom, because $p_{2}$ then vanishes and becomes a \textit{constraint} and equation of motion for $q_{2}$ does not exist, thus reducing the number of degrees of freedom to two.
This observation is crucial to remember if one would like to suppose that this higher-derivative theory has a small $g$ limit equivalent to the simple harmonic oscillator,
because from eq.~\eqref{H_alho_q1b} and eq.~\eqref{H_alho_q2b} it can be seen that such a limit, taken \textit{after} the velocities are inverted, renders these equations ill-defined.
This means that current treatment of the theory is incompatible with the perturbative approach, as expected based on the discussion so far in this chapter.

One now proceeds to define the total Hamiltonian via the Legendre transform, substituting all velocities for the momenta,
    \begin{align}
    \label{H_alho_ham}
        \mathcal{H} = p_{1}q_{2} + p_{2}\dot{q}_{2} - \mathcal{L}(q_{1},q_{2},p_{1},p_{2}) = -\frac{p_{2}^2}{2g} + f q_{1}p_{2} + q_{2} p_{1} - \frac{m}{2}q_{2}^2 + \frac{k}{2} q_{1}^2\ ,
    \end{align}
where $\dot{q}_{1}=q_{2}$ was used in the first term in the first equality. 
It is clear that $g\rightarrow 0$ limit is meaningless, i.e. the theory and its solutions are not perturbatively expandable in $g$. But that is alright since we have assumed from the beginning that $g/m > 1$.
If one would like to use this theory to describe a classical system, then one derives Hamilton's equations of motion from the Poisson brackets as usual,
    \begin{align}
    \label{H_alho_eom1}
        \dot{q}_{1} & = \dd{\mathcal{H}}{p_{1}} = q_{2}\ ,\quad & \dot{p}_{1} & = - \dd{\mathcal{H}}{q_{1}} = - k q_{1}\ ,\\[12pt]
    \label{H_alho_eom2}
        \dot{q}_{2} & = \dd{\mathcal{H}}{p_{2}} = - \frac{p_{2}}{g} + f q_{1}\ ,\quad & \dot{p}_{2} & = - \dd{\mathcal{H}}{q_{2}} = m q_{2} - p_{1}\ .
    \end{align}
The first equation in \eqref{H_alho_eom1} is just the definition of the new variable $q_{2}$.
The second equation in eq.~\eqref{H_alho_eom1} is the Euler-Lagrange equation of motion, cf. eqs.~\eqref{H_alho_q1} and \eqref{ALHOEOMEL}.
The first equation in \eqref{H_alho_eom2} is just the inverted velocity, eq.~\eqref{H_alho_q2b}.
The second equation in \eqref{H_alho_eom2} is equivalent to the second equation in \eqref{H_alho_q1}.

There is another approach to the Hamiltonian formulation, which gives the same results and this is the approach we are going to use in the subsequent sections for the Hamiltonian formulation of gravity.
Namely, instead of introducing the new variables after the variation to find the momenta, one could introduce the new variables already at th elevel of the Lagrangian, i.e. before the variation, in order to find the momenta.
This is done by introducing a constraint $\lambda\left(Y - \dot{x}\right)$ into the Lagrangian, where $\lambda$ is a Lagrange multiplier, and substituting all $\dot{x}$ for $Y$ in it, thus obtaining the \textit{constrained Lagrangian},
    \begin{equation}
    \label{H_alho_Lagcost}
        \mathcal{L}_{c}=\frac{1}{2}\left( 
        m Y^{2} - k x^2 - g \left(\dot{Y} - f x\right)^2
        \right) - \lambda\left(Y - \dot{x}\right)\ .
    \end{equation}
There are now actually three variables in the system, but $\lambda$ does not have any kinetic term, nor velocities, so its momentum vanishes (it will turn out that this is just the statement of $Y = \dot{x}$).
The conjugate momenta are
    \begin{align}
    \label{H_alho_px}
        p_{x} & = \dd{\mathcal{L}}{\dot{x}} \deq \lambda\quad\Rightarrow\quad p_{x} - \lambda \deq 0\ ,\\[6pt]
    \label{H_alho_pY}
        P_{Y} & = \dd{\mathcal{L}}{\dot{Y}} = - g \left(\dot{Y} - f x\right)\ ,\\[6pt]
    \label{H_alho_pl}
        p_{\lambda} & = \dd{\mathcal{L}}{\dot{\lambda}} \deq 0\ .
    \end{align}
Equation \eqref{H_alho_pl} is a \textit{primary constraint} and ``$\deq$'' is Dirac's ``weak equality'', which we rename here as the ``delayed equality'', that delays setting $p_{\lambda}$ to zero until all Poisson brackets have been calculated; Appendix \ref{app_const} should be consulted for the details on the constraint analysis and the used definitions.
It just says that $\lambda$ is an arbitrary variable.
Equation \eqref{H_alho_px} is also a primary constraint. It is an interesting contrast to the momentum of the first-order theory, $p = m\dot{x}$, since it says that $\dot{x}$ cannot be inverted from it.
This should not be alarming, since $\dot{x}$ has been moved into the new variable $Y$, so $p_{x}$ has only an auxiliary meaning, until one decides to restore to the original variables when the Hamiltonian formulation is complete.
The total Hamiltonian is formed by the Legendre transform
    \begin{align}
    \label{Ham_osc_Ham}
        \mathcal{H} & = p_{x}\dot{x} + P_{Y}\dot{Y} + p_{\lambda}\dot{\lambda} - \mathcal{L}_{c}\nl
        & = -\frac{P_{Y}^2}{2g} + f x P_{Y} + Y p_{x} - \frac{m}{2}Y^2 + \frac{k}{2}x^2 + p_{\lambda}\dot{\lambda}\ ,
    \end{align}
where $p_{\lambda}$ is not yet set to zero, as Dirac's ``delayed equality'' $\deq$ implies.
This is the only difference compared to the Hamiltonian in eq.~\eqref{H_alho_ham} (apart from a trivial relabeling of variables), so there has to be a way to safely set $p_{\lambda}=0$.
The time preservation of primary constraint in eq.~\eqref{H_alho_pl} must be required and leads to
    \begin{equation}
        \dot{p}_{\lambda} = \PB{p_{\lambda}}{\mathcal{H}} = Y - \dot{x} \deq 0\ ,
    \end{equation}
which is just the constraint we added to the Lagrangian. 
The important thing here is that $p_{x}-\lambda\deq 0$ and $p_{\lambda}\deq 0$ are \textit{second-class} constraints (cf. eq.~\eqref{EM_1st2nd}) and the discussion around eq.~\eqref{app_DB_osc} in appendix~\ref{app_dirb} shows that these constraints can be set strongly to zero, thus eliminating $\lambda$ explicitly from the theory.
This makes the last term in eq.~\eqref{Ham_osc_Ham} vanish, leaving us with
    \begin{align}
    \label{Ham_osc_Ham1}
        \mathcal{H} & = -\frac{P_{Y}^2}{2g} + f x P_{Y} + Y p_{x} - \frac{m}{2}Y^2 + \frac{k}{2}x^2\ ,
    \end{align}
which coincides with eq.~\eqref{H_alho_ham}.

Let us now take a closer look at the Hamiltonian in eq.~\eqref{Ham_osc_Ham1} to consider its features.
First of all, we see that there is no kinetic term of $q_{1}$; the only kinetic term is the one of $q_{2}$. 
The negative sign in front of it does not matter, because one could simply change the sign of $g$, if the theory allows it.
If one imagines a theory with even higher order derivative terms, the corresponding Hamiltonian would always contain only one kinetic term, corresponding to the highest order variable, unless one mixes the momenta and coordinates via special canonical transformation \cite{Wood}.
But the bigger problem is the term linear in $p_{1}$.
Namely, as explained in \cite{Wood}, according to the theorem of Ostrogradski any higher-derivative theory exhibits runaway solutions because of this type of terms. (The other term $f q_{1} p_{2}$ is just the artefact of the particular model we are considering; this term could be eliminated by setting $f=0$.) 
Tracing the steps backwards, it can be concluded that this term arises \textit{always} if one wishes to do the Hamiltonian formulation of a higher-derivative theory.
It allows to counter the kinetic term and drive the energy of the system (which is essentially the value of $\mathcal{H}$) arbitrarily high positive or low negative values.
This property is imprinted on the solutions as well, in accordance with our discussion in section \ref{subs_HDLHO}.
It can be shown that a specific canonical transformation and quantization of the system implies that the system is equivalent to two coupled harmonic oscillators whose total energy is indefinite \cite{Wood}, because there appears another pair of creation and annihilation operators which turn out to act on a state in such a way to give a negative energy spectrum.
The associated particle excitations are named ``ghosts'' (not to be confused with Faddeev-Popov ghosts) and they can be shown to break unitarity \cite{HH}, if one opts for keeping the positive energy interpretation.
There are ways to tackle this problem by a variety of alternative quantization procedures \cite{BendMan, HH, Salv} but they are aimed at systems whose Hamiltonian does not vanish, such as the one discussed in our toy model.
However, we shall see in the next chapter that quantization procedure of theories with vanishing Hamiltonian, followed by a carefully tailored semiclassical approximation could suggest that ghosts remain in the realm of quantum gravity and beyond Planck energies, which are practically unobservable.
As a hint of how does this happen, notice that since the only kinetic term is the one of the highest order variable, the dynamics of the wave function is established by the evolution of that variable.
Moreover, note that if one would set $m=0$, no issues would arise and one would still have the mentioned kinetic term in the Hamiltonian, thus ensuring that the evolution with respect to the higher-order variable continues.
In short, Emmy's theory cannot be considered as an exact classical theory whose limit $g\rightarrow 0$ is a well defined low-energy limit and thus we can expect that Emmy's theory makes sense only as a full quantum theory. Thus we will not be interested in the corresponding equations of motion and its classical interpretation.

\subsection{Perturbative interpretation of the higher derivative theory}
\label{subs_HD_pert}

For Richard's theory, on the other hand, we claim that it makes no sense to be interpreted as a quantum theory while $g/m \ll 1$ if the higher-derivative term is expected to become important at high energies (which we have shown in section \ref{subs_HDLHO} that it does).
Rather, his theory is \textit{semiclassical}, in analogy to the SEE.
The only way Richard could make the Hamiltonian formulation of his theory legal is to implement the perturbative nature in eqs.~\eqref{H_alho_q1a} and \eqref{H_alho_q2a} \textit{before} inverting these equations for the velocities.
This is done in the similar way as imposing the Einstein equations at the first order of perturbation in eqs.~\eqref{SEE_full_TlessB1} and \eqref{SEE_full_TrB1}: multiply the first equalities in eqs.~\eqref{H_alho_q1a} and \eqref{H_alho_q2a} by $g$ and neglect all terms of $\mathcal{O}(g^2)$ order.
The result is:
    \begin{align}
    \label{H_alhoR_q1a}
        g p_{1} & = g m \dot{x} + \mathcal{O}(g^2)\ ,\\[12pt]
    \label{H_alhoR_q2a}
        g p_{2} & = 0 + \mathcal{O}(g^2)\ .
    \end{align}
As one can see, $p_{2}=0$ has to be imposed at order $g$ at each step of the derivation, thus ensuring that the extra degree of freedom (i.e. Emmy's $q_{2}$) is excluded from the theory.
Equation \eqref{H_alhoR_q1a} is just another way of saying the same: it means ``impose the definition of the  'classical' momentum $p_{0} = m \dot{x}$ at order $\mathcal{O}(g)$''. 
The already known Hamilton equation of motion for the `classical' momentum is also imposed at this order,
    \begin{equation}
    \label{zerothmom}
        g \dot{p}_{1} = g m\ddot{x} = - g m\omega_{m}^2 x\ ,
    \end{equation}
and one can take as many derivatives of this equation as necessary to eliminate the higher derivatives from the Lagrangian, which then takes the form we met before in eq.~\eqref{ALHOactionSubs}.
Only from that ``perturbatively reduced'' Lagrangian can one derive the correct momentum, as pointed out by Mazzitelli \cite{Mazzit}, and this momentum coincides with the unperturbed case in eq.~\eqref{zerothmom} (in a more general case of a higher-derivative theory it is possible the momentum differs from the zeroth order form).
This is the meaning of \textit{the method of perturbative constraints}.
The Hamiltonian is found by a Legendre transform,
    \begin{align}
    \label{Ham_osc_HamRic}
        \mathcal{H} = p_{1}\dot{x} - \mathcal{L}(x,p) = \frac{p_{1}^2}{2m} + \frac{m\omega^2_{m}}{2}\left(1 + \frac{g}{m}\frac{\left(f + \omega^2_{m}\right)^2}{\omega^2_{m}}\right)x^2\ .
    \end{align}
The most important point is that it can be written as a sum of the ``classical'', zeroth order Hamiltonian (evaluated for $g=0$ in the above equation) and the perturbation term
    \begin{equation}
        \mathcal{H}_{\ssst pert} = \frac{g}{2}\left(f + \omega^2_{m}\right)^2x^2\ ,
    \end{equation}
which vanishes in $g\rightarrow 0$ limit, validating the method of perturbative constraints.
The Hamilton equations of motion derived from this perturbative approach agree with the Euler-Lagrange equation of motion we derived in eq.~\eqref{ALHOpertsolfin}. 

Note the drastic difference between Richard's (eq.~\eqref{Ham_osc_HamRic}) and Emmy's (eq.~\eqref{Ham_osc_Ham1}) Hamiltonians. These are essentially two \textit{distinct} theories.
If Richard would like to quantize this theory, there has to be a good reason to do so, which in this case would lie in the assumption that $g$ is unrelated to a \textit{quantum} correction.
There could be systems in which that might indeed be the case, but if we suppose that $g\sim\hbar$ then the higher derivative term is interpreted as a quantum correction and mimics the role of the $H_{\mu\nu}$ and $B_{\mu\nu}$ tensors in the SEE (valid under the assumption of $l^2\gg 1$),
implying that there is no much sense in assuming that the \textit{full} quantum theory would be obtained by quantizing what is supposed to be its semiclassical limit.
In other words, ``perturbation before quantization'' (PbQ) is not a meaningful way to proceed in constructing the full quantum theory if the higher-derivative terms are the large length scale \textit{perturbative corrections} of the small-scale \textit{quantum} effects.
Hence, we conclude that it makes sense to quantize only Emmy's version of the theory, i.e. to approach with ``quantization before perturbation'' (QbP).
The classical treatment of the Lagrangian valid at Emmy's energies is disregarded and is reserved for the classical perturbative treatment of the Lagrangian at Richard's energies.

\section{Hamiltonian formulation of General Relativity}
\label{sec_HEH}

Before we go into Hamiltonian formulation of theories with higher derivatives we make a short detour in order to present the Hamiltonian formulation of GR \cite[chapter 20]{Blau},\cite[chapter 12]{Pad}.
In its Hamiltonian formulation, GR is a theory that describes how three-dimensional spatial hypersurface, described by the three-metric metric, evolves in time.
Even though things are a bit more subtle than, say, a particle in spacetime with a potential, one could think of the dynamics of the three-metric in an analogous way.
Instead of just restating the Hamiltonian formulation of GR here, we shall employ our unimodular-conformal variables defined in section \ref{sec_31conf} and present thus resulting Hamiltonian formulation.
In that way, the evolution of the three-metric is split into the evolution of its conformally invariant part $\hb_{ij}$ and its scale-full part $a$.
We shall also add a non-minimally coupled scalar field (cf. section \ref{sec_31chi}) to the EH action.

\subsection{Unimodular-conformal variables} 

The EH action without the cosmological constant and with a scalar (density) field reads
    \begin{align}
    \label{HEHchi_action}
        S^{\srm E\chi} = \inttx \left(\mathcal{L}^{\srm E}_{\srm{ADM}} + \mathcal{L}^{\srm \chi}\right) - l^2\inttx a^2\Nb\left(\nabla_{\mu}\left(\bar{n}^{\mu}\kb\right)
	    -\frac{2}{\Nb}\mathbf{D}\cdot\mathbf{D}\Nb\right)
    \end{align}
where
    \begin{align}
    \label{HEH_Lag}
        \mathcal{L}^{\srm E}_{\srm{ADM}} = \frac{l^2 \hbar\Nb a^2}{2}\left(a^2\,^{\ssst (3)}\! R + \bktb\cdot\bktb-6\kb^2\right)
    \end{align}
is called the ADM Lagrangian \cite{ADM} and is based on eq.~\eqref{Rdec31umod2}, while $\mathcal{L}^{\srm \chi}$ is given by eq.~\eqref{chi31Lagdec}. 
The last term in eq.~\eqref{HEHchi_action} is the boundary term \cite[section 12.4]{Pad} and will be disregarded (i.e. we assume no issue with boundaries of spacetime and space), along with all divergences in the matter field Lagrangian in eq.~\eqref{chi31Lagdec}.
This boundary term effectively eliminates the second derivatives of the metric from the Lagrangian and shows that the Lagrangian of GR is not a genuine higher derivative theory.
Note that since GR is a first order theory there is no need to consider the extrinsic curvature components as independent variables, so in this section they are treated merely as labels in order to simplify notation.

At first glance, one notices that there are no velocities $\dot{\Nb}$ and $\dot{N}^{i}$ in the Lagrangian. 
According to the constraint analysis (cf. appendix \ref{app_const}), one should expect constraints, because this implies that the lapse density $\Nb$ and shift vector $N^{i}$ are arbitrary.
The momenta are defined as\footnote{Note that the all momenta have dimensions of $[\hbar]$, except $p_{\chi}$, whose dimension is $[\hbar]^{1/2}$.}
    \begin{subequations}
    \begin{align}
	\label{GRchi_pchi}
    	p_{\chi}=\frac{\del\mathcal{L}^{{\srm E}\chi}}{\del \dot{\chi}}&=
	    \bar{n}^{\mu}\del_{\mu}\chi + 6\xi_{c}\kb\chi-\frac{\del_{i}N^{i}}{3\Nb}\chi\nonumber\\[12pt]
	    &\Rightarrow \quad
	    \dot{\chi}=\Nb\left(p_{\chi} - 6\xi_{c}\kb\chi\right)+\frac{\del_{i}N^{i}}{3}\chi+N^{i}\del_{i}\chi\ ,\\[18pt]
	\label{GRchi_pam}
	    p_{a}=\frac{\del\mathcal{L}^{{\srm E}\chi}}{\del\dot{a}}
	    &=-\frac{6l^2 \hbar a}{\tilde{l}^2}\kb + 6\xi_{c}\frac{\chi}{a}p_{\chi}\\[12pt]
	\label{GRchi_pa}
	    &\Rightarrow \quad
	    \kb=-\frac{\tilde{l}^2}{6l^2\hbar a}\left(p_{a} - \xi_{c}\frac{\chi}{a}\, p_{\chi}\right)\\[18pt]
	\label{GRchi_padot}
	    &\Rightarrow \quad
	    \dot{a} = -\frac{\Nb\tilde{l}^2}{6l^2\hbar }\left(p_a - 6\xi_{c}\frac{\chi}{a} p_{\chi}\right)+\frac{1}{3}a D_{i}N^{i}\ ,\\[18pt]
	\label{GRchi_phdot}
	    \bpb=\frac{\del\mathcal{L}^{{\srm E}\chi}}{\del \dot{\bhb}}
	    & = \frac{l^2\hbar a^2}{2}\left(1 - \xi \frac{\chi^2}{l^2\hbar a^2}\right)\bktb_{\sharp}\nonumber\\[12pt]
	    &\Rightarrow \quad
	    \bktb=\frac{2}{l^2\hbar a^2}\left(1 - \xi \frac{\chi^2}{l^2\hbar a^2}\right)^{-1}\bpb_{\flat}\\[12pt]
	    &\Rightarrow \quad
	\label{GRchi_hdot}
	    \dot{\bhb}=\frac{4}{l^2\hbar a^2}\Nb\left(1-\xi \frac{\chi^2}{l^2\hbar a^2}\right)^{-1}\bpb_{\flat}+2\left[\bar{D}_{(i}\Nb_{j)}\right]^{\srm T}\ ,
	\end{align}
	\end{subequations}
where we have introduced a $\chi$- and $a$-dependent dimensionless coupling	
	\begin{equation}
    \label{GR_kt}
	    \tilde{l}^2 := \frac{1}{1 + 6\xi \xi_{c}\dfrac{\chi^2}{l^2\hbar a^2}}\ .
	\end{equation}
The bold notation is defined as follows.
Objects such as $\bpb$ and $\bktb$ stand for $\bpb := \pb^{ij}$ and $\bktb:=\ktb_{ij}$; then $\bpb\cdot\bpb\equiv \hb_{ik}\hb_{jl}\pb^{ij}\pb^{kl}$ and $\bktb\cdot\bktb\equiv \hb_{ik}\hb_{jl}\kb^{ij}\kb^{kl}$.
The musical notation designates that object's indices are lowered (${\flat}$) or raised (${\sharp}$) by $\hb_{ij}$'s: $\bpb_{\flat}:=\pb_{kl}=\hb_{ik}\hb_{jl}\pb^{ij}$ and $\bktb_{\sharp}:=\kb^{{\srm T}kl}=\hb^{ik}\hb^{jl}\ktb_{ij}$.
We allow to mix the bold notation with the index notation, since the use of the bold notation is just a matter of convenience.
Note that $p_{\chi}$, $p_{a}$ and $\pb^{ij}$ are tensor densities of scale weight two, two and five, respectively.

We also have two vanishing momenta, i.e. two primary constraints (see appendix~\ref{app_const}) due to Lagrangian's independence of $\dot{\Nb}$ and $\dot{N}^{i}$,
    \begin{align}
    \label{HEH_pN}
        p_{\ssst \Nb} = \frac{\del\mathcal{L}^{{\srm E}\chi}}{\del\dot{\Nb}} \deq 0\ ,\qquad
        p_{i} = \frac{\del\mathcal{L}^{{\srm E}\chi}}{\del\dot{N}^{i}} \deq 0\ .
    \end{align}
The Poisson bracket defined in eq.~\eqref{PB_definition} in unimodular-conformal variables reads
    \begin{align}
    \label{eqn:PB}
        &\left\lbrace A(\bx), B({\mathbf y})\right\rbrace \nld
        &=\!\int{\rm d}^3z\Biggl(\frac{\delta A(\bx)}{\delta
          \bar{h}_{ij}({\mathbf z})}\frac{\delta B({\mathbf y})}{\delta
          \bar{p}^{ij}({\mathbf z})}
          +\frac{\delta A}{\delta
          a}\frac{\delta B}{\delta p_{a}}
        +\frac{\delta A}{\delta
          \chi}\frac{\delta B}{\delta p_{\chi}}
          %
        +\frac{\delta A}{\delta \Nb}\frac{\delta B}{\delta
          p_{\ssst\Nb}}
          +\frac{\delta A}{\delta N^{i}}\frac{\delta B}{\delta
          p_i} - A(\bx)\leftrightarrow B(\by)\Biggr) \ .
    \end{align}
Then the canonical pairs obey the following equal-time Poisson brackets,
    \begin{align}
    \label{HEH_PBvars1}
        \PB{\hb_{ij}(\bx)}{\pb^{ab}(\by)}
        & =\mathbb{1}^{{\srm{T}}ab}_{(ij)}\delta
        (\bx,\by)\ , \\[6pt]
    \label{HEH_PBvars2}
        \PB{q^{A}(\bx)}{\Pi_{B}(\by)}
        & =\delta_{B}^{A}\delta (\bx,\by)\ ,
    \end{align}
where $q^{A}=(a,\chi)$ and $\Pi_{B}=(p_{a},p_{\chi})$. Note that in eq.~\eqref{HEH_PBvars1} the result is the \textit{traceless} identity $\mathbb{1}^{ab\srm T}_{ij}$ because the variation of the shape parts of the metric and its momentum are traceless, cf. eq.~\eqref{tracelessvar}.
Lapse density and shift vector obey analogous Poisson brackets to eq.~\eqref{HEH_PBvars2}, while all other Poisson brackets vanish.

The preservation of primary constraints in eq.~\eqref{HEH_pN} in time will give two more constraints as we shall soon see.
First one needs to find the total Hamiltonian via the Legendre transform by expressing the velocities in terms of the momenta with eqs.~\eqref{GRchi_pchi}-\eqref{GRchi_hdot}. 
To that purpose we need to apply the product rule for derivatives in three terms: the next-to-last term in eq.~\eqref{GRchi_pchi} multiplied by $p_{\chi}$, the last term in eq.~\eqref{GRchi_padot} multiplied by $p_{a}$ and the last term in eq.~\eqref{GRchi_hdot} contracted with $\pb^{ij}$.
The first two cases are trivially treated and they produce the following surface terms,
    \begin{equation}
    \label{GRchi_Hsurf1}
        \frac{1}{3}\intx\,\del_{i}\left(N^{i}\chi p_{\chi} + N^{i} a p_{a}\right)\ .
    \end{equation}
We focus on the third case now.
Recall that symmetrization on $ij$ indices and subtraction of trace ensures that $\left[\bar{D}_{(i}\Nb_{j)}\right]^{\srm T}$ is scale-less (cf. eq.~\eqref{delVtless}).
For this reason we can drop the symmetrization and traceless notation if this term is contracted with $\pb^{ij}$, which is symmetric and traceless, i.e. we have the following,
    \begin{align}
    \label{GRchi_pDN}
        \intx\,\pb^{ij}\left[\bar{D}_{(i}\Nb_{j)}\right]^{\srm T} & = \intx\,\pb^{ij}\bar{D}_{i}\Nb_{j} = \intx\,\bar{D}_{i}\left(N^{k}\hb_{kj}\pb^{ij}\right) - \intx\,N^{k}\bar{D}_{i}\left(\hb_{kj}\pb^{ij}\right) \nld
        & = \intx\,\del_{i}\left(N^{k}\hb_{kj}\pb^{ij}\right) - \intx\,N^{k}\bar{D}_{i}\left(\hb_{kj}\pb^{ij}\right)\ .
    \end{align}
Substituting velocities in eqs.~\eqref{GRchi_pchi}-\eqref{GRchi_hdot} into the following Legendre transform and using eq.~\eqref{GRchi_pDN} and other product rules, the resulting total Hamiltonian is given by
    \begin{align}
    \label{GR_Htot}
        H^{\srm{E\chi}} & = \intx\left( \dot{a}p_{a} + \dot{\hb}_{ij}\pb^{ij} + \dot{\chi}p_{\chi} + \lambda_{\ssst \Nb}p_{\ssst \Nb} + \lambda_{i}p^{i} - \mathcal{L}_{\srm{ADM}}\right)\nld
        & = \intx\left\lbrace \Nb\Ho{E\chi} + N^{i}\Hi{E\chi} + \lambda_{\ssst\Nb}p_{\ssst\Nb} + \lambda^{i}p_{i}\right\rbrace + H^{\srm{ E\chi}}_{\ssst surf}\ ,
    \end{align}
where the surface terms 
    \begin{equation}
    \label{GRchi_HsurfEchi}
        H^{\srm{E\chi}}_{\ssst surf} = \intx\,\del_{i}\left(2 N^{k}\hb_{kj}\pb^{ik} + \frac{1}{3}N^{i}\chi p_{\chi} + \frac{1}{3}N^{i} a p_{a}\right)\ ,
    \end{equation}
which arise from product rules will be left out. Note that we have already named $\lambda_{\ssst \Nb}\equiv\dot{\Nb}$ and $\lambda^{i}\equiv\dot{N}^{i}$ as Lagrange multipliers.
Expressions $\Ho{E\chi}$ and $\Hi{E\chi}$ are independent of $\Nb$ and $N^{i}$ which follow from the demand that primary constraints in eq.~\eqref{HEH_pN} are to be preserved in time,
    \begin{subequations}
    \begin{align}
    \label{GRxi_H0}
    & \dot{p}_{\ssst \Nb} = \PB{p_{\ssst \Nb}}{H} \deq 0\nld
       &\Rightarrow\quad\Ho{E\chi}  =
	    -\frac{\tilde{l}^2}{12 l^2\hbar}\left(p_{a}-6\xi_{c}\frac{\chi}{a}p_{\chi}\right)^2
	    +\frac{1}{2}p_{\chi}^2
        +\frac{2}{l^2\hbar\, a^2}\frac{1}{\left(1-\xi \dfrac{\chi^2}{l^2\hbar a^2}\right)}\bpb\cdot\bpb\nld
	    &\qquad\qquad\quad 
	    - \frac{l^2\hbar a^4}{2}\,^{\ssst (3)}\! R
        + \frac{1}{2}U^{\ssst\chi}
        \deq 0\ ,\\[18pt]
    \label{GRxi_Hi}
       &  \dot{p}_{i} = \PB{p^{i}}{H} \deq 0 \nld
       &\Rightarrow\quad\Hi{E\chi}  =
        -2\bar{D}_{j}\left(\hb_{ik}\pb^{kj}\right)
        -\frac{1}{3}D_{i}\left(a\,p_{a}\right) - \frac{1}{3}\left(\chi\del_{i}p_{\chi} - 2\del_{i}\chi \, p_{\chi}\right)
        \deq 0\ .
    \end{align}
    \end{subequations}
where $U^{\ssst\chi}$ was defined in eq.~\eqref{Uchi_def}.
Constraint in eq.~\eqref{GRxi_H0} is called \textit{the Hamiltonian constraint} and \eqref{GRxi_Hi} is called \textit{the momentum constraint}\footnote{Technically speaking, there are three constraints in what is called the ``momentum constraint'', i.e. one for each value of the index.}.
The constraints represent relations among the phase space variables which are to hold at every moment in time.

\subsection{Original, ADM variables}
\label{subs_HEADM}

Before we say more on these constraints, let us consider the constraints in the original variables in vacuum GR \cite[chapter 12]{Pad}, \cite[chapter 1]{Thiemann}, \cite[chapter 4]{OUP}, \cite[chapter 20]{Blau},
    \begin{subequations}
    \begin{align}
	\label{GR_Hoold}
	    \mathcal{H}^{\srm E}_{\bot} & = 2\kappa \mirl{G}_{ikjl}p_{\srm{ADM}}^{ij}p_{\srm{ADM}}^{kl}-\frac{\h}{2\kappa}\,^{\srm (3)}\! R\deq 0\ ,\\[6pt]
	\label{GR_Hiold}
	    \mathcal{H}_{i}^{\srm E} & = -2D_{j}\left(h_{ik}p_{\srm{ADM}}^{kj}\right)\deq 0 \ ,
	\end{align}
	\end{subequations}
where
    \begin{equation}
    \label{GR_pADM}
        p_{\srm{ADM}}^{ij} =\frac{1}{2\kappa}G^{ikjl}K_{kl} = \frac{\h}{2\kappa}\left(K^{ij}-h^{ij}K\right)\ ,
    \end{equation}
is called ``the ADM momentum'' and $G^{ikjl}$ is called \textit{DeWitt supermetric} and is defined by
    \begin{equation}
    \label{HEH_DWmet}
        G^{ikjl} := \frac{\h}{2}\left(h^{ik}h^{jl} + h^{il}h^{jk} - 2 h^{ij}h^{kl}\right)\ .
    \end{equation}
It is the metric on the space of symmetric rank-2 tensors with an inverse
    \begin{equation}
    \label{HEH_DWmetin}
        \mirl{G}_{ikjl} := \frac{1}{2\h}\left(h_{ik}h_{jl} + h_{il}h_{jk} - h_{ij}h_{kl}\right)\ ,\qquad G^{ikjl}\mirl{G}_{kmln} = \mathbb{1}_{(mn)}^{ij}\ .
    \end{equation}
It is important to note that simple lowering of indices with $h_{ij}$ does not define the inverse of the DeWitt metric from eq.~\eqref{HEH_DWmet},
    \begin{equation}
    \label{HEH_DeWittNotInv}
        h \mirl{G}_{ikjl} \neq h_{ia}h_{kc}h_{jb}h_{ld}G^{acbd} = \frac{\h}{2}\left(h_{ac}h_{bd} + h_{ad}h_{bc} - 2 h_{ab}h_{dc}\right)\ .
    \end{equation}
In terms of the DeWitt supermetric the kinetic term of the ADM Lagrangian can be written as
    \begin{equation}
        K_{ij}K^{ij} - K^2 = \frac{1}{\h}G^{ikjl}K_{ik}K_{jl}\ .
    \end{equation}
    
Now, the constraints obey what is called ``the hypersurface foliation (or deformation) algebra''.
Namely, if one defines a smeared version of a constraint $\mathcal{C}^{A}(\bx)$ as a functional of a smearing function $\eta(\bx)$,
    \begin{equation}
    \label{HEH_smearC}
        \mathcal{C}^{A}[\eta]=\intx \,\eta (\bx)\cdot\mathcal{C}^{A}(\bx)\ ,
    \end{equation}
the Hamiltonian and the momentum constraints close the following algebra\
    \begin{align}
    \label{HEH_AlgHH}
        \left\lbrace \mathcal{H}_{\bot}^{\srm E}[\varepsilon_1],\mathcal{H}_{\bot}^{\srm E}[\varepsilon_2]\right\rbrace &=\mathcal{H}_{||}^{\srm E}[\varepsilon_1 \del^{i}\varepsilon_2-\varepsilon_2 \del^{i}\varepsilon_1]\,,\\[6pt]
    \label{HEH_AlgHm}
        \left\lbrace \mathcal{H}_{||}^{\srm E}[\vec{\eta}],\mathcal{H}_{\bot}^{\srm E}[\varepsilon]\right\rbrace &=\mathcal{H}_{\bot}^{\srm E}[\mathcal{L}_{\vec{\eta}}\varepsilon]\,,\\[6pt]
    \label{HEH_Algmm}
        \left\lbrace \mathcal{H}_{||}^{\srm E}[\vec{\eta}_1],\mathcal{H}_{||}^{\srm E}[\vec{\eta}_2]\right\rbrace & = \mathcal{H}_{||}^{\srm E}[\mathcal{L}_{\vec{\eta}_{1}}\vec{\eta}_2]\ ,
    \end{align}
where $\mathcal{H}_{\bot}^{\srm E}[\varepsilon_1]$ is the smeared version of the Hamiltonian constraint and $\mathcal{H}_{||}^{\srm E}[\vec{\eta}]$ is the smeared version of the momentum constraint.
Addition of matter contribution to $\mathcal{H}_{\bot}^{\srm E}$ and $\mathcal{H}_{||}^{\srm E}$ does not spoil the algebra.
The meaning of this algebra is that GR is a reparametrization-invariant theory.
In particular, eq.~\eqref{HEH_Algmm} says that the theory is invariant under three-dimensional diffeomorphisms and it is usually said that the momentum constraint is the generator of spatial coordinate transformations.
As for the Hamiltonian constraint, it is related to the reparametrizations of time coordinate and is usually referred to as the generator of time translations.
However, eq.~\eqref{HEH_AlgHH} and eq.~\eqref{HEH_AlgHm} show that one cannot simply transform the time coordinate without affecting the way the spatial hypersurfaces have been chosen --- which is expected because the hypersurfaces are defined \textit{in terms of} the time function $t$. 
For this reason one must think of these four constraints not as separate generators but as $3+1$ decomposition of some generator of four-dimensional coordinate transformations related to $GL(4,\mathbb{R})$ group.
This was clarified by Castellani \cite{C}, Pons et al. \cite{PSS}, Pitts \cite{PittsEM} and others; we come back to this in section~\ref{sec_Gencf31}.
On the other hand, in relation to the reparametrization invariance, it was shown by Hojman et al. \cite{HKT76}~(cf.~\cite[chapter 4]{OUP}) that the form of the Hamiltonian and the momentum constraint of GR can be \textit{derived} from dynamics of three-hypersurfaces if one starts from an
assumption that the three-metric and its conjugate momentum is the only gravitational pair of canonical variables defined on the three-dimensional hypersurface. 
As remarked in a textbook by Thiemann \cite[section 1.5]{Thiemann} and shown by Deruelle et al. \cite{DerrHD} for a class of actions whose Lagrangian is a general function of the Riemann tensor $f(R^{\mu}{}_{\alpha\nu\beta})$, any reparametrization-invariant theory of spacetime obeys such an algebra, regardless of the specific form of the Hamiltonian and momentum constraints\footnote{This general result implies that there might be a possibility that more general theories than only GR could be derived from the dynamics of three-hypersurfaces if one negates the condition of \cite{HKT76} that the three-metric and its
conjugate momentum are the only gravitational pair of canonical variables defined on the three-dimensional hypersurface.}.

Coming back to our formulation, let us establish the relationship with the usual ADM formulation.
There are three things to consider: the relationship between the momenta, the comparison of the constraints and their algebra, and the DeWitt supermetric.
Let us write down constraints in eqs.~\eqref{GRxi_H0} and \eqref{GRxi_Hi} for $\chi=0$ and $p_{\chi}=0$, i.e. for vacuum,
    \begin{subequations}
    \begin{align}
    \label{GRxi_H0v}
       \Ho{E}  & =
	    -\frac{1}{12 l^2\hbar}p_{a}^2
        +\frac{2}{l^2\hbar\, a^2}\bpb\cdot\bpb -
        \frac{l^2\hbar a^4}{2}\,^{\ssst (3)}\! R
        \deq 0\ ,\\[18pt]
    \label{GRxi_Hiv}
        \Hi{E} & =
        -2\bar{D}_{j}\left(\hb_{ik}\pb^{kj}\right)
        -\frac{1}{3}D_{i}\left(a\,p_{a}\right)
        \deq 0\ .
    \end{align}
    \end{subequations}
The relationship between the momenta can be found by making use of the decomposed extrinsic curvature given by eq.~\eqref{Kalldec}, $\h = l_{0}^3a^3$, $h^{ij}=l_{0}^{-2}a^{-2}\hb^{ij}$ and $\kappa = l_{p}^2/\hbar$ in eq.~\eqref{GR_pADM}, which results in
    \begin{align}
    \label{GRxi_padmcan1}
        p_{\srm{ADM}}^{ij} & = a^{-2}\pb^{ij} + \frac{a^{-1}}{6}\hb^{ij}p_{a} \ ,\\[12pt] 
    \label{GRxi_padmcan}
        p_a & = 2a^{-1}p_{\srm{ADM}} = - 6l^2\hbar a\kb\ ,\qquad \pb^{ij}=a^2 p_{\srm{ADM}}^{ij \srm T} = \frac{l^2\hbar a^2}{2}\hb^{ik}\hb^{jl}\ktb_{kl}\ ,
    \end{align}
i.e. $p_a$ and $\pb^{ij}$ are rescaled trace $p_{\srm{ADM}}$ and traceless $p_{\srm{ADM}}^{ij \srm T}$ parts of the ADM momentum.
Comparison of eq.~\eqref{GRxi_H0v} with eq.~\eqref{GR_Hoold} shows that they differ by $\mathcal{H}_{\bot}^{\srm E}=a^{-1}\Ho{E}$, but since it must be that $N\mathcal{H}_{\bot}^{\srm E}=\Nb\Ho{E}$ so the total Hamiltonian does not change in transition to the unimodular-conformal variables.
This is related to the fact that the unimodular-conformal variables can be derived by a \textit{canonical} transformation from the usual ADM variables, as we prove in appendix \ref{app_canon}.
Because of this, we claim without proof that the hypersurface algebra in eqs.~\eqref{HEH_AlgHH}-\eqref{HEH_Algmm} holds for the constraints in unimodular-conformal variables in eqs.~\eqref{GRxi_H0} and \eqref{GRxi_Hi}, or eqs.~\eqref{GRxi_H0v} and \eqref{GRxi_Hiv}.
This should not change if the matter is present, as in eqs.~\eqref{GRxi_H0} and \eqref{GRxi_Hi}.
Therefore, the Hamiltonian formulation of GR in unimodular-conformal variables is equivalent to the ADM formulation of GR.

\subsection{DeWitt supermetric} 
\label{subs_DeWittmet}

The final note should be on the DeWitt supermetric.
First of all, note that $\hb_{ij}$ and $\pb^{ij}$ have five independent components.
This means that the second term in eq.~\eqref{GRxi_H0v} can be transformed into a sum of five terms.
In relation to this, DeWitt \cite{Witt67} has shown that the supermetric has a signature $(-,+,+,+,+,+)$, i.e. as if the supermetric describes the line element in a six-dimensional pseudo-Riemannian space where one direction is the direction of pure dilations (i.e. conformal transformations) --- the direction of the scale density --- and other five orthogonal directions are the directions of shear (volume-preserving) deformations of the three-metric.
This space is the space of all three-geometries, whose evolution can be described by the evolution of each of the six three-metric components. Such space is usually referred to as the \textit{superspace} and one can study its geometry \cite{Witt67, Giulini}.
Now, denote with another index $I,J=a,1,...,5$ the scale component and the five shape components of the three-metric and with index $\bar{I},\bar{J}=1,...,5$ only the five shape components.
Then we can define the inverse of the DeWitt metric as
    \begin{equation}
    \label{HEH_DeWittgb}
        \mirl{\mathscr{G}}_{\ssst IJ} := 
            \Big(
            \begin{matrix}
                -\frac{1}{12} & 0 \\
                0 & 2a^{-2}\bar{\mirl{\mathscr{G}}}_{\ssst \bar{I}\bar{J}}
            \end{matrix}
            \Big)\ ,
    \end{equation}
where $\mirl{\mathscr{G}}_{aa}=- 1/12$, and write the kinetic term in eq.~\eqref{GRxi_H0v} as
    \begin{equation}
    \label{HEH_DWittgbKin}
        \frac{1}{l^2 \hbar}\mirl{\mathscr{G}}_{\ssst IJ}p^{\ssst I}p^{\ssst J}\ .
    \end{equation}
Note that formulation of the theory in the unimodular-conformal variables automatically exposes the minus sign in front of the kinetic term of the scale density.
DeWitt \cite{Witt67} has done a similar unimodular transformation\footnote{In \cite{Witt67, Giulini} a variable $\tau$ defined as $\tau:=4\vert\zeta-1/3\vert^{1/2} (\sqrt{h})^{1/2}$ was used instead of the scale density $a:= (\sqrt{h})^{1/3}$ as we do here.} which does the same.
The ``timelike'' direction of the scale density should not be understood as having something to do with the timelike direction of the spacetime.
This minus sign in the DeWitt supermetric in eq.~\eqref{HEH_DWmet} is just a simple geometrical consequence which is \textit{independent of the dimension}. 
It has to do with the minus sign in the term $ K^2 - K_{ij}K^{ij}$ in the ADM Lagrangian in eq.~\eqref{HEH_Lag}.
Namely, by considering a $D+1$ formulation of the $d=D+1$ dimensional spacetime it can be shown \cite[eq. (23a)]{Witek} that the form of the $D+1$ Hamiltonian constraint
remains unchanged; in the vacuum case it takes the following form
    \begin{equation}
        K_{ij}K^{ij} - K^2 -{}^{\ssst (D)}\! R = 0\ , \qquad i,j=1,...,D\,\ .
    \end{equation}
where $K_{ij}$ is the extrinsic curvature of the $D$-dimensional spatial hypersurface and ${}^{\ssst (D)}\! R$ is its intrinsic Ricci scalar.
To an intuitive eye this should not be surprising because the form $ K^2 - K_{ij}K^{ij}$ is nothing other than the second scalar invariant of matrix $K_{ij}$:
the term $K^2 - K_{ij}K^{ij}$ for $D = 2$ dimensions is just the determinant of the extrinsic curvature, while in higher dimensions is always the coefficient in front of the $D-2$-th order term in the latter's characteristic polynomial for the eigenvalue problem for $K_{ij}$.
This is a consequence of Gauss' \textit{theorema egregium} that relates the intrinsic with extrinsic curvature of a $D$-dimensional hypersurface embedded in a $D+1$-dimensional space.
Hence, the DeWitt supermetric in eq.~\eqref{HEH_DWmet} is unchanged in $D$-dimensions and it is important to understand that this is just a geometrical consequence of the generalization of Gauss' \textit{theorema egregium}.
However, a better insight is gained if one interprets the DeWitt supermetric as the metric on superspace: one can define a line element on this space.
Before we show this line element, for the purposes of later discussions, it is also instructive to define a more general supermetric and its inverse
    \begin{align}
    \label{HEH_DWmetB}
        G^{ikjl}_{\zeta} &:= \frac{\h}{2}\left(h^{ik}h^{jl} + h^{il}h^{jk} - 2\zeta h^{ij}h^{kl}\right)\nl
        & = \frac{1}{2a}\left(\hb^{ik}\hb^{jl} + \hb^{il}\hb^{jk} - 2\zeta \hb^{ij}\hb^{kl}\right)\ ,\\[12pt]
    \label{HEH_DWmetBin}
        \mirl{G}_{ikjl}^{\zeta} &:= \frac{1}{2\h}\left(h_{ik}h_{jl} + h_{il}h_{jk} - \frac{2\zeta}{\zeta D - 1} h_{ij}h_{kl}\right)\nl
        & = \frac{a}{2}\left(\hb_{ik}\hb_{jl} + \hb_{il}\hb_{jk} - \frac{2\zeta}{\zeta D - 1} \hb_{ij}\hb_{kl}\right)
    \end{align}
where in the second lines in both equations above we exposed the scale and shape parts of the three-metric. 
If $\zeta = 1$, one recovers the DeWitt supermetric of GR. One may investigate the consequences of other values of $\zeta$, as done in \cite{Witt67, GiuKief, Giulini}.
The most important point is that for the critical value $\zeta_{c} = 1/D$ the inverse supermetric cannot be defined.
Now, the line element on superspace --- the distance between two three-metrics --- is defined as follows \cite{Witt67, Giulini}
    \begin{align}
    \label{DeWittMet_lineel}
       \d S^{2} := G^{ikjl}_{\zeta}\d h_{ij}\otimes\d h_{kl} & = G^{ikjl}_{\zeta}\Bigpar{
       4a^{2}\hb_{ij} \hb_{kl}\d a\otimes\d a  + a^{4} \d \hb_{ij}\otimes\d \hb_{kl} \nld
       &\quad + 2 a^{3} \hb_{ij} \d a \otimes \d \hb_{kl} + 2 a^3 \hb_{kl} \d \hb_{ij} \otimes\d a 
       }\ ,\nld
       & = a^3 \left[- 4 D (\zeta D - 1) \frac{\d a}{a}\otimes \frac{\d a}{a} + \hb^{ik}\hb^{jl}\d \hb_{ij}\otimes\d \hb_{kl}\right]\ ,\nld
       & = a^3 \left[- 4 D (\zeta D - 1) \frac{\d a}{a}\otimes \frac{\d a}{a} +  \bar{\mathscr{G}}^{\ssst\bar{I}\bar{J}}\d b_{\ssst \bar{I}}\otimes \d b_{\ssst \bar{J}}\right]
       \ ,
    \end{align}
where $b_{\ssst \bar{I}}$ are the five independent shape components of $\hb_{ij}$ and $\bar{\mathscr{G}}^{\ssst\bar{I}\bar{J}}$ is the ``shape part'' of the supermetric (whose inverse appears in eq.~\eqref{HEH_DeWittgb}), which depends only on $\hb_{ij}$.
In terms of $\hb_{ij}$, one can define the \textit{traceless DeWitt supermetric} as
    \begin{equation}
    \label{DeWittmet_shape}
        \bar{G}^{ikjl} := \frac{1}{2}\left(\hb^{ik}\hb^{jl} + \hb^{il}\hb^{jk}\right) - \frac{1}{D}\hb^{ij}\hb^{kl}\ ,
    \end{equation}
which may also be called the \textit{shape DeWitt supermetric}. Note that the last term above vanishes identically when contracted with $\d \hb_{ij}\d \hb_{kl}$.
We shall hear more about it in the next section.
One can clearly identify what we shall from now on call the \textit{scale-like} and the \textit{shape-like} direction in superspace, which are the analogues of the timelike and the spacelike directions in spacetime.
The shape ``subspace'' on which the shape part $\bar{\mathscr{G}}^{\ssst\bar{I}\bar{J}}$ of the supermetric defines distances can be shown to be an Einstein space with a negative constant scalar curvature whose Ricci curvature is proportional to the negative of the shape part of the supermetric \cite[eq. (5.15)]{Witt67}.
This space is inert to the spatial conformal transformations as the shape supermetric and the ``coordinates'' on this space are $SL(3,\mathbb{R})$ tensors.
Note that the shape subspace is independent of $\zeta$.
But we again see that the critical value $\zeta_{c} = 1/D$ plays an important role in the scale-like direction: for $\zeta > \zeta_{c}$ the supermetric is indefinite, while for $\zeta < \zeta_{c}$ the sign of the scale-like direction becomes positive\footnote{The value of $\zeta$ can be studied in the context of theories generalizing GR \cite{GiuKief} and it also has consequences on the geometry of superspace \cite{Giulini}.}.
For the critical case we see that the scale-like direction drops out and one has a singularity there.
DeWitt supermetric has a more complicated form in the presence of matter. This can be seen on an example of the non-minimally coupled scalar (density) field that we used at the beginning of this section, i.e. from eq.~\eqref{GRxi_H0}.
From there one can clearly see that the signature depends on the evolution of the matter field and the scale density, and on the value of involved coupling constants.
This was investigated by Kiefer in \cite{KiefNM} in the context of the initial value problem\footnote{There the scalar field is rescaled as $\varphi = a^{-6\xi}\chi$ so the equations are different and simpler, but $\tilde{l}^2$ in eq.~\eqref{GR_kt} is of the same form up to differences in notation and the critical value of $\chi$ is the same up to an appropriate rescaling by $a$.}: 
for a critical value $\chi_{\ssst crit}$ for which eq.~\eqref{GR_kt} vanishes, the scale-like direction disappears and the DeWitt metric becomes positive-definite, whereas for values $\tilde{l}^2 > 0$ or $\tilde{l}^{2} < 0$ the DeWitt metric has an indefinite signature.
However, we are not interested in the features of the DeWitt supermetric appearing in GR, nor its general extensions such as $\zeta\neq 1$. 
We are interested in an object of similar role and features as the DeWitt supermetric that could appear in higher-derivative theories and the identification of the scale-like part of such a supermetric. 
There, other parameters than $\zeta$ or dimension $D$ could conspire to change the signature of such a supermetric and the behavior of the scale-like direction.
The important thing to keep in mind from the present discussion is that the existence of the scale-like direction in superspace is related to the fact that the theory (in this case GR) is not conformally invariant.

It was already mentioned that the scale-like direction is analogous to the timelike direction in classical relativistic mechanics.
The scale density as an evolution parameter is referred to as the \textit{intrinsic time} \cite{Witt67}.
This becomes obvious if one compares the Hamiltonian constraint of GR with the Hamiltonian of the relativistic particle with mass $m$:
    \begin{equation}
    \label{KGHam}
        H = - p_{t}^2 + p^{2}_{i} + m^2 = 0\ .
    \end{equation}
In GR's Hamiltonian constraint in eq.~\eqref{GRxi_H0v} the potential (the Ricci scalar and the cosmological constant) would be a kind of a ``mass'' term in the relativistic particle language, but it would be a ``space''- and ``time''-dependent mass term because it depends on the shape and the scale parts of the three-metric.
But because of this indefinite signature it is tempting to think of the scale density as directly related to the notion of coordinate time $t$ itself, but even though there are implications of the hyperbolic nature of the kinetic term in the Hamiltonian constraint of GR to the observed dynamics, one should distance oneself from direct identification of the scale density being the clock with respect to which we measure coordinate time \cite{OUP}.
It is more appropriate to think of it as one of many choices for an evolution parameter with respect to which the dynamics of variables \textit{within the configuration space itself} may be expressed.
To connect the notion of time with the notion of intrinsic time (or any other evolution parameter defined in terms of the configuration space variables of an underlying theory of gravity) one must address the contradiction that quantum field theory on curved spacetime refers to the spacetime as \textit{a fixed background}, while in GR the spacetime itself is a dynamical object \cite{KieferEss}.
Therefore, one needs to be careful what one means by ``$t$''.
It is hoped that quantum gravity may address this issue, and indeed the notion of \textit{the problem of time} arises there \cite{IshamTime} as one of the most important unsolved questions about the observable Universe.

\section{Hamiltonian formulation of a general quadratic curvature theory}
\label{sec_HHDall}

With this section we start the Hamiltonian formulation of higher-derivative theories of gravity that we discuss in this thesis.
Kaku \cite{Kaku1982} was first to open this field on the example of an action in which only the Weyl-tensor term is present.
Later, the same theory was addressed by Boulware \cite{Blw}.
Hamiltonian formulation of the most general quadratic curvature theory akin to the one discussed in this thesis was analysed by Szczyrba \cite{Sz} using the symplectic formalism; Szczyrba's work seems to be the first account of its kind and is rather detailed on the matter of features and number of degrees of freedom of the theory.
Another work which shows a great detail into features and symmetries of the theory is by Odintsov et al. \cite{Odints}.
Some exact solutions of this theory were obtianed by Demaret et al. \cite{DemQ,Dem} using the Hamiltonian formulation of quadratic curvature gravity \cite{Quer}.
More recently, Deruelle et al. \cite{DerrHD} discussed the Hamiltonian formulation of a general $f(Riemann)$ theory of gravity and pointed out some important features that were not mentioned elsewhere.
Kluso\u n et al. \cite{Kluson2014} have presented the Hamiltonian formulation of a theory based on the same action as we are using here, except with a conformally coupled scalar field added to the case of the Weyl-tensor gravity only.
The author's Master thesis \cite{MSc} covers the Weyl-tensor (W) and Weyl-Einstein (WE) gravity with their canonical quantization.
The case of the W gravity was further considered in \cite{ILP} where the notion of the generator of conformal transformations was introduced for the first time correctly.
These are the most important examples of Hamiltonian formulation of generic higher-derivative theories of gravity.
Other examples of the Hamitlonian formulation of higher-derivative theories mainly deals with specific models, see e.g. \cite{Hwz,Schm} and comprehensive list of references in \cite[chapter 4]{Quer}.

In none of the aforementioned works except \cite{Odints,Sz} the idea to use variables similar to the unimodular-conformal variables has appeared.
The fact that the conformal invariance of the $3+1$-decomposed $C^2$ term should be manifest in terms of the absence of $K$ and $\h$ from the Hamiltonian and momentum constraints was left unnoticed except in the two cited references (to our knowledge).
Szczyrba \cite{Sz} has noticed that the extrinsic curvature separates naturally into its traceless and trace parts if one uses what we call here the shape density $\hb_{ij}$ and $\h$ as independent variables.
They have also analysed the constraints and the number of degrees of freedom for various special combinations of the terms in the action.
Odintsov et al. \cite{Odints} have used the Hamiltonian formulation for the purpose of the path integral formulation of higher-derivative theories (covered in greater detail in \cite{BOS}).
The recent work which is closest to suggesting that another set of variables can reveal the conformal features of the $C^2$ theory is \cite{DerrHD}, who mentioned at the end of their analysis that a reformulation of their approach to the Hamiltonian formulation of the $C^2$ theory could show that $\h$ could be eliminated from the constraints due to its conformal invariance.
Kluso\u n et al. \cite{Kluson2014}, despite their very detailed constraint analysis, have only realized that the velocities $h^{ij}\dot{K}_{ij}$ cannot be inverted for in the $C^2$ theory.
Their result --- and likewise the results of Kaku \cite{Kaku1982} and Boulware \cite{Blw} --- fails to recognize that $\h$ has to be absent from the theory.
They do, however, notice that the traceless part of the extrinsic curvature $K_{ij}^{\srm T}$ and its conjugate momentum appears to be the only dynamical variable from the extrinsic curvature sector, although their form of constraints still depended on the trace $K$.
In the author's Master thesis \cite{MSc} the Hamiltonian formulation of the W gravity was achieved by using $K_{ij}^{\srm T}$ and $K$ as independent canonical variables, but the Hamiltonian and momentum constraints of the theory still had the form that depends on $\h$.
The author was unaware of the results of \cite{ Odints,Sz} at the time of working on \cite{MSc} and \cite{MSc}. 
With the introduction of the unimodular-conformal variables that we presented in chapter \ref{ch:umtocf1} we are able to formulate a Hamiltonian version of the W and WE theories in which the constraints manifestly exhibit conformal properties \cite{KN17}, such that $a$ and $\kb$ are completely eliminated from the Hamiltonian and momentum constraints.
In this section we aim to extend the application of the unimodular-conformal variables employed in \cite{KN17} to the more general action given by eq.~\eqref{SEE_TheAction}.
We show that the choice of unimodular-conformal variables --- as motivated in chapters \ref{ch:umtocf}-\ref{ch:defcf} --- can completely and clearly separate the degrees of freedom which are introduced by the $R^2$ and the $C^2$ terms into conformally invariant and conformally non-invariant ones.
The upcoming sections will focus on particular cases, one of which is covered in \cite{KN17}, such that features of the $R^2$ and the $C^2$ terms are presented in a manner not yet encountered in the literature.

\subsection{Hamiltonian formulation in unimodular-conformal variables}
\label{subs_HDall_HamForm}

The Lagrangian we are working with in this section is based on eq.~\eqref{SEE_TheAction} plus the action for a non-minimally coupled scalar field whose Lagrangian was derived in eq.~\eqref{chi31Lagdec}.
In unimodular-conformal variables the action takes the following form,
    \begin{align}
    \label{HDall_LagAll}
        S^{\srm{ERW\chi}} = \inttx  \mathcal{L}^{\srm{ERW\chi}} \equiv \inttx \left(\mathcal{L}^{\srm E} + \mathcal{L}^{\srm R} + \mathcal{L}^{\srm W} + \mathcal{L}^{\srm \chi}\right)\ ,
    \end{align}
where
    \begin{align}
    \label{HDall_LagADM}
        \mathcal{L}^{\srm E} & = \frac{l^2 \hbar\Nb a^2}{2}\left(a^2\left(\,^{\ssst (3)}\! R 
        - 2\bar{\Lambda}\right) 
        + \bktb\cdot\bktb 
        - 6\kb^2\right)\ ,\\[12pt]
    \label{HDall_LagR2}
        \mathcal{L}^{\srm R} & = \frac{\br\hbar}{72} \Nb \left[6 \mathcal{L}_{\bar{n}}\kb 
        + 6 \kb^2 
        + \bktb\cdot\bktb
        + a^2 \,^{\ssst (3)}\! R 
        - \frac{2}{\Nb}\mathbf{D}\cdot\mathbf{D}\Nb\right]^2\ ,\\[12pt]
    \label{HDall_LagW}
        \mathcal{L}^{\srm W} & = -\aw\hbar\Nb\left( \frac{1}{2}\bar{\mathbf{C}}^{\srm E}\cdot\bar{\mathbf{C}}^{\srm E} 
        - \bar{\mathbf{C}}^{\srm B}\cdot\bar{\mathbf{C}}^{\srm B}  \right)\ .
    \end{align}
The $R^2$ term is derived using eq.~\eqref{Rdec31umod2}, the Weyl-tensor term comes from eq.~\eqref{W2EB}, while the scalar density field Lagrangian is given by eq.~\eqref{chi31Lagdec}.
For convenience, we have redefined the coupling of the $R^2$ term as $\br\rightarrow \br/18$.
For simplicity we shall assume that all divergences giving rise to boundary terms are subtracted.

There are two important features of this higher-derivative Lagrangian. Firstly, unlike in GR, this Lagrangian \textit{depends} on velocities $\dot{\Nb},\dot{N}^{i}$, through $\dot{\kb}$ and $\dot{\kb}^{\srm{T}}_{ij}$.
One might be tempted to think that this fact prevents one from deriving the corresponding Hamiltonian and momentum constraints, thereby obscuring the diffeomorphism invariance of the theory, but this is not the case, as we shall soon see.
Secondly, the Lagrangian depends not only on the first but also on the second time derivatives of the three-metric\footnote{It depends on the first and second space derivatives as well, but for the statement in the text only time derivatives are relevant so we suppress the notation of explicit dependence on the former.}, and so one can write
    \begin{equation}
    \label{HDall_LagDep}
        \mathcal{L}^{\srm{ERW\chi}} = \mathcal{L}^{\srm{ERW\chi}} \left(\Nb,N^{i}, a,\hb_{ij},\chi,\dot{\Nb},\dot{N}^{i},\dot{a},\dot{\chi},\dot{\hb}_{ij},\ddot{a},\ddot{\hb}_{ij}\right)\ .
    \end{equation}
The second time derivatives cannot be partially integrated away (since, unlike the EH Lagrangian, the Lagrangian satisfies eq.~\eqref{ALHO_HDcond}).

Now, since we are dealing with a higher-derivative Lagrangian, one needs to reduce the order of the theory in order to arrive at the Hamiltonian formulation.
The method for doing this was explained on an example of a simple harmonic oscillator with a higher derivative term in section \ref{subs_Ham_osc}: 
define a new set of variables such that all first derivatives are the new \textit{independent} variables themselves and add the necessary constraints to the Lagrangian which ensure that the new variables are treated independently only until the constraints are enforced\footnote{One may also choose the second derivatives of the three-metric components as the new variables, as in \cite{Blw,Hwz}, for example.
The difference between the two sets of variables amounts to a canonical transformation of exchanging the variables with its conjugate momenta.}.
In the present case, one uses the components of the extrinsic curvature as the new variables, which ``hide'' the velocities of the components of the three-metric.
The constraints can be introduced by the following ``delayed equalities'' (cf. appendix \ref{app_const}),
    \begin{align}
    \label{HDall_KtlessC}
        \bar{\mathcal{K}}^{\srm{T}}_{ij} & := 2\Nb\ktb_{ij} - \dot{\hb}_{ij} + 2\left[\bar{D}_{(i}\Nb_{j)}\right]^{\srm T} \deq 0 \ ,\\[12pt]
    \label{HDall_KtC}
        \mathcal{K} & := \Nb \kb - \frac{\dot{a}}{a} + \frac{D_{i}N^{i}}{3} \deq 0\ ,
    \end{align}
which are based on definitions of the traceless and trace extrinsic curvature density in eqs.~\eqref{Kbardef2} and \eqref{Ktbardef2}, respectively.
In analogy to eq.~\eqref{H_alho_Lagcost}, these constraints are added to the initial Lagrangian in eq.~\eqref{HDall_LagAll},
    \begin{align}
    \label{HDall_Lagconst}
        &\mathcal{L}^{\srm{ERW\chi}}_{c} \left(\Nb,N^{i},a,\hb_{ij},\chi,\kb,\ktb_{ij},\dot{\chi}; \dot{a},\dot{\hb}_{ij}, \bar{\lambda}, \bar{\lambda}^{ij\srm{T}}\right) :=\nld &\mathcal{L}^{\srm{ERW\chi}}\left(\Nb,N^{i}, a,\hb_{ij},\chi,\kb,\ktb_{ij},\dot{\chi},\dot{\kb},\dot{\kb}^{\srm{T}}_{ij}\right)
        - \bar{\lambda}^{ij\srm{T}} \bar{\mathcal{K}}^{\srm{T}}_{ij} 
        - a\lambda \mathcal{K}\ ,
    \end{align}
where $\bar{\lambda}^{ij\srm{T}}$ and $\bar{\lambda}$ are Lagrange multipliers which are tensor densities of scale weight five and two, respectively.
One can see now that all the dependence on $\dot{a}$ and $\dot{\hb}_{ij}$ in eq.~\eqref{HDall_Lagconst} comes only through the constraints.
Furthermore, the first term on the right hand side of eq.~\eqref{HDall_Lagconst} is the same Lagrangian as in eq.~\eqref{HDall_LagAll}, except that $\kb$ and $\ktb_{ij}$ are not just mere labels but the actual \textit{independent}, but auxiliary variables.
They are the analog of $q_{2}$ in the first equation in \eqref{H_alho_q2}.
The additional degrees of freedom are thus made explicit.
Moreover, observe that all time derivatives of the lapse density and the shift have been absorbed into the new variables; 
in conclusion, reformulating a higher-derivative gravity theory as a first order theory eliminates explicit dependence of the Lagrangian on the first time derivatives $\dot{\Nb}$ and $\dot{N}^{i}$.

The conjugate momenta are now derived from the constrained Lagrangian in eq.~\eqref{HDall_Lagconst} and they take the following form
    \begin{align}
    \label{HDall_pchi}
        p_{\chi} & =\frac{\del\mathcal{L}^{\srm{ERW\chi}}}{\del \dot{\chi}} =
	    \bar{n}^{\mu}\del_{\mu}\chi + 6\xi_{c}\kb\chi-\frac{\del_{i}N^{i}}{3\Nb}\chi\ ,\\[12pt]
    \label{HDall_pNb}
        p_{\ssst \Nb} & = \dd{\mathcal{L}^{\srm{ERW\chi}}_{c}}{\dot{\Nb}} \deq 0  \ ,\qquad
        p_{i} = \dd{\mathcal{L}^{\srm{ERW\chi}}_{c}}{\dot{N}^{i}} \deq 0   \ ,\\[12pt]
    \label{HDall_pa}
        p_{a} & = \dd{\mathcal{L}^{\srm{ERW\chi}}_{c}}{\dot{a}} = a\bar{\lambda}  \ ,\\[12pt]
    \label{HDall_ph}
        \pb^{ij} & = \dd{\mathcal{L}^{\srm{ERW\chi}}_{c}}{\dot{\hb}_{ij}} = \bar{\lambda}^{ij\srm{T}}  \ ,\\[12pt]
    \label{HDall_Pk}
        \Pb & = \dd{\mathcal{L}^{\srm{ERW\chi}}_{c}}{\dot{\kb}} = \br \hbar R \nld
        & = \frac{\br\hbar}{6} \left[6 \mathcal{L}_{\bar{n}}\kb 
        + 6 \kb^2 + \bktb\cdot\bktb
        + a^2 \,^{\ssst (3)}\! R 
        - \frac{2}{\Nb}\mathbf{D}\cdot\mathbf{D}\Nb\right]\ ,\\[16pt]
    \label{HDall_Pkt}
        \Pb^{ij} & = \dd{\mathcal{L}^{\srm{ERW\chi}}_{c}}{\dot{\kb}_{ij}^{\srm{T}}}= -\aw\hbar \hb^{ik}\hb^{jl}\bar{C}^{\srm E}_{kl} \nld
        & = -\aw \hbar \hb^{ia}\hb^{jb}\left[\mathcal{L}_{\bar{n}}\kb_{ij}^{\srm T}
          - \frac{2}{3}\hb_{ij}\bktb \cdot \bktb 
          - \,^{\ssst (3)}\!\bar{R}_{ij}^{\srm T}
          - \frac{1}{\Nb}\left[\bar{D}_{i}\del_{j}\Nb\right]^{\srm
          T}\right]  \ ,\\[12pt]
    \label{HDall_pL}
        p_{\lambda} & = \dd{\mathcal{L}^{\srm{ERW\chi}}_{c}}{\dot{\bar{\lambda}}} = 0  \ ,\qquad \pb^{\lambda}_{ij} = \dd{\mathcal{L}^{\srm{ERW\chi}}_{c}}{\dot{\bar{\lambda}}^{ij\srm{T}}} = 0   \ ,
    \end{align}
where $p_{a}$, $\pb^{ij}$, $\Pb$ and $\Pb^{ij}$ are tensor densities of scale weight three, five, two and four; note that $p_{\chi}$ is the same as eq.~\eqref{GRchi_pchi} derived in GR, except that there $\kb$ was expressed in terms of $p_{a}$ and $p_{\chi}$.
Note that we have $2 \times 6 = 12$ additional variables compared to the original theory: the Lagrange multipliers and their conjugate momenta.
But as explained in section \ref{subs_Ham_osc} and appendix \ref{app_dirb}, equalities in eqs.~\eqref{HDall_pa}, \eqref{HDall_ph} and \eqref{HDall_pL} can already be set to zero without delays, since the introduction of $\bar{\lambda}^{ij\srm{T}}$ and $\bar{\lambda}$ does not interfere with the dynamics of the original theory \cite{Kluson2014}.
This is expected because they are introduced via a simple internal relabeling of objects which does not give rise to any additional structure in the theory.
This means that these 12 additional variables can be eliminated even before the calculation of the Poisson brackets.
This leaves one with only seven invertible velocities --- $\dot{\chi}$, which is the same as eq.~\eqref{GRchi_pchi}:
    \begin{align}
    \label{HDall_Ktvel}
        \dot{\kb} & = \Nb\left[\frac{\Pb}{\br\hbar} 
        - \frac{1}{6}\left( 
        6\kb^2
        + \bktb\cdot\bktb 
        + a^2 \,^{\ssst (3)}\! R - \frac{2}{\Nb}\mathbf{D}\cdot\mathbf{D}\Nb
         \right)
         \right] 
         + \mathcal{L}_{\vec{N}}\kb\ ,\\[12pt]
    \label{HDall_Kvel}
        \dot{\kb}_{ij}^{\srm{T}} & = - \Nb\left[\frac{\hb_{ia}\hb_{jb}\Pb^{ab}}{\aw \hbar}
        - \frac{2}{3}\hb_{ij}\bktb \cdot \bktb
        - \,^{\ssst
          (3)}\!\bar{R}_{ij}^{\srm T}
        - \frac{1}{\Nb}\left[\bar{D}_{i}\del_{j}\Nb\right]^{\srm T}
        \right]
        + \mathcal{L}_{\vec{N}}\ktb_{ij}\ ,
    \end{align}
where in eq.~\eqref{HDall_Kvel} we used eq.~\eqref{LieKtdec1}.
It is now obvious that the dynamics will be contained in the extrinsic curvature sector.
Note in passing that the trace of eq.~\eqref{HDall_Kvel} does not vanish, because the fourth term in there survives upon contraction with $\hb^{ij}$.
Furthermore, observe that each term in eq.~\eqref{HDall_Ktvel} has a pair in eq.~\eqref{HDall_Kvel}: the first terms in eq.~\eqref{HDall_Ktvel} and eq.~\eqref{HDall_Kvel} are the trace and traceless part of what would be the momentum conjugate to $K_{ij}$ up to scaling with $a$;
the second term in eq.~\eqref{HDall_Kvel} represents subtraction of the trace of $\mathcal{L}_{\bar{n}}K_{ij}^{\srm T}$ which is just the third term in eq.~\eqref{HDall_Ktvel}; the fourth term in eq.~\eqref{HDall_Ktvel} is the trace of the Ricci tensor, while the latter's traceless piece is the third term in eq.~\eqref{HDall_Kvel};
the fifth term in eq.~\eqref{HDall_Ktvel} and the fourth term in eq.~\eqref{HDall_Kvel} are the traceless and trace pieces of $D_{i}D_{j}\Nb$;
the last term in eq.~\eqref{HDall_Ktvel} together with the second term, correspond to the last term in eq.~\eqref{HDall_Kvel}.

We would like to make an important observation at this point.
It was mentioned earlier in this chapter that $R^2$ and $C^2$ (or $R_{\mu\nu}R^{\mu\nu} - R^2 /3$, cf. footnote~\ref{SchWsquared} on page~\pageref{SchWsquared}) terms are the only quadratic curvature terms in four dimensions which can appear as the counter-terms, up to a reformulation done in transition from eq.~\eqref{SEE_effAct} to eq.~\eqref{SEE_TheAction} due to identity in eq.~\eqref{GaussBonnet} valid only in four dimensions.
Using unimodular-conformal decomposition in this section, these two terms may be interpreted as two independent kinetic terms of the conformally invariant and conformally variant part of the extrinsic curvature.
Namely, observe that the $R^2$ term hosts the time derivative of the expansion density $\kb$ only, while the $C^2$ term hosts the time derivative of the expansion density $\ktb_{ij}$ only.
This is related to the fact that the Weyl tensor and the Ricci scalar are orthogonal pieces of the Riemann tensor.
An analogy can be drawn with the kinetic term of the EH action: there the kinetic term splits into the scale part and the shape part in an orthogonal way, because the expansion density $\kb$ and the shear density $\ktb_{ij}$ are orthogonal pieces of the extrinsic curvature.
Furthermore, we can observe a certain asymmetry between the two quadratic curvature kinetic terms: the term $C^2$ is completely deprived of $a$ and $\kb$, while the term $R^2$ necessarily contains $\ktb_{ij}$.
In other words, this asymmetry shows that the kinetic term of $\ktb_{ij}$ is independent of the scale, but the kinetic term of $\kb$ is \textit{not} independent of the shape.
We have already met this asymmetry in section \ref{subsec_curvAsh}. There we have seen that the Riemann tensor cannot be split into scale-independent and shape-independent pieces in an orthogonal way and that this fact is reflected in the behavior of the shape and scale parts of the metric in a small neighbourhood of a geodesic;
the asymmetry in the mentioned kinetic terms --- which are the independent pieces of the Riemann tensor --- follows from this.
This has certain implications to the dynamics of the higher-derivative theory that we shall come back to in the following sections.

Compared to GR, the phase space of the presently discussed higher-derivative theory is extended by six canonical pairs of the extrinsic curvature sector.
The Poisson brackets in eq.~\eqref{eqn:PB} therefore contain six more terms and their antisymmetrized counterpart.
The canonical pairs thus obey the following Poisson brackets,
    \begin{align}
    \label{HDall_PBvars1}
        \PB{\hb_{ij}(\bx)}{\pb^{ab}(\by)}
        & = \mathbb{1}^{{\srm{T}}ab}_{ij}\delta
        (\bx,\by)\ ,\qquad 
        &\PB{\ktb_{ij}(\bx)}{\Pb^{ab}(\by)}
        & = \mathbb{1}^{{\srm{T}}ab}_{ij}\delta
        (\bx,\by)\ , \\[6pt]
    \label{HDall_PBvars2}
        \PB{a(\bx)}{p_{a}(\by)}
        & = \delta (\bx,\by)\ , \qquad 
        &\PB{\kb(\bx)}{\Pb(\by)}
        & = \delta (\bx,\by)\ , \\[6pt]
    \label{HDall_PBvars3}
        \PB{\chi(\bx)}{p_{\chi}(\by)}
        & = \delta (\bx,\by)\ , & &
    \end{align}
which is similar to eq.~\eqref{HEH_PBvars1} and eq.~\eqref{HEH_PBvars2}, except that now we have six additional pairs in the extrinsic curvature sector.
Lapse density and shift vector again obey Poisson brackets analogous to eq.~\eqref{HDall_PBvars2}.
All other Poisson brackets vanish.

For the Legendre transform, which takes the following form,
    \begin{align}
    \label{HDall_LegTransf}
        H^{\srm{ERW\chi}} & = \intx\,\Bigg(  \dot{a}p_{a} + \dot{\hb}_{ij}\pb^{ij} 
        + \dot{\kb}_{ij}^{\srm T}\Pb^{ij} + \dot{\kb}\Pb 
        + \dot{\chi}p_{\chi} + \lambda_{\ssst\Nb}p_{\ssst\Nb} + \lambda_{i}p^{i} - \mathcal{L}^{\srm{ERW\chi}}_{c}\Bigg)\ ,
    \end{align}
we need to deal with a few partial integrations on the last two terms in eq.~\eqref{HDall_Ktvel} multiplied by $\Pb$ and the last two terms in eq.~\eqref{HDall_Kvel} contracted by $\Pb^{ij}$.
Using eq.~\eqref{LieKbarN} and partial integration we can deal with the following term,
    \begin{align}
    \label{HDall_LegTrLiePbar}
        \intx\,
        \Pb\mathcal{L}_{\vec{N}}\kb
        & = \intx\,\left(\Pb N^{k}\del_{k}\kb + \frac{1}{3}\kb\Pb\del_{k}N^{k}\right) \nld
        & = \intx\,\left(\Pb N^{i}\del_{k}\kb 
        + \frac{1}{3}\del_{k}\left(N^{k}\kb\Pb\right)
        - \frac{1}{3}\del_{k}\left(\kb\Pb\right)
        \right)\nld
        & = \frac{1}{3}\intx\,N^{k}\left(\Pb \del_{k}\kb 
        -  \del_{k}\left(\kb\Pb\right)
        \right)
        + \frac{1}{3}\intx\,\del_{i}\left(N^{i}\kb\Pb\right)\ .
    \end{align}
The last term in the above equation is a boundary term.
Using the expression for $\mathcal{L}_{\vec{N}}\ktb_{ij}$ in eq.~\eqref{LieKtdec} and partial integration in a similar way as above, the following term is treated as well,
    \begin{align}
    \label{HDall_LegTrLiePt}
        \intx\,
        \Pb^{ij}\mathcal{L}_{\vec{N}}\ktb_{ij}
        & = \intx\,\Bigg(\Pb^{ij} N^{k}\del_{k}\ktb_{ij} + 2\del_{i}\left(N^{k}\ktb_{jk}\Pb^{ij}\right)
        - 2 N^{k}\del_{i}\left(\ktb_{jk} \Pb^{ij} \right)\nld
        & \qquad\qquad
        - \frac{1}{3}\del_{k}\left(N^{k}\ktb_{ij}\Pb^{ij}\right)
        + \frac{1}{3}N^{k}\del_{k}\left(\ktb_{ij}\Pb^{ij}\right)
        \Bigg) \nld
        & = \intx\,N^{k}\Bigg(\Pb^{ij} \del_{k}\ktb_{ij} 
        - 2 \del_{i}\left(\ktb_{jk} \Pb^{ij} \right)
        + \frac{1}{3}\del_{k}\left(\ktb_{ij}\Pb^{ij}\right)
        \Bigg)  \nld
        & \quad
        + \intx\,\left(2\del_{i}\left(N^{k}\ktb_{jk}\Pb^{ij}\right)
        - \frac{1}{3}\del_{k}\left(N^{k}\ktb_{ij}\Pb^{ij}\right)
        \right)\ ,
    \end{align}
where the symmetrization on lower indices $ij$ is dropped because $\Pb^{ij}$ is symmetric. The same contraction picks up only the traceless parts of the objects contracted with it.
The last two terms in the above equation are boundary terms.
Furthermore, the fifth term in eq.~\eqref{HDall_Ktvel} multiplied with $\Pb$ can be partially integrated as follows,
    \begin{align}
    \label{HDall_DDPN}
        \intx\,\Pb D^{i}D_{i}\Nb & = \intx\,\left( 
        D^{i}\left(\Pb D_{i}\Nb \right)
        - D^{i}\Pb D_{i}\Nb\right)\nld
        & = \intx\,\left( 
        D^{i}\left(\Pb D_{i}\Nb \right)
        - D_{i}\left(\Nb D^{i}\Pb \right)
        + \Nb D_{i}D^{i}\Pb  \right)
        \nld
        & = \intx\,\left( 
        \del_{i}\left(\Pb D^{i}\Nb 
        - \Nb D^{i}\Pb \right)
        \right)
        + \intx\,\Nb D_{i}D^{i}\Pb\ .
    \end{align}

A similar partial integration can be done with the fourth term in eq.~\eqref{HDall_Kvel} contracted with $\Pb^{ij}$, but it is easier to do it if we undo the cancellation of the scale-dependent terms by $\,^{\ssst (3)}\!R_{ij}$ due to the unimodular-conformal decomposition for a moment.
We then have the following,
    \begin{align}
    \label{HDall_DDPtN}
        \intx\,\Pb^{ij} D_{i}D_{j}\Nb & =  \intx\,\bigg(
        D_{i}\left(\Pb^{ij} D_{j}\Nb\right)
        - D_{i}\Pb^{ij} D_{j}\Nb
        \bigg)\nld
        & = \intx\, \bigg(D_{i}\left(\Pb^{ij} D_{j}\Nb\right)
        - D_{j}\left(\Nb D_{i}\Pb^{ij} \right)\bigg)
        + \intx\,\Nb D_{j}D_{i}\Pb^{ij}\nld
        & = \intx\, \del_{i}\bigg(\Pb^{ij} D_{j}\Nb
        - \Nb D_{j}\Pb^{ij} \bigg)
        + \intx\,\Nb D_{j}D_{i}\Pb^{ij}\ ,
    \end{align}
where in the last equality we used renaming of indices in the second term and the fact that $\Pb^{ij} D_{j}\Nb$ and $\Nb D_{j}\Pb^{ij}$ are both vector densities of weight $1$ (corresponding to scale weight 3) to turn the covariant derivative into the partial one.
Those two terms are thus just boundary terms.
But now it is not obvious that the sum of $\,^{\ssst (3)}\bar{R}_{ij}\Pb^{ij}$ and eq.\eqref{HDall_DDPtN} is conformally invariant.
However, we can prove that in the following indirect way.
Expand the derivatives in the total divergence in the first two terms in eq.~\eqref{HDall_DDPtN} and observe that the scale $a$ cancels out,
    \begin{align}
        \Pb^{ij} D_{j}\Nb
        - \Nb D_{j}\Pb^{ij} & = 
        \Pb^{ij} \del_{j}\Nb 
        + \Nb \Pb^{ij} \del_{j}\log a
        - \Nb \bar{D}_{j}\Pb^{ij}\nld
        & \quad
        - \Nb \Sigma^{j}{}_{jk}\Pb^{ik}
        - \Nb \Sigma^{i}{}_{kj}\Pb^{kj}
        + 4\Nb \Pb^{ij}\del_{j}\log a\nld
        & = \Pb^{ij} \del_{j}\Nb 
        + \Nb \Pb^{ij} \del_{j}\log a
        - \Nb \bar{D}_{j}\Pb^{ij}\nld
        & \quad
        - \Nb \Sigma^{j}{}_{jk}\Pb^{ik}
        - \Nb \Sigma^{i}{}_{kj}\Pb^{kj}
        + 4\Nb \Pb^{ij}\del_{j}\log a\nld
        & = \Pb^{ij} \del_{j}\Nb 
        - \Nb \bar{D}_{j}\Pb^{ij}\ .
    \end{align}
The sum of the second and the last term above cancels with all terms containing the scale connection due to eq.~\eqref{Sigma} and eq.~\eqref{Clogaprop} applied to three dimensions and using the fact that $\hb_{ij}\Pb^{ij}$ vanishes.
Therefore, the boundary term in eq.~\eqref{HDall_DDPtN} is conformally invariant.
Now, adding $\intx\,^{\ssst (3)}\!R_{ij}\Pb^{ij}$ to the integrand in both sides of eq.~\eqref{HDall_DDPtN},
    \begin{equation}
        ^{\ssst (3)}\!R_{ij}^{}\Pb^{ij} + \left[D_{i}D_{j}\Nb\right]^{\srm T}\Pb^{ij}  =\,^{\ssst (3)}\!R_{ij}^{}\Pb^{ij} 
        + \Nb D_{j}D_{i}\Pb^{ij}
        + \del_{i}\left(\Pb^{ij} \del_{j}\Nb
        - \Nb \bar{D}_{j}\Pb^{ij}\right)\ ,
    \end{equation}
we can conclude that since the LHS of the equation is conformally invariant (cf. appendix \ref{App_DD}) and the divergence term on the RHS is also conformally invariant, then the first two terms together on the RHS must be conformally invariant as well and we can simply relabel $\,^{\ssst (3)}\!R_{ij}\Pb^{ij}\rightarrow \,^{\ssst (3)}\!\bar{R}_{ij}\Pb^{ij}$, $D_{j}D_{i}\rightarrow \bar{D}_{j}\bar{D}_{i}\Pb^{ij}$.
Furthermore, since $\bar{D}_{i}\Pb^{ij}$ is a vector density, $\bar{D}_{j}\bar{D}_{i}\Pb^{ij} = \del_{j}\bar{D}_{i}\Pb^{ij}$ because $\bar{D}_{j}$ derivative does not recognize the non-zero weight of a tensor density it acts on and we also have $\bar{\Gamma}^{i}{}_{ij}=0$ (cf. eq.~\eqref{Gammabarprop}).
Finally, we have
    \begin{equation}
    \label{HDall_DDPtN1}
        \left(\Nb\,^{\ssst (3)}\!R_{ij}^{\srm T} + \left[D_{i}\del_{j}\Nb\right]^{\srm T} \right)\Pb^{ij} = \Nb\left(\,^{\ssst (3)}\!\bar{R}_{ij}^{\srm T} 
        +  \del_{j}\bar{D}_{i}\right)\Pb^{ij}
        + \del_{i}\left(\Pb^{ij} \del_{j}\Nb
        - \Nb \bar{D}_{j}\Pb^{ij}\right)\ .
    \end{equation}

Plugging eq.~\eqref{HDall_Ktvel} and eq.~\eqref{HDall_Kvel} into the Legendre transform in eq.~\eqref{HDall_LegTransf}, using eq.~\eqref{HDall_LegTrLiePbar}, eqs.~\eqref{HDall_LegTrLiePt}, \eqref{HDall_DDPN} and \eqref{HDall_DDPtN1} in it, and substituting all velocities and Lagrange multipliers, we obtain the total Hamiltonian of the quadratic curvature higher-derivative theory of gravity with a non-minimally coupled scalar (density) field,
    \begin{equation}
        H^{\srm{ERW\chi}} = 
        \intx\left\lbrace \Nb\Ho{ERW\chi}
        + N^{i}\Hi{ERW\chi}
        + \lambda_{\ssst\Nb}p_{\ssst\Nb}
        + \lambda^{i}p_{i}\right\rbrace
        + H^{\srm{ERW}\chi}_{\ssst surf}\ ,
    \end{equation}
with a rather different Hamiltonian and momentum constraints compared to the case of GR (cf. eqs.~\eqref{GRxi_H0} and \eqref{GRxi_Hi}),
    \begin{align}
    \label{HDall_Hofin}
        \Ho{ERW\chi} & = 
        \frac{1}{2\br\hbar}\Pb^2
        - \frac{1}{2\aw\hbar}\bPb\cdot\bPb
        - \mathcal{D}_{\srm R}^2\Pb
        + \bm{\mathcal{D}}^2_{\srm W}\cdot\bPb
        + a\kb p_{a}
        + 2\bktb\cdot\bpb\nld
        &\quad 
        - \aw\hbar\bar{\mathbf{C}}^{\srm B}\cdot\bar{\mathbf{C}}^{\srm B} 
        - \frac{l^2 \hbar a^2}{2}\Big(a^2\left(\,^{\ssst (3)}\! R 
        - 2\bar{\Lambda}\right) 
        + \bktb\cdot\bktb 
        - 6\kb^2\Big)\nld
        & \quad
        + \frac{1}{2}p_{\chi}^2
        - 6\xi_{c}\kb\chi p_{\chi}
        + \frac{1}{2}V^{\chi}
        \deq 0\ ,\\[20pt]
    \label{HDall_Hifin}
        \Hi{ERW\chi} & = 
        - 2\bar{D}_{j}\left(\hb_{ik}\pb^{kj}\right)
        -\frac{1}{3}D_{i}\left(a\,p_{a}\right)
        - \frac{1}{3}\left(\chi\del_{i}p_{\chi}
        - 2\del_{i}\chi \, p_{\chi}\right)\nld
        & \quad
        + \Pb^{jk} \bar{D}_{i}\ktb_{jk} 
        - 2 \bar{D}_{j}\left(\ktb_{ik} \Pb^{jk} \right)
        + \frac{1}{3}\del_{i}\left(\ktb_{jk}\Pb^{jk}\right)
        + \Pb \del_{i}\kb 
        - \del_{i}\left(\kb\Pb\right)
    \end{align}
where we defined
    \begin{align}
    \label{HDall_HamDDR}
        \mathcal{D}_{\srm R}^2\Pb & := 
        \frac{1}{6}\Big( 
        6\kb^2
        + \bktb\cdot\bktb 
        + a^2 \,^{\ssst (3)}\! R - 2\mathbf{D}\cdot\mathbf{D}
         \Big)\Pb
         \ ,\\[12pt]
    \label{HDall_HamDDW}
        \bm{\mathcal{D}}^2_{\srm W}\cdot\bPb & := \Big(\,^{\ssst
          (3)}\!\bar{R}_{ij}^{\srm T}
        + \left[\del_{j}\bar{D}_{i}\right]^{\srm T}\Big)\Pb^{ij}
    \end{align}
in order to simplfy the equations and recall that $V^{\chi}$ is given by eq.~\eqref{Vchi_def}.
The term $H^{\srm{ERW\chi}}_{\ssst surf}$ contains, in addition to the surface term appearing in the total Hamiltonian of GR with $\chi$-field given by eq.~\eqref{GRchi_HsurfEchi}, surface terms from eqs.~\eqref{HDall_LegTrLiePbar}, \eqref{HDall_LegTrLiePt} and \eqref{HDall_DDPtN1}, and is given by
    \begin{align}
        H^{\srm{ERW\chi}}_{\ssst surf} & =
        H^{\srm{E\chi}}_{\ssst surf}
        + \frac{1}{3}\intx\,\del_{i}\Big(
        N^{i}\kb\Pb
        + 2 N^{k}\ktb_{jk}\Pb^{ij}
        - N^{i}\ktb_{jk}\Pb^{jk}\nld
        &\quad 
        + \Pb D^{i}\Nb 
        - \Nb D^{i}\Pb
        + \Pb^{ij} \del_{j}\Nb
        - \Nb \bar{D}_{j}\Pb^{ij}
        \Big)\ ,
    \end{align}
which, when integrated, shows that not only canonical variables in the three-metric and scalar (density) field sector contribute to the surface term, but the extrinsic curvature sector contributes as well.
Before we go into details about similarities and differences between the constraints of this theory and the constraints of GR, we would like to mention that the constraint algebra of the quadratic curvature higher-derivative theory or gravity with matter discussed here should be the same as in GR, given by eqs.~\eqref{HEH_AlgHH}-\eqref{HEH_Algmm}.
This expectation is based on the fact that we are dealing with a reparametrization-invariant theory \cite[section 1.5]{Thiemann} and is further supported by the results of Deruelle et al.~\cite{DerrHD} who showed that a generic metric theory of gravity whose Lagrangian is an arbitrary function of the Riemann curvature tensor obeys the hypersurface foliation algebra, in accordance to its reparametrization invariance.
The addition of matter and the formulation in another set of \textit{canonical} variables (such as the unimodular-conformal variables in our case) should not change this outcome.
Therefore, we think that there is enough evidence to claim without proof that the constraints in eqs.~\eqref{HDall_Hofin} and \eqref{HDall_Hifin} are first class constraints and that they satisfy 
    \begin{align}
    \label{HDall_AlgHH}
        \left\lbrace \mathcal{H}_{\bot}^{\srm{ERW\chi}}[\varepsilon_1],\mathcal{H}_{\bot}^{\srm{ERW\chi}}[\varepsilon_2]\right\rbrace &=\mathcal{H}_{||}^{\srm{ERW\chi}}[\varepsilon_1 \del^{i}\varepsilon_2-\varepsilon_2 \del^{i}\varepsilon_1]\ ,\\[6pt]
    \label{HDall_AlgHm}
        \left\lbrace \mathcal{H}_{||}^{\srm{ERW\chi}}[\vec{\eta}],\mathcal{H}_{\bot}^{\srm{ERW\chi}}[\varepsilon]\right\rbrace &=\mathcal{H}_{\bot}^{\srm{ERW\chi}}[\mathcal{L}_{\vec{\eta}}\varepsilon]\ ,\\[6pt]
    \label{HDall_Algmm}
        \left\lbrace \mathcal{H}_{||}^{\srm{ERW\chi}}[\vec{\eta}_1],\mathcal{H}_{||}^{\srm{ERW\chi}}[\vec{\eta}_2]\right\rbrace & = \mathcal{H}_{||}^{\srm{ERW\chi}}[\mathcal{L}_{\vec{\eta}_{1}}\vec{\eta}_2]\ .
    \end{align}

Let us now take a closer look at the Hamiltonian and the momentum constraints.
The momentum constraint contains in the second line of eq.~\eqref{HDall_Hifin} contributions from the extrinsic curvature sector.
This is expected because the theory must ensure that not only the metric components but also the extrinsic curvature components are allowed to transform under spatial coordinate transformations, since the latter are treated as auxiliary \textit{independent} variables.
However, because of the constraints in eqs.~\eqref{HDall_KtlessC} and \eqref{HDall_KtC}, the spatial coordinate transformation of the extrinsic curvature components is \textit{induced} and not truly independent.
The aspect of $K_{ij}$'s independence is encoded through the phase space and dynamics.

On the other hand, there is no little difference between the Hamiltonian constraint in eq.~\eqref{HDall_Hofin} and its counterpart in GR given by eq.~\eqref{GRxi_H0} or in vacuum by eq.~\eqref{GRxi_H0v}.
The most striking departure from GR is that the presence of the EH term in a higher-derivative theory induces only a potential term and does not give rise to any kinetic term.
It shows that the Hamiltonian formulation of a theory of gravity based on higher-derivative extensions of the EH action does not ``add corrections'' to the Hamiltonian of pure GR but completely alters the theory, making lower-order contributions --- the EH action --- playing the role of a potential.
Related to this is the fact that the limit $\aw,\br\rightarrow 0$ in eq.~\eqref{HDall_Hofin} in hope of recovering the pure GR makes no meaning unless the momenta $\bPb$ and $\Pb$ are set to zero \textit{as constraints}.
This is in accordance with what we learned form the Hamiltonian formulation of a higher-derivative toy model in section~\ref{subs_Ham_osc}.
Furtehrmore, if one started with this higher-derivative theory without the EH term, i.e. $l=0$, the kinetic terms of this theory would not change and the matter part of the Hamiltonian constraint (the third line in eq.~\eqref{HDall_Hofin}) would remain unchanged.
Contrast this with the case of the non-minimally coupled scalar (density) field in GR: as can be seen from eq.~\eqref{GRxi_H0}, the kinetic terms of the scalar (density) field and the three-metric are entwined in a non-trivial way.
The reason for this simplification compared to GR is that $\bktb$ and $\kb$ are not conjugate momenta of the three-metric field components in a higher-derivative theory and thus the second term in the third line of eq.~\eqref{HDall_Hofin} does not represent mixing between $p_{a}$ and $p_{\chi}$, as is the case in the kinetic term of GR in eq.~\eqref{GRxi_H0}.
This term is one of the signatures that the conformal symmetry is broken, since it contains $\kb$; we shall revisit the importance of this term in the next section.

\subsection{DeWitt supermetric on the extended superspace}
\label{subs_HDall_DeWittMet}

Let us now inspect the gravitational kinetic term in eq.~\eqref{HDall_Hofin} itself. 
It consists of a conformally invariant part $\bPb\cdot\bPb$ from the Weyl-tensor term in the action and the conformally non-invariant part $\Pb^2$ --- the scale part --- arising from the conformally non-invariant $R^2$ term in the action.
The signs in front of these two terms are opposite, but $\aw$ and $\br$ might have negative values in general, since these are unknown couplings.
Yet, as mentioned in the beginning of this chapter, we choose these couplings to be strictly positive.
Nevertheless, the values of these couplings determine ``the relative strength'' of the negative-definite and positive-definite terms and it is interesting to draw an analogy with the DeWitt supermetric in GR.
Namely, the $\Pb^2$ term is analogous to $p_{a}^2$ term in eq.~\eqref{GRxi_H0}, while the $\bPb\cdot\bPb$ term is analogous $a^{-2}\bpb\cdot\bpb$ term in eq.~\eqref{GRxi_H0}.
The former could be called \textit{the expansion-like} direction and the latter could be called \textit{the shear-like} direction in the extended superspace.
The big difference is that these expansion density and shear density kinetic terms have \textit{different} coupling constants, whereas in GR the scale and the shape kinetic terms come with the same coupling constants.
Because of this it is more appropriate to draw analogy with eq.~\eqref{HEH_DWmetB}, i.e. with a generalized DeWitt supermetric whose parameter $\zeta$ is now a function of $\aw,\br$ in a fixed dimension of three.
To see this clearly, we state here the form of the DeWitt supermetric and its inverse in original variables, which was derived in \cite[eqs. (4.76) and (4.77)]{Kluson2014}, but which we state here based on eqs.~\eqref{HEH_DWmetB} and \eqref{HEH_DWmetBin}, with $\zeta = (3\aw + \br)/9$,
    \begin{align}
    \label{HDall_DeWittMet}
        G^{ikjl}_{\aw,\br} & = 
        \frac{a^3}{2}\left(\aw h^{ik}h^{jl} + \aw h^{il}h^{jk} - \frac{3\aw + \br}{9} h^{ij}h^{kl}\right)\ ,\\[12pt]
    \label{HDall_DeWittMetInv}
        \mirl{G}_{ikjl}^{\aw,\br} & = 
        \frac{1}{2\aw a^3}\left(h_{ik}h_{jl} + h_{il}h_{jk} - \frac{3\aw + \br}{3\br} h_{ij}h_{kl}\right)\ ,
    \end{align}
where we have adopted our notation.
The kinetic term that appears in \cite[eq. (4.81)]{Kluson2014} has the following form
    \begin{equation}
    \label{HDall_KinDW}
        - \mirl{G}_{ikjl}^{\aw,\br} P^{ij}P^{kl}\ ,
    \end{equation}
where $P^{ij}$ is the notation of \cite{Kluson2014} for the momentum conjugate to $K_{ij}$.
Terms proportional to $\aw$ come from the $C^2$ term and terms proportional to $\br$ come from the $R^2$ term.
Equation \eqref{HDall_DeWittMetInv} was derived assuming $\aw \neq 0$ and $\br \neq 0$.
The case $\br = 0$ achieves in eq.~\eqref{HDall_DeWittMet} the elimination of trace because one is then left with $\aw/3$.
This case is problematic for the definition of the inverse supermetric for the same reason that $\zeta_c = 1/D$ is problematic for the definition of the DeWitt supermetric in eq.~\eqref{HEH_DWmetBin}.
Namely, we see that $\zeta = (3\aw + \br)/9 = \aw/3$ for $\br = 0$ which is just $\zeta_c = 1/3$ up to a redefinition by a coupling constant. 
Therefore, by this reasonining, we see that in a theory with the Weyl-tensor term alone in the gravitational sector would prevent one from defining the inverse of the DeWitt metric and thus would suggest that the velocities $\bktb$ are not invertible.
However, we have just derived these velocities in eq.~\eqref{HDall_Kvel} so it should be possible to reconcile these apparently contradicting results.
Furthermore, the authors of \cite{Kluson2014} have claimed (without explanation) in equations (4.9) and (4.10) in their section 4.1, that the inverse of the traceless DeWitt supermetric \textit{is} well-defined,
    \begin{align}
    \label{HDall_DeWittproj}
         \mirl{G}_{akbl}^{\aw,0}G^{ikjl}_{\aw,0} 
        = \left( h^{i(k}h^{l)j} - \frac{1}{3} h^{ij}h^{kl}\right)\left(h_{k(a}h_{b)l} - \frac{1}{3} h_{ij}h_{kl}\right)
        = \mathbb{1}_{(ab)}^{{\srm T}ij}\ ,
    \end{align}
but note that on the right-hand side one does not have an identity rank-2 tensor, yet its \textit{traceless} version.
Similarly, in present author's master thesis \cite{MSc} the same conclusion was reached.
How can one make sense of these apparently contradicting results?

The proper way of interpreting the DeWitt supermetric in the case $\br = 0$ is to say that such DeWitt supermetric lives in a space of \textit{traceless} rank-2 symmetric tensors.
How does one reduce the space of all symmetric rank-2 tensors to the space of \textit{traceless} symmetric rank-2 tensors? --- with two steps: \textbf{one}, by interpreting $\mirl{G}_{akbl}^{\aw,0}$ as the \textit{projector}, since eq.~\eqref{HDall_DeWittproj} can be interpreted as the \textit{idempotency relation};
this is possible because --- unlike in the case of the DeWitt supermetric in GR, cf. eq.~\eqref{HEH_DeWittNotInv} --- here lowering the indices of $G^{ikjl}_{\aw,0}$ by $h_{ij}$ \textit{does} give the ``inverse'' $\mirl{G}_{akbl}^{\aw,0}$;
\textbf{two}, by noticing that $\mathbb{1}_{(ab)}^{{\srm T}ij}$ must be the unit element in the space of all \textit{traceless} symmetric rank-2 tensors.
The situation can be compared to that of the $3+1$ decomposition of the metric, in which one defines the spatial metric from the projector onto the three-hypersurface, cf. eqs.~\eqref{4met31} and \eqref{31proj}.
This becomes much more clear if one uses unimodular-conformal decomposition and refers to section \ref{subs_DeWittmet}. 
Namely, recall from eq.~\eqref{DeWittMet_lineel}, which defines the line element between two points in the superspace of GR, that one loses the scale-like direction for the critical value of $\zeta_c = 1/3$ in three dimensions and the six-dimensional supermetric has a singularity at that point. 
But one could simply interpret this as a \textit{restriction to the shape-like superspace} as the space of all unimodular metrics which are positive definite and five-dimensional.
In a similar way, one can interpret $\mirl{G}_{akbl}^{\aw,0}$ as the \textit{projector onto the five dimensional sub-superspace of all shear densities $\bktb$}.
The direction of the expansion density $\kb$ in the superspace would be analogous to the direction of the scale density in the superspace of GR.
Then one might imagine investigations of the geometry of the \textit{extended} superspace, which was discussed in \cite{MSc}, in a similar way as was done in \cite{Giulini} and mentioned in the end of section \ref{subs_DeWittmet}.
We shall not go into such discussions here; we only want to emphasize that the dynamics in this extended superspace that appears in higher-derivative theories bears some similarities with the superspace of GR but also brings novelties that so far do not seem to have been explored.
This conclusion is of relevance mainly for the quantum gravity context, but not only in canonical approach such as geometrodynamics:
it could be of importance to keep this in mind even if one would like to study the non-perturbative behavior of higher-derivative theories of gravity in the context of the program of asymptotic safety for gravity \cite{Bened}.
There, the couplings $\aw$ and $\br$ are redefined to depend on the energy scale and could affect the signature of the DeWitt supermetric, which could in turn dictate which degrees of freedom introduced by the higher-derivative terms appear at high energies.

\subsection{The significance of terms linear in momenta}
\label{subs_LinTerms}

Recall that a higher-derivative theory suffers from instabilities. In the Hamiltonian formulation, these instabilities manifest themselves as terms linear in momenta in the Hamiltonian constraint.

In the first line of eq.~\eqref{HDall_Hofin}, there are four terms linear in momenta that can be divided in two groups. One consists of the first two terms and another consists of the other two terms.
The first group follows simply because of the non-trivial form of the higher-derivative terms in the action --- these are analogues of the term proportional to $f$ in the Hamiltonian of our toy model given by eq.~\eqref{Ham_osc_Ham}.
The second group is the one which is responsible for the instabilities in a generic case. Those terms are the analogues of the third term in eq.~\eqref{Ham_osc_Ham}. 
In our toy model these terms can go to arbitrarily negative values and thus drive the energies to its negative values without bounds.
However, in geometrodynamics of higher-derivative theories the Hamiltonian constraint vanishes at each moment in time and because of that whichever term evolves towards negative energies, the rest of the terms necessarily ``keep the balance'' by countering with positive values.
The situation become more clear if one demands that the matter part of the Hamiltonian $\Ho{\chi}$ is positive (corresponding to the positive energy density). 
If this demand is taken seriously then the rest of the terms must add up to its negative, no more and no less.
This is of course, very loose argumentation, but we do not intend to go deeper into it because the instabilities at the classical level are of no importance to us, as we interpret the higher-derivative theory in the quantum context only.
In the quantum context, however, one must be more careful, since the problem of instability might reflect badly on the nature of the quantum state and give rise to a negative norm, thus forcing one to integrate away the additional degrees of freedom \cite{HH}. 
However, we are not aware of such a discussion in the context of constrained, reparametrization invariant theories of gravity, where the Hamiltonian is constrained to vanish.
The vanishing of the Hamiltonian constraint might have different consequences to the notion of the norm of a quantum state in a higher-derivative theory of gravity compared to quantum theory based on classical models in which no such constraints exist.

The linear terms arise due to introduction of the additional degrees of freedom, which can be seen by inspecting the constrained Lagrangian in eq.~\eqref{HDall_Lagconst}.
One might be tempted to say that $p_{a}$ and $\pb^{ij}$ are arbitrary since the Lagrange multipliers $\bar{\lambda}$ and $\bar{\lambda}^{ij\srm{T}}$ are arbitrary.
But it turns out --- as shown by Kluso\u n et al. \cite{Kluson2014} --- that one ends up with equations of motion for $p_{a}$ and $\pb^{ij}$. 
It is not clear from their result that one would end up with the equations of motion for the ADM momentum in the case of $\aw,\br\rightarrow 0$, which is expected because such limit is impossible without further restrictions.
Since we are not interested in the equations of motion, we do not pursue the possibility to derive the equations of motion for the ADM variables from the higher-derivative theory of gravity. 
However, we would like to point out that in this thesis we are ultimately seeking a way of deriving the Einstein equations from a semiclassical approximation in a quantized theory, in which case the linear terms play crucial role and therefore are not to be dismissed or sought to be eliminated for any reason, as we shall see in the next chapter.

\section{Hamiltonian formulation of Weyl-Einstein and Weyl-tensor theory}
\label{sec_HWE}

An important special case of the theory covered in the previous section is the case of Weyl-tensor (W) gravity $(\br = l^2 = 0)$ and Weyl-Einstein (WE) gravity $(\br = 0)$, both supplemented by a non-minimaly coupled scalar field.
Compared to the general theory discussed in the previous section, the novelty about the W and the WE theory is that two additional constraints appear due to the absence of the velocities $\dot{\kb}$ from the Lagrangian.
As emphasized in the Introduction and throughout the present chapter, we do not take W and WE theories seriously as classical theories.
We are merely interested in the possibility of exploring conformal symmetry in quantum gravity and W and WE theories are suitable for this.
Moreover, on the example of these theories one can motivate the $3+1$ version of the generator of conformal transformations, which we introduced in the full $d$-dimensional spacetime in chapter \ref{ch:defcf}.
Most of the material from this section is a central topic of \cite{KN17} --- which represent a significant improvement compared to \cite{MSc} --- but contains a few minor corrections which were missed there.

\subsection[Weyl-Einstein theory, Weyl-tensor theory \texorpdfstring{\\}{} and conformal symmetry]{Weyl-Einstein theory, Weyl-tensor theory and conformal symmetry}
\label{HWE_W}

We would first like to understand the conformal properties of the Weyl-Einstein gravity with non-minimally coupled scalar field from a convariant perspective.
The most important property of the \textit{pure} Weyl-tensor theory coupled with a non-minimally coupled scalar field is that the trace of the equations of motion demands that the trace of the energy-momentum tensor of the matter (in our case the non-minimally coupled scalar field) vanishes.
If the EH term is added to the theory, the trace of the equations of motion is the same as the trace of the Einstein equations.
From the variation of the covariantly written Weyl-Einstein action with respect to the four-dimensional scale density $A$ we have (cf. appendix~\ref{varprincdec} and eq.~\eqref{ttREE} in there)
    \begin{equation}
    \label{HWchi_tracezero}
        - R + 4\bar{\Lambda} = l^2\hbar T = l^2\hbar A^{-4}\mathcal{T}
    \end{equation}
where $\mathcal{T}$ is given by eq.~\eqref{traceT}.
This equation is the same as in GR because the Bach tensor $B_{\mu\nu}$ is identically traceless and contains no scale density $A$.
From chapter~\ref{ch:defcf} we have learned that vanishing of $\mathcal{T}$ has something to do with the matter action being conformally invariant and that some caution must be taken with such a claim.
The usual conclusion in Weyl-tensor theory of classical gravity, see e.g.~\cite{Mann1}, is that the energy-momentum tensor must be \textit{identically} traceless, i.e. that only conformal matter can be allowed in the Weyl-tensor theory, if the latter is to be interpreted as a classical theory.
We think that there are some issues with this conclusion and we shall explain below why.
We shall find some evidence for challenging this conclusion in the Hamiltonian formulation of the theory.

Suppose now that we are dealing with a pure Weyl-tensor theory so $R=0, \bar{\Lambda} = 0$.
In chapter \ref{ch:defcf} we have explained the difference between $\mathcal{T} = 0$ holding for an arbitrary variation $\delta A$ and $\mathcal{T} = 0$ holding for an arbitrary scale density itself.
We think that interpreting eq.~\eqref{HWchi_tracezero} correctly in terms of this difference reveals a problem with claims of \cite{Mann1}.
Namely, condition in eq.~\eqref{HWchi_tracezero} is \textit{on-shell}, i.e. it determines the nature of the solutions and holds for arbitrary \textit{variations} $\delta A$.
According to our discussion in chapter \ref{ch:defcf}, this would correspond to eq.~\eqref{confinvS1}, not  eq.~\eqref{confinvS2}.
Therefore, the condition for conformal invariance in eq.~\eqref{confinvS2} \textit{is not implied by the equation of motion} \eqref{HWchi_tracezero}.
For the same reason one does not say that solutions to the Einstein equations having a vanishing Ricci-scalar, $R = 0$, imply conformal invariance of the EH action; 
or equivalently, one never encounters a demand ``conformal matter is not allowed in GR because it implies identical vanishing of the Ricci scalar, which is not possible''\footnote{A more careful investigation of such a statement could actually lead to some interesting implications for the meaning of interaction between matter and spacetime which deserves further inquiry.
For example, one may start with a hypothesis \textit{the scale density and the notion of length are impossible to define if the only matter considered is the conformal matter}, relying on a postulate \textit{points of spacetime have no meaning without interacting matter}.}. 
However, $R = 0$ in Einstein equations is usually interpreted as a demand that the matter must be conformally invariant and therefore independent of $A$.
Hence, one again makes an ambiguity between the conditions in eqs.~\eqref{confinvS1} and \eqref{confinvS2}. 
But if one understands that there is no ambiguity between the conditions in eq.~\eqref{confinvS1} (which holds on-shell) and eq.~\eqref{confinvS2} (which holds off-shell) themselves, then one has to accept that $R = 0$ (which follows as an on-shell condition) does not exclusively necessitate conformal matter in Einstein equations but demands that $\mathcal{T} = 0$ exists as a condition (an on-shell condition!) between the scale density $A$ and the rest of the variables for \textit{any kind} of matter.
In some cases, such as non-minimally but not conformally coupled scalar field, $\mathcal{T} = 0$ actually has the meaning of an equation of motion for the scale density $A$, as can be seen from~eq.~\eqref{traceT}, which contains a second time derivative of $A$.
That means that even though the scale density is absent from the geometry side of the equations of motion the scale density does become dynamical by the non-minimally (but not conformally) coupled scalar (density) field.
The consequence of this kind of reasoning is that non-conformal matter may be allowed to be coupled to the conformally invariant gravity sector, in this case determined by the Weyl-tensor term.
This is in contrast to the usual conclusions, which say that only conformal matter can be allowed in Weyl-tensor theory of classical gravity.
(The conclusion is generalized for the case of Weyl-Einstein theory, in which eq.~\eqref{HWchi_tracezero} holds on-shell as well.)
The point is that one may or may not \textit{require} conformal invariance of the total action based on the Weyl-tensor term; only in the case where one does require it should one restrict the form of matter in Weyl-tensor theory to only conformal one.

We think that our reasoning can be justified by the evidence arising from the Hamiltonian formulation of the theory: as a consequence of constraint analysis, an equation arises which determines the second time derivative of the scale density $a$, cf. eq.~\eqref{HWchi_KdotLambda}.
This result was not realized at the time of writing of \cite{MSc} and \cite{KN17}.

\subsection{Hamiltonian formulation}
\label{subs_HWEchiH}

The Hamiltonian formulation shall be based on the following Lagrangian,
    \begin{align}
    \label{HWchi_Lag}
        S^{\srm{WE\chi}} = \inttx  \mathcal{L}^{\srm{WE\chi}} \equiv
        \inttx \left(\mathcal{L}^{\srm W}
        + \mathcal{L}^{\srm E}
        + \mathcal{L}^{\srm \chi}\right)\ .
    \end{align}
All conjugate momenta are the same as eqs.~\eqref{HDall_pchi}-\eqref{HDall_pL}, except the momentum $\Pb$, conjugate to the expansion density $\kb$. 
The theory is characterized by the vanishing of this momentum,
    \begin{equation}
        \label{HWchi_Pk}
        \Pb = \dd{\mathcal{L}^{\srm{WE\chi}}_{c}}{\dot{\kb}} \deq 0\ ,
    \end{equation}
which means that $\dot{\kb}$ cannot be determined from it and one has to consider eq.~\eqref{HWchi_Pk} as a constraint.
Note that $\mathcal{L}^{\srm \chi}$ depends on $\kb$, except in the conformally coupled case $\xi_{c}=0$.
In deriving the Hamiltonian constraint from the total Hamiltonian one needs to take some care.
The total Hamiltonian can \textit{preliminary} be written as
    \begin{align}
    \label{HWchi_HamWchi1}
        H^{\srm{WE\chi}} = 
        & \intx\bigg\lbrace \Nb
        \bigg[
        - \frac{1}{2\aw\hbar}\bPb\cdot\bPb
        + \bm{\mathcal{D}}^2_{\srm W}\cdot\bPb
        + a\kb p_{a}
        + 2\bktb\cdot\bpb
        - \aw\hbar\bar{\mathbf{C}}^{\srm B}\cdot\bar{\mathbf{C}}^{\srm B}\nld
        &\quad
        - \frac{l^2 \hbar a^2}{2}\Big(a^2\left(\,^{\ssst (3)}\! R 
        - 2\bar{\Lambda}\right) 
        + \bktb\cdot\bktb 
        - 6\kb^2\Big)
        + \frac{1}{2}p_{\chi}^2
        - 6\xi_{c}\kb\chi p_{\chi}
        + \frac{1}{2}V^{\chi}
        \bigg]\nld
        &\quad
        + N^{i}\Hi{WE\chi}
        + \lambda_{\ssst\Nb}p_{\ssst\Nb}
        + \lambda^{i}p_{i}
        + \lambda_{\ssst\kb}\Pb\bigg\rbrace
        + H^{\srm{WE}\chi}_{\ssst surf}\ .
    \end{align}
First of all, note the new term $\lambda_{\ssst\kb}\Pb$. 
It appears because $\dot{\kb}$ cannot be inverted from eq.~\eqref{HWchi_Pk}, so it is then rewritten as $\lambda_{\ssst\kb}$, a Lagrange multiplier.
The term multiplying $\Nb$ is \textit{not} the Hamiltonian constraint \cite{MSc}, in contrast to the claim of \cite{Kluson2014}.
The reason is that within this term another constraint hides.
It appears as a demand that $\Pb$ is preserved in time,
    \begin{align}
    \label{HWchi_Pdot}
        \dot{\Pb} = \PB{\Pb}{H^{\srm{WE\chi}}} = - \dd{H^{\srm{WE\chi}}}{\kb}
        & = - \Nb\left(a p_{a} 
        - 6\xi_{c}\chi p_{\chi}
        + 36\xi\xi_{c}\kb\chi^2
        + 6l^2\hbar a^2\kb\right)\nld
        & = - \Nb\left(a p_{a} 
        - 6\xi_{c}\chi p_{\chi}
        + \frac{6l^2\hbar a^2}{\tilde{l}^2}\kb
        \right)
        \deq 0\ ,
    \end{align}
where $\tilde{l}$ was defined in eq.~\eqref{GR_kt}.
The last term in the above equation results from the EH term and from the only $\kb$-dependent term in the potential $V^{\chi}$ given by eq.~\eqref{Vchi_def}.
Note that this is a new secondary constraint. 
It is usually called \textit{the conformal constraint} \cite{Blw, ILP, Kaku1982, Kluson2014, MSc}, since it is claimed that it generates conformal transformations \cite{Blw, Kluson2014}.
We shall keep the name, but we must point out that it is incorrect to claim that it is a generator of conformal transformations because it generates only a part of conformal transformation \cite{ILP, KN17, MSc}; we shall give more detail on this in section \ref{sec_Gencf31}.
The conformal constraint\footnote{We use an overbar to distinguish our result from the result of the previous works for the conformal constraint, since it does have a different form in unimodular-conformal variables.}
    \begin{equation}
    \label{HWchi_Q}
        \Q{WE\chi} := a p_{a} 
        - 6\xi_{c}\chi p_{\chi}
        + \frac{6l^2\hbar a^2}{\tilde{l}^2}\kb \deq 0
    \end{equation}
and $\Pb$ are of \textit{second-class} because they do not commute;
using the smeared version of these constraints (cf. eq.~\eqref{HEH_smearC}), one obtains
    \begin{equation}
    \label{HWchi_QP}
        \PB{\Pb[\epsilon]}{\Q{WE\chi}[\omega]} = -\frac{6l^2\hbar }{\tilde{l}^2}\intx\,\epsilon\omega a^2\ .
    \end{equation}
One further has to demand that $\Q{WE\chi}$ is preserved in time, but before we look for $\dot{\bar{\mathcal{Q}}}^{\srm WE\chi}$, we can see that eq.~\eqref{HWchi_Q} can be found in the term multiplied by $\Nb$ in eq.~\eqref{HWchi_HamWchi1}, which is just the condition for preservation of the first constraint in eq.~\eqref{HDall_pNb},
    \begin{align}
    \label{HWchi_pNdot}
        \dot{p}_{\ssst \Nb} & = \PB{p_{\ssst \Nb}}{H^{\srm{WE\chi}}}\nld
        & = - \Nb\bigg[ - \frac{1}{2\aw\hbar}\bPb\cdot\bPb
        + \bm{\mathcal{D}}^2_{\srm W}\cdot\bPb
        + 2\bktb\cdot\bpb
        - \aw\hbar\bar{\mathbf{C}}^{\srm B}\cdot\bar{\mathbf{C}}^{\srm B}\nld
        &\quad
        + \frac{1}{2}p_{\chi}^2
        + \frac{1}{2}\tilde{V}^{\chi}
        - \frac{l^2 \hbar a^2}{2}\Big(a^2\left(\,^{\ssst (3)}\! R 
        - 2\bar{\Lambda}\right) 
        + \bktb\cdot\bktb 
        + 6\kb^2\Big)\bigg]\nld
        &\quad
        - \Nb\kb\Big( a p_{a} 
        - 6\xi_{c}\chi p_{\chi}
        + \frac{6l^2\hbar a^2}{\tilde{l}^2}\kb\Big)
         \deq 0\ .
    \end{align}
In the above equation the term $2\cdot 36\kb^2\chi^2$ has been subtracted from the potential $V_{\chi}$ and the term $6l^2\hbar a^2\kb^2 $ has been subtracted from the last term in the second line in order to form the conformal constraint in the parentheses $\kb\left(...\right)$, which vanishes upon releasing the delayed equality.
What remains is
    \begin{equation}
        \tilde{V}^{\chi} := V^{\chi} - 2 \cdot 36\kb^2\chi^2\ ,
    \end{equation}
which effectively means that $\tilde{V}^{\chi}$ is equal to $V^{\chi}$ with an opposite sign in front of the $\kb^2\chi^2$ term.
Also note the sign change in the last term in the third line of eq.~\eqref{HWchi_pNdot}.
Therefore, one only needs to demand that terms in $\left[...\right]$ in the first line in eq.~\eqref{HWchi_pNdot} vanish with delayed equality; \textit{this} is the Hamiltonian constraint in unimodular-conformal variables,
    \begin{align}
    \label{HWchi_HamCnst}
        \Ho{WE\chi} & = - \frac{1}{2\aw\hbar}\bPb\cdot\bPb
        + \bm{\mathcal{D}}^2_{\srm W}\cdot\bPb
        + 2\bktb\cdot\bpb
        - \aw\hbar\bar{\mathbf{C}}^{\srm B}\cdot\bar{\mathbf{C}}^{\srm B}\nld
        &\quad
        + \frac{1}{2}p_{\chi}^2
        + \frac{1}{2}\tilde{V}^{\chi}
        - \frac{l^2 \hbar a^2}{2}\Big(a^2\left(\,^{\ssst (3)}\! R 
        - 2\bar{\Lambda}\right) 
        + \bktb\cdot\bktb 
        + 6\kb^2\Big)
        \deq 0\ .
    \end{align}
Note that this equation does not correspond to the usual definition of the Hamiltonian constraint, which would be given by eq.~\eqref{HWchi_pNdot} and which is usually found (in original variables) in the literature, e.g. \cite{ILP,Kluson2014}.
It should be kept in mind then that the matter terms in the second line in eq.~\eqref{HWchi_HamCnst} do not correspond to the matter Hamiltonian.
This is because we choose to demand the delayed vanishing of only those terms which do not already vanish according to the other constraints in the theory.
That this makes sense, we draw attention to the fact that the authors of \cite{Kluson2014} have \textit{added} the conformal constraint\footnote{Even though we refer to the particular case of the vacuum WE theory in their work, the properties of the conformal constraint are the same in the general non-vacuum case since conformal symmetry is in that case broken as well, see further below the case of vacuum WE theory.}
to the total Hamiltonian with a new Lagrange multiplier and found that this Lagrange multiplier vanishes, thus eliminating the constraint they had just added to the total Hamiltonian.
They did not notice that the conformal constraint is already present within what they derived to be the Hamiltonian constraint.
Based on our result in eq.~\eqref{HWchi_pNdot} and claim that the conformal constraint is \textit{already} in the total Hamiltonian, hidden in what \cite{Kluson2014} call the Hamiltonian constraint, it is not surprising that the Hamiltonian formulation of the theory did not let the authors of \cite{Kluson2014} add the additional conformal constraint.
This situation is similar to the case of a massive vector field whose constraint analysis we presented in appendix \ref{app_const}.
As we shall see soon below, there are no further constraints and this will allow us to write the total Hamiltonian in the following form,
    \begin{equation}
    \label{HWchi_Htot}
        H^{\srm{WE\chi}} = 
        \intx\left\lbrace \Nb\Ho{WE\chi}
        + N^{i}\Hi{WE\chi}
        + \left(\Nb\kb\right)\bar{\mathcal{Q}}^{\srm{WE\chi}}
        + \lambda_{\ssst\Nb}p_{\ssst\Nb}
        + \lambda^{i}p_{i}
        + \lambda_{\kb}\Pb\right\rbrace
        + H^{\srm{WE}\chi}_{\ssst surf}\ ,
    \end{equation}
where
    \begin{align}
    \label{HWchi_Hicns}
        \Hi{WE\chi} & = 
        - 2\bar{D}_{j}\left(\hb_{ik}\pb^{kj}\right)
        -\frac{1}{3}D_{i}\left(a\,p_{a}\right)
        - \frac{1}{3}\left(\chi\del_{i}p_{\chi}
        - 2\del_{i}\chi \, p_{\chi}\right)\nld
        & \quad
        + \Pb^{ij} \bar{D}_{k}\ktb_{ij} 
        - 2 \bar{D}_{i}\left(\ktb_{jk} \Pb^{ij} \right)
        + \frac{1}{3}\del_{k}\left(\ktb_{ij}\Pb^{ij}\right)
    \end{align}
is the momentum constraint.
Note that the momentum constraint does not contain $\kb$ and $\Pb$ terms, in contrast to eq.~\eqref{HDall_Hifin}.
Equation \eqref{HWchi_Htot} is written as a sum of pairs of primary constraints (last three terms) plus the corresponding secondary constraints (first three terms).

Now we come back to the demand that eq.~\eqref{HWchi_QP} is preserved in time.
It is enough to assume for a moment that we are dealing with spatially homogeneous variables.
Then the Poisson bracket of $\bar{\mathcal{Q}}^{\srm{WE}\chi}$ with the total Hamiltonian reduces to
    \begin{align}
    \label{HWchi_Qdot}
        \dot{\bar{\mathcal{Q}}}^{\srm{WE}\chi} & = \PB{\bar{\mathcal{Q}}^{\srm{WE}\chi}}{H^{\srm{WE\chi}}} = \PB{\bar{\mathcal{Q}}^{\srm WE\chi}}{\Ho{WE\chi}[\Nb]}\nld
        & = \PB{a p_{a}}{\Ho{WE\chi}[\Nb]}
        - 6\xi_{c}\chi\PB{p_{\chi}}{\Ho{WE\chi}[\Nb]}
        - 6\xi_{c}\PB{\chi}{\Ho{WE\chi}[\Nb]}p_{\chi}\nld
        & \quad 
        + \frac{6l^2\hbar a^2}{\tilde{l}^2}\PB{\kb}{\lambda_{\ssst\kb}\Pb}
        + 72\xi\xi_{c}\kb\chi\PB{\chi}{\Ho{WE\chi}[\Nb]}\nld
        & = 
        l^2 \hbar \Nb a^2\Big(a^2\,^{\ssst (3)}\! R 
        + \bktb\cdot\bktb 
        + 6\kb^2
        - 4a^2\bar{\Lambda}\Big)\nld
        &\quad 
        + 12\xi_{c}\Nb\left(\frac{1}{2}p_{\chi}^2
        - 6\xi\chi\kb p_{\chi}
        + \frac{1}{2}\tilde{V}^{\chi}\right)
        + \frac{6l^2\hbar a^2}{\tilde{l}^2}\lambda_{\ssst\kb} \deq 0\ .
    \end{align}
The first line in the above equation is due to $\bar{\mathcal{Q}}^{\srm{WE}\chi}$ commuting with itself.
The first term in the second line produces the first term in the last equality --- arising from the derivative of the EH potential term with respect to $a$, where $a^2\,^{\ssst (3)}\! R = \,^{\ssst (3)}\! \bar{R} $ for homogeneous case and no additional terms appear in the general case other than complete $a^2\,^{\ssst (3)}\! R$ term, up to surface terms which we disregarded in the calculation.
Note that the cosmological constant term contributes twice as much compared to other terms from the EH potential.
The same first term in the second line contributes with the inhomogeneous terms in the potential $\tilde{V}^{\chi}$ in the general case, up to surface terms.
The Poisson brackets in the second and third term in the second line evaluate to $-\Nb\tilde{V}^{\chi}/\chi$ and $-\Nb p_{\chi}$ in homogeneous case, or the same up to surface terms in the inhomogeneous case.
This condition determines Lagrange multiplier $\lambda_{\ssst\kb}$ which is, as we mentioned before, velocity $\dot{\kb}$,
    \begin{align}
    \label{HWchi_KdotLambda}
        \frac{6l^2\hbar a^2}{\tilde{l}^2}\frac{1}{\Nb}\lambda_{\ssst\kb} = \frac{6l^2\hbar a^2}{\tilde{l}^2}\frac{1}{\Nb}\dot{\kb}
        & = - l^2 \hbar a^2\Big(a^2\,^{\ssst (3)}\! R 
        + \bktb\cdot\bktb 
        + 6\kb^2
        - 4a^2\bar{\Lambda}\Big)\nld
        &\quad 
        - 12\xi_{c}\left(\frac{1}{2}p_{\chi}^2
        - 6\xi\chi\kb p_{\chi}
        + \frac{1}{2}\tilde{V}^{\chi}\right)\ .
    \end{align}
Now, from this equation one can see that $\dot{\kb}$ cannot be determined if
    \begin{equation}
        \frac{l^2}{\tilde{l}^2} = l^{2} + 6\xi\xi_{c}\frac{\chi^2}{\hbar a^2} = 0\ ,
    \end{equation}
which leads to four possible cases: 
    \begin{enumerate}
        \item Vacuum Weyl-tensor gravity: $l=0$ and $\chi = 0$ (which also implies $p_{\chi} = 0$)
        \item Weyl-tensor gravity with conformally coupled scalar field: $l=0$ and $\xi_{c} = 0$
        \item Weyl-tensor gravity with minimally coupled scalar field: $l=0$ and $\xi = 0$
        \item Weyl-Einstein gravity with non-minimally coupled scalar field $\frac{\chi^2}{a^2} = - \frac{l^2\hbar}{6\xi\xi_{c}}$, for some critical value of the ratio $\chi/a$ if $\xi$ is fixed.
    \end{enumerate}
Also note that eq.~\eqref{HWchi_QP} vanishes in these cases, since the $\kb$-dependent term in the conformal constraint in eq.~\eqref{HWchi_Q} disappears.
In each of these cases the constraint analysis must be repeated if one is to completely understand the details of their implications. 
We shall not do so here.
We shall only point out the differences compared to the key equations in the general case, because most of the derivations are the same.
To this purpose it is important and also interesting to ask, what is the meaning of eq.~\eqref{HWchi_KdotLambda}?
This equation has been derived in a vacuum WE theory by \cite{Kluson2014}, but they did not notice its importance nor have interpreted it, which we think is a crucial step in the light of our discussion in the previous subsection.
It has also been derived in \cite{KN17}, equation (106), but its meaning was not understood in there at the time.
Namely, the equivalence between Hamiltonian and Lagrangian formulation implies that they contain the same information in the equations they consist of.
This means that there has to be an equation in the covariant, Lagrangian formulation which corresponds to eq.~\eqref{HWchi_KdotLambda}.
The correct equation is the trace of the equations of motion, given by eq.~\eqref{HWchi_tracezero} in the previous subsection.
Hence, the trace of the covariant equations of motion in a theory based on the Weyl-tensor term emerges from Lagrange multiplier $\lambda_{\ssst\kb}$.
This is most easily seen in the case of vacuum WE gravity ($\chi = 0$), as we shall see further below.

We turn now to Dirac brackets of the theory.
Since the theory contains second-class constraints, one needs to substitute Poisson brackets with Dirac brackets after implementing the second-class constraints strongly (i.e. the delayed equality ``$\deq$'' is set to strong equality), if one would like to proceed to find equations of motion once the Lagrange multiplier has been determined.
Using Dirac brackets instead of Poisson brackets is equivalent to using Poisson brackets after the second-class constraints have been implemented.
Substituting the second-class constraints means eliminating the canonical pair of variables which one thought was arbitrary but which turned out that one can express them as a function of other canonical variables.
This pair $\kb,\Pb$, so one can expect non-trivial Dirac brackets of $\kb,\Pb$ with other canonical variables.
Dirac brackets in the presently discussed theory were derived in \cite{KN17} using the recipe from appendix \ref{app_dirb}.
But here we would like to show that one can derive Dirac brackets more intuitively, by directly translating the following sentence we stated above
    \begin{quote}
        \textit{Substituting the second-class constraints means eliminating the canonical pair related to an apparent arbitrary variable},
    \end{quote}
 in the sense of nomenclature introduced in appendix \ref{app_const}.
 Namely, the conformal constraint in eq.~\eqref{HWchi_Q} set strongly to zero means that $\kb$ is not an independent and true arbitrary variable but is given as a function of other canonical variables by
    \begin{equation}
    \label{HWchi_QderK}
        \kb = \frac{\tilde{l}^2}{6l^2\hbar a^2}\left( - a p_{a} + 6\xi_{c} \chi p_{\chi}\right)\ .
    \end{equation}
The primary constraint $\Pb\deq 0$ trivially becomes $\Pb = 0$.
Then one can implement this information by substituting the following Poisson brackets involving $\kb$ and $\Pb$ by Poisson brackets with eq.~\eqref{HWchi_QderK} implemented\footnote{In what follows we suppress the Dirac delta function and the explicit dependence on spatial coordinates},
    \begin{align}
    \label{HWchi_DBKP}
        \PB{\kb}{\Pb} \rightarrow \PB{\kb}{\Pb}_{\ssst D} & = \PB{\frac{\tilde{l}^2}{6l^2\hbar a^2}\left( - a p_{a} + 6\xi_{c} \chi p_{\chi}\right)}{0} = 0\ ,\\[12pt]
        \PB{\kb}{a} \rightarrow \PB{\kb}{a}_{\ssst D} & = \PB{\frac{\tilde{l}^2}{6l^2\hbar a^2}\left( - a p_{a} + 6\xi_{c} \chi p_{\chi}\right)}{a} \nld
        & = - \ddel{}{p_a}\frac{\tilde{l}^2}{6l^2\hbar a^2}\left( - a p_{a} + 6\xi_{c} \chi p_{\chi}\right)
        = \frac{\tilde{l}^2}{6l^2\hbar a}
        \ ,\\[12pt]
    \label{HWchi_DBKpa}
        \PB{\kb}{p_{a}} \rightarrow \PB{\kb}{p_{a}}_{\ssst D} & = \ddel{\kb}{a} = \frac{\tilde{l}^2}{6l^2\hbar a^2} p_{a} - \frac{2\tilde{l}^2\xi_{c}}{l^2\hbar a^3}\chi p_{\chi} \deq - \frac{\kb}{a} - \frac{\tilde{l}^2\xi_{c}}{l^2\hbar a^3}\chi p_{\chi}\ ,\\[12pt]
        \PB{\kb}{\chi} \rightarrow \PB{\kb}{\chi}_{\ssst D} & = - \ddel{\kb}{p_{
        \chi}} = - \frac{\tilde{l}^2\xi_{c}}{l^2\hbar a^2}\chi \ ,\\[12pt]
    \label{HWchi_DBKpchi}
        \PB{\kb}{p_{\chi}} \rightarrow \PB{\kb}{p_{\chi}}_{\ssst D} & = \ddel{\kb}{
        \chi} \deq \frac{\tilde{l}^2\xi_{c}}{l^2\hbar a^2}\left( p_{\chi} + 12\xi\chi \kb\right) \ ,
    \end{align}
which are then just the Dirac brackets.
Up to notation differences the Dirac brackets above are equal to the ones derived in \cite{KN17} in equations (154).
Note that in eqs.~\eqref{HWchi_DBKpa} and \eqref{HWchi_DBKpchi} we used the conformal constraint to introduce $\kb$, which is marked by using the delayed equality, but this is not necessary.
All other Dirac brackets are equal to the corresponding Poisson brackets.

It is interesting to observe the conformal constraint n eq.~\eqref{HWchi_Q} in a little bit more detail.
According to the interpretation from the constraint analysis perspective, this equation is a second-class constraint and says that one of the variables $p_{a},p_{\chi},a$ and $\kb$ is not an independent variable.
But there is another way of looking at this equation: as the definition of $p_{a}$. Namely, recall that $p_{a}$ is just a Lagrange multiplier $\bar{\lambda}$ --- there is no information about $\dot{a}$ that could be retrieved from it.
But we see that the conformal constraint has the role of determining this Lagrange multiplier.

For the simplest case of vacuum Weyl-tensor gravity (which we shall visit soon below in more detail), $\chi=0$ and $l=0$ in eq.~\eqref{HWchi_Q} gives trivially $\bar{\mathcal{Q}}^{\srm{W}} = a p_{a} = 0$.
This means that $a$ is arbitrary, which makes sense because this theory is conformally invariant.
Constraints $\bar{\mathcal{Q}}^{\srm{W}}=0$ and $\Pb=0$ are in this case of first class and there are trivially no further constraints.

If conformal matter $\xi_{c}=0$ is present in the Weyl-tensor gravity, from  eq.~\eqref{HWchi_Q} we again have $a p_{a} = 0$, meaning again that there is no scale in the theory and conformal symmetry holds in this case as well, with no further constraints.

However, if a minimally coupled scalar field is present in the Weyl-tensor gravity, even though eq.~\eqref{HWchi_QP} vanishes, there appear further constraints which could severely constrain the scalar field, but we do not calculate them here.
Note that in that case the trace of the energy-momentum tensor of the non-minimally coupled scalar is required to vanish (so we expect that the additional constraint shall eventually lead to the another equation that determines the Lagrange multiplier $\lambda_{\ssst \kb}$), thus putting a condition on the allowed solutions for the scalar field.

In the non-vacuum case of the WE theory we see that eq.~\eqref{HWchi_Q} defines $p_{a}$, since the conformal symmetry is broken and the scale density $a$ is not arbitrary anymore, becoming dynamical.
One could then look at eq.~\eqref{HWchi_Q} as the definition of momentum $p_{a}$.
The resulting equation is remarkably nothing other than the ADM momentum in non-vacuum GR, cf. eq.~\eqref{GRchi_pam}.
This outcome is independent on whether or not one has conformal coupling $\xi_{c}=0$.
This is a very interesting observation because one could imagine a theory in which pure Weyl-tensor gravity is valid at high energies, and then as the energies become lower (through the change of balance of the respective couplings) the EH term starts being important, breaks the conformal symmetry of the theory which generates the dynamical scale.
With 
Of course, we do not claim that this is necessarily so but we rather point out the ``big picture'' that the conformal constraint paints.

It is the conformal constraint that could play a crucial role in Hamilton-formulated theories in determining whether a dynamical scale could emerge from a broken conformal symmetry of a theory.
In relation to this, since the conformal constraint plays a crucial role in the definition of the generator of conformal transformations \ref{sec_Gencf31}, further studies of this generator could provide some novel tools for studying the generation of dynamical scale in conformally invariant theories, both classical and quantum.

\subsection{Vacuum Weyl-Einstein gravity}
\label{subs_WEvac}

In this case the Hamiltonian constraint is given by
    \begin{align}
    \label{HWchi_HamCnstWE}
        \Ho{WE} & = - \frac{1}{2\aw\hbar}\bPb\cdot\bPb
        + \bm{\mathcal{D}}^2_{\srm W}\cdot\bPb
        + 2\bktb\cdot\bpb
        - \aw\hbar\bar{\mathbf{C}}^{\srm B}\cdot\bar{\mathbf{C}}^{\srm B}\nld
        &\quad
        - \frac{l^2 \hbar a^2}{2}\Big(a^2\left(\,^{\ssst (3)}\! R 
        - 2\bar{\Lambda}\right) 
        + \bktb\cdot\bktb 
        + 6\kb^2\Big)
        \deq 0\ ,
    \end{align}
the momentum constraint is given by eq.~\eqref{HWchi_Hicns} with $\chi=0$.
The conformal constraint in eq.~\eqref{HWchi_Q} reduces to
    \begin{equation}
    \label{HWchi_QWE}
        \Q{WE} := a p_{a} 
        + 6l^2\hbar a^2\kb \deq 0\ ,
    \end{equation}
while the Poisson bracket stating that in vacuum WE theory conformal constraint and its primary ancestor are second-class constraints is given by eq.~\eqref{HWchi_QP} with $\tilde{l} = 1$, 
    \begin{equation}
    \label{HWchi_QPWE}
        \PB{\Pb[\epsilon]}{\Q{WE}[\omega]} = - 6 l^2\hbar \intx\,\epsilon\omega a^2\ .
    \end{equation}
Assuming spatial homogeneity of the theory, the equation for Lagrange multiplier $\lambda_{\ssst \kb}$ derived in eq.~\eqref{HWchi_KdotLambda} reduces to
    \begin{align}
    \label{HWchi_KdotLambdaWE}
        \frac{6}{\Nb}\dot{\kb}
        & = - \Big(a^2\,^{\ssst (3)}\! R 
        + \bktb\cdot\bktb 
        + 6\kb^2\Big)
        + 4a^2\bar{\Lambda}\ ,
    \end{align}
which can be recognized as the four-dimensional Ricci scalar in unimodular-conformal $3+1$ variables (cf. eqs.~\eqref{LieKdecn} and \eqref{Rdec31umod} without spatial derivatives) plus the cosmological constant,
    \begin{equation}
        - R + 4\bar{\Lambda} = 0\ ,
    \end{equation}
which is precisely the covariant equation of motion for the trace density stated in the previous subsection given by eq.~\eqref{HWchi_tracezero}.
This conclusion should hold even if spatial homogeneity requirement is relaxed, but this we claim without pursuing a proof.

Dirac brackets in the vacuum WE theory can be derived from the general case given by eqs.~\eqref{HWchi_DBKP}-\eqref{HWchi_DBKpa} by setting $\chi = 0$, $\tilde{l} = 1$, which results in the following,
    \begin{align}
    \label{HWchi_WEDBKP}
        \PB{\kb}{\Pb} \rightarrow \PB{\kb}{\Pb}_{\ssst D} & = - \PB{\frac{1}{6l^2\hbar a} p_{a}}{0} = 0\ ,\\[12pt]
        \PB{\kb}{a} \rightarrow \PB{\kb}{a}_{\ssst D} & = - \PB{\frac{1}{6l^2\hbar a} p_{a}}{a} = \frac{1}{6l^2\hbar a}
        \ ,\\[12pt]
    \label{HWchi_WEDBKpa}
        \PB{\kb}{p_{a}} \rightarrow \PB{\kb}{p_{a}}_{\ssst D} & = \frac{p_{a}}{6l^2\hbar a^2} \deq - \frac{\kb}{a}  \ ,
    \end{align}
and these Dirac brackets are equal to the ones derived in \cite[eq. (118)]{KN17} using the recipe from appendix \ref{app_dirb}.

As mentioned in the previously discussed general theory, in the vacuum WE theory one can interpret the conformal constraint in eq.~\eqref{HWchi_QWE} as the defining equation of the ADM scale momentum.
Thus we see that the term linear in momentum $p_{a}$ has indeed a significant role, as anticipated in section \ref{subs_LinTerms}.
Nothing less is to be expected in the quantum version of the theory.

\subsection{Weyl-tensor gravity with matter}
\label{subs_Wmat}

If the EH term is absent ($l = 0$) but matter is present we have the Weyl-tensor gravity with matter.
The Hamiltonian constraint is in this case given by
    \begin{align}
    \label{HWchi_WHamC}
        \Ho{W\chi} & = - \frac{1}{2\aw\hbar}\bPb\cdot\bPb
        + \bm{\mathcal{D}}^2_{\srm W}\cdot\bPb
        + 2\bktb\cdot\bpb
        - \aw\hbar\bar{\mathbf{C}}^{\srm B}\cdot\bar{\mathbf{C}}^{\srm B}
        + \frac{1}{2}p_{\chi}^2
        + \frac{1}{2}\tilde{V}^{\chi}
        \deq 0\ .
    \end{align}
The momentum constraint is the same as eq.~\eqref{HWchi_Hicns}. The conformal constraint is obtained by setting $l^2 /\tilde{l}^2 = 6\xi\xi_{c}\chi^2\hbar a^2$ in eq.~\eqref{HWchi_Q}, leaving
    \begin{equation}
    \label{HWchi_WchiQ}
        \bar{\mathcal{Q}}^{\srm W\chi} := a p_{a} 
        - 6\xi_{c}\chi p_{\chi}
        + 36\xi\xi_{c}\chi^2\kb \deq 0\ .
    \end{equation}
As mentioned earlier above, this equation may be rewritten as a definition of the momentum $p_{a}$
    \begin{equation}
    \label{HWchi_Wchipa}
        p_{a} := 
        6\xi_{c}\frac{\chi}{a} p_{\chi}
        + 36\xi\xi_{c}\frac{\chi^2}{a}\kb \ ,
    \end{equation}
which tells us that the scale density $a$ is dynamical in this theory.
The preservation of $\bar{\mathcal{Q}}^{\srm W\chi}$ in time gives an equation for $\dot{\kb}$ that we obtained in eq.~\eqref{HWchi_KdotLambda} with the first line eliminated and $l^2 /\tilde{l}^2 = 6\xi\xi_{c}\chi^2/\hbar a^2$ substituted in there.
As we explained earlier, this is just the trace of the equations of motion.

The Dirac brackets in this theory can be found from eqs.~\eqref{HWchi_DBKP}-\eqref{HWchi_DBKpchi} by setting $l^2 /\tilde{l}^2 = 6\xi\xi_{c}\chi^2/\hbar a^2$ in eq.~\eqref{HWchi_QderK} before starting their calculation.
The result is given by the following set of equations
    \begin{align}
    \label{HWchi_WchiDBKP}
        \PB{\kb}{\Pb} \rightarrow \PB{\kb}{\Pb}_{\ssst D} & = 0\ ,\\[12pt]
        \PB{\kb}{a} \rightarrow \PB{\kb}{a}_{\ssst D} & = - \ddel{\kb}{p_a} = \frac{a}{36\xi\xi_{c}\chi^2}
        \ ,\\[12pt]
    \label{HWchi_WchiDBKpa}
        \PB{\kb}{p_{a}} \rightarrow \PB{\kb}{p_{a}}_{\ssst D} & = \ddel{\kb}{a} = - \frac{p_a}{36\xi\xi_{c}\chi^2}\ ,\\[12pt]
        \PB{\kb}{\chi} \rightarrow \PB{\kb}{\chi}_{\ssst D} & = - \ddel{\kb}{p_{
        \chi}} = - \frac{1}{6\xi\chi} \ ,\\[12pt]
    \label{HWchi_WchiDBKpchi}
        \PB{\kb}{p_{\chi}} \rightarrow \PB{\kb}{p_{\chi}}_{\ssst D} & = \ddel{\kb}{
        \chi} = - \frac{p_{\chi}}{6\xi\chi^2} \ .
    \end{align}

For conformal coupling the whole action is conformally invariant.
The scale density $a$ and the expansion density $\kb$ completely disappear from the theory (as we learned in chapter \ref{ch:defcf}). 
Setting $\xi_{c}=0$ in eq.~\eqref{HWchi_WchiQ} completely eliminates $\kb$ from constraints and implies the vanishing of the momentum $p_{a}$, agreeing with the absence of the scale density from a conformally invariant theory.
Moreover, this makes $\Pb$ and $\bar{\mathcal{Q}}^{\srm W\chi}$ first-class constraints, since eq.~\eqref{HWchi_QP} vanishes.
Dirac brackets are thus equal to the Poisson brackets and $\lambda_{\ssst \kb}$ remains undetermined. The latter means that the trace of the equations of motion is \textit{identically} zero.

Of course, if mass term $m^2a^2\chi^2$ were present in the potential of the conformally coupled scalar (density) field the conformal symmetry would have been broken by the appearance of the scale density.
However, something interesting happens in that case.
The constraint $\bar{\mathcal{Q}}^{\srm W\chi}$ would not change (since the mass term does not depend on $\kb$), but $\dot{\bar{\mathcal{Q}}}^{\srm W\chi}$ would give a further secondary constraint.
Assumming homogeneous case, this new constraint is $\Nb \bar{\mathcal{Q}}^{\srm W\chi}_1 :=  - \dot{\bar{\mathcal{Q}}}^{\srm W\chi}= \Nb m^2 a^2 \chi^2 \deq 0$.
Its preservation in time gives\footnote{Recall that we are using conformal coupling $\xi_{c} = 0$.}
    \begin{equation}
        \dot{\bar{\mathcal{Q}}}^{\srm W\chi}_1 = 2m^2 \PB{a^2\chi^2}{H^{\srm{W}\chi}} = 2\Nb m^2 a^2\chi^2\left( \kb + \frac{p_{\chi}}{\chi}\right) \deq 0\ .
    \end{equation}
But we see that $\bar{\mathcal{Q}}^{\srm W\chi}_1\deq 0$ implies $m = 0$ or $a = 0$, which produces no further constraints from above.
Furthermore, $m = 0$ simply eliminates the mass term and thus forbids it in the Weyl-tensor theory with conformally coupled scalar (density) field.
This invites a curious question: why is the mass term forbidden, but non-conformal coupling is allowed in the Weyl-tensor theory, if they both break conformal symmetry?
We think that the question is only obscured and its answer might be straightforward: the condition $m=0$ in the conformal but massive case is the same as requiring that the trace of the corresponding energy-momentum tensor vanishes (cf. chapter \ref{ch:defcf}).
This condition on trace is already achieved upon derivation of $\bar{\mathcal{Q}}^{\srm W\chi}_1\deq 0$ constraint.
In the massless but non-conformally coupled case the trace $\mathcal{T}$ contains $\kb$ in the non-minimal coupling term and in the kinetic term.
This makes eq.~\eqref{HWchi_KdotLambda} for the Weyl-tensor gravity to pick up the Lagrange multiplier $\lambda_{\ssst \kb} = \dot{\kb}$ which necessarily appears in the trace of the equations of motion.
The point is that in both case the trace of the equations of motion is recovered, but from two different terms, since the trace itself is different.
In the massive conformal case the trace does not have any other terms to help $a$ be determined so the only remaining possibility is $m=0$.
It is a coincidence that in this case the on-shell and off-shell conditions in eqs.~\eqref{confinvS1} and \eqref{confinvS2} mean the same thing: that $\mathcal{T}$ must vanish identically.
We expect that in the inhomogeneous case the conclusion is the same but we do not test that claim here.
A similar discussion is expected in the case of the Weyl-tensor gravity with minimally coupled scalar field but we do not pursue it here.

\subsection{Vacuum Weyl-tensor gravity}
\label{subs_Wvac}

The situation in vacuum Weyl-tensor is very similar to the case of the Weyl-tensor gravity with a conformally coupled massless scalar (density) field described above, except that Hamitlonian and momentum constraints do not have any matter terms.
All constraints are trivially of first class: there is no $a$ or $p_{a}$ or $\kb$ or $\Pb$ in the Hamiltonian constraint, which is given by
    \begin{align}
    \label{HWchi_W0HamC}
        \Ho{W} & = - \frac{1}{2\aw\hbar}\bPb\cdot\bPb
        + \bm{\mathcal{D}}^2_{\srm W}\cdot\bPb
        + 2\bktb\cdot\bpb
        - \aw\hbar\bar{\mathbf{C}}^{\srm B}\cdot\bar{\mathbf{C}}^{\srm B}
        \deq 0\ ,
    \end{align}
so the conformal constraint
    \begin{equation}
    \label{HWchi_WQ}
        \bar{\mathcal{Q}}^{\srm W} := a p_{a} \deq 0
    \end{equation}
trivially commutes with it.
The momentum constraint is given by
    \begin{align}
    \label{HWchi_W0Hicns}
        \Hi{W} & = 
        - 2\bar{D}_{j}\left(\hb_{ik}\pb^{kj}\right)
        - \frac{1}{3}\left(\chi\del_{i}p_{\chi}
        - 2\del_{i}\chi \, p_{\chi}\right)\nld
        & \quad
        + \Pb^{ij} \bar{D}_{k}\ktb_{ij} 
        - 2 \bar{D}_{i}\left(\ktb_{jk} \Pb^{ij} \right)
        + \frac{1}{3}\del_{k}\left(\ktb_{ij}\Pb^{ij}\right)\ ,
    \end{align}
from which we have excluded and partially integrated the term $-D_{i}\left(a\,p_{a}\right)/3$ within the total Hamiltonian, because this term can be written as $- D_{i}\left(\bar{\mathcal{Q}}^{\srm W}\right)/3$, which vanishes with delayed equality.
Therefore, writing 
    \begin{equation}
    \label{HWchi_WDap}
        N^{i}D_{i}\bar{\mathcal{Q}}^{\srm W} = D_{i}\left(N^{i}\bar{\mathcal{Q}}^{\srm W}\right) - D_{i}N^{i}\bar{\mathcal{Q}}^{\srm W}\ ,
    \end{equation}
we can put this term with the conformal constraint into the total Hamiltonian, which can be written as follows
    \begin{align}
    \label{HWchi_WHtot}
        H^{\srm{W}} & = 
        \intx\bigg\lbrace \Nb\Ho{W}
        + N^{i}\Hi{W}
        + \left(\Nb\kb + \frac{1}{3}D_{i}N^{i}\right)\bar{\mathcal{Q}}^{\srm{W}}
        + \lambda_{\ssst\Nb}p_{\ssst\Nb}
        + \lambda^{i}p_{i}
        + \lambda_{\kb}\Pb\bigg\rbrace
        + H^{\srm{W}}_{\ssst surf},
    \end{align}
where the surface term is given by
    \begin{equation}
    \label{HWchi_WHsurf}
        H^{\srm{W}}_{\ssst surf} = \intx\,\del_{i}\Big(2 N^{k}\hb_{kj}\pb^{ik}
        + 2 N^{k}\ktb_{jk}\Pb^{ij}
        - N^{i}\ktb_{jk}\Pb^{jk}
        + \Pb^{ij} \del_{j}\Nb
        - \Nb \bar{D}_{j}\Pb^{ij}
        \Big)\ ,
    \end{equation}
which is missing the term $D_{i}\left(N^{i}a p_{c}\right)$, because it cancels with the first term in eq.~\eqref{HWchi_WDap}.
But doesn't the term in front of $\bar{\mathcal{Q}}^{\srm{W}}$ look familiar? It is equal to $\dot{a}/a$, according to eq.~\eqref{Ktbardef2}, which defines the expansion density $\kb$.
Things now fall into place like the few last missing pieces of a puzzle by interpreting the term in front of $p_{a}$ (which vanishes, just as $\Pb$ vanishes and $\dot{\kb}$ is its Lagrange multiplier) as a Lagrange multiplier: 
    \begin{equation}
        \left(\Nb\kb + \frac{1}{3}D_{i}N^{i}\right)\bar{\mathcal{Q}}^{\srm{W}} = \dot{a} p_{a} \equiv \lambda_{a}p_{a}\ ,
    \end{equation}
which accompanies the term $\lambda_{\kb}\Pb$.
(This could have been done in the case of Weyl-tensor gravity with conformally coupled scalar field as well.)
Let us rewrite the total Hamiltonian with this new notation,
    \begin{align}
    \label{HWchi_WHtot1}
        H^{\srm{W}} & = 
        \intx\bigg\lbrace \Nb\Ho{W}
        + N^{i}\Hi{W}
        + \lambda_{a}p_{a}
        + \lambda_{\ssst\Nb}p_{\ssst\Nb}
        + \lambda^{i}p_{i}
        + \lambda_{\kb}\Pb\bigg\rbrace
        + H^{\srm{W}}_{\ssst surf}\ .
    \end{align}
Reading the above equation, whose simple and straightforward form we remind is a result of the use of the unimodular-conformal variables in the Hamiltonian formulation, it is clear that the scale density $a$ and the expansion density $\kb$ are true arbitrary variables, their velocities being Lagrange multipliers.

We finish this section by stating the algebra of constraints. Namely, in previous works, e.g. \cite{Kluson2014, ILP}, the algebra of constraints for the vacuum Weyl-tensor theory was rather involved.
This is due to the use of the original variables and the fact that the conformal constraint had a more complicated form compared to the one in the present work.
In the original variables the conformal constraint is given by the following expression \cite{Kluson2014}
    \begin{equation}
    \label{HWchi_Qkluson}
        \mathcal{Q} = 2 h_{ij}p^{ij} + P^{ij}K_{ij} \deq 0\ ,
    \end{equation}
which cannot be reduced to our form in eq.~\eqref{HWchi_WQ} by a direct change of variables. 
This constraint contains a relationship between 24 canonical variables! If one were to substitute one of these variables in terms of the others in the rest of the equations upon implementing the conformal constraints one would end up with a rather complicated expression.
Contrast this to our form of the conformal constraint given by eq.~\eqref{HWchi_WQ}: a single variable is constrained to vanish.
This statement refers to a single degree of freedom, i.e. the scale density $a$.
Similar simplification is seen in $\Pb \deq 0$ constraint, which in original variables reads $h_{ij}P^{ij} \deq 0$, thus relating 12 canonical variables.
In contrast, $\Pb \deq 0$ is a constraint for a single variable.
The use of unimodular-conformal variables thus significantly simplifies the form and improves the interpretation of constraints.
It also simplifies the constraint algebra, which is given by the following,
    \begin{align}
    \label{HWchi_AlgHH}
        \left\lbrace\Ho{W}[\varepsilon_1],\Ho{W}[\varepsilon_2]\right\rbrace &=\bar{\mathcal{H}}_{||}^{\srm W}[\varepsilon_1 \del^{i}\varepsilon_2-\varepsilon_2 \del^{i}\varepsilon_1]\,,\\[6pt]
    \label{HWchi_AlgHm}
        \left\lbrace\bar{\mathcal{H}}_{||}^{\srm W}[\vec{\eta}],\Ho{W}[\varepsilon]\right\rbrace &=\Ho{W}[\mathcal{L}_{\vec{\eta}}\varepsilon]\,,\\[6pt]
    \label{HWchi_Algmm}
        \left\lbrace\bar{\mathcal{H}}_{||}^{\srm W}[\vec{\eta}_1],\bar{\mathcal{H}}_{||}^{\srm W}[\vec{\eta}_2]\right\rbrace &=\bar{\mathcal{H}}_{||}^{\srm W}[\mathcal{L}_{\vec{\eta}_{1}}\vec{\eta}_2]\,,\\[6pt]
    \label{HWchi_AlgHP}
        \left\lbrace\Ho{W}[\varepsilon],\Pb[\omega]\right\rbrace &=0\,,\\[6pt]
    \label{HWchi_AlgmP}
        \left\lbrace\bar{\mathcal{H}}_{||}^{\srm W}[\vec{\eta}],\Pb[\omega]\right\rbrace &=0\,,\\[6pt]
    \label{HWchi_AlgHQ}
        \left\lbrace\Ho{W}[\varepsilon],\bar{\mathcal{Q}}^{\srm W}[\omega]\right\rbrace &=0\,,\\[6pt]
    \label{HWchi_AlgHmQ}
        \left\lbrace\bar{\mathcal{H}}_{||}^{\srm W}[\vec{\eta}],\bar{\mathcal{Q}}^{\srm W}[\omega]\right\rbrace &=0\,,\\[6pt]
    \label{HWchi_AlgPQ}
        \PB{\Pb[\omega_{1}]}{\bar{\mathcal{Q}}^{\srm W}[\omega_{2}]} &=0.
    \end{align}
Equations \eqref{HWchi_AlgHH}-\eqref{HWchi_Algmm} are given without proof because they should be equivalent to eqs.~\eqref{HDall_AlgHH}-\eqref{HDall_Algmm} and eqs.~\eqref{HEH_AlgHH}-\eqref{HEH_Algmm}, which all express the hypersurface foliation algebra.
Note, however, that in the case of a general higher-derivative theory and pure GR the Hamiltonian and momentum constraints contain the conformally non-invariant terms depending on the scale density $a$ and the expansion density $\kb$, so the foliation algebra takes into account the freedom to perform both spatial conformal and spatial shear transformations.
In contrast, the hypersurface foliation algebra of the pure Weyl theory refers only to $SL(3,\mathbb{R})$ transformations because the Hamiltonian and momentum constraints are already conformally invariant.
The remaining information about conformal invariance must be accounted for in some way and it indeed is, in the through eqs.~\eqref{HWchi_AlgHP}-\eqref{HWchi_AlgPQ} in a rather trivial way.
That these equations convey the meaning of conformal invariance has to with the interpretation of the first-class constraints (in this case $\Pb$ and $\bar{\mathcal{Q}}^{\srm W}$) as the generators of symmetry transformations (ini this case the conformal transformation), as noted by Dirac \cite[page 21]{Dirac}.
However, as we shall review in the following section, this interpretation needs more rigor.
For now, it is enough to take this information as it is and conclude that conformal invariance of the pure Weyl-tensor gravity is conveyed by commutation of $\Pb$ and $\bar{\mathcal{Q}}^{\srm W}$ with the Hamiltonian and momentum constraints.

As a last note, which will visually sum up the strength of using the unimodular-conformal variables, we state here the same algebra of constraints of the pure Weyl-tensor theory as derived in \cite[eq. (23)-(30)]{ILP} (using our notation and sorting the equations in parallel to the above),
    \begin{align}
    \label{HWchi_AlgHHILP}
        \left\lbrace\mathcal{H}_{\bot}^{\srm W}[\varepsilon_1],\mathcal{H}_{\bot}^{\srm W}[\varepsilon_2]\right\rbrace & =\mathcal{H}_{||}^{\srm W}[\varepsilon_1 \del^{i}\varepsilon_2-\varepsilon_2 \del^{i}\varepsilon_1]\nld
        &\quad
        + P[(\varepsilon_1 D^{i}\varepsilon_2 - \varepsilon_2 D^{i}\varepsilon_1)(D_{j}K^{j}{}_{i} - D_{i}K)]\,,\\[12pt]
    \label{HWchi_AlgHmILP}
        \left\lbrace\mathcal{H}_{||}^{\srm W}[\vec{\eta}],\mathcal{H}_{\bot}^{\srm W}[\varepsilon]\right\rbrace & =
        \mathcal{H}_{\bot}^{\srm W}[\mathcal{L}_{\vec{\eta}}\varepsilon]\,,\\[12pt]
    \label{HWchi_AlgmmILP}
        \left\lbrace\mathcal{H}_{||}^{\srm W}[\vec{\eta}_1],\mathcal{H}_{||}^{\srm W}[\vec{\eta}_2]\right\rbrace & =
        \mathcal{H}_{||}^{\srm W}[\mathcal{L}_{\vec{\eta}_{1}}\vec{\eta}_2]\,,\\[12pt]
    \label{HWchi_AlgHPILP}
        \left\lbrace\mathcal{H}_{\bot}^{\srm W}[\varepsilon],P[\omega]\right\rbrace & =
        \mathcal{Q}^{\srm W}[\varepsilon\omega] + P[\varepsilon\omega K]\,,\\[12pt]
    \label{HWchi_AlgmPILP}
        \left\lbrace\mathcal{H}_{||}^{\srm W}[\vec{\eta}],P[\omega]\right\rbrace &= P[\mathcal{L}_{\vec{\eta}}\omega]\,,\\[12pt]
    \label{HWchi_AlgHQILP}
        \left\lbrace\mathcal{H}_{\bot}^{\srm W}[\varepsilon],\mathcal{Q}^{\srm W}[\omega]\right\rbrace & =
        \mathcal{H}_{\bot}^{\srm W}[\varepsilon\omega] + P[D_{i}D^{i}(\varepsilon\omega) + \omega D_{i}D^{i}\varepsilon - D_{i}\varepsilon D^{i}\omega]\,,\\[12pt]
    \label{HWchi_AlgHmQILP}
        \left\lbrace\mathcal{H}_{||}^{\srm W}[\vec{\eta}],\mathcal{Q}^{\srm W}[\omega]\right\rbrace & = \mathcal{Q}^{\srm W}[\vec{\eta}\omega]\,,\\[12pt]
    \label{HWchi_AlgPQILP}
        \PB{P[\omega_1]}{\mathcal{Q}^{\srm W}[\omega_2]} &=P[\omega_1\omega_2].
    \end{align}
Compared to eqs.~\eqref{HWchi_AlgHH}-\eqref{HWchi_AlgPQ}, the additional terms in eqs.~\eqref{HWchi_AlgHHILP}, \eqref{HWchi_AlgHPILP}, \eqref{HWchi_AlgmPILP} and \eqref{HWchi_AlgHQILP} are due to the fact that \cite{ILP} --- and the same is with \cite{Kluson2014} --- did not isolate the conformal constraint from what they call the Hamiltonian constraint and due to the fact that they did not use the unimodular-conformal variables, as mentioned earlier in this section.
It is obvious that unimodular-conformal variables reveal manifest conformal invariance of the Weyl-tensor gravity.
The same is expected for other conformally invariant theories.

\subsection{DeWitt supermetric in Weyl-tensor theory}
\label{subs_HWDeWitt}

In sections \ref{subs_DeWittmet} and \ref{subs_HDall_DeWittMet} we discussed the DeWitt supermetric in the superspace (of GR) and in the extended superspace (of a general higher-derivative theory).
As a particular case, we mentioned the DeWitt metric which is missing the trace term, i.e. eq.~\eqref{HDall_DeWittproj}, which introduces the DeWitt metric in the pure Weyl-tensor gravity as discussed in \cite{Kluson2014}.
This supermetric arises if $\br=0$ in eq.~\eqref{HDall_DeWittMet}.

From the previous subsection --- the kinetic term in eq.~\eqref{HWchi_W0HamC} --- it can be deduced that using unimodular-conformal variables reveals that DeWitt supermetric and its inverse in Weyl-tensor theory could simply be defined as
    \begin{align}
    \label{HWchi_DeWitt}
        \bar{G}^{ikjl}_{\aw} & = \frac{\aw}{2}\left(\hb^{ik}\hb^{jl} + \hb^{il}\hb^{jk}\right) - \frac{\aw}{3}\hb^{ij}\hb^{kl} \ ,\\[12pt]
        \bar{\mirl{G}}_{ikjl}^{\aw} & = \frac{1}{2\aw}\left(\hb_{ik}\hb_{jl} + \hb_{il}\hb_{jk}\right) - \frac{1}{3\aw}\hb_{ij}\hb_{kl}  \ ,\\[12pt]
        \bar{\mirl{G}}_{akbl}^{\aw}\bar{G}^{ikjl}_{\aw}  & = \mathbb{1}_{(ab)}^{{\srm{T}}ij}\ .
    \end{align}
Note that there is no problem with defining the inverse supermetric because this metric is defined on the space of all \textit{traceless} rank-2 symmetric tensors.
Now, it is interesting to observe that the traceless DeWitt metric is actually the same as the shape part of the DeWitt supermetric discussed in GR, cf. eq.~\eqref{DeWittmet_shape}.
Moreover, the scale part, i.e. the trace part, of the DeWitt supermetric in the superspace is related to the scale-like part of the DeWitt supermetric in the extended superspace.
This is expected because the kinetic term of GR splits in a similar way as the kinetic term of a general quadratic higher-derivative theory in unimodular-conformal variables: 
the scale part of the DeWitt supermetric in GR determines the scale-like direction in superspace, while the scale part of the DeWitt supermetric in the higher-derivative theory determines the expansion-like direction in the extended superspace;
the shape part of the DeWitt supermetric in GR determines the shape-like direction in superspace, while the shape part of the DeWitt supermetric in the higher-derivative theory determines the shear-like direction in the extended superspace.
The two supermetrics have the exact same properties, the only difference being the factor of $a^{2}$ in the shape part of the DeWitt supermetric, which arises because the shape momenta and shear momenta have different scale weight.
This is expected since the expansion density is built from the scale density and the shear density is built from the shape density.
But what is the metric of the complete \textit{extended} superspace?
Such a metric should have 12 independent elements.
But one may then wonder, where is the DeWitt supermetric part which defines distances in the three-metric sector of the extended superspace?
It would have been obvious that there is such a part if the Hamiltonian constraint in eq.~\eqref{HDall_Hofin} had a kinetic term of $a$ and $\pb^{ij}$ as well.
But we do think that the same supermetric is hiding in the last two terms in the first line of eq.~\eqref{HDall_Hofin} --- the terms linear in momenta.
To see this, recall the form of the ADM momenta in unimodular-conformal variables given by eq.~\eqref{GRxi_padmcan} and express the expansion density and the shear density from there; then using the DeWitt metric in unimodular-conformal variables given by eq.~\eqref{HEH_DeWittgb}, which gives
    \begin{equation}
        a\kb p_{a} + 2\bktb \cdot\bpb = - \frac{1}{l^2\hbar }p_{a}^2 + \frac{4}{l^2\hbar a^2} \bpb\cdot\bpb\ ,
    \end{equation}
which is nothing else than twice the kinetic term of vacuum GR, cf. eq.~\eqref{GRxi_H0v}.
Of course, there is no justification to substitue the extrinsic curvature in eq.~\eqref{HDall_Hofin}.
But it is interesting to see that there is some relationship between these terms linear in momenta and the kinetic term of GR (which also appears in the EH potential in the second line of eq.~\eqref{HDall_Hofin}).
These linear terms are of crucial importance for the quantum theory and its semiclassical approximation, as we shall see in the next chapter.

\section{Generator of conformal transformations in \texorpdfstring{$3+1$}{3+1} formulation}
\label{sec_Gencf31}

In several works over the past few decades \cite{AB,C,PittsEM,Pons88,PSS,PSS00} it has been pointed out and proven that first-class constraints are not \textit{each by themselves} generators of symmetry transformations in a theory (as was proposed by Dirac \cite[page 21]{Dirac}), but that only a ``tuned sum'' \cite{PittsEM} of them forms the correct generator.
This has been shown on examples of both GR and Yang-Mills theories \cite{C,PSS00} and also on the example of electromagnetism \cite{PittsEM}.

Now let's think about the Hamiltonian and momentum constraints.
Their meaning is usually interpreted as: the momentum constraints generate spatial coordinate transformations, while the Hamiltonian constraint generates time transformation.
From the algebra given by eq.~\eqref{HDall_Algmm}, the former is true if taken by itself. However, GR and other reparametrizaton invariant theories of spacetime are \textit{four-dimensionally covariant} theories, which implies that separating spatial from temporal coordinate transformations is artificial and is bound to lead to inconsistencies.
It can be seen from eq.~\eqref{HDall_AlgHH} that two ``temporal'' transformations mix into a spatial coordinate transformation, if the constraints are interpreted as generators of symmetry transformations.
Therefore, the spatial and temporal coordinate transformations mix, but this is expected since it is a $3+1$ decomposition of a full four-dimensional diffeomorphisms.
That the interpretation of each individual first class constraint as a generator of a gauge symmetry leads to inconsistencies in GR can be found in the work of Pitts \cite{PittsGR}.
This inconsistency can be observed also in the case of vacuum electromagnetism~\cite{PittsEM}, where it can be shown that Gauss' constraint in eq.~\eqref{EM_GaussC}, with $m=0$, by itself and its primary constraint $\Pi^{t}$ by itself generate a wrong gauge transformation because they only picks up the spatial part and temporal part, respectively, of the full gauge transformation $A_{\mu}\rightarrow A_{\mu} + \del_{\mu} f$.
The result is that $F_{\mu\nu}$ is \textit{not} invariant under the action of individual primary and secondary constraints, which can be seen on the example of the Gauss' constraint (using its smeared version with an arbitrary function $f(t,\vec{x})$)
    \begin{align}
        \delta A_{\mu} & = \PB{A_{\mu}}{\intx\,\del_{i}\bar{\Pi}^{i}f} = -\delta_{\mu}^{i}\del_{i}\epsilon + surf.\nld
        \Rightarrow & \qquad \delta F_{\mu\nu} 
        = \del_{\mu}\delta A_{\nu} - \del_{\nu}\delta A_{\mu} 
        = - \delta_{\nu}^{i}\del_{\mu}\del_{i}f - \delta_{\mu}^{i}\del_{\nu}\del_{i}f\ ,
    \end{align}
where ``$surf$'' denotes surface terms\footnote{All definitions further below related to the generators are valid up to surface terms.}.
The magnetic field $\sim F_{ij}$ is unchanged, but the electric field $\sim F_{0j}$ obviously is not invariant under the action of the Gauss' constraint alone
    \begin{equation}
        \delta F_{0j} = - \delta_{j}^{i}\del_{0}\del_{i}\epsilon = - \del_{0}\del_{j}\epsilon \neq 0\ .
    \end{equation}
Similar result is obtained for the action of the primary constraint.
The issue is resolved if a particular linear combination of primary and secondary constraints is ``tuned'' such that it gives the correct gauge transformation.
Namely, if one considers
    \begin{align}
        \delta A_{\mu} & = \PB{A_{\mu}}{\intx\,\left(- \Pi^{t}\del_{t}\epsilon + \del_{i}\bar{\Pi}^{i}f \right)} = - \delta_{\mu}^{t}\del_{t}\epsilon - \delta_{\mu}^{i}\del_{i}\epsilon + surf. = -\del_{\mu}f + surf.
    \end{align}
The electromagnetic field strength is of course invariant under this transformation.
Hence, one can define the generator of $U(1)$ transformation for vacuum electromagnetism as
    \begin{align}
    \label{G_GEM}
        \mathcal{G}_{\ssst U(1)}[f] & := \intx\,\left(- \Pi^{t}\del_{t}f - \bar{\Pi}^{i}\del_{i}f \right) + surf.\nld
        & = -\intx\,\Pi^{\mu}\del_{\mu}f + surf.\ ,
    \end{align}
which commutes with the field strength,
    \begin{equation}
    \label{G_GFmn}
        \delta F_{\mu\nu} = \PB{F_{\mu\nu}}{\mathcal{G}_{\ssst U(1)}[f]} = 0\ .
    \end{equation}
Note that in the second line of eq.~\eqref{G_GEM} a partial integration is used to write the generator in a more intuitive, covariant form, which to our knowledge is not often met in the literature.
In fact, we think that attempting to rewrite generators in $3+1$ formulation into their covariant form is a good exercise towards the definition of generators in other decompositions of spacetime than $3+1$ decomposition, starting from their covariant form.
A covariant notation of generators related to reparametrization invariance was to some extent achieved by \cite{C,PSS}, see below.

But we think that definitions of this and other generators can be generalized to be independent of the theory from which it was derived, which is something we already attempted and succeded with the generator of conformal transformation in chapter~\ref{ch:defcf} but in the full covaraint formalism.
In other words, suppose that the form of eq.~\eqref{G_GEM} were \textit{given} and one would like to define an object, say a tensor field, which is invariant under the action of this generator.
Then one would use eq.~\eqref{G_GEM} as a \textit{condition}, from which it would follow that $F_{\mu\nu}$ must be antisymmetric and therefore expressible as a curl of a covector.
In order to have such a generator at one's disposal independent of a theory in question, one needs to start by an attempt to understand the meaning of the known generators without the bias of a particular theory within which the generator is considered.
So let us see what is the $U(1)$ generator doing without reference to the theory of electromagnetism.
What this generator is doing is that it makes a shift of a given four-vector field by a gradient of a scalar function.
It seems at first that all components of the vector field change.
But let us use the wisdom accompanying the search for a set of variables which separate into those that change and those that do not change under this transformation.
This is already known as the transverse-longitudinal decomposition of the electromagnetic potential, see e.g. \cite{PittsEM, Sunder}, and it is the decomposition that we used in chapter \ref{ch:umtocf} to understand the difference between those transformations which change lengths and those that do not.
This guiding principle is the same as the one used to formulate the unimodular-conformal variables, by looking for those variables which change under a conformal transformation and isolating them from those variables which do not.
Splitting a vector field as $V_{\mu} = V_{\mu}^{\bot} + V_{\mu}^{\parallel}$ with a condition that\footnote{We are assuming Minkowski metric here.} $\del_{\mu}V^{\mu}_{\bot} = 0$ and $V^{\mu}_{\bot}V_{\mu}^{\parallel}=0$, which defines the transversality of this component, it follows that the longitudinal component has only one degree of freedom in four dimensions.
This component can also be defined as the curl-free part of $V_{\mu}$ and therefore can be written as $V_{\mu}^{\parallel} = \del_{\mu}\phi$, $\phi$ being a scalar function which carries a single degree of freedom.
The conjugate momenta to these components may be found by splitting $\Pi^{\mu} = \Pi^{\mu}_{\bot} + \Pi^{\mu}_{\parallel}$, whose transversal and longitudinal components obey the same relations as their corresponding configuration variables.
Now, let us define a transformation which makes a shift of the longitudinal component as $\phi\rightarrow \phi + f$ by a function $f$, while it leaves the transversal component invariant.
Using that a Fourier-dual version of the condition $\del_{\mu}V^{\mu}_{\bot} = 0$ is $k_{\mu}V^{\mu}_{\bot} = 0$, where $k_{\mu}$ is the momentum along the direction of propagation of $V^{\mu}$, it follows that the integrand $\Pi^{\mu}\del_{\mu}f$ in eq.~\eqref{G_GEM} reduces to $\Pi^{\mu}k_{\mu}f = \Pi^{\mu}_{\parallel}k_{\mu}f$, i.e. it does not depend on the transversal component of the momentum.
Therefore, a $U(1)$ transformation of a vector field is just a shift in its longitudinal component, leaving the transversal component invariant, so $\delta V_{\mu} = \delta V_{\mu}^{\parallel}  \sim k_{\mu} f$ under a $U(1)$ transformation.
This is a heuristic way of explaining why the generator of $U(1)$ transformation in the second line of eq.~\eqref{G_GEM} has such a form; the Poisson bracket for $V^{\mu}_{\bot}$ component vanishes.
But the point is that we think that this generator could be defined by itself, based on the underlying Lie group itself.
Just as the generator of rotations in space exists by itself, so does any other generator which is related to some Lie group.
Then such generator could be used in \textit{any} theory, independently of whether a theory is invariant under its action or not.
One only needs to find a suitable new set of variables which exposes the relevant variables that are affected by this transformation, such as the longitudinal component of each starting variable, in the case of $U(1)$ transformation.
In the case of vacuum electromagnetism, the action of $U(1)$ on the Lagrangian gives a zero change because of eq.~\eqref{G_GFmn}, but if a mass term is present, as is in Proca field theory (cf. appendix~\ref{app_Proca}), the action of the $U(1)$ generator gives
    \begin{equation}
        \frac{1}{2}\PB{m^2\eta^{\mu\nu}A_{\mu}A_{\nu}}{G_{\ssst U(1)}[f]} = - m^2\eta^{\mu\nu}A_{\mu}\PB{A_{\nu}}{\intx\,\Pi^{\mu}\del_{\mu}f} = - m^2A^{\mu}\del_{\mu}f
    \end{equation}
which means that Proca field theory is not invariant under a $U(1)$ transformation.
Hence, the longitudinal component in the Proca field theory does not vanish due to the mass term and shows that the generator in eq.~\eqref{G_GEM} makes sense to be defined in a theory with second-class constraints as well, which does not enjoy the $U(1)$ symmetry. 

An algorithm for constructing the generator of a symmetry transformation for a given Hamiltonian formulation of a particular theory has been developed by Castellani \cite{C}, but the idea of generators constructed from first-class constraints was initiated by Anderson and Bergmann \cite{AB}; see also historical remarks in \cite{PittsEM, PittsGR}.
We refer to this algorithm as the ``ABC algorithm'' and it is important to keep in mind that this algorithm works with first-class constraints only.
Let us review it.
The ABC algorithm consists of the following steps \cite{C}:
    \begin{enumerate}
        \item Define the following sum
            \begin{equation}
            \label{G_GenSum}
                \mathcal{G}[\epsilon] = \intx\,\sum_{k = 0}^{m}\epsilon^{(k)}\mathcal{G}_{k}
            \end{equation}
        where $m$ is the total number of first-class constraints appearing in a chain starting with a particular primary constraint; $\epsilon^{(k)}:=\d^{k}\epsilon/\d t^{k}$ is the $k$-th time derivative of an arbitrary scalar function $\epsilon$ which is the parameter of the symmetry transformation\footnote{It may happen that this parameter is a tensor density of any rank and weight but that does not affect the essence of the algorithm.}.
        This sum defines the generator of a symmetry transformation related to a set of first-class constraints in a given theory.
        \item Identify all primary first class constraints and form the following sum:
            \begin{equation}
            \label{G_PFCs}
                PFC \equiv \intx\,\sum_{k=1}^{n'}\rho_{k}\mathcal{P}^{k}\ ,
            \end{equation}
        where $\mathcal{P}^{k}$ are primary constraints (not necessarily momenta) and $\rho_{k}$ are arbitrary functions to be determined.
        Choose one primary first-class constraint or a linear combination given by the above equation and set it equal to $\mathcal{G}_{m}$.
        The generator that is based on this constraint gives one particular symmetry transformation: one primary first-class constraint leads to one generator of symmetry transformation.
        If we think of each primary constraint as an equation telling us that a certain variable is a true arbitrary variable, then each symmetry generator arises due to appearance of one true arbitrary variable in a theory.
        \item Apply the following iteration procedure:
            \begin{align}
            \label{G_ABCalgor}
                \mathcal{G}_{m} & = PFC\ ,\nld
                \mathcal{G}_{m-1} + \PB{\mathcal{G}_{m}}{H_{\srm{TOT}}} & = PFC\ ,\nl
                & \vdots\nl
                \mathcal{G}_{0} + \PB{\mathcal{G}_{1}}{H_{\srm{TOT}}} & = PFC\ ,\nl
                \PB{\mathcal{G}_{0}}{H_{\srm{TOT}}} & = PFC\ ,
            \end{align}
            where $H_{\srm{TOT}}$ is a total Hamiltonian of the theory in question.
            \item From the resulting set of equations the coefficients in eq.~\eqref{G_PFCs} can be determined and $\mathcal{G}_{k}$ found. 
    \end{enumerate}
This procedure works for any kind of symmetry --- intrinsic or extrinsic, see e.g. the example of Einstien-Yang-Mills theory \cite{PSS00} ---  as long as these symmetries can be found in a given theory.
Using this procedure Castellani \cite{C} has constructed generators of four-dimensional diffeomorphisms in their $3+1$ decomposed version within the Hamiltonian formulation of GR.
Later, Pons et al. \cite{PSS} have polished the ABC procedure and have given more details on the construction of these generators based on the idea that gauge transformations of configuration variables in Lagrangian formalism should induce a particular transformation of the phase space variables in the Hamiltonian formalism.
They derived a concise version of the generator of the four-dimensional coordinate transformations within the $3+1$ Hamiltonian formulation of any four-dimensional reparametrization invariant theory of spacetime in the following form
    \begin{equation}
    \label{G_GdiffPSS}
        \mathcal{G}[\xi^{\nu}] = \intx \left(\dot{\xi}^{\mu}p_{\mu} + \left(\mathcal{H}_{\mu} + N^{\alpha}{C^{\nu}}_{\mu\alpha}p_{\nu}\right)\xi^{\mu}\right)\ ,
    \end{equation}
where $\xi^{\mu}$ are arbitrary functions of space and time that describe the transformation \textit{in the phase space} and are called ``descriptors'';
one also writes concisely $N^{\alpha}=(N,N^{i})$, $p_{\nu} = (p_{\ssst N}, p_i)$, $\mathcal{H}_{\mu} = (\mathcal{H}_{\bot},\mathcal{H}_{i})$, while ${C^{\nu}}_{\mu\alpha}$ are the structure functions (``coefficients'') of the hypersurface foliation ``algebra'', see e.g. \cite{PSS} and \cite[section 20.7]{Blau},
    \begin{subequations}
    \begin{align}
    \label{G_Calgcoeff1}
        {C^{i}}_{00} & =  \left(h^{ij}(\bx) + h^{ij}(\bx ') \right)\del_{j}\delta(\bx,\bx ')\ ,\\[12pt]
    \label{G_Calgcoeff2}
        {C^{0}}_{i0} & =\del_{i}\delta(\bx,\bx ') = - {C^{0}}_{0i}\ , \\[12pt]
    \label{G_Calgcoeff3}
        {C^{i}}_{jk} & = \left(\delta^{i}_{j}\del_{k} + \delta^{i}_{k}\del_{j}\right)\delta(\bx,\bx ')\ ,
    \end{align}
    \end{subequations}
where derivatives are with respect to $\bx$ and all other components vanish. 
These structure functions would appear if one would have derived eqs.~\eqref{HEH_AlgHH}-\eqref{HEH_Algmm} in terms of the constraints themselves instead of their smeared versions.
This is in agreement with the claims about the general validity of hypersurface foliation algebra for any reparametrization-invariant theory of spacetime \cite{Derr,Thiemann}, because this algebra arises from the algebra of generators, as shown in \cite{C,PSS}.
The gauge generator generates any coordinate transformation and is given for each diffeomorphism class of metrics.
The explicit form of the generators is achieved by plugging eqs.~\eqref{G_Calgcoeff1}-\eqref{G_Calgcoeff3} into eq.~\eqref{G_GdiffPSS} and it has the following form \cite{C,PSS00},
    \begin{subequations}
    \begin{align}
    \label{G_G0}
        \mathcal{G}_{\ssst\bot}[\xi^{0}] & =
        \intx\left(\xi^{0}\left(\mathcal{H}_{\ssst\bot} + h^{ij}p_{i}\del_{j}N + \del_{i}\left(N h^{ij}p_{j}\right) + \del_{i}\left(p_{\ssst N}N^{i}\right)\right) + \dot{\xi}^{0}p_{\ssst N} \right)\ ,\\[12pt]
    \label{G_Gi}
        \mathcal{G}_{\ssst\parallel}[\xi^{i}] & = 
        \intx \left( \xi^{i}\left(\mathcal{H}_{i} + p_{j}\del_{i}N^{j} + \del_{j}\left(N^{j}p_{i}\right) + \del_{i}N p_{\ssst N}\right) + \dot{\xi}^{i}p_{i}
        \right)\ .
    \end{align}
    \end{subequations}
Then the sum of these generators $\mathcal{G}[\xi^{\mu}] = \mathcal{G}_{\ssst\bot}[\xi^{0}] + \mathcal{G}_{\ssst\parallel}[\xi^{i}]$ generates a general coordinate transformation $x^{\mu}\rightarrow x^{\mu} + \epsilon^{\mu}$ of the four-dimensional metric components, i.e. it gives a Lie derivative of $g_{\mu\nu}$ along $\epsilon^{\mu}$,
    \begin{equation}
        \delta g_{\mu\nu} = \mathcal{L}_{\epsilon} g_{\mu\nu} = \PB{g_{\mu\nu}}{\mathcal{G}[\xi^{\mu}]} = \epsilon^{\alpha}\del_{\alpha}g_{\mu\nu} + g_{\alpha\nu}\del_{\mu}\epsilon^{\alpha} + g_{\mu\alpha}\del_{\nu}\epsilon^{\alpha}\ ,
    \end{equation}
if arbitrary functions $\epsilon^{\mu}$ and $\xi^{\mu}$ are related by
    \begin{equation}
        \epsilon^{\mu} = \delta^{\mu}_{i}\xi^{i} + n^{\mu}\xi^{0}\ ,
    \end{equation}
as shown in \cite{C,PSS00}.
The meaning of the above equation is just that $\epsilon^{\mu}$ is decomposed into a piece parallel to the three-hypersurface (the first term) and a piece which is orthogonal to the hypersurface (the second term).
For a special case of translations in time (only the second term above is present and $\epsilon^{\mu} = \delta^{\mu}_{0}$) one expects that this generator coincides with the total Hamiltonian, the latter being the generator of a global evolution in time; indeed, as shown by \cite{PSS}, for a specific form of descriptors $\xi^{\mu} = N^{\mu}$ one recovers the total Hamiltonian
    \begin{equation}
        \mathcal{G}[N^{\nu}] = H_{\srm{TOT}} = \intx \left(\dot{N}^{\mu}p_{\mu} + N^{\mu}\mathcal{H}_{\mu}\right)\ ,
    \end{equation}
which is just a sum of primary-secondary pairs of first-class constraints (as derived in the original variables).

From all this we see that the generators as derived by the ABC algorithm have a more important fundamental role than each individual first-class constraints --- the first class constraints are just pieces of generators which do not have a clear meaning on their own, in terms of symmetry transformations.
Now, the novelty that we propose here is that one could look for generators of various transformations \textit{outside} the ABC algorithm, because the group of transformations that they belong to exists independently of the theory in question --- as we argued, the existence of the $U(1)$ generator given by eq.~\eqref{G_GEM} is independent of the formulation of electromagnetism.
The diffeomorphism generator in eq.~\eqref{G_GdiffPSS} is also defined for any theory in $3+1$ formalism with first-class constraints\footnote{An interesting side quest would be to look for these generators in other types of spacetime decompositions, such as e.g. the double-null decomposition \cite{Vick}}.
In a similar way, we can look at the generator of conformal transformations that arises in the pure Weyl-tensor theory (or the same with conformally coupled scalar field).
It was claimed in \cite{Blw,Kaku1982, Kluson2014} that the conformal constraint in eq.~\eqref{HWchi_Qkluson} by itself is the generator of conformal transformations.
However, this is not correct. The conformal constraint alone does not give the correct conformal transformation, as shown in \cite{MSc} and \cite{KN17}.
It is a simple matter to prove this by commuting  eq.~\eqref{HWchi_Qkluson} with the extrinsic curvature,
    \begin{equation}
        \PB{K_{ij}}{\mathcal{Q}[\omega]} = \omega K_{ij}\ .
    \end{equation}
Comparing with the actual conformal transformation of $K_{ij}$ given by eq.~\eqref{31_KconfTransfDec} it is obvious that the inhomogeneous part is missing.
If instead of $K_{ij}$ one considers the action $\mathcal{Q}$ on its traceless part only, then the result is correct.
But the trace $K$ cannot be correctly transformed using only $\mathcal{Q}$.
This problem is the direct analog of the case of Gauss' constraint in electromagnetism.
Furthermore, the authors of \cite{Kluson2014} wondered what is the physical interpretation of the ``generator'' $h_{ij}P^{ij}$ and left the question open.
The reason why they did not notice the relevance of this primary first-class constraint is that they followed Dirac's definition of gauge transformation generators, which is incorrect, as explained above.
Therefore, in the light of the present discussion, the only physical interpretation that could be found is the one that lies in a particular \textit{linear combination} of the primary-secondary pair of constraints $h_{ij}P^{ij}$ and $\mathcal{Q}$.
It was in \cite{ILP} that the correct generator of conformal transformation has been derived using the ABC algorithm.
Its form is given by \cite[eq. (39)]{ILP}
    \begin{equation}
    \label{G_WILPGconf}
        \mathcal{G}_{\srm{ ILP}}[\omega] = \intx\,\left(\frac{\dot{\omega}}{N}P + \omega\left(\mathcal{Q} + N p_{\ssst N} + \mathcal{L}_{\vec{N}}\frac{P}{N}\right)\right)\ .
    \end{equation}
It can be checked easily that a Poisson bracket of eq.~\eqref{G_WILPGconf} with $K_{ij}$ gives the correct infinitesimal conformal transformation.
The inhomogeneous part is generated by the term with the primary constraint $P$ in eq.~\eqref{G_WILPGconf}.
On the other hand, in \cite{KN17} it was shown that the correct generator can be \textit{guessed} in unimodular-conformal variables by ``tuning the sum'' of $\Pb$ and $\bar{\mathcal{Q}^{\srm W}}$ such that its Poisson bracket with $a$ and $\kb$ gives eqs.~\eqref{31_confa} and \eqref{31_confK} the correct conformal transformation.
This can be done by asking for the following action of the generator of conformal transformation $\mathcal{G}[\omega]$,
    \begin{align}
        \delta_{\omega} a & = \PB{a}{\mathcal{G}[\omega]} \shalleq \omega a\ ,\\[6pt]
        \delta_{\omega} \kb & = \PB{\kb}{\mathcal{G}[\omega]} \shalleq \bar{n}^{\mu}\del_{\mu}\omega\ ,
    \end{align}
and noticing that from the first equation it must be that $\mathcal{G}[\omega]\sim \intx\, \omega a p_{a}$ and that from the second equation it must be that $\mathcal{G}[\omega]\sim \intx\, \bar{n}^{\mu}\del_{\mu}\omega\,\Pb$. Since there is no other variables which transform under conformal transformation, the generator of conformal transformation can be written as a sum of these
    \begin{equation}
    \label{G_GconfUmod}
        \mathcal{G}[\omega] = \intx\, \left(\omega a p_{a} + \bar{n}^{\mu}\del_{\mu}\omega\,\Pb\right)\ ,
    \end{equation}
where $a p_{a} \equiv \mathcal{Q}^{\srm W}$ is the secondary, conformal constraint.
We think that this generator can be derived using the ABC algorithm as well, but we were unable to show that.
The problem is that the ABC algorithm can generate only $\omega a p_{a} + \dot{\omega}\Pb$, if $\mathcal{G}_{1}=\Pb/\Nb$ is used, where the spatial derivatives of $\omega$ are missing.
This could be remedied perhaps by using $\mathcal{L}_{\bar{n}}\omega$ in eq.~\eqref{G_GenSum} instead of $\dot{\omega}$  but we leave this problem open in this thesis and take eq.~\eqref{G_GconfUmod} for granted as it is.


As can be seen, if unimodular-conformal variables are used, the form of the generator of conformal transformations is rather trivial and intuitive: it is built from the only two variables that are affected by a conformal transformation.
Its action on all other unimodular-conformal variables vanishes.
This is why this generator can be used in any other theory as well, including GR and a general higher-derivative theory that we described in the previous two sections.
Despite the fact that the generator of conformal transformations cannot be derived in a general higher-derivative theory or pure GR using the ABC algorithm because there are no first class constraints $\Pb \deq 0$ or $p_{a} \deq 0$, one could easily study Poisson brackets of the Hamiltonian and momentum constraints in those theories with the generator in eq.~\eqref{G_GconfUmod}.
The result will be non-vanishing and would give the outcome of an infinitesimal conformal transformation due to the presence of $a$ and $\kb$.
There is no reason why the notion of such generators would not exist in a theory which does not possess the corresponding symmetries.

If one were to combine the information from this section with the results of chapter~\ref{ch:defcf}, where we defined the covariant generator of conformal transformation, one would be tempted to attempt to derive eq.~\eqref{G_GconfUmod} from the covariant form of the generator of conformal transformation in eq.~\eqref{GconfdefA} using $3+1$ decomposition and unimodular-conformal variables.
We believe that this is indeed possible. One would have to make the following substitutions
    \begin{equation}
        p_{a}\rightarrow \ddel{}{a}\,\qquad \Pb\rightarrow \ddel{}{\kb}\ ,
    \end{equation}
such that eq.~\eqref{G_GconfUmod} would have the form
    \begin{equation}
    \label{G_GconfUmodD}
        \hat{\mathcal{G}}_{\omega}\cdot := \intx\, \left(\omega a \ddel{\,\cdot}{a} + \bar{n}^{\mu}\del_{\mu}\omega\,\ddel{\,\cdot}{\kb}\right)\ ,
    \end{equation}
in \textit{any} theory formulated with the $3+1$ decomposition (not as a Hamiltonian theory).
Then one would have to show that variational derivative with respect to $A$ gives rise to variational derivative with respect to $a$ and variational derivative with respect to $\kb$, using a chain rule due to the change of variables from $A$ to $a$ and $\dot{a}$ to $\kb$.
We shall not pursue the investigation of this hypothesis, but we do think this would be an interesting and important line of research that could be applied to other generators as well.
The result should be the following. Let $S[q^{I}]$ be a functional (say, an action) of fields $q^{I}$ defined on spacetime and $S[z^{I}]$ is the same functional but expressed in $3+1$ decomposition formalism where $z^{I}$ are the $3+1$ configuration variables defined on spatial hypersurface $\Sigma_{t}$ parametrized by a time function $t$.
Then the definition of conformal invariance of this functional should ensure that
    \begin{align}
        \hat{\mathcal{G}}_{\omega}S[q^{I}] & = \int\d^4 x\, \omega (x) A(x)\ddel{S[q^{I}]}{A(x)} \nld
        & \shalleq \int\d t\intx\, \left(\omega(t,\vec{x}) a(t,\vec{x}) \ddel{S[z^{I}]}{a(t,\vec{x})} + \bar{n}^{\mu}\del_{\mu}\omega(t,\vec{x})\,\ddel{S[z^{I}]}{\kb(t,\vec{x})}\right) \nld
        & = \int\d t\hat{\mathcal{G}}_{\omega} S[z^{I}] = 0\ .
    \end{align}
The additional integral over $t$ is necessary because eq.~\eqref{G_GconfUmodD} is defined on the spatial hypersurface at each instant of parameter $t$, but it acts on a functional of fields $z^{I}(t,\vec{x})$ which are evaluated at a particular $t$.
In other words, eq.~\eqref{G_GconfUmodD} should be derivable from eq.~\eqref{GconfdefA} using the $3+1$ formalism and unimodular-conformal variables, since the two equations should convey the same information.

We mention an interesting observation as a final note of this section. Since the generator of conformal transformation in its covariant form eq.~\eqref{GconfdefA} can be interpreted as the generator of changes along a single direction in the configuration space of all metrics $g_{\mu\nu}$, in the $3+1$ formulation of higher-derivative theories of gravity the generator in its $3+1$ form eq.~\eqref{G_GconfUmodD} should express changes along directions of $a$ and $\kb$ in the \textit{extended} superspace.
The dynamics of a conformally invariant theory can then be thought to take place in the hypersurface of the extended superspace which is orthogonal to these two directions.
Since the dynamics of the Weyl-tensor theory takes place on this hypersurface, we can think of the traceless DeWitt supermetric in eq.~\eqref{HWchi_DeWitt} as the projection of the dynamics of the general higher-derivative theory in the entire extended superspace onto the hypersurface which is invariant under conformal transformations.
Hence, using the notion of the traceless DeWitt supermetric and the notion of the generator of conformal transformations one could have a way of constructing the dynamics of three-hypersurfaces whose volume is preserved in time.
Now, recall that we mentioned in section \ref{subs_HEADM} that the Hamiltonian formulation of GR could be derived as a unique outcome of the assumption that the three-hypersurface is completely described by the canonical pair $h_{ij},p^{ij}_{\srm{ADM}}$.
If this assumption were relaxed, would it be possible to derive other theories that describe dynamics of three-hypersurfaces? We think so, because: 1.) \textit{any} reparametrization-invariant theory obeys the hypersurface foliation algebra, cf.~eqs.~ \eqref{HDall_AlgHH}-\eqref{HDall_Algmm}; 2.) higher derivative theories of gravity are mathematically without issues, the only difference is that they have a richer structure than GR and ``live'' on an extended superspace where the extrinsic curvature sector adds six more directions; 3.) one is able to specify additional symmetries, such as conformal symmetry, using the algebra of the generator of conformal transformations with other generators in the theory.
Taking this into account, it is natural to attempt to derive a dynamics of three-hypersurfaces starting from an assumption that hypersurfaces are described not only by the pair $(h_{ij}, p^{ij})$ but also $(K_{ij},P^{ij})$.
This could be worth investigating, but is beyond the aims of this thesis.

\section{Einstein-Hilbert action as a higher-derivative theory without higher derivatives}
\label{sec_HEHHD}

The EH Lagrangian does not contain second order time derivatives of the three-metric (after the partial integration), and it is not necessary to introduce the extrinsic curvature components as independent canonical variables. But it is not harmful, either.
Let us therefore make a short excursion and ask what can one learn if one treats the EH theory as if it were a higher-derivative theory.
The contents of this section are slight revision, reformulation and extension of~\cite{EHHD}, with few corrected typos.

Starting from the ADM Lagrangian in eq.~\eqref{HEH_Lag}, we add to it constraints in eq.~\eqref{HDall_KtlessC} and eq.~\eqref{HDall_KtC} with Lagrange multipliers $\bar{\lambda}^{ij\srm{T}}$ and $a\lambda$ to obtain a constrained Lagrangian similar to eq.~\eqref{HDall_Lagconst} in a general higher-derivative theory,
    \begin{align}
    \label{EHHD_Lagconst}
        &\mathcal{L}^{\srm{EH}}_{c} \left(\Nb,N^{i},a,\hb_{ij},\chi,\kb,\ktb_{ij},\dot{\chi}; \dot{a},\dot{\hb}_{ij}, \lambda, \bar{\lambda}^{ij\srm{T}}\right) :=\nld 
        &\quad \mathcal{L}^{\srm{EH}}\left(\Nb,N^{i}, a,\hb_{ij},\chi,\kb,\ktb_{ij},\dot{\chi}\right)
        - \bar{\lambda}^{ij\srm{T}} \bar{\mathcal{K}}^{\srm{T}}_{ij} 
        - a\lambda \mathcal{K}\ .
    \end{align}
Note that since there is no $\dot{\kb}$ or $\dot{\kb}^{\srm{T}}_{ij}$ in the Lagrangian, the corresponding momenta both vanish as primary constraints, while the rest of the momenta are as in a higher-derivative theory,
    \begin{align}
    \label{EHHD_p}
        p_{\ssst \bar{N}} \deq 0\ ,\qquad p_{i} \deq 0\ ,&\qquad p_{a} = \bar{\lambda}\ ,\qquad \pb^{ij} = 2\bar{\lambda}^{ij\srm T}\ ,\\[12pt]
    \label{EHHD_bP}
        \Pb & \deq 0\ ,\\[12pt]
    \label{EHHD_Pt}
        \bP & \deq 0 \ .
    \end{align}
It is obvious that something strange is happening here, since neither of the momenta seems to be invertible in terms of their velocities. Nevertheless, let us proceed.
Legendre transform is similar to the one used in a higher-derivative theory in eq.~\eqref{HDall_LegTransf} except that we have Lagrange multipliers instead of the exstrinsic curvature velocities,
    \begin{align}
    \label{EHHD_LegTransf}
        H^{\srm{EH}} & = \intx\,\Bigg(  \dot{a}p_{a} + \dot{\hb}_{ij}\pb^{ij} 
        + \bar{\lambda}_{ij}^{\srm T}\Pb^{ij} + \bar{\lambda}_{\ssst \kb}\Pb 
        + \lambda_{\ssst\Nb}p_{\ssst\Nb} + \lambda_{i}p^{i} - \mathcal{L}^{\srm{EH}}_{c}\Bigg)\ .
    \end{align}
The result is the following total Hamiltonian,
    \begin{align}
        \label{EHHD_Htot}
        H^{\srm{EH}} = &\intx\biggbr{ \bar{N}\Ho{EH} + N^{i}\Hi{EH}
        + (2\bar{N}\bar{K}_{ij}^{\ssst\rm T})\mathcal{Y}^{ij}_{\srm{EH}}+\left(\bar{N}\bar{K}\right)\Q{EH}\\[12pt]
        & +\lambda_{\ssst\bar{N}}p_{\ssst\bar{N}}+\lambda^{i}p_{i}+\lambda_{ij}\bar{P}^{ij}+\lambda_{\ssst\kb}\bar{P}}\ ,
    \end{align}
where the secondary constraints $\mathcal{Y}^{ij}_{\srm{EH}}$ and $\Q{EH}$ follow from the preservation of constraints in eq.~\eqref{EHHD_bP} and eq.~\eqref{EHHD_Pt} and their form is given by
    \begin{align}
    \label{EHHD_Psec}
        \dot{\Pb}&\deq 0 \quad\Rightarrow\quad  \Q{EH}\equiv a p_{a} + 6 l^2\hbar a^2\kb\deq 0\ ,\\[12pt]
    \label{EHHD_Ptsec}
        \dot{\bPb} & \deq 0 \quad\Rightarrow\quad \mathbf{\mathcal{Y}}_{\srm{EH}}\equiv 2\bpb - l^2\hbar a^2\bktb_{\sharp}\deq 0\ .
    \end{align}
Preservation in time of the first equation in eq.~\eqref{EHHD_p} --- after extracting eq.~\eqref{EHHD_Psec} and eq.~\eqref{EHHD_Ptsec} in a very similar way that Hamiltonian constraint in the WE theory is derived in eq.~\eqref{HWchi_HamCnst} from eq.~\eqref{HWchi_pNdot} --- gives the Hamiltonian constraint,
    \begin{align}
    \label{EHHD_Ho}
        \dot{p}_{\ssst \bar{N}}&\deq 0 \quad\Rightarrow\quad\Hi{EH}=-\frac{l^2\hbar a^2}{2}\biggl(a^2\,^{\ssst (3)}R - \bktb\cdot\bktb
            + 6\bar{K}^2\biggr)
        \deq 0\ .
    \end{align}
The momentum constraint follows from the time derivative of the second equation in eq.~\eqref{EHHD_p} and is the same as eq.~\eqref{GRxi_Hi} without $\chi$-dependent terms, since we are considering the vacuum case.
 
The interesting fact about this formulation is that it seem that there is no kinetic term in eq.~\eqref{EHHD_Ho}.
Extrinsic curvature is only in the original ADM formulation related to the ADM momenta and makes up the kinetic term, but here things seem to be ``frozen''.
But let us not jump into conclusions.
Now, didn't we meet eq.~\eqref{EHHD_Psec} earlier? In the WE theory, in eq.~\eqref{HWchi_QWE} we met the exact same constraint. 
There the constraint led to eq.~\eqref{HWchi_QPWE} because it is a second-class constraint.
In the present case of the EH gravity treated in a strange, higher-derivative way, the same is true; in fact, both
eq.~\eqref{EHHD_Psec} and eq.~\eqref{EHHD_bP} are second-class constraints becasue they do not commute with their primary constraint pair,
    \begin{align}
    \label{eqn:E-PB2nd}
        \left\lbrace \bar{P},\,\Q{EH}\right\rbrace &  = - 6l^2 \hbar a^2\ ,\\[12pt]
        \left\lbrace \bar{P}^{ij},\,\mathcal{Y}^{mn}_{\srm{EH}}\right\rbrace & = l^2\hbar a^2\left(\bar{h}^{im}\bar{h}^{jn}-\frac{1}{3}\bar{h}^{ij}\bar{h}^{mn}\right)\ .
    \end{align}
What is the meaning of eq.~\eqref{EHHD_Psec} and eq.~\eqref{EHHD_bP}? First of all, since they are second-class constraints, they fix certain canonical variables to make them a function of the other canonical variables which remain independent.
Rewriting the equations to express the extrinsic curvature leads to\footnote{In \cite{EHHD} in equation (32) the first expression should contain ``$a$'' instead of $a^2$ on the RHS and the second expression should contain $a^2$ instead of $a$ onthe RHS, as in equation (19).}
    \begin{align}
    \label{EHHD_ADMpa}
        \kb & = - \frac{1}{6 l^2\hbar a }p_{a} \ ,\\[12pt]
    \label{EHHD_ADMpb}
        \bktb_{\sharp} & = \frac{2}{l^2\hbar a^2}\bpb\ ,
    \end{align}
which we can recognize from eq.~\eqref{GRxi_padmcan} as none other than ADM momenta.
Secondly, since they are second-class constraints, their preservation in time should eventually lead to fixing of Lagrange multipliers $\lambda_{\kb}$ and $\bar{\lambda_{ij}^{\srm T}}$.
Indeed, $\lambda_{\kb}$ is fixed in the same way is in the WE theory in eq.~\eqref{HWchi_KdotLambda} and eq.~\eqref{HWchi_KdotLambdaWE}, which we showed (assuming spatial homogeneity) that it is just the trace of the EE.
In the present case the calculation is the same (up to neglecting the cosmological constant from the start) and eq.~\eqref{HWchi_KdotLambdaWE} as the trace of the EE follows --- the calculation is exactly the same, because the Weyl-tensor part of the WE theory has a vanishing trace.
What is the meaning of the other second-class constraint, in eq.~\eqref{EHHD_Ptsec}?
Since that constraint contains $\bktb$, it will pick up the Lagrange multiplier $\bar{\lambda}_{ij}^{\srm T}$ form the following Poisson bracket
    \begin{equation}
    \label{EHHD_TlessEE}
        \PB{2\bpb - l^2\hbar a^2\bktb_{\sharp}}{H^{\srm{EH}}_{\srm{T}}} \sim \bar{\lambda}_{ij}^{\srm T} + ... \deq 0\ ,
    \end{equation}
whose vanishing stops the constraint analysis and determines the $\bar{\lambda}_{ij}^{\srm T}$, which is just the velocity $\dot{\kb}_{ij}^{\srm T}$.
Which part of the EE might eq.~\eqref{EHHD_TlessEE} be? The only remaining equation is the traceless part of the Ricci tensor, but we only claim this as an educated guess without calculation and leave it as it is.

Finally, to recover the ADM Hamiltonian constraint, use eq.~\eqref{EHHD_ADMpa} and eq.~\eqref{EHHD_ADMpb} in eq.~\eqref{EHHD_Ho}.
The result is given by
    \begin{equation}
        \Ho{EH}  =
	    -\frac{1}{12 l^2\hbar}p_{a}^2
        +\frac{2}{l^2\hbar\, a^2}\bpb\cdot\bpb -
        \frac{l^2\hbar a^4}{2}\,^{\ssst (3)}\! R
        \deq 0\ ,
    \end{equation}
which is exactly the same as eq.~\eqref{GRxi_H0v}.

It is interesting to wonder about a ``broken symmetry'' behind the second-class constraint $\mathcal{Y}^{mn}_{\srm{EH}}\deq 0$. The question is motivated from what we learned about $\Q{W}\deq 0$, $\Q{WE}\deq 0$ and $\Q{WE\chi}\deq 0$ constraints --- these are related to an established or broken conformal symmetry.
We can therefore definitely say that the conformal symmetry is broken in the EH theory, even in vacuum.
This is in accordance with our discussion in section~\ref{subsec_EH} and section~\ref{sec_Gencf31}: there exists a theory-independent generator of conformal transformation which can tell if an object it acts on is conformally invariant or not.
It is thus natural to ask if there a generator related to $\bpb$ and $\mathcal{Y}^{mn}_{\srm{EH}}$ which is obviously broken in any theory which contains velocities of $\dot{\bhb}$.
Note that this hypothetical generator is expected to be related to all transformations except conformal transformations.
So the question can be reformulated as: is there a theory which is independent of the shape density?
As we have discussed in chapter~\ref{ch:umtocf1}, there exists an asymmetry between the scale and the shape degrees of freedom, such that one can find spacetimes in which the scale density is unaffected by curvature, but one cannot ever find spacetimes in which the shape density is unaffected by curvature while the volume is.
The tensorial degrees of freedom of the metric in order to related the points in space --- without it there is no meaning to distance.
So the answer to the question is no.
It would still be interesting to see how can one go about investigating the nature of transformations related to $\bpb$ and $\mathcal{Y}^{mn}_{\srm{EH}}$.

\section{Final remarks}

Three main messages are to be remembered from this section. 
Firstly, exact classical higher-derivative theories introduce additional degrees of freedom which in general exhibit a runaway behavior.
Secondly, the interpretation of the higher-derivative terms in the context of their role at different energy/length scales determines whether a theory containing such terms shall be treated as an exact classical higher-order theory or a classical first-order theory with the higher-derivative terms reduced pertrubatively to a form of perturbative corrections.
This also depends heavily on the context.
The Hamiltonian formulation of these two cases differs significantly because in the latter case one introduces the additional degrees of freedom, while in the former case the degrees of freedom is unchanged compared to the first order theory but their dynamics is affected by the perturbative corrections due to the higher-derivative terms.
For our purposes it is necessary to Hamilton-formulate the exact higher-derivative theory, because we are guided by the principle of \textit{quantization before perturbation}, as explained in the Introduction.
Therefore, our main interest was to explore the form of the constraints and not to derive classical equations of motion because the latter make no sense within the proposed context of this thesis.
Thirdly, despite our omission of a lot of details of the exact classical higher-derivative theory of gravity, we think that this chapter also serves as a testimony of the power of the unimodular-conformal variables.
It is clear that without the application of the results in chapter \ref{ch:umtocf1} to the Hamiltonian-formulation of the higher-derivative theories of gravity would not illuminate the subtleties of conformal features of the theory. 
Using unimodular-conformal variables has proven of crucial significance in identifying the conformal degrees of freedom and interpreting the roles of the $R^2$ and $C^2$ terms in the action, but also in the GR itself.
We hope that these results will motivate further applications of unimodular-conformal variables in other fields of classical and quantum gravity.


Of particular importance is the generator of conformal transformations formulated in $3+1$ formalism.
It was already introduced in \cite{ILP}, but our formulation presented in section \ref{sec_Gencf31} is in unimodular-conformal variables, which makes the interpretation and the action of the generator much more clear compared to its formulation in \cite{ILP}.
Since the derivation of the generator of conformal transformation using the ABC algorithm is tied to the \textit{first-class} theory in question, it becomes impossible to talk about the derivation of the generator of conformal transformation in a theory whose conformal invariance is broken, such as the WE theory. 
However, we have argued already in chapter \ref{ch:defcf} that the existence of the generator should be independent of the theory and the same should be true for the $3+1$ version of the generator appearing in constrained systems.
For this reason we think it is important to understand if there is a possibility to generalize the ABC algorithm to theories which contain second-class constraints.
Moreover, we think that it is worth pursuing the derivation of the $3+1$ form of the generator of conformal transformation from its covariant form defined in \ref{ch:defcf}.

As a final note we would like to encourage the use of the method of perturbative constraints on classical and semiclassical considerations of theories of gravity.
With the ongoing observations through the lens of gravitational waves one might hope for signatures of semiclassical gravity in the gravitational wave signals.
There are already proposals for testing the higher-derivative theories of gravity as exact theories, but we think one should in parallel consider tests of such theories treated perturbatively.

{\centering \hfill $\infty\quad$\showclock{0}{10}$\quad\infty$ \hfill}

\chapter[Quantum geometrodynamics of higher derivative theories]{Quantum geometrodynamics \texorpdfstring{\\}{} of higher derivative theories}
    \label{ch:Quant}
    
Quantum geometrodynamics of General Relativity (QGDGR) is an approach to quantum gravity which is based on the canonical (Dirac) quantization of the Hamiltonian formulation, which we reviewed in the previous chapter.
The main questions about this approach to quantum gravity revolve around the resulting \textit{Wheeler-DeWitt equation}, which is a dynamical equation for the \textit{wave functional} that formally describes a quantum state of an entire universe, including both gravity and matter.
The results of the full theory are rather formal and there are several issues that are still unresolved.
One of those issues concerns the semiclassical approximation and the derivation of the renormalized semiclassical Einstein equations (SEE).
Although methods of semiclassical approximations have been established in the past, it is of our interest in this chapter to question the absence of the quadratic curvature terms in the SEE which are otherwise necessary for the renormalization of the expectation value of the energy-momentum tensor operator can take place.
After presenting the QGDGR in unimodular-conformal variables --- a mere reformulation of the already established results in the literature in terms of unimodular-conformal variables --- we shall address the issue of the unrenormalized SEE and what one should expect from it.
We suggest a way of dealing with this by formulating QGD of a higher-derivative theory instead of the sole EH term.
It will be shown what one can expect if such an approach is adopted and what are the properties of such a quantum gravity theory.
Moreover, independently of the theory in question, a quantization procedure based on the generators of diffeomorphisms shall be argued for.

\newpage
\section{Quantum geometrodynamics of General Relativity}

QGDGR\footnote{Since this quantization procedure is quite general and independent on the theory in question, we choose to use ``QGD'' as the name of a method, while we reserve ``QGDGR'' for quantum geometrodynamics of General Relativity in particular.} was introduced by DeWitt~\cite{Witt67} and it is the most conservative approach to direct quantization of gravity~\cite{OUP}.
It is formulated as a Dirac quantization of the constrained theory of GR (presented in section~\ref{sec_HEH}).
The main consequence of canonical quantization of gravity is that the three-metric field itself is quantized and one talks about superpositions of states that refer to different \textit{three-geometries}.
In such a theory space and time seize to exists, as we shall review.
That means that the notions of space and time need to emerge from the quantum gravity theory, as the energies become lower and lower.
Since one does not know which quantum gravity theory is the correct one and whether QGD based on GR, in particular, makes sense (as straightforward as it appears to be), one needs to investigate various semiclassical approximation schemes with an aim to derive a meaningful low-energy limit to quantum gravity in which the classical spacetime described by GR emerges.
In doing this, one meets several issues that may be relevant for the interpretation of the results of a semiclassical approximation.
We shall point out some existing problems which we think are relevant for the aim of this thesis.

\subsection{Wheeler-DeWitt equation}

We shall quantize GR in a similar way as is presented elsewhere (see~\cite[eq. (5.21)]{OUP} for vacuum case and~\cite{Witt67,KiefNM, KieferBR} for the case with non-minimally coupled scalar field).
The difference will be that we use the theory formulated in the unimodular-conformal variables in this thesis.
The difference with our work is that instead of the three-volume element as a variable we use the scale density $a$ and in our treatment the scalar density field is defined in a different way, cf. eq.~\eqref{chidef3}.
Moreover, in our treatment the use of the lapse density $\Nb$ brings a certain extra factor of $a$ in the Hamiltonian constraint, as compared to the usual formulation.

The central object in canonical quantum gravity is the wave functional
    \begin{equation}
    \label{QGR_Psi}
        \Psi \equiv \Psi\left[a,\bhb,\chi\right] \ ,
    \end{equation}
which is a functional of both gravitational and non-gravitational (matter) fields defined on the three-dimensional space.
The wave functional should in principle also have dependence on $\Nb$ and $N^{i}$, but it turns out that it is independent of them due to their arbitrary nature (cf. eq.~\eqref{QGR_qP} below).
One then adopts the following Dirac quantization rules,
    \begin{align}
    \label{QGR_qNp}
        \hat{\Nb}(\bx)\Psi & =
        \Nb(\bx)\cdot\Psi , & \hat{p}_{\ssst \Nb}(\bx)\Psi & = \frac{\hbar}{i}
        \frac{\delta}{\delta \Nb (\bx)}\Psi,\\[6pt]
    \label{QGR_qNip}
        \hat{N}^i(\bx)\Psi & =
        N^i(\bx)\cdot\Psi , & \hat{p}_{i}(\bx)\Psi & = \frac{\hbar}{i}
        \frac{\delta}{\delta N^i (\bx)}\Psi,\\[6pt]
    \label{QGR_qhp}
        \hat{\bhb}(\bx)\Psi & =
        \bhb(\bx)\cdot\Psi , & \hat{\bpb}(\bx)\Psi & = \frac{\hbar}{i}
        \frac{\delta}{\delta \bhb (\bx)}\Psi,\\[6pt]
    \label{QGR_qap}
        \hat{a}(\bx)\Psi & = a(\bx)\cdot\Psi  , & \hat{p}_{a}(\bx)\Psi & = \frac{\hbar}{i}
        \frac{\delta}{\delta a(\bx)}\Psi , \\[6pt]
    \label{QGR_qxp}
        \hat{\chi}(\bx)\Psi & =
        \chi(\bx)\cdot\Psi , & \hat{p}_{\chi}(\bx)\Psi & = \frac{\hbar}{i}
        \frac{\delta}{\delta \chi (\bx)}\Psi
    \end{align}
such that Poisson brackets in eqs. \eqref{HEH_PBvars1} and \eqref{HEH_PBvars2} are promoted to the following commutators
    \begin{align}
    \label{QGR_CM1}
        \CM{\hat{\hb}_{ij}(\bx)}{\hat{\pb}^{ab}(\by)}\Psi
        & = i\hbar\,\mathbb{1}^{{\srm{T}}ab}_{(ij)}\,\delta
        (\bx,\by)\Psi\ , \\[6pt]
    \label{QGR_CM2}
        \CM{\hat{q}^{A}(\bx)}{\hat{\Pi}_{B}(\by)}\Psi
        & = i\hbar\,\delta_{B}^{A}\,\delta (\bx,\by)\Psi\ ,
    \end{align}
where $\hat{q}^{A} = (\hat{a},\hat{\chi})$ and $\hat{\Pi}_{B} = (\hat{p}_{a}, \hat{p}_{\chi})$.
From now on we suppress labeling the dependence on space coordinates, unless an explicit need arises.
Then the wave functional is defined as the state which is annihilated by the total Hamiltonian,
    \begin{equation}
    \label{QGR_HPsi} 
        \hat{H}^{\srm{E\chi}}\Psi = \intx\,\left(\Nb\qHo{E\chi}\Psi + N^{i}\qHi{E\chi}\Psi + \lambda_{\ssst \Nb}\hat{p}_{\ssst\Nb}\Psi + \lambda^{i}\hat{p}_{i}\Psi\right) + \hat{H}_{\ssst sruf}^{\srm{E\chi}}\Psi= 0\ ,
    \end{equation}
for which could be argued that is equivalent to the statement that each constraint by itself annihilates the wavefunction as $\qHo{E\chi}\Psi = 0$ and $\qHi{E\chi}\Psi = 0$.
This is the usual assumption in the canonical quantization procedure \cite{Witt67,Dirac,OUP}, but there are several important remarks regarding this procedure that one must be at least aware of and which we shall breifly discuss further below.
Let us first write out each term in eq.~\eqref{QGR_HPsi},
    \begin{align}
    \label{QGR_qP}
        \intx\,\lambda_{\ssst \Nb}\hat{p}_{\ssst \Nb}\Psi & \quad\Rightarrow\quad
        \ddel{\Psi}{ \Nb } = 0\qquad \wedge \qquad \intx\,\lambda^{i}\hat{p}_{i}\Psi = 0 \quad\Rightarrow\quad
        \ddel{\Psi}{N^i } = 0\ ,\\[18pt]
    \label{QGR_qHo}
        \intx\,\Nb\qHo{E\chi}\Psi & = \intx\,\Nb\Biggbr{
	    \frac{\hbar^2}{12 \left(\Lr^2\hbar a^2 + 6\xi \xi_{c}\chi^2\right)}\left(a\ddel{}{a} - 6\xi_{c}\chi\ddel{}{\chi}\right)^2
	    -\frac{\hbar^2}{2}\ddel{^2}{\chi^2}\nld
        &\qquad\qquad
        -\frac{2\hbar^2}{\left(\Lr^2 \hbar a^2 - \xi\chi^2\right)}\ddel{}{\bhb}\cdot\ddel{}{\bhb}
	    - \frac{\Lr^2\hbar a^4}{2}\,^{\ssst (3)}\! R
        +\frac{1}{2}U^{\chi}
        }\Psi= 0\ ,\\[12pt]
    \label{QGR_qHi}
        \intx\,N^{i}\qHi{E\chi}\Psi & = i\hbar \intx\,N^{i}\Biggbr{
        2\bar{D}_{j}\left(\hb_{ik}\ddel{\Psi}{\hb_{kj}}\right)
        +\frac{1}{3}D_{i}\left(a\,\ddel{\Psi}{a}\right)\nld
        &\qquad\qquad\quad
        +\frac{1}{3}\left(\chi\del_{i}\ddel{\Psi}{\chi} - 2\del_{i}\chi \, \ddel{\Psi}{\chi}\right)
        }
        = 0\ .
    \end{align}
The first term in the second line of eq. \eqref{QGR_qHo} is understood as
    \begin{equation}
        \ddel{}{\bhb}\cdot\ddel{}{\bhb} \equiv \hb^{ik}\hb^{jl}\ddel{}{\hb_{ij}}\ddel{}{\hb_{kl}}\ .
    \end{equation}
We color the relative dimensionless coupling $\Lr$ in red in order to be able to later keep track of equations with ease.
Equation \eqref{QGR_HPsi} is equivalent to stating that the wavefunction is invariant under \textit{spacetime} reparametrizations, because $\hat{H}^{\srm{E\chi}}$ is just the quantized generator of four-dimensional diffeomorphisms derived by Pons et al. \cite{PSS} that we reviewed in section~\ref{sec_Gencf31}.
Interpretation of each of the above individual equations is as follows: equations~\eqref{QGR_qP} express the independence of $\Psi$ on the true arbitrary variables --- the lapse density and the shift density --- meaning that the quantum state should not depend on the way the three-hypersurfaces are defined;
eq.~\eqref{QGR_qHi} expresses the independence of the wave functional on the choice of spatial coordinates, i.e. the wave functional is three-diffeomorphism invariant;
eq.~\eqref{QGR_qHo} is the \textit{dynamical} equation for $\Psi$ and is called \textit{the Weeler-DeWitt equation} (WDW equation).
According to Dirac \cite{Dirac}, the reason to impose the quantization conditions as annihilation of $\Psi$ by each \textit{individual} first class constraint is that each of these constraints has the meaning of a gauge generator of a symmetry transformation --- so the interpretation of the above equations is that each symmetry generator produces a vanishing change of the wave functional.
Such wave functional (in analogy to wave functions in ordinary quantum mechanics) is usually referred to as ``the physical state''.
However, recall from section~\ref{sec_Gencf31} that it is incorrect to state that \textit{each individual first class constraint is a generator of a symmetry transformation}.
Instead, as we have reviewed there, the true symmetry generators are a specific linear combination of first class constraints.
Practically, this does not matter for a system such as GR because, as follows from eq.~\eqref{QGR_HPsi}, the result is the same.
It is, however, misleading and care should be taken in more complicated systems.

Equation \eqref{QGR_qHo} plays the central role in QGDGR.
The WDW equation is an equation that resembles the Klein-Gordon equation in its form, the scale density direction being analogous to the time direction; the scalar (density) field behaves like an additional ``spatial'' direction.
The hyperbolic form of this equation is reflected in the opposite signs of the first kinetic terms.
By inspection of eq.~\eqref{QGR_qHo} one can see that the addition of non-minimally coupled scalar matter gives rise to an ambiguity in the signature of the kinetic term because $\xi$ and $\xi_{c}$ may in principle be negative and overcome $\Lr^2\hbar a^2$ term in the kinetic term, as studied by Kiefer~\cite{KiefNM}.
Note that the situation becomes more complicated if factor ordering ambiguity (see further below) were taken into account.
In any case, it is not possible to solve this equation except in some very special cases~\cite{Witt67,OUP}, such as the minisuperspace, where one deals with spatially homogeneous fields and imposes the homogeneity conditions \textit{before} quantization, avoiding functional derivative altogether.
Therefore, all further discussion concerns only the equation itself, not the solutions.

As mentioned in section~\ref{subs_DeWittmet}, \textit{the problem of time} arises in quantum gravity and one is able to see that from the WDW equation: this equation does not resemble the usual differential equations in quantum mechanics and quantum field theory which contain derivatives with respect to space and time.
Time and space are meaningless concepts and the evolution of $\Psi$ is with respect to the changes of the three-metric in the directions of scale density and shape density, but also in the direction of the non-gravitational fields, such as $\chi$ in our treatment.
The wave functional thus lives on the configuration space of gravitational and matter fields.
All derivatives are with respect to fields which are \textit{functions} of space and these fields play the role of ``coordinates''.
Since these fields are tensors and tensor densities, a change of coordinates affects the form of components of these fields but does not affect the wave functional itself, due to eq.~\eqref{QGR_qHi} --- the wave functional is a \textit{timeless} and \textit{spaceless} object and thus provides no information on them.
Therefore, with respect to the spacetime parameters, eq.~\eqref{QGR_HPsi} expresses the so-called ``static'' nature of the quantum state, as in a time-independed Schr\"odinger equation.
But this is where the hyperbolicity of the WDW equation becomes important: it tells one that it is possible to talk about the initial value problem and express the evolution of $\Psi$ in terms of the scale density.
This discussion becomes non-trivial if the matter is coupled to $\dot{a}$.
Namely, such coupling leads to a much more complicated kinetic term as in eqs.~\eqref{QGR_qHo} whose signature depends on the value of the involved fields \cite{KiefNM}, i.e. only for $\Lr^2\hbar a^2 + 6\xi\xi_{c}\chi^2 > 0$ and $\Lr^2\hbar a^2 - \xi\chi^2 > 0$ the WDW equation is of hyperbolic nature (cf. equations (2.16a) and (2.16b) in \cite{KiefNM}), which could then be achieved only in certain regions of the configuration space.
We mentioned in section~\ref{subs_DeWittmet} that the absence of the scale-like direction in the superspace would imply conformal invariance, but that this is not possible in GR because GR by itself is not conformally invariant.
But we have seen in chapter~\ref{ch:HDclass} that higher-derivative theories offer the possibility of having conformal invariance within the gravitational sector itself.
We shall see in the remaining of this chapter that this raises very interesting questions in the corresponding quantum gravity theory and the hyperbolic nature of the kinetic terms in the respective equations.

It was mentioned earlier that there are some important issues to be aware of. One is the problem of the definition of the Hilbert space.
Namely, since there is no time in the usual sense in quantum gravity, the usual notion of probability is ill defined, because there is no time evolution parameter with respect to which the measure in a potential Hilbert space could be conserved.
But it may be argued that it is also unnecessary since there is also no space in which potential observers would sit and measure this probability.
(However, this is just one point of view.)
In relation to this, one also speaks about the issue of unitarity. One may try to find a way to impose a Hilbert space structure and a well-defined measure, but there are many ways to do this~\cite{IshamTime} and we shall not go into it.
The approach that we take is that the concept of time and Hilbert space as we use in ordinary quantum mechanics is understood to be only of an \textit{emergent}, approximate nature.
The implications of the WDW equation --- if its form is taken literally --- are observationally inaccessible due to its timeless and spaceless nature.
One is then led to seek a meaningful semiclassical approximation scheme from which one could recover the concepts of space and time, as well as quantum field theory on a fixed background, in a meaningful way.
Only then can one hope to have some observational signatures of quantum gravity at one's disposal.
The semiclassical approximation is the problem of our concern here.

Other problems are related to the factor ordering ambiguity --- the problem of non-commutation of momenta and ``coordinate'' operators.
There are at least two aspects of this problem. One is that there is no empirical indication which factor ordering should one choose, in contrast to quantum mechanics where different factor ordering choices can be distinguished by the experimental results.
The simplest and most naive choice is ``momenta to the right'', so that they are the first objects to act on the wave functional in a sequence of operator actions.
One may instead opt for the Laplace-Beltrami-like factor ordering, in which case one speaks of the notion of covariance in the superspace~\cite{Barvinsky} or for the conformal factor ordering~\cite{RyanConf}.
In both cases it is important to understand the implications to the quantization of the hypersurface foliation algebra, because not only the constraints themselves but also the structure functions play a role in factor ordering ambiguity and may affect the formulation of the quantum theory.
In relation to this, we think that the notion of generators of symmetry transformations as discussed by \cite{PSS} (cf. section~\ref{sec_31conf}) might be a better starting point to tackle this problem.
By including $\Nb$ and $N^{i}$ in eqs.~\eqref{QGR_qHo}-\eqref{QGR_qHi} we wanted to emphasize that point, as well as to encourage one to be aware of the subtleties that are easily obscured if one writes down only the integrand in eq.~\eqref{QGR_HPsi}.
An example of the factor order ambiguity can be observed already by comparing the WDW equation derived in the original variables with the WDW equation in the unimodular-conformal variables, if one recalls that the Hamiltonian constraints in the two approaches differ by a factor of $a$;
the change of variables in the quantum theory would produce inconsistencies of the two approaches if this rescaling is not taken into account.
More generally, one could take into account that $\Nb$ can be rescaled by an arbitrary function of configuration space variables in which case one could use the conformal factor ordering\footnote{A non-minimal coupling term proportional to the Ricci scalar of the configuration space appears, akin to the KG equation for a conformally coupled scalar field.} \cite{RyanConf} that ensures (in some limiting and simplified scenarios) that the difference disappears.
In spite of these issues, we proceed with naive factor ordering and are aware of the possible limitations of the results.

Another issue is that one often neglects (as we do) the surface terms  arising form the Hamiltonian formulation, cf. eq.~\eqref{GRchi_HsurfEchi}.
These are again easily overlooked if one writes only the integrand in eq.~\eqref{QGR_HPsi}.

The last issue with QGD that one has to bear in mind is the fact that the second functional derivatives in the kinetic terms in the WDW equation are evaluated \textit{at the same point in space}, which gives rise to some terms proportional to delta function of vanishing argument $\delta(0)$, which is ill-defined even under an integral.
This has been rather recently addressed by Feng~\cite{Feng}, who hinted that regularization of the three-volume (which formally diverges) apparently may lead to the elimination of $\delta(0)$ problem of the second functional derivatives.
This problem persists independently of factor ordering ambiguity.
One consequence of their work which we find interesting and possibly relevant in relation to the topic of the thesis is that the regularized form of the WDW equation gives rise to quadratic curvature terms in the WDW equation itself.
This may be an important topic of an interesting, alternative and less ``artificial'' line of inquiry to address the problem of the missing quadratic curvature terms that the renormalization of the backreaction requires one to introduce in the SEE at this point by hand, cf. section~\ref{subs_HDSEE}.

In the remainder of this whole chapter we assume a naive factor ordering (momenta act first), neglect the possibility of rescaling $\Nb$ (or changing $N^i$ in any way) and neglect the surface terms. 
We focus on the structure of the WDW equation and general, although formal, implications of the semiclassical apporximation.

\subsection{Semiclassical approximation: general remarks}
\label{subs_QGR_SCbr1}

The semiclassical approximation is a topic in quantum gravity of a particular interest because it is the means by which one can obtain the observable classical universe with classical theory of gravity.
It is reasonable to expect that the SEE given by eqs.~\eqref{SEE_full_Tless} and \eqref{SEE_full_Tr} have to arise in the semiclassical approximation to QGDGR.
What kind of approximation scheme should one employ?
There are few steps and assumptions that are made in the preparation of the semiclassical approximation, which are taken by analogy to the quantum mechanics with atoms and molecules \cite[section 5.4]{OUP}.
Let us briefly sketch it here and leave more details for the upcoming subsection.
The standard approach is to combine the Born-Oppenhemer (BO) type of approximation with the WKB-like approximation.
This has been studied on a number of occasions, e.g. in~\cite{Banks, Witt67, KiefSing, PadSing}, see also an overview in~\cite{KieferSC} and~\cite[section 5.4]{OUP}.
We proceed by using the existing methods but using our own notation.

The BO part of the approximation consists of writing the quantum system as a product state of a part which dominates at scales $\Lr^2 \gg 1$ and a part which is suppressed at these scales.
The WKB-like part of the approximation is an ansatz for $\Psi$ expressed in terms of a rapidly oscillating phase and a slowly oscillating amplitude $\Psi \sim A \exp(i S/\hbar)$, which both are then expanded in a series of inverse powers of $\Lr^2 \gg 1$, $S = \Lr^2 S_{0} + S_{1} + \Lr^{-2} S_{2} + ...$ and similarly for the amplitude.
The point is to notice that $\Lr^2 \gg 1$ diminishes the kinetic term in the WDW equation, compared to the gravitational potential and to the kinetic and potential terms of the matter:
    \begin{align}
    \label{QGR_SC_sk1}
        \ddel{^2\Psi}{\chi^2} &\sim \left(\Lr^2\right)^2 \left(\ddel{S_{0}}{\chi}\right)^2 \Psi + \mathcal{O}\left(\Lr^2\right)\ ,\\[12pt]
    \label{QGR_SC_sk2}
        \frac{1}{\Lr^2}\ddel{^2\Psi}{a^2} & \sim \Lr^2 \left(\ddel{S_{0}}{a}\right)^2 \Psi + \mathcal{O}\left(\Lr^0\right)\ ,\\[12pt]
    \label{QGR_SC_sk3}
         \frac{1}{\Lr^2}\ddel{}{\bhb}\cdot\ddel{\Psi}{\bhb} & \sim \Lr^2\ddel{S_{0}}{\bhb}\cdot\ddel{S_{0}}{\bhb} \Psi + \mathcal{O}\left(\Lr^0\right)\ .
    \end{align}
From the above equations and upon inspection of the WDW equation~\eqref{QGR_qHo} one can see that eq.~\eqref{QGR_SC_sk1} is the only term that survives at $\left(\Lr^2\right)^2$ order.
Since the RHS of the WDW equation equates to zero, this implies that $S_{0}$ is independent of $\chi$ and this information is used at each subsequent order of the approximation.
With this, the part of the quantum system which dominates at the $\Lr^2\gg 1$ scales is recognized and referred to as the ``heavy'' part. 
The ``heavy'' part is determined only by the gravitational background.
The ``light'' part is significant only at orders lower than $\mathcal{O}(\Lr^2)$ and is determined by both matter and gravitational background.
Taking into account this discrepancy in orders of magnitude, the BO$+$WKB approximation scheme can be employed. 
But the result is that in the highest order of the approximation (where the ``heavy'' part dominates), one obtains \textit{vacuum} Einstein-Hamilton-Jacobi (EHJ) equation, which is equivalent to the EE, not to SEE.
Then in the subsequent order, $\Lr^0$, one obtains the functional Schr\"odinger equation for the ``light'' part, along with the \textit{backreaction contribution} to the EHJ --- which is equivalent to the SEE \textit{without classical matter and counter-terms}, i.e. to eq.~\eqref{SEE_full} with $\aw=\br=0$ and $T_{\mu\nu}^{\ssst cl} = 0$.
This establishes the quantum field theory on a classical \textit{vacuum} curved spacetime, without the counter-terms (which is indeed consistent).
At even lower orders of $\Lr^{2}$ the quantum gravitational corrections to the functional Schr\"odinger equation are derived.

In order to obtain a more general, non-vacuum result for the EHJ and therefore the SEE given by eq.~\eqref{SEE_full}, additional matter field action needs to be added to the theory.
But this must be done in such a way that this additional matter is not suppressed at the order $\Lr^2$.
It has to enter the ``heavy'' part at order $\Lr^2$ because it needs to contribute to the EHJ.
Assuming one has introduced an additional matter field designated by $\chi_{0}$, the situation in eqs.~\eqref{QGR_SC_sk1}-\eqref{QGR_SC_sk3} should look like this:
    \begin{subequations}
    \begin{align}
    \label{QGR_SC_sk1a}
        \ddel{^2\Psi}{\chi^2} &\sim \left(\Lr^2\right)^2 \left(\ddel{S_{0}}{\chi}\right)^2 \Psi + \mathcal{O}\left(\Lr^2\right)\ ,\\[12pt]
    \label{QGR_SC_sk1ab}
        \frac{1}{\Lr^2}\ddel{^2\Psi}{\chi_{0}^2} &\sim \Lr^2 \left( \ddel{S_{0}}{\chi_0}\right)^2 \Psi + \mathcal{O}\left(\Lr^0\right)\ ,\\[12pt]
    \label{QGR_SC_sk2a}
        \frac{1}{\Lr^2}\ddel{^2\Psi}{a^2} & \sim \Lr^2 \left(\ddel{S_{0}}{a}\right)^2 \Psi + \mathcal{O}\left(\Lr^0\right)\ ,\\[12pt]
    \label{QGR_SC_sk3a}
         \frac{1}{\Lr^2}\ddel{}{\bhb}\cdot\ddel{\Psi}{\bhb} & \sim \Lr^2\ddel{S_{0}}{\bhb}\cdot\ddel{S_{0}}{\bhb} \Psi + \mathcal{O}\left(\Lr^0\right)\ .
    \end{align}
    \end{subequations}
How can a contribution in eq.~\eqref{QGR_SC_sk1ab} be implemented in the WDW equation? 
Comparing with eq.~\eqref{QGR_SC_sk1a}, we see that a formal replacement of the form $\chi\rightarrow\Lr \chi$ eliminates the $\Lr^4$ order, preventing one to conclude that $S_{0}$ is independent of $\chi$ and thus putting the field $\chi$ into the ``heavy'' part, side by side with $a$ and $\bhb$. 
Therefore, adding another copy of the scalar density field Lagrangian and changing $\chi\rightarrow\Lr \chi_{0}$ would do the trick.
But this needs to be done with a good enough justification at the level of the action before the quantization. 
Authors of \cite{KieferBR}, which deals with a background scalar field \textit{and} its pertubration in the EH theory, have done this by simply rescaling the background scalar field by the Planck mass to make it dimensionless (in $\hbar=1$ units, which they use), $\varphi\rightarrow m_{p}^{-1}\varphi$.
This produces a coupling constant $1/\kappa$ of the background scalar field action --- the same coupling as the EH action.
The consequence is that the kinetic terms of the gravitational and matter $\varphi$ sectors in the Hamiltonian constraint (and therefore in the WDW equation) appear at the same order in $m_{p}^2$, leading to terms similar to eq.~\eqref{QGR_SC_sk1ab}.
This redefinition seems a bit ad-hoc assumption but it achieves the goal.
But in the context of this thesis where fields are deprived of their lenght/mass dimensions and we deal with dimensionless scales, we think that such rescalings could be safely reformulated using the dimensionless parameter $\Lr$.
Hence, we give here an alternative justification for such rescaling.
Namely, let us imagine a system consisting of the EH action plus an action for a scalar (density) field $\mathscr{X}$ and before quantization let us \textit{perturb}\footnote{Note that in a realistic scenario perturbations of matter induce perturbations in spacetime.
In~\cite{KieferBR} this was taken into account. Here we do not take this into account (which is unrealistic) and claim that it does not affect the discussion in an essential way.} the field $\mathscr{X}$ with respect to $\Lr$:
    \begin{equation}
    \label{QGR_SC}
        \mathscr{X} = \Lr \chi_{0} + \chi\ ,
    \end{equation}
interpreting $\chi_{0}$ as the background and $\chi$ as an independent perturbation.
Compared to \cite{KieferBR}, we have simply explicitly stated at which order of $\Lr$ do $\chi_{0}$ and $\chi$ appear, without rescaling the fields into their dimensionless versions.
If one demands that the classical EE make sense at the order of $\Lr^2$, the above ansatz achieves this explicitly, because it generates a ``coupling constant'' $\Lr^2$ in the kinetic term of the $\chi_{0}$ Lagrangian.
Recalling eqs.~\eqref{SEE_full_TlessB}-\eqref{SEE_full_TrB1}, which argue that $T_{\mu\nu}^{\ssst cl}$ is of the order of $\Lr^2\hbar$,
this implies that $T_{\mu\nu}^{\ssst cl}$ is determined solely by $\chi_{0}$, thereby pushing dependence on $\chi$ into $\left\langle \hat{T}_{\mu\nu} \right\rangle$.
We see that eq.~\eqref{QGR_SC} is compatible with that claim and makes it explicit.
Let us therefore implement an additional scalar density field $\chi_{0}$ into the classical GR Lagrangian in eq.~\eqref{HEHchi_action} described in section~\ref{sec_HEH} and then repeat the quantization.
This consists of adding another copy of $\mathcal{L}^{\ssst \chi}$ to eq.~\eqref{HEHchi_action} with $\chi\rightarrow\Lr \chi_{0}$, changing $p_{a}$ from eq.~\eqref{GRchi_pam} into\footnote{In principle, one could attribute independent non-minimal couplings $\xi$ (and therefore $\xi_{c}$) for $\chi_{0}$ and $\chi$.
This depends on a physical situation one has at hand and presents a separate question that we shall not pursue here.
Therefore, for simplicity we assume that both fields have the same non-minimal coupling constant.}
    \begin{align}
        p_{a} & = -\frac{6\left(\Lr^2\hbar a^2 + 6\Lr^2\xi\xi_{c}\chi_{0}^2 + 6\xi\xi_{c}\chi^2\right)}{a}\kb 
        + 6\xi_{c}\frac{\chi_{0}}{a}p_{\chi_{0}} 
        + 6\xi_{c}\frac{\chi }{a}p_{\chi}\ ,\\[12pt]
    \label{QGR_Knew}
        \Rightarrow\quad \kb & = -\frac{a}{6\left(\Lr^2\hbar a^2 + 6\Lr^2\xi\xi_{c}\chi_{0}^2 + 6\xi\xi_{c}\chi^2\right)}\left(p_{a} 
        - 6\xi_{c}\frac{\chi_{0}}{a}p_{\chi_{0}} 
        - 6\xi_{c}\frac{\chi }{a}p_{\chi}\right)
    \end{align}
changing $\bpb$ from eq.~\eqref{GRchi_phdot} to
    \begin{align}
        \bpb & = \frac{1}{2}\left(\Lr^2\hbar a^2  -  \Lr^2\xi\chi_{0}^2 - \xi\chi^2 \right)\bktb_{\sharp}\ ,\\[12pt]
    \label{QGR_Ktlessnew}
        \Rightarrow\quad \bktb & = \frac{2}{\left(\Lr^2\hbar a^2  -  \Lr^2\xi\chi_{0}^2 - \xi\chi^2 \right)}\bpb_{\flat}
    \end{align}
and adding another copy of eq.~\eqref{GRchi_pchi} with $p_{\chi}\rightarrow p_{\chi_{0}}/\Lr^2$ in it,
    \begin{align}
        p_{\chi_{0}} & = \Lr^2\left(\bar{n}^{\mu}\chi_{0} + 6\xi_{c}\kb\chi_{0} - \frac{\del_{i}N^{i}}{3\Nb}\chi_{0}\right)\ ,\\[12pt]
    \label{QGR_dotchinew}
        \Rightarrow\quad \dot{\chi}_{0} & = \Nb\left( \frac{1}{\Lr^2}p_{\chi_{0}} - 6\xi_{c}\kb\chi_{0}\right) + \frac{\del_{i}N^{i}}{3}\chi_{0} + N^{i}\del_{i}\chi_{0}\ .
    \end{align}
With these additions, the new Hamiltonian constraint can be derived in the following form,
    \begin{align}
    \label{QGR_Hchichi}
        \Ho{E\chi_{o}\chi} & = -\frac{\left(a p_{a} - 6\xi_{c}\chi_{0} p_{\chi_{0}} - 6\xi_{c}\chi p_{\chi}\right)^2}{12 \left(\Lr^2\hbar a^2 + 6\Lr^2\xi\xi_{c}\chi_{0}^2 + 6\xi\xi_{c}\chi^2\right)}
	    +\frac{1}{2\Lr^2}p_{\chi_{0}}^2  + \frac{1}{2}p_{\chi}^2
        +\frac{2\bpb\cdot\bpb}{\Lr^2\hbar\, a^2 - \Lr^2\xi\chi_{0}^2- \xi\chi^2}\nld
	    &\qquad\qquad\quad - \frac{\Lr^2\hbar a^4}{2}\,^{\ssst (3)}\! R
        + \frac{\Lr^2}{2}U^{\chi_{0}}
        + \frac{1}{2}U^{\chi}
        \deq 0\ .
    \end{align}
Note how more complicated the kinetic term --- and therefore the DeWitt metric, cf. eq.~\eqref{HEH_DeWittgb} --- is now.
It shows that there is certain mixing between the two matter fields because both fields are coupled to the scale density and its time derivative ($\kb$).
In quantum theory, this adds even more drastic ambiguities of factor ordering, but we shall ignore that.
The important thing is that we have one matter field $(\chi_{0})$ at the same order as the gravitational fields (both of which now comprise the ``heavy'' part) and another matter field at one order lower than that (the ``light'' part), so that the expansion scheme sketched with eqs.~\eqref{QGR_SC_sk1a}-\eqref{QGR_SC_sk3a} is now achievable.
It has to be emphasized that the sum of the second and the next-to-last term in eq.~\eqref{QGR_Hchichi} are not the only parts of the $\chi_{0}$ Hamiltonian, because the non-minimal coupling term in $V^{\ssst \chi}$ \textit{mixes} the potential of the $\chi_{0}$ field with the kinetic term of GR, which is reflected in the first term and the fourth term.
The same can be said for the terms related to the $\chi$ field.
This is the reason why we refrain from writing $\Ho{\chi_{0}}$ and $\Ho{\chi}$ in eq.~\eqref{QGR_Hchichi}.

Let us now jump back to quantization. There is now another copy of eq.~\eqref{QGR_qxp} with $\chi\rightarrow\chi_{0}$, and similar additions to eqs.~\eqref{QGR_CM1} and \eqref{QGR_CM2}.
The wave functional obtains an additional dependence on $\chi_{0}$,
    \begin{equation}
    \label{QGR_Psichi0}
        \Psi \equiv \Psi [q^{\ssst A},\chi]\ ,
    \end{equation}
where we define $q^{\ssst A}:= \{ a,\bhb,\chi_{0} \}$, ${\ssst A} = \{a, \bhb, \chi_{0}\}$ the set of ``heavy'' fields.
Having presented the general formalism (with all its problems) of discussing non-minimally coupled fields in unimodular-conformal variables in GR, there is no need for us to keep things as general any further.
We shall therefore assume that we are dealing with conformal coupling, in order to simplify the discussion and prevent obscuring the main point.
Setting $\xi=1/6$ and $\xi_{c} = 0$, eq.~\eqref{QGR_Hchichi} results in
    \begin{align}
    \label{QGR_Hconfchichi}
        \Ho{E\chi_{o}\chi} & = 
        - \frac{ p_{a}^2}{12 \Lr^2\hbar }
	    + \frac{1}{2\Lr^2}p^2_{\chi_{0}}
	    + \frac{1}{2}p_{\chi}^2
        + \frac{2\bpb\cdot\bpb}{\Lr^2\hbar\, a^2 - \Lr^2\frac{\chi_{0}^2}{6}- \frac{\chi^2}{6}}\nld
	    &\quad - \frac{\Lr^2\hbar a^4}{2}\,^{\ssst (3)}\! R
        + \frac{\Lr^2}{2}U^{\chi_{0}}_{c}
        + \frac{1}{2}U^{\chi}_{c}
        \deq 0\ ,
    \end{align}
where $U^{\chi_{0}}_{c} = U^{\chi_{0}}(\xi=1/6)$ and $U^{\chi}_{c} = U^{\chi}(\xi=1/6)$.
We shall define the inverse metric of the ``heavy'' part using the concept of the inverse DeWitt metric as follows,
    \begin{equation}
    \label{QGR_DeWittMet}
        \tilde{\mirl{\mathscr{G}}}^{\ssst AB} := 
            \left(
            \begin{matrix}
                -\frac{1}{12} & 0 & 0\\
                0 & \dfrac{2\hb^{ik}\hb^{jl}}{ a^2 - \frac{\chi_{0}^2}{6\hbar}- \frac{\chi^2}{6\Lr^2 \hbar}} & 0\\
                0 & 0 & \frac{\hbar}{2}
            \end{matrix}
            \right)\ ,
    \end{equation}
such that it defines the kinetic term proportional to $\Lr^{-2}$,
    \begin{equation}
        \frac{1}{\Lr^2 \hbar}\tilde{\mirl{\mathscr{G}}}^{\ssst AB}p_{\ssst A}p_{\ssst B}\ ,
    \end{equation}
where $p_{\ssst A} = \{p_a, p_{\bhb}, p_{\chi_{0}\}}$, in analogy to eq.~\eqref{HEH_DWittgbKin}.
Note that \textit{upper} indices of the inverse DeWitt metric are merely labels, which we choose to employ because of convenience in writing the momenta and the functional derivatives with the lower index.
Repeating the quantization procedure, thereby focusing only on the Hamiltonian constraint itself by dropping the spatial integral, we have the following WDW equation
    \begin{align}
    \label{QGR_WDWconf}
        \qHo{E\chi_{o}\chi}\Psi & = \Biggsq{ - \frac{\hbar}{\Lr^2}\tilde{\mirl{\mathscr{G}}}^{\ssst AB}\delta^{2}_{\ssst AB}
        + \frac{\Lr^2\hbar}{2}U^{\ssst q}
        - \frac{\hbar^2}{2}\delta^2_{\chi\chi}
        + \frac{1}{2}U^{\chi}_{c}
        }\Psi[q^{\ssst A},\chi]
        = 0
    \end{align}
where we use a short-hand notation $\delta_{\ssst AB}^{2}\equiv \delta^{2}/\delta q^{\ssst A}\delta q^{\ssst B}$ and $\delta^2_{\chi\chi}\equiv\delta^{2}/\delta \chi\delta \chi$ for second functional derivatives and where we defined the (dimensionless) ``heavy'' potential by
    \begin{equation}
    \label{QGR_Vheavy}
        U^{\ssst q} := - a^4 \,^{\ssst (3)}\! R + \frac{1}{\hbar}U^{\chi_{0}}_{c}\ .
    \end{equation}
That $\chi_{0}$ appears at the same order as $a$ and $\bhb$ is now even more apparent.
However, it should be noted that the introduction of the DeWitt metric by eq.~\eqref{QGR_DeWittMet} was possible because the matters were simplified by considering only the conformally coupled fields --- in a more general case there would be cross terms between ``heavy'' and ``light'' kinetic terms, as is apparent from eq.~\eqref{QGR_Hchichi}.
One should also keep in mind that these cross terms emerge in the way they do because the unimodular-conformal variables were used. 
In order to take into account different choices of variables while at the same time having a somewhat clearer definition of the DeWitt supermetric in quantum theory it is crucial to take into account the factor ordering.
But as we said before, we stick to a simple factor ordering choice.
In order to bypass this ambiguity in the definition of the DeWitt supermetric, we have used the ``tilde'' notation in eq.~\eqref{QGR_DeWittMet} as a \textit{temporary} notation, because the \textit{pure} ``heavy'' or \textit{classical background} DeWitt supermetric shall be defined only at the classical level.
The latter should emerge at the highest order of the semiclassical approximation, to which we turn in the next subsection.

\subsection{Semiclassical approximation: the Born-Oppenheimer type and the WKB-like approach}
\label{subs_QGR_SCbr2}

One of the main motivations for introducing the dimensionless coupling $\Lr$ in chapter~\ref{ch:umtocf1} was to obtain a suitable dimensionless parameter with respect to which one could formulate the semiclassical approximation and talk about different scales in a units-independent way.

As hinted in the previous subsection, the approximation scheme consists of two parts: one, separating the ``heavy'' part from the ``light'' part in the quantum state (the BO-type approximation) based on the asymmetry of the kinetic terms in the WDW equation with respect to $\Lr^2$; two, expanding the quantum state using the WKB-like expansion in appropriate powers of $\Lr^2$.

The BO ansazt applied to the quantum state in eq.~\eqref{QGR_Psichi0} reads as follows,
    \begin{align}
    \label{QGR_SC_BOans}
        \Psi[q^{\ssst A},\chi] = \Phi[q^{\ssst A}] \psi[q^{\ssst A},\chi] = \Phi[q^{\ssst A}] e^{\phi} e^{-\phi} \psi[q^{\ssst A},\chi] = \Phi'[q^{\ssst A}] \psi'[q^{\ssst A},\chi]\ ,
    \end{align}
where $\Phi[q^{\ssst A}]$ is referred to as the ``heavy'' part, which is independent of $\chi$, and $\psi[q^{\ssst A},\chi]$ is referred to as the ``light'' part of the wave functional.
The second and the last equality convey the fact that this separation into ``heavy'' and ``light'' parts is actually arbitrary \cite{Leo19}, because one can make an appropriate rescaling of the parts using the complex functional $\phi\equiv \phi[q^{\ssst A}]$, which depends only on the set of ``heavy'' variables and behaves as a gauge.
The choice of $\phi$ affects all subsequent equations unless they are written in a gauge-independent form.
We shall not go into such details but simply assume a choice of $\phi$ has been made such that the notation of the first equality in eq.~\eqref{QGR_SC_BOans} is adopted and certain conditions on $\psi[q^{\ssst A},\chi]$ imposed which we shall come to shortly.
This will be enough for achieving the aim of the thesis.
We emphasize, however, that the work of this thesis should be revisited in the light of Chataignier's work~\cite{Leo19}. 

The second step is to employ a WKB-like approximation in the following form,
    \begin{equation}
    \label{QGR_WKBans}
        \Psi[q^{\ssst A},\chi] = \mathcal{A}[q^{\ssst A}]\exp\left(\frac{i}{\hbar}\Lr^2 S^{\srm{E}\chi_{0}}[q^{\ssst A}]\right)\psi[q^{\ssst A},\chi]\ ,
    \end{equation}
where $\mathcal{A}[q^{\ssst A}]$ is the ``slowly changing amplitude'' and $\Lr^2 S^{\srm{E}\chi_{0}}/\hbar$ is the ``rapidly oscillating phase''.
The ``slow'' and ``rapid'' refer to the fact that derivatives of $\mathcal{A}$ are neglected compared to the derivatives of $S^{\srm{E}\chi_{0}}$ at the order $\Lr^2$.
The amplitude and the phase in eq.~\eqref{QGR_WKBans} are assumed to be expanded in power series in $\Lr^{-2}\rightarrow 0$, as $\Lr\rightarrow \infty$,
    \begin{align}
    \label{QGR_SC_Sexp}
        S^{\srm{E}\chi_{0}}[q^{\ssst A}] & = S^{\srm{E}\chi_{0}}_{0}[q^{\ssst A}] + \Lr^{-2}S^{\srm{E}\chi_{0}}_{1}[q^{\ssst A}] + \mathcal{O}(\Lr^{-4})\ ,\\[12pt]
    \label{QGR_SC_Aexp}
        \mathcal{A}[q^{\ssst A}] & = \mathcal{A}_{0}[q^{\ssst A}] + \Lr^{-2}\mathcal{A}_{1}[q^{\ssst A}]+ \mathcal{O}(\Lr^{-4})\ ,
    \end{align}
while $\psi$ is considered to be determined at the order $\Lr^{0}$, as we shall see.
The remaining steps consist of plugging into the WDW equation and equating to zero all terms coming with the same power of $\Lr^2$.

First one plugs eq.~\eqref{QGR_SC_BOans} into eq.~\eqref{QGR_WDWconf} and obtains\footnote{We stress that a different factor ordering (e.g. Laplace-Beltrami) would yield a more complicated equation involving derivatives of the DeWitt metric.
Even if the most general factor ordering is considered, all these equations suffer from ill-defined delta functions evaluated at zero.
Therefore we stress that all semiclassical approximation schemes in full canonical quantum gravity must be revisited to deal with these issues, see e.g. Feng~\cite{Feng}.}
    \begin{align}
    \label{QGR_SC_BO1}
        \frac{\hbar}{\Lr^2}\tilde{\mirl{\mathscr{G}}}^{\ssst AB}\biggsq{
        \psi\delta^{2}_{\ssst AB}\Phi + 2\delta_{\ssst A}\Phi\delta_{\ssst B}\psi
        + \Phi\delta^{2}_{\ssst AB}\psi
        }
        - \frac{\Lr^2\hbar}{2}U^{\ssst q}\Phi\psi
        = - \Phi\frac{\hbar^2}{2}
        \delta^2_{\chi\chi}\psi
        + \frac{1}{2}U^{\chi}_{c}\Phi\psi\ ,
    \end{align}
which was rewritten in a more convenient form.
The aim is to multiply the above equation from left by $\psi^{*}$, which is a complex-conjugate of $\psi$, and perform a functional integration.
To do so one has to impose an appropriate inner product in the ``light'' sector, $\vert\psi\vert^2$, and then divide eq.~\eqref{QGR_SC_BO1} by $\vert\psi\vert^2$ after the functional integration, in order to normalize.
One can \textit{demand} $\vert\psi\vert^2$ to be
    \begin{equation}
    \label{QGR_SC_inner}
        \vert\psi\vert^2 : = \int \mathcal{D}[\chi]\psi^{*}[q^{\ssst A},\chi]\psi[q^{\ssst A},\chi]\ .
    \end{equation}
By doing this one also says that $\psi$ lives in a (Hilbert) space in which it is possible to define such a measure. (Note that we have avoided claiming the same for the total wave-functional $\Psi$.)
The integration over matter fields only is related to the choice of $\chi$ being the only ``light'' variable.
It can be shown (see e.g.~\cite{Leo19}) that eq.~\eqref{QGR_SC_inner} arises from the lowest non-trivial order of $\Lr^{-2}$ expansion of the Klein-Gordon inner product of the total wave function $\Psi$, for a given choice of gauge $\phi$.
Nevertheless, we assume (as is usually done) that $\psi$ obeys eq.~\eqref{QGR_SC_inner} up to the order to which we confine our discussion here, without referring to the Klein-Gordon inner product.
One is then able to introduce the following definition of the expectation value of an arbitrary operator $\hat{O}$,
    \begin{equation}
    \label{QGR_SC_paravg}
        \left\langle \hat{O}\right\rangle := \frac{1}{\vert\psi\vert^2}\int \mathcal{D}[\chi]\psi^{*}\hat{O}\psi\ .
    \end{equation}
These expectation values are called \textit{partial averages} because they are calculated with respect to the $\chi$-subspace of the total configuration space \cite[eq.~(90)]{Leo19}.
One is now able to introduce the following two definitions,
    \begin{align}
    \label{QGR_SC_d}
        \left\langle \delta_{\ssst A}\right\rangle & := \frac{1}{\vert\psi\vert^2}\int \mathcal{D}[\chi]\psi^{*}\delta_{\ssst A}\psi\ ,\\[12pt]
    \label{QGR_SC_dd}
        \left\langle \delta^2_{\ssst AB}\right\rangle & := \frac{1}{\vert\psi\vert^2}\int \mathcal{D}[\chi]\psi^{*}\delta_{\ssst A}\delta_{\ssst B}\psi\ ,
    \end{align}
which are in general complex functionals of the ``heavy'' variables.

One now has a choice how to normalize the inner product in eq.~\eqref{QGR_SC_inner}. 
It could be assumed that it is just constant, i.e. independent of $q^{\ssst A}$, and let us choose $\vert\psi\vert^2$.
It can be shown that such an assumption is compatible with the demand that the real part of eq.~\eqref{QGR_SC_d} is zero~\cite{Leo19}, because~\cite[eq.~(13)]{Leo19}
    \begin{equation}
    \label{QGD_SC_Redel}
        \rm{Re}\left\langle \delta_{\ssst A}\right\rangle = \frac{1}{2}\delta_{\ssst A}\int \mathcal{D}[\chi]\log \vert\psi\vert^2\ ,
    \end{equation}
which vanishes for constant norm $\vert\psi\vert^2$. 
It can be shown that by appropriately choosing the real part of the gauge $\phi$ one achieves $\vert\psi\vert^2 = 1$, which eliminates eq.~\eqref{QGD_SC_Redel}.
This is an example of utilizing the freedom in choosing $\psi$ and $\Phi$ in the BO ansatz in eq.~\eqref{QGR_SC_BOans}.
With these assumptions and definitions, eq.~\eqref{QGR_SC_BO1} can be integrated over $\chi$, assuming $\vert\psi\vert^2 = 1$. (If the latter assumption were relaxed, one would simply divide by $\vert\psi\vert^2$ without affecting the derivation.)
The result is given by
    \begin{align}
    \label{QGR_SC_BO2}
        \frac{\hbar}{\Lr^2}\tilde{\mirl{\mathscr{G}}}^{\ssst AB}\biggsq{
        \delta^{2}_{\ssst AB}\Phi + 2\delta_{\ssst A}\Phi\left\langle \delta_{\ssst B}\right\rangle
        + \Phi\left\langle \delta^{2}_{\ssst AB}\right\rangle
        }
        - \frac{\Lr^2\hbar}{2}U^{\ssst q}\Phi
        = \left\langle - \frac{\hbar^2}{2}
        \delta^2_{\chi\chi}
        + \frac{1}{2}U^{\chi}_{c}\right\rangle  \Phi\ ,
    \end{align}
where on the RHS is the partial average of the operator in the angled brackets.
The RHS of the above equation is usually written as the expectation value of the $\chi$ Hamiltonian constraint operator, but in our case it is not (yet) so because the part of the non-minimally coupled term is stuck inside the kinetic term.
This is actually a feature of using the unimodular-conformal variables.
Therefore, it is important to emphasise that at this stage of derivation one cannot, in general, identify the RHS of eq.\eqref{QGR_SC_BO2} with $\left\langle\qHo{\chi}\right\rangle$.

Coming back to the derivations, eq.~\eqref{QGR_SC_BO2} can be thought of as an equation for $\Phi$ component sourced by the second, third and the term in the RHS of the equation.
Now one multiplies eq.~\eqref{QGR_SC_BO2} by $\psi$ and subtracts it from eq.~\eqref{QGR_SC_BO1}, then divides the whole result by $\Phi$ and gets
    \begin{align}
        \label{QGR_SC_BO3}
        \frac{2\hbar}{\Lr^2}\tilde{\mirl{\mathscr{G}}}^{\ssst AB}\Biggsq{
        \frac{1}{\Phi}\delta_{\ssst A}\Phi\Big\lbrace
        \delta_{\ssst B}
        -
        \left\langle \delta_{\ssst B}\right\rangle
        \Big\rbrace\psi
        }
        & = \biggsq{
        - \frac{\hbar^2}{2}
        \delta^2_{\chi\chi}
        + \frac{1}{2}U^{\chi}_{c}
        - \left\langle - \frac{\hbar^2}{2}
        \delta^2_{\chi\chi}
        + \frac{1}{2}U^{\chi}_{c}\right\rangle 
        }\psi\nld
        &\quad - \frac{\hbar}{\Lr^{2}}\tilde{\mirl{\mathscr{G}}}^{\ssst AB}\Big\lbrace
        \delta^{2}_{\ssst AB}
        - \left\langle \delta^{2}_{\ssst AB}\right\rangle\Big\rbrace\psi\ .
    \end{align}
Equations~\eqref{QGR_SC_BO2} and \eqref{QGR_SC_BO3} are still just intermediate equations because we still need to expand $\Phi$ and $\tilde{\mirl{\mathscr{G}}}^{\ssst AB}$ in $\Lr^{-2}$.

Expanding $\tilde{\mirl{\mathscr{G}}}^{ \bhb\bhb}$ element in eq.~\eqref{QGR_DeWittMet} up to $\mathcal{O}(\Lr^{-2})$ order we obtain
    \begin{align}
    \label{QGR_SC_DeWittMet_App}
        \frac{\hbar}{\Lr^2}\tilde{\mirl{\mathscr{G}}}^{\ssst AB} & \approx 
        \frac{\hbar}{\Lr^2}\left(\mirl{\mathscr{G}}^{\ssst AB}_{0}
            +
            \Lr^{-2}\tilde{\mirl{\mathscr{G}}}^{\ssst AB}_{1}
        \right)
            \ , \\[12pt]
    \label{QGR_SC_DeWittMet_App0}
        \mirl{\mathscr{G}}^{\ssst AB}\equiv\tilde{\mirl{\mathscr{G}}}^{\ssst AB}_{0} & := \left(
            \begin{matrix}
                -\frac{1}{12} & 0 & 0\\
                0 & \frac{2\hb^{ik}\hb^{jl}}{ a^2 - \frac{\chi_{0}^2}{6\hbar}}  & 0\\
                0 & 0 & \frac{\hbar}{2}
            \end{matrix}
            \right)\ ,\qquad
        \tilde{\mirl{\mathscr{G}}}^{\ssst AB}_{1}  :=    
            \left(
            \begin{matrix}
                0 & 0 & 0\\
                0 & \frac{2\hb^{ik}\hb^{jl} \chi^2}{6\hbar \left(a^2 - \frac{\chi_{0}^2}{6\hbar}\right)^2}  & 0\\
                0 & 0 & 0
            \end{matrix}
            \right)\ .
    \end{align}
The object $\mirl{\mathscr{G}}^{\ssst AB}$ defined in eq.~\eqref{QGR_SC_DeWittMet_App0} depends only on the ``heavy'' fields and will turn out to be the classical DeWitt supermetric in the configuration space of $q^{\ssst A}$ variables.
The other object $\tilde{\mirl{\mathscr{G}}}^{\ssst AB}_{1}$ can be considered as a $\Lr^{-2}$ correction to the classical DeWitt supermetric.
Now, using the WKB ansatz given in eqs.~\eqref{QGR_WKBans}-\eqref{QGR_SC_Aexp} and eq.~\eqref{QGR_SC_DeWittMet_App}, we determine the following terms up to $\mathcal{O}(\Lr^{0})$ order,
    \begin{align}
    \label{QGR_SC_delPhi}
        \frac{\hbar}{\Lr^2}\frac{1}{\Phi}\tilde{\mirl{\mathscr{G}}}^{\ssst AB}\delta_{\ssst A}\Phi  & = 
        \frac{\hbar}{\Lr^2}\tilde{\mirl{\mathscr{G}}}^{\ssst AB}\left(\delta_{\ssst A}\log \mathcal{A} + \frac{i\Lr^{2}}{\hbar}\delta_{\ssst A}S^{\srm{E}\chi_{0}}\right)\Phi\nld
        & \approx i\mirl{\mathscr{G}}^{\ssst AB}\delta_{\ssst A}S^{\srm{E}\chi_{0}}_{0}
        + \mathcal{O}(\Lr^{-2})
        ,\\[18pt]
    \label{QGR_SC_del2Phi}
        \frac{\hbar}{\Lr^2}\frac{1}{\Phi}\tilde{\mirl{\mathscr{G}}}^{\ssst AB}\delta_{\ssst AB}^2\Phi & =
        \frac{\hbar}{\Lr^2}\tilde{\mirl{\mathscr{G}}}^{\ssst AB}
        \Biggpar{
        \frac{\delta_{\ssst AB}^2 \mathcal{A}}{\mathcal{A}}
        + \frac{2i\Lr^2}{\hbar}\delta_{\ssst A}\log\mathcal{A}\,\,\delta_{\ssst B}S^{\srm{E}\chi_{0}} \nld
        & \qquad\qquad 
        + \frac{i\Lr^2}{\hbar} \delta_{\ssst AB}^2 S^{\srm{E}\chi_{0}}
        - \frac{\Lr^{4}}{\hbar^2}\delta_{\ssst A}S^{\srm{E}\chi_{0}}\delta_{\ssst B}S^{\srm{E}\chi_{0}}
        }\nld
        & \approx 
        i \left(
        2\mirl{\mathscr{G}}^{\ssst AB}\delta_{\ssst A}\log\mathcal{A}_{0}\,\,\delta_{\ssst B}S^{\srm{E}\chi_{0}}_{0}
        + \mirl{\mathscr{G}}^{\ssst AB}\delta_{\ssst AB}^2 S^{\srm{E}\chi_{0}}_{0}
        \right)\
        - \frac{\Lr^{2}}{\hbar}\mirl{\mathscr{G}}^{\ssst AB}\delta_{\ssst A}S^{\srm{E}\chi_{0}}_{0}\delta_{\ssst B}S^{\srm{E}\chi_{0}}_{0}\nld
        & \quad
        - \frac{2}{\hbar}\mirl{\mathscr{G}}^{\ssst AB}\delta_{\ssst A}S^{\srm{E}\chi_{0}}_{0}\delta_{\ssst B}S^{\srm{E}\chi_{0}}_{1}
        - \frac{1}{\hbar}\tilde{\mirl{\mathscr{G}}}^{\ssst AB}_{1}\delta_{\ssst A}S^{\srm{E}\chi_{0}}_{0}\delta_{\ssst B}S^{\srm{E}\chi_{0}}_{0}
        + \mathcal{O}(\Lr^{-2})\ ,
    \end{align}
where we used $\delta_{\ssst AB}^2\log \mathcal{A} + \delta_{\ssst A}\log \mathcal{A} \, \, \delta_{\ssst B}\log \mathcal{A} = \mathcal{A}^{-1}\delta_{\ssst AB}^2 \mathcal{A}$.
Note that the last term in eq.~\eqref{QGR_SC_del2Phi} comes from the $\Lr^{-2}$ correction to the DeWitt supermetric in eq.~\eqref{QGR_SC_DeWittMet_App}; this term must not be neglected because it obviously contributes to the WDW equation at the relevant order $\Lr^{0}$.
This term is the missing non-minimal coupling term in the potential for the $\chi$ field in the RHS of eq.~\eqref{QGR_SC_BO2}.

Plugging eq.~\eqref{QGR_SC_delPhi} and eq.~\eqref{QGR_SC_del2Phi} into the equation for $\Phi$ given by eq.~\eqref{QGR_SC_BO2}, neglecting all terms of $\mathcal{O}(\Lr^{-2})$ order and lower, we end up with
    \begin{align}
    \label{QGR_SC_BO2a}
        & - \frac{\Lr^{2}}{\hbar}\mirl{\mathscr{G}}^{\ssst AB}\delta_{\ssst A}S^{\srm{E}\chi_{0}}_{0}\delta_{\ssst B}S^{\srm{E}\chi_{0}}_{0}
        - \frac{\Lr^2\hbar}{2}U^{\ssst q}
        + i \left(
        2\mirl{\mathscr{G}}^{\ssst AB}\delta_{\ssst A}\log\mathcal{A}_{0}\,\,\delta_{\ssst B}S^{\srm{E}\chi_{0}}_{0}
        + \mirl{\mathscr{G}}^{\ssst AB}\delta_{\ssst AB}^2 S^{\srm{E}\chi_{0}}_{0}
        \right)
        \nld
        & \qquad\qquad\qquad
        - \frac{2}{\hbar}\mirl{\mathscr{G}}^{\ssst AB}\delta_{\ssst A}S^{\srm{E}\chi_{0}}_{0}\delta_{\ssst B}S^{\srm{E}\chi_{0}}_{1}
        + 2i\mirl{\mathscr{G}}^{\ssst AB}\delta_{\ssst A}S^{\srm{E}\chi_{0}}_{0}
        \left\langle \delta_{\ssst B}\right\rangle\nld
        & = \left\langle - \frac{\hbar^2}{2}
        \delta^2_{\chi\chi}
        + \frac{1}{2}U^{\chi}_{c}\right\rangle
        + \frac{1}{\hbar}\tilde{\mirl{\mathscr{G}}}^{\ssst AB}_{1}\delta_{\ssst A}S^{\srm{E}\chi_{0}}_{0}\delta_{\ssst B}S^{\srm{E}\chi_{0}}_{0}
        \ .
    \end{align}
Note that the last term in the RHS of the above equation is equal to its partial average and comes at the same order as the first term on the same side. 
Hence, it can be included into this first term, which adds to the potential $U^{\ssst \chi}_{c}$.
This hints that the mentioned last term is the missing non-minimal coupling, but one can show that only after a few more steps.
Namely, there are three things to observe.
First, recall that $\left\langle \delta_{\ssst B}\right\rangle$ is purely imaginary because of the demand that $\vert\psi\vert^2 = 1$;
this means that the last term in the second line in eq.~\eqref{QGR_SC_BO2a} is real.
Second, the parentheses containing the last two terms in the second line is purely imaginary.
Thirdly, and by taking the previous two points into account, one can take the real part and imaginary part of the equation and separate the orders $\Lr^{2}$ and $\Lr^{0}$.

Terms at order $\Lr^{2}$ are all real and they equate to
    \begin{equation}
    \label{QGR_SC_EHJ}
        \frac{1}{\hbar}\mirl{\mathscr{G}}^{\ssst AB}\delta_{\ssst A}S^{\srm{E}\chi_{0}}_{0}\delta_{\ssst B}S^{\srm{E}\chi_{0}}_{0}
        + \frac{\hbar}{2}U^{\ssst q} = 0\ .
    \end{equation}
This is the EHJ equation anticipated earlier in this section and it was shown by Gerlach~\cite{Gerlach} in vaccum case to be equivalent to the EE.
Compare the above equation with the Hamiltonian constraint given by eq.~\eqref{QGR_Hconfchichi}: if one writes the ``heavy'' momenta via the HJ method,
    \begin{equation}
    \label{QGR_SC_momenta}
        p_{\ssst A} \rightarrow \Lr^{2}\ddel{S^{\srm{E}\chi_{0}}_{0}}{q^{\ssst A}}\ ,
    \end{equation}
and expands that equation in descending powers of $\Lr^{2}$, neglecting terms of order $\mathcal{O}(\Lr^{0})$ and lower just gives eq.~\eqref{QGR_SC_EHJ}, taking into account definitions in eq.~\eqref{QGR_Vheavy} and eq.~\eqref{QGR_SC_DeWittMet_App0}.
Hence, the classical \textit{non-vacuum} GR has been recovered and its solution is the highest order contribution to the phase of $\Phi$, $S^{\srm{E}\chi_{0}}_{0}$.

Taking the imaginary part of eq.~\eqref{QGR_SC_BO2a} we have
    \begin{equation}
    \label{QGR_SC_ContA}
        2\mirl{\mathscr{G}}^{\ssst AB}\delta_{\ssst A}\log\mathcal{A}_{0}\,\,\delta_{\ssst B}S^{\srm{E}\chi_{0}}_{0}
        =
        - \mirl{\mathscr{G}}^{\ssst AB}\delta_{\ssst AB}^2 S^{\srm{E}\chi_{0}}_{0} \quad\Rightarrow\quad 
        \mirl{\mathscr{G}}^{\ssst AB}\delta_{\ssst A}\left( \mathcal{A}_{0}^2 \delta_{\ssst B}S^{\srm{E}\chi_{0}}_{0}\right) = 0\ .
    \end{equation}
This equation determines $\mathcal{A}_{0}$, \textit{given} the solution to the EHJ equation, $S^{\srm{E}\chi_{0}}_{0}$.
$\mathcal{A}_{0}$ is related to the van Vleck determinant which describes the density of classical trajectories in the configuration space.
It should resemble a continuity equation for a conserved ``current'' $\mirl{\mathscr{G}}^{\ssst AB}\mathcal{A}_{0}^2 \delta_{\ssst B}S^{\srm{E}\chi_{0}}_{0}$ describing the flow of points on classical trajectories, but the reason why it does not lies in the fact that we do not work with Laplace-Beltrami factor ordering.
If such factor ordering were used then eq.~\eqref{QGR_SC_ContA} would take on the following form
    \begin{equation}
        \delta_{\ssst A}\left(\frac{1}{\sqrt{\mirl{\mathscr{G}}}}
        \mirl{\mathscr{G}}^{\ssst AB} \mathcal{A}_{0}^2 \delta_{\ssst B}S^{\srm{E}\chi_{0}}_{0}\right) = 0\ ,
    \end{equation}
where $\sqrt{\mirl{\mathscr{G}}}$ is the square root of the determinant of the inverse DeWitt supermetric.
Then this equation could be interpreted as the continuity equation.

Lastly, defining 
    \begin{equation}
    \label{QGR_SC_BerryConn}
        \mathcal{B}_{\ssst B} := \rm{Im}\left\langle \delta_{\ssst B}\right\rangle\ ,
    \end{equation}
which is called \textit{the Berry connection},
and taking the real part of eq.~\eqref{QGR_SC_BO2a} at order $\Lr^{0}$ one obtains
    \begin{equation}
        \label{QGR_SC_S1andBR}
        \frac{2}{\hbar} \mirl{\mathscr{G}}^{\ssst AB}\delta_{\ssst A}S^{\srm{E}\chi_{0}}_{0}
       \left(\delta_{\ssst B}S^{\srm{E}\chi_{0}}_{1}
       + \hbar\mathcal{B}_{\ssst B}\right)
        =
        - \left\langle \mathcal{\qHo{\chi}}
        \right\rangle\ ,
    \end{equation}
where we have absorbed the missing potential term into 
    \begin{align}
    \label{QGR_SC_Backreact}
        \left\langle \mathcal{\qHo{\chi}}
        \right\rangle
        & \equiv 
        \left\langle - \frac{\hbar^2}{2}
        \delta^2_{\chi\chi}
        + \frac{1}{2}V^{\chi}_{c}
        \right\rangle\ ,\\[12pt]
    \label{QGR_SC_Vrecovered}
        V^{\chi}_{c} & =  
        \frac{1}{2}U^{\chi}_{c}
        + \frac{1}{\hbar}\tilde{\mirl{\mathscr{G}}}^{\ssst AB}_{1}\delta_{\ssst A}S^{\srm{E}\chi_{0}}_{0}\delta_{\ssst B}S^{\srm{E}\chi_{0}}_{0}\ .
    \end{align}
By including the non-minimal coupling term into the potential we have recovered the Hamiltonian constraint operator of the scalar density field $\chi$.
Equation~\eqref{QGR_SC_Backreact} represents what is called \textit{backreaction} and we shall come back to it shortly in more detail.
It should be noted that in a more general case of non-conformal coupling few other terms contribute to recover the correct $\mathcal{\qHo{\chi}}$; these additional terms are all dependent on $\delta_{a}S^{\srm{E}\chi_{0}}_{0}$ because precisely those are eliminated by conformal coupling in the present case.
Hence we claim without proof that in the more general case one can still recover eq.~\eqref{QGR_SC_Backreact}.

Let us now turn to eq.~\eqref{QGR_SC_BO3}, i.e. the equation for $\psi$.
Using eqs.~\eqref{QGR_SC_DeWittMet_App}-\eqref{QGR_SC_delPhi} and \eqref{QGR_SC_BerryConn} in there and neglecting all terms of order $\mathcal{O}(\Lr^{-2})$ and lower, we have
    \begin{align}
    \label{QGR_SC_Sch0}
        2i\mirl{\mathscr{G}}^{\ssst AB}\delta_{\ssst A}S^{\srm{E}\chi_{0}}_{0}
        \Bigbr{
        \delta_{\ssst B}
        -
        i\mathcal{B}_{\ssst B}
        }\psi
        & = \left(
        \mathcal{\qHo{\chi}}
        - \left\langle \mathcal{\qHo{\chi}}
        \right\rangle
        \right)\psi\ ,
    \end{align}
where we have also used eqs.~\eqref{QGR_SC_Backreact} and \eqref{QGR_SC_Vrecovered}, by adding and subtracting the missing non-minimal term in the RHS of the equation in order to complete the potential $V^{\ssst\chi}$.
Note that eq.~\eqref{QGR_SC_Sch0} is invariant under the following phase transformation of $\psi$
    \begin{equation}
    \label{QGR_SC_unitTr}
        \psi \rightarrow e^{i\phi[q^{\ssst A}]}\psi\ ,
    \end{equation}
where $\phi$ is real, because the Berry connection transforms as (cf. eq.~\eqref{QGR_SC_BerryConn})
    \begin{equation}
    \label{QGR_SC_BerryTr}
        \mathcal{B}_{\ssst A} \rightarrow \mathcal{}B_{\ssst A} + \delta_{\ssst A}\phi\ ,
    \end{equation}
thereby ensuring that curly brackets in eq.~\eqref{QGR_SC_Sch0} are unchanged.
This motivates one to treat the combination in the curly brackets in eq.~\eqref{QGR_SC_Sch0} as a kind of a covariant derivative~\cite{KiefWich}.
Now, observe that the Berry connection term $\mathcal{B}_{\ssst B}$ and the backreaction term in eq.~\eqref{QGR_SC_Sch0} precisely add up to eq.~\eqref{QGR_SC_S1andBR}, which can be used to give
    \begin{align}
    \label{QGR_SC_Sch1}
        2i\mirl{\mathscr{G}}^{\ssst AB}\delta_{\ssst A}S^{\srm{E}\chi_{0}}_{0}
        \Bigbr{
        \delta_{\ssst B}
        +
        \frac{i}{\hbar}\delta_{\ssst B}S^{\srm{E}\chi_{0}}_{1}
        }\psi
        & = 
        \mathcal{\qHo{\chi}}
        \psi\ .
    \end{align}
Hence, if eq.~\eqref{QGR_SC_Sch0} is invariant under eq.~\eqref{QGR_SC_unitTr}, then eq.~\eqref{QGR_SC_Sch1} is also invariant.
But if that is so, then it follows that $S^{\srm{E}\chi_{0}}_{1}$ must transform as
    \begin{equation}
    \label{QGR_SC_S1Tr}
        S^{\srm{E}\chi_{0}}_{1} \rightarrow S^{\srm{E}\chi_{0}}_{1} - \hbar\phi\ .
    \end{equation}
This in turn implies that eq.~\eqref{QGR_SC_S1andBR} stays invariant under a unitary transformation in eq.~\eqref{QGR_SC_unitTr}, because eq.~\eqref{QGR_SC_BerryTr} induces eq.~\eqref{QGR_SC_S1Tr} such that $\phi$ cancels.
Looking closely, one deduces that such a unitary transformation is just shifting a phase of $\psi$ at the expense of the phase of $\Phi$ in eq.~\eqref{QGR_SC_BOans}.
It is interesting that the choice of $q^{\ssst A}$-dependent phase of $\psi$ requires $S^{\srm{E}\chi_{0}}_{1}$ to change but not $S^{\srm{E}\chi_{0}}_{0}$.
This shows that $S^{\srm{E}\chi_{0}}_{1}$ is directly related to the presence of quantum matter.
We now turn to the interpretation of the main equations obtained in this subsection.

\subsection{The Hamilton-Jacobi equation, the functional Schr\"odinger equation and the WKB-evolution paramter}
\label{subs_HJSchWKBtime}

The semiclassical approximation leads to two equations relevant for the description of quantum matter fields propagating on a curved spacetime background.

As already mentioned, eq.~\eqref{QGR_SC_EHJ} is the EHJ equation, whose solution $S^{\srm{E}\chi_{0}}_{0}$ is related to the action consisting of the EH action and $\chi_{0}$-matter action.
Quantum effects of the quantized matter $\chi$ do not contribute here.
The classical Hamiltonian constraint is recovered if one identifies
    \begin{equation}
    \label{QGR_SC_momenta0}
        p_{\ssst A}^{(0)} :=  \Lr^{2}\delta_{\ssst A}S^{\srm{E}\chi_{0}}_{0}
    \end{equation}
with the \textit{classical} momenta, as hinted around eq.~\eqref{QGR_SC_momenta}, leading to
    \begin{equation}
    \label{QGR_SC_EHJm}
        \frac{1}{\Lr^{2}\hbar}\mirl{\mathscr{G}}^{\ssst AB}p_{\ssst A}^{(0)}p_{\ssst B}^{(0)}
        + \frac{\Lr^2\hbar}{2}U^{\ssst q}
        = 0
        \ .
    \end{equation}
This in principle recovers the classical GR.
How does backreaction change the classical momenta? 
Using eqs.~\eqref{QGR_SC_ContA} and~\eqref{QGR_SC_Backreact} in eq.~\eqref{QGR_SC_BO2a}, adding and subtracting 
    \begin{equation}
    \label{QGR_S1Ba_add}
        \frac{\hbar}{\Lr^{2}}\mirl{\mathscr{G}}^{\ssst AB}\left(\delta_{\ssst A}S^{\srm{E}\chi_{0}}_{1} + \hbar\mathcal{B}_{\ssst A}\right)
        \left(\delta_{\ssst B}S^{\srm{E}\chi_{0}}_{1} + \hbar\mathcal{B}_{\ssst B}\right)
    \end{equation}
from it and aiming to complete the square, one can rewrite eq.~\eqref{QGR_SC_BO2a} as
    \begin{equation}
    \label{QGR_SC_BO2b}
        \frac{\Lr^{2}}{\hbar}\mirl{\mathscr{G}}^{\ssst AB}\Bigbr{\delta_{\ssst A}
        S^{\srm{E}\chi_{0}}_{(1)}
        +\hbar\mathcal{B}_{\ssst A}
        }
        \Bigbr{
        \delta_{\ssst B}S^{\srm{E}\chi_{0}}_{(1)}
        +\hbar\mathcal{B}_{\ssst B}
        }
        + \frac{\Lr^2\hbar}{2}U^{\ssst q}
        = - \left\langle \qHo{\chi}\right\rangle
        \ ,
    \end{equation}
where
    \begin{equation}
    \label{QGR_SC_S1}
        S^{\srm{E}\chi_{0}}_{(1)} := S^{\srm{E}\chi_{0}}_{0} + \Lr^{-2}S^{\srm{E}\chi_{0}}_{1}\ .
    \end{equation}
One would be tempted to identify the \textit{corrected} momenta as $\Lr^{2}\delta_{\ssst A}S^{\srm{E}\chi_{0}}_{(1)}$, but this momenta is not invarint under a phase transformation of $\psi$, see eq.~\eqref{QGR_SC_S1Tr}.
However, since the Berry connection also transforms according to eq.~\eqref{QGR_SC_BerryTr}, the entire curly bracket in eq.~\eqref{QGR_SC_BO2b} should be identified as the \textit{corrected momenta},
    \begin{equation}
    \label{QGR_SC_momenta1}
        p_{\ssst A}^{(1)} :=  \Lr^{2}\delta_{\ssst A}S^{\srm{E}\chi_{0}}_{(1)} + \hbar\mathcal{B}_{\ssst A}\ ,
    \end{equation}
so that $p_{\ssst A}^{(1)} = p_{\ssst A}^{(0)} + \delta_{\ssst A}S^{\srm{E}\chi_{0}}_{1} + \hbar\mathcal{B}_{\ssst A}$ is invariant under the $q^{\ssst A}$-dependent phase transformation of $\psi$.
Equation~\eqref{QGR_SC_BO2b} is correct if terms of order $\mathcal{O}(\Lr^{-2})$ are neglected, and can be written in terms of the corrected momenta as
    \begin{equation}
    \label{QGR_SC_HcBR}
        \frac{1}{\Lr^{2}\hbar}\mirl{\mathscr{G}}^{\ssst AB}p_{\ssst A}^{(1)}p_{\ssst B}^{(1)}
        + \frac{\Lr^2\hbar}{2}U^{\ssst q}
        = - \left\langle \qHo{\chi}\right\rangle
        \ .
    \end{equation}
This equation is the EHJ equation \textit{with backreaction} $\left\langle \qHo{\chi}\right\rangle$.
It should correspond (up to certain rescalings) to the $00$ component of the SEE given by eq.~\eqref{SEE_full}.

But it does not. Namely, if the methods of regularization are applied to $\left\langle \qHo{\chi}\right\rangle$ in order to isolate the divergent quantities as explained in section~\ref{subs_HDSEE}, these divergent quantities cannot be absorbed into the corresponding counter-terms because eq.~\eqref{QGR_SC_HcBR} \textit{does not contain higher-derivative counter-terms}.
The problem stems from the same issue that was encountered in eq.~\eqref{SEE_eqn} and now we have seen how it carries over into the semiclassical approximation of the quantized theory.
We see that counter-terms still need to be \textit{added by hand} into otherwise (formally) consistent derivation of the semiclassical EHJ equation, in order to absorb the divergencies.
Furthermore, divergences appear also in eq.~\eqref{QGR_SC_S1andBR} and eq.~\eqref{QGR_SC_Sch0}.
These two equations have the Berry connection in common, so it would be plausible that the Berry connection has something to do with counter-terms.
Since $S^{\srm{E}\chi_{0}}_{1}$ cannot be determined without the Berry connection and the backreaction, and since it represents the correction to $S^{\srm{E}\chi_{0}}_{0}$ due to the backreaction, it seems that counter-terms could also be sought in $S^{\srm{E}\chi_{0}}_{1}$ too.

The other important equation is eq.~\eqref{QGR_SC_Sch0}, or equivalently eq.~\eqref{QGR_SC_Sch1}; this is the equation for $\psi$.
It can be read as: the rate of change of $\psi$ with respect to $q^{\ssst A}$ variables projected along $\delta_{\ssst A}S^{\srm{E}\chi_{0}}_{0}$ is proportional to the Hamiltonian acting on $\psi$.
(Let us not forget that the derivations discussed here take place under the integral in eq.~\eqref{QGR_HPsi}.
Therefore, eq.~\eqref{QGR_SC_Sch1} should be accompanied by the contributions from the momentum constraint, eq.~\eqref{QGR_qHi}.
In order to keep things simple, we shall proceed as if $N^{i}=0$, so that the additional terms are not included explicitly.
This won't affect the main point of this review discussion.)
This \textit{evolution} of $\psi$ is interpreted as the \textit{functional Schr\"odinger equation} if one defines a functional $\bar{\tau}$ in one of the following two ways,
    \begin{align}
    \label{QGR_SC_TauBR}
        \mathscr{D}_{\bar{\tau}'} & := \frac{2\Nb}{\hbar}\mirl{\mathscr{G}}^{\ssst AB}\delta_{\ssst A}S^{\srm{E}\chi_{0}}_{0}
        \Bigbr{
        \delta_{\ssst B}
        -
        i\mathcal{B}_{\ssst B}
        } \ ,\\[12pt]
    \label{QGR_SC_Tau}
        \ddel{}{\bar{\tau}} & := \frac{2\Nb}{\hbar}\mirl{\mathscr{G}}^{\ssst AB}\delta_{\ssst A}S^{\srm{E}\chi_{0}}_{0}
        \delta_{\ssst B}\ ,
    \end{align}
with which eq.~\eqref{QGR_SC_Sch0} and eq.~\eqref{QGR_SC_Sch1} are rewritten as
    \begin{align}
    \label{QGR_SC_SchTauBR}
        i\hbar\mathscr{D}_{\bar{\tau}'}\psi & = \Nb\left(\qHo{\chi} - \left\langle\qHo{\chi}\right\rangle\right)\psi\ ,\\[12pt]
    \label{QGR_SC_SchTau}
        i\hbar\ddel{\psi}{\bar{\tau}} - \ddel{S^{\srm{E}\chi_{0}}_{1}}{\bar{\tau}}\psi & = \Nb\qHo{\chi}\psi\ .
    \end{align}
Note that eq.~\eqref{QGR_SC_SchTauBR} contains what may be called a ``covariant derivative'' defined by eq.~\eqref{QGR_SC_TauBR}.
Also note that the extra term in eq.~\eqref{QGR_SC_SchTau} can be eliminated by defining
    \begin{equation}
    \label{QGR_SC_PsiTild}
        \tilde{\psi} := e^{\frac{i}{\hbar}S^{\srm{E}\chi_{0}}_{1}}\psi\ ,
    \end{equation}
which leads to
    \begin{equation}
    \label{QGR_SC_SchPlain}
        i\hbar\ddel{\tilde{\psi}}{\bar{\tau}} = \Nb\qHo{\chi}\tilde{\psi}\ .
    \end{equation}
Both eq.~\eqref{QGR_SC_SchTauBR} and eq.~\eqref{QGR_SC_SchTau} are invariant under the $q^{\ssst A}$-dependent phase transformation of $\psi$ if one recalls that $S^{\srm{E}\chi_{0}}_{1}$ is also required to change accordingly, such that the total wave functional in eq.~\eqref{QGR_SC_BOans} remains unchanged.
Moreover, $\tilde{\psi}$ itself is invariant under the phase transformation of $\psi$, so one must be careful in interpreting it as the wave functional on the same footing as $\psi$.
The $\bar{\tau}$ functional is interpreted as the evolution parameter along a classical trajectory described by $S^{\srm{E}\chi_{0}}_{0}$ and is usually called ``WKB time'' or ``bubble time''.
It has nothing to do with the coordinate time and one must be careful not to mix the two before some more considerations have been made.
This evolution parameter is determined by the background (``heavy'') variables, intrinsic to the hypersurface itself.
It is important to note that each observer on the hypersurface has their own $\bar{\tau}$ (i.e. their own ``bubble'' in which they write their own evolutions of $\psi$).
Since there are infinitely many observers related by diffeomorphisms, there are infinitely many equations of the form of eq.~\eqref{QGR_SC_SchTauBR}, eq.~\eqref{QGR_SC_SchTau} and eq.~\eqref{QGR_SC_SchPlain}.
Only upon integration of eq.~\eqref{QGR_SC_SchTauBR} or eq.~\eqref{QGR_SC_SchTau} one obtains the following forms of the Schr\"odinger equation
    \begin{align}
    \label{QGR_SC_SchInt}
        i\hbar\dd{\psi}{t'} & := i\hbar\intx\, \,\mathscr{D}_{\bar{\tau}'}\psi = \intx \Nb\left(\qHo{\chi} - \left\langle\qHo{\chi}\right\rangle\right)\psi\ ,\\[12pt]
    \label{QGR_SC_SchIntS1}
        i\hbar\dd{\psi}{t} & := i\hbar\intx\,\left(\ddel{\psi}{\bar{\tau}} - \ddel{S^{\srm{E}\chi_{0}}_{1}}{\bar{\tau}}\right)\psi  =  \intx\,\Nb\qHo{\chi}\psi
    \end{align}
where one defines $t'$ or $t$ to be the usual coordinate time, \textit{after} fixing the coordinate gauge by choosing $\Nb$ (and $N^{i}$, if the contribution from the momentum constraint is properly included, as it should be).
It is in this way that one recovers the notion of time from a timeless quantum gravity theory.
Time emerges from a semiclassical approximation to QGDGR.
It should be emphasized that dependency of definitions of the WKB evolution parameter eq.~\eqref{QGR_SC_TauBR} and eq.~\eqref{QGR_SC_Tau} on $S_{0}^{\srm{E\chi_0}}$ implies that each classical solution to the EHJ equation~\eqref{QGR_SC_EHJm} gives rise to its own time evolution.
These are then called the ``WKB branches''.
Furthermore, the full wave functional $\Psi$ is then a superposition of components such as eq.~\eqref{QGR_WKBans}.
That means there is, in principle, interference between different WKB branches, which then raises the question ``how does this interference disappear to give the single observable classical Universe?''.
The answer can be given using the program of \textit{decoherence}~\cite{Joos,Schloss}, which explains the emergence of a classical world from a quantum world in a continuous manner.
We shall not go into details of decoherence here.

It is not important whether equation eq.~\eqref{QGR_SC_SchTauBR} or eq.~\eqref{QGR_SC_SchTau} is integrated, because it is $t$ that one ultimately uses as an evolution parameter.
As for eq.~\eqref{QGR_SC_SchPlain}, it is not clear to us what would be the meaning of $\tilde{\psi}$ in it and whether such a wavefunctional conveys the same information as $\psi$ in the integrated Schr\"odinger equation.
The important point is that the same problem appears here as in eq.~\eqref{QGR_SC_HcBR}:
eq.~\eqref{QGR_SC_SchInt} gives rise to \textit{divergences} on the RHS once one tries to evaluate the backreaction.
In order to absorb these divergencies using the methods of renormalization, one deduces that counter-terms have to be contained in $\mathcal{B}_{\ssst B}$ in eq.~\eqref{QGR_SC_TauBR} or, equivalently, in $S^{\srm{E}\chi_{0}}_{1}$ in eq.~\eqref{QGR_SC_Tau}.
The current state of matters in the canonical quantum gravity and the semiclassical approximation scheme does not offer means of formulating the EHJ and the Schr\"odinger equation with counter-terms\footnote{We note again that Feng's work~\cite{Feng} offers one way of staying with canonical quantum gravity and still finding the counter-terms by dealing with the yet unsolved problem of regularizing the second functional derivatives evaluated at the same point.}.

The backreaction-corrected EHJ equation given by eq.~\eqref{QGR_SC_HcBR} and the Schr\"odinger equation given by eq.~\eqref{QGR_SC_SchInt} constitute the two equations of quantum field theory on curved spacetimes, i.e. the semiclassical theory of gravity and quantum fields.
Since we have not found the results of this semiclassical approximation scheme satisfactory due to the absence of counter-terms, we would like to offer a way to address this issue in the remainder of this chapter.


\section{Quantum geometrodynamics of a general quadratic curvature theory}
\label{QGDHD}

In this section we shall apply canonical quantization on a general quadratic curvature gravity in unimodular-conformal variables formulated as a Hamiltonian theory in sections~\ref{sec_HHDall} and~\ref{sec_HWE}.
It was argued in the previous chapter that a higher-derivative theory which is first perturbatively constrained at the classical level and then quantized is not satisfactory.
The reason is, we recall, that a quantum theory is a high-energy entity while a perturbatively constrained theory is a low-energy entity and the two approaches conceptually contradict each other.
Recalling sections~\ref{subs_HDLHO} and~\ref{subs_Ham_osc}, Richard cannot quantize his Lagrangian because it is perturbatively interpreted --- on the other hand, Emmy can only make sense of her Lagrangian if it describes a full quantum theory.
To make a transition from Emmy's to Richard's theory, a careful construction of semiclassical approximation must take place.
We are in Richard's shoes and in this section we are trying to make sense of the exact (i.e. not perturbatively constrained) higher-derivative theory given by eq.~\eqref{SEE_TheAction} as a quantum gravity theory.
Therefore, we shall present what we'll call \textit{quantization before perturbation} (QbP) method of formulating a quantum higher-derivative theory.
This is in high contrast to what Mazzitelli~\cite{Mazzit} did: he used \textit{perturbation before quantization} (PbQ) approach to quantize Emmy's higher-derivative theory. 
His result --- which he obtained directly after quantizing the already perturbatively constrained theory --- resembled the form of the WDW equation (i.e. eq.~\eqref{QGR_qHo}) \textit{corrected} by contributions from the quadratic curvature terms.
These corrections were non-linear in $p_{\ssst A}$ momenta (of third and fourth power), which means that they turned into third and fourth order derivatives of the wave functional after the quantization.
He then applied a WKB-type approximation (of the form of eq.~\eqref{QGR_SC_Sexp}) but in \textit{three perturbation parameters}: $\Lr^{-2},\aw$ and $\br$, and recovered the correct form of the SEE \textit{with} perturbatively reduced counter-terms.
It is important to note that Mazzitelli applied perturbative order reduction in $\aw$ and $\br$ \textit{two} times: once before and once after the quantization.
Nevertheless, on one hand, this is a remarkable result and to our knowledge one of a kind in the literature.
But on the other hand, this results is expected because his WKB approximation simply follows what has already been implemented in the unquantized classical theory.
Hence, our alternative approach of QbP avoids saying anything about the perturbative nature of the higher-derivative terms before the quantization.
We shall see that only a semiclassical approximation (which also uses the same three perturbation parameters $\Lr^{-2},\aw$ and $\br$) then has the necessary power to tell us the meaning and rule of the higher-derivative terms.

\subsection{Higher-derivative Wheeler-DeWitt equation}
\label{subs_HDWDW}

Canonical quantization of the higher-derivative theory proceeds much in the same way as QGDGR.
One may choose to quantize the constraints or to quantize the diffeomorphism generators given by eq.~\eqref{G_GdiffPSS}.
These are equivalent procedures, as it is the case with quantization of GR described in the previous section.
However, the result of the quantization of higher-derivative theories is quite different compared to QGDGR.
The most important fact, which underlays all of the differences, is that one is dealing with an \textit{extended} configuration space.
This means that the wave functional contains additional dependence on $\kb$ and $\bktb$, compared to the wave functional in QGDGR:
    \begin{equation}
        \Psi \equiv \Psi[a,\bhb,\kb,\bktb,\chi]\ .
    \end{equation}
But before we continue with quantization, we shall modify the higher-derivative theory slightly by adding another matter action, which will play the role of a background matter field as one of the ``heavy'' variables.
This is just the same thing we did in QGDGR in section~\ref{subs_QGR_SCbr1} in order to prepare the grounds for having a non-vacuum ``heavy'' sector.
The drastic difference with the cases of GR and QGDGR is that in QGDHD there is no cross term that mixes the geometric and matter momenta.
This fact holds because $\ktb_{ij}$ and $\kb$ are \textit{not related} to $\pb^{ij}$ and $p_{a}$, so terms such as $\sim\kb \dot{\chi}$ turn into $\sim\kb p_{\chi}$, instead into $\sim p_{a} p_{\chi}$ as in the case of GR (cf. kinetic terms in eq.~\eqref{QGR_Hchichi}).
It is for this reason that one has $V^{\chi}$, instead of $U^{\chi}$, in eq.~\eqref{HDall_Hofin}.
Hence, we simply add the following term
    \begin{equation}
    \label{QHD_Hchi0}
        \frac{1}{2\Lr^2}p_{\chi_{0}}^2
        - 6\xi_{c}\kb\chi_{0} p_{\chi_{0}}
        + \frac{\Lr^{2}}{2}V^{\chi_{0}}
    \end{equation}
to extend eq.~\eqref{HDall_Hofin} with a background matter field Hamiltonian, and the following term
    \begin{equation}
    \label{QHD_Hichi0}
        - \frac{1}{3}\left(\chi\del_{i}p_{\chi_{0}}
        - 2\del_{i}\chi_{0} \, p_{\chi_{0}}\right)
    \end{equation}
to extend eq.~\eqref{HDall_Hifin} with the corresponding contribution to the momentum constraint.
We shall not assume conformal coupling for any of the matter fields in the present case and this will allow for more general conclusions.
Taking into account this additional matter field, the wave functional can be written as
    \begin{equation}
        \Psi \equiv \Psi[q^{\ssst A},Q^{\ssst I},\chi]\ .
    \end{equation}
where $q^{\ssst A} := \{a,\bhb,\chi_{0}\}$, $Q^{\ssst I} := \{\kb,\bktb\}$ and indices ${\ssst A}=\{a,\bhb,\chi_{0}\}$, ${\ssst I}=\{\kb,\bktb\}$.
Variables $Q^{\ssst I}$ are the components of the extrinsic which extend the ``heavy'' configuration space spanned by $q^{\ssst A}$.    
    
Due to extended configuration space and the additional matter field $\chi_{0}$, in addition to eqs.~\eqref{QGR_qNp}-\eqref{QGR_CM2}, one has the following quantization rules
    \begin{align}
       \label{QHD_qKPt}
        \hat{\bar{\mathbf{K}}}^{\srm T}(\bx)\Psi & =
        \bktb(\bx)\cdot\Psi , & \hat{\bPb}(\bx)\Psi & = \frac{\hbar}{i}
        \frac{\delta}{\delta \bktb (\bx)}\Psi,\\[6pt]
    \label{QHD_qKP}
        \hat{\kb}(\bx)\Psi & = \kb(\bx)\cdot\Psi  , & \hat{\Pb}(\bx)\Psi & = \frac{\hbar}{i}
        \frac{\delta}{\delta \kb(\bx)}\Psi , \\[6pt]
    \label{QHD_qchip}
        \hat{\chi}_{0}(\bx)\Psi & =
        \chi_{0}(\bx)\cdot\Psi , & \hat{p}_{\chi_{0}}(\bx)\Psi & = \frac{\hbar}{i}
        \frac{\delta}{\delta \chi_{0} (\bx)}\Psi
    \end{align}
and the following commutation relations
    \begin{align}
    \label{QHD_CMKP}
        \CM{\hat{\bar{K}}^{\srm T}_{ij}(\bx)}{\hat{\Pb}^{ab}(\by)}\Psi
         = i\hbar\,\mathbb{1}^{{\srm{T}}ab}_{(ij)}\,\delta
        (\bx,\by)\Psi\ ,&\qquad \CM{\hat{\kb}(\bx)}{\hat{\Pb}(\by)}\Psi
         = i\hbar\,\delta (\bx,\by)\Psi\ ,\\[12pt]
    \label{QHD_CMchi0}
        \CM{\hat{\chi}_{0}(\bx)}{\hat{p}_{\chi_{0}}(\by)}\Psi &= i\hbar\,\delta (\bx,\by)\Psi\ ,
    \end{align}
all other commutators vanishing.
It should be noted that the commutators in quantized higher-derivative theory are just promoted Poisson brackets in eq.~\eqref{HDall_PBvars1}-eq.~\eqref{HDall_PBvars3} \textit{if} the theory contains first-class constraints only.
If, however, a theory contained second-class constraints, such as the WE theory described in section~\ref{sec_HWE}, then one must promote Dirac brackets in eqs.~\eqref{HWchi_DBKP}-\eqref{HWchi_DBKpchi} to commutators.

In the higher-derivative theory described in section~\ref{sec_HHDall} which contains both the $R^2$ and the $C^2$ term the quantization proceeds by quantizing the constraints in the same way as in QGDGR.
The Hamiltonian and momentum constraints given by eq.~\eqref{HDall_Hofin} and eq.~\eqref{HDall_Hifin}, extended by eq.~\eqref{QHD_Hchi0} and eq.~\eqref{QHD_Hichi0} respectively give rise to the following two quantum equations,
    \begin{align}
    \label{QHD_WDW}
        \qHo{ERW\chi_{0}\chi}\Psi & := \left[
        \qHo{Q}
        + \qHo{q}
        + \qHo{\chi}
        \right]\Psi = 0\ ,\\[12pt]
    \label{QHD_WDWm}
        \qHi{ERW\chi_{0}\chi}\Psi & := \left[
        \qHi{Q}
        + \qHi{q}
        + \qHi{\chi}
        \right]\Psi = 0\ ,
    \end{align}
where
    \begin{subequations}
    \begin{align}
    \label{QHD_H0Q}
        \qHo{Q}\Psi & :=
        \left[-\hbar\mathring{\mirl{\mathscr{G}}}^{\ssst IJ}_{\ssst{\aw,\br}}\delta^2_{\ssst IJ}
        - \frac{\hbar}{i}\mathcal{D}^{\ssst I}\delta_{\ssst I}
        - \aw\hbar\bar{\mathbf{C}}^{\srm B}\cdot\bar{\mathbf{C}}^{\srm B}\right]\Psi
        \ ,\\[12pt]
    \label{QHD_H0met}
        & \qquad \mathring{\mirl{\mathscr{G}}}^{\ssst IJ}_{\ssst{\aw,\br}} :=\left(
            \begin{matrix}
                \frac{1}{\br} & 0 \\
                0 & -\frac{1}{2\aw}\hb^{ik}\hb^{jl} 
            \end{matrix}
            \right)\ ,\\[12pt]
    \label{QHD_H0D}
        & \qquad \mathcal{D}^{\ssst I} := \left(\mathcal{D}_{\srm R}^2\quad
        - \bm{\mathcal{D}}^2_{\srm W}\right)\ ,\\[18pt]
    \label{QHD_H0q}
        \qHo{q}\Psi & := \left[
        \frac{\hbar}{i}\left(a\kb \delta_{a}
        + 2\bktb\cdot\delta_{\bhb}\right)
        - \frac{\hbar^2}{2\Lr^{2}}\delta^2_{\chi_{0}\chi_{0}}
        - \frac{6\hbar}{i}\xi_{c}\kb\chi_{0} \delta_{\chi_{0}}
        + \frac{\Lr^{2}\hbar}{2}V^{q}\right]\Psi
        \ ,\\[12pt]
    \label{QHD_Vq}
        & \qquad V^{\ssst q}  := - V^{\srm{E}} + \frac{1}{\hbar}V^{\ssst \chi_{0}}\ ,\\[12pt]
    \label{QHD_Ve}
        & \qquad V^{\srm E} := a^2\Big(a^2\left(\,^{\ssst (3)}\! R 
        - 2\bar{\Lambda}\right) 
        + \bktb\cdot\bktb 
        - 6\kb^2\Big)\ ,\\[18pt]
    \label{QHD_H0chi}
        \qHo{\chi}\Psi& := \left[- 
        \frac{\hbar^2}{2}\delta^2_{\chi\chi}
        - \frac{6\hbar}{i}\xi_{c}\kb\chi \delta_{\chi}
        + \frac{1}{2}V^{\chi}
        \right]\Psi\ ,
    \end{align}
    \end{subequations}
and
    \begin{subequations}
    \begin{align}
    \label{QHD_HiQ}
        \qHi{Q}\Psi & := \frac{\hbar}{i}\Biggpar{
        \ddel{\Psi}{\ktb_{jk}} \bar{D}_{i}\ktb_{jk} 
        - 2 \bar{D}_{j}\left(\ktb_{ik} \ddel{\Psi}{\ktb_{jk}} \right)
        + \frac{1}{3}\del_{i}\left(\ktb_{jk}\ddel{\Psi}{\ktb_{jk}}\right)\nld
        &\quad
        + \ddel{\Psi}{\kb} \del_{i}\kb 
        - \del_{i}\left(\kb\ddel{\Psi}{\kb}\right)
        }\ , \\[18pt]
    \label{QHD_Hiq}
        \qHi{q}\Psi & := \frac{\hbar}{i}\Biggpar{
        2\bar{D}_{j}\left(\hb_{ik}\ddel{\Psi}{\hb_{kj}}\right)
        +\frac{1}{3}D_{i}\left(a\,\ddel{\Psi}{a}\right)
        +\frac{1}{3}\left(\chi_{0}\del_{i}\ddel{\Psi}{\chi_{0}}
        - 2\del_{i}\chi_{0} \, \ddel{\Psi}{\chi_{0}}\right)
        }\ ,\\[18pt]
    \label{QHD_Hichi}
        \qHi{\chi}\Psi & := \frac{\hbar}{3i}\left(\chi\del_{i}\ddel{\Psi}{\chi} - 2\del_{i}\chi \, \ddel{\Psi}{\chi}\right)\ .
    \end{align}
    \end{subequations}

Equation~\eqref{QHD_WDW} will be referred to as \textit{the higher-derivative Wheeler-DeWitt} (HDWDW) equation.
We separated the equation into three parts: eq.~\eqref{QHD_H0Q} contains the derivatives with respect to $\kb$ and $\bktb$ only and the Weyl-tensor potential.
Note that we have introduced in eq.~\eqref{QHD_H0met} an upper-index inverse DeWitt supermetric of the higher-derivative sector.
The terms linear in momenta (here derivatives with respect to $a, \bhb$) --- whose importance at the classical level has been emphasized in section~\ref{subs_LinTerms} --- are grouped together with the Hamiltonian of the ``heavy'' $\chi_{0}$ field in eq.~\eqref{QHD_H0q}.
Even though these terms are generated by constraints which came to aid to Hamilton-formulate the higher-derivative theory (cf. eqs.~\eqref{HDall_KtlessC}-\eqref{HDall_KtC}), the crucial role that they play in semiclassical approximation will justify grouping them in eq.~\eqref{QHD_H0q}, as we shall see in the following subsections.
Also note that $V^{\srm E}$ --- which is proportional to the ADM Lagrangian of GR --- is contained in the same equation.
The last component of the HDWDW equation is given by eq.~\eqref{QHD_H0chi}, which is just the Hamiltonian constraint operator of the ``light'' sector $\chi$ field.
Note that both matter fields have the full potential introduced in eq.~\eqref{Vchi_def}, because the non-minimally coupled extrinsic curvature terms do not participate in the formulation of the $p_{a}$ and $\bpb$ momenta since they are treated as independent variables.
Contrast this to the case of GR, eq.~\eqref{QGR_Hchichi}, where these non-minimal coupling terms are migrated to the kinetic term, leaving eq.~\eqref{Uchi_def} instead of eq.~\eqref{Vchi_def} for the potential of the matter fields.
In short --- and this is important for later discussion --- the presence of linear terms made $\kb$ and $\bktb$ explicitly appear in potentials $V^{\ssst \chi_{0}}, V^{\ssst \chi}$ and $V^{\srm E}$, denying their relation to the momenta $p_{a}$ and $\bpb$.

Equation eq.~\eqref{QHD_WDWm}, with individual terms in eqs.~\eqref{QHD_HiQ}-\eqref{QHD_Hichi}, is different from eq.~\eqref{QGR_qHi} only in that it contains additional terms referring to the higher-derivative degrees of freedom $\kb$ and $\bktb$.
Its interpretation is just a generalization of diffemorphic invariance of $\Psi$ from QGDGR, namely, that $\Psi$ is invariant under the spatial coordinate transformations not only in variables $a,\bhb,\chi_{0}$ and $\chi$ but now also in $\kb$ and $\bktb$. 

QGDHD suffers from the same problems as QGDGR.
However, the ordering ambiguity has a slightly different flavor not ony because of the presence of terms linear in momenta but also because the DeWitt supermetric does not depend on the extrinsic curvature so things seem a bit simpler as far as the kinetic term in the HDWDW equation is concerned.
Perhaps the most important ``problem'' is the interpretation of dependence of $\Psi$ on the additional degrees of freedom carried by $\kb,\bktb$.
These are true dynamical degrees of freedom in the full quantum gravity theory, but what does this mean for the evolution of $\Psi$?
Does it mean that this quantum gravity theory does not ``know'' of the relationship between the intrinsic metric of the hypersurface and its first order change in the timelike direction (as interpreted by a classical observer)? 
We shall not go into this question, but we think that it could be worth investigating the quantized version of the hypersurface algebra in order to gain some additional insight.

We now turn to the formulation of an appropriate semiclassical scheme with an aim to recover the semiclassical gravity and QFT in curved spacetime.

\subsection{Semiclassical approximation: the Born-Oppenheimer type ansatz}

The most important difference between the HDWDW and the WDW is that there is no $\Lr^{-2}$ parameter in the kinetic term of the HDWDW.
In other words, the presence of higher-derivative terms occurs at the same order as the matter field $\chi$, i.e. their kinetic terms enter the HDWDW at the same order.
This would suggest one to separate the part of the wave functional which depends on $q^{\ssst A}$ from a part which depends on all variables, as one does in the case of QGDGR, cf.~eq.~\eqref{QGR_SC_BO2}.
However, we have concluded in the previous chapter (and the previous section would suggests the same) that the HD terms act to correct the ``heavy'' part of the system.
This means that one cannot separate out the contributions of $Q^{\ssst I}$ neither from the ``heavy'' nor from the ``light'' sector --- the higher-derivative terms play the role in both parts because the backreaction appears in both eq.~\eqref{QGR_SC_BO2} and eq.~\eqref{QGR_SC_BO3}.
Hence, we do not put restrictions on separating the dependency on $Q^{\ssst I}$ --- we shall see that certain restrictions naturally follow from the semiclassical approximation we aim to construct.

That being said, we can only use $\Lr^{2}$ as the scale separation parameter at first, since we still cannot say anything certain about the appearance of $Q^{\ssst I}$-dependent terms with respect to $\aw$ and $\br$. (This important point had not yet been observed in the author's Master thesis~\cite{MSc}.)
Therefore, we assume the following BO-type ansatz,
    \begin{align}
    \label{QHD_SC_BOans}
        \Psi[q^{\ssst A}, Q^{\ssst I},\chi] = \Phi[q^{\ssst A}, Q^{\ssst I}] \psi[q^{\ssst A},Q^{\ssst I},\chi]\ ,
    \end{align}
Compared to eq.~\eqref{QGR_SC_BOans}, it is quite a similar ansatz and it also holds that there is freedom to choose this separation by choosing a rescaling factor $\phi\equiv\phi[q^{\ssst A},Q^{\ssst I}]$.

Using eq.~\eqref{QHD_SC_BOans} into the HDWDW equation given by eq.~\eqref{QHD_WDW}, i.e. into eqs.~\eqref{QHD_H0Q}, \eqref{QHD_H0q} and \eqref{QHD_H0chi}, we obtain the following equations
    \begin{subequations}
    \begin{align}
    \label{QHD_WDWQ_BO}
        \qHo{Q}\Psi & =
        -\hbar\mathring{\mirl{\mathscr{G}}}^{\ssst IJ}_{\ssst{\aw,\br}}
        \Bigsq{\psi\delta^2_{\ssst IJ}\Phi
        + 2\delta_{\ssst I}\Phi\delta_{\ssst J}\psi
        + \Phi\delta^2_{\ssst IJ}\psi
        }\nld
        &\quad - \frac{\hbar}{i}\mathcal{D}^{\ssst I}\Bigsq{
        \psi\delta_{\ssst I}\Phi
        + \Phi\delta_{\ssst I}\psi
        }
        - \aw\hbar\bar{\mathbf{C}}^{\srm B}\cdot\bar{\mathbf{C}}^{\srm B}\Phi\psi
        \ ,\\[18pt]
    \label{QHD_WDWq_BO}
        \qHo{q}\Psi & = 
        \frac{\hbar}{i}\Bigsq{
        \psi\left(
        a\kb \delta_{a}\Phi
        + 2\bktb\cdot\delta_{\bhb}\Phi
        \right)
        +
        \Phi\left(
        a\kb \delta_{a}\psi
        + 2\bktb\cdot\delta_{\bhb}\psi
        \right)
        }\nld
        &\quad 
        - \frac{\hbar^2}{2\Lr^{2}}\Bigsq{
        \psi\delta^2_{\chi_{0}\chi_{0}}\Phi
        + 2 \delta_{\chi_{0}}\Phi\delta_{\chi_{0}}\psi
        + \Phi\delta^2_{\chi_{0}\chi_{0}}\psi
        }\nld
        &\quad - \frac{6\hbar}{i}\xi_{c}\kb\chi_{0} \Bigsq{
        \psi\delta_{\chi_{0}}\Phi
        + \Phi\delta_{\chi_{0}}\psi
        }
        + \frac{\Lr^{2}\hbar}{2}V^{q}\Phi\psi
        \ ,\\[18pt]
    \label{QHD_WDWchi_BO}
        \qHo{\chi}\Psi & = \Phi\qHo{\chi}\psi\ .
    \end{align}
    \end{subequations}
We shall continue to analyize the HDWDW equation by analyzing their three parts separately.

In order to take a partial average of the HDWDW equation with respect to the $\chi$-field --- in analogy to what was done in the previous section to obtain eq.~\eqref{QGR_SC_BO2} --- we assume that $\psi$ is normalized as
    \begin{equation}
    \label{QHD_SC_inner}
        \vert\psi\vert^{2} = \int\mathcal{D}[\chi]\psi^{*}[q^{\ssst A},Q^{\ssst A},\chi]\psi[q^{\ssst A},Q^{\ssst A},\chi] = 1\ ,
    \end{equation}
in analogy to eq.~\eqref{QGR_SC_inner}.
This assumption holds as long as we neglect any contributions of order $\mathcal{O}(\Lr^{-2})$ and lower, because all manipulations from now on shall hold at $\mathcal{O}(\Lr^{0})$ order.
Moreover, it should be emphasized that eq.~\eqref{QHD_SC_inner} is also assumed to hold as a \textit{perturbative approximation} up to order $\mathcal{O}(\aw)$ and $\mathcal{O}(\br)$, i.e. terms with higher powers of $\aw$ and $\br$ are excluded from eq.~\eqref{QHD_SC_inner}.
We shall soon define more clearly what does this mean.
Furthermore, in direct analogy to eqs.~\eqref{QGR_SC_paravg}, \eqref{QGR_SC_d} and \eqref{QGR_SC_dd}, we define the same partial averages but with respect to $\psi$ corresponding to the HDWDW equation.
Therefore, we shall simply borrow those three definitions while keeping in mind that $\psi\equiv\psi[q^{\ssst A},Q^{\ssst A},\chi]$ and $\vert\psi\vert^{2}=1$ there.
In addition to those partial averages, we have to introduce the following two definitions
    \begin{align}
    \label{QHD_SC_d}
        \left\langle \delta_{\ssst I}\right\rangle & := \int \mathcal{D}[\chi]\psi^{*}\delta_{\ssst I}\psi\ ,\\[12pt]
    \label{QHD_SC_dd}
        \left\langle \delta^2_{\ssst IJ}\right\rangle & := \int \mathcal{D}[\chi]\psi^{*}\delta_{\ssst I}\delta_{\ssst J}\psi\ ,
    \end{align}
recalling that indices $ I,J = \{ \kb,\bktb \}$.
These terms will appear in what follows.

We now take a partial average of eq.~\eqref{QHD_WDW}, i.e. of eqs.~\eqref{QHD_WDWQ_BO}-\eqref{QHD_WDWchi_BO}, divide by $\Phi$ and obtain
    \begin{subequations}
    \begin{align}
    \label{QHD_WDWQ_BOavg}
        \frac{1}{\Phi}\left\langle\qHo{Q}\Phi\right\rangle & =
        -\hbar\mathring{\mirl{\mathscr{G}}}^{\ssst IJ}_{\ssst{\aw,\br}}
        \Biggsq{\frac{\delta^2_{\ssst IJ}\Phi}{\Phi}
        + 2\frac{\delta_{\ssst I}\Phi}{\Phi}\left\langle\delta_{\ssst J}\right\rangle
        + \left\langle\delta^2_{\ssst IJ}\right\rangle
        }\nld
        &\quad 
        - \frac{\hbar}{i}\mathcal{D}^{\ssst I}\biggsq{
        \frac{\delta_{\ssst I}\Phi}{\Phi}
        + \left\langle\delta_{\ssst I}\right\rangle
        }
        - \aw\hbar\bar{\mathbf{C}}^{\srm B}\cdot\bar{\mathbf{C}}^{\srm B}
        \ ,\\[18pt]
    \label{QHD_WDWq_BOavg}
        \frac{1}{\Phi}\left\langle\qHo{q}\Phi\right\rangle & =
        \frac{\hbar}{i}\biggsq{
        \frac{1}{\Phi}\left(
        a\kb \delta_{a}\Phi
        + 2\bktb\cdot\delta_{\bhb}\Phi
        \right)
        +
        \left(
        a\kb \left\langle\delta_{a}\right\rangle
        + 2\bktb\cdot\left\langle\delta_{\bhb}\right\rangle
        \right)
        }\nld
        &\quad 
        - \frac{\hbar^2}{2\Lr^{2}}\biggsq{
        \frac{\delta^2_{\chi_{0}\chi_{0}}\Phi}{\Phi}
        + 2 \frac{\delta_{\chi_{0}}\Phi}{\Phi}\left\langle\delta_{\chi_{0}}\right\rangle
        + \left\langle\delta^2_{\chi_{0}\chi_{0}}\right\rangle
        }\nld
        &\quad - \frac{6\hbar}{i}\xi_{c}\kb\chi_{0} \biggsq{
        \frac{\delta_{\chi_{0}}\Phi}{\Phi}
        + \left\langle\delta_{\chi_{0}}\right\rangle
        }
        + \frac{\Lr^{2}\hbar}{2}V^{q}
        \ ,\\[18pt]
    \label{QHD_WDWchi_BOavg}
        \frac{1}{\Phi}\left\langle \qHo{\chi} \Phi\right\rangle & = \left\langle  - 
        \frac{\hbar^2}{2}\delta^2_{\chi\chi}
        - \frac{6\hbar}{i}\xi_{c}\kb\chi \delta_{\chi}
        + \frac{1}{2}V^{\chi}
        \right\rangle = \left\langle \qHo{\chi}\right\rangle\ .
    \end{align}
    \end{subequations}
The sum of the three terms above is the equation for $\Phi$, which we write in the following concise form
    \begin{equation}
    \label{QHD_SC_BO2}
        \frac{1}{\Phi}\left\langle\qHo{Q}\Phi\right\rangle + \frac{1}{\Phi}\left\langle\qHo{q}\Phi\right\rangle = - \left\langle \qHo{\chi}\right\rangle\ .
    \end{equation}
This equation is analogous to eq.~\eqref{QGR_SC_BO2} in QGDGR.
Note that we have obtained --- without any additional assumptions --- the expectation value of the $\chi$-Hamiltonian operator, $\left\langle \qHo{\chi}\right\rangle$.
In the current HD quantum theory this expectation value is not of the form in eq.~\eqref{QGR_SC_Backreact}, because of the explicit appearance of the non-minimally coupled terms with extrinsic curvature instead of $\delta_{\bhb}S_{0}^{\srm{E\chi_{0}}}$ (and $\delta_{a}S_{0}^{\srm{E\chi_{0}}}$ for non-conformal coupling).
Apart from this observation, the two equations differ drastically and leave one to wonder if it is at all possible that they be related to each other.

The next step is to find an equation for $\psi$, the analog of eq.~\eqref{QGR_SC_BO3}.
Multiplying eq.~\eqref{QHD_SC_BO2} by $\psi$ and subtracting the result from eq.~\eqref{QHD_WDWQ_BO}, which is first divided by $\Phi$, we obtain
    \begin{subequations}
    \begin{align}
    \label{QHD_WDWQ_BOsch}
        \frac{1}{\Phi}\qHo{Q}\Psi - \frac{1}{\Phi}\left\langle\qHo{Q}\Phi\right\rangle\psi & =
        -\hbar\mathring{\mirl{\mathscr{G}}}^{\ssst IJ}_{\ssst{\aw,\br}}
        \Biggsq{
        2\frac{\delta_{\ssst I}\Phi}{\Phi}
        \Bigbr{
        \delta_{\ssst J}
        -\left\langle\delta_{\ssst J}\right\rangle
        }
        + \Bigbr{
        \delta^2_{\ssst IJ}
        - \left\langle\delta^2_{\ssst IJ}\right\rangle
        }
        }\psi   \nld
        &\quad 
        - \frac{\hbar}{i}\mathcal{D}^{\ssst I}\Bigbr{
        \delta_{\ssst I}
        - \left\langle\delta_{\ssst I}\right\rangle
        }\psi
        \ ,\\[18pt]
    \label{QHD_WDWq_BOsch}
        \frac{1}{\Phi}\qHo{q}\Psi - \frac{1}{\Phi}\left\langle\qHo{q}\Phi\right\rangle\psi
        & = \frac{\hbar}{i}\Biggsq{
        a\kb \Bigbr{
        \delta_{a}
        - \left\langle\delta_{a}\right\rangle
        }
        + 2\bktb\cdot\Bigbr{
        \delta_{\bhb}
        - \left\langle\delta_{\bhb}\right\rangle
        }
        }\psi   \nld
        &\quad 
        - \frac{\hbar^2}{2\Lr^{2}}\Biggsq{
        2 \frac{\delta_{\chi_{0}}\Phi}{\Phi}\Bigbr{
        \delta_{\chi_{0}}
        - \left\langle\delta_{\chi_{0}}\right\rangle
        }
        +
        \Bigbr{
        \delta^2_{\chi_{0}\chi_{0}}
        - \left\langle\delta^2_{\chi_{0}\chi_{0}}\right\rangle
        }
        }\psi     \nld
        &\quad - \frac{6\hbar}{i}\xi_{c}\kb\chi_{0}
        \Bigbr{
        \delta_{\chi_{0}}
        - \left\langle\delta_{\chi_{0}}\right\rangle
        }\psi\ ,\\[18pt]
    \label{QHD_WDWchi_BOsch}
        \frac{1}{\Phi}\qHo{\chi}\Psi - \frac{1}{\Phi}\left\langle \qHo{\chi}\Psi\right\rangle\psi & = \left(\qHo{\chi} - \left\langle \qHo{\chi}\right\rangle\right)\psi\ .
    \end{align}
    \end{subequations}
Adding the above three equations and putting eq.~\eqref{QHD_WDWchi_BOsch} and all second derivatives of $\psi$ to the opposite side of the resulting equation, we obtain the following result
    \begin{align}
    \label{QHD_WDWQ_BOschF}
        & 2\hbar\mathring{\mirl{\mathscr{G}}}^{\ssst IJ}_{\ssst{\aw,\br}}
        \frac{\delta_{\ssst I}\Phi}{\Phi}
        \Bigbr{
        \delta_{\ssst J}
        -\left\langle\delta_{\ssst J}\right\rangle
        }\psi
        + \frac{\hbar}{i}\mathcal{D}^{\ssst I}\Bigbr{
        \delta_{\ssst I}
        - \left\langle\delta_{\ssst I}\right\rangle
        }\psi
        \nld
        & \qquad\qquad - \frac{\hbar}{i}
        \Biggsq{
        a\kb \Bigbr{
        \delta_{a}
        - \left\langle\delta_{a}\right\rangle
        }
        + 2\bktb\cdot\Bigbr{
        \delta_{\bhb}
        - \left\langle\delta_{\bhb}\right\rangle
        }
        }\psi
        \nld
        &\qquad\qquad\qquad\qquad 
        + \frac{\hbar^2}{\Lr^{2}}
        \frac{\delta_{\chi_{0}}\Phi}{\Phi}\Bigbr{
        \delta_{\chi_{0}}
        - \left\langle\delta_{\chi_{0}}\right\rangle
        }\psi
        + \frac{6\hbar}{i}\xi_{c}\kb\chi_{0}
        \Bigbr{
        \delta_{\chi_{0}}
        - \left\langle\delta_{\chi_{0}}\right\rangle
        }\psi    \nld
        &\qquad 
        =  \left(\qHo{\chi} - \left\langle \qHo{\chi}\right\rangle\right)\psi 
        -
        \frac{\hbar^2}{2\Lr^{2}}\Bigbr{
        \delta^2_{\chi_{0}\chi_{0}}
        - \left\langle\delta^2_{\chi_{0}\chi_{0}}\right\rangle
        }\psi
        - \hbar\mathring{\mirl{\mathscr{G}}}^{\ssst IJ}_{\ssst{\aw,\br}}
        \Bigbr{
        \delta^2_{\ssst IJ}
        - \left\langle\delta^2_{\ssst IJ}\right\rangle
        }\psi\ .
    \end{align}
Compare this equation with eq.~\eqref{QGR_SC_BO3} in the case of QGDGR.
The term on the LHS of eq.~\eqref{QHD_WDWQ_BOschF} contains terms of order higher than the LHS of eq.~\eqref{QGR_SC_BO3}, which is strange because such terms are also of higher order than the quantum Hamiltonian on the RHS, indicating that $\psi$ and its evolution are not entirely determined by the matter quantum Hamiltonian.
These are very interesting observations which will shall address in the following subsections in more detail.

\subsection{Semiclassical approximation: the WKB-type expansion}

In this subsection we shall merely derive the expanded equations order by order and give some remarks and comparisons with the corresponding case in QGDGR.
Then in the following subsection, we shall engage into further formulation of the semiclassical approximation and actual interpretation of the derived equations.

We proceed by applying the WKB approximation in terms of $\Lr^2$:
    \begin{equation}
    \label{QHD_WKBans}
        \Psi[q^{\ssst A},Q^{\ssst I},\chi] = \mathcal{A}[q^{\ssst A},Q^{\ssst I}]\exp\left(\frac{i}{\hbar}\Lr^2 S^{\srm{HD}}[q^{\ssst A},Q^{\ssst I}]\right)\psi[q^{\ssst A},Q^{\ssst I},\chi]\ ,
    \end{equation}
where the superscript ``HD'' stands for $\rm{HD} = \rm{ERW\chi_{0}} $. 
We shall first calculate general forms of the derivatives appearing in the equations of concern.
To this purpose, let us introduce general indices $X,Y$ which can represent either $X,Y = I,J = \{\kb,\bktb\}$ or $X,Y = A,B = \{ a, \bhb, \chi_{0}\}$.
This will help us manage the variety of terms in the main equations.
We are interested in the following derivatives expanded up to order $\mathcal{O}(\Lr^{0})$,
    \begin{subequations}
    \begin{align}
    \label{QHD_SC_delPhi}
        \frac{\delta_{\ssst X}\Phi}{\Phi}  & = 
        \delta_{\ssst X}\log \mathcal{A} + \frac{i\Lr^{2}}{\hbar}\delta_{\ssst X}S^{\srm{HD}}\Phi\nld
        & \approx \frac{i\Lr^{2}}{\hbar}\delta_{\ssst X}S^{\srm{HD}}_{0}
        + \frac{i}{\hbar}\delta_{\ssst X}S^{\srm{HD}}_{1}
        + \delta_{\ssst X}\log \mathcal{A}_{0}
        + \mathcal{O}(\Lr^{-2})
        ,\\[18pt]
    \label{QHD_SC_del2Phi}
        \frac{\delta_{\ssst XY}^2\Phi}{\Phi} & =
        \frac{\delta_{\ssst XY}^2 \mathcal{A}}{\mathcal{A}}
        + \frac{i\Lr^2}{\hbar}\left(2\delta_{\ssst X}\log\mathcal{A}\,\,\delta_{\ssst Y}S^{\srm{HD}} 
        + \delta_{\ssst XY}^2 S^{\srm{HD}}
        \right)
        - \frac{\Lr^{4}}{\hbar^2}\delta_{\ssst X}S^{\srm{HD}}\delta_{\ssst Y}S^{\srm{E}\chi_{0}}\nld
        & \approx 
        \frac{\delta_{\ssst XY}^2 \mathcal{A}_{0}}{\mathcal{A}_{0}}
        - \frac{1}{\hbar^{2}}\delta_{\ssst X}S^{\srm{HD}}_{1}   \delta_{\ssst Y}S^{\srm{HD}}_{1}
        \nld
        &\quad + \frac{i}{\hbar}\left( 2\delta_{\ssst X}\log\mathcal{A}_{0} \,\,  \delta_{\ssst Y}S^{\srm{HD}}_{1}
        + 2\delta_{\ssst X} \left( \frac{\mathcal{A}_{1}}{\mathcal{A}_{0}}\right) \delta_{\ssst Y}S^{\srm{HD}}_{0}
        + \delta^2_{\ssst XY}S^{\srm{HD}}_{1}
        \right)
        \nld
        & \quad 
        - \frac{2\Lr^{2}}{\hbar^{2}}\delta_{\ssst X}S^{\srm{HD}}_{0}   \delta_{\ssst Y}S^{\srm{HD}}_{1}
        + \frac{i\Lr^2}{\hbar}\left(
        2\delta_{\ssst X}\log\mathcal{A}_{0}\,\,   \delta_{\ssst Y}S^{\srm{HD}}_{0}
        + \delta_{\ssst XY}^2 S^{\srm{HD}}_{0}
        \right)
        \nld
        &\quad - \frac{\Lr^{4}}{\hbar^{2}}\delta_{\ssst X}S^{\srm{HD}}_{0}    \delta_{\ssst Y}S^{\srm{HD}}_{0}
        + \mathcal{O}(\Lr^{-2})\ ,
    \end{align}
    \end{subequations}
where we again used $\delta_{\ssst XY}^2\log \mathcal{A} + \delta_{\ssst X}\log \mathcal{A} \, \, \delta_{\ssst Y}\log \mathcal{A} = \mathcal{A}^{-1}\delta_{\ssst XY}^2 \mathcal{A}$ in the second equation.
These are very similar to eqs.~\eqref{QGR_SC_delPhi} and~\eqref{QGR_SC_del2Phi}, except that we are focusing on the derivatives only, without any pre-factors or coefficients.
This is because not all terms --- corresponding to the above expressions for different values of indices $X,Y$ --- come with the same coefficients of certain power of $\Lr^{2}$ and this leads to non-trivial structure of expanded HDWDW equation.

Let us first look at eq.~\eqref{QHD_SC_BO2}, whose terms are given by eqs.~\eqref{QHD_WDWQ_BOavg}-\eqref{QHD_WDWchi_BOavg}.
If eqs.~\eqref{QHD_SC_delPhi} and \eqref{QHD_SC_del2Phi} with indices $X,Y = I,J = \{\kb,\bktb\}$ are used in eq.~\eqref{QHD_WDWQ_BOavg}, none of the derivative terms is suppressed.
This implies one of the most important points of the semiclassical approximation to HDWDW: the term $\sim\Lr^{4}$, i.e. the first term in the last line of eq.~\eqref{QHD_SC_del2Phi}, is \textit{the only} highest order surviving term in the \textit{entire} equation~\eqref{QHD_SC_BO2}.
We can confirm that by noting that this term comes from the \textit{second} functional derivative of $\Phi$ and the only other place in eq.~\eqref{QHD_SC_BO2} where the second functional derivative of $\Phi$ appears is the first term in the middle line of eq.~\eqref{QHD_WDWq_BOavg}, corresponding to $X,Y = \chi_{0}$; but this term is suppressed by $\Lr^{-2}$ which means that it reduces the order of each term in eq.~\eqref{QHD_SC_del2Phi} by one.
Hence, the highest order term from the only other second functional derivative of $\Phi$ is only of the order $\Lr^{2}$ (same order as the potential $V^{\ssst q}$!).
This further implies that we can already deduce something without plugging everything we calculated so far into the equations: we have that the following holds,
    \begin{equation}
    \label{QHD_SC_PhiL4}
        \mathcal{O}_{\Phi}(\Lr^{4}): \quad  \mathring{\mirl{\mathscr{G}}}^{\ssst IJ}_{\ssst{\aw,\br}}\delta_{\ssst I}S^{\srm{HD}}_{0}  \delta_{\ssst J}S^{\srm{HD}}_{0} = 0\ ,
    \end{equation}
where $\mathcal{O}_{\Phi}$ stands for the order of the equation for $\Phi$.
The DeWitt supermetric in eq.~\eqref{QHD_SC_PhiL4} is indefinite, so at first, it seems that one cannot conclude much from the above equation.
So let us keep this equation in mind and we shall come back to it when we start dealing with perturbative interpretation of the final equations.

The next order brings the following.
We need $X,Y = \{I,J\}$ versions of terms of order $\mathcal{O}(\Lr^{2})$ in eq.~\eqref{QHD_SC_del2Phi} and $X,Y = I,J$ versions of terms of order $\mathcal{O}(\Lr^{2})$ in eq.~\eqref{QHD_SC_delPhi}; these are to be used in the first and second term in the first line of eq.~\eqref{QHD_WDWQ_BOavg}, as well as in the first term in the second line of the same equation.
We also need $X, Y = \chi_{0}$ versions of the same terms in eq.~\eqref{QHD_SC_delPhi} and eq.~\eqref{QHD_SC_del2Phi} as with $X,Y = \{I,J\}$ versions;
these are used in the first term in th emiddle line of eq.~\eqref{QHD_WDWq_BOavg} and in the first term in the last line of the same equation.
Lastly, we need $X, Y = \{a, \bhb\}$ versions of order $\mathcal{O}(\Lr^{2})$ term in eq.~\eqref{QHD_SC_delPhi}; these appear in the first two terms in eq.~\eqref{QHD_WDWQ_BOavg}.
Note, in passing, that there are no contributions to the second order derivative of $\Phi$ for index values $X, Y = \{a, \bhb\}$, because there is no kinetic term with respect to the metric variables.
This is in drastic contrast to the case of QGDGR and the semiclassical approximation in there, cf.~eq.~\eqref{QGR_SC_BO2}.
In QGDGR, it was precisely the second order derivative term which gave rise to the kinetic term of the classical EHJ equation, cf.~eq.~\eqref{QGR_SC_del2Phi} and eq.~\eqref{QGR_SC_EHJ}.
It thus seems at the moment that there is no hope of recovering the classical momenta in eq.~\eqref{QGR_SC_momenta} conjugate to $a$ and $\bhb$ variables.
Since there are no other terms of order $\mathcal{O}(\Lr^{2})$, the resulting equation at this order is given by
    \begin{align}
    \label{QHD_SC_PhiL2}
        \mathcal{O}_{\Phi}(\Lr^{2}):\quad & 2\mathring{\mirl{\mathscr{G}}}^{\ssst IJ}_{\ssst{\aw,\br}}\delta_{\ssst I}S^{\srm{HD}}_{0}  \delta_{\ssst J}S^{\srm{HD}}_{1}\nld
        &\quad - i\mathring{\mirl{\mathscr{G}}}^{\ssst IJ}_{\ssst{\aw,\br}}\left( 2\delta_{\ssst I}\log\mathcal{A}_{0}\delta_{\ssst J}S^{\srm{HD}}_{0}
        + \delta_{\ssst IJ}^{2}S^{\srm{HD}}_{0}
        + 2\delta_{\ssst I}S^{\srm{HD}}_{0}\left\langle\delta_{\ssst J}\right\rangle
        + \mathcal{D}^{\ssst I}\delta_{\ssst I}S^{\srm{HD}}_{0}\right)
        = 0\ .
    \end{align}
We can further split this equation into its real and imaginary parts:
    \begin{subequations}
    \begin{align}
    \label{QHD_SC_PhiL2Re}
        \rm{Re}\mathcal{O}_{\Phi}(\Lr^{2}): \quad &  2\mathring{\mirl{\mathscr{G}}}^{\ssst IJ}_{\ssst{\aw,\br}}\delta_{\ssst I}S^{\srm{HD}}_{0} \biggsq{ \delta_{\ssst J}S^{\srm{HD}}_{1} 
        + \rm{Im}\left\langle\delta_{\ssst J}\right\rangle
        }
        + a\kb\delta_{a}S^{\srm{HD}}_{0}
        + 2\bktb\cdot\delta_{\bhb}S^{\srm{HD}}_{0}\nld
        &\quad
        + \frac{1}{2}\delta_{\chi_{0}}S^{\srm{HD}}_{0}\delta_{\chi_{0}}S^{\srm{HD}}_{0}
        -6\xi_{c}\kb\chi_{0}\delta_{\chi_{0}}S^{\srm{HD}}_{0}
        + \frac{\hbar}{2}V^{q} = 0\ , \\[18pt]
    \label{QHD_SC_PhiL2Im}
        \rm{Im}\mathcal{O}_{\Phi}(\Lr^{2}): \quad & \mathring{\mirl{\mathscr{G}}}^{\ssst IJ}_{\ssst{\aw,\br}}\left( 2\delta_{\ssst I}\log\mathcal{A}_{0}\delta_{\ssst J}S^{\srm{HD}}_{0}
        + \delta_{\ssst IJ}^{2}S^{\srm{HD}}_{0}
        + \delta_{\ssst I}S^{\srm{HD}}_{0}\rm{Re}\left\langle\delta_{\ssst J}\right\rangle
        + \mathcal{D}^{\ssst I}\delta_{\ssst I}S^{\srm{HD}}_{0}\right) = 0\ .
    \end{align}
    \end{subequations}
We can see that the real part does not only contain $S^{\srm{HD}}_{0}$, but also $S^{\srm{HD}}_{1}$ and $\rm{Im}\left\langle\delta_{\ssst J}\right\rangle$, so it seems that one needs the knowledge of both of the latter terms in order to determine $S^{\srm{HD}}_{0}$ from that equation.
Furthermore,  the first term structurally reminds of eq.~\eqref{QGR_SC_S1andBR}, if $\rm{Im}\left\langle\delta_{\ssst J}\right\rangle$ were interpreted as $J$-components of what we may call the \textit{extended Berry connection},
    \begin{equation}
    \label{QHD_SC_Berry}
        \mathcal{B}_{\ssst X} = \rm{Im}\left\langle\delta_{\ssst X}\right\rangle\ .
    \end{equation}
This equation is thus invariant with respect to a phase transformation of $\psi$.
Therefore, solving eq.~\eqref{QHD_SC_PhiL2Re} seems impossible because one must know both the next order phase $S^{\srm{HD}}_{1}$ \textit{and} the matter wave functional $\psi$.
As for the imaginary part, we have only temporarily written the term $\rm{Re}\left\langle\delta_{\ssst J}\right\rangle$ as non-vanishing, only to remind that it plays a role in these equations, even though it is eliminated for the same reasons as eq.~\eqref{QGD_SC_Redel}, i.e. due to the choice $\vert\psi\vert^2 = 1$.
Lastly, note that through $\mathcal{D}^{\ssst I}\delta_{\ssst I}S^{\srm{HD}}_{0}$ both extrinsic and intrinsic curvature appear in eq.~\eqref{QHD_SC_PhiL2Im}, cf.~eqs.~\eqref{HDall_HamDDR} and~\eqref{HDall_HamDDW}.

The remaining equations of interest follow from the $\mathcal{O}(\Lr^{0})$ order of eqs.~\eqref{QHD_WDWQ_BOavg}-\eqref{QHD_SC_BO2}.
We use the $\mathcal{O}(\Lr^{0})$ order terms in eq.~\eqref{QHD_SC_del2Phi} and the $\mathcal{O}(\Lr^{0})$ order terms in eq.~\eqref{QHD_SC_delPhi} for $X,Y = I, J$ in eq.~\eqref{QHD_WDWQ_BOavg}.
Furthermore, the $\mathcal{O}(\Lr^{0})$ order terms in eq.~\eqref{QHD_SC_delPhi} for $X,Y = \{ a,\bhb\}$ are used in the first line of eq.~\eqref{QHD_WDWq_BOavg}.
Lastly, we use the $X,Y = \chi_{0}$ versions of the $\mathcal{O}(\Lr^{0})$ order terms in eq.~\eqref{QHD_SC_delPhi} in the third line of eq.~\eqref{QHD_WDWq_BOavg} and the $\mathcal{O}(\Lr^{2})$ order terms in eq.~\eqref{QHD_SC_delPhi} and eq.~\eqref{QHD_SC_del2Phi} in the second line of eq.~\eqref{QHD_WDWq_BOavg}.
Note that eq.~\eqref{QHD_WDWchi_BOavg} is already of the $\mathcal{O}(\Lr^{0})$ order.
Putting all these terms into eq.~\eqref{QHD_SC_BO2}, taking the real and imaginary parts of the resulting equation, using eq.~\eqref{QHD_SC_Berry} and $\rm{Re}\left\langle\delta_{\ssst J}\right\rangle = 0$, we obtain the following two equations
    \begin{subequations}
    \begin{align}
    \label{QHD_SC_PhiL0Re}
        \rm{Re}\mathcal{O}_{\Phi}(\Lr^{0}): \quad &  -\mathring{\mirl{\mathscr{G}}}^{\ssst IJ}_{\ssst{\aw,\br}}
        \left[ 
        \hbar\frac{\delta_{\ssst IJ}^2 \mathcal{A}_{0}}{\mathcal{A}_{0}}
        - \frac{1}{\hbar}\delta_{\ssst I}S^{\srm{HD}}_{1}   \delta_{\ssst J}S^{\srm{HD}}_{1}
        - 2\delta_{\ssst I}S^{\srm{HD}}_{1}\mathcal{B}_{\ssst J}
        + \hbar\rm{Re}\left\langle\delta^2_{\ssst IJ}\right\rangle
        \right]\nld
        &- \mathcal{D}^{\ssst I}\left(\delta_{\ssst I}S^{\srm{HD}}_{1} + \hbar\mathcal{B}_{\ssst I}\right) 
        - \aw\hbar\bar{\mathbf{C}}^{\srm B}\cdot\bar{\mathbf{C}}^{\srm B}\nld
        &\quad + a\kb\left(\delta_{a}S^{\srm{HD}}_{1} + \hbar\mathcal{B}_{a}\right) 
        + 2\bktb\cdot \left(\delta_{\bhb}S^{\srm{HD}}_{1} + \hbar\mathcal{B}_{\bhb}\right) \nld
        &\quad\quad + \delta_{\chi_{0}}S^{\srm{HD}}_{0}\left(\delta_{\chi_{0}}S^{\srm{HD}}_{1} + \hbar\mathcal{B}_{\chi_{0}}\right) 
        - 6\xi_{c}\kb\chi_{0}\left(\delta_{\chi_{0}}S^{\srm{HD}}_{1} + \hbar\mathcal{B}_{\chi_{0}}\right) \nld
        & \qquad = - \left\langle\qHo{\chi}\right\rangle\ ,\\[12pt]
    \label{QHD_SC_PhiL0Im}
        \rm{Im}\mathcal{O}_{\Phi}(\Lr^{0}): \quad & 
        -\mathring{\mirl{\mathscr{G}}}^{\ssst IJ}_{\ssst{\aw,\br}}
        \biggsq{ 
         2\delta_{\ssst I}\log\mathcal{A}_{0} \,\, \left( \delta_{\ssst J}S^{\srm{HD}}_{1}  + \hbar\mathcal{B}_{\ssst J}     \right)
        + 2\delta_{\ssst I} \left( \frac{\mathcal{A}_{1}}{\mathcal{A}_{0}}\right) \delta_{\ssst J}S^{\srm{HD}}_{0}
        \nld
        & \quad 
        + \delta^2_{\ssst IJ}S^{\srm{HD}}_{1}
        + \rm{Im}\left\langle\delta^2_{\ssst IJ}\right\rangle
        }
        + \hbar\mathcal{D}^{\ssst I}\delta_{\ssst I}\log\mathcal{A}_{0}
        \nld
        &\qquad - \hbar a\kb\delta_{a}\log\mathcal{A}_{0} - 2\hbar\bktb\cdot\delta_{\bhb}\log\mathcal{A}_{0}\nld
        &\qquad\quad - \hbar\delta_{\chi_{0}}\log\mathcal{A}_{0}\delta_{\chi_{0}}S^{\srm{HD}}_{0}  -  \frac{\hbar}{2}\delta_{\chi_{0}\chi_{0}}^{2}S^{\srm{HD}}_{0} = 0 \ .
    \end{align}
    \end{subequations}
Equation~\eqref{QHD_SC_PhiL0Re} is analogous to eq.~\eqref{QGR_SC_S1andBR}, while eq.~\eqref{QHD_SC_PhiL0Im} is analogous to eq.~\eqref{QGR_SC_ContA}.
Note the drastic difference due to the presence of terms from higher-derivative contributions.

We now turn to implementation of the WKB expansion in the equation for $\psi$, given by eq.~\eqref{QHD_WDWQ_BOschF}.
Unlike in the case of the semiclassical approximation to QGDGR, in the present case we have contributions of order $\mathcal{O}(\Lr^{2})$ in the equation for $\psi$, which we get using the first term in eq.~\eqref{QHD_SC_delPhi} for $X,Y = I, J$ in eq.~\eqref{QHD_WDWQ_BOschF}.
    \begin{subequations}
    \begin{align}
    \label{QHD_SC_psiL2}
        \mathcal{O}_{\psi}(\Lr^{2}): \quad & 2i \mathring{\mirl{\mathscr{G}}}^{\ssst IJ}_{\ssst{\aw,\br}}
        \delta_{\ssst I}S^{\srm{HD}}_{0}
        \Bigbr{
        \delta_{\ssst J}   -  i \mathcal{B}_{\ssst J}
        }\psi = 0\ ,\\[12pt]
    \label{QHD_SC_psiL0}
        \mathcal{O}_{\psi}(\Lr^{0}): \quad &
        2i\mathring{\mirl{\mathscr{G}}}^{\ssst IJ}_{\ssst{\aw,\br}}
        \biggpar{
        \delta_{\ssst I}S^{\srm{HD}}_{1}
        - i\hbar\delta_{\ssst I}\log\mathcal{A}_{0}
        }
        \Bigbr{
        \delta_{\ssst J}   -  i \mathcal{B}_{\ssst J}
        }\psi
        - i \hbar\mathcal{D}^{\ssst I}\Bigbr{
        \delta_{\ssst I}   -  i \mathcal{B}_{\ssst I}
        }\psi
        \nld
        & \quad + i\hbar
        \Biggsq{
        a\kb \Bigbr{
        \delta_{a}
        - i \mathcal{B}_{a}
        }
        + 2\bktb\cdot\Bigbr{
        \delta_{\bhb}
        - i \mathcal{B}_{\bhb}
        }
        }\psi
        \nld
        &\qquad
        + i\hbar
        \delta_{\chi_{0}}S^{\srm{HD}}_{0}\Bigbr{
        \delta_{\chi_{0}}
        - i \mathcal{B}_{\chi_{0}}
        }\psi
        - 6i\hbar\xi_{c}\kb\chi_{0}
        \Bigbr{
        \delta_{\chi_{0}}
        - i \mathcal{B}_{\chi_{0}}
        }\psi    \nld
        &\qquad 
        =  \left(\qHo{\chi} - \left\langle \qHo{\chi}\right\rangle\right)\psi 
        - \hbar\mathring{\mirl{\mathscr{G}}}^{\ssst IJ}_{\ssst{\aw,\br}}
        \Bigbr{
        \delta^2_{\ssst IJ}
        - \left\langle\delta^2_{\ssst IJ}\right\rangle
        }\psi\ . 
    \end{align}
    \end{subequations}
Equation \eqref{QHD_SC_psiL2} can be seen as a parallel transport of $\psi$ along the direction $\delta_{\ssst I}S^{\srm{HD}}_{0}$; in other words, $\psi$ \textit{does not evolve along the changes of} $S^{\srm{HD}}_{0}$ \textit{with respect to the extrinsic curvature}.
There are two other possibilities for that equation to be satisfied automatically.
One is that eq.~\eqref{QHD_SC_PhiL4} gives $\delta_{\ssst I}S^{\srm{HD}}_{0} = 0$ trivially.
Another possibility is that $\psi$ is independent of $\kb$ and $\bktb$.
Of course, since extrinsic curvature is definitely present in the Hamiltonian operator for the $\chi$ field, cf. eq.~\eqref{QHD_H0chi}, one would at first think that $\delta_{\ssst I}\psi = 0$ is too restrictive and unfounded an assumption.
But given what we just discussed above and the fact that one should expect that $\psi$ evolves along the classical background determined either by the EHJ equation or by the EHJ equation corrected by \textit{perturbatively reduced} counter-terms, there is no reason to expect that $\psi$ explicitly depends on the additional degrees of freedom.
After all, the whole pint of this thesis is to find a semiclassical limit to a QGDHD leading to a QFT on curved spacetime \textit{with} counter-terms but \textit{without} additional degrees of freedom.
This is what we aim to finally achieve in the following subsection.

\subsection{Semiclassical approximation: implementation of perturbative constraints}

In QGDGR it is shown that the EHJ equation given by eq.~\eqref{QGR_SC_EHJm} and eq.~\eqref{QGR_SC_momenta0} follows from a semiclassical approximation scheme based on the expansion of the WDW equation in powers of $\Lr^{-2}$.
It is also shown that this equation can be corrected by the backreaction, resulting in eq.~\eqref{QGR_SC_HcBR}, with momenta corrected according to eq.~\eqref{QGR_SC_momenta1}.
The matter wave functional is determined at order $\mathcal{O}(\Lr^{0})$ by the functional Schr\"odinger equation given by eq.~\eqref{QGR_SC_SchTauBR} or eq.~\eqref{QGR_SC_SchTau}.
However, we argued that the problem is that divergences which appear once the backreaction is calculated are ``naked'' because there are no counter-terms which could absorb them.
As we have argued in chapter~\ref{ch:HDclass}, a classical theory of gravity based on a higher-derivative extension of the EH action must be interpreted as the EH theory plus perturbations.
These perturbations are in principle determined by making use of the first-order usual non-vacuum EE without counter-terms, which lead to their consistent role as an integral part of the SEE in absorbing the divergencies from the backreaction.
The aim of the QGDHD is to recover the SEE with counter-terms, as well as the functional Schr\"odinger equation, and we present here few steps which could be taken in order to achieve that goal.
There are several difficulties that we encounter and these are left open, apart from some suggestions and educated guesses.
The main lines of thought that we are led by are: that there has to be a way to derive the EHJ equation and, if possible, to determine the form of the EHJ equation corrected by the counter-terms; that implementation of the perturbative nature of the higher-derivative contributions plays a crucial role in this endeavour.
Only then one could address the equation for $\psi$ which is expected to lead to a functional Schr\"dinger equation.

The following important assumptions and observations are our starting point:
    \begin{enumerate}
        \item Parameters $\aw$ and $\br$ are \textit{bare} coupling constants and thus observationally meaningless until all divergences from the backreaction are absorbed, which is a meaningful method only within the BO-WKB approximation with respect to $\Lr^{2}$.
        In other words, treating parameters $\aw$ and $\br$ as ``small'' cannot be done until the the BO-WKB semiclassical approximation with respect to $\Lr^{2}$ has been employed.
        \item Parameters $\aw$ and $\br$ are \textit{independent}. They correspond to two different and \textit{independent} terms in the original action --- $C^2$ and $R^2$ --- which is a fact we emphasized in the previous chapter.
        Recall that $C^2$ does not depend on $\dot{\kb}$ which leads to eq.~\eqref{HWchi_Pk}. Suppose a HJ functional $S^{\srm W}$ were introduced for the W or WE theory: then it would follow that $\Pb=\ddel{S^{\srm W}}{\kb} = 0$ at order $\aw$, because no contribution from $\dot{\kb}$ enters the theory.
        Furthermore, the surviving momenta in this theory is eq.~\eqref{HDall_Pk} so $\ddel{S^{\srm W}}{\bktb}\sim \aw$ \textit{even in the presence of $R^2$ term.}
        A similar argument can be made for the momenta in a theory with the $R^2$ term but without the $C^2$ term.
        Therefore, setting $\ddel{S^{\srm HD}}{\kb} = 0$ to zero and $\br=0$ leaves contributions of the $C^2$ term only;
        setting $\ddel{S^{\srm HD}}{\bktb} = 0$ to zero and $\aw=0$ leaves contributions of the $R^2$ term only.
        \item It can be deduced from the previous statement and from the derivation of the momenta $\Pb$ and $\bPb$ in eq.~\eqref{HDall_Pk} and eq.~\eqref{HDall_Pkt} that the part of the action coupled with $\aw$ gives terms $\sim\aw$ as the highest non-vanishing contribution to $\Pb$ and that the part of the action coupled with $\br$ gives terms $\sim\br$ as the highest non-vanishing contribution to $\bPb$.
        It is also important to emphasize that these contributions appear at the order $\mathcal{O}(\Lr^{0})$ in the action.
        \item Given the above and the fact that the WKB phase $\Lr^{2}S^{\srm{HD}}_{0}$ is of order $\mathcal{O}(\Lr^{2})$ (recalling that $\Lr^{2}$ was factored out from $S^{\srm{HD}}_{0}$ in eq.~\eqref{QHD_WKBans} only for convenience) and that $S^{\srm{HD}}_{1}$ is of order $\mathcal{O}(\Lr^{0})$, it follows that any dependence of the WKB phase of on $\kb$ and $\bktb$ can enter only in $S^{\srm{HD}}_{1}$.
        This conclusion is exactly compatible with our expectation from eq.~\eqref{QGR_SC_S1andBR} that any counter-terms that may be needed for absorbing the divergences from the backreaction must appear either through $S^{\srm{E\chi_{0}}}_{1}$ or through $\mathcal{B}_{\ssst A}$.
        Therefore, we may assume that $S^{\srm{HD}}_{0} = S^{\srm{HD}}_{0}[q^{\ssst A}]$ only, i.e. that the highest order WKB phase contribution \textit{is not generated by the higher-order contributions}.
        \item Perhaps the two most important technical difficulties encountered in the subsequent analysis are found in the fact that the Hamiltonian formulation of the classical (and therefore quantized) higher-derivative theory rests entirely on the Legendre transform given by eq.~\eqref{HDall_LegTransf}, which introduces the additional degree of freedom into the root of the canonical and canonically quantized theory.
        Because of this the resulting Hamiltonian is \textit{not equivalent} to the perturbatively constrained Hamiltonian.
        Yet we are trying to \textit{perturb} the \textit{exact higher-derivative theory} in the semiclassical approximation.
        This seems contradictory and in fact it is, given what we have learnt from section~\ref{subs_Ham_osc}: perturbatively reduced theory given by the Hamiltonian in eq.~\eqref{Ham_osc_HamRic} cannot be obtained by $g\rightarrow 0$ perturbation of the exact higher-derivative Hamiltonian in eq.~\eqref{Ham_osc_Ham1}.
        The only way one could avoid blowing up the kinetic term is to assume in the latter equation that $P_{Y}$ is of the order $\sim g$.
        This is indeed compatible with the definition of momentum $P_{Y}$ given in eq.~\eqref{H_alho_pY}, where it can be clearly seen that $P_{Y}\sim g$
        The situation with that toy model is in direct analogy with the situation we have at hand here, as explained in the previous two points.
        Is then an expansion of $S^{\srm{HD}}_{1}$ in terms of $\aw,\br$ enough? No, it is not, because such an expansion cannot undo what a Legendre transform did: introduction of \textit{additional terms} $\dot{\kb}\Pb$ and $\dot{\kb}^{\srm T }_{ij}\Pb^{ij}$ to form the total Hamiltonian gives a different result as they would in a perturbatively reduced case.
        In the perturbatively reduced theory modeled by the Hamiltonian in eq.~\eqref{Ham_osc_HamRic} the higher-order contributions \textit{transform} (i.e. they are not added through a Legendre transform) into additional \textit{potential terms} in the Hamiltonian, not kinetic terms.
        Hence, if perturbatively reduced higher-derivative gravity were formulated canonically, the $\dot{\kb}\Pb$ terms would not even enter the Legendre transform --- the higher-order derivative terms would transform into lower-order terms and woudl contribute to the kinetic term of pure Hamiltonian GR and its potential.
        From this important observation it follows that the Hamilton-Jacobi formulation of the exact higher-derivative theory \textit{is not equivalent to the Hamilton-Jacobi formulation of the perturbatively reduced theory}.
        Unfortunately, this is an inevitable problem that we are encountering in the very derivation this section is devoted to.
        Is there a way to turn the additional kinetic terms of the exact theory into potential terms of the perturbatively constrained theory we would like to have at the end?
        Since we are hopeful of the possibility that the answer is positive, we shall give possible directions and pitfalls that would be important to be aware of in a future work.
    \end{enumerate}
With these observations and assumptions we are ready to analyze eqs.~\eqref{QHD_SC_PhiL4}, \eqref{QHD_SC_PhiL2Re}, \eqref{QHD_SC_PhiL2Im}, \eqref{QHD_SC_PhiL0Re}, \eqref{QHD_SC_PhiL0Im}, \eqref{QHD_SC_psiL2} and \eqref{QHD_SC_psiL0}.

Let us start with eq.~\eqref{QHD_SC_PhiL4}.
Since we have argued above that contributions to the WKB phase of order $\aw$ and $\br$ are to be sought in $S^{\srm{HD}}_{1}$, it follows that
    \begin{equation}
    \label{QHD_SC_S0noQ}
        S^{\srm{HD}}_{0}[q^{\ssst A},Q^{\ssst I}] = S^{\srm{HD}}_{0}[q^{\ssst A}] \equiv S^{\srm{HD}}_{0}\ .
    \end{equation}
Hence all its derivatives with respect to $I,J$ vanish identically.
Alternatively, we could have deduced the same if we noticed in another way: observe that appearance of $\aw$ and $\br$ in the \textit{denominator} inside the DeWitt supermetric prevents one from taking a limit $\aw,\br\rightarrow 0$; then one could apply observation point 2. from the above, i.e. that the contributions from $C^2$ and $R^2$ terms, proportional to $\aw$ and $\br$ respecitvely, are \textit{independent};
this would lead to the following two expansions of $S^{\srm{HD}}_{0}$,
    \begin{align}
    \label{QHD_SC_S0expa}
        S^{\srm{HD}}_{0}[q^{\ssst A},Q^{\ssst I}] & =  \prescript{\ssst  0}{}{S}^{\srm{HD}}_{0}[q^{\ssst A}] + \aw \prescript{\ssst \alpha}{}{S}^{\srm{HD}}_{0}[q^{\ssst A},\bktb] \ ,\\[12pt]
    \label{QHD_SC_S0expb}
        S^{\srm{HD}}_{0}[q^{\ssst A},Q^{\ssst I}] & =  \prescript{\ssst  0}{}{S}^{\srm{HD}}_{0}[q^{\ssst A}] + \br \prescript{\ssst \beta}{}{S}^{\srm{HD}}_{0}[q^{\ssst A},\kb]\ .
    \end{align}
Equation \eqref{QHD_SC_S0expa} is plugged into eq.~\eqref{QHD_SC_PhiL4} with a condition $\ddel{S^{\srm{HD}}_{0}[q^{\ssst A},Q^{\ssst I}]}{\kb} = 0$.
Equation \eqref{QHD_SC_S0expb} is plugged into eq.~\eqref{QHD_SC_PhiL4} with a condition $\ddel{S^{\srm{HD}}_{0}[q^{\ssst A},Q^{\ssst I}]}{\bktb} = 0$.
Collecting the powers of $\aw$ and $\br$ in the respective equations leads to
    \begin{equation}
        \prescript{\ssst \alpha}{}{S}^{\srm{HD}}_{0}[q^{\ssst A},Q^{\ssst I}] = \prescript{\ssst \alpha}{}{S}^{\srm{HD}}_{0}[q^{\ssst A}]\ ,\quad  \prescript{\ssst \beta}{}{S}^{\srm{HD}}_{0}[q^{\ssst A},Q^{\ssst I}] =\prescript{\ssst \beta}{}{S}^{\srm{HD}}_{0}[q^{\ssst A}] \ ,
    \end{equation}
i.e. both expressions in eq.~\eqref{QHD_SC_S0expa} and eq.~\eqref{QHD_SC_S0expb} are independent of $Q^{\ssst I}$ and thus deprived of higher-derivative degrees of freedom.
We can now use information from eq.~\eqref{QHD_SC_S0noQ} in the subsequent orders of the WKB expansion.
Note that we do not write $\approx$ in eq.~\eqref{QHD_SC_S0expa} and eq.~\eqref{QHD_SC_S0expb} and we neglect contributions of the higher powers of couplings such as $\aw^2,\,\aw\br,\,\br^2\, ...$ because this expansion is \textit{exact} and compatible with the fact that there are only $\sim\aw$ and $\sim\br$ contributions in the starting theory (action).
This is in accordance with the method of perturbative constraints and is just an application of discussion in section~\ref{subs_HDLHO}, page~\pageref{p_pertg}.

Let us next look at eqs.~\eqref{QHD_SC_PhiL2Re} and~\eqref{QHD_SC_PhiL2Im}.
We do not need to expand anything there because each term is proportional to $\delta_{\ssst I}S^{\srm{HD}}_{0}[q^{\ssst A}] = 0$ and no information is obtained from this equation.
The real counterpart of that equation given by eq.~\eqref{QHD_SC_PhiL2Re} contains a derivative $\delta_{\ssst I}S^{\srm{HD}}_{0}$ in the first term only and thus this term is gone.
The remaining terms amount to (using eq.~\eqref{QHD_Vq} and eq.~\eqref{QHD_Ve}) the following equation
    \begin{align}
    \label{QHD_SC_PhiL2ReSab}
        \rm{Re}\mathcal{O}_{\Phi}(\Lr^{2}): \quad &
        a\kb\delta_{a}S^{\srm{HD}}_{0}
        + 2\bktb\cdot\delta_{\bhb}S^{\srm{HD}}_{0} 
        - \frac{\hbar a^2}{2}\Big(a^2\left(\,^{\ssst (3)}\! R 
        - 2\bar{\Lambda}\right) 
        + \bktb\cdot\bktb 
        - 6\kb^2\Big)\nld
        &\quad
        + \frac{1}{2}\delta_{\chi_{0}}S^{\srm{HD}}_{0}\delta_{\chi_{0}}S^{\srm{HD}}_{0}
        -6\xi_{c}\kb\chi_{0}\delta_{\chi_{0}}S^{\srm{HD}}_{0}
        + \frac{1}{2}V^{\ssst \chi_{0}}
        = 0\ ,
    \end{align}
recalling that $V^{\ssst \chi_{0}}$ is defined by eq.~\eqref{Vchi_def}.
Equation~\eqref{QHD_SC_PhiL2ReSab} is one of the most important points in this thesis.
There are terms explicitly depending on $\kb$ and $\bktb$ but the functional $S^{\srm{HD}}_{0}$ is independent of them and the equation seems to be an equation that, in principle, determines $S^{\srm{HD}}_{0}$.
It seems that one needs to somehow fix $\kb$ and $\bktb$ in order to solve for $S^{\srm{HD}}_{0}$.
But it was shown already in~\cite{MSc} by the author that (in the case of the WE theory without $\chi_{0}$ field) equation such as eq.~\eqref{QHD_SC_PhiL2ReSab} itself contains enough information to fix $\kb$ and $\bktb$. 
Actually, it would be more precise to say that eq.~\eqref{QHD_SC_PhiL2ReSab} is a \textit{constraint equation} representing a relationship between $\kb,\bktb$ and $\delta_{\ssst A}S^{\srm{HD}}_{0}$.
To see this, we use the same procedure as we did in \cite{MSc}: act with a functional derivative $\delta_{\ssst I}$ on eq.~\eqref{QHD_SC_PhiL2ReSab} and use the result back in it.
Taking into account eq.~\eqref{QHD_SC_S0noQ}, for $I = \kb$ we have the following result of a functional derivation\footnote{We are being quite imprecise here with functional differentiation. Firstly, one must recall that all equations that we are discussing in this section are under an integral $\intx\Nb$.
Secondly, acting on a functional derivative on a functional involves another integration which is cancelled once the functional derivative has produced a delta function.
If one then looks into the resulting integrand, one has what we otherwise immediately write out in our derivations.},
    \begin{equation}
    \label{QHD_SC_EHJK}
        a\delta_{a}S^{\srm{HD}}_{0}
        + 6\hbar a^2\kb
        -6\xi_{c}\chi_{0}\delta_{\chi_{0}}S^{\srm{HD}}_{0}
        + 36\xi\xi_{c}\kb\chi^2_{0}
        = 0\ ,
    \end{equation}
where we have used (cf. eq.~\eqref{Vchi_def})
    \begin{equation}
        \delta_{\kb}V^{\ssst\chi_{0}} = 
        72\xi\xi_{c}\kb\chi^2_{0}\ .
    \end{equation}
For $I = \bktb$, the functional derivative of eq.~\eqref{QHD_SC_PhiL2ReSab} yields
    \begin{equation}
    \label{QHD_SC_EHJKtless}
        2\delta_{\bhb}S^{\srm{HD}}_{0} 
        - a^2 \hbar \bktb_{\sharp}
        + \xi\bktb_{\sharp}\chi_{0}^{2} = 0\ ,
    \end{equation}
where we have used (cf. eq.~\eqref{Vchi_def})
    \begin{equation}
        \delta_{\bktb}V^{\ssst\chi_{0}} = 
        2\xi\bktb_{\sharp}\chi_{0}^{2}\ .
    \end{equation}
Don't eqs.~\eqref{QHD_SC_EHJK} and \eqref{QHD_SC_EHJKtless} remind us of something we have seen earlier?
Let us rewrite them in a slightly different form:
    \begin{align}
    \label{QHD_SC_KFix}
        \kb & 
        = - \frac{a}
        {6\left(\Lr^{2}\hbar a^2 + 6\Lr^{2}\xi\xi_{c}\chi_{0}^{2}\right)}\left(
        \Lr^{2}\delta_{a}S^{\srm{HD}}_{0} - 6\xi_{c}\frac{\chi_{0}}{a}\Lr^{2}\delta_{\chi_{0}}S^{\srm{HD}}_{0}
        \right)\ ,\\[12pt]
    \label{QHD_SC_KtlessFix}
        \bktb_{\sharp} & 
        = \frac{2}{\Lr^{2}\hbar a^2 - \Lr^{2}\xi\chi_{0}^{2}} \Lr^{2}\delta_{\bhb}S^{\srm{HD}}_{0} \ ,
    \end{align}
where we multiplied and divided both equations with $\Lr^{2}$.
Given the following definitions
    \begin{equation}
    \label{QGDHD_SC_pHJfix}
        p_{a}^{(0)} := \Lr^{2}\delta_{a}S^{\srm{HD}}_{0}\ ,\quad \bpb^{(0)} := \Lr^{2}\delta_{\bhb}S^{\srm{HD}}_{0}\ ,\quad p_{\chi_{0}}^{(0)} := \Lr^{2}\delta_{\chi_{0}}S^{\srm{HD}}_{0}\ ,
    \end{equation}
eq.~\eqref{QHD_SC_KFix} is nothing other than eq.~\eqref{QHD_SC_KFix}, while eq.~\eqref{QHD_SC_KtlessFix} is nothing other than eq.~\eqref{GRchi_phdot}, if appropriate substitution $\chi\rightarrow \Lr\chi_{0}$ is used.
Equivalently, comparison of eq.~\eqref{QHD_SC_KFix} with eq.~\eqref{QGR_Knew} and comparison of eq.~\eqref{QHD_SC_KtlessFix} with eq.~\eqref{QGR_Ktlessnew} leads to the same conclusion if only terms of order $\mathcal{O}(\Lr^{-2})$ are kept.
This means that we have just recovered $\kb$ and $\bktb$ in terms of the momenta for a classical non-vacuum GR theory!
If that is the case, then let us use eq.~\eqref{QHD_SC_KFix}, eq.~\eqref{QHD_SC_KtlessFix}, eq.~\eqref{QGDHD_SC_pHJfix} and eq.~\eqref{Vchi_def} back into eq.~\eqref{QHD_SC_PhiL2ReSab}.
After a straightforward algebra and multiplication of the whole equation by $\Lr^{2}$, we obtain
    \begin{align}
    \label{QHD_SC_HAMGR}
        & -\frac{\left(a p_{a}^{\ssst (0)} - 6\xi_{c}\chi_{0} p^{\ssst (0)}_{\chi_{0}} \right)^2}{12 \Lr^2\left(\hbar a^2 + 6\xi\xi_{c}\chi_{0}^2\right)}
	    +\frac{1}{2\Lr^2}\left(p^{\ssst (0)}_{\chi_{0}}\right)^2
        +\frac{2\bpb^{\ssst (0)}\cdot\bpb^{\ssst (0)}}{\Lr^2\left(\hbar\, a^2 - \xi\chi_{0}^2\right)}\nld
	    &\qquad\qquad - \frac{\Lr^2\hbar a^4}{2}\left(\,^{\ssst (3)}\! R 
        - 2\bar{\Lambda}\right) 
        + \frac{\Lr^2}{2}U^{\chi_{0}} = 0\ ,
    \end{align}
which is just the Hamiltonian constraint of GR given by eq.~\eqref{QGR_Hchichi} with $\chi=p_{\chi}=0$, or equivalently by eq.~\eqref{QGR_SC_EHJm}.
If we use eq.~\eqref{QGDHD_SC_pHJfix} we obtain the EHJ equation, meaning that $S^{\srm{HD}}_{0} \equiv S^{\srm{E\chi_{0}}}_{0}$.
Hence, we derive the following important conclusion:
    \begin{quote}
        \textit{The non-vacuum EHJ equation corresponding to the classical non-vacuum GR arises not only in the semiclassical approximation to QGDGR but also in a semiclassical approximation to a quantum gravity theory based on a general quadratic curvature gravity with the EH term.}
    \end{quote}
The correct classical gravity limit is therefore not unique to QGDGR (as anticipated in~\cite{MSc}).
It must be kept in mind that it is the nature of the approximation that allows one to recover the first order classical theory, rather than a specific action for gravity that we have chosen to work with.
Precisely such generality of this result encourages one to look for other theories where a similar semiclassical approximation scheme may perturbatively exclude unwanted contributions which would otherwise spoil the classical description of gravity in terms of GR, if GR is taken as a valid first order classical theory.
An example that immediately comes to mind is to investigate the action containing terms related to conformal anomaly (cf. discussion following eq.~\eqref{confanom}) within quantum geometrodynamics approach.
Another example could be a quantized truncated infinite-derivative theory of gravity based on~\cite{Maz, Mazum}.

Next we address eqs.~\eqref{QHD_SC_PhiL0Re}, \eqref{QHD_SC_PhiL0Im}, \eqref{QHD_SC_psiL2} and \eqref{QHD_SC_psiL0}. It is, however, not clear how to proceed from here.
The difficulties arise mainly from point 5. stated at the beginning of the current subsection.
Here we only suggest certain directions and explain our educated guess towards a Hamilton-Jacobi equation that would correspond to the SEE with counter-terms.

First of all, it is important to understand the results obtained at order $\mathcal{O}(\Lr^{2})$ given by eq.~\eqref{QHD_SC_KFix} and eq.~\eqref{QHD_SC_KtlessFix}.
What is their meaning? 
We think the proper way to think about these two equations is to see them as \textit{constraint} equations. 
In fact, an equation similar to eq.~\eqref{QHD_SC_KFix} arises as a second-class constraint in the WE theory, i.e. eq.~\eqref{HWchi_Q}.
The latter equation could be interpreted within the Dirac constraint analysis as a second-class constraint that eliminates $\kb$, as discussed in section~\ref{subs_HWEchiH}.
We saw in the said section that this ``elimination'' has a price: the need for Dirac brackets arises because Poisson brackets between $\kb$ and other canonical variables do not necessarily vanish.
Hence, eq.~\eqref{QHD_SC_KFix} could be treated in analogy with eq.~\eqref{HWchi_Q}, i.e. as a statement of a broken conformal symmetry in classical GR which says that $\kb$ is not an arbitrary independent variable but a \textit{function of other canonical variables} which arises as a solution to the EHJ equation.
We therefore should continue with $\kb\rightarrow\kb(q^{\ssst A})$.
A similar interpretation of equation eq.~\eqref{QHD_SC_KtlessFix} is to be made, except that we have not included in this thesis a Hamiltonian formulation of a EH$+R^2$ gravity.
The EH$+R^2$ theory would lead to $\bPb = 0$ (cf. eq~\eqref{HDall_Pkt}) and an associated second-class secondary constraint $\dot{\bPb} \deq 0$.
This second-class constraint would be equivalent to eq.~\eqref{QHD_SC_KtlessFix}.
Thus, eq.~\eqref{QHD_SC_KtlessFix} means that $\bktb$ is not an independent canonical variable but is now fixed as a \textit{function of other canonical variables}.
Hence, we should think of $\bktb$ as $\bktb\rightarrow\bktb(q^{\ssst A})$.
Then one of the main problems in the derivation of the semiclassical approximation concerns with finding a consistent way of including eq.~\eqref{QHD_SC_KFix} and eq.~\eqref{QHD_SC_KtlessFix} into subsequent orders of the approximation.
Since $S^{\srm{HD}}_{1}$ is a functional of $q^{\ssst A}$ and $Q^{\ssst I}$, one has to implement the following\footnote{
We could expect that this substitution corresponds to using the EE in eq.~\eqref{SEE_full_TlessB2} and eq.~\eqref{SEE_full_TrB1} in the next order, i.e. in eq.~\eqref{SEE_full_TlessB2} and eq.~\eqref{SEE_full_TrB2}.}
    \begin{equation}
    \label{QHD_SC_S1_red}
        S^{\srm{HD}}_{1} = S^{\srm{HD}}_{1}[q^{\ssst A},Q^{\ssst I}(q^{\ssst B})]
    \end{equation}
in the equations.
This implies that all derivatives $\delta_{\ssst I}S^{\srm{HD}}_{1}$ have to be evaluated at $Q^{\ssst I} = Q^{\ssst I}(q^{\ssst B})$,
    \begin{equation}
    \label{QHD_SC_setDelS1}
        \delta_{\ssst I}S^{\srm{HD}}_{1} \rightarrow\delta_{\ssst I}S^{\srm{HD}}_{1}  \bigg\rvert_{Q^{\ssst I} = Q^{\ssst I}(q^{\ssst B})}\ ,
    \end{equation}
while all derivatives $\delta_{\ssst A}S^{\srm{HD}}_{1}$ are still to be evaluated while $Q^{\ssst I}$ is held fixed.

Secondly, we can write~$S^{\srm{HD}}_{1}$ as
    \begin{align}
    \label{QHD_SC_S1expa}
        S^{\srm{HD}}_{1}[q^{\ssst A},Q^{\ssst I}] & =  \prescript{\ssst  0}{}{S}^{\srm{HD}}_{1}[q^{\ssst A}] + \aw \prescript{\ssst \alpha}{}{S}^{\srm{HD}}_{1}[q^{\ssst A},\bktb] + \br \prescript{\ssst \beta}{}{S}^{\srm{HD}}_{1}[q^{\ssst A},\kb]\ ,
    \end{align}
and consider the contributions of $\aw$ and $\br$ separately. $\prescript{\ssst  0}{}{S}^{\srm{HD}}_{1}[q^{\ssst A}]$ is separated out such that it corresponds to ${S}^{\srm{E\chi_{0}}}_{1}[q^{\ssst A}]$ in eq.~\eqref{QGR_SC_S1andBR}.
Since the extended Berry connection $\mathcal{B}_{\ssst I}$ appears at the same order $\aw$ and $\br$, it is reasonable to expect that it too can be written as
    \begin{align}
    \label{QHD_SC_Bexpa}
        \mathcal{B}_{\ssst X}[q^{\ssst A}, Q^{\ssst I}] & = \prescript{\ssst  0}{}{\mathcal{B}}_{\ssst X}[q^{\ssst A}] + \aw\prescript{\ssst \alpha}{}{\mathcal{B}}_{\ssst X}[q^{\ssst A}, \bktb]
        +
        \br \prescript{\ssst \beta}{}{\mathcal{B}}_{\ssst X}[q^{\ssst A},\kb]\ .
    \end{align}
Furthermore, we introduce
    \begin{equation}
    \label{QHD_SC_Pidef}
        \Pi_{\ssst X} : = \delta_{\ssst X} S^{\srm{HD}}_{1} + \hbar \mathcal{B}_{\ssst X}\ ,
    \end{equation}
such that
    \begin{align}
    \label{QHD_SC_Piexpa}
        \Pi_{\ssst X}[q^{\ssst A}, Q^{\ssst I}] & = \prescript{\ssst  0}{}{\Pi}_{\ssst X}[q^{\ssst A}] + \aw\prescript{\ssst \alpha}{}{\Pi}_{\ssst X}[q^{\ssst A}, \bktb]
        + \br \prescript{\ssst \beta}{}{\Pi}_{\ssst X}[q^{\ssst A}, \kb]\ .
    \end{align}

Now we look at eq.~\eqref{QHD_SC_PhiL0Re} and for simplicity let us assume $\xi_{c} = 0$. 
We shall make an assumption that $\delta_{\ssst I}\mathcal{A}_{0} = 0$.
This assumption could be justified by demanding that the amplitude of the wave function $\Psi$ does not contribute to the SEE.
Let us add and subtract $\hbar \mathcal{B}_{\ssst I} \mathcal{B}_{\ssst J}$ inside the first square bracket in eq.~\eqref{QHD_SC_PhiL0Im} and use the first order solutions eq.~\eqref{QHD_SC_KFix} and eq.~\eqref{QHD_SC_KtlessFix}.
Then, using eq.~\eqref{QHD_SC_Pidef}, eq.~\eqref{QHD_SC_PhiL0Re} becomes
    \begin{align}
    \label{QHD_SC_PhiL0Re1}
        \rm{Re}\mathcal{O}_{\Phi}(\Lr^{0}): \quad &  \mathring{\mirl{\mathscr{G}}}^{\ssst IJ}_{\ssst{\aw,\br}}
        \left[
         \frac{1}{\hbar}\Pi_{\ssst I}\Pi_{\ssst J}- 
        \hbar\mathcal{B}_{\ssst I}\mathcal{B}_{\ssst J}
        - \hbar\rm{Re}\left\langle\delta^2_{\ssst IJ}\right\rangle
        \right]
        - \mathcal{D}^{\ssst I}\Pi_{\ssst I} 
        - \aw\hbar\bar{\mathbf{C}}^{\srm B}\cdot\bar{\mathbf{C}}^{\srm B}\nld
        & \quad
        - \frac{1}{6\hbar}\delta_{a} S^{\srm{HD}}_{0} \Pi_{a} 
        + \frac{4}{\hbar a^2 - \frac{\chi_{0}^2}{6}}\delta_{\bhb}S^{\srm{HD}}_{0}\cdot \Pi_{\bhb}
        + \delta_{\chi_{0}}S^{\srm{HD}}_{0}\Pi_{\chi_{0}} 
        = - \left\langle\qHo{\chi}\right\rangle\ .
    \end{align}
Next step could be to compress the last line using the DeWitt supermetric given in the first of eq.~\eqref{QGR_SC_DeWittMet_App0},
    \begin{align}
    \label{QHD_SC_PhiL0Re2}
        \rm{Re}\mathcal{O}_{\Phi}(\Lr^{0}): \quad &  \mathring{\mirl{\mathscr{G}}}^{\ssst IJ}_{\ssst{\aw,\br}}
        \left[
         \frac{1}{\hbar}\Pi_{\ssst I}\Pi_{\ssst J}- 
        \hbar\mathcal{B}_{\ssst I}\mathcal{B}_{\ssst J}
        - \hbar\rm{Re}\left\langle\delta^2_{\ssst IJ}\right\rangle
        \right]
        - \mathcal{D}^{\ssst I}\Pi_{\ssst I} 
        - \aw\hbar\bar{\mathbf{C}}^{\srm B}\cdot\bar{\mathbf{C}}^{\srm B}\nld
        & \quad
        + \frac{2}{\hbar}\mirl{\mathscr{G}}^{\ssst AB}\delta_{\ssst A} S^{\srm{HD}}_{0} \Pi_{\ssst B}
        = - \left\langle\qHo{\chi}\right\rangle\ .
    \end{align}
This concise result should be compared to eq.~\eqref{QGR_SC_S1andBR} in the semiclassical approximation to QGDGR --- the difference is the entire first line of eq.~\eqref{QHD_SC_PhiL0Re1}.
But we still have a problem with $\aw,\, \br \rightarrow 0$ limit. 

Let us now turn to making the limit $\aw, \br \rightarrow 0$ possible.
Firstly, it should be noted that $\rm{Re}\left\langle\delta^2_{\ssst IJ}\right\rangle$ should in principle also be expanded, because $\mathcal{B}_{\ssst I}\mathcal{B}_{\ssst J} - \rm{Re}\left\langle\delta^2_{\ssst IJ}\right\rangle $ should be invariant under a phase transformation of $\psi$.
So far we have assumed no constraints on $\psi$ regarding its dependence on $Q^{\ssst I}$.
(Perhaps it could be possible to set $\hbar\mathcal{B}_{\ssst I}\mathcal{B}_{\ssst J} = \hbar\rm{Re}\left\langle\delta^2_{\ssst IJ}\right\rangle $ but we cannot find a reasonable motivation, other than just a convenient choice of gauge.)
But from the first line of eq.~\eqref{QHD_SC_PhiL0Re2} it can be seen that $\psi$ must contain a phase dependent on $Q^{\ssst I}$ in order to keep that line invariant under a phase transformation.
Both $\hbar\mathcal{B}_{\ssst I}\mathcal{B}_{\ssst J}$ and $\hbar\rm{Re}\left\langle\delta^2_{\ssst IJ}\right\rangle $ contain two derivatives of $\psi$, which means that they contain two derivatives of the phase (recall that $\psi$ is normalized to one so its amplitude is independent on $q^{\ssst A}$ and $Q^{\ssst I}$); this implies that the phase $\theta [q,Q]$ of $\psi$ can be expanded to isolate the contribution at order $\aw$ and $\br$.
In fact, this was implicitly assumed in eq.~\eqref{QHD_SC_Bexpa}.
So one could imagine that the phase of $\psi$ can be separated as
    \begin{equation}
    \label{QHD_SC_thetaexp}
        \theta [q^{\ssst A},Q^{\ssst I}] = \prescript{\ssst  0}{}{\theta} [q^{\ssst A}] + \aw \prescript{\ssst  \alpha}{}{\theta} [q^{\ssst A},\bktb]\ ,
    \end{equation}
and similarly for $\br$ contribution.
But this is where one needs to be cautious.
Namely, eq.~\eqref{QHD_SC_thetaexp} implies that also the backreaction is expanded in $\aw$ and $\br$.
However, constants $\aw$ and $\br$ are \textit{bare} at this point, which means that regularization and renormalization is yet to take place.
Only after the divergences that appear from the backreaction upon its evaluation are taken care of can one make an expansion of the backreaction in terms of renormalized $\aw$ and $\br$ in order to solve the SEE equation.
But renormalization cannot yet be done because we do not have a correct form of the EHJ at order $\mathcal{O}_{\Phi}(\Lr^{0})$.
Hence, assuming $\aw \ll 1$ and $\br\ll 1$  or that $\aw/\Lr^{2}\ll 1$ and $\br/\Lr^{2}\ll 1$ has no meaning before the renormalization has taken place.
So it seems that this is a problem.
A possible way out would be to assume $\delta_{\ssst I}\psi = 0$, which could be justified by thinking of $\psi$ as evolving on a classical background described only by $q^{\ssst A}$ degrees of freedom.
This means $\mathcal{B}_{\ssst I} = 0$ and $\left\langle\delta_{\ssst IJ}^{2}\right\rangle = 0$ as well.
However, such assumptions put constraints on the $Q^{\ssst I}$-dependent part of $S^{\srm{HD}}_{1}$; moreover, they eliminate any interesting contribution from the higher-derivative terms in the equation for $\psi$, i.e. the entire first line of eq.~\eqref{QHD_SC_psiL0} as well as the last term on the RHS of the equation vanish.
We shall assume $\delta_{\ssst I}\psi = 0$, only because we would like to simplify things and focus on remedying the issue of treating $\aw$ and $\br$ as expansion parameters before renormalization --- an issue which would persist even if $\delta_{\ssst I}\psi = 0$ were not assumed.

Recall that the whole point of treating $\aw$ and $\br$ as perturbation parameters is to solve the SEE as a second-order differential equation instead of the fourth order one.
This was achieved in e.g. \cite{ParSim} by treating $\hbar$ as an expansion parameter, as we mentioned in the previous chapter, or in \cite{Mazzit} in order to reduce the order of equations \textit{before} the quantization.
But since $\aw$ and $\br$ appear at the order $\Lr^{0}$ in a WKB expansion with $\Lr^{-2}$ as the expansion parameter, the WKB approximation has already achieved the order reduction: as we have shown by deriving eqs.~\eqref{QHD_SC_KFix}-\eqref{QHD_SC_HAMGR}, the classical non-vacuum GR arises without the need of approximations in terms of $\aw$ and $\br$.
This is a very important distinction between our derivation and treatments in \cite{Mazzit, Simon1,Simon2,ParSim}.
Therefore, let us change the way we think about parameters $\aw$ and $\br$ and treat them simply as bare couplings, without any assumptions on their size yet.
We have already noted with eqs.~\eqref{QHD_SC_S1_red} and~\eqref{QHD_SC_setDelS1} that one should implement the $\mathcal{O}(\Lr^{2})$ solution (i.e. eqs.~\eqref{QHD_SC_KFix}-\eqref{QHD_SC_HAMGR} ) in the subsequent orders --- without any perturbation in $\aw$ and $\br$ parameters.
Namely, let us think of eqs.~\eqref{QHD_SC_S1expa},\eqref{QHD_SC_Bexpa} and \eqref{QHD_SC_Piexpa} only for indices $X = I$ as sort of \textit{separation Ans\"atze}, which tells us that $\kb$ and $\bktb$ appear within contributions that are coupled to $\aw$ and $\br$, respectively.
These Ans\"atze simply demand that $\aw = 0$ and $\br = 0$ switches the dependence on $\kb$ and $\bktb$ off.
If we thought of eqs.~\eqref{QHD_SC_S1expa},\eqref{QHD_SC_Bexpa} and \eqref{QHD_SC_Piexpa} in that way and if we recall that $\kb$ and $\bktb$ in those equations are functions of $q^{\ssst A}$ due to the classical EE (cf.  eqs.~\eqref{QHD_SC_KFix}-\eqref{QHD_SC_HAMGR} ), then there is no need to talk about perturbative constraints in terms of $\aw$ and $\br$ parameters --- the theory is already formally perturbatively reduced.

Given this conclusion, we can demand eqs.~\eqref{QHD_SC_S1expa}, \eqref{QHD_SC_Bexpa} and \eqref{QHD_SC_Piexpa} for index $X = I$ only, while leaving $X = A$ components unperturbed. 
Then we plug $X = I$ versions of these equations into eq.~\eqref{QHD_SC_PhiL0Re2} and obtain the following,
    \begin{align}
    \label{QHD_SC_PhiL0Re3}
        & 
         \frac{\br}{2\hbar}\prescript{\ssst \beta}{}{\Pi}_{\ssst \kb}\prescript{\ssst \beta}{}{\Pi}_{\ssst \kb}
         - \frac{\aw}{2\hbar}\prescript{\ssst \alpha}{}{\Pi}_{\ssst \bktb}\cdot \prescript{\ssst \alpha}{}{\Pi}_{\ssst \bktb}
        - \br\mathcal{D}^{2}_{\srm{R}}\prescript{\ssst \beta}{}{\Pi}_{\ssst \kb} +  \aw\mathcal{D}^{2}_{\srm{W}}\cdot\prescript{\ssst \alpha}{}{\Pi}_{\ssst \bktb}
        - \aw\hbar\bar{\mathbf{C}}^{\srm B}\cdot\bar{\mathbf{C}}^{\srm B}
        \nld
        & \quad
        + \frac{2}{\hbar}\mirl{\mathscr{G}}^{\ssst AB}\delta_{\ssst A} S^{\srm{HD}}_{0} \Pi_{\ssst B}
        = - \left\langle\qHo{\chi}\right\rangle\ ,
    \end{align}
where we remind that the entire equation is evaluated at the solution to the EHJ equation obtained at the previous order $\mathcal{O}(\Lr^{2})$ which determines the extrinsic curvature in each term above through eq.~\eqref{QHD_SC_KFix} and eq.~\eqref{QHD_SC_KtlessFix} in terms of $a,\bhb$ and $\chi_{0}$. 
It can finally be seen that coupling constants $\aw$ and $\br$ are ready to absorb divergences that might appear in $\left\langle\qHo{\chi}\right\rangle$. 
Equation~\eqref{QHD_SC_PhiL0Re3} at the same time represents an equation for $\Pi_{\ssst B}$, but one can solve for it (in principle) only after the renormalization has been done, because only then one can employ the perturbative constraints approach in terms of $\aw^{\ssst phys}$ and $\br^{\ssst phys}$ \textit{running} couplings.

Let us only mention that eq.~\eqref{QHD_SC_psiL0} resembles the functional Schr\"odinger equation eq.~\eqref{QGR_SC_Sch0} after implementation of eq.~\eqref{QHD_SC_KFix} and eq.~\eqref{QHD_SC_KtlessFix} and using the assumption that $\delta_{\ssst I}\psi = 0$.
As we said earlier, the latter assumption might be a severe restriction.
We do not go further into the details here.

Equation \eqref{QHD_SC_PhiL0Im} contains ``corrections'' to the amplitude even if assumption $\delta_{\ssst I}\psi = 0$ is implemented. However, these corrections cannot be determined before eq.~\eqref{QHD_SC_PhiL0Re3} is solved.

There is one more issue that is encountered here and is perhaps the most important technical one. As already mentioned in point 5. at the beginning of this subsection, even though the first line of eq.~\eqref{QHD_SC_PhiL0Re3} resembles the higher-derivative part of the Hamiltonian constraint in eq.~\eqref{HDall_Hofin}, one must not identify $\Pb$ with $\prescript{\ssst \beta}{}{\Pi}_{\ssst \kb}$ and $\bPb$ with $\prescript{\ssst \alpha}{}{\Pi}_{\ssst \bktb}$.
The reason is the that $\prescript{\ssst \beta}{}{\Pi}_{\ssst \kb}$ and $\prescript{\ssst \alpha}{}{\Pi}_{\ssst \bktb}$ do not make sense as momenta because there are no additional variables they are conjugate to --- any additional variables are removed through eq.~\eqref{QHD_SC_KFix} and eq.~\eqref{QHD_SC_KtlessFix}.
More importantly, the reason for the first line in eq.~\eqref{QHD_SC_PhiL0Re3} resembling the part of the Hamiltonian constraint in eq.~\eqref{HDall_Hofin} is the Legendre transform, which is rooted in the Hamiltonian theory and therefore in the quantized theory.
Because of this, one must find a way to relate $\prescript{\ssst \beta}{}{\Pi}_{\ssst \kb}$ and $\bPb$ with $\prescript{\ssst \alpha}{}{\Pi}_{\ssst \bktb}$ (or $S^{\srm{HD}}_{1}$) to a HJ functional that corresponds to a theory which is \textit{first} perturbatively constrained and \textit{after that} Hamilton-formulated.
Even though a Hamiltonian formulation of perturbatively constrained theories exists in \cite{Mazzit}, the relationship between an exactly formulated Hamiltonian theory and the perturbatively constrained Hamiltonian theory does not seem to exist, to our knowledge. We do think it is possible to achieve it through the Hamilton-Jacobi approach, but we are unsure how to proceed.
The most important point here that is certain is that relationship of the higher-derivative momenta and the perturbatively constrained HJ functional must ensure that the kinetic term in the higher-derivative Hamiltonian is turned into the potential, i.e that the Legendre transform in the sector of additional degrees of freedom is undone.
We hope that manipulations presented here may serve as a guideline to achieve such a goal in the future.

\section{Final remarks}

This chapter has shown opportunities and difficulties in QGDHD as compared to QGDGR.
The formulation of both theories was achieved in terms of unimodular-conformal variables which to our knowledge does not exist in the literature to the extent presented in here.
The main concern was the semiclassical approximation to the HDWDW equation and the derivation of the EHJ equation and its corrections which take into account the backreaction and the possibility for absorbing the divergences into couplings $\aw$ and $\br$.
It was shown that --- at least in principle --- that it is possible to achieve this by performing a BO-WKB type of approximation to the HDWDW equation in terms of inverse powers of $\Lr^{2}$.
The result is the non-vacuum EHJ equation obtained at the highest order of the expansion, given by eq.~\eqref{QGDHD_SC_pHJfix} and~\eqref{QHD_SC_PhiL0Re3}.
This proves that a canonically quantized general quadratic curvature gravity with higher-derivative terms and the EH term gives a valid classical theory given by GR without any contributions of the higher-derivative terms and the plague of additional degrees of freedom they carry.
The additional degrees of freedom were eliminated by eq.~\eqref{QHD_SC_KFix} and eq.~\eqref{QHD_SC_KtlessFix}, which determine (or fix) $\kb$ and $\bktb$ in terms of the first-order configuration variables $a,\bhb$ and $\chi_{0}$.
This information must be implemented in all subsequent orders of the semiclassical approximation.
This is enough to achieve the order reduction which was argued for in the previous chapter.

However, unlike the usual entirely classical perturbative order reduction of the SEE reviewed in the previous chapter, we have shown that order reduction happens already with the BO-WKB expansion in terms of inverse powers of $\Lr^{2}$ from a \textit{quantum gravity} theory.
Moreover, the approximation puts counter-terms automatically at the same order as the backreaction, making them ready to absorb the divergences that stem form the evaluation of the latter into couplings $\aw$ and $\br$.
This tells in favor of interpreting the higher-derivative terms as being relevant at high energies (or relatively small length scales) instead of being genuine classical entities which introduce additional degrees of freedom and spacetime instabilities.

The crucial role in deriving the semiclassical limit is played by the terms in the Hamiltonian formulation of a higher-derivative theory which are linear in momenta.
The presence of these terms has led to eq.~\eqref{QHD_SC_PhiL2ReSab}, from which the EHJ equation stems. 
This equation resembles the Legendre transform in the usual Hamiltonian formulation of GR --- which is why the method works.
However, of crucial significance was to apply \textit{derivatives} with respect to $\kb$ and $\bktb$ to this equation, in order to fix the extrinsic curvature.
We stress that this could be a clear example of Dirac second-class constraint, whose preservation in time eventually determines a Lagrange multiplier.

There were several pitfalls which we tried to point out. The most important one is how to relate the Hamilton-Jacobi treatment of an exact higher-derivative theory and its perturbatively constrained version.
The problem revolves around expressing the higher-derivative momenta in terms of the first-order HJ functional.
We hope we gave some pathways on how to approach this problem.

The insights gained from our treatment might motivate a search of connections of quantum geometrodynamics with other approaches to quantum gravity\footnote{E.g. the program of \textit{asymptotic safety} \cite{Bened, OUP, ReuSauer}, where one investigates the possibility of finding an effective action which in the limit of high energies has finite number of terms and non-divergent couplings; moreover, the so-called ``infinite-derivative'' theories \cite{Maz} where one considers a non-local theory of gravity valid at high-energies.
In both of these approaches it seems that additional and runaway degrees of freedom are avoided.}.
We think that this question is important and we hope this thesis encourages its investigation.

{\centering \hfill $\infty\quad$\showclock{0}{40}$\quad\infty$ \hfill}

\chapter{Conclusions}
    \label{ch:time}
    The beauty of using the unimodular-conformal variables is that one can clearly identify a single degree of freedom with a single variable.
Its choice is driven by carefully listening to the symmetry features inscribed in a given theory and then transforming one ``coordinate system'' in the configuration space into another such that the new ``directions'' are suitably tuned to these symmetry features.

In the case of metric theories of gravity which may or may not have conformal symmetry, the suitable choice is the set of unimodular-conformal variables, because it aligns the direction of conformal transformation with the ``axis'' of the scale density (which is related to a volume) and the expansion density (which is related to volume's timelike evolution), while other directions are orthogonal to it and aligned to the shape density (which encodes the conformally invariant metric degrees of freedom) and the shear density (which represents the timelike evolution of the conformal degrees of freedom).
This split into conformal and non-conformal degrees of freedom is motivated not only by examining various coordinate transformations of the $GL(4,\mathbb{R})$ group but also by questioning the meaning of attributing units to coordinates themselves. The latter led us to motivate a dimensionless relative measure of a length scale $\Lr$ which encodes how large is the area of spacetime measured by the scale density compared to the Planck length scale.

Introduction of the dimensionless relative length scale and formulation of geometric objects (Christoffel symbols, curvature tensors) in terms of unimodular-conformal variables has allowed us to examine the conformal properties of any theory that lives on Riemannian geometry.
It has further led to a formulation of a generator of conformal transformation of fields (local Weyl rescaling) and the definition of conformal invariance in terms of it.
Furthermore, using unimodular-conformal variables in $3+1$ decomposition of spacetime, we have shown why the Weyl-tensor squared and the $R^2$ terms independently contribute with conformal and non-conformal degrees of freedom, which itself represents significant improvement compared to the previous works on the topic.
Moreover, the Hamiltonian formulation --- and later the quantization --- is also made more clear using our methods, as compared to the existing formulations in the literature.
These tools have proven invaluable in understanding the higher-derivative theories and they deserved a significant part of the thesis.

The higher-derivative theories are usually sought as alternatives to classical theory of gravity described by GR, but they suffer from instabilities and increasing number of degrees of freedom.
They are motivated by requirements of semiclassical gravity, which necessitates the introduction of quadratic curvature terms in Einstein equations in order to absorb divergent terms that appear in the expectation value of the energy-momentum tensor operator (the backreaction).
The counter-terms of most significance can be described by two pieces: the Weyl-tensor squared and the $R^2$ term.
In spite of the usual discussion of these theories as exact classical theories of gravity, we have embarked on an attempt to make sense of the quadratic curvature terms as \textit{perturbations} relevant at higher energies. 
Our work was inspired by \cite{Simon1, Simon2, ParSim} and \cite{Mazzit}, who favoured the perturbative nature of the counter-terms over their interpretation as an exact contribution to the theory of gravity.
The main purpose of the thesis was to quantize a higher-derivative theory of gravity and investigate the possibility to have a meaningful semiclassical approximation where such a theory naturally gives rise to the mentioned counter-terms --- an outcome which is not met in QGDGR.
We have reviewed the quantization and the semiclassical approximation in canonical GR using the unimodular-conformal variables and the dimensionless relative length scale $\Lr$.
The latter is used as an expansion parameter and thus we avoided usual issues with limits of dimensionful parameters.
The formulation of a QGDHD was shown to be rather similar to the formulation of the QGDGR: canonical quantization of the constraints derived in the Hamiltonian formulation.
The semiclassical approximation to QGDHD was based on the same ansatz as in QGDGR, with the exception that the wave functional lives on an extended configuration space which includes the extrinsic curvature as additional degrees of freedom.
The approximation itself is a combination of the Born-Oppenheimer type and a WKB-type of approximation.
Two of the main questions of the approach were how to eliminate the additional degrees of freedom in the semiclassical approximation to a higher-derivative quantum gravity and how to recover non-vacuum GR.
The answers turned out to follow without any additional assumptions because the higher-derivative terms appear only at order $\mathcal{O}(\Lr^{0})$, while the highest order of approximation $\mathcal{O}(\Lr^{2})$ produced equations which fixed the extrinsic curvature in terms of the first-order variables.
Manipulating this information we showed that classical non-vacuum GR in the form of non-vacuum Einstein-Hamilton-Jacobi equation emerges from a QGDHD.
This proves that non-vacuum GR is not a classical limit unique to canonically quantized GR.
Furthermore, the higher-derivative terms appeared at the same order as the backreaction, thus introducing the correction to the EHJ equation due to the presence of counter-terms.
This was achieved without any use of the perturbative approach in terms of couplings of the higher-derivative terms.
This is important to emphasize because these couplings may only be constrained to be ``small'' if they have already been redefined by absorbing the divergences.
It is thus necessary to perturb the equations in terms of these parameters only once the couplings are renormalized and one wishes to solve the SEE with a regularized backreaction.

There are two main issues with the approach we pursued. One is that it involves a few unclear assumptions which have left the question of deriving the functional Schr\"odinger equation unresolved. 
The other is the lack of proof that our result corresponds to the classical perturbatively constrained higher-derivative theory.
The former is an issue that requires more analysis of the semiclassical approximation.
The latter is an issue which would have to be addressed in a less complicated context, on a toy model of a constrained system or a minisuperspace model.

In spite of the issues, we think that our work investigated promising possibilities for considering QGDHD at least as seriously as QGDGR.
Furthermore, we think that quantum gravity community lacks investigations of interconnections among different approaches to quantum gravity.
Our work opens some doors in addressing this gap, because the questions raised in this thesis may relate to infinite derivative theories \cite{Mazum} and asymptotic safety approach to gravity \cite{Bened,ReuSauer}.
We therefore hope that our work will inspire further investigations in various directions.

A single answer does not always correspond to a single question.


\begin{appendix}
    \renewcommand\thechapter{}
    \chapter*[Appendix]{Appendix}
    \addcontentsline{toc}{chapter}{Appendix}
    \renewcommand\thechapter{A}
    \section{Coordinate variation of Christoffel symbols}
\label{App_VarGamma}

In order to derive the so-called integrability condition for the Killing vector field $\xi^{\mu}$, that is, equation \eqref{CT_2} or \eqref{CT_2} set to zero, one is usually referred (see e.g. Appendix C.3 in \cite{Wald}) to use a specific sum of cyclic permutation of indices of the definition of the Riemann tensor via commutator of covariant derivatives of that Killing vector,
    \begin{equation}
        \left[\nabla_{\mu},\nabla_{\nu}\right]\xi^{\alpha}={R^{\alpha}}_{\beta\mu\nu}\xi^{\beta}\ ,
    \end{equation}
and Bianchi identity for the Riemann tensor. However, one should be able to derive such integrability condition from some principle which can be invoked in order to use the specific sum of cyclic index permutations of the above expression. Of course, if one is familiar with the Bianchi identity of the Riemann tensor (which one should be on any course on General Relativity), an idea to use it may come to one's mind and after some trial and error, a correct answer is obtainable.
But it is more satisfactory to know the reason why this works.

The reason is simply the demand that the variation of curvature and the variation of connection vanish under along a Killing vector. It is actually enough to demand that the variation of the connection along the killing vector vanish. One can show that, if a metric is present, the Lie derivative of the connection can be expressed as
    \begin{align}
    \label{deltaGammaCalc}
        \delta {\Gamma^{\alpha}}_{\mu\nu}&=
        \frac{1}{2}g^{\alpha\beta}
        \Biggpar{
        \reduline{{\nabla_{\mu}\nabla_{\beta}\xi_{\nu}}}
        +\nabla_{\mu}\nabla_{\nu}\xi_{\beta}
        +\gruline{\nabla_{\nu}\nabla_{\beta}\xi_{\mu}}
        +\nabla_{\nu}\nabla_{\mu}\xi_{\beta}
        -\reduline{\nabla_{\beta}\nabla_{\mu}\xi_{\nu}} 
        -\gruline{ \nabla_{\beta}\nabla_{\nu}\xi_{\mu}}
        }\nonumber\\[6pt]
        &=g^{\alpha\beta}
        \Biggpar{
        \nabla_{(\mu}\nabla_{\nu)}\xi_{\beta}
        +\reduline{{\nabla_{[\mu}\nabla_{\beta]}\xi_{\nu}}}
        +\gruline{\nabla_{[\nu}\nabla_{\beta]}\xi_{\mu}}
        }\nonumber\\[6pt]
        &=g^{\alpha\beta}
        \Biggpar{
        \nabla_{(\mu}\nabla_{\nu)}\xi_{\beta}
        +\frac{1}{2}\xi_{\rho}{R^{\rho}}_{\nu\beta\mu}
       +\frac{1}{2}\xi_{\rho}{R^{\rho}}_{\mu\beta\nu}
        }
        \nonumber\\[6pt]
        &=g^{\alpha\beta}
        \Biggpar{
        \nabla_{(\mu}\nabla_{\nu)}\xi_{\beta}
        -\frac{1}{2}\xi_{\rho}{R^{\rho}}_{\nu\mu\beta}
       -\frac{1}{2}\xi_{\rho}{R^{\rho}}_{\mu\nu\beta}
        }=g^{\alpha\beta}
        \Biggpar{
        \nabla_{(\mu}\nabla_{\nu)}\xi_{\beta}
        -\xi_{\rho}{R^{\rho}}_{(\mu\nu)\beta}
        }\ .
    \end{align}
Setting this expression to zero gives the integrability condition. The meaning behind this requirement is understood if one thinks of what should happen to the curvature along a motion in the direction of a Killing vector: the geometry of space does not change under a symmetry transformation and all geometric objects should acquire a zero physical change when evaluated at points along the direction of a Killing vector.

The above equation has a meaning even if it does not vanish. If $\xi^{\mu}$ is not a Killing vector, then the above equation expresses the change of the Christoffel symbols along the corresponding direction. For example, we have studied the form of this change under a general infinitesimal coordinate transformation in section \ref{sec_genCT} and showed in equation \eqref{varGamma} that it can be split into two parts --- the one due to the shear (volume-preserving) transformations 
and the one due to the scale (shape-preserving) transformations.
Imposing this explicit change of the Christoffel symbol on one side of the equation and equating it with the result of \eqref{deltaGammaCalc} results in an expression which one may call ``non-isometry integrability condition'', which can (in principle) be used to find transformation vectors in any geometry.
This is how we found the relevant equations for finding vectors that generate conformal transformations in the Minkowski spacetime in section \ref{conftrM}.

\section{3+1 decomposition of spacetime}
\label{App_31}

In this Appendix we briefly sketch the $3+1$ decomposition of spacetime. It is the basis of canonical quantization of theories of gravity \cite{OUP} as well as numerical relativity \cite{EricG, Pad}.

In this formalism the four-dimensional spacetime is described by three-dimensional spacelike hypersurfaces $\Sigma_t$ embedded in four-dimensional spacetime as evolving in time. Therefore a four-dimensional metric shall be descomposed into three-dimensional metric parametrized by a scalar
function $t$ that governs distances on the three-hypersurface and the rest of the components, which describe one's choice of orienting this hypersurface with respect to a defined timelike direction.
This timelike direction is defined as a covariant derivative of the time function $t$
    \begin{equation}
    \label{ntime}
        n_{\mu}=-N\nabla_{\mu}t
    \end{equation}
where $N>0$ is the \textit{lapse function}, and this vector ($n^{\mu}=g^{\mu\nu}n_{\nu}$) is normalized by $g_{\mu\nu}n^{\mu}n^{\nu}=-1$. This vector is orthogonal to $\Sigma_t$ at each point on it and one could imagine that as one is walking along $\Sigma_t$ the orthogonal vector (which therefore extends into the time dimension)
changes its orientation depending on how the hypersurface curves into the time dimension due to embedding. In other words, variation of $n^{\mu}$ will describe the rate of change of the three-dimensional metric.
A particular choice of normalized vector \eqref{ntime} is given in Arnowitt-Deser-Misner (ADM)
variables as
    \begin{equation}
        n_{\mu}=\left(-N,0,0,0\right)\ ,\qquad n^{\mu}=\left(\frac{1}{N},-\frac{N^{i}}{N}\right)\ ,
    \end{equation}
where $N^{i}$ is called the shift vector. The metric is then decomposed as
    \begin{equation} 
    \label{4met31}
        g_{\mu\nu}=h_{\mu\nu}-n_{\mu}n_{\nu}\ ,
    \end{equation}
where $h_{\mu\nu}$ is the metric induced on $\Sigma_t$, such that
    \begin{equation}
    \label{31proj}
        h_{\mu\nu}n^{\mu}=0\ ,\quad  h^{\mu}{}_{\alpha}h^{\alpha}{}_{\nu}=h^{\mu}{}_{\nu}\ ,\quad g^{\mu\nu}h_{\mu\nu}=3\ .
    \end{equation}
Therefore, the $h^{\mu}{}_{\nu}$ and $n^{\mu}n_{\nu}$ are just projection operators: they project any four-dimensional index onto spacelike hypersurface and timelike orthogonal direction. 
Using these projections, a four-tensor $T_{\mu\nu}$, for example, can be
decomposed in the following way: 
    \begin{align}
    \label{Tproj}
        T_{\mu\nu}&=\left(h^{\alpha}_{\mu}-n^{\alpha}n_{\mu}\right)\left(h^{\beta}_{\nu}-n^{\beta}n_{\nu}\right)T_{\alpha\beta} \nonumber\\
        &=\,_{\ssst\parallel}\!T_{\mu\nu}-\,_{\ssst\parallel}\!T_{\mu\ssst\bot}-\,_{\ssst\parallel}\!T_{{\ssst\bot}\nu}+T_{\ssst\bot\bot},
    \end{align}
where ``$\ssst\parallel$'' denotes that the greek indices are projected to the
hypersurface using  $h^{\alpha}_{\mu}$, while ``$\bot$'' denotes the
position of an index that has been projected along the orthogonal
vector $n^{\mu}$. 

The four functions, the lapse $N$ and the shift $N^{i}$, describe the mentioned choice of coordinates. This is seen explicitly from the decomposition \eqref{4met31} which implies that the four-metric and its determinant decompose as
    \begin{equation}
    \label{g31mat}
        g_{\mu\nu}=\left(
        \begin{array}{ccc}
        -N^{2}+N_{i}N^{i}& N_{i}\\[6pt]
        N_{i} & h_{ij} 
        \end{array}
        \right)\ ,\quad \sqrt{g}=N\sqrt{h},
    \end{equation} 
where $h_{ij}$ is now the three-metric as directly formulated with
spatial indices, which is used to raise and lower spatial indices; we also defined $h:={\rm det}\,h_{ij}$.
The inverse of the four-metric has the form
    \begin{equation}
    \label{g31invmat}
        g^{\mu\nu}=\left(
        \begin{array}{ccc}
        -\dfrac{1}{N^2}& \dfrac{N^{i}}{N^2}\\[12pt]
        \dfrac{N^{i}}{N^2} & h^{ij}-\dfrac{N^{i}N^{j}}{N^2} 
        \end{array}
        \right)\ .
    \end{equation}
With these definitions, the time components of objects projected onto
the hypersurface vanish; in \eqref{Tproj}, for example, all
components with ``$\ssst\parallel$'' are now spatial, and the ``$\ssst\parallel$'' can be
dropped with the understanding that greek indices can there be turned
into latin ones $i,j$, etc.: $\,_{\ssst\parallel}\!T_{\mu\nu}\rightarrow \,^{\ssst
  (3)}\! T_{ij}$, $\,_{\ssst\parallel}\!T_{{\ssst\bot}\nu}\rightarrow T_{{\ssst\bot} j}$, etc.,
where objects denoted with a left superscript ``$(3)$'' are intrinsic
to the hypersurface.

We mentioned that the variation of $n^{\mu}$ along the hypersurface will describe the rate of change of the three-dimensional metric. More precisely this means that one forms an object defined by $K_{\alpha\beta}:=h^{\mu}{}_{\alpha}h^{\nu}{}_{\beta}\nabla_{(\alpha}n_{\beta)}$ and uses the above-mentioned fact that now the indices can be considered as spatial ones. The result can be shown \cite[eq. (12.14)-(12.21)]{Pad}  to be\footnote{Note, however the difference in the sign convention in the definition of the extrinsic curvature as compared to \cite{Pad}.}
    \begin{align}
    \label{Kijdef}
        K_{ij} & : =N\Gamma^{0}{}_{ij}=\frac{1}{2}\mathcal{L}_{n}h_{ij}=\frac{1}{2N}\left(\dot{h}_{ij}-2D_{(i}N_{j)}\right),\\[12pt]
    \label{Ktrdef}
        K & : = h^{ij}K_{ij}=\frac{\mathcal{L}_{n}\sqrt{h}}{\sqrt{h}}=\frac{1}{N}\left(\frac{\dot{\sqrt{h}}}{\sqrt{h}}-D_{i}N^{j}\right)\ ,
    \end{align}
where $D_{i}$ denotes the covariant derivative with respect to the three-metric $h_{ij}$, i.e. the covariant derivative which strictly speaks of parallel transport on $\Sigma_{t}$.
Object $K_{ij}$ is called \textit{the extrinsic curvature tensor} and $K$ is its trace. Therefore, the extrinsic curvature is the Lie derivative of the three-metric along the timelike orthogonal vector $n^{\mu}$.
One must be careful to keep in mind that $\mathcal{L}_{n}h_{ij}$ is derived via \textit{projection} of $\mathcal{L}_{n}g_{\mu\nu}$ onto the hypersurface, which eliminates certain terms such as spatial derivatives of $N$.
It is important to note that the trace of the extrinsic curvature involves the three-volume $\sqrt{h}$ and thus can be interpreted as the rate of change of the three-volume.

The Riemann tensor, Ricci tensor, and Ricci scalar can be decomposed in a manner similar to \eqref{Tproj}, but we are interested in the Ricci scalar only, for which one obtains~\cite[chapter 12]{Pad},
    \begin{subequations}
    \begin{align}
    \label{Rdec31}
        R & =\,^{\ssst (3)}R+K_{ij}K^{ij}+K^{2}+2\mathcal{L}_{n}K-\frac{2}{N}D^{i}D_{i}N\,\\[6pt]
    \label{Rdec31a}
        & =\,^{\ssst (3)}R+K_{ij}K^{ij}-K^{2}+2\nabla_{\mu}\left(n^{\mu}K\right)-\frac{2}{N}D^{i}D_{i}N \ ,
    \end{align}
    \end{subequations}
where $\,^{\ssst (3)}R$ is the \textit{intrinsic Ricci scalar curvature} formed from traces of the three-dimensional Ricci tensor\footnote{In three dimensions Weyl tensor identically vanishes so Ricci tensor components are the only remaining non-zero set of components of the Riemann tensor. }, describing the curvature of $\Sigma_{t}$.

The two versions of the Ricci scalar are equivalent but their use depends on the context. For example, \eqref{Rdec31a} is more suitable for calculations in classical and quantum GR because the second-to-last term manifestly represents a boundary term when put into an action. However \eqref{Rdec31} may be more useful for discussions
based on the action of the non-minimally coupled scalar field or $R^2$ gravity. Thus it is important to be aware of both forms and how can one switch from one to the other and this is done by a simple manipulation
$\nabla_{\mu}\left(n^{\mu}K\right)=K\nabla_{\mu}n^{\mu}+n^{\mu}\nabla_{\mu}K=K^2+n^{\mu}\nabla_{\mu}K$, which explains the change of the sign in front of $K^2$ in \eqref{Rdec31a} compared to \eqref{Rdec31}.

The Weyl tensor has two relevant components: $C_{i{\ssst\bot} j{\ssst\bot}}$ and $C^{kl}{}_{j{\ssst\bot}}$. It has been derived in 
\cite{Kluson2014} and \cite{ILP} and already used in author's Master thesis \cite{MSc}. Here we only state the final expressions that are relevant for this thesis given by
    \begin{subequations}
    \begin{align}
    \label{elW}
        C_{ij}^{\srm E}:= -2C_{i{\ssst\bot} j{\ssst\bot}}&=\mathbb{1}_{(ij)}^{\srm Tab}\left(\mathcal{L}_{n}K_{ab}-\!\,^{\ssst (3)}\!R_{ab}-K_{ab}K-\frac{1}{N}D_{a}D_{b}N\right)\ ,\nonumber\\[6pt]
        &=\left(\mathcal{L}_{n}K_{ij}\right)^{\srm T}-\!\,^{\ssst
          (3)}\!R_{ij}^{\srm T}-K_{ij}^{\srm
          T}K-\frac{1}{N}(D_{i}D_{j})^{\srm T}N\ ,\\[6pt]
    \label{magW}
        C_{ij}^{\srm B}:= \varepsilon_{ikl}C^{kl}{}_{j{\ssst\bot}}&=\varepsilon^{kl}{}_{(i}D_{k}K_{j)l}\ .
    \end{align}
    \end{subequations}
Note that $\mathcal{L}_{n}K_{ab}$ is an object that does not correspond to the Lie derivative of $K_{ab}$ along the four-vector $n^{\mu}$ because its derivation involves projection onto the hypersurface which eliminates some terms \cite[section 3.4]{EricG}; this object has the following form:
    \begin{subequations}
    \begin{align}
    \label{LieKdef1}
        \mathcal{L}_{n}K_{ab} &: = \frac{1}{N}\left(\dot{K}_{ab}^{\srm T} - \mathcal{L}_{\vec{N}}K_{ab}\right)\ ,\\[6pt]
    \label{LieKdef2}
        \mathcal{L}_{\vec{N}}K_{ab} & = N^{i}\del_{k}K_{ab} + K_{aj}\del_{b}N^{j} + K_{bj}\del_{a}N^{j}\ ,
    \end{align}
    \end{subequations}
where $\mathcal{L}_{\vec{N}}K_{ab}$ is the Lie derivative of $K_{ab}$ along the three-dimensional vector $N^{i}$.
The two objects in \eqref{elW} and \eqref{magW} are the ``electric'' and ``magnetic'' parts of the Weyl tensor\footnote{Note that in this thesis we choose to work with $C_{ij}^{\srm B}$ instead of $C_{ijk{\ssst\bot}}$ as we did in \cite{MSc} and \cite{KN17} due to its simpler and more intuitive form. 
Also, the notation in definitions of electric and magnetic parts may differ only up to a constant factor from the ones in the literature.} \cite{ILP,xact}, where $\mathbb{1}_{(ij)}^{{\srm T}ab}$ is ``three-dimensional traceless rank-two identity'' defined as
    \begin{align}
    \label{eqn:oneT}
        \mathbb{1}_{(ij)}^{{\srm T}ab}:= \delta_{(i}^{a}\delta_{j)}^{b}-\frac{1}{3}h_{ij}h^{ab},
    \end{align}
rendering each term traceless, $(D_{i}D_{j})^{\srm T}\equiv\mathbb{1}_{(ij)}^{{\srm T}ab}D_{a}D_{b}$ and $\varepsilon_{ikl}$ is the three-dimensional Levi-Civita tensor density. Both the electric and magnetic parts of the Weyl tensor are traceless and they carry five degrees of freedom each, agreeing with a total of 10 for the Weyl tensor. Also, each of these components is conformally invariant but this is obvious only after applying unimodular-conformal decomposition, see subsection \ref{EBwumod}.
Finally, the Weyl invariant, constructed from the Weyl tensor contracted with itself, then takes the form similar to the electromagnetic invariant $F_{\mu\nu}F^{\mu\nu}\sim E^{2}-B^{2}$ and is given \cite{Kluson2014, ILP} by 
    \begin{align}
    \label{C2dec}
        C_{\mu\nu\lambda\rho}C^{\mu\nu\lambda\rho}=2C_{ij}^{\srm E}C^{ij}_{\srm E}-4C_{ij}^{\srm B}C^{ij}_{\srm B}\ .
    \end{align}
Note that the term $C_{ij}^{\srm E}C_{\srm E}^{ij}$ 
in \eqref{C2dec} contains only traceless quantities and does not contain velocities of the trace $K$, but seems to contain the trace $K$ itself. As stated in the main text
of this thesis, if the Weyl tensor is invariant under conformal transformations the trace $K$ should not appear in it due to its inhomogeneous transformation. Moreover, the magnetic part $C_{ij}$ should also not contain the trace $K$. All this is made manifest in section \ref{sec_31conf} by utilizing the unimodular-conformal decomposition.






\section{Various proofs}

\subsection{Conformally invariant expressions with differential operators}
\label{App_DD}

We are interested in investigating conformal properties of the following expression
    \begin{equation}
    \label{RDDN}
        \left(\,^{\ssst (3)}\! R_{ij} + \frac{1}{N}D_{j}\del_{j}N\right)^{\srm T}
    \end{equation}
which appears in \eqref{elWsmpl}, studied in subsection \ref{EBwumod}. 
Let us study the two terms separately and it is instructive to do the calculation in $d$ dimensions and thus switch to greek indices and a $d$-dimensional metric and covariant derivatives.

Consider a scalar field $\phi$ of conformal weight $n_{\phi}=1$. Then according to \eqref{fieldresc} the corresponding conformally invariant scalar density is of the scale weight $\wb=-1$ and is defined by $\bar{\phi}:=A^{-1}\phi$. Then the second term in \eqref{RDDN} evaluates to
    \begin{align}
    \label{DDphi}
        \frac{\nabla_{\mu}\del_{\nu}\phi}{\phi}&=\frac{\nabla_{\mu}\nabla_{\nu}\bar{\phi}}{\bar{\phi}}\nonumber\\
        &=\frac{1}{\bar{\phi}}\bar{\nabla}_{\mu}\del_{\nu}\bar{\phi}
        +\bar{\nabla}_{(\mu}\del_{\nu)}\log A
        +\gb_{\mu\nu}\gb^{\alpha\beta}\del_{\alpha}\log A\,\del_{\beta}\bar{\phi}\nonumber\\
        &\quad
        -\left(2\delta_{(\mu}^{\alpha}\delta_{\nu)}^{\beta}
        -\gb_{\mu\nu}\gb^{\alpha\beta}\right)\del_{\alpha}\log A\,\del_{\beta}\log A \ .
    \end{align}
Now observe that the second and fourth terms appear (up to
$d$-dependent coefficients) in $R_{\mu\nu}$ in \eqref{riccidec}
with an opposite sign. However, the third term contains\footnote{If one generalizes the calculation to an arbitrary scale weight $\wb$ then additional terms proportional to $(1+\wb)\del_{\mu}\log A\,\del_{\nu}\bar{\phi}$ would appear. Then, for example, for the Klein-Gordon scalar field of conformal weight $n_{\phi}=-\wb=-1$ these terms cannot be eliminated unless one takes the trace and subtracts a certain multiple of $R$, as in section \ref{sec_xiphi}.}
$\del_{\beta}\bar{\phi}$ and cannot be found in there, and is present without an opposite-signed pair in $R_{\mu\nu}$ to be canceled with. But this term is in its totality a part of the trace of expression \eqref{DDphi},
    \begin{align}
        \gb^{\mu\nu}\frac{\nabla_{\mu}\nabla_{\nu}\bar{\phi}}{\bar{\phi}}&=\frac{1}{\bar{\phi}}\gb^{\mu\nu}\bar{\nabla}_{\mu}\del_{\nu}\bar{\phi}+\gb^{\mu\nu}\bar{\nabla}_{\mu}\del_{\nu}\log A + d\,\gb^{\alpha\beta}\del_{\alpha}\log A\,\del_{\beta}\bar{\phi}\nonumber\\
        &\quad-\left(d-2\right)\gb^{\alpha\beta}\del_{\alpha}\log A\,\del_{\alpha}\log A \,,
    \end{align}
which means that the \textit{traceless} part of \eqref{DDphi} does not contain it. Therefore, this ``coincidence'' can be used to form a traceless operator from \textit{traceless parts} $\,^{\ssst (3)}\! R_{\mu\nu}^{\srm T}$ and $(\nabla_{\mu}\del_{\nu}\phi)^{\srm T}/\phi$,
    \begin{align}
    \label{DDRT}
        R_{\mu\nu}^{\srm T} + \left(d-2\right)\frac{1}{\phi}(\nabla_{\mu}\nabla_{\nu})^{\srm T}\phi &= \bar{R}_{\mu\nu}^{\srm T}
        -\left(d-2\right)\left(\bar{\nabla}_{(\mu}\del_{\nu)}\log A-\del_{\mu}\log A\,\del_{\nu}\log A\right)^{\srm T}\nonumber\\[12pt]
        &\quad + \left(d-2\right)\left(\frac{1}{\bar{\phi}}\bar{\nabla}_{\mu}\del_{\nu}\bar{\phi}
        +\bar{\nabla}_{\mu}\del_{\nu}\log a -\del_{\mu}\log A\,\del_{\nu}\log A\right)^{\srm T}\nonumber\\[12pt]
        &= \bar{R}_{\mu\nu}^{\srm T}+\left(d-2\right)\frac{1}{\bar{\phi}}\left[\bar{\nabla}_{\mu}\del_{\nu}\bar{\phi}\right]^{\srm T},
    \end{align}
which is indeed manifestly conformally invariant. In $d=3$ dimensions and setting $\bar{\phi}=\Nb$ we obtain the last two terms in \eqref{elWumod}, completing the proof of its manifest conformal invariance.

Let us multiply the above result by $(d-2)^{-1}\phi$ and switch the order of terms to get
    \begin{align}
    \label{DDRTeq}
        \left((\nabla_{\mu}\nabla_{\nu})^{\srm T}+\frac{1}{d-2}R_{\mu\nu}^{\srm T}\right)\phi= A\left(\bar{\nabla}_{\mu}\del_{\nu}+\frac{1}{d-2}\bar{R}_{\mu\nu}\right)^{\srm T}\bar{\phi}\ .
    \end{align}
This equation is to some extent analogous to the Klein-Gordon operator for the conformally coupled scalar field of conformal weight $n_{\ssst KG}=(2-d)/2$ that is studied in Section \eqref{sec_xiphi} with an exception that it is not derived from any Lagrangian. It does testify, however, that not only the conformal weight of a field it
acts on but also the property of tracelessness is relevant to the notion of conformal invariance of differential operators.

The same operator with $d=3$ appears in the Weyl-tensor part of the Hamiltonian constraint \eqref{HDall_Hofin}, cf. \eqref{HDall_HamDDW}. Namely,
the traceless momentum density $\Pb^{ij}$ of scale weight
$\omega_{a}=4$, contracted with $\left(D_{i}D_{j}+\frac{1}{d-2}R_{ij}\right)$ ensures that 
    \begin{align}
    \label{eDDRTPeq}
        \left((D_{i}D_{j})^{\srm T}+R_{ij}^{\srm T}\right)\Pb^{ij}= \left(\del_{i}\bar{D}_{j}+\bar{R}_{ij}\right)^{\srm T}\Pb^{ij}\,,
    \end{align}
is conformally invariant. These derivations
add to the power of the method of using the unimodular-conformal decomposition.

We finally make the interesting observation that the very same operator
considered above is precisely the one that appears in the Bach
equations, which are conformally invariant. Setting $d=4$, contraction
of $\left(\nabla_{\mu}\nabla_{\nu}+\frac{1}{d-2}R_{\mu\nu}\right)$ with the Weyl tensor
ensures that the operator is traceless, thus eliminating all the
scale-dependent terms from it. That is why the Bach tensor
\eqref{C2eom} can be simplified to 
    \begin{equation}
    \label{BachSmpl}
        A^{2}\left(\nabla_{\alpha}\nabla_{\beta}+\frac{1}{2}R_{\alpha \beta}\right)C^{\alpha}{}_{\mu}{}^{\beta}{}_{\nu}=
        \gb^{\beta\rho}\left(\bar{\nabla}_{\alpha}\bar{\nabla}_{\beta}+\frac{1}{2}\bar{R}_{\alpha \beta}\right)C^{\alpha}{}_{\mu\rho\nu}\ ,
    \end{equation}
which is manifestly $A$-independent and thus conformally invariant.

\subsection{3+1 decomposition of the non-minimally coupled scalar field}
\label{app_31chi}

In subsection \ref{sec_31chi} we use the results of the current Appendix, where we derive in detail the Lagrangian in unimodular-conformal variables in $3+1$ formalism, using results from section \ref{sec_31conf}. Note that we shall use
    \begin{equation}
    \label{phitochi3}
        \varphi = a^{s}\chi
    \end{equation}
decomposition, $a$ is the three-dimensional scale density. This is because we want this decomposition to be independent of the choice of lapse density and we put a general scale weight for now.

First we prepare the Lagrangian of the non-minimally coupled scalar field in the following way:
    \begin{align}
    \label{eqn:nonx-Lag31}
        \mathcal{L}^{\varphi}&=-\frac{1}{2}\sqrt{-g}\Biggl[g^{\mu\nu}\del_{\mu}\varphi\del_{\nu}\varphi+\xi R\varphi^2+V(\varphi)\Biggr]\nonumber\\
        &=\frac{1}{2}N\sqrt{h}\Biggl[\underbrace{\left(n^{\mu}\del_{\mu}\varphi\right)^2
        + \xi\left(\frac{2}{3}K^{2}-2\nabla_{\mu}\left(n^{\mu}K\right)\right)\varphi^2}_{\rm I}\nonumber\\[12pt]
        &\quad \underbrace{- h^{ij}\del_{i}\varphi\,\del_{j}\varphi-\xi\left( \,^{\ssst (3)}R - \frac{2}{N}D^{i}D_{i}N\right)\varphi^2}_{\rm{II}}\nonumber\\
        &\quad-\xi K_{ij}^{\srm T}K^{ij\srm T}\varphi^2-V(\varphi)\Biggr]\ ,
    \end{align}
where I and II are useful designations.
Now we proceed with calculation of each term. First we have the kinetic term and the last term in I:
    \begin{align}
    \label{eqn:nonx_Ia}
        N\sqrt{h}\left(n^{\mu}\del_{\mu}\varphi\right)^2 &=\Nb a^{2(1+s)}\Bigl[\bar{n}^{\mu}\del_{\mu}\chi+s\, \bar{n}^{\mu}\del_{\mu}\log a\chi\Bigr]^2\\[12pt]
    \label{eqn:nonx_Ik1}
        \nabla_{\mu}\left(n^{\mu}K\right)&=K^2+n^{\mu}\del_{\mu}K
    \end{align}
The last term in \eqref{eqn:nonx_Ik1} can be partially integrated to extract $\kb$ from under the derivative:
    \begin{align}
       \label{eqn:nonx_Ik}
        -6\xi \Nb a^{2(1+s)}\bar{n}^{\mu}\del_{\mu}\bar{K}\chi^2&=-6\xi \del_{\mu}\left(\Nb a^{2(1+s)}\bar{n}^{\mu}\bar{K}\chi^2\right)
        + 6\xi\bar{K}\del_{\mu}\left(\bar{n}^{\mu}\Nb a^{2(1+s)}\chi^2\right)\nonumber\\[12pt]
        & = -2\xi\del B + 12\xi(1+s)\Nb a^{2(1+s)}\bar{K}^2\chi^2\nonumber\\[12pt] 
        &\quad + 6\xi(2s-1)\Nb a^{2(1+s)}\frac{\del_{i}N^{i}}{3\Nb}\bar{K}\, \chi^2 
        + 6\xi\Nb a^{2(1+s)}\bar{n}^{\mu}\del_{\mu}\chi^2\ ,\nonumber\\[12pt]
        \del B &\equiv \frac{6}{2}\del_{\mu}\left(\Nb a^{2(1+s)}\bar{n}^{\mu}\bar{K}\chi^2\right)
    \end{align}
Note that this partial integration eliminates the second time derivative of $a$ from the Lagrangian. This is necessary only in GR in order to eliminate the second time derivatives. But if the scalar field is considered within a higher derivative theory of gravity one could leave this term alone and do the partial integration in the kinetic term in order to \textit{generate} the second time derivative of $a$ (equivalently the first time derivative of $\kb$) such that the expressions are simplified in another way. For the purposes of this thesis, the former is more convenient.

The expressions above are needed in order to calculate the following term:
    \begin{align}
    \label{eqn:nonx_Ik2}
        \xi N\sqrt{h}\left(\frac{2}{3}K^{2}-2\nabla_{\mu}\left(n^{\mu}K\right)\right)\varphi^2&=\xi N\sqrt{h}\left(\frac{2}{3}K^{2}-2K^2-2n^{\mu}\del_{\mu}K\right)\varphi^2\nonumber\\[12pt]
        &=-\xi N\sqrt{h}\left(\frac{4}{3}K^{2}+2n^{\mu}\del_{\mu}K\right)\varphi^2\nonumber\\[12pt]
        &=-\xi \Nb a^{2(1+s)}\Biggpar{
        12\bar{K}^2-6\bar{n}^{\mu}\del_{\mu}\log a\bar{K}+6\bar{n}^{\mu}\del_{\mu}\bar{K}
        }\chi^2\nonumber\\[12pt]
        &=6\xi \Nb a^{2(1+s)}\Biggpar{(1+2s)\bar{K}^2\chi^2\nonumber\\[12pt]
        &\quad\qquad+2\, s\,\bar{K}\frac{\del_{i}N^{i}}{3\Nb}+ a^{2(1+s)}\bar{n}^{\mu}\del_{\mu}\chi^2
        } - 2\xi\del B 
    \end{align}
Finally, putting together \eqref{eqn:nonx_Ia}-\eqref{eqn:nonx_Ik2}, we obtain
    \begin{align}
    \label{eqn:nonx_I}
        {\rm I}&=N\sqrt{h}\left(\left(n^{\mu}\del_{\mu}\varphi\right)^2+\xi\left(\frac{2}{3}K^{2}
        -2\nabla_{\mu}\left(n^{\mu}K\right)\right)\varphi^2\right)\nonumber\\[6pt]
        &=\Nb a^{2(1+s)}\left[\left(\bar{n}^{\mu}\del_{\mu}\chi+(s+6\xi)\bar{K}\chi+s\frac{\del_{i}N^{i}}{3\Nb}\chi\right)^2+6\xi(1-6\xi)\bar{K}^2\chi^2\right]-2\xi\del B\ .
    \end{align}
For expression II we will need the following expressions:
    \begin{align}
    \label{eqn:nonx_IIa}
        N\sqrt{h}\, h^{ij}\del_{i}\varphi\,\del_{j}\varphi &=\Nb a^{2(1+s)}\bar{h}^{ij}\left(\del_{i}\chi\,\del_{j}\chi+s\,\del_{i}\log a\,\del_{j}\chi^2+s^2\,\del_{i}\log a\,\del_{j}\log a\,\chi^2\right)\ ,
    \end{align}
    \begin{align}
    \label{eqn:nonx_IIb}
        \sqrt{h}\,h^{ij}D_{i}\del_{j}N\,\varphi^2&=D_{i}\left(\sqrt{h}\,h^{ij}\del_{j}N\right)\varphi^2=\del_{i}\left(\sqrt{h}\,h^{ij}\del_{j}N\right)\varphi^2\nonumber\\[12pt]
        &=\del_{i}\left(\sqrt{h}\,h^{ij}\del_{j}N\varphi^2\right)-\sqrt{h}h^{ij}\del_{j}N\del_{i}\varphi^2\nonumber\\[12pt]
        &=\rm{ BT1}-\rm{BT2}+N\del_{j}\left(\sqrt{h}h^{ij}\del_{j}\varphi^2\right)\nonumber\\[12pt]
        &=\rm{ BT1}-\rm{BT2}+\Nb a^{2(1+s)}\Big[2s\,\del_{i}\left(\bar{h}^{ij}\del_{j}\log a\right)\chi^2\nonumber\\[12pt]
        &\quad + 2s(1+2s)\bar{h}^{ij}\del_{i}\log a\,\del_{j}\log a\, \chi^2\nonumber\\[12pt]
        &\quad +2\chi\del_{i}\bar{h}^{ij}\del_{j}\chi +2\,\bar{h}^{ij}\del_{i}\chi\del_{j}\chi+(1+4s)\bar{h}^{ij}\del_{i}\log a\,\del_{j}\chi^2\Big]\ ,\\[12pt]
    \label{eqn:nonx_IIbBT}
        \rm{ BT1}-\rm{BT2}&=\del_{i}\left(\sqrt{h}h^{ij}\del_{j}N\varphi^2\right)-\del_{i}\left(\sqrt{h}h^{ij}N\del_{i}\varphi^2\right)\nonumber\\[12pt]
        &=(1-2s)\del_{i}\left(\Nb a^{2(1+s)}\bar{h}^{ij}\del_{j}\log a\chi^2\right)\nonumber\\[12pt]
        &\quad + \del_{i}\left(a^{2(1+s)}\bar{h}^{ij}\left(\del_{j}\Nb\chi^2-\Nb\del_{j}\chi^2\right)\right)\ .
    \end{align}
Recalling the unimodular decomposition of the Ricci scalar in three dimensions \eqref{Rdec}, we have
    \begin{align}
    \label{eqn:nonx_Ricdec}
        a^2 h^{ij}R_{ij}=\bar{h}^{ij}R_{ij}&=\bar{R}-4\bar{h}^{ij}\biggl[\bar{D}_{i}\del_{j}\log a+\frac{1}{2}\del_{i}\log a \,\del_{j}\log a\biggr]\nonumber\\[12pt]
        &=\bar{R}-4\del_{i}\left(\bar{h}^{ij}\del_{j}\log a\right)-2\bar{h}^{ij}\del_{i}\log a \,\del_{j}\log a\ .
    \end{align}
Collecting \eqref{eqn:nonx_IIa}-\eqref{eqn:nonx_Ricdec}, we can calculate the second contribution,
    \begin{align}
    \label{eqn:nonx_II}
        \rm{II}&=-N\sqrt{h}\left(h^{ij}\del_{i}\varphi\,\del_{j}\varphi-\xi\left( \,^{\ssst (3)}R-\frac{2}{N}D^{i}D_{i}N\right)\varphi^2\right)\nonumber\\[12pt]
        &=\Nb a^{2(1+s)}\Bigg[4\xi\del_{i}\left(\chi\bar{h}^{ij}\del_{j}\chi\right)-\bar{h}^{ij}\del_{i}\chi\del_{j}\chi-\xi\bar{R}\chi^2+S\left(a; s,\xi\right)\Bigg]\nonumber\\[12pt]
        &\quad + 2\xi\rm{BT1}-2\xi\rm{BT2}
    \end{align}
where $S\left(a; s,\xi\right)$ is the collection of $a-$dependent terms,
    \begin{align}
    \label{eqn:nonx_S}
        S\left(a; s,\xi\right)&\equiv 4\xi(1+s)\del_{i}\left(\bar{h}^{ij}\del_{j}\log a\right)\chi^2-\left[s^{2}(1-6\xi)-2\xi(1+s)^2\right]\bar{h}^{ij}\del_{i}\log a\,\del_{j}\log a\,\chi^2\nonumber\\[12pt]
        &\quad-\left[s(1-6\xi)-2\xi(1+s)\right]\bar{h}^{ij}\del_{i}\log a\,\del_{j}\chi^2\ .
    \end{align}
The Lagrangian in its final form is then given by
    \begin{align}
    \label{eqn:nonx_Lagdec}
        \mathcal{L}^{\varphi}=\mathcal{L}^{\chi}&=\frac{1}{2}\Nb a^{2(1+s)}\Bigg[\left(\bar{n}^{\mu}\del_{\mu}\chi
        + (s+6\xi)\bar{K}\chi+s\frac{\del_{i}N^{i}}{3\Nb}\chi\right)^2+6\xi(1-6\xi)\bar{K}^2\chi^2 \nonumber\\[12pt]
        &\quad + 4\xi\del_{i}\left(\chi\bar{h}^{ij}\del_{j}\chi\right)-\bar{h}^{ij}\del_{i}\chi\del_{j}\chi
        -\xi\bar{R}\chi^2-\xi\bar{K}_{ij}^{{\srm T}2}\,\chi^2+S\left(a; s,\xi\right)  \Bigg]\nonumber\\[12pt]
        &\quad-\xi\del B + 2\xi\rm{BT1}-2\xi\rm{BT2}
    \end{align}

Now we can choose\footnote{
One could have also chosen $s=-6\xi$ as in \cite{KiefNM} which is suitable if one is dealing with conformally coupled scalar field because
the cross term $\kb\dot{\chi}\sim \dot{a}\dot{\chi}$ term is gone and this eliminates mixing between the momenta with respect to $a$ and $\chi$ in GR. 
However, the price that one has to pay is that $\chi$ is no longer conformally invariant for a general non-minimal coupling and the length dimension does not coincide with the scale weight. 
This inconsistency is not what we want in this thesis, even though this choice might have some calculational advantages. } 
$s=-1$ as motivated in the previous subsection which sets the scaling of the scalar field to be $\varphi = a^{-1}\chi$ and the Lagrangian now reads
    \begin{align}
    \label{eqn:nonx_LagdeSone}
        \mathcal{L}^{\varphi}=\mathcal{L}^{\chi}&=\frac{1}{2}\Nb\Bigg[\left(\bar{n}^{\mu}\del_{\mu}\chi-(1-6\xi)\bar{K}\chi-\frac{\del_{i}N^{i}}{3\Nb}\chi\right)^2+6\xi(1-6\xi)\bar{K}^2\chi^2 \nonumber\\[6pt]
        &\quad + 4\xi\del_{i}\left(\chi\bar{h}^{ij}\del_{j}\chi\right)-\bar{h}^{ij}\del_{i}\chi\del_{j}\chi-\xi\bar{R}\chi^2-\xi\bar{K}_{ij}^{{\srm T}2}\,\chi^2+S\left(a; \xi\right)  \Bigg]\nonumber\\[12pt]
        &\quad-\xi\del B+\xi\rm{BT1}-\xi\rm{BT2}
    \end{align}
with
    \begin{align}
    \label{eqn:nonx_Sf}
    S\left(a; \xi\right)&\equiv (1-6\xi)\left[\bar{h}^{ij}\del_{i}\log a\,\del_{j}\chi^2-\bar{h}^{ij}\del_{i}\log a\,\del_{j}\log a\,\chi^2\right]\ ,
    \end{align}
while the total divergences reduce to
	\begin{align}
	\label{eqn:nonx_delB}
        \del B &= \frac{6}{2}\del_{\mu}\left(\Nb\bar{n}^{\mu}\bar{K}\chi^2\right)\\[12pt]
	\label{eqn:nonx_BT12}
        \rm{BT1}-\rm{BT2}&=3\del_{i}\left(\Nb\bar{h}^{ij}\del_{j}\log a\chi^2\right)+\del_{i}\left(\bar{h}^{ij}\left(\del_{j}\Nb\chi^2-\Nb\del_{j}\chi^2\right)\right)\ .
	\end{align}
Note that qualitatively only one term has dropped from the Lagrangian, namely the one containing second spatial derivative of the scale density $4\xi(1+s)\del_{i}\left(\bar{h}^{ij}\del_{j}\log a\right)$ from \eqref{eqn:nonx_S}. The rest of the terms have remained with simplified coefficient. These coefficients now depend only on $\xi$, whose choice controls whether one will deal with conformally coupled, minimally coupled, or general non-minimally coupled scalar field.

Observe now that for conformal coupling $\xi=1/6$ Lagrangian \eqref{eqn:nonx_LagdeSone} reduces to
    \begin{align}
    \label{eqn:nonx_Lagdecf}
        \mathcal{L}^{\varphi}=\mathcal{L}^{\chi}&=\frac{1}{2}\Nb\Bigg[\left(\bar{n}^{\mu}\del_{\mu}\chi-\frac{\del_{i}N^{i}}{3\Nb}\chi\right)^2 \nonumber\\[6pt]
        &\quad +\frac{2}{3}\del_{i}\left(\chi\bar{h}^{ij}\del_{j}\chi\right)-\bar{h}^{ij}\del_{i}\chi\del_{j}\chi-\frac{1}{6}\bar{R}\chi^2-\frac{1}{6}\bar{K}_{ij}^{{\srm T}2}\,\chi^2  \Bigg]\nonumber\\[12pt]
        &\quad-\frac{1}{6}\del B+\frac{1}{6}\rm{BT1}
    \end{align}
since $S\left(a; 1/6\right)=0$. Note that no $a$ or $\kb$ appear in here and thus we have shown that the Lagrangian is manifestly conformally invariant.

\subsection[Canonical transformation from the ADM variables \texorpdfstring{\\}{} to the unimodular-conformal variables]{Canonical transformation from the ADM to the unimodular-conformal variables}
\label{app_canon}

We prove here that a general Poisson bracket defined by \eqref{PB_definition} with respect to the ADM variables $h_{ij},p_{\srm{ADM}}^{ij}$, $N,p_{\ssst N}$ and $N^{i},p_{i}$ gives rise to canonical pairs $(a,p_{a})$ and $(\hb_{ij},\pb^{ij})$, $\Nb,\pb_{\ssst N}$, $N^{i},p_{i}$ in transition to the unimodular-conformal variables.
For this proof we shall suppress the coordinate and time dependence and consider all components and functions evaluated at the same point (thereby formally substituting the functional with partial derivatives).

Let us consider only the first term in the Poisson bracket involving the pair $h_{ij},p_{\srm{ADM}}^{ij}$.
We would like to see how should the ADM momentum transform in order for \eqref{3unimoddec} to be a canonical transformation.
Let us first define the traceless $p^{{\srm{T}}ij}_{\srm{ADM}}$ and trace $p$ parts of $p^{ij}_{\srm{ADM}}$,
    \begin{equation}
        p^{{\srm{T}}ij}_{\srm{ADM}} := \mathbb{1}_{kl}^{{\srm T}ij}p^{kl}_{\srm{ADM}}\ ,\qquad p_{\srm{ADM}} := h_{ij}p^{ij}_{\srm{ADM}}\ ,
    \end{equation}
where $\mathbb{1}_{kl}^{{\srm T}ij}$ is defined in \eqref{eqn:oneT}.
Then we use the unimodular decomposition of the three-metric given by \eqref{3unimoddec} along with the three-dimensional version of \eqref{vardecD1} and work out the following expression
    \begin{align}
    \label{PB_canonProof}
        \ddel{F}{h_{ij}}\dd{G}{p^{ij}_{\srm{ADM}}} & = \left(a^{-2}\mathbb{1}_{kl}^{{\srm T}ij}\dd{F}{\hb_{kl}} + \frac{a}{6}h^{ij}\dd{F}{a}\right)\dd{G}{p^{ij}_{\srm{ADM}}}\nld
        & = a^{-2}\dd{F}{\hb_{kl}}\mathbb{1}_{kl}^{{\srm T}ij}\dd{G}{p^{ij}_{\srm{ADM}}} + \frac{a}{6}\dd{F}{a}h^{ij}\dd{G}{p^{ij}_{\srm{ADM}}}\ ,
    \end{align}
where now we see that the first term picks up only the traceless part while the second term picks up only the trace part of the derivative with respect to $p^{ij}_{\srm{ADM}}$.
Using the chain rule
    \begin{align}
    \label{PB_canon_Proof1}
        \dd{G}{p^{ij}_{\srm{ADM}}} &= \dd{p^{{\srm T}mn}_{\srm{ADM}}}{p^{ij}_{\srm{ADM}}}\dd{G}{p^{{\srm T}mn}_{\srm{ADM}}} + \dd{p}{p^{ij}_{\srm{ADM}}}\dd{G}{p_{\srm{ADM}}} = \mathbb{1}_{ij}^{{\srm T}mn}\dd{G}{p^{{\srm T}mn}_{\srm{ADM}}} + h_{ij}\dd{G}{p_{\srm{ADM}}} \nld
        & = \dd{G}{p^{{\srm T}ij}_{\srm{ADM}}} + h_{ij}\dd{G}{p_{\srm{ADM}}}\ .
    \end{align}
Plugging \eqref{PB_canon_Proof1} into \eqref{PB_canonProof} we obtain
    \begin{align}
    \label{PB_canonProof3}
        \ddel{F}{h_{ij}}\dd{G}{p^{ij}_{\srm{ADM}}} & = a^{-2}\dd{F}{\hb_{kl}}\dd{G}{p^{{\srm T}ij}_{\srm{ADM}}} + \frac{a}{2}\dd{F}{a}\dd{G}{p_{\srm{ADM}}}\ ,
    \end{align}
from which we see that the correct canonical transformation of the ADM momentum's pieces is
    \begin{equation}
        \pb^{ij} = a^{2} p^{{\srm T}ij}_{\srm{ADM}}\ ,\qquad p_{a} = \frac{2}{a} p_{\srm{ADM}}\ ,
    \end{equation}
which agrees with \eqref{GRxi_padmcan}.


\section{Variational principle in terms of the scale and the shape}
\label{varprincdec}

Based on \eqref{gsplit1}, the variation of the metric decomposes into variations of the scale $\delta A$ and variations of the shape $\delta \gb_{\mu\nu}$:
    \begin{subequations}
    \begin{align}
    \label{varg}
        \delta g_{\mu\nu}& = A^{2}\delta \gb_{\mu\nu} + 2  \gb_{\mu\nu} A\delta A\ ,\\[6pt]
    \label{varg2}
        \delta g^{\mu\nu}& = A^{-2}\delta \gb^{\mu\nu} - 2  \gb^{\mu\nu} A^{-3}\delta A \ .
    \end{align}
    \end{subequations}
An important property of the above decomposition is that the variation of the shape is \textit{traceless},
    \begin{equation}
    \label{tracelessvar}
        g^{\mu\nu}\delta\gb_{\mu\nu}=A^{-2}\gb^{\mu\nu}\delta\gb_{\mu\nu}=0\ ,
    \end{equation}
meaning that the two pieces of variation in \eqref{varg} are orthogonal to each other. This is just another way of saying that scale and shape are \textit{orthogonal} ``directions'' in the configuration space of metric components.
Based on the above decomposition the variational derivative with respect to the metric can be decomposed as follows:
    \begin{subequations}
    \begin{align}
    \label{vardecUP}
        \frac{\delta }{\delta g^{\mu\nu}}
        =\frac{\delta A}{\delta g^{\mu\nu}}\frac{\delta }{\delta A}
        +\frac{\delta \gb^{\alpha\beta}}{\delta g^{\mu\nu}}\frac{\delta }{\delta \gb^{\alpha\beta}} 
        & = - \frac{A}{2d}\, g_{\mu\nu}\frac{\delta }{\delta A}+A^2\mathbb{1}^{{\srm{T}}\alpha\beta}_{\mu\nu}\frac{\delta }{\delta \gb^{\alpha\beta}}\\[12pt]
    \label{vardecUP1}
        & = - \frac{A^3}{2d}\, \gb_{\mu\nu}\frac{\delta }{\delta A}+A^2\mathbb{1}^{{\srm{T}}\alpha\beta}_{\mu\nu}\frac{\delta }{\delta \gb^{\alpha\beta}}\\[12pt]
    \label{vardecD}
        \frac{\delta }{\delta g_{\mu\nu}}
        =\frac{\delta A}{\delta g_{\mu\nu}}\frac{\delta }{\delta A}
        +\frac{\delta \gb_{\alpha\beta}}{\delta g_{\mu\nu}}\frac{\delta }{\delta \gb_{\alpha\beta}}
        & = \frac{A}{2d}\, g^{\mu\nu}\frac{\delta }{\delta A}+A^{-2}\mathbb{1}^{{\srm{T}}\alpha\beta}_{\mu\nu}\frac{\delta }{\delta \gb_{\alpha\beta}}\\[12pt]
    \label{vardecD1}
        & = \frac{A^{-1}}{2d}\, \gb^{\mu\nu}\frac{\delta }{\delta A}+A^{-2}\mathbb{1}^{{\srm T}\alpha\beta}_{\mu\nu}\frac{\delta }{\delta \gb_{\alpha\beta}} \ ,
    \end{align}
    \end{subequations}
where we write several forms of equations with and without completely exposing the shape and scale densities and where $\mathbb{1}^{{\srm T}\alpha\beta}_{\mu\nu}$ makes the contracted variational derivative explicitly traceless. The most important equation to keep in mind throughout this work is \eqref{tracelessvar}.

These are the variational tools for unimodular-conformal formulation.  
They are used to re-derive equations of motion and energy-momentum tensor for various theories in Chapter \ref{ch:defcf}.
We can give a small example here to show how can this tool be used to look at the equations of motion for gravitational actions in a different way.

Let us take an example of the EH theory with a cosmological constant and some matter described by action $S^{m}$,
    \begin{equation}
    \label{appEH}
        S = \frac{1}{2\kappa}\int\d^{d}x\, \sqrt{g}(R-2\Lambda)+S^{m}\ .
    \end{equation}
Using \eqref{varg2}, variation of the EH term with respect to $g_{\mu\nu}$ leads to
    \begin{align}
        \delta S^{\srm{ EH}} & = \frac{1}{2\kappa}\int\d^{d}x\, \sqrt{g}
        \left(R_{\mu\nu}-\frac{1}{2}g_{\mu\nu}R + g_{\mu\nu}\Lambda\right)\delta g^{\mu\nu} \nonumber\\[12pt]
        & = \frac{1}{2\kappa}\int\d^{d}x\, \sqrt{g}
        \biggsq{
        A^{-2}\left(R_{\mu\nu}-\frac{1}{2}g_{\mu\nu}R + g_{\mu\nu}\Lambda\right)\delta \gb^{\mu\nu}\nld
        &\qquad\qquad
        - 2\left(R_{\mu\nu}-\frac{1}{2}g_{\mu\nu}R + g_{\mu\nu}\Lambda\right)\gb^{\mu\nu}A^{-3}\delta A
        }\ ,
    \end{align}
up to a boundary term. On the other hand, based on \eqref{Tmn_dec} the variation induces a split of the energy-momentum tensor into trace and traceless components because of \eqref{tracelessvar}. For the same reason, the two terms above become the traceless and trace parts of the Einstein tensor. Putting all this information together, we arrive at
    \begin{align}
    \label{ttREE}
        R_{\mu\nu}^{\srm T}=\kappa T_{\mu\nu}^{\srm{T}}\ ,\qquad - R + 4\Lambda = \kappa T\ ,
    \end{align}
which are just Einstein equations split into traceless and trace parts. The traceless part has been shown to arise in unimodular gravity \cite{Fink} by a variation with respect to the metric with a determinant constrained to be unity. As pointed out in \cite{PadSal}, this constraint is just a particular gauge fixing within the GR and we tend to agree with their claim. 
What we have in \eqref{ttREE} are the equations of motion for the shape density and the scale density, respectively.
The approach to variation with respect to the scale and shape can be applied to any theory.

\section{Constraint analysis}
\label{app_const}

\subsection{Example: a massive vector field}
\label{app_Proca}

In this Appendix we briefly introduce what is known as Dirac or Dirac-Bergmann constraint analysis \cite{Dirac}, although it is a collection of results by Rosenfeld, Anderson, Bergmann and Dirac, see \cite{Sali} and \cite[Appendix C]{Sunder}.
The procedure presented here on an example of a massive vector field (so-called \textit{Proca field}) theory on a general curved spacetime.
The treatment is reformulated in the unimodular-conformal variables, introduced in section \ref{sec_31conf}.

The Lagrangian density for a massive vector field is given by
    \begin{equation}
        \mathcal{L}^{\ssst\rm A} = -\frac{1}{2}\sqrt{g}\left(\frac{1}{2} F_{\mu\nu}F^{\mu\nu} + m^2A_{\mu}A^{\mu}\right)\ ,
    \end{equation}
where $F_{\mu\nu}=\del_{\mu}A_{\nu} - \del_{\nu}A_{\mu}$ and $m$ mass parameter of the vector field $A^{\mu}$.
Only in the special case $m=0$ (which describes vacuum electromagnetism) is the theory invariant under gauge transformations $A_{\mu}\rightarrow A_{\mu} + \del_{\mu}f$, with $f$ and arbitrary function on spacetime.

Let us first use the $3+1$ splitting of spacetime (cf. Appendix \ref{App_31}) with unimodular-conformal variables (cf. subsection \ref{subs_umoddec31}) to decompose the Lagrangian and expose its conformal properties. We shall assume that the vector field lives on a curved fixed spacetime whose Lagrangian is of no interest here and does not interfere with the derivations.
The Lagrangian of a massive vector field is worked out to be
    \begin{align}
	\label{EMm_Lag}
    	\mathcal{L}^{\ssst\rm A} & = -\frac{1}{4}\Nb a^4F_{\mu\nu}a^{-4}\left(\bar{h}^{\mu\alpha}\bar{h}^{\nu\beta}+\bar{n}^{\mu}\bar{n}^{\alpha}\bar{n}^{\nu}\bar{n}^{\beta}-\bar{h}^{\mu\alpha}\bar{n}^{\nu}\bar{n}^{\beta}-\bar{n}^{\mu}\bar{n}^{\alpha}\bar{h}^{\nu\beta}\right)F_{\alpha\beta}\nl
    	&\quad  - \frac{\Nb a^4}{2}m^2a^{-2}A_{\mu}h^{\mu\nu}A_{\nu}+\frac{\Nb  a^4}{2}m^2a^{-2}\bar{n}^{\mu}\bar{n}^{\nu}A_{\mu}A_{\nu}\nl
    	& = \frac{1}{2}\bar{N}\left(\bar{F}_{\bot i}\bar{h}^{ij}\bar{F}_{\bot j} -\frac{1}{2}F_{ij}\bar{h}^{ia}\bar{h}^{jb}F_{ab}
    	- m^2a^2A_{i}h^{ij}A_{j} + m^2a^2A_{\bot}^2\right)\nl
    	&=\frac{1}{2}\bar{N}\left(\bar{\mathbf{F}}_{\bot}\cdot \,\bar{\mathbf{F}}_{\bot}-\frac{1}{2}\mathbf{F}\cdot \mathbf{F} - m^2a^2\mathbf{A}\cdot \mathbf{A}+m^2a^2\mathbf{A}_{\bot}^2\right)\ ,
	\end{align}
where $\bar{\mathbf{F}}_{\bot}:=\bar{n}^{\mu}F_{\mu i}$.
The dot notation designates contraction of all indices with $\bar{h}_{ij}$ and its inverse. 
In the dot product of a vector and a 2nd rank tensor it matters if the vector is on the left or on the right of the tensor. 
On the left side it is contracted with the left index of the tensor, and if it is on the right side then with the right index of the tensor, i.e. $\mathbf{N}\cdot\mathbf{F}:=N^{i}F_{ij}$ and $\mathbf{F}\cdot\mathbf{N}:=F_{ij}N^{j}=N^{j}F_{ij}=-N^{j}F_{ji}$. Thus one has to be careful with the position of indexes and the relative position of the object in this simplified notation of contraction.
Note that $A_{\mu}$ is already conformally invariant, so $A_{\mu}=\bar{A}_{\mu}$ and $F_{\mu\nu}=\bar{F}_{\mu\nu}$. 
These objects are given by
	\begin{align}
	\label{EM_Fuc}
	&\bar{\mathbf{F}}_{\bot}:=\bar{F}_{\bot\, i}=\bar{n}^{\mu}\del_{\mu}A_{i}-\bar{n}^{\mu}\del_{i}A_{\mu}=\frac{1}{\bar{N}}\left(\dot{A}_{i}-N^{j}F_{ji}-\del_{i}A_{t}\right)\ ,\\[6pt]
	&\mathbf{F}:=F_{ij}=\del_{i}A_{j}-\del_{j}A_{i}\ ,\\[6pt]
	&\mathbf{A}=A_{i}\ ,\qquad \bar{\mathbf{A}}_{\bot}:=\bar{A}_{\bot}
	=\bar{n}^{\mu}A_{\mu}=\frac{1}{\bar{N}}\left(A_{t}-N^{i}A_{i}\right)\ .
	\end{align}
Note that in \eqref{EMm_Lag} only the last two terms --- those with dimensionful coupling constant $m$ --- depend on the scale. Therefore these terms break not only gauge but also conformal symmetry of the Lagrangian.

Conjugate momenta are defined as
	\begin{align}
	\label{EM_puc}
    	\bar{\Pi}^{i}=\frac{\del \mathcal{L}^{\ssst\rm A}}{\del \dot{A}_i}&=\bar{h}^{ij}\bar{F}_{\bot j}\equiv \bar{\mathbf{\Pi}}=\frac{\del \mathcal{L}^{\ssst\rm A}}{\del \dot{\mathbf{A}}}\nonumber\\[6pt]
    	&\Rightarrow\quad \bar{F}_{\bot i}=\bar{h}_{ij}\bar{\Pi}^{j}\equiv\bar{\mathbf{F}}_{\bot}=\bar{\mathbf{h}}\cdot\bar{\mathbf{\Pi}}\\[6pt]
    	&\Rightarrow \quad\dot{A}_i=\bar{N}\bar{h}_{ij}\bar{\Pi}^{j}+\del_{i}A_{t}+N^{j}F_{ji}\nonumber\\[6pt]
    \label{EM_Ai}
    	\equiv\dot{\mathbf{A}} &=\bar{N}\bar{\mathbf{h}}\cdot\bar{\mathbf{\Pi}}+\mathbf{\del}A_{t}-\mathbf{F}\cdot\mathbf{N}\ ,\\[6pt]
    \label{EM_Pt}
    	&\bar{\Pi}^{t}=\frac{\del \mathcal{L}^{\ssst\rm A}}{\del \dot{A}_{t}}\deq 0\ .
	\end{align}
These canonical pairs form the following equal-time Poisson brackets
    \begin{equation}
        \PB{A_{\mu}(\bx,t)}{\bar{\Pi}^{\nu}(\by,t)} = \delta^{\nu}_{\mu}\delta(\bx,\by)\ ,
    \end{equation}
where $\delta(\bx,\by)$ is the three-dimensional delta distribution. For a general set of canonical pairs $q_{\ssst A}(\bx,t)$ and $p^{\ssst A}(\bx,t)$, the Poisson bracket is defined as
    \begin{equation}
    \label{PB_definition}
        \PB{F(\bx)}{G(\by)}: = \int\d ^3 z\left(\ddel{F(\bx)}{q_{\ssst A}(\bz)}\ddel{G(\bx)}{p^{\ssst A}(\bz)} - \ddel{G(\bx)}{q_{\ssst A}(\bz)}\ddel{F(\bx)}{p^{\ssst A}(\bz)}\right)\ ,
    \end{equation}
for two functions $F(\bx),G(\bx)$ on phase space and index $\ssst A$ designates a phase space variable, and is assumed to be summed over.
In the presently discussed case $q_{\ssst A}(\bx,t)=A_{\mu}(\bx,t)$ and $p^{\ssst A}(\bx,t)=\Pi^{\mu}(\bx,t)$.
From now on the time and space dependence shall be implicitly assumed and only in Poisson brackets will the latter be recovered

Note that $\dot{A}_{t}$ is missing from the theory, which is why its momentum vanishes --- the corresponding velocity cannot be inverted for.
In order to study properties of this theory within the Hamiltonian formulation, Dirac introduced a ``weak equality'', which in this thesis we denote as ``$\deq$'' and rename it as the ``delayed equality'', whose purpose is to delay setting the expression ``strongly'' to zero until all Poisson brackets have been calculated. 
This prevents inconsistencies in Poisson brackets such as $\PB{\bar{\Pi}^{t}}{.} = \PB{0}{.} \neq 0$, in cases where such a bracket is indeed not zero under ``$\deq$'' sign.
A constraint which directly follows from the Lagrangian and relates momenta with coordinates is called a \textit{primary constraint}.
This usually point to an arbitrary degree of freedom, in this case $A_{t}$, which does not have its own kinetic term and thus no equation of motion.
Using the above definitions the Lagrangian can then be written as
	\begin{align}
	\mathcal{L}^{\ssst\rm A}&=\frac{1}{2}\bar{N}\left(\bar{\mathbf{\Pi}}\cdot \bar{\mathbf{\Pi}}-\frac{1}{2}\mathbf{F}\cdot \mathbf{F}- m^2a^2\mathbf{A}\cdot \mathbf{A}+m^2a^2\bar{\mathbf{A}}_{\bot}^2\right)\ ,
	\end{align}
which is manifestly conformally invariant. The total Hamiltonian is defined via the Legendre transform,
	\begin{align}
	\label{EM_Huc}
    	H & = \intx\left(\dot{A}_{i}\bar{\Pi}^{i}+\dot{A}_{t}\bar{\Pi}^{t}-\mathcal{L}^{\ssst\rm A}\right)\nl
    	& = \intx\Bigg\lbrace
    	\frac{\bar{N}}{2}\left[\bar{\mathbf{\Pi}}\cdot \bar{\mathbf{\Pi}}+\frac{1}{2}\mathbf{F}\cdot \mathbf{F} - m^2a^2\mathbf{A}\cdot \mathbf{A}+m^2a^2\bar{\mathbf{A}}_{\bot}^2 \right]
    	+\mathbf{N}\cdot\left[\mathbf{F}\cdot\bar{\mathbf{\Pi}} \right]\nonumber\\[12pt]
    	&\qquad\qquad - A_{t}\del\cdot\bar{\mathbf{\Pi}}+\lambda_{t}\Pi^{t}
    	+\del\cdot\left(A_{t}\bar{\mathbf{\Pi}}\right)\Bigg\rbrace\ .
	\end{align}
Now, we have included $\dot{A}_{t}\bar{\Pi}^{t}$ in the Legendre transform, which is not the usual procedure. One usually starts without this term --- since the Lagrangian does not depend on velocity $\dot{A}_{t}$ --- and then defines another Hamiltonian (``primary Hamiltonian'') with $\lambda_{t}\bar{\Pi}^{t}$ term added, where $\lambda_{t}$ is the Lagrange multiplier.
We find such a procedure unnecessary, because if one starts as we did in the above equation one is lead naturally to the conclusion that  $\lambda_{t}:=\dot{A}_{t}$ is a Lagrange multiplier. Hence, all information about the theory is already contained in the theory itself and there is no need to add things to it.
Furthermore, from definition \eqref{EM_Ai} the term $\bar{\Pi}^{i}\del_{i}A_{t}$ is partially integrated to produce the first and the third term (this is a total divergence) in the last line of the above equation.
Now, the primary constraint has to be preserved in time. In order to simplify showing the point of this discussion, we shall choose $\Nb=1,\mathbf{N}=0$ without harm.
So we look for time derivative $\dot{\bar{\Pi}}^{i}$ and obtain
    \begin{align}
        \dot{\Pi}^{t} & = \PB{\Pi^{t}}{H} = - \dd{H}{A_{t}} = \left( \del\cdot\bar{\mathbf{\Pi}} - m^2 a^2A_{t}\right) \deq 0\ ,
    \end{align}
where again one demands the ``delayed equality''. This expression obviously does not vanish automatically and therefore represents another constraint,
    \begin{equation}
    \label{EM_GaussC}
        \mathcal{G}_c:=\del\cdot\bar{\mathbf{\Pi}} - m^2 a^2A_{t} \deq 0\ .
    \end{equation}
Such constraints --- derived from conditions for the time preservation of the primary constraints --- are called \textit{secondary constraints}.
The meaning of constraint \eqref{EM_GaussC} is recognized in the case of electromagnetism when $m=0$: this is the Gauss constraint.
So we see that breaking the conformal and gauge symmetry manifests itself as a source term in the Maxwell equation for the divergence of the electric field.
Now, there is a way to tell that a theory enjoys some symmetry or if that symmetry is broken.
Since \eqref{EM_GaussC} is a condition that needs to hold at each moment in time, Dirac-Bergmann procedure requires that one demands its time derivative to vanish as well,
    \begin{align}
    \label{EM_GaussCt}
        \dot{\mathcal{G}_c}=\PB{\mathcal{G}_c}{H} &= \PB{\del\cdot\bar{\mathbf{\Pi}}}{H} 
        - m^2 a^2\PB{A_{t}}{H}\nonumber\\[12pt]
        & = - \frac{m^2 a^2}{2}\hb^{ij}\PB{\del\cdot\bar{\mathbf{\Pi}}}{A_{i}A_{j}} - \lambda_{t}\PB{A_{t}}{\Pi^{t}}\nonumber\\[12pt]
        &= - m^2 a^2\del_{i}A^{i} - m^2 a^2\lambda_{t}\deq 0\quad\Rightarrow\quad \del_{t}A_{t} = -\del_{i}A^{i}\ .
    \end{align}
We see that by this last equation the Lagrange multiplier is not actually arbitrary but is \textit{determined}. Why is this so? 
Note that both surviving terms in the above calculation are proportional to mass $m$.
So in the case of electromagnetism $\dot{\mathcal{G}_c}\equiv 0$ and $\mathcal{G}_c:=\del\cdot\bar{\mathbf{\Pi}}\deq 0$ and there are no more constraints, leaving $\lambda_{t}$ undetermined.
The information about whether or not a Lagrange multiplier is determined is inscribed the the Poisson bracket of the primary and secondary constraints (in the present case \eqref{EM_Pt} and \eqref{EM_GaussC}, respectively),
    \begin{equation}
    \label{EM_PB2ndclass}
        \PB{\bar{\Pi}^t(\by)}{\mathcal{G}_{c}(\by)} = m^2 a^2\delta(\bx,\by)\ ,
    \end{equation}
which vanishes only if $m=0$, as in electromagnetism.
In general, for the Poisson bracket between two constraints $\phi_{I}(\bx)$ and $\phi_{J}(\bx)$ we have the following cases, nomenclature and meaning,
    \begin{equation}
    \label{EM_1st2nd}
        \PB{\phi_{I}(\bx)}{\phi_{J}(\by)}\Biggl\lbrace
        \begin{matrix}
            \deq 0 \quad\lor\quad \equiv0\,\, &\Rightarrow\,\, \phi_{I}(\bx),\phi_{J}(\by) \text{ \small ``\textbf{1st class}''}  \Rightarrow &\text{ \small symmetry }\\[12pt]
            \stackrel{\ssst D}{\neq}0 \,\, &\Rightarrow\,\, \phi_{I}(\bx),\phi_{J}(\by) \text{ \small ``\textbf{2nd class}''}  \Rightarrow & \text{ \small broken symmetry }
        \end{matrix}
    \end{equation}
    
Let us explain this. It could happen that the Poisson bracket between $\phi_{I}(\bx)$ and $\phi_{J}(\bx)$ gives a linear combination of already existing constraints, in which case the Poisson bracket vanishes once ``$\deq$'' is promoted to ``$=$''.
Or it could happen that the bracket vanishes identically (as is the case with \eqref{EM_PB2ndclass} for $m=0$).
In both of these cases the involved constraints are called the \textit{first-class constraint} and they are related to a symmetry of the theory.
In electromagnetism \eqref{EM_PB2ndclass} vanishes, the Gauss constraint is first-class and the gauge symmetry holds, while $A_{t}$ completely disappears from the theory (one may call this a ``true arbitrary variable'').
In the second case in \eqref{EM_1st2nd} the Poisson bracket does not vanish even after all delayed equalities are set to strong equalities; in this case the constraints are called the \textit{second-class constraints} and are a signal of a broken symmetry (either a gauge is fixed or a symmetry-breaking term appears in a symmetric Lagrangian).
This is the case with \eqref{EM_PB2ndclass} because $m^2a^2$ is not a constraint --- a consequence of the symmetry breaking term $m^2A_{\mu}A^{\mu}$.
In this case one may call $A_{t}$ ``an apparent arbitrary variable'', since it only seems arbitrary but it turns out it can be fixed in terms of other variables.
An important consequence of the appearance of the second-class constraints in a theory is that Poisson brackets have to be modified in order to accommodate the fact that a variable which was initially undetermined turns out to be fixed in terms of other variables.
The modified brackets are called \textit{Dirac brackets} but we postpone their calculation for the next subsection.
Once all Dirac brackets are calculated all second-class constraints can be strongly set to zero and if one wishes to quantize the theory, then it is the Dirac brackets which are quantized instead of the Poisson brackets.


\subsection{Dirac brackets}
\label{app_dirb}

Since we are dealing in this thesis with theories that have both first- and second-class constraints 
the Poisson brackets should be replaced by Dirac brackets in order to make equations of motion consistent. For a
general function $F({\mathbf x})$ and $G({\mathbf x})$, and a system with two second-class constraints the Dirac
bracket reads \cite{Dirac}  
\begin{align}
\label{eqn:DB}
&\left\lbrace F({\mathbf x}),G({\mathbf
  y})\right\rbrace_{D}=\left\lbrace F({\mathbf x}),G({\mathbf
  y})\right\rbrace - \int {\rm d}^{3}z\,{\rm d}^{3}z'\left\lbrace F({\mathbf
  x}),\phi_{I}({\mathbf z})\right\rbrace\mathcal{M}^{IJ}\left\lbrace
  \phi_{J}({\mathbf z}'),G({\mathbf y})\right\rbrace,
\end{align}
where the sum is understood as running over the second-class
constraints here labelled by $I,J=(1,2)$ and
$\mathcal{M}^{IJ}$ is the inverse matrix to 
    \begin{align}
    \label{Dir_Mat}
    \mathcal{M}_{IJ}&=
    \begin{pmatrix}
    \left\lbrace \phi_{1}({\mathbf z}),\phi_{1}({\mathbf z}')\right\rbrace& \left\lbrace \phi_{1}({\mathbf z}),\phi_{2}({\mathbf z}')\right\rbrace\\[6pt]
    \left\lbrace \phi_{2}({\mathbf z}),\phi_{1}({\mathbf z}')\right\rbrace& \left\lbrace \phi_{2}({\mathbf z}),\phi_{2}({\mathbf z}')\right\rbrace
    \end{pmatrix}\ .
    \end{align}

\textbf{Harmonic oscillator with higher derivatives}. A simple example demonstrating how Dirac brackets are calculated is met in subsection~\ref{subs_Ham_osc}. 
It is at the same time an explanation of why could those constraints be set to strongly vanish from the start\footnote{This explanation is a simplified version of that in Appendix C of~\cite{Kluson2014}.}.
The above matrix and its inverse for constraints $\phi_{1} = p_{x} - \lambda $ and $\phi_{1} = p_{\lambda}$ derived there reads
    \begin{align}
    \mathcal{M}_{IJ} =
    \begin{pmatrix}
    0 & -1\\[6pt]
    1 & 0
    \end{pmatrix}\ ,\qquad \mathcal{M}^{IJ} =
    \begin{pmatrix}
    0 & 1\\[6pt]
    -1 & 0
    \end{pmatrix}\ .
    \end{align}
The Dirac bracket then reads
    \begin{equation}
    \label{app_DB_osc}
        \PB{F}{G}_{D} = \PB{F}{G} - \PB{F}{p_{x} - \lambda}\PB{p_{\lambda}}{G} + \PB{G}{p_{\lambda}}\PB{p_{x} - \lambda}{F}\ ,
    \end{equation}
and it can be seen that only those Dirac brackets in which one of the $F$ and $G$ functions depends on $\lambda$ and the other depends on $x$ or $p_{\lambda}$ is distinct from the corresponding Poisson bracket.
But after setting $p_{x} = \lambda$ and $p_{\lambda}=0$ strongly, no function can depend on $\lambda$ so the Dirac bracket is the same as the Poisson bracket.
Therefore, including $\lambda$ and its conjugate momentum is unnecessary. This is expected because adding a constraint that simply relabels what is meant by velocity in a higher-derivative theory should not affect the physics that theory describes.

The second-class constraints in the above example are not related to any broken symmetry; they are demands put in by hand outside of theory. 
But more generally, second-class constraints and Dirac brackets appear in a more fundamental context, such as broken conformal symmetry or broken gauge invariance.

\textbf{Massive vector field}. In the previous subsection we discussed the example of a massive vector field which turned out to be a system with second-class constraints.
The following Poisson bracket
    \begin{equation}
    \label{PBAA}
        \PB{A_{t}(\bx)}{A_{i}(\by)}
    \end{equation}
is expected to vanish, but $A_{t}$ is a function of the momentum $\Pi^{i}$ if the second-class constraint given by \eqref{EM_GaussC} is set to zero, which means that the above bracket actually does not vanish and $A_{t}$ is not an independent canonical variable.
That is where Dirac brackets come to help resolve the contradiction.
Namely, the matrix inverse to \eqref{Dir_Mat} for the case of constraints obeying \eqref{EM_PB2ndclass} are
    \begin{equation}
    \mathcal{M}^{IJ} =
        -\frac{1}{m^2 a^2}\begin{pmatrix}
        0 & 1\\[6pt]
        -1 & 0
        \end{pmatrix}\delta^{3}(\bx,\by)
    \end{equation}
and using this result in \eqref{eqn:DB} with $\phi_{1}=\Pi^{t}$ and $\phi_{2}=\mathcal{G}_{c}$, the Dirac bracket version of \eqref{PBAA} is straightforwardly calculated to be
    \begin{equation}
        \PB{A_{t}(\bx)}{A_{i}(\by)}_{D} = - \frac{1}{m^2 a^2}\del_{i}\delta^{3}(\bx,\by)\ .
    \end{equation}

\end{appendix}


\chapter*[Acknowledgments]{Acknowledgments}
\addcontentsline{toc}{chapter}{Acknowledgments}
\thispagestyle{empty}
I would like to thank Prof. Dr. Claus Kiefer for giving me the opportunity to work on this thesis and pursue my own interest. I am also grateful for discussions, patience and understanding over the last four years. I am thankful to Prof. Dr. Domenico Giulini for agreeing to be the second reviewer of my thesis and for patience regarding the organization of the final date of defense.

I would also like to thank Friedrich Hehl for several discussions and for drawing my attention to the literature on shear current and the work of Ogievetski.

I am indebted to Dr. J. Brian Pitts for his help, discussions, patience and encouragement for pursuing the consequences of unimodular decomposition to extrinsic curvature and conformal gauge generators. These were some of the most important ingredients that laid the foundations of the methods used in this thesis.

It was pleasure to unexpectedly meet an artist, teacher, physicist and a philosopher in Ka\'ca Bradonji\'c from Hampshire College, Boston, USA, and share the passion for research on projective and conformal structures through several hours long discussions. The mind-connection was immediate and I am grateful for it. I received such a boost for research from you!

I think my experience from the first years of being a PhD student couldn't have been any better thanks to Nick. You have never made me feel like it is a bad thing that I do not know things someone says I should know. It was so fun to correct sheets with you and get distracted by cracking the pottery or talking about guitar pedals. Your spirit is priceless and I am grateful that we have met.

The last months of writing my thesis were quite dark in trying to figure out the semiclassical approximation. Leonardo has found time to explain certain details to me which were crucial for my understanding of the bigger picture of my own work. Thank you Leo!

Doing a PhD is not a stress-free thing, to say the least. I could go on and on about it, but let's just say that it's good to have a professional that is actually listening to you and whoc can offer coping tools especially in times of great anxiety, worry and thought dead-ends. I am grateful to have B. G. for my therapist, who has made the last month of writing my thesis much more manageable than it would have been otherwise.

As far as overall atmosphere during my PhD research, I am so grateful that I have been surrounded by my groupmates Nick Kwidzinski, Leonardo Chataignier, Tim Schmitz, Yi-Fan Wang, Dimitris Gkiatas, Dennis Piontek, Anirudh Ghundi, Sebastian Arenas and Christina Koliofoti. Countless lunchkins and puns and complaining about movies or how some student's handwriting is untidy --- which has another level to it because some of you were my students as well!
\\

Speaking of my students, Lucas, Alex, Christoph, Jana, Marcel, Tim, Edmond, Basel, Ali, Sandeep, Mario, Mateo, Nicola, Mugdha, Ben, Bj\"orn, Atefeh, Karthik, Lukas, Guido, Dominik, Daniel, Christina, Julius, Ghalib, Mahsa, Valentin, Anirudh --- and I wish I could name all of you, but some of your names I forgot simply because we haven't seen each other anymore. If I saw you on the street, I would surely remember you and I guess that's most important. In any case, to all of you who were my students in courses from 2015-2018: there is nothing more enjoyable than sharing excitement, understanding, a-ha! moments and knowing that I have made a difference. Know that you have made a great difference for me as well and that goes for anything you'll do in your life. If you are my student and reading this, well hello from 22. July 2019! :) Thank you so much.

Some of my students have become friends and that's wonderful to experience. My heart is so much bigger since I met Ali, Sandeep, Mugdha, Marcel and Nicola. Their friendship and support came unexpected and wonderful. There were moments when you just know that you share them. 

Indeed, nice things usually happen when you least expect them. It is especially nice when they keep happening in spite of certain miseries of life. Meeting Nick Nu\ss baum and getting to know him was (and continues to be) an unexpectedly fulfilling and joyful experience.

I have also had the privilege to meet \"Ozkan in my dormitory who has become a close friend and whose deep conversations and chess games and emotions made my room feel not that small after all. It was a pleasure to witness important milestones in each other's lives!
Bonn has also brought me Bahadir and Shishir as well and I am glad I had even a few moments to share with them.

I would like to mention that Marija and Ljubi\v sa were with me in spirit and memory, as a glimpse of peace as well as inspiration at times where I did not know I needed it. Grl.

Where do I begin, Pranjal and Armin? I guess in that moment where we all always meet, when we realize time hasn't passed at all as soon as we hear from each other.

The typos and other corrections would have been left unnoticed in the final version of the manuscript without the sharp eyes of Prof. Dr. Claus Kiefer, Pranjal, Ali, Nicola, Sandeep, Mugdha and Antonela.

There are things beyond time and space and words and numbers and equations: it's that silence shared with someone you love. A thread that has made the fabric of the ground I learned to stand on today, a whisper from the past that turned into echo which turned into a breeze, a music of all kinds of colors and dreams, some waypoints after a longest journey... You have been so patient with me and always had support when I needed it, even if I didn't ask. Thank you, Antonela. This thesis is dedicated to you.

\thispagestyle{empty}
\newpage
\thispagestyle{empty}
\thispagestyle{empty}
{\centering \LARGE \textbf{Erkl\"arung}}
\\
\\
\noindent Ich versichere, dass ich die von mir vorgelegte Dissertation selbst\"andig angefertigt, die  benutzten  Quellen  und  Hilfsmittel  vollst\"andig  angegeben  und  die  Stellen  der  Arbeit -- einschlie{\ss}lich Tabellen, Karten und Abbildungen --, die anderen Werken im Wortlaut  oder  dem  Sinn  nach  entnommen  sind,  in  jedem  Einzelfall  als  Entlehnung  kenntlich  gemacht  habe;  dass  diese  Dissertation  noch  keiner  anderen  Fakult\"at  oder  Universität zur Pr\"ufung vorgelegen hat; dass sie -- abgesehen von unten angegebenen Teilpublikationen -- noch nicht ver\"offentlicht worden ist, sowie, dass ich eine solche Ver\"offentlichung  vor  Abschluss  des  Promotionsverfahrens  nicht  vornehmen  werde.  Die Bestimmungen der Promotionsordnung sind mir bekannt. Die von mir vorgelegte Dissertation ist von Hernn Prof. Dr. Claus Kiefer betreut worden.
\\
\\
\\
K\"oln, den \rule{3cm}{0.5pt}   \hspace*{3.5cm}  \rule{5.5cm}{0.5pt}\\
\hspace*{10cm}                              Branislav Nikoli\'c

\vspace{2cm}
{\centering \Large \textbf{Teilpublikationen}}
\\
\\
1. Claus Kiefer und \underline{Branislav Nikoli\'c},\\
\textit{Conformal and Weyl-Einstein gravity: Classical geometrodynamics},\\
Physics Review D \textbf{95}, 084018.
\\
\\
2. Claus Kiefer und \underline{Branislav Nikoli\'c},
\textit{Quantum geometrodynamics of Einstein and conformal (Weyl-squared) gravity},\\
Journal of Physics: Conference Series \textbf{880}, (2017) no.1, 012002.
\\
\\
3. Branislav Nikoli\'c,\\
\textit{Treating the Einstein-Hilbert action as a higher derivative Lagrangian: revealing the missing information about conformal non-invariance},\\
Journal of Physics: Conference Series \textbf{880}, (2017) no.1, 012027.
\\
\\
4. Claus Kiefer und \underline{Branislav Nikoli\'c},\\
\textit{Notes on Semiclassical Weyl Gravity}, In:
Gravity and the Quantum. Fundamental Theories of Physics, vol 187, pp 127-143. Springer, Cham.

\thispagestyle{empty}

\end{document}